\documentclass[11pt,dutch,british]{book}
\usepackage{amsfonts,amsmath,amssymb} 
\usepackage{comment} 
\excludecomment{comment}  

\usepackage{color} 

\usepackage{pslatex}
\usepackage[section]{placeins} 
\usepackage{float}
\usepackage{subfigure}
\usepackage{graphicx,wrapfig,hyperref}

\usepackage[small,bf]{caption}
\usepackage{slashed}
\usepackage{cite}

\numberwithin{equation}{section}
\definecolor{Darkgreen}{RGB}{0,100,0}
\definecolor{Forestgreen}{RGB}{34,139,34}
\definecolor{Mediumblue}{RGB}{0,0,205} 

\newdimen\mylength

\begin{document} 
\pagenumbering{Roman} 
\thispagestyle{empty}   
%
 \noindent
 \begin{minipage}{3cm}%
   \includegraphics*[width=2cm]{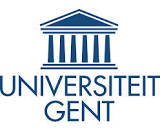}
 \end{minipage}\hfill
 \begin{minipage}{8cm}
 \raggedleft
 \textsf{Ghent University\\
 Faculty of Sciences\\
 Department of Physics and Astronomy}
 \end{minipage}
%
\vspace{4cm}
\bigskip
   \begin{flushleft}
 \centerline{\LARGE \textsf{Hagedorn String Thermodynamics}}
 \vspace{0.5cm}
 \centerline{\LARGE \textsf{in Curved Spacetimes and near Black Hole Horizons}}
   \end{flushleft}
%
 \bigskip
\centerline{\LARGE\noindent \textsf{- Thomas Mertens -}} \hfill
 \bigskip
%
%
 \normalsize
 
 \vspace{3cm}
 \centerline{Promotor: Prof. Dr. Henri Verschelde}
%
 \vspace{3cm}
\centerline{Thesis submitted in fulfillment of the requirements for the degree of}
\centerline{\textbf{Doctor In Sciences: Physics}}
\centerline{at Ghent University}
\vspace{2cm}
\centerline{ Academic Year 2014-2015}
\newpage
\thispagestyle{empty}
\mbox{}
\newpage
\newpage

\thispagestyle{empty}

\centerline{\LARGE Acknowledgements}
\vspace{1.5cm}

\noindent Firstly, I would like to thank Henri Verschelde and Valentin I. Zakharov for the collaboration on which this work is based. \\
I am grateful to Ben Craps, David Dudal, Oleg Evnin, Amit Giveon, Diego R. Granado, Nissan Itzhaki, Igor Justo, Hampus Linander, Lihui Liu, Shiraz Minwalla, Blagoje Oblak, Jo\~{a}o Penedones, Waldemar Schulgin, Chung-I Tan, Karel Van Acoleyen, Joris Vanhoof, Herman Verlinde and Frank Verstraete for many interesting discussions on the topics presented here at various stages during the completion of this work. \\

\noindent I thank the UGent Special Research Fund for financial support. \\

\noindent Finally, I would like to thank my friends, my parents, my grandparents and my sister for the support during the years.

\thispagestyle{empty}
\mbox{}
\newpage

\pagenumbering{roman} 
\tableofcontents
\newpage
\pagenumbering{arabic}

\chapter{General Introduction}

This work concerns the study of high-temperature string theory on curved backgrounds, generalizing the notions of Hagedorn temperature and thermal scalar to general backgrounds. The quantum model we are working with is the so-called non-linear sigma model describing string propagation on a curved background. The results we will present focus almost exclusively on non-interacting string thermodynamics. Hence our methods are firmly rooted in the perturbative formulation of string theory. More elaborate discussions will follow in the first few chapters. \\

\noindent Our main goal of this curved space formalism is to apply it to black hole horizons, i.e. studying string thermodynamics near the horizon of a black hole. \\

\noindent There are several reasons for studying this problem. Firstly, in holography it is known that the deep infrared regime of the boundary (gauge) theory is dual to the near-horizon degrees of freedom of a higher-dimensional black hole. String thermodynamics close to black hole horizons hence seems to have some influence on deep IR features of gauge theories, perhaps even the quark-gluon plasma. Secondly, the recent firewall paradox urges us to have a better understanding of black hole horizons, especially in a consistent theory of quantum gravity. Thirdly, string thermodynamics has some puzzling features even in flat space that are ill-understood. Perhaps studying the general problem might shed some light on the nature of string thermodynamics. \\

\noindent This work is based on the following research papers \cite{Mertens:2013pza}\cite{Mertens:2013zya}\cite{Mertens:2014nca}\cite{Mertens:2014cia}\cite{Mertens:2014dia}\cite{Mertens:2014saa}\cite{Mertens:2015hia}. \\
Let me first briefly mention how these papers fit together in the research program outlined above. \\
In \cite{Mertens:2013pza} we proposed a formalism to deal with the curved space generalization of the random walk picture of the near-Hagedorn string gas. We provided a symbiosis of first-quantized (path integral) and second-quantized (field theory) methods, both of which have pros and cons. Combined however, a full picture emerges. We tested this on several elementary examples in flat space. \\
In \cite{Mertens:2013zya} we provided the application of this formalism to Rindler spacetime, as the near-horizon approximation to uncharged black holes. We analyzed the different string types (bosonic, type II and heterotic) and found that the Hagedorn temperature of such a geometry is precisely equal to the Hawking temperature of the original black hole. This has profound consequences of which some are given here. \\
In \cite{Mertens:2014nca} we studied the 3d $AdS_3$ WZW model as another application. We presented several methods to deduce the thermal spectrum of such spaces and the resulting thermal scalar and random walk picture. BTZ black holes are closely related to this model and we discussed these as well, leading to some puzzling results on the absence of the thermal scalar. \\
In \cite{Mertens:2014cia} some other basic examples on the formalism were provided. These are simple enough to be exactly solvable, yet they are complicated enough to teach us something. \\
In \cite{Mertens:2014dia} we studied whether the thermal scalar tells us something more about the high-temperature string gas besides the thermal partition function. The answer is in the affirmative as the near-Hagedorn stress tensor and string charge of the gas are both given by the same quantity of the thermal scalar field. We apply the formulas to some of the examples that were studied in our earlier papers. \\
In \cite{Mertens:2014saa} we revisited Rindler space string thermodynamics, tying up some loose ends and providing some additional interpretations and computations on the results we previously obtained. Also higher genus corrections were analyzed. \\
Finally, in \cite{Mertens:2015hia} we provided a simple argument to show that long strings close to the black hole horizon are capable of yielding the tree-level black hole entropy, precisely because of the equality between Hawking and Hagedorn temperatures. \\

\noindent This work combines these results with additional material taken from the existing literature to, hopefully, create a single consistent story on string thermodynamics. To that end, the results of the above papers have been reordered in a way that, in my opinion, reflects more their logical entanglement than their chronological order of appearance. The structure of the remaining chapters is illustrated in the scheme below.
\begin{figure}[h]
\centering
\includegraphics[width=0.7\textwidth]{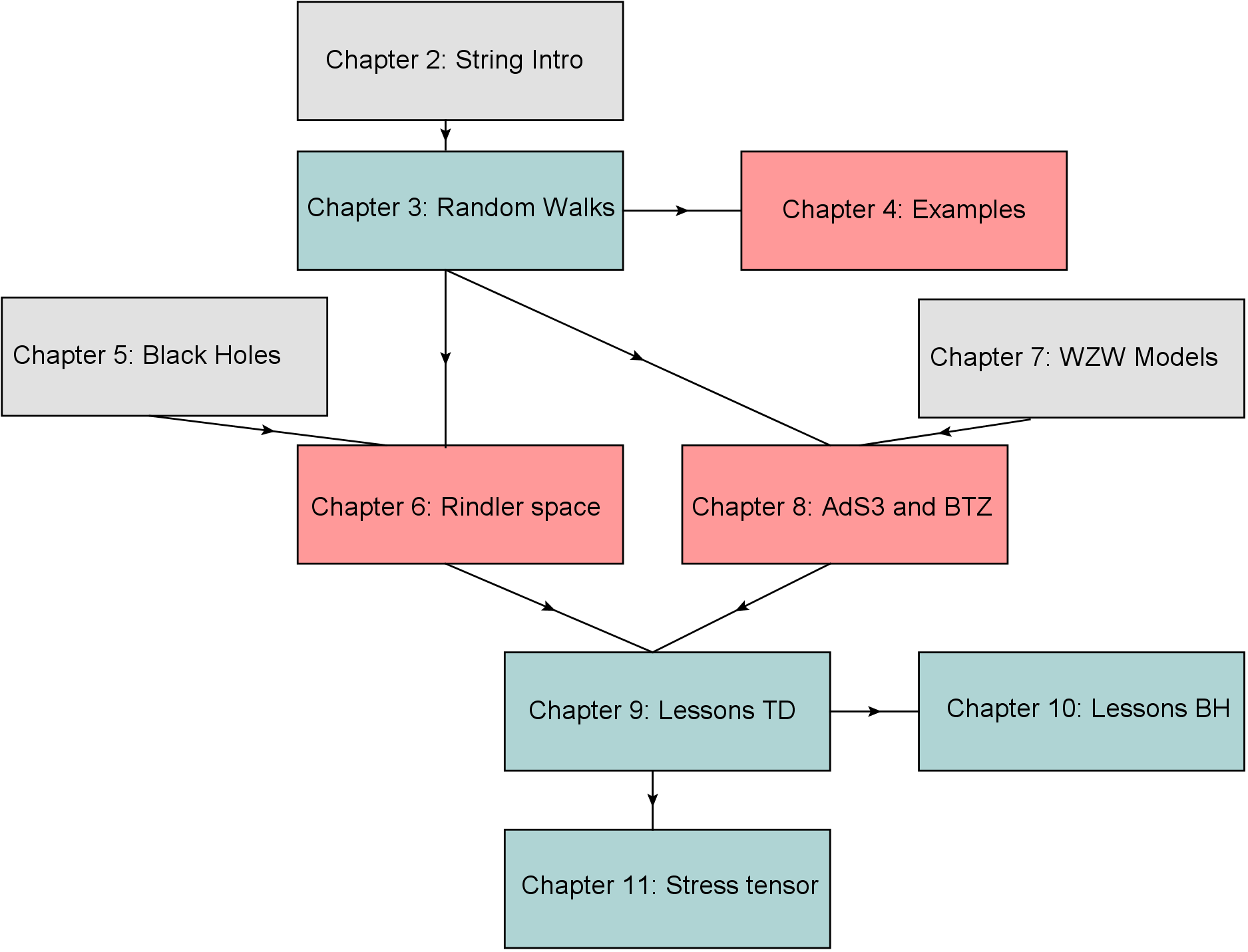}
\caption{Scheme of the chapters with the logical connections shown. Gray boxes indicate introductory chapters. Green boxes indicate formalism developing chapters. Red boxes indicate example chapters.}
\label{sche}
\end{figure}

\noindent Sections labeled by a $*$ correspond to appendices in the original papers and should be viewed as supplementary (or technical) material. We follow the conventions of Polchinski's textbooks throughout. \\

\noindent During this time, the author also contributed to the following research paper \cite{Dudal:2014jfa} in collaboration with Prof. Dr. David Dudal. This paper concerns with the application of holographic methods to study the melting of the $J/\psi$ vector meson in a strong background magnetic field, an effect that is important in the quark-gluon plasma phase of QCD where strong magnetic fields are indeed produced in off-axis collisions of nuclear matter. In the paper we proposed a DBI extension of the famous soft-wall model to account for magnetic field influence on the charged constituents of the charmonium bound state. \\

\part{String Thermodynamics}

\chapter{String thermodynamics: a general introduction}
\label{introTD}
This chapter contains an introduction to string thermodynamics and its strange features as it has been known since the '80s \cite{Mitchell:1987hr}\cite{Mitchell:1987th}\cite{Bowick:1989us}\cite{Deo:1989bv}\cite{Atick:1988si}\cite{Horowitz:1997jc}\cite{Barbon:2004dd}. Most of the discussions are standard textbook material, although some material has, to the best of my knowledge, not been discussed before in the literature. It is my hope that even the more seasoned string theorist might learn something new in this chapter.  
\section{Intro to string theory}
The quantitive treatment of string theory (see e.g. \cite{Polchinski:1998rq}) starts with the generalization of the worldline action of a relativistic point particle \begin{equation}
S = - m \int ds = - m \int d\tau \sqrt{-\partial_\tau X^{\mu} \partial_\tau X_{\mu}}
\end{equation}
(with $s$ the invariant distance and $\tau$ the proper time used to parameterize the worldline of the particle) to the worldsheet action of a relativistic string
\begin{equation}
\label{NambuGoto}
S_{NG} = - \frac{1}{2\pi\alpha'}\int d\sigma^1 d\sigma^2 \sqrt{-\det(\partial_a X^{\mu}\partial_{b} X_{\mu})},
\end{equation}
where $\sigma^1$ and $\sigma^2$ denote coordinates used to parameterize the 2-dimensional worldsheet of the string. We assume spacetime to be flat at this stage. The prefactor $T = \frac{1}{2\pi\alpha'}$ is the analog of the point particle mass $m$ and is called the \emph{string tension}. This action, the Nambu-Goto action, can be used to quantize the string in the so-called \emph{light-cone gauge}. This gauge can be compared with the Coulomb gauge in QED and leads to a manifestly unitary theory, sacrificing manifest Lorentz invariance in the process. Though we will not need much of this gauge, it is instructive to skim through the procedure quickly. \\ 
In light-cone gauge, one gauge-fixes the theory at the classical level. Then one solves the constraints and quantizes only the truly independent degrees of freedom. This procedure does not require the introduction of ghosts and unitarity issues related to this. As a gauge choice, one sets the light-cone coordinate $X^+$ proportional to worldsheet time $\tau$ as
\begin{equation}
X^+ = \frac{1}{\sqrt{2}}(X^0 + X^1) = \alpha' p^+ \tau,
\end{equation}
for worldsheet coordinates $\tau$ and $\sigma$ and conserved momentum $p$.\footnote{This choice of prefactor is for closed strings. In addition to this, an appropriate gauge choice for the $\sigma$-coordinate needs to be made.} The constraints of the theory then determine $X^- = \frac{1}{\sqrt{2}}(X^0 - X^1)$ in terms of the transverse remaining coordinates $X^I$. In the end, only the transverse coordinates $X^I$ are independent degrees of freedom. For a very elaborate treatment on string theory in this gauge, we refer to the book \cite{Zwiebach:2004tj}. We will utilize this gauge only in our discussion on Susskind's long string arguments in chapter \ref{chBH}. \\

\noindent A classically equivalent action can be written down by introducing a metric on the string worldsheet $h_{ab}$ and leads to
\begin{equation}
S_P = - \frac{1}{4\pi\alpha'}\int d\sigma^1 d\sigma^2 \sqrt{-h} h^{ab}\partial_a X^{\mu}\partial_b X_{\mu},
\end{equation}
the Polyakov action, to be viewed as a functional of both $X$ and $h_{ab}$. After eliminating $h_{ab}$ through its equations of motion, one recovers the Nambu-Goto action (\ref{NambuGoto}), proving the classical equivalence of both actions. This action however, has the virtue of not containing a square root. This allows a more direct quantization approach. The Polyakov action has one additional symmetry over the Nambu-Goto action: local worldsheet rescalings. This symmetry, called Weyl invariance, changes $h_{ab} \to \exp(2\omega(\tau,\sigma)) h_{ab}$ while keeping $X$ fixed. The local (on the worldsheet) symmetries of this action are hence both the worldsheet diffeomorphism invariance and the Weyl invariance, denoted diff $\times$ Weyl. \\
Upon Euclideanization of the worldsheet, string theory can be perturbatively defined as a series over random surfaces, suitably weighted in a path integral:
\begin{equation}
Z = \int \left[\mathcal{D}X\right] \left[\mathcal{D} h\right]e^{-S_P + \lambda \chi},
\end{equation}
where $\chi$ is the Euler characteristic of the surface and $\lambda$ an arbitrary constant. Path integration is performed over a suitable set of worldsheet metrics on each surface and on the embedding coordinates in the ambient spacetime. The above local symmetries show that these surfaces (equiped with a Riemannian metric) should be identified under diffeomorphisms and Weyl rescalings. It can be shown that such 2d Riemannian manifolds when identified under Weyl rescalings are equivalent to 2d Riemann surfaces (2d manifolds equiped with a complex structure). In fact the gauge fixing of diff $\times$ Weyl can be made explicit using the Faddeev-Popov procedure. \\
The strategy of this procedure is to use the diff $\times$ Weyl gauge symmetry to fix the worldsheet metric to a prescribed fiducial metric. Actually, there is too much gauge symmetry for this and a residual subgroup is left: holomorphic diffeomorphisms that can be compensated by Weyl transformations to preserve the fiducial metric. In the gauge-fixed theory these can be interpreted as providing a conformal invariance. Thus \emph{the gauge-fixed diff $\times$ Weyl theory gives a conformal field theory (CFT)}. The above analysis was done locally. Globally however (on the entire Riemann surface), only a subgroup of the conformal group survives, this is called the conformal Killing group (CKG).\\
From here on, we focus exclusively on closed strings. \\
The defining property of a 2d CFT is the presence of a \emph{Virasoro algebra}:\footnote{Actually two commuting Virasoro algebras for closed strings.}
\begin{equation}
[L_m,L_n] = (m-n)L_{m+n} + \frac{c}{12}(m^3-m)\delta_{m,-n}.
\end{equation}
Physical states in closed string theory $\left|\psi\right\rangle$ can then be defined by
\begin{equation}
(L_0 +\bar{L}_0)\left|\psi\right\rangle = 0 ,\quad L_{n}\left|\psi\right\rangle = 0, \, n>0, \quad \bar{L}_{n}\left|\psi\right\rangle = 0, \, n>0.
\end{equation}
The first equation provides the on-shell equation: it imposes the Klein-Gordon equation on the physical state. The remaining relations are non-trivial constraints for excited string states. These correspond to gauge choices: for instance Lorenz gauge for the open string massless gauge field. Hence from a spacetime point of view, this set of conditions corresponds to a specific gauge choice for all of the higher spin string states such that the remaining equation of motion reduces to a Klein-Gordon equation. \\

\begin{figure}[h]
\centering
\includegraphics[width=0.8\textwidth]{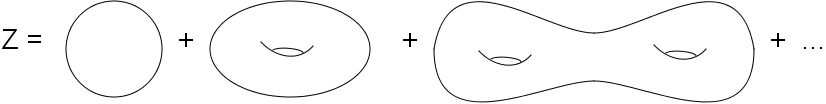}
\caption{Genus expansion of the string path integral.}
\label{genusexp}
\end{figure}

\noindent A technical complication when dealing with gauge-fixed string theory is that the complex structure on such Riemann surfaces is not necessarily unique. The set of inequivalent complex structures is called the \emph{moduli space} of the surface and it can be viewed as the set of surfaces that can not be transmuted in each other using diff $\times$ Weyl operations. \\
Thus the full perturbative string amplitude reduces to a sum over all different Riemann surfaces, which breaks up in a sum over the discrete genus and an integral (with a suitable measure) over the moduli space for each genus (figure \ref{genusexp}).
The string coupling arises from the Euler character of the different surfaces. \\
A crucial point is that interactions are inherently present in this construction: no local interaction terms should (and can) be added to the Polyakov action.\\
To go beyond perturbation theory, different methods are available: the discovery of D-branes as supergravity p-branes \cite{Polchinski:1995mt} paved the way for (strong-weak) S-dualities, the holographic correspondence \cite{Maldacena:1997re} provides a definition of the strongly coupled string theory, matrix models (M-theory) also provide non-perturbative definitions of string theory \cite{Witten:1995ex}\cite{Banks:1996vh}\cite{Dijkgraaf:1997vv}. In this work we will not use any of these developments and we are firmly fixed in the perturbative regime. In fact, much of the results presented here could have been studied several years ago. \\

\noindent To proceed with the construction of string theory, one introduces \emph{vertex operators}. These are local (on the worldsheet) operators that provide punctures on the Riemann surface. Using Weyl rescalings, it is easy to see that they can be interpreted as providing asymptotic states for scattering amplitudes. Thus worldsheets with local punctures can be interpreted as providing scattering amplitudes. For instance for a graviton with momentum $k^\mu$, the vertex operator is given by
\begin{equation}
V(\sigma^1, \sigma^2) = \frac{g_c}{\alpha'}\int d^2 \sigma \sqrt{h}h^{ab}s_{\mu\nu}\partial_a X^{\mu} \partial_b X^{\nu} e^{i k \cdot X},
\end{equation}
with $s_{\mu\nu}$ a symmetric traceless tensor. If we then try to look at a coherent state of such gravitons, we can try to exponentiate this vertex operator. This leads to an additional term in the Polyakov action, which can be combined with the Polyakov action into
\begin{equation}
S = - \frac{1}{4\pi\alpha'}\int d\sigma^1 d\sigma^2 \sqrt{h} h^{ab}\left(\delta_{\mu\nu} - 4 \pi g_c e^{i k \cdot X} s_{\mu\nu}\right) \partial_a X^{\mu}\partial_b X^{\nu}.
\end{equation}
It is hence clear that this provides a curved space generalization of the Polyakov action. Generalizing this argument to include also a Kalb-Ramond and dilaton field, we obtain the non-linear sigma model \cite{Callan:1985ia}:
\begin{equation}
S = - \frac{1}{4\pi\alpha'}\int d\sigma^1 d\sigma^2 \sqrt{h} \left[\left(h^{ab}G_{\mu\nu}(X) +i \epsilon^{ab} B_{\mu\nu}(X)\right)\partial_a X^{\mu}\partial_b X^{\nu} + \alpha' R \Phi(X)\right],
\end{equation}
providing the dynamics of string propagation in a background $G_{\mu\nu}$, $B_{\mu\nu}$, $\Phi$. The worldsheet theory is interacting in this case and the quantum theory becomes much more difficult to develop. \\
An important feature in a quantum theory is the possibility of anomalies in symmetries. It turns out that the Weyl invariance symmetry can be anomalous. If this is the case, our interpretation in terms of strings breaks down. This condition is so fundamental to string theory that it can be viewed as a definition of a (quantum) string theory. For the flat space string theory, the absence of a Weyl anomaly leads to the requirement of vanishing total central charge of the worldsheet CFT. \\
For a curved background, the restrictions are more elaborate and can be split into three separate sets of equations, which are given by (to lowest order in $\alpha'$): 
\begin{align}
\beta^G_{\mu\nu} &= \alpha' R_{\mu\nu} + 2\alpha' \nabla_\mu \nabla_\nu \Phi - \frac{\alpha'}{4} H_{\mu\lambda\omega}H_{\nu}^{\,\,\lambda\omega} + \mathcal{O}(\alpha'^2), \\
\beta^B_{\mu\nu} &= -\frac{\alpha'}{2} \nabla^{\omega}H_{\omega \mu\nu} + \alpha'\nabla^{\omega}\Phi H_{\omega \mu\nu} + \mathcal{O}(\alpha'^2), \\
\beta^{\Phi} &= \frac{d-26}{6} - \frac{\alpha'}{2}\nabla^2 \Phi + \alpha'\nabla_{\omega}\Phi\nabla^{\omega}\Phi - \frac{\alpha'}{24}H_{\mu\nu\lambda}H^{\mu\nu\lambda} + \mathcal{O}(\alpha'^2).
\end{align}
The vanishing of these three beta-functionals ensures a Weyl-invariant quantum theory. These equations encapsulate Einstein's equations and the equations of motion of $B_{\mu\nu}$ and $\Phi$. An important feature is that if $\beta^G_{\mu\nu} = \beta^B_{\mu\nu} = 0$, the worldsheet theory admits the construction of a Virasoro algebra and is hence conformally invariant. These conditions ensure that $\beta^{\Phi}$ becomes independent of $x^{\mu}$ and proportional to the central charge of the resulting CFT \cite{Callan:1985ia}. The vanishing of the central charge of this CFT then finally ensures the ability to couple this CFT in a Weyl-invariant way to a curved (worldsheet) metric. \\
It is immediately clear from this that flat space (with all other background fields turned off) is a solution only when $d=26$. To get to a viable 4d theory, one would then need to compactify 22 dimensions. This immediately incorporates higher dimensions and Kaluza-Klein theory in string theory. \\
The most succesful way of obtaining valid string backgrounds (to all order in $\alpha'$) is by considering group manifolds (and the resulting coset manifolds) as backgrounds. As will be discussed further on, these backgrounds automatically have a conformal algebra and hence satisfy $\beta^G_{\mu\nu} = \beta^B_{\mu\nu} = 0$ to all orders in $\alpha'$. To have a valid string theory, one then needs to combine several of these models together to have a vanishing total central charge (and hence satisfy $\beta^{\Phi}=0$) and obtain a valid string background.  \\

\noindent Bosonic string theory as defined above cannot be the real theory of nature: it has a tachyon at the lowest excitation level and it does not include fermions. Superstring theory arises by including worldsheet fermionic degrees of freedom in the theory. On a cylindrical worldsheet (closed superstrings), these fermions can satisfy either periodic or anti-periodic boundary conditions along the circular dimension. These are called, respectively, Ramond (R) and Neveu-Schwarz (NS) boundary conditions. Combining left-movers and right-movers, one obtains four sectors: NS-NS, NS-R , R-NS and R-R. It turns out that this theory should include a further projection of the allowed states down to a subspace, the so-called \emph{GSO projection}. This is needed for quantum consistency of the theory and projects out the tachyonic state. Moreover, the resulting theory (called \emph{type II superstring theory}) turns out to be spacetime supersymmetric as well after this projection. Different choices of projection lead to type IIA and type IIB theories.\footnote{There also exist type 0A and 0B theories without any spacetime supersymmetry and with a tachyon. These are less useful but can occasionally be used as laboratories to learn lessons on general string theory.} The absence of the Weyl anomaly leads in this case to the critical number of dimensions $d=10$ for flat space. \\

\noindent Another approach to fermionic theories is to combine the, say, left-moving side of the bosonic string with the right-moving side of the superstring on the worldsheet. These are the \emph{heterotic} theories and these have originally been strong candidates for phenomenology. Nowadays however these are less popular and type II phenomenology is more promising. Nonetheless, they are part of the large web of consistent string theories and we will discuss these as well at some points during this work.

\section{Busher Rules and T-duality}
\label{bushe}
A fundamental property of strings is their so-called T-duality. In the simplest setting, one finds an equivalence between theories of closed strings defined on very small and on very large circles. For instance, considering flat space with one dimension periodically identified (with radius $R$) leads to the conformal weights (for the matter sector)
\begin{align}
L_0 &= \frac{\alpha'p^2}{4} + \frac{\alpha'}{4}\left(\frac{n}{R} + \frac{wR}{\alpha'}\right)^2 + N, \\
\bar{L}_0 &= \frac{\alpha'p^2}{4} + \frac{\alpha'}{4}\left(\frac{n}{R} - \frac{wR}{\alpha'}\right)^2 + \bar{N},
\end{align}
where $p$ labels the spacetime momentum of the state (excluding the compact dimension) and $N$ is the oscillator number of the excited states. Also, $n$ denotes the \emph{discrete momentum} of the string along the compact dimension and $w$ denotes the \emph{winding} of the string along the compact dimension. One immediately observers that this spectrum is retained upon simultaneously switching $R \leftrightarrow \alpha'/R$ and $n\leftrightarrow w$. T-duality swaps discrete momentum modes and stringy winding modes in the non-interacting spectrum. \\
On a general curved background, T-duality acts \emph{to lowest order in $\alpha'$} by the so-called \emph{Busher rules} as \cite{Buscher:1987sk}\cite{Buscher:1987qj}
\begin{align}
G_{00} \to \frac{1}{G_{00}}&,\quad G_{0i} \to \frac{B_{0i}}{G_{00}},\quad G_{ij} \to G_{ij} - \frac{G_{0i}G_{0j}}{G_{00}} + \frac{B_{0i}B_{0j}}{G_{00}}, \nonumber \\
B_{0i} \to \frac{G_{0i}}{G_{00}} &,\quad B_{ij} \to B_{ij} - \frac{G_{0i}B_{0j}}{G_{00}} + \frac{B_{0i}G_{0j}}{G_{00}}, \nonumber \\
&\Phi \to \Phi - \frac{1}{2}\ln\left(G_{00}\right),\quad T_{n} \leftrightarrow T_{w}.
\end{align}
The final formula here denotes swapping of momentum with winding modes. \\
It is known that this symmetry is an exact symmetry of the non-linear sigma model \cite{Rocek:1991ps}, although the concrete transformation formulas could receive $\alpha'$ corrections. However, for type II superstrings, these corrections seem absent and the above rules are likely $\alpha'$-exact.\footnote{At least the two-loop order contribution vanishes (unlike for bosonic and heterotic strings) and concrete examples discussed further show this explicitly to any order. For WZW models (both gauged and ungauged) this can be shown explicitly in general. In somewhat more detail, it is known that for type II superstrings, integrating out the worldsheet gauge field can be done exactly at one loop (including the dilaton contribution) \cite{Tseytlin:1992ri}\cite{Tseytlin:1993my}, both for the axially and vectorially gauged models. Since at lowest order these backgrounds are related by the above Busher rules, it immediately follows that for these models, the Busher rules are $\alpha'$-exaxt.} This is not the case for bosonic and heterotic string theories \cite{Kaloper:1997ux}\cite{Garousi:2013gea}.

\section{Highly excited strings}
Throughout this entire work, we denote $d$ as the total spacetime dimension and $D$ as the number of noncompact spatial dimensions. The quantity $d-1$ is the total number of spatial dimensions (compact or not).  \\
The string spectrum in flat space can be computed and leads to the conformal weights (for the matter sector)
\begin{equation}
L_0 = \alpha'p^2 + N, 
\end{equation}
for open strings and 
\begin{align}
L_0 &= \frac{\alpha'p^2}{4} + N, \\
\bar{L}_0 &= \frac{\alpha'p^2}{4} + \bar{N},
\end{align}
for closed strings. The on-shell condition of a string translates into $L_0 =1$ for open strings and $L_0 + \bar{L}_0 = 2$ for closed strings. This leads to the mass spectrum
\begin{align}
\alpha' m^2 &= -1 + N , \\
\alpha' m^2 &= -4 + 2(N+\bar{N}),
\end{align}
for open and closed strings respectively. \\
The early successes of string theory are explained by this spectrum: the closed string has a massless state at the first excited level $N=\bar{N}=1$. It turns out that this state is described by a spin-2 field and can be decomposed into a symmetric traceless part (the graviton), an antisymmetric part (the Kalb-Ramond field) and a trace part (the dilaton). String theory includes gravitation and hence provides an interesting candidate for a consistent theory of quantum gravity. \\
For the purposes of this work, we are interested in a different feature of the string spectrum: the behavior of the high-energy part. Highly excited strings have a large oscillator number $N$ (and $\bar{N}$) and the number of states for a fixed $N$ increases dramatically for large $N$. The number of such states is given by the number of partitions of the integer $N$. Its asymptotic form is given by the Hardy-Ramanujan formula and the resulting density of states is of the form:
\begin{equation}
p(N) \approx \alpha N^{-\gamma}\exp(\delta \sqrt{N})
\end{equation}
for large $N$ and with positive $\alpha$, $\gamma$ and $\delta$. The precise values of these coefficients depend on the type of string theory one studies. For instance for bosonic strings, one obtains
\begin{equation}
\boxed{p(N) \approx \alpha N^{-\frac{d+1}{4}}\exp\left(2\pi \sqrt{\frac{N(d-2)}{6}}\right)}
\end{equation}
for $d$ spacetime dimensions ($d-2$ bosonic oscillators). For closed bosonic strings, the level-matching condition $N=\bar{N}$ and $\alpha' m^2 \approx 4N$, then lead to the mass density of states
\begin{equation}
\rho(M) \sim M^{-d}\exp\left(2\pi \sqrt{\alpha'} M \sqrt{\frac{d-2}{6}}\right),
\end{equation}
upon including the anti-holomorphic sector and taken the transformation of the measure into account. \\
\noindent Next we need to go from this mass density to the energy density of states and both expressions are not trivially related to each other.\footnote{Note that thinking in terms of the energy is more convenient and more general than thinking in terms of the mass. Energy, as the eigenvalue of the temporal derivative of a state, is generally defined also in curved (stationary) spacetimes and in compact spaces, whereas mass requires a separate definition that is adapted to each of these situations. For instance in the prototypical Kaluza-Klein reduction of a Klein-Gordon field, one has both the original (5d) mass and the lower-dimensional effective (4d) mass, differing by the discrete momentum around the compact dimension. Energy on the other hand is the same throughout.} 
We first generalize the situation to the case where a number of compact dimensions are present. Consider a space of $d$ flat spacetime dimensions of which $\tilde{d}$ compact flat dimensions and $D$ non-compact flat dimensions, such that $d=D+\tilde{d}+1$. The weights are given by
\begin{align}
L_0 &= \frac{\alpha'p^2}{4} + \frac{\alpha'}{4}\left(\frac{n_i}{R_i} + \frac{w_iR_i}{\alpha'}\right)^2 + N, \\
\bar{L}_0 &= \frac{\alpha'p^2}{4} + \frac{\alpha'}{4}\left(\frac{n_i}{R_i} - \frac{w_iR_i}{\alpha'}\right)^2 + \bar{N},
\end{align}
with level-matching $\bar{L}_0 - L_0 = N - \bar{N} + n_iw_i = 0$. One finds on-shell:
\begin{align}
N &= \frac{\alpha'}{4}\left[E^2-\mathbf{p}^2 -\left(\frac{n_i}{R_i} + \frac{w_iR_i}{\alpha'}\right)^2\right]+1, \\
\bar{N} &= \frac{\alpha'}{4}\left[E^2-\mathbf{p}^2 -\left(\frac{n_i}{R_i} - \frac{w_iR_i}{\alpha'}\right)^2\right]+1.
\end{align}
The energy of an on-shell string state satisfies
\begin{equation}
\alpha' E^2 = \alpha' \mathbf{p}^2 + \alpha'\frac{n_i^2}{R_i^2} + \frac{w_i^2R_i^2}{\alpha'} - 4 + 2(N+\bar{N}),
\end{equation}
and at first sight, high energy can arise by either high oscillator number or high momentum (or a combination). However, the density of states as a function of $\left|\mathbf{p}\right|$ has a power-law rise, whereas the density of states as a function of $N$ rises exponentially as we saw above. This means that the dominant part of the high energy regime arises by taking $N$ very large and $\mathbf{p}$ arbitrarily low: $\left|\mathbf{p}\right| \ll E$. The same holds for the compact dimensions. The energy density of states is given by
\begin{equation}
\omega (E) dE \sim EdE \frac{V_{D}}{(2\pi)^D} \sum_{n_i w_i} \int d^{D}\mathbf{p} p(N)p(\bar{N}).
\end{equation}
For high energy $E$, and using the Taylor expansion $\sqrt{N} \approx \frac{\sqrt{\alpha'}}{2}\left[E - \frac{1}{2E}\left(\mathbf{p}^2 +\left(\frac{n_i}{R_i} + \frac{w_iR_i}{\alpha'}\right)^2\right)\right]$, one can approximate the partitions of $N$ further as:
\begin{align}
p(N) &\sim N^{-\frac{d+1}{4}}\exp\left(2\pi \sqrt{\frac{N(d-2)}{6}}\right) \nonumber \\
&\sim E^{-\frac{d+1}{2}} \exp\left(\pi \sqrt{\alpha'} E \sqrt{\frac{d-2}{6}}\right) \exp\left(-\frac{\sqrt{\alpha'}\pi}{2E}\left(\mathbf{p}^2 +\left(\frac{n_i}{R_i} + \frac{w_iR_i}{\alpha'}\right)^2\right)\sqrt{\frac{d-2}{6}}\right),
\end{align}
and hence 
\begin{equation}
p(N)p(\bar{N}) \sim \sim E^{-(d+1)} \exp\left(2\pi \sqrt{\alpha'} E \sqrt{\frac{d-2}{6}}\right) \exp\left(-\frac{\sqrt{\alpha'}\pi}{E}\left(\mathbf{p}^2 +\left(\frac{n_i}{R_i}\right)^2 + \left(\frac{w_iR_i}{\alpha'}\right)^2\right)\sqrt{\frac{d-2}{6}}\right).
\end{equation}
For momenta $\left|\mathbf{p}\right| \ll E$, one can use the saddle point approximation on the integrals over $\mathbf{p}$. The summations over the compact dimensions can be approximated by integrals when $E \gg 1/R_i$ and $E \gg R_i/\alpha'$, both of which are satisfied in the high energy regime. The first inequality allows the treatment of $n_i$ as a continuous quantum number. The second equality allows the winding numbers to also be treated as a continuous quantum number. The latter condition is the difference between a non-compact dimension (for which $E<R/\alpha'$ by definition) and a compact dimension. The salient part of these integrals is the generation of a factor $\sqrt{E}$ per integral. In the end, one obtains 
\begin{equation}
\omega (E) \sim V_{D} E^{-d} \exp\left(2\pi \sqrt{\alpha'} E\sqrt{\frac{d-2}{6}}\right)E^{D/2+\tilde{d}} \sim V_{D} \frac{\exp\left(2\pi E \sqrt{\alpha'}\sqrt{\frac{d-2}{6}}\right)}{E^{D/2+1}}.
\end{equation}
And indeed, the high energy density of states $\omega(E)$ is sensitive to the number of \emph{non-compact} dimensions. The general expression we have derived is 
\begin{equation}
\boxed{\omega(E) \sim V \frac{e^{\beta_H E}}{E^{D/2+1}}}.
\end{equation}

\noindent In a canonical ensemble, the large energy contribution to the single-string thermal trace can hence be rewritten as
\begin{equation}
\label{lapll}
z = \text{Tr}e^{-\beta H} \approx \int^{+\infty} dE \frac{e^{\beta_H E}}{E^{D/2+1}} e^{-\beta E}
\end{equation}
and it is clear that this quantity diverges as soon as $\beta < \beta_H$. This upper temperature $T_H$ is called the \emph{Hagedorn temperature} and the non-interacting canonical partition function does not exist beyond it \cite{Hagedorn:1965st}\cite{Frautschi:1971ij}\cite{Carlitz:1972uf}. \\
The lower boundary of the above integral (say $E_0$) should be chosen such that the asymptotic formula is valid throughout the interval $[E_0, +\infty[$. Then one can evaluate the above integral in the $\beta\to\beta_H$ limit as
\begin{equation}
\int_{E_0}^{+\infty} dE \frac{e^{\beta_H E}}{E^{D/2+1}} e^{-\beta E} \approx (\beta-\beta_H)^{D/2}\Gamma\left(-D/2,E_0(\beta-\beta_H)\right),
\end{equation}
in terms of the incomplete Gamma function. This leads to
\begin{equation}
z \propto 
\left\{
    \begin{array}{ll}
        \left(\beta - \beta_{H}\right)^{D/2}\ln\left(\beta - \beta_{H}\right), \quad D\text{ even}, \\
        \left(\beta - \beta_{H}\right)^{D/2}, \quad D\text{ odd},
    \end{array}\right.
\end{equation}
for the most non-analytic contribution. \\
\noindent In particular, when $D=0$, one obtains $z \approx -\ln(\beta-\beta_H)$. The proportionality factor does not have a fixed sign as $D$ changes; only when $D=0$ does this expression really dominate and one has the requirement that $z>0$, which is indeed satisfied. \\ 

\noindent In this work we are interested in the behavior of the string ensemble at temperatures close to this temperature but still below it. We will see that strings behave rather differently at these temperatures (than their zero-temperature counterparts). \\

\noindent From the above manipulations, it is clear that determining the high energy density of states in the microcanonical ensemble is not without subtleties. A different method of determining this consists in computing the thermal partition function and then reading off the density of states by writing this as a Laplace transform as in (\ref{lapll}). We will illustrate this further on. 

\section{Random walk model of a highly excited string}
\begin{figure}[h]
\centering
\includegraphics[width=0.2\textwidth]{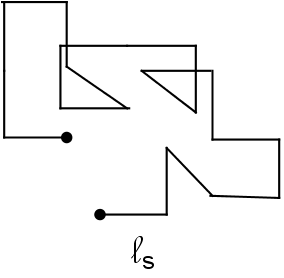}
\caption{Random walk string segment model of open string. Each segment is of length $\ell_s$.}
\label{orw}
\end{figure}

The random walk model of a highly excited string has proven to be a useful mental picture of a string with high excitation number, although the arguments are a bit heuristic \cite{Mitchell:1987hr}\cite{Mitchell:1987th}\cite{Bowick:1989us}\cite{Deo:1989bv}\cite{Horowitz:1997jc}\cite{Barbon:2004dd}. 
A highly excited string is very massive, implying classical arguments should be a good guide. Classically, the energy (or mass) of the string arises from its length, so
\begin{equation}
E \sim \frac{L}{\ell_s^2}. 
\end{equation}
The string is hence long. By looking at the string as arising from string segments of length $\ell_s$ each of which having a random orientation, one can account for the entropy of the highly excited string as follows (figure \ref{orw}). For simplicity we first focus on an open string.
The total number of microstates with fixed energy $E$, is given by
\begin{equation}
\omega(E) \sim n^{L/\ell_s} \sim e^{\ell_s E \text{ln}(n)},
\end{equation}
with $n$ twice the number of accessible space dimensions. This leads to an entropy $S \sim E \ell_s$, which is indeed the correct entropy for a highly excited long string. \\
For a closed string (and a closed random walk), some correction factors are in order. Up to this point, the density of single string states derived from the random walk model is of the form
\begin{equation}
\omega(E) \sim e^{\beta_H E}
\end{equation}
for some constant $\beta_H$. The spatial extent of a random walk is $R \sim E^{1/2}$, which for large $E$ is much smaller than its length $E$. Hence its volume for a well-contained walk in $D$ spatial dimensions is of the order $V_{walk} \sim E^{D/2}$. For a walk that reaches the boundary however, we have $V_{walk} \sim V$. \\
For a closed random walk, we should correct the number of microstates of the walk by dividing by the volume of the walk (correcting for its closedness)\footnote{The author finds this result non-trivial: it is illuminating to know that mathematicians computed the probability for a random walk to return to its origin after $N$ steps to behave as $\sim 1/N^{D/2}$ for large $N$ and for $D$ non-compact dimensions.}, multiply by the spatial volume available (correcting for the arbitrary location), and finally divide by its length $E$ (correcting for the arbitrary starting point of the walk along its perimeter). This leads to 
\begin{equation}
\omega(E) \sim V \frac{e^{\beta_H E}}{E^{1+D/2}},
\end{equation}
which indeed coincides with the high-energy density of states as determined above from the Hardy-Ramanujan formula and the ensuing momentum integrals. \\

\noindent It has been shown that for open random walks between Dp- and Dq-branes, the number of random walks again correctly reproduces the high-energy density of states \cite{Abel:1999rq}. \\

\noindent An additional piece of evidence for the random walk picture is to look at the transverse spread of a highly excited string from its center of mass \cite{Mitchell:1987th}. The transverse coordinate $X^i$ of a string in light-cone gauge is given by: 
\begin{equation}
X^i(\tau,\sigma) = \underbrace{x_0^i + \alpha' p^{i} \tau}_{X_{CM}^i(\tau)} + i \sqrt{\frac{\alpha'}{2}}\sum_{n\neq0}\frac{e^{-in\tau}}{n}\left(\alpha_n^i e^{in\sigma} + \bar{\alpha}_n^i e^{-in\sigma}\right) 
\end{equation}
where $\left[\alpha_m^\mu , \alpha_n^{\nu}\right] = m\delta_{m,-n}\eta^{\mu\nu}$. The spread of a string state $\left|n_k\right\rangle$ in the transverse direction $i$ can then be computed directly and yields\footnote{No sum over $i$ here.}
\begin{equation}
\label{transverseloc}
R_i^2 = \frac{1}{2\pi} \int_{0}^{2\pi} d\sigma \left\langle n_k\right|\left|X^i - X^i_{CM}\right|^2\left|n_k\right\rangle = \frac{\alpha'}{2} \left(2R_{0pt}^2 + f_i(n_k) + f_i(\bar{n_k})\right)
\end{equation}
where $R_{0pt}^2$ is the zero-point fluctuation that we will neglect here.\footnote{It will play a prominent role in chapter \ref{chBH}. Alternatively, one can define the quadratic spread of the string by including a normal ordering prescription as in \cite{Mitchell:1987th}.} The function $f_i$ sums the inverses of the oscillator numbers in the $i$-direction. For instance, a state $\alpha_{-5}^{i}\alpha_{-4}^{j}\alpha_{-2}^{i}\left|0\right\rangle$ would give $1/5 + 1/2$ as the value for $f_i$. \\
For fixed level $N$ in the $i$-direction, it is clear that the state $\left(\alpha_{-1}^i\right)^N\left|0\right\rangle$ has the longest spread $\sim N$ in the $i$-direction, whereas the state $\alpha_{-N}^i\left|0\right\rangle$ has the shortest spread $\sim 1/N$. Obviously, exciting no oscillators at all in the $i$-direction leads to zero size in this direction. \\
This holds for any fixed transverse direction $i$. When averaging over all states with fixed level $N$, full rotational invariance in the $d-1$ spatial dimensions should be visible. This then ultimately leads to the spread $R^2_{av} \sim (d-1)\sqrt{N}$ \cite{Mitchell:1987th}. Since $\sqrt{N} \sim L$, we find again a characteristic feature of a random walk: its size is proportional to the square root of its length: $R_{av} \sim \sqrt{L}$. For large $L$, this implies it will form a highly tangled ball of string. \\

\noindent Finally, we refer the reader to \cite{Manes:2004nd} for scattering amplitude arguments in favor of a random walk interpretation of the highly excited string.

\section{Multi-string coalescence}
One of the earliest discoveries that was made for such highly excited strings was their tendency to elongate. Up to this point, only a single string was studied. The effect we wish to highlight here is inherently multi-string and hence we will make the transition to multiple strings and their interactions. The treatment we will present is sufficiently general to apply to curved spaces as well: the only input required is the single-string density of states $\omega(E)$. The multi-string density of states $\Omega(E)$ is constructed by summing over all possible number of strings. Within string theory, interactions allow strings to split and join, meaning keeping fixed the number of strings in an ensemble is nonsensical. Instead one treats the gas of strings in a manner similar to the photon gas studied in elementary statistical physics. \\

\noindent To further place this section in this work, I point out that almost everything that follows in other chapters concerns the single-string density of states. After the analysis we will present there, one should come back to this section to find out how the full string gas behaves in terms of number of long strings. \\ 
An elementary property of highly excited strings is that they tend to join together and elongate further. We will present details on this strange effect from three different perspectives. First, we present a crude argument from a 2-string perspective, as described in the textbook \cite{Zwiebach:2004tj}. Secondly, we will present a more elaborate microcanonical computation, based on \cite{Bowick:1989us}. Finally, we show that the canonical ensemble also contains this information. This final perspective has, as far as I know, not been treated before in the existing literature. 

\subsection{Crude argument}
For simplicity we focus on open bosonic strings in flat space. Highly excited strings have an exponential degeneracy approximated by 
\begin{equation}
p(N) \approx \alpha N^{-\gamma}\exp(\delta \sqrt{N})
\end{equation}
for large $N$ and with positive $\alpha$, $\gamma$ and $\delta$. Let us consider the entropy change for the process of joining two highly excited strings together, where total energy is conserved. We neglect momentum of the strings in this paragraph. For a highly excited open bosonic string in flat space we have: $\alpha' E^2 \approx N$. The change in entropy in this process of joining strings 1 and 2 and obtaining string 3 (using $E_1+E_2 = E_3$) is given by
\begin{equation}
\Delta S \approx \gamma \ln\left(\frac{N_1N_2}{N_3}\right) - \ln(\alpha).
\end{equation}
Assuming $N_1 \geq N_2$, we can approximate this further as
\begin{equation}
\Delta S \approx \gamma \ln\left(N_2\right) - C
\end{equation}
for some positive constant $C$. It is clear that for large enough $N_2$ (and hence $N_1$ and $N_3$), the entropy change will always be positive. We conclude that \emph{long strings are entropically favored}. Of course this argument is a bit crude, and a more detailed exposition is required. \\
The reason is that the microcanonical ensemble only fixes the total energy in the box: the energy of each individual string is not kept fixed. This leads to an appreciable modification of the $D=0$ case as is illustrated in the next section.

\subsection{Microcanonical treatment}
The single-string density of states is related to the single-string partition function $z$ by a Laplace transform
\begin{equation}
z = \int_0^{+\infty}dE \omega(E)e^{-\beta E}.
\end{equation}
Likewise the multi-string density of states and the multi-string partition function $Z_{\text{mult}}$ are related as 
\begin{equation}
Z_{\text{mult}} = \int_0^{+\infty}dE \Omega(E)e^{-\beta E}.
\end{equation}
In the Maxwell-Boltzmann approximation (which is valid at high temperatures), one has $Z_{\text{mult}}=e^{z}$ and this can be used to relate the densities of states as \cite{Bowick:1989us}\footnote{One writes $\Omega$ in terms of $Z_{\text{mult}}$ with an inverse Laplace transform. $Z_{\text{mult}}$ is related to $z$ by exponentiation and this can be Taylor-expanded. Finally using the relation between $z$ and $\omega$ gives the desired result. The $n=0$ term is simply $\delta(E)$.}
\begin{equation}
\Omega(E) = \sum_{n=0}^{+\infty}\frac{1}{n!}\int \prod_{i=1}^{n} dE_i \omega(E_i)\delta\left(E-\sum_{i=1}^{n} E_i\right).
\end{equation}
For a single string density of states of the form (at high energies)
\begin{equation}
\omega(E) \approx \frac{e^{\beta_H E}}{E^{D/2+1}},
\end{equation}
one can readily see that the most dominant configuration in the multi-string density of states is when all energy is in one string. The asymptotic approximation for the single string density of states breaks down at intermediate energies and this can be taken care of by cutting off the integrals at $E = E_0$. In this approximation, one chooses one $i$ for which $E_i= E$. Upon dropping the vacuum contribution ($n=0$),\footnote{The $n=0$ contribution provides an additive contribution to $\Omega$ as $\delta(E)$, a delta-peak at zero energy. Obviously in the large $E$ regime that we care about here, this vanishes.} and integrating the remaining $E_i$, one obtains \cite{Bowick:1989us}\cite{Athanasiu:1988st}
\begin{align}
\Omega(E) &\sim \sum_{n=1}^{+\infty}\frac{1}{(n-1)!} \frac{e^{\beta_H E}}{E^{D/2+1}}\left(\frac{2}{DE_0}\right)^{n-1}, \quad D>0, \\
\Omega(E) &\sim \sum_{n=1}^{+\infty}\frac{1}{(n-1)!} \frac{e^{\beta_H E}}{E}\left(\ln(E/E_0)\right)^{n-1}, \quad D=0,
\end{align}
and upon summation:
\begin{align}
\Omega(E) &\sim \frac{e^{\beta_H E}}{E^{D/2+1}}e^{\frac{2}{DE_0}}, \quad D>0, \\
\Omega(E) &\sim \frac{e^{\beta_H E}}{E_0}, \quad D=0,
\end{align}
showing that quite generally the multi-string density of states is of the same form as the single-string density of states. The major discrepancy is when $D=0$, where there is a difference in the denominator. Hence the density of multi-string states $\Omega(E)$ is dominated for $D\neq0$ by a single long string. Modifications on this result for $D=1,2$ are known (they come from subleading contributions), but will not be considered here \cite{Deo:1989bv}.

\subsection{Canonical treatment}
It is interesting to realize that the above microcanonical approach to single long strings, can be just as well studied in the canonical ensemble. The criterion is dominance (or non-analyticity) of the partition function $Z_{\text{mult}}$ as a function of the temperature. \\

\noindent The multi-string partition function is (near $T_H$) related to the single-string partition function as $Z_{\text{mult}} = e^{z}$, where $z$ has a random walk interpretation. The multi-string partition function hence has the form of multiple random walks (with an unspecified amount). Using the general expression for the single-string partition function\footnote{The logarithm is absent for $D$ odd. The constant $C$ changes sign as $D$ changes. For $D=0$, $C=1$.}
\begin{equation}
z = - C(\beta-\beta_H)^{D/2}\ln(\beta-\beta_H),
\end{equation}
one can write down the following expressions. For $D$ odd, one finds
\begin{equation}
Z_{\text{mult}} = 1 - C(\beta-\beta_H)^{D/2} + \frac{C^2}{2}(\beta-\beta_H)^{D} - \frac{C^3}{6}(\beta-\beta_H)^{3D/2} + \hdots
\end{equation}
Dropping the no-string contribution, it is clear that the single-string contribution carries the most dominant non-analytic contribution to $Z_{\text{mult}}$. For $D$ even and non-zero, one finds
\begin{equation}
Z_{\text{mult}} = 1 - C(\beta-\beta_H)^{D/2}\ln(\beta-\beta_H) + \frac{C^2}{2}(\beta-\beta_H)^{D}\ln(\beta-\beta_H)^2 - \frac{C^3}{6}(\beta-\beta_H)^{3D/2}\ln(\beta-\beta_H)^3 + \hdots,
\end{equation}
and again the single-string part $z$ contains the dominant non-analytic contribution. For $D=0$, this is absolutely not the case:
\begin{equation}
Z_{\text{mult}} = \frac{1}{\beta-\beta_H} = 1 - \ln(\beta-\beta_H) + \frac{C^2}{2}\ln(\beta-\beta_H)^2 - \frac{C^3}{6}\ln(\beta-\beta_H)^3 + \hdots,
\end{equation}
and higher multi-string contributions carry even more important divergences. Concluding, for $D\neq0$ and dropping the vacuum contribution, one finds
\begin{equation}
Z_{\text{mult}} \approx z
\end{equation}
where $z$ is the single-string partition function. The multi-string partition function gets its main non-analytical contribution from single strings.

\section{String Thermodynamics in flat space}
In this section we present the main ideas related to string thermodynamics in flat space. This section is very important for the remainder of this work, as our main goal is to seek an extension of these results to curved backgrounds.
\subsection{From particles to strings}
In this section we introduce string thermodynamics in the canonical ensemble in flat space. Let us start by considering a single bosonic degree of freedom. Its free energy is of the following form
\begin{align}
F &= \frac{V}{\beta} \int \frac{d^{d-1}k}{(2\pi)^{d-1}}\ln\left(1-\exp(- \beta E)\right) \\
&= -\frac{V}{\beta} \sum_{r=1}^{+\infty}\frac{1}{r}\int \frac{d^{d-1}k}{(2\pi)^{d-1}}\exp(- r\beta E).
\end{align}
To proceed, one uses the identity
\begin{equation}
\frac{1}{r}\exp(-\beta r E) = \frac{\beta}{\sqrt{2\pi}}\int_{0}^{+\infty}\frac{ds}{s^{3/2}}\exp\left(-\frac{E^2s}{2}-\frac{r^2\beta^2}{2s}\right),
\end{equation}
and we obtain
\begin{equation}
F = -V \int_{0}^{+\infty}\frac{ds}{s(2\pi s)^{d/2}} \sum_{r=1}^{+\infty}\exp\left(-\frac{m^2 s}{2} - \frac{r^2\beta^2}{2s}\right).
\end{equation}
This can be interpreted in a first-quantized path integral language as
\begin{equation}
-\beta F = \int_{0}^{+\infty}\frac{ds}{2s}\sum_{w=-\infty}^{'+\infty}\int_{\substack{X^0(0) = X^{0}(s) + r\beta \\ X^{i}(0) = X^{i}(s)}}\left[\mathcal{D}X\right]\exp\left(-\frac{1}{2}\int_{0}^{s}d\tau\left[\left(\frac{\partial X^{\mu}}{\partial t}\right)^{2} + m^2\right]\right)
\end{equation}
and the prime indicates that $w=0$ has been excluded in the sum. The integer $w$ can be interpreted as the winding number of the particle worldline around the compact time dimension. The path integral is over all paths that are periodic (up to the possible winding in the $X^0$ dimension). The interpretation of the integral over $s$ is to sum over all proper times where an overcounting has been divided out (corresponding to an arbitrary starting point on the loop and an arbitrary orientation). The zero-winding contribution does not give a temperature-dependent effect and is unimportant for thermodynamics. \\

\noindent This rewriting of the free energy shows that one can just as well compute thermodynamical quantities by computing the \emph{vacuum amplitude} on the \emph{thermal manifold}, the latter being obtained by Wick-rotating the temporal dimension and periodically identifying this with period $\beta$. \\

\noindent To get to string theory with $d=26$, one can sum this expression over the entire bosonic string spectrum, with masses
\begin{equation}
m^2 = \frac{2}{\alpha'}(N+\bar{N}-2),
\end{equation}
where $N=\bar{N}$ for level-matching. The constraint can be imposed by including the integral
\begin{equation}
\int_{-1/2}^{1/2}d\tau_1 \exp(2\pi i \tau_1(N-\bar{N}))
\end{equation}
and, after changing $s=2\pi\alpha'\tau_2$, the resulting free energy can be written as
\begin{align}
F &= -V \sum_{r=-\infty}'^{+\infty} \int_{0}^{+\infty}\frac{d\tau_2}{2\tau_2}\int_{-1/2}^{1/2}d\tau_1\frac{1}{(4\pi^2 \alpha'\tau_2)^{d/2}} \sum_{i \in \mathcal{H}}e^{-2\pi\tau_2(N_i+\bar{N_i}-2)}e^{2\pi i \tau_1(N_i-\bar{N_i})}\exp\left( - \frac{r^2\beta^2}{4\pi\alpha'\tau_2}\right) \\
&= -V \sum_{r=-\infty}'^{+\infty} \int_{0}^{+\infty}\frac{d\tau_2}{2\tau_2} \int_{-1/2}^{1/2}d\tau_1\frac{1}{(4\pi^2 \alpha'\tau_2)^{d/2}}\left|\eta(\tau)\right|^{-2d+4}\exp\left( - \frac{r^2\beta^2}{4\pi\alpha'\tau_2}\right),
\end{align}
where we have set $q=e^{2\pi i \tau}$ where $\tau= \tau_1+ i \tau_2$, and $\mathcal{H}$ denotes the physical Hilbert space. The prime denotes the omission of the $r=0$ term. In these expressions the Dedekind $\eta$ function is given by
\begin{equation}
\eta(\tau) = q^{1/24}\prod_{n=1}^{+\infty}(1-q^n).
\end{equation}
Polchinski \cite{Polchinski:1985zf} proved that this quantity can also be obtained by computing the torus path integral on the modular strip domain:
\begin{equation} 
Z_{T_2} = \int_0^\infty \frac{d\tau_2}{2\tau_2} \int_{-1/2}^{1/2} d\tau_1 \Delta_{FP} \int \left[\mathcal{D}X\right]\exp -\frac{1}{4\pi\alpha'} \int d^2\sigma \sqrt h h^{\alpha \beta} \partial_\alpha X^\mu \partial_\beta X_\mu,
\end{equation}
and by considering strings that only wrap one of the torus cycles ($\sigma_1 \sim \sigma_1 + 1 \,\, , \,\, \sigma_2 \sim \sigma_2 + 1$) around the time dimension:
\begin{align}
X^\mu(\sigma_1+1,\sigma_2) & =  X^\mu(\sigma_1,\sigma_2), \quad \mu = 0\hdots d-1,\nonumber \\
X^i(\sigma_1,\sigma_2+1) & =  X^i(\sigma_1,\sigma_2), \quad i=1 \hdots d-1,\nonumber \\
X^0(\sigma_1,\sigma_2+1) & =  X^0(\sigma_1,\sigma_2) + r \beta,
\end{align}
where $\sigma_2$ is interpreted as the worldsheet time coordinate. The interpretation is just like in the particle case: one considers trajectories that wind around the temporal dimension on the thermal manifold. In this case, the string winds along its temporal worldsheet dimension, which can be interpreted as a single closed string making $r$ loops around the $X^0$ dimension.\\

\subsection{Modular invariance and the thermal scalar}
\begin{figure}[h]
\centering
\includegraphics[width=0.5\textwidth]{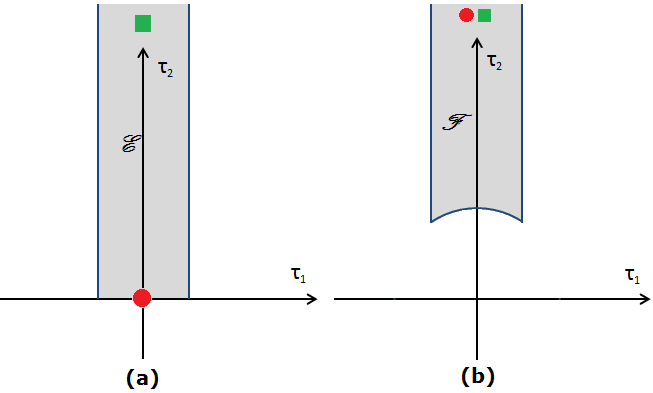}
\caption{(a) Modular strip region where the thermodynamic interpretation is manifest. The thermal divergence arises from the small $\tau_2$ region. Other non-thermal divergences appear in the large $\tau_2$ region. (b) Modular fundamental domain where modular invariance is manifest. The thermal divergence can only arise from a tachyonic divergence at large $\tau_2$, being the thermal scalar.}
\label{modreg}
\end{figure}

The above results are not manifestly modular invariant. This can be achieved by performing two steps. Firstly one restricts the region of integration of $\tau$ to the modular fundamental domain. Secondly, an extra quantum number is introduced to be summed over. This technical result was obtained independently by McClain and Roth \cite{McClain:1986id} and O'Brien and Tan \cite{O'Brien:1987pn}. The result is given by (for $d=26$)
\begin{align}
\label{freeee}
F = -V \sum_{m,w=-\infty}^{+\infty} \int_{\mathcal{F}}\frac{d\tau_1 d\tau_2}{2\tau_2} \frac{1}{(4\pi^2 \alpha'\tau_2)^{13}}\left|\eta(\tau)\right|^{-48}\exp\left( - \frac{\beta^2\left|m-w\tau\right|^2}{4\pi\alpha'\tau_2}\right).
\end{align}
This is to be interpreted as the genus one path integral of string theory on the thermal manifold, with wrapping numbers around \emph{both} torus cycles and a modular integral over the fundamental domain. The modular regions are shown in figure \ref{modreg}. 
A further Poisson resummation on the quantum number $m$ into $n$ allows us to give a Hamiltonian interpretation of this path integral as
\begin{equation}
-\beta F = Z = \int_{\mathcal{F}}\frac{d\tau_1 d\tau_2}{2\tau_2} \text{Tr}\left(q^{L_0}\bar{q}^{\bar{L}_0}\right),
\end{equation}
with 
\begin{align}
h &= \alpha'\frac{p^2}{4} + \frac{\alpha'}{4}\left(\frac{2\pi n}{\beta} + \frac{w\beta}{2\pi \alpha'}\right)^2 + N, \\
\bar{h} &= \alpha'\frac{p^2}{4} + \frac{\alpha'}{4}\left(\frac{2\pi n}{\beta} - \frac{w\beta}{2\pi \alpha'}\right)^2 + \bar{N}. 
\end{align}
It is instructive to think about which states are present in the trace. The intermediate states are required to satisfy $L_n = \bar{L}_n = 0$ for $n>0$. They do not satisfy $L_0=\bar{L}_0=0$ on the other hand: off-shell states propagate in virtual channels. However, when looking into the most divergent state (for thermodynamic purposes), the $\tau_1$ integral reaches over the entire interval $\left[-1/2, 1/2\right]$ and this projects the state onto a level-matched state. Hence divergences can only come from level-matched states. \\

\noindent In the strip modular domain, the Hagedorn divergence can be found by looking into the region $\tau_2 \approx 0$. The $r=\pm 1$ terms have the leading behavior and the exponents behave as
\begin{equation}
e^{\frac{4\pi}{\tau_2}}e^{-\frac{\beta^2}{4\pi\alpha'\tau_2}},
\end{equation}
causing a divergence of the integral in the $\tau_2 \to 0$ region when $\beta \leq \beta_H = 4\pi\sqrt{\alpha'}$. \\
In the fundamental domain, this same divergence arises from the large $\tau_2$ region as a tachyonic excitation. This can be appreciated as follows. Since the fundamental domain excludes the small $\tau_2$ region, one has to look for the divergence elsewhere. In fact, since the interpretation is that of string theory on the thermal manifold, the only place where a divergence can appear is in the large $\tau_2$ region. Thus a tachyon of the thermal theory should encode the Hagedorn divergence. And indeed, the stringy state with $n=0$, $w=\pm1$ and $N=\bar{N}=0$ has the following conformal weight:
\begin{equation}
h= - \frac{\alpha'm^2}{4} + \frac{\beta^2}{16\pi^2\alpha'}.
\end{equation} 
Setting $h=1$ implies $m^2<0$ at $\beta < \beta_H$. Note that this argument uses the on-shell condition ($h=1$). \\
In fact, this condition is not required: for convergence in the partition function one only requires the matter conformal weight to be larger than 1. The momentum quantum number can be integrated out harmlessly and the resulting condition reduces to 
\begin{equation}
\frac{\beta^2}{16\pi^2\alpha'} \geq 1,
\end{equation}
which is the equation we need. \\
This thermal state that is singly wound around the compact Euclidean time dimension is called the \emph{thermal scalar}. It encodes the Hagedorn divergence and the long string dominance. The entire remainder of this work is focused around generalizing this argument to curved backgrounds. \\

\noindent The critical free energy (as determined by the thermal scalar) can be used to correctly determine the (single-string) high-energy density of states as follows. For $\beta \approx \beta_H$ and large $\tau_2$, expression (\ref{freeee}) approximates to\footnote{The sector $w=\pm1$ is dominant. We also use the Maxwell-Boltzmann approximation which can be shown to apply as soon as the thermal scalar state dominates on the thermal manifold. We will present this argument in chapter \ref{chgentd}.}
\begin{align}
z \approx -\beta F &\approx 2\beta V \int^{+\infty}\frac{d\tau_2}{2\tau_2} \frac{1}{(4\pi^2 \alpha'\tau_2)^{13}}\exp\left(4\pi\tau_2\right)\exp\left(-\frac{\beta^2\tau_2}{4\pi\alpha'}\right)\sqrt{\frac{4\pi^2\alpha'\tau_2}{\beta^2}} \\
&\approx 2V \int^{+\infty}\frac{d\tau_2}{2\tau_2} \frac{1}{(4\pi^2 \alpha'\tau_2)^{25/2}}\exp\left(\frac{2}{\sqrt{\alpha'}}(\beta_H-\beta)\tau_2\right).
\end{align}
The final factor in the first line originates from Poisson resumming the summation over $m$. \\
Upon rescaling the $\tau_2$ variable, one obtains the standard high energy Laplace transform of the density of states (for $D=d-1$ non-compact spatial directions):
\begin{equation}
\omega(E) \sim V \frac{e^{\beta_H E}}{E^{D/2+1}}
\end{equation}
and this route provides a quicker way to obtain the high-energy density of states.

\subsection{Superstring thermodynamics}
For superstrings, analogous discussions can be made, which are obviously more involved than for bosonic strings. In this subsection, we will only touch upon two particular aspects of superstring thermodynamics: the modified GSO projection on the thermal manifold and then the thermal sign factors. The reader interested in the detailed computations of the partition functions can consult some of the early literature \cite{Atick:1988si}\cite{Alvarez:1986sj} or the recent detailed review \cite{Liu:2014nva}.

\subsubsection{Modified GSO projection}
The major new feature when considering superstring thermodynamics in the fundamental domain is an interplay between worldsheet fermionic boundary conditions (spin structure) and bosonic boundary conditions (winding numbers). Following \cite{Giveon:2013ica}, we can write the mutual locality condition of two vertex operators as
\begin{equation}
(F_1\alpha_2 - F_2\alpha_1) - (\bar{F}_1\bar{\alpha}_2-\bar{F}_2\bar{\alpha}_1) + 2(n_1w_2+n_2w_1) \in 2 \mathbb{Z}.
\end{equation}
The two quantum numbers $n$ and $w$ are along the thermal direction. $F$ denotes the worldsheet fermion number and $\alpha$ implements the spin structure: $\alpha=0$ is the NS-sector, whereas $\alpha=1$ corresponds to the R-sector. To implement thermal boundary conditions, we should choose the discrete momentum to be
\begin{equation}
n \in \mathbb{Z} + \frac{\alpha-\bar{\alpha}}{2},
\end{equation}
which implies periodic boundary conditions for the NS-NS and R-R sectors, but antiperiodic boundary conditions for the NS-R and R-NS sectors, which agrees with their spacetime statistics. Implementing this thermal condition leads to
\begin{equation}
((F_1+w_1)\alpha_2 - (F_2+w_2)\alpha_1) - ((\bar{F}_1+w_1)\bar{\alpha}_2-(\bar{F}_2+w_2)\bar{\alpha}_1) \in 2 \mathbb{Z},
\end{equation}
causing $F+w$ to play the role of the worldsheet fermion number in the GSO-projections. \\
We want to remark here that this type of reasoning seems insufficient to conclude we have obtained a consistent string theory: one needs to make sure that a modular invariant partition function can be constructed. For type II superstrings, this has been done more generally in \cite{Atick:1988si}. \\

\noindent For instance for type IIB this leads to
\begin{equation}
(-)^{F+w} = (-)^{\bar{F}+w} = +1.
\end{equation}
The zero-winding state in the NS-NS sector is projected out (due to $(-)^F \left|NS\right\rangle = - \left|NS\right\rangle$), but odd winding states are projected in. In particular the $w=\pm1$ state survives the GSO projection. \\
For type IIA, the projection is instead
\begin{equation}
(-)^{F+w} = 1, \quad (-)^{\bar{F}+w} = (-)^{\bar{\alpha}}.
\end{equation}
In the NS-NS sector, the projection acts in the same way as for type IIB. We conclude that the thermal scalar and its dominant thermodynamics is equal for both type IIA and type IIB superstrings. \\
As a last example, the GSO projection for type 0 superstrings is given by
\begin{align}
\alpha &= \bar{\alpha}, \quad (-)^{F+w} = (-)^{\bar{F}+w}, \quad \text{0B}, \\
\alpha &= \bar{\alpha}, \quad (-)^{F+w} = (-)^{\bar{F}+w+\bar{\alpha}}, \quad \text{0A},
\end{align}
which means the thermal winding $w$ is irrelevant. In particular, the zero-winding tachyon is left untouched. At the same time, the thermal scalar is present as well. We will not have much to say on this type of string theory. 

\subsubsection{Thermal sign factors}
Atick and Witten \cite{Atick:1988si} showed that, compared to the non-thermal case, thermal one-loop amplitudes include the additional sign factors for superstrings:
\begin{align}
U_1 &= \frac{1}{2}\left(-1 + (-)^w + (-)^m + (-)^{w+m}\right), \\
U_2 &= \frac{1}{2}\left(1 - (-)^w + (-)^m + (-)^{w+m}\right), \\
U_3 &= \frac{1}{2}\left(1 + (-)^w + (-)^m - (-)^{w+m}\right), \\
U_4 &= \frac{1}{2}\left(1 + (-)^w - (-)^m + (-)^{w+m}\right),
\end{align}
where $U_3$ is the NS-sector $(-,-)$, $U_4$ corresponds to the $\widetilde{\text{NS}}$-sector $(-,+)$. $U_2$ corresponds to the R-sector $(+,-)$ and finally $U_1$ is the $\widetilde{\text{R}}$-sector $(+,+)$. These formulas can be simplified into
\begin{align}
U_1 &= (-)^{wm}(-)^{w+m}, \\
U_2 &= (-)^{wm}(-)^{m}, \\
U_3 &= (-)^{wm}, \\
U_4 &= (-)^{wm}(-)^{w}.
\end{align}
While originally proved for flat space, these formulas are quite general and follow purely from the thermal boundary conditions for fermions combined with modular invariance. \\

\noindent For type II superstrings, for which one combines holomorphic and antiholomorphic sectors each including the above sign factors, the overall phase factor $(-)^{wm}$ disappears in the modulus squared in the end and one finds
\begin{align}
U_1 &= (-)^{w+m}, \\
U_2 &= (-)^{m}, \\
U_3 &= 1, \\
U_4 &= (-)^{w}.
\end{align}
These factors have a clear physical origin. \\ 
The factors of $(-)^m$ originate from the antiperiodic boundary conditions on the spacetime fermions. On the worldsheet, this can be achieved by inserting $(-)^m$ in the R-sectors ($U_1$ and $U_2$) for both left- and right-movers. This ensures only the R-NS and NS-R sectors obtain these factors in the end. The $(-)^m$ factor appearing here signals antiperiodic fields. Upon Poisson resummation $m \to n$, one finds half-integral values of the discrete momentum $n$. \\
The factors of $(-)^w$ originate from the modified GSO-projection, necessary to achieve mutual locality of the vertex operators. These should be inserted in the twisted sectors $\widetilde{\text{NS}}$ ($U_4$) and $\widetilde{\text{R}}$ $(U_1$). \\


\noindent For heterotic strings, the overall phase factor $(-)^{wm}$ does not cancel and is important. For odd $w$, one obtains a further extra sign $(-)^m$ in all sectors. This means the sectors change spacetime periodicity: the NS sector becomes antiperiodic whereas the R sector becomes periodic. This implies that the thermal scalar \emph{must} have discrete momentum as well with $n=1/2$. 

\subsection{Summary}
Since the treatment presented in this section is very important for the remainder of this work, let us repeat the important points. \\
In field theory, one can consider thermal quantities either from the Lorentzian perspective (as $Z=\text{Tr}e^{-\beta H}$) or from the thermal manifold perspective, on which the vacuum amplitude contains the same information. From a field theory perspective, this arises from the interpretation of $Z$ as propagation over imaginary time $i\beta$ and the trace identifies initial and final states. This procedure can be reinterpreted as computing a vacuum amplitude on the thermal manifold. \\
In string theory, one can repeat the above procedure for each string field and then sum over the fields in the string spectrum. This immediately leads to the modular strip region and closed strings propagating $r$ times around the Euclidean time direction. An added feature is that modular invariance can be made manifest by including a second set of wrapping numbers. This latter perspective has the full interpretation of a vacuum amplitude on the thermal manifold. \\
In the strip domain, the dominant near-Hagedorn behavior comes from the $r=1$ part in the small $\tau_2$ region, whereas in the fundamental domain the $w=\pm1$ state at large $\tau_2$ is a thermal string state, the thermal scalar, which contains the same information.

\section{Thermodynamics philosophy}
In non-interacting thermodynamics, one starts with a weakly interacting gas of matter. Then one adiabatically tunes down the coupling to reach the non-interacting gas. It is vital to start with the interacting gas since equilibration processes cannot occur otherwise. It is clear that the time-scale of variation of the coupling should be much longer than the interaction and dissipation times (mean free path) in the interacting gas. \\
For string theory, the same strategy should be used. However, a varying coupling constant seems problematic at first sight.\footnote{We thank O. Evnin for bringing this issue to our attention.} One should always consider an on-shell background and the time-variation of the coupling constant causes backreaction on the geometry (and possible other background fields). The solution is of course to take the time-variation very small, such that corrections to Einstein's equations coming from this time-variation are parametrically smaller than the other terms in the action. Consider for instance the lowest-order dilaton field equation in bosonic string theory:
\begin{equation}
\frac{d-26}{6} - \frac{\alpha'}{2}\nabla^2\Phi - \alpha'\nabla_{\mu}\Phi \nabla^{\mu} \Phi = 0.
\end{equation}
If we take $\nabla_0 \Phi \sim \frac{1}{T}$ where $T^2 \gg \alpha'$, then the two $\alpha'$ terms are very small compared to the $d-26$ term, and one can safely choose $d=26$ and retain flat 26-dimensional Minkowski space throughout this adiabatic process. The same reasoning holds for the other equations of motion and their $\alpha'$-corrections. \\
So for string theory, actually two conditions are required for the adiabatic process: $T$ should be much longer than the interaction and dissipation time scales and also $T$ must be much longer than the string scale. Of course, both of these conditions are actually the same since in string theory dissipation processes are also string scale. \\

\noindent Both from this perspective and from the specific holographic interval for $g_s$ we will present further on (formula (\ref{holoint})), we see that the string gas may be chosen to be arbitarily weakly coupled, but this is always a limiting process: the coupling constant $g_s$ is not strictly zero in thermodynamics. This is important for our discussion on the importance of higher genus corrections.

\section{Discrepancy microcanonical and canonical ensemble}
A peculiar and ill-understood feature of near-Hagedorn string thermodynamics is that the microcanonical and canonical ensemble disagree. The reason is readily appreciated. Normally, in statistical systems the link between both ensembles is made as a saddle point approximation in the partition function. The saddle point of 
\begin{equation}
Z(\beta) = \int dE \Omega(E)e^{-\beta E}
\end{equation}
is found by setting 
\begin{equation}
\frac{\partial S}{\partial E} = \beta.
\end{equation}
As long as the fluctuations in energy in the canonical ensemble are small, both ensembles agree. Unfortunately this is not the case for a Hagedorn system. One can check that different formulas are obtained by computing the internal energy in both ensembles. Let us focus on a fully compact space ($D=0$) where $\Omega(E) = e^{\beta_H E}$. In the microcanonical picture, one finds then
\begin{equation}
\frac{\partial S}{\partial E} = \beta = \beta_H,
\end{equation}
and the temperature is fixed for any energy: the thermodynamics is \emph{degenerate}. In the canonical ensemble on the other hand, one computes the internal energy as
\begin{equation}
E = \frac{\partial (\beta F)}{\partial \beta} = \frac{1}{\beta-\beta_H}
\end{equation}
and one does obtain a non-trivial relation between energy and temperature. For non-compact spaces, one also finds formulas that do not agree. \\

\noindent One way of dealing with this nuisance is by using the canonical ensemble as a tool to distill the free energy and then the single-string density of states. From there on, one proceeds in the microcanonical picture. Above we have indeed shown that computing the density of states is much simpler in the canonical ensemble. \\
Alternatively, it is interesting to study this discrepancy further.

\section{Another perturbative instability: the Jeans effect}
\subsection{Jeans instability}
Thermal physics including gravity is in general not well defined. The reason is that matter tends to collapse and form heterogeneous clumps of matter instead. This effect is called the \emph{Jeans effect} and was discovered in 1902 for fluids in classical Newtonian gravity. In general, hot matter in gravity is unstable towards long wavelength density fluctuations. For a system of mass density $\rho$, the classical Jeans length is given by
\begin{equation}
L_J = v_{s}\left(\frac{\pi}{G \rho}\right)^{1/2},
\end{equation}
with $v_s$ the speed of sound in the medium. \\
In a gravitational system, the relevant speed is $v_s \sim c$ and in natural units we have
\begin{equation}
L_J \sim \left(\frac{1}{G \rho}\right)^{1/2}.
\end{equation}
This can be interpreted as the instability of a system under gravitational collapse when it becomes smaller than its own Schwarzschild radius. Indeed, for a system with a fixed mass density $\rho$ and radius $R$, the relevant quantities are related as
\begin{equation}
r_s^{d-3} \sim G M , \quad M \sim \rho R^{d-1}.
\end{equation}
Demanding $R<r_s$ to obtain an instability, we obtain indeed
\begin{equation}
R > \left(\frac{1}{G \rho}\right)^{1/2} = L_J.
\end{equation}
In \cite{Gross:1982cv} this effect was studied in the semiclassical Euclidean quantum gravity regime, where it was shown that this instability arises from higher loop corrections to the graviton propagator (from either the graviton itself or other matter present in the system). This instability causes a negative $m^2$ for the graviton and a non-real free energy.\\
To mitigate this instability, we have to surround our thermal system by some sort of box of volume $V < L_J^{d-1}$. This can be realized by an actual box or by curvature effects (such as in Anti-de Sitter space as we will study further on). \\

\noindent An analogous formula can be written down in the canonical ensemble \cite{Barbon:2001di}. The (internal) energy density of a (massless) gas equals $E/V \sim T^d$, so that we obtain
\begin{equation}
L_J(T) \sim \left(\frac{1}{G T^d}\right)^{1/2}.
\end{equation}
An alternative argument for this formula is given by looking at the formula for the one-loop tachyonic mass of the ($G_{\tau\tau}$ component) thermal graviton:
\begin{equation}
m^2 \sim - GT^d
\end{equation}
and it is then natural to associate the length scale $L_J(T)$ to this mass.

\subsection{AdS container}
A first way of circumventing the Jeans effect is to embed the thermal system in a box of length $L$ such that $\ell_s \ll L \ll L_J$. In this way, we ensure having a sufficient number of degrees of freedom on the one hand but still not enough to be unstable towards gravitational collapse. \\

\noindent A more sophisticated, alternative method of mitigating the Jeans instability can be found by using $AdS$ space as a container \cite{Barbon:2001di}. \\
The (Euclidean) $AdS$ metric is of the form
\begin{equation}
ds^2 = \left(1+\frac{r^2}{R^2}\right)d\tau^2 + \frac{1}{1+\frac{r^2}{R^2}}dr^2 + r^2d\Omega_{d-2}^2,
\end{equation}
with $\tau \sim \tau + \beta$. For $r<R$, the thermal circle is almost of constant size $\beta$. For $r>R$, the thermal circle increases (asymptotically) linear with $r$. Hence in the interior $r<R$, $AdS$ looks like flat space. Outside, the gas is forced back inwards by the gravitational potential. Hence this space works as a large container of thermal radiation. To disable the Jeans instability one needs simply $R < L_J$. \\

\noindent Demanding both a large number of excitations inside the container and also the absence of the Jeans instability leads to
\begin{equation}
\ell_s \ll R \ll \frac{\ell_s}{g_s},
\end{equation}
where the second inequality is found by evaluating $L_J(\ell_s^{-1})$, which is the smallest Jeans length possible without encountering the Hagedorn divergence. Taking $g_s \ll 1$ is hence sufficient. \\

\noindent In a holographic system, one can use the $AdS/CFT$ duality to describe more concretely what is needed when both $R \gg \ell_s$ and $R \ll L_J$ in terms of the boundary data. It turns out one needs\footnote{One uses the dictionary $R = \ell_s \lambda^{1/4}$, $\lambda = g_s N$ and $g_{YM}^2 = g_s$ for a dual Yang-Mills theory with $N$ colors.}
\begin{equation}
\label{holoint}
\frac{1}{N} \ll g_s  \ll \frac{1}{N^{1/5}}.
\end{equation}
As $N$ is taken to infinity, the string coupling must scale to zero respecting these inequalities.

\subsection{Maxwell construction in the canonical ensemble}
\begin{figure}[h]
\centering
\includegraphics[width=0.5\textwidth]{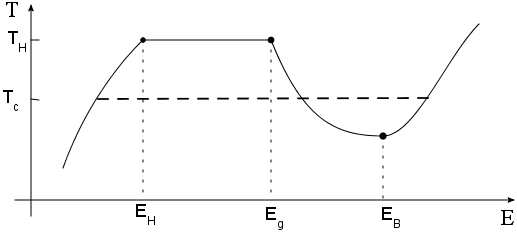}
\caption{Temperature versus energy diagram of the stringy matter in $AdS$. Four different phases can be discerned in a microcanonical approach: a graviton phase, a long string phase (and the corresponding Hagedorn plateau), a small black hole phase and a large black hole phase. Three energy scales are defined in the text: $E_H$, $E_g$ and $E_B$. In a canonical picture, both middle phases are missed completely due to the presence of a first-order phase transition in between. This implies the Hagedorn phase is a superheated phase.}
\label{phasediagram}
\end{figure}

Since $AdS$ space turns out to be a convenient IR regulator of the Jeans instability, it is mandatory to obtain a better understanding of the additional structure provided by the $AdS$ length $R$ to string thermodynamics. \\
It can be shown \cite{Barbon:2004dd}\cite{Abel:1999dy} that the phase diagram of stringy matter in $AdS$ space has the following temperature versus energy behavior (figure \ref{phasediagram}).
For low energies, the massless gas of gravitons dominates thermodynamics and causes the characteristic $T \sim E^{\frac{1}{d}}$ behavior (with $d$ the number of spacetime dimensions). For high energies (higher than $E_H$), long string behavior starts appearing and the temperature becomes effectively constant $T=T_H$, the Hagedorn temperature. At even higher energies $E_g \sim \frac{1}{\ell_s g_s^2}$, small $AdS$ black holes start dominating.\footnote{This energy scale is found by equating the entropy of a Hagedorn gas $S \propto \ell_s E$ and the entropy of a Schwarzschild black hole $S = \frac{A}{4G} \propto \frac{r_s^{d-2}}{\ell_s^{d-2}g_s^2} \propto E \left(g_s^2E\right)^{1/(d-3)}\ell_s^{(d-2)/(d-3)}$. The last equality is found by relating the black hole mass $M=E$ to the Schwarzschild radius as $M \propto \frac{r_s^{d-3}}{g_s^2\ell_s^{d-2}}$. Note that a small $AdS$ black hole behaves precisely the same as a Schwarzschild black hole in flat space. } These are however thermodynamically unstable leading to the declining behavior of the curve. Finally, large $AdS$ black holes (which are thermodynamically stable) take over. The fact that at large energies, black holes dominate thermal systems, was called \emph{asymptotic darkness} in \cite{Barbon:2004dd} \\
We will not go into the details of all of these phases. We only wish to highlight several features that are important for our work. Since we work in the canonical ensemble, the presence of such an unstable phase in the microcanonical ensemble requires the utilization of a Maxwell construction to obtain the canonical phase diagram. Let us therefore first give a small digression on this construction in the canonical ensemble. \\

\begin{figure}[h]
\centering
\includegraphics[width=0.5\textwidth]{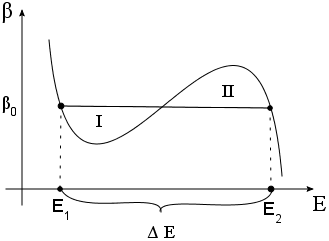}
\caption{Maxwell construction in a thermal system. The phase transition temperature $\beta_0^{-1}$ is determined by demanding the area of region I and II to match. The latent heat is $\Delta E$.}
\label{Maxwell}
\end{figure}

\noindent We illustrate the Maxwell construction in the canonical ensemble on a general thermal system whose $\beta(E)$ curve in the microcanonical ensemble is of form as illustrated in figure \ref{Maxwell}.
A first order canonical phase transition is characterized as a transition with $\Delta F=0$ occuring at a fixed (inverse) temperature $\beta_0$ with a latent heat $\Delta E$ and an entropy difference $\Delta S$. Using $F= E-TS$ and $\frac{\partial S}{\partial E} = \beta(E)$, one infers that 
\begin{equation}
\beta_0\Delta F = 0 = \int_{E_1}^{E_2}\left(\beta_0 - \beta(E)\right) dE.
\end{equation}
The interpretation is that given in figure \ref{Maxwell}. The area of region I and II should be equal, which determines $\beta_0$. \\

\noindent Applying this to the $AdS$ phase diagram, we conclude that both the Hagedorn phase and the small black hole phase are missed entirely in the canonical picture (as shown by the dashed line in figure \ref{phasediagram}). The Hagedorn dominance regime is a \emph{superheated} configuration, meaning it has become meta-stable already at a lower temperature. This phenomenon was called \emph{Hagedorn censorship} in \cite{Barbon:1998ix}\cite{Barbon:1998cr}\cite{Barbon:2001di}. \\
A second remark that we wish to make is that the transition temperature $T_c \sim 1/R$. This is the stringy ancestor of the Hawking-Page phase transition in gravity. \\
Thirdly, we note that the black hole phases are dominant at energies at least as high as $\frac{1}{\ell_s g_s^2}$. Focusing on weakly coupled strings ($g_s \ll 1$) hence ensures these phases are only reached at a very high energy. Also the latent heat is of this same scale and is very high in the coupling regime we are interested in. \\

\noindent Let us finally look at the flat space limit ($R\to\infty$). The energy scale $E_B \sim \frac{R^{d-3}}{g_s^2 \ell_s^{d-2}}$ which can be deduced from the fact that this occurs for a black hole whose size is the $AdS$ radius. Hence, as $R$ increases, $E_B$ moves away to infinity and $T(E_B) \to 0$. The Maxwell construction then implies that $T_c \to 0$ as well. The stringy Hawking-Page transition is possible hence for any non-zero temperature and is in agreement with the fact that hot flat space is unstable towards black hole nucleation at any non-zero temperature \cite{Gross:1982cv}. The stable phase seems to be an infinitely large black hole which needs an infinite latent heat to produce. \\

\noindent Note that this story has a peculiarity for $AdS_3$, since no small (unstable) black holes exist: the BTZ black hole is stable. This apparantly implies the absence of an intermediate unstable phase, necessary for the application of the Maxwell construction and the ensuing Hawking-Page transition. However, recently in \cite{Kleban:2013wba} it was argued that there is in fact an unstable phase consisting of self-gravitating radiation in this case.

\section{Non-perturbative instabilities}
Besides the above (perturbative) instability, there are also several non-perturbative instabilities that can be expected and one expects these to be linked in the manner we now describe.\\
Firstly, in \cite{Gross:1982cv} Gross, Perry and Yaffe found that hot flat space is non-perturbatively unstable towards black hole nucleation, for any non-zero temperature. \\
This instability is related to the Hawking-Page transition in $AdS$ space. This is a non-perturbative (tunneling) transition between a thermal gas in $AdS$ and a black hole in $AdS$. The critical temperature is called the Hawking-Page temperature \cite{Hawking:1982dh}. Taking the flat-space limit of this transition, one finds that the Hawking-Page temperature goes to zero. This corresponds to the above flat space instability of a thermal gas. The Gross-Perry-Yaffe black hole nucleation is a special case of the one-parameter family of Hawking-Page transitions. \\
Going to string theory, Atick and Witten suggested that thermal string theory in flat space undergoes a first-order Hagedorn transition at temperatures smaller than the Hagedorn temperature \cite{Atick:1988si}. The reason is the coupling to the dilaton in the effective action of the thermal scalar. This first order transition is again non-perturbative and takes the hot string gas to some phase that is not known explicitly. However, comparing this story to that above, it is strongly believed that the high-temperature phase of string theory is in fact a black hole phase. \\
Combining these features, the string gas in $AdS$ is expected to have a string-corrected Hawking-Page transition, which in the flat limit reduces to the Atick-Witten transition. On the other hand, taking the low curvature limit, the original Hawking-Page transition is found with the corresponding Gross-Perry-Yaffe instability in the flat limit. These relations are shown schematically below in figure \ref{diagram}.
\begin{figure}[h]
\centering
\includegraphics[width=0.7\textwidth]{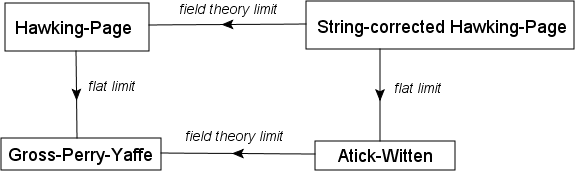}
\caption{Diagram showing the tentative links between the non-perturbative effects discussed in the text. The effects are linked through flat limits and/or dropping all higher stringy fluctuations.}
\label{diagram}
\end{figure}
The precise quantitive link between (most of) these effects is an open problem in the literature that we will have nothing more to say about.

\subsection{Atick-Witten phase transition}
\begin{figure}[h]
\centering
\includegraphics[width=0.8\textwidth]{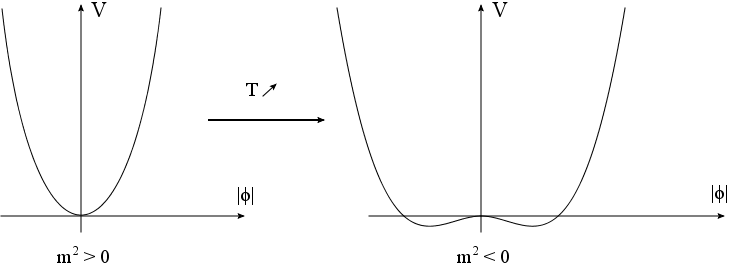}
\caption{$\lambda>0$ behavior of the effective potential. For $m^2>0$, the system is stable (left figure). For $m^2<0$, the system has a perturbative tachyonic instability.}
\label{AtWitten}
\end{figure}

\begin{figure}[h]
\centering
\includegraphics[width=0.3\textwidth]{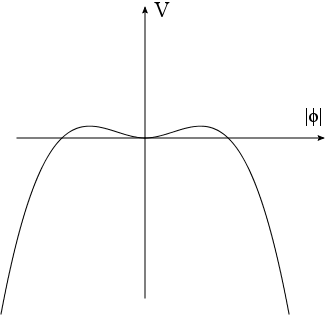}
\caption{$\lambda<0$ behavior of the effective potential. The system is unstable for evolving to larger values of $\left|\phi\right|$ even though no tachyon is present.}
\label{AtWitten2}
\end{figure}

To appreciate the Atick-Witten phase transition, consider the following effective potential for the thermal scalar field $\phi$ for small values of $\phi$:
\begin{equation}
V(\phi,\phi^*) = m^2 \phi \phi^* + \lambda g_s^2 T (\phi \phi^*)^2.
\end{equation}
The behavior depends crucially on the sign of $\lambda$. Consider first the case $\lambda > 0$. For $m^2 >0$, the potential has a positive curvature throughout. For $m^2<0$, the potential at the origin (in field space) has a negative curvature, and the real minimum is located at nonzero $\phi$. For small $\phi$, one finds the minimum at $\left\langle \phi\phi^*\right\rangle = \frac{\left|m^2\right|}{2\lambda g^2 T}$. This behavior is consistent with a second order phase transition in which the order parameter changes continuously. The behavior is shown in figure \ref{AtWitten}.
In the other case ($\lambda < 0$), the potential always has values lower than zero for larger field values (assuming the potential is not substantially corrected at such field vevs). This implies a more stable configuration can be achieved by passing through (or over) the potential barrier, characteristic of a first order phase transition. Since this behavior occurs for $m^2 > 0$, the possibility opens up of a first order phase transition at a temperature lower than the Hagedorn temperature. This is illustrated in figure \ref{AtWitten2}. We note that in this work, if such a phase transition indeed occurs below $T_H$, we will study superheated spaces, where the space is non-perturbatively unstable (as shown here), but perturbatively stable.
In fact, as far as this simple analysis is realistic, any non-zero temperature seems to suffice. If this is the case, the link with the flat limit of the Hawking-Page transition (being the Gross-Perry-Yaffe instability) seems immediate, were indeed hot flat space at \emph{any} non-zero temperature was shown to be unstable. 

\section{Summary}
We have shown how the high-energy density of states arises in several ways.
\begin{itemize}
\item Using the Hardy-Ramanujan formula and some delicate computations on the momentum integrals.
\item Using the thermal manifold to compute $F$ and its inverse Laplace transform then gives $\omega(E)$. One approximates the full free energy by the thermal scalar contribution to obtain the near-Hagedorn contribution.
\item Using the geometric random walk picture. It is very general and gives the correct result in every known case.
\end{itemize}
We have also provided a detailed discussion on how the multi-string density of states is obtained from this and how (and when) it encodes long string dominance. 
String theory allows two different treatments of the thermal manifold: where the modular domain is chosen to be the strip or the fundamental domain. Both descriptions are equivalent and have their merits. \\

\noindent All of these computations will be relevant for the curved space generalizations presented in the remainder of this work. \\

\noindent Finally, we have pointed out several problems with string thermodynamics and how we handle them in this work. 
\begin{itemize}
\item Discrepancy between microcanonical and canonical ensemble.\\
We use the canonical ensemble as a tool to get the density of states and the random walk, from there on we can proceed in the microcanonical picture.
\item Jeans instability. \\
We embed the thermal system in a box or in $AdS$ space.
\item The Hagedorn transition might happen at a lower temperature than $T_H$. \\
If this happens, we will study superheated configurations.
\end{itemize}


\chapter{Random Walks - Formalism in Curved Backgrounds}
\label{chth}
%


In this chapter we discuss the generalization of the random walking picture to a general curved spacetime based on \cite{Mertens:2013pza}. The first few sections treat the first-quantized description where an explicit random walk picture will be obtained. After that, a second quantized formalism is used which complements the previous description. The next chapter continues this study by applying the developed formalism to some simple examples and providing some mild extensions. 
\section{Introduction}

It is an old fact that the behavior of string theory at high temperatures is rather different than that of a conventional thermodynamic system \cite{Atick:1988si}\cite{Horowitz:1997jc}\cite{Barbon:2004dd}. \\
Consider a gas of strings in a box with a fixed total amount of energy (microcanonical ensemble) \cite{Mitchell:1987hr}\cite{Mitchell:1987th}\cite{Bowick:1989us}\cite{Deo:1989bv}. As one increases the energy gradually, nothing special happens until suddenly most of the strings in the gas coalesce and form highly excited long strings. All energy pumped into the system goes into the excitation modes of the long string(s) and not in increasing the temperature. Actually the story is a bit more involved: a single string at high energy behaves as a long random walk. Whether the entire string gas is dominated by one or many long strings depends on the number of non-compact dimensions as is extensively discussed in \cite{Deo:1989bv} and in the previous chapter.\footnote{For 1 or 2 spatial non-compact dimensions, the gas is not dominated by long strings at all.} No further discussion on this will be provided in the remainder of this work. 
Important to note is that the single string density of states always shows long random walk behavior and it is this aspect that we will consider. \\
Now consider the same story, but from a canonical ensemble point of view. Increasing the temperature to a critical value, causes the partition function to diverge due to the high density of highly excited string modes. This ultimate temperature is the so-called \emph{Hagedorn temperature} \cite{Hagedorn:1965st} and the canonical ensemble is not useable at higher temperatures. \\
A few years ago, the authors of \cite{Kruczenski:2005pj} made an explicit derivation of the single string random walk picture directly from the canonical ensemble using the string path integral at genus one. \\
This phenomenon also has a different manifestation. Thermodynamics on any spacetime can be calculated on the so-called `thermal manifold' by Wick rotating the time coordinate and periodically identifying this coordinate. For particles nothing dramatic happens when doing this, strings however can wrap this Euclideanized time direction. The divergence manifests itself here by the masslessness of a winding string state at the critical temperature (and it becomes tachyonic when further heating the system). This string field is what is known as the \emph{thermal scalar} and when nearing the Hagedorn temperature, this field dominates the thermodynamics in the same way that the single string dominates the microcanonical picture \cite{Atick:1988si}. This field effectively represents the large density of states of highly excited string states. One should remark that this field is not a real field corresponding to physical particles but is an effective field theory degree of freedom which dominates the string thermodynamics at high energy. \\

\noindent This random walk picture also arises in black hole geometries \cite{Susskind:1993ws}\cite{Susskind:2005js}. The long string surrounds the event horizon and forms the stretched horizon (or the black hole membrane as it is called in the earlier literature\footnote{See e.g. \cite{Thorne:1986iy} and references therein.}). This work started with the question: `Can we apply the methods developed in \cite{Kruczenski:2005pj} to the black hole case?'
In the case of a black hole, the local temperature increases as one approaches the horizon and at a distance of the order of the string length $\sqrt{\alpha'}$, it exceeds the flat space Hagedorn temperature. One expects a condensate of winding tachyons close to the horizon which in the Lorentzian case is responsible for the appearance of a stretched horizon. Strong evidence for this scenario was shown using the exactly solvable 2D black hole or cigar \cite{Kutasov:2005rr} appearing in Little String Theory \cite{Kutasov:2000jp}.
A relation between condensed winding modes and black hole entropy has also been discussed in \cite{Dabholkar:2001if} using $\mathbb{C}/\mathbb{Z}_n$ orbifolds. The idea \cite{Adams:2001sv} is that the tachyons condense at the tip of the cone and relax the cone to flat space. In this case, closed string field theory can be used because of the localized nature of the winding tachyons \cite{Okawa:2004rh}. For Schwarzschild black holes, the situation is less clear \cite{Kutasov:2005rr}.
Recently, research on the nature of the stretched horizon has been rekindled in \cite{Almheiri:2012rt} who propose the existence of a firewall.\footnote{See \cite{Braunstein:2009my} for an earlier development in this direction.} As pointed out in \cite{Giveon:2012kp}\cite{Giveon:2013ica}, the existence of a winding string zero mode close to the horizon is a possible stringy realization of this idea and deserves further study. Our endeavor is therefore to develop from the string path integral, a general framework for the thermal scalar in curved spacetime backgrounds. \\

\noindent Before we arrive there however, we will first reanalyze the derivation of \cite{Kruczenski:2005pj} and extend and test their result in several situations that are easier to understand than black hole horizons. We discuss the application to black hole horizons themselves in chapter \ref{chri}. Several other examples will be given in the next chapter. 
We want to analyze the random walk picture of highly excited strings in general backgrounds from the canonical ensemble and see if we can get aspects of the above picture out of it. \\

\noindent This chapter is organized as follows.\\
In section \ref{pathderiv1} we review and extend the path integral derivation of the random walk behavior in general backgrounds as was put forward by \cite{Kruczenski:2005pj}. The beauty of this path integral approach is the physical picture of a random walk that clearly emerges once the dust settles. We comment on several of the difficulties that appear in our derivation. The most important of these is that we seem to miss several terms in the resulting particle action, indicating that we did not take the near-Hagedorn limit correctly.\\
In section \ref{compa} we compare the results from the second section to some explicitly known flat spacetime results. We will see in these explicit examples that we do indeed reproduce the expected results if we include a correction corresponding to the flat space tachyon mass in the action. \\
In section \ref{corrections} we will calculate the correction terms explicitly in flat spacetime while taking a path integral perspective (i.e. without comparing to known results). This will demonstrate where the above term comes from. \\
We take a different point of view in section \ref{alternative} and try to see whether we can make contact with one-loop results of field theory actions. The reason we take this approach is because in this case it is computationally easier to deal with field theory actions than to manipulate path integral expressions. We will find a match for flat backgrounds but for general backgrounds we find other terms as well in the action. These are terms arising from the $\sqrt{G_{00}}$ metric component in the measure in the field theory action. We interpret these as other terms we missed in the derivation of the second section. \\
Several technical computations are given in the supplementary sections.
Examples of these methods will be presented further on in subsequent chapters.

\section{Path integral analysis for dominance by singly wound strings}
\label{pathderiv1}
The authors of \cite{Kruczenski:2005pj} have given an explicit path integral picture of the thermal scalar. In this section we review and extend their derivation of the random walk picture of highly excited strings (while making some modifications near the end). \\
The goal is to derive the free energy of a gas of non-interacting strings in a curved background in the limit where highly excited strings dominate (in the microcanonical ensemble). In the canonical ensemble this corresponds to temperatures near the Hagedorn temperature of the specific background. Let us first remark that the relation between the microcanonical and canonical ensemble is not entirely clear in string theory: several conceptual problems arise due to the asymptotic exponential degeneracy of states \cite{Mitchell:1987th}. In what follows we will perform our computations in the canonical ensemble and leave further study of this issue in our case to future work. A partial motivation for this is that the canonical ensemble is often used as a starting point to compute the relevant microcanonical quantities \cite{Brandenberger:1988aj}\cite{Deo:1988jj}.

\subsection{The thermal manifold to calculate string thermodynamics}

Thermodynamics in a general background depends obviously on the choice of time variable. So to describe e.g. the free energy we first have to choose a preferred time coordinate and then calculate the free energy associated to that specific time coordinate.\\
The starting point of the derivation is the \emph{assumption} that the free energy of a non-self-interacting string gas in a certain time-independent background (defined as the sum of the free energies of the individual particle states in the string spectrum) is proportional to the torus partition function of a single string on the thermal manifold (same background, but with $X^0$ Wick-rotated and compactified with length $\beta$, the inverse temperature). So
\begin{equation}
\label{cruc}
Z_{1 string} = -\beta F_{string gas}.
\end{equation}
We say non-self-interacting since the string gas does interact with the background, but it does not interact with itself (reflected in the fact that we only consider the torus amplitude). \\
We make the following comments regarding this assumption:
\begin{itemize}
\item{It holds for the flat bosonic case as proven by Polchinski \cite{Polchinski:1985zf}.}
\item{It holds also for toroidal compactifications of the flat bosonic string. As an example, we will prove such a statement in the next chapter.}
\item{In \cite{Maldacena:2000kv} this equality was used in $AdS_3$ to identify the string spectrum with the proposed spectrum obtained from harmonic analysis \cite{Maldacena:2000hw}.}
\item{The authors of \cite{Ferrer:1990na} prove that the analogous statement holds for open strings on the cylinder worldsheet in a constant background electromagnetic field.}
\item{For flat space superstrings and heterotic strings, such a statement also holds \cite{Alvarez:1986sj} but one needs to be careful in the interplay between the bosonic boundary conditions and the fermionic boundary conditions (spin structure), i.e. the GSO projection.}
\end{itemize}

\subsection{Deriving the thermal scalar}
\label{deriv}

We will now explore the thermodynamics of closed strings at the one loop level (genus 1). \\

\noindent We must first be a bit more precise on the relation (\ref{cruc}). The modular integration of the torus amplitude is chosen to be the entire strip \cite{Polchinski:1985zf} and we restrict the Euclidean time coordinate to winding around only one torus cycle. In a second stage, in flat space, one can use the theorem by \cite{McClain:1986id}\cite{O'Brien:1987pn} to relate this to a modular integral over the fundamental domain, while replacing the zero-mode sum over a single quantum number by a double sum over both momenta and winding. For superstrings, an extension of this theorem needs to be used. The logic is basically the same: if we start with the fundamental domain, we restrict the double sum to a single sum and extend the modular domain to the entire strip (see e.g. \cite{Kutasov:2000jp} for an example of such a procedure). 
In what follows we view the strip domain as the relevant one for thermodynamics and restrict the Euclidean time coordinate to a single torus cycle. For now, we also assume that no other coordinates are compactified. We will extend this assumption in the next chapter.\\

\noindent We are interested in the dominant contribution and so we restrict ourselves to winding $\pm 1$ around the Euclidean time direction. At least in spaces where the thermal circle is topologically stable, we expect strings that are wrapped multiple times to be more massive. Indeed, in the flat space string spectrum, the winding $\pm 1$ mode becomes massless at the Hagedorn temperature \cite{Atick:1988si} and strings with higher winding numbers are massive. The zero-winding modes correspond to the zero-temperature vacuum energy and we are not interested in this here. \\
This intuition is well-founded for spaces with topologically stable thermal circles, but what about other spaces? We know Euclidean black hole backgrounds are cigar-shaped and the thermal circle shrinks to zero size at the horizon. Is there still a dominating winding mode present? 
For now we will \emph{assume} that indeed winding $\pm 1$ modes are dominant and we focus on them. We present examples of these phenomena in the remainder of this work.\\

\begin{figure}[h]
\centerline{\includegraphics[width=0.8\textwidth]{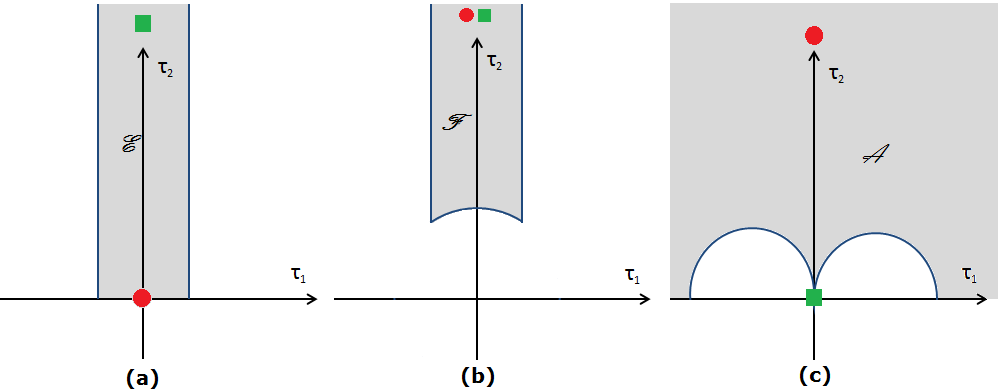}}
\caption{A circle represents the winding tachyon divergence. A square represents the closed string tachyon divergence (or the massless states for superstrings). (a) Free energy evaluated in the strip. (b) Free energy as a partition function of a single string on the thermal manifold. The equivalence with (a) follows from the theorem in \cite{McClain:1986id}\cite{O'Brien:1987pn}. (c) Free energy after the modular transformation $\tau \to -\frac{1}{\tau}$ on (a).}
\label{fund}
\end{figure}

\noindent We start from the following torus path integral in an external field $G_{\mu\nu}$ in $d$ spacetime dimensions: 
\begin{equation} 
Z_{T_2} = \int_0^\infty \frac{d\tau_2}{2\tau_2} \int_{-1/2}^{1/2} d\tau_1 \Delta_{FP} \int \left[\mathcal{D}X\right]\sqrt{G}
\exp -\frac{1}{4\pi\alpha'} \int d^2\sigma \sqrt h h^{\alpha \beta} \partial_\alpha X^\mu \partial_\beta X^\nu G_{\mu\nu}(X).
\end{equation}
where $\Delta_{FP}$ denotes the Faddeev-Popov determinant from the (Diff $\times$ Weyl) gauge-fixing procedure. We choose Euclidean signature for both the worldsheet and the target space manifold. The worldsheet metric has been fixed to
\begin{equation}
h_{\alpha\beta} = \left[\begin{array}{cc} 
1 & \tau_1  \\
\tau_1 & \tau_1^2+\tau_2^2 \end{array}\right],
\end{equation}
and the torus worldsheet is represented by a square with sides equal to 1.
We consider strings that are singly wound around the Euclidean time direction:
\begin{align}
X^\mu(\sigma_1+1,\sigma_2) & =  X^\mu(\sigma_1,\sigma_2), \quad \mu = 0\hdots d-1,\nonumber \\
X^i(\sigma_1,\sigma_2+1) & =  X^i(\sigma_1,\sigma_2), \quad i=1 \hdots d-1,\nonumber \\
X^0(\sigma_1,\sigma_2+1) & =  X^0(\sigma_1,\sigma_2) \pm \beta.
\end{align}
where the wrapping is around the temporal worldsheet coordinate.
The Hagedorn divergence is due to the $\tau_2 \rightarrow 0$ (ultraviolet) behavior of the torus path integral (as shown in figure \ref{fund}(a)). 
The essential idea of the thermal scalar is that this ultraviolet divergence for $\tau_2 \rightarrow 0$ can be described through a UV/IR connection as an infrared divergence for $\tau_2 \rightarrow \infty$. 
In standard approaches \cite{Atick:1988si}, one uses the link between the strip and the fundamental domain to make this correspondence. The divergence appears there as a string state that becomes massless at precisely the Hagedorn temperature (see figure \ref{fund}(b)). However, we follow \cite{Kruczenski:2005pj} and instead use a modular transformation on the strip domain (figure \ref{fund}(c)). This will enable us to deal with the integral over $\tau_1$ later on.
So, qualitatively, the ultraviolet divergence becomes an infrared divergence (due to the massless thermal scalar). More precisely, we perform the transformation $\tau \to -\frac{1}{\tau}$ in the modular integral and then swap the roles of $\sigma_1$ and $\sigma_2$. As a consequence, the string wraps the thermal circle along the spatial worldsheet coordinate $\sigma_1$: 
\begin{align}
X^\mu(\sigma_1,\sigma_2+1) & = X^\mu(\sigma_1,\sigma_2), \quad \mu = 0\hdots d-1,\nonumber\\
X^i(\sigma_1+1,\sigma_2) & = X^i(\sigma_1,\sigma_2), \quad i =1\hdots d-1,\nonumber\\
X^0(\sigma_1+1,\sigma_2) & = X^0(\sigma_1,\sigma_2) \pm \beta.
\end{align}

\noindent In figure \ref{fund}(a), there is a UV divergence caused by the exponentially growing string density. This corresponds to long string dominance. For bosonic strings, after using the theorem from \cite{McClain:1986id}\cite{O'Brien:1987pn}, we get the picture of figure \ref{fund}(b). Here we interpret this divergence as the tachyonic character of the singly wound string. To extract solely the winding tachyon contribution, we use a modular transformation in the strip (a). This displaces the winding tachyon divergence to $\tau_2 \to \infty$ (figure \ref{fund}(c)).\footnote{Strictly speaking, this is only valid for $\tau_1=0$. Nevertheless, we will obtain the expected results.} \\

\noindent Let us now consider this dual (in the modular sense $\tau \rightarrow -\frac{1}{\tau}$) path integral for $\tau_2 \rightarrow \infty$. 
We can use the reparametrization invariance of the path integral to define new worldsheet coordinates $(\sigma,\tau)$:
\begin{equation}
\left\{ \begin{array}{l} \sigma= \frac{\sigma_1}{\tau_2}, \tau = \sigma_2 \\ X(\sigma_1,\sigma_2) \rightarrow X(\sigma,\tau). \end{array} \right.
\end{equation}
The worldsheet action now becomes using the torus metric: 
\begin{eqnarray}
\label{action11}
S = \frac{1}{4\pi \alpha'} \left[ \left(1 + \frac{\tau_1^2}{\tau_2^2}\right) \int_0^{1/\tau_2} d\sigma \int_0^1 d\tau G_{\mu\nu} \partial_\sigma X^\mu \partial_\sigma X^\nu \right. \nonumber \\
\left.
+ 2 \frac{\tau_1}{\tau_2} \int_0^{1/\tau_2} d\sigma \int_0^1 d\tau G_{\mu\nu} \partial_\sigma X^\mu \partial_\tau X^\nu 
+ \int_0^{1/\tau_2} d\sigma \int_0^1 d\tau G_{\mu\nu} \partial_\tau X^\mu \partial_\tau X^\nu \right].
\end{eqnarray} 
We next consider a Fourier series expansion in the $\sigma$ worldsheet coordinate. 
\begin{align}
X^i(\sigma,\tau) & = \sum_{n=-\infty}^{+\infty} e^{i(2\pi n \tau_2) \sigma} X_n^i(\tau), \\
X^0(\sigma,\tau) & = \pm \beta \tau_2 \sigma +  \sum_{n=-\infty}^{+\infty} e^{i(2\pi n \tau_2) \sigma} X_n^0(\tau).
\end{align}
In the $\tau_2 \rightarrow \infty$ limit, only the $n=0$ mode survives ($\langle (X_n^\mu)^2\rangle \sim 1/\tau_2$ for $n\neq 0$) and we get a dimensional reduction from a two dimensional non-linear $\sigma$-model on the worldsheet to quantum mechanics on the worldline and the string theory reduces to a particle theory (of the thermal scalar). Physically this corresponds to neglecting the temporal worldsheet dependence of the string which we started with.\\
Defining 
\begin{equation}
X_0^i(\tau) = X^i (\tau), \quad X_0^0(\tau) = X^0 (\tau),
\end{equation}
the particle action becomes: 
\begin{eqnarray} 
S_{part} =  \frac{1}{4\pi \alpha' \tau_2} \left[ \beta^2(\tau_1^2 + \tau_2^2)  \int_0^1 d\tau G_{00} \pm 2 \tau_1 \beta \int_0^1 d\tau G_{00} \partial_\tau X^0 \right. \nonumber \\
\qquad \qquad\qquad\qquad
\left.\pm 2 \tau_1 \beta \int_0^1 d\tau G_{0i} \partial_\tau X^i +
\int_0^1 d\tau G_{\mu\nu} \partial_\tau X^\mu \partial_\tau X^\nu \right]
\end{eqnarray}
for winding number $\pm1$. 

\noindent Of course, in dimensional reduction, one always loses information about the high energy degrees of freedom on the worldsheet, which could be important. The situation is similar to dimensional reduction in high temperature field theory where one loses the perturbative Stefan-Boltzmann result coming from the high energy degrees of freedom \cite{Braaten:1995jr}\cite{Braaten:1995cm}. In a sense, one can view $\tau_2$ as a `spatial' worldsheet temperature.
The dimensional reduction has replaced a string path integral with a particle path integral. 
So the lost degrees of freedom are the orthogonal oscillations of the string. \\

\noindent From a more technical perspective, what is neglected is the determinant of quadratic fluctuations. These give corrections to the classical part. \\

\noindent In the case of bosonic strings in flat space ($G_{\mu\nu} = \delta_{\mu\nu}$), the orthogonal oscillations (including the Faddeev-Popov factor $\Delta_F \tau_2^{-1}$) give a factor: 
\begin{equation}
 |\eta(\tau)|^{-48} \exp -\frac{\pi R^2}{\tau_2 \alpha'} |n\tau - m|^2 ,
\end{equation}
for one compact dimension with radius $R$. The symbol $\eta$ denotes the Dedekind $\eta$-function.
Using the modular transformation and taking the limit $\tau_2 \to \infty$ yields
\begin{equation}
e^{4\pi \tau_2} e^{-\frac{\beta^2}{4\pi \alpha' \tau_2} (\tau_1^2+\tau_2^2)}.
\end{equation}
Comparing with $S_{part}$, we find that we have to add 
\begin{equation}
\label{bosAdd}
\Delta S = -4\pi \tau_2
= -\frac{\tau_2^2 \beta_{H,\text{flat}}^2}{4\pi \alpha' \tau_2}.
\end{equation}
This correction term was only calculated for flat spacetime. We expect other corrections when we consider a generally curved background. One of the goals of this work is precisely to get a handle on these corrections. We will have more to say about this further on. \\

\begin{figure}[h]
\centering
\includegraphics[width=0.35\textwidth]{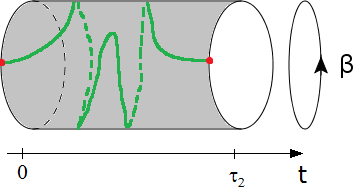}
\caption{An arbitrary integration path of $\tilde{X}_0$ is shown. The red dots are starting and ending point and these are identified. Intermediate values are allowed to range over the entire real axis to allow for configurations such as the one shown here where the path winds and unwinds again as it traverses the interval $\left[0,\tau_2\right]$.}
\label{cylconfig}
\end{figure}

\noindent Introducing the parameter $t=\tau_2 \tau$ and adding $\Delta S$, the particle action is finally 
\begin{equation}
\label{particleaction}
S_{part} =  \frac{1}{4\pi \alpha' } \left[ \beta^2\frac{\left|\tau\right|^2}{\tau_2^2} \int_0^{\tau_2} dt G_{00} - \beta_{H,\text{flat}}^2 \tau_2 \pm 2 \frac{\tau_1}{\tau_2} \beta \int_0^{\tau_2} dt G_{0\mu} \partial_t X^\mu 
+\int_0^{\tau_2} dt G_{\mu\nu} \partial_t X^\mu \partial_t X^\nu \right].
\end{equation}
If the metric is time independent, whatever the corrections are, the $X^0$ integration is Gaussian in this case and can be integrated out exactly. 
From now on we set $G_{0i}=0$. For the Gaussian $X^0$ integration ($G_{00}$ is independent of $X^0$) we first solve the classical equation: 
\begin{equation}
\partial_t \left[ \left( G_{00}(\vec X) \partial_t X^{0,cl}(t) \right) \pm \frac{\tau_1}{\tau_2} \beta G_{00}(\vec X) \right] = 0
\end{equation} 
or 
\begin{equation} 
G_{00}(\vec X) \partial_t X^{0,cl} \pm \frac{\tau_1}{\tau_2} \beta G_{00}(\vec X) = C.
\end{equation} 
The constant $C$ is determined by periodicity: 
\begin{equation}
\int_0^{\tau_2} \partial_t X^0 dt = 0 
\end{equation}
so that 
\begin{equation}
C = \pm \frac{ \tau_1 \beta}{\left \langle 1/G_{00} \right \rangle},
\end{equation}
where we denoted $\langle A \rangle = \int_{0}^{\tau_2}A dt.$
The classical action (of the $X^{0}$-dependent contributions) is
\begin{equation} 
S\left(X^{0,cl}\right) =  \frac{1}{4\pi \alpha' } \left[ \frac{ \tau_1^2 \beta^2}{\left \langle 1/G_{00} \right \rangle} - \frac{\tau_1^2}{\tau_2^2}\beta^2 \langle G_{00} \rangle \right].
\end{equation}
The second term in the classical action cancels the $\tau_1^2$ term in $S_{part}$.
Therefore, putting $X^0 = X^{cl} + \tilde X^0$, we find: 
\begin{equation} 
S_p =  \frac{1}{4\pi \alpha'} \left[ \frac{ \tau_1^2 \beta^2}{\left \langle 1/G_{00} \right \rangle} + \beta^2 \int_0^{\tau_2} dt G_{00} - \beta_{H,\text{flat}}^2 \tau_2 +  \int_0^{\tau_2} dt G_{00} (\partial_t \tilde X^0)^2 + \int_0^{\tau_2} dt G_{ij} \partial_t X^i \partial_t X^j \right].
\end{equation}
The $\tilde X^0$ path integral is: 
\begin{equation} 
Z_0 = \int \left[ \mathcal{D}\tilde X^0 \right] \prod_t \sqrt{G_{00}(\vec X(t))} \exp - \frac{1}{4\pi \alpha'} \int_0^{\tau_2} dt G_{00}(\vec X) (\partial_t \tilde X^0)^2.
\end{equation}
Using the following identity for one timestep $\epsilon$:
\begin{align} 
\int_{-\infty}^{+\infty} du \sqrt{\frac{a_1}{\pi}} &\exp \left(- \frac{a_1}{\epsilon}(x-u)^2\right) \sqrt{\frac{a_2}{\pi}} \exp \left(- \frac{a_2}{\epsilon}(u-y)^2\right) \nonumber \\
&=
\sqrt \epsilon \sqrt{\left(\frac{a_1 a_2}{a_1+a_2}\right) \frac{1}{\pi}} \exp \left(- \frac{a_1 a_2}{a_1+a_2} \frac{(x-y)^2}{\epsilon}\right),
\end{align}
we find after $n$ intermediate timesteps: 
\begin{equation} 
\left(\sqrt \epsilon\right)^n \sqrt{\left(\frac{1}{\sum_{i=1}^n 1/a_i}\right) \frac{1}{\pi}} 
\exp \left(- \frac{1}{\sum_{i=1}^n 1/a_i} \frac{(x-y)^2}{\epsilon}\right).
\end{equation}
The path $\tilde{X}_0(t)$ is periodic whereby each intermediate value of the path takes values in $\mathbb{R}$. The final endpoints however only take values in an interval of size $\beta$. Else one would overcount several configurations obtained by rigid translation of multiples of $\beta$ along the circle. This is shown in figure \ref{cylconfig}.
\noindent Using 
\begin{equation}
a_i =\frac{1}{4\pi \alpha'} G_{00}(t_i),
\end{equation}
and performing the final integration over $\tilde X_0(0) =\tilde X_0(\tau_2)$ which gives a factor $\beta$, we find 
\begin{equation} 
Z_0 = N \beta \sqrt{\frac{1}{\langle G_{00}^{-1}\rangle}},
\end{equation}
where $N$ is a normalization factor given by\footnote{By carefully taking the measure into account, we can determine the normalization factor $N$. We can determine this constant equally easy from the flat space limit (since the constant does not depend on the background) where the path integral is given by
\begin{equation}
Z_0 = \int \left[dX_0\right]\exp{-\frac{1}{4\pi\alpha'}}\int_{0}^{\tau_2}dt (\partial_t X_0)^{2} = \frac{\beta}{\sqrt{4\pi^2\alpha'\tau_2}}.
\end{equation}
Comparing the results immediately gives the above value for $N$.
}
\begin{equation}
N = \sqrt{\frac{1}{4\pi^2\alpha'}}.
\end{equation}
Note that the appearance of $\prod_t \sqrt{G_{00}}$ in the measure is essential for this simple result. \\

\noindent The $\tau_1$ integration can also be performed exactly for $\tau_2 \rightarrow \infty$ because the dual domain of integration for $\tau_1$ is $]-\infty,+\infty[$. 
This gives the factor 
\begin{equation} 
\frac{2\pi \sqrt{\alpha'} \sqrt{\langle G_{00}^{-1}\rangle}}{\beta}.
\end{equation}
Putting everything together, the $\langle G_{00}^{-1}\rangle$ factors cancel and we finally obtain the following particle path integral after integrating out $X_0$: 
\begin{equation} 
Z_p = 2\int_0^\infty \frac{d \tau_2}{2\tau_2} \int \left[ \mathcal{D}X \right] \sqrt{\prod_{t} \det G_{ij}} \exp - S_p(X) 
\end{equation}
where 
\begin{equation}
S_p = \frac{1}{4\pi \alpha'}\left[ \beta^2 \int_0^{\tau_2} dt G_{00} - \beta_{H,\text{flat}}^2 \tau_2 
+\int_0^{\tau_2} dt G_{ij} \partial_t X^i \partial_t X^j\right].
\end{equation}
The prefactor $2$ is the result of summing over both windings $\pm 1$, since these give equal contributions.
Note that we do not find the path integral measure $\sqrt{G_{00}  \det G_{ij}}$ in contrast to the authors of \cite{Kruczenski:2005pj}. \\

\noindent In all, we have reduced the full string partition function to a partition function for a non-relativistic particle moving in the purely spatial curved background. Because of the swapping of worldsheet coordinates, the time evolution of the particle in its random walk is equal to the spatial form of the long highly excited string. The long string in real spacetime has a shape described by the above random walk.\\

\noindent Before moving on, let us briefly recapitulate several of the delicate points in the previous derivation:
\begin{itemize}
\item{We assumed that the free energy of a string gas can be described by a path integral on the thermal manifold over the modular strip. As we already discussed, this is a very plausible statement. But, as far as we know, it has not been proven in general.}

\item{This picture provides an explicit realization of the correspondence between the long highly excited string and the dominant behavior of the canonical (single-string) partition function. The Euclidean time coordinate has been integrated out to explicity show the random walk behavior in the spatial submanifold. If we are interested in the configurations of the entire string gas, we should look at the multi-string partition function which is given by
\begin{equation}
Z_{\text{mult}} = e^{-\beta F}
\end{equation}
and it is clear that this has contributions from multiple random walks. Whether this is dominated by single random walks then requires an analysis analogous to \cite{Deo:1989bv} that was done in the previous chapter.}

\item{Note that since the final result does not depend on $X^{0}$ anymore, the Wick rotation (naively) appears trivial in this case. To put it another way, we end up with a random walk in the spatial submanifold and it appears irrelevant whether we view this as the spatial submanifold of the Lorentzian or Euclidean background. There is however one important influence of this Wick rotation: if one chooses a black hole background with $G_{00} \to 0$ as $r \to r_{H}$ (for instance a Schwarzschild black hole in Schwarzschild coordinates), the radial coordinate only lives outside the event horizon. When viewing this random walk as a walk in Lorentzian signature spacetime, this means the random walk cannot intersect the horizon and go into the interior of the black hole. This makes sense, since we are discussing thermodynamics as observed by fiducial observers and spacetime effectively ends at the horizon for these observers (cfr. the membrane paradigm \cite{Thorne:1986iy}). The concrete application to random walks near black hole horizons is discussed in \cite{Mertens:2013zya} and will be considered in chapter \ref{chri}.}

\item{We stated that the $\tau_2 \to$ 0 limit corresponding to the UV divergence, is mapped to the $\tau_2 \to \infty$ limit as a IR divergence. This only holds for $\tau_1 = 0$. We did not consider the effect of other values of $\tau_1$ on the derivation. This is a subtle point, but we will find further on that nevertheless we reproduce expected results whenever we can compare with other results.}

\item{In the worldsheet dimensional reduction, we ignored all higher order terms (we set them to zero) because they were not dominant in the large $\tau_2$ limit. A good approach would be to integrate all of these out and get corrections for the lowest Fourier mode. This would then give us the correction term that we missed. We previously found this by comparing our result with the known exact result. In what follows we will compare the previous result with several exact results from type II and heterotic superstrings and we will find again the same type of correction. We will also exactly integrate out all higher Fourier modes in flat spacetime and indeed see that these provide us with the missed correction term in the action.}

\item{There is a divergence for $\tau_2 \to 0$ in this result. This was not present at first and was introduced `by hand' when we chose the lower boundary for the $\tau_2$ integral. In principle, we should \emph{not} take $0$ as the lower limit, since we assumed that $\tau_2$ was large. We chose this to find agreement with the field theory picture (see further) that indeed has lower value $0$. This is actually the same reasoning as discussed in \cite{Polchinski:1998rq} (chapter 7): if we rewrite the path integral to explicitly display all string contributions and then drop all but one of these states, we would get the same result as a field theory first-quantized vacuum loop (provided the integral boundaries are kept the same).
This divergence is not a consequence of a tachyon, but instead follows from the fact that we only retain one string field. The divergence that sets in is the usual field theory UV divergence. By dropping all other string excitations, we have lost the finiteness of the amplitudes. We are however only interested here in the $\tau_2 \to +\infty$ limit so this will not bother us. Also note that in the bosonic string case, the full string amplitude has a tachyon in the small $\tau_2$ limit, but when doing the manipulations in the large $\tau_2$ limit, we lose this tachyon divergence but instead get the field theory divergence.}
\end{itemize}

\subsection{Some extensions}
\label{exten}
Let us now make two extensions to the method discussed above: a background Kalb-Ramond field and general stationary spacetimes (so $G_{0i} \neq 0$).
\subsubsection*{Adding a Kalb-Ramond background}
We extend the previous result to contain a background Kalb-Ramond field. The starting point is the following addition to the string action
\begin{equation}
S_{extra} = \frac{1}{4\pi\alpha'}\int d^{2}\sigma \sqrt{h}i \epsilon^{ab}B_{\mu\nu}(X)\partial_{a}X^{\mu}\partial_{b}X^{\nu}.
\end{equation}
Since $\sqrt{h}\epsilon^{12} = 1$, the modular transformation has no effect on this term except the swapping of the roles of $\sigma_1$ and $\sigma_2$. We obtain
\begin{equation}
S_{extra} = -\frac{1}{2\pi\alpha'}\int d\sigma_1 d\sigma_2 i B_{\mu\nu}(X)\partial_{1}X^{\mu}\partial_{2}X^{\nu}.
\end{equation}
Doing the substitution to $\sigma$ and $\tau$ as before yields
\begin{equation}
\label{Bfield}
S_{extra} = -\frac{1}{2\pi\alpha'}\int_{0}^{1/\tau_2} d\sigma \int_{0}^{1}d\tau i B_{\mu\nu}(X)\partial_{\sigma}X^{\mu}\partial_{\tau}X^{\nu}.
\end{equation}
The next step is the Fourier expansion. There is only one non-zero possibility here and it finally gives the following addition to the particle action
\begin{equation}
S_{extra} = \mp i\frac{\beta}{2\pi\alpha'}\int_{0}^{\tau_2}dt B_{0i}(X)\partial_{t}X^{i}.
\end{equation}
This can be interpreted as the minimal coupling of a non-relativistic particle (with a suitably normalized charge) to a vector potential $A_{i} = B_{0i}$. So the resulting particle that lives in one dimension less than the original string, is also minimally coupled to an electromagnetic field. \\
Note that such a term is no longer symmetric under the positive and negative windings (as is obvious from the orientation reversal symmetry breaking property of the Kalb-Ramond field). In the particle path integral, the interpretation is that the particles are oppositely charged under the electromagnetic field.

\subsubsection*{Non-static spacetimes}
\label{nonstatt}
One can readily extend the previous calculation to the case where $G_{0i} \neq 0$. So we consider stationary but non-static spacetimes (e.g. a string-corrected version of the Kerr black hole). We present the derivation itself as a supplementary section \ref{AppAA}. One arrives at
\begin{equation} 
Z_p = 2\int_0^\infty \frac{d \tau_2}{2\tau_2} \int \left[ d\vec X \right] \sqrt{\prod \det \left(G_{ij} - \frac{G_{0i}G_{0j}}{G_{00}}\right)} \exp - S_p(\vec X) 
\end{equation}
where 
\begin{equation}
S_p = \frac{1}{4\pi \alpha'}\left[ \beta^2 \int_0^{\tau_2} dt G_{00} - \beta_{H,\text{flat}}^2 \tau_2 
+\int_0^{\tau_2} dt \left(G_{ij} - \frac{G_{0i}G_{0j}}{G_{00}}\right) \partial_t X^i \partial_t X^j\right].
\end{equation}
Thus the only modification is a change in the spatial metric $G_{ij} \to G_{ij} - \frac{G_{0i}G_{0j}}{G_{00}}$. We will find this action again in section \ref{dimred} using dimensional reduction and T-duality.

\section{Comparison with exact results}
\label{compa} 
In the previous section we saw that for bosonic strings we needed to include a correction term in the action proportional to the Hagedorn temperature. We now look into the other types of strings. Just like we did for the bosonic string in the previous section, our strategy will be to compare the path integral result with the known exact expression for the free energy. 

\subsection{Flat space superstring}
\label{supers}
For the superstring, one might be initially worried about the effects of the worldsheet fermions on the previous derivation and especially their interplay with the worldsheet bosons (the GSO projection). In \cite{Alvarez:1986sj} it is shown that in thermal path integrals with the strip as the modular integration domain, one must choose only one spin structure for the fermions. 
We thus expect that the fermionic contribution will only be a modification of the Hagedorn temperature (at least for flat space). We now show that this is indeed the case.
The free energy for superstrings is given by the following expression \cite{Alvarez:1986sj}\footnote{One readily checks that this is the same expression as that obtained in \cite{Atick:1988si} when one restricts the expression for the free energy from \cite{Atick:1988si} to a single sum and extends the domain of integration to the entire strip.}
\begin{equation}
F = -2 V_9\int_{0}^{+\infty}d\tau_2\int_{-1/2}^{1/2}\frac{d\tau_1}{\tau_2^{6}(2\pi^2\alpha')^5}\left[\vartheta_3\left(0,\frac{i\beta^2}{4\pi^2\alpha'\tau_2}\right)-\vartheta_4\left(0,\frac{i\beta^2}{4\pi^2\alpha'\tau_2}\right)\right]\left|\vartheta_4(0,2\tau)\right|^{-16}.
\end{equation}
In supplementary section \ref{appsuper} we prove that in the limit of large $\tau_2$ (after the modular transformation) this is equal to
\begin{equation}
\label{freeEnsuper}
F = -2 V_9\iint_{\mathcal{A}} \frac{d\tau_2d\tau_1}{2\tau_2}\frac{1}{(4\pi^2\alpha'\tau_2)^5}e^{-\frac{\beta^2}{4\pi\alpha'}\frac{\tau_1^2+\tau_2^2}{\tau_2}}e^{2\pi\tau_2}.
\end{equation}
where the integration region $\mathcal{A}$ was shown in figure \ref{fund}(c).\\
The path integral from the previous section (\emph{without} a correction) is given by 
\begin{equation}
Z = 2\iint_{\mathcal{A}}\frac{d\tau_1d\tau_2}{2\tau_2}\int\left[\mathcal{D}X\right]e^{-\frac{\tau_1^2\beta^2}{4\pi\alpha'\tau_2}}e^{-\frac{\beta^2\tau_2}{4\pi\alpha'}}e^{-\frac{1}{4\pi\alpha'}\int_{0}^{\tau_2}dt\left(\partial_tX^\mu\right)^2}.
\end{equation}
Note that each flat space integral is given by
\begin{equation}
\int \left[\mathcal{D}X\right]e^{-\frac{1}{4\pi\alpha'}\int_{0}^{\tau_2}dt\left(\partial_tX^\mu\right)^2} = \frac{L}{\sqrt{4\pi^2\alpha'\tau_2}},
\end{equation}
where $L$ is the length of the space. So the path integration over the free coordinate fields reproduces precisely the $\frac{\beta V_9}{(4\pi^2\alpha'\tau_2)^5}$ factor and the factor $2$ in the beginning is the result of the sum over both windings. Considering finally the relation $Z = -\beta F$, we conclude that this is exactly the same as equation (\ref{freeEnsuper}) if we include a factor $e^{2\pi\tau_2}$. This corresponds to the superstring Hagedorn temperature $\beta_{H} = \pi \sqrt{8\alpha'}$. \\

So we learn that also the flat spacetime superstring gets the same type of correction as the flat spacetime bosonic string.

\subsection{Flat space heterotic string}
\label{heteroticz}
We now look into heterotic strings in flat spacetime. For concreteness we focus on the $E_8 \times E_8$ string but the result is more general (see remarks further). The general expression for the free energy of the heterotic string is given by \cite{Alvarez:1986sj}
\begin{equation}
F = -2 V_9\int_{0}^{+\infty}d\tau_2\int_{-1/2}^{1/2}\frac{d\tau_1}{16\tau_2^{6}(2\pi^2\alpha')^5}\left[\vartheta_3\left(0,\frac{i\tilde{\beta}^2}{\tau_2}\right)-\vartheta_4\left(0,\frac{i\tilde{\beta}^2}{\tau_2}\right)\right]\frac{\Theta_{E_8 \oplus E_8}(-\overline{\tau})}{\vartheta_4(0,2\tau)^{8}\eta(-\overline{\tau})^{24}}.
\end{equation}
where we denoted $\tilde{\beta}^2 = \frac{\beta^2}{4\pi^2\alpha'}$. \\
The large $\tau_2$ limit is again deferred to section \ref{appheterotic}. The result is
\begin{equation}
F = -2 V_9\iint_{\mathcal{A}} \frac{d\tau_2d\tau_1 }{2\tau_2(4\pi^2\alpha'\tau_2)^5}e^{-\frac{\beta^2}{4\pi\alpha'}\frac{\tau_1^2+\tau_2^2}{\tau_2}}e^{\pi i \tau_1} e^{3\pi \tau_2}.
\end{equation}
We clearly see that the correction is given by $e^{\pi i \tau_1} e^{3\pi \tau_2}$. The second factor looks like a Hagedorn-like correction, but there is an extra $e^{\pi i \tau_1}$ factor. Intuitively, the heterotic string is left-right asymmetric, so it is not totally unexpected to find such a factor. Factors of this type are also seen in other limits \cite{Alvarez:1986sj}, where they lead to (unexpected) infrared convergence. \\
If we would ignore this $\tau_1$ correction, we would have the prediction that the Hagedorn temperature is equal to 
\begin{equation}
\beta_H = \sqrt{12}\pi\sqrt{\alpha'}
\end{equation}
which is the value predicted by \cite{Alvarez:1986sj}. However, this value is inconsistent with other values in the literature \cite{Gross:1985fr} and with the expected `thermal duality' of the heterotic string \cite{O'Brien:1987pn}. \\
If we proceed correctly and integrate over $\tau_1$, we get
\begin{equation}
\int_{-\infty}^{+\infty}d\tau_1\exp\left(-\frac{\beta^2}{4\pi\alpha'\tau_2}\tau_1^2 + \pi i \tau_1\right) = \sqrt{\frac{4\pi^2\alpha'\tau_2}{\beta^2}}\exp\left(-\frac{\pi^3\alpha'}{\beta^2}\tau_2\right).
\end{equation}
We see that we get a correction to the Hagedorn temperature from this integration. \\
To ensure convergence in the large $\tau_2$ limit, we should have
\begin{equation}
\frac{\beta^2}{4\pi\alpha'} + \frac{\pi^3\alpha'}{\beta^2} \geq 3\pi. 
\end{equation}
At first sight, it seems rather strange to have a $\beta$ factor in the denominator, but we now argue that indeed this must be the case.\\
Solving the previous equation gives \emph{two} critical temperatures
\begin{eqnarray}
\beta_{H1} = (2+\sqrt{2})\pi\sqrt{\alpha'},\quad T_{H1} = \frac{1}{\pi\sqrt{\alpha'}}\left(1-\frac{1}{\sqrt{2}}\right), \\
\beta_{H2} = (2-\sqrt{2})\pi\sqrt{\alpha'},\quad T_{H2} = \frac{1}{\pi\sqrt{\alpha'}}\left(1+\frac{1}{\sqrt{2}}\right).
\end{eqnarray}
The partition function converges for $T \leq T_{H1}$ or $T \geq T_{H2}$. The first temperature is the physical one we encounter in heating up a gas of strings. It is necessary that we have two solutions since the heterotic string has a duality symmetry \cite{O'Brien:1987pn} under 
\begin{equation}
\beta \leftrightarrow \frac{2\pi^2\alpha'}{\beta},
\end{equation}
and the previous convergence condition indeed has this symmetry.

\section{Flat spacetime corrections}
\label{corrections}
In the previous sections, we retained only the lowest Fourier mode in the path integral. We saw that we missed certain contributions and we corrected for them by comparing with the known results in flat spacetime. These correction terms obviously have to correspond to the higher Fourier modes that we neglected. In this section we explicitly show that this is indeed the case for strings in flat spacetime. We will determine the exact correction term in the particle path integral result starting from the string path integral alone and by relating the latter to those for particles in uniform magnetic fields. One can presumably also obtain these correction terms by using a double Fourier mode decomposition and regulating several infinite products via zeta-regularization \cite{Alvarez:1985fw}\cite{Alvarez:1986sj}. \\

\noindent The starting point is the string action after the modular transformation
\begin{eqnarray}
S = \frac{1}{4\pi\alpha'}\left[\left(1+\frac{\tau_1^2}{\tau_2^2}\right)\int_{0}^{1/\tau_2}d\sigma\int_{0}^{1}d\tau G_{\mu\nu}\partial_{\sigma}X^{\mu}\partial_{\sigma}X^{\nu}\right. \nonumber \\
\left. + 2\frac{\tau_1}{\tau_2}\int_{0}^{1/\tau_2}d\sigma\int_{0}^{1}d\tau G_{\mu\nu}\partial_{\sigma}X^{\mu}\partial_{\tau}X^{\nu}+\int_{0}^{1/\tau_2}d\sigma\int_{0}^{1}d\tau G_{\mu\nu}\partial_{\tau}X^{\mu}\partial_{\tau}X^{\nu}\right].
\end{eqnarray}
In the previous derivation we only kept the lowest worldsheet mode in $\tau$. The expansion of the target coordinate fields is given by
\begin{eqnarray}
X^{i}(\sigma,\tau) =& \sum_{n=-\infty}^{+\infty}e^{2\pi i n \tau_2 \sigma}X^{i}_{n}(\tau), \quad i=1\hdots D-1,\nonumber\\
X^{0}(\sigma,\tau) =& \pm \beta \tau_2 \sigma + \sum_{n=-\infty}^{+\infty}e^{2\pi i n \tau_2 \sigma}X^{0}_{n}(\tau).
\end{eqnarray}
If we keep all of these terms and plug it into the action, we get a total action (in flat space) given by
\begin{equation}
S = S_{0} + \sum_{n=1}^{+\infty}S_{n},
\end{equation}
where the $S_{n}$ are particle actions that combine the modes $\pm n$ and are given by\footnote{A sum over $\mu$ is implied; the metric is flat Euclidean space here.}
\begin{equation}
S_{n} = \frac{1}{4\pi\alpha'}\int_{0}^{1}\frac{d\tau}{\tau_2}\left[2\dot{X}^{\mu}_{n}\dot{X}^{\mu}_{-n} + 4\pi i \tau_1n\left(X^{\mu}_{n}\dot{X}^{\mu}_{-n}-\dot{X}^{\mu}_{n}X^{\mu}_{-n}\right) + 8\pi^2n^2\left(\tau_1^2+\tau_2^2\right)X^{\mu}_{n}X^{\mu}_{-n}\right].
\end{equation}
To see this, note that the integral over $\sigma$ is only non-zero if the two contributing factors have opposite $n$.
The reality of the target fields requires that $X^{\mu}_{-n} = X^{\mu*}_{n}$. Setting
\begin{equation}
X^{\mu}_{n} = A_{n} + i B_{n}, \quad X^{\mu*}_{n} = A_{n} - i B_{n}
\end{equation}
for two real scalar fields $A_{n}$ and $B_{n}$, gives
\begin{align}
S_{n} = \frac{1}{4\pi\alpha'}\int_{0}^{1}\frac{d\tau}{\tau_2}\left[2\dot{A}_{n}\dot{A}_{n} + 2\dot{B}_{n}\dot{B}_{n} + 8\pi\tau_1n\left(A_{n}\dot{B}_{n}-\dot{A}_{n}B_{n}\right) \right. \nonumber \\
\left. + 8\pi^2n^2\left(\tau_1^2+\tau_2^2\right)\left(A_{n}^{2}+B_{n}^2\right)\right]
\end{align}
for one target space component field $X^{\mu}$. 
These fields, that only depend on time, need to be path-integrated over the entire two-dimensional plane (in field space). 
We notice that:
\begin{enumerate}
\item{the corrections are independent of the $n=0$ contribution, so they will generate a term independent of the $X_{0}^{i}$ and $X_{0}^{0}$ fields,}
\item{in the end we will multiply this action by 26 (or 10 for the superstring) since each target space coordinate $X^{\mu}$ yields the same action,}
\item{the actions corresponding to different $n$ are also decoupled and of identical form, so if we can compute the path integral for one of them, we can compute all of them.}
\end{enumerate}
The way to solve the path integral corresponding to $S_n$ is to reinterpret it as a path integral for a 2D particle moving in a harmonic oscillator potential and interacting with a (imaginary) magnetic field. Exact results are known for such systems \cite{Khandekar:1986ib}\cite{Jones:1971kk}. 
We rename $A_{n} = x$ and $B_{n} = y$. \\
The Hamiltonian corresponding to the Lagrangian above is given by (see e.g. \cite{Gao-Feng:2008})
\begin{equation}
H = \frac{1}{2m}\left[\left(p_{x}+\frac{qBy}{2}\right)^2 + \left(p_{y}-\frac{qBx}{2}\right)^2\right] + \frac{1}{2}m\omega^2\left(x^2+y^2\right)
\end{equation}
where 
\begin{equation}
m = \frac{1}{\pi \alpha' \tau_2}, \quad \frac{qB}{2} = i\frac{2\tau_1 n}{\alpha' \tau_2}, \quad \omega = \sqrt{4\pi^2n^2\left(\tau_1^2+\tau_2^2\right)}.
\end{equation}
The eigenstates of this Hamiltonian are known exactly:
\begin{equation}
\psi_{N,m_{\ell}}(x,y) \propto \rho^{\left|m_{\ell}\right|}F\left(-N,\left|m_{\ell}\right|+1,\gamma^2\rho^2\right)e^{-\gamma^2\rho^2/2}e^{im_{\ell}\phi}
\end{equation}
where $N = 0,1,2, \hdots$ and $m_{\ell} = 0, \pm1, \pm2, \hdots$ and $\gamma^2 = m\omega_0 = \frac{2n}{\alpha'}$.\footnote{In writing this expression, we have corrected a typo in \cite{Gao-Feng:2008} and \cite{Mertens:2013pza}.} $F$ is the confluent hypergeometric function and $\rho^2 = x^2+y^2$.
The energies are given by
\begin{equation}
\label{geomsum}
E_{N,m_{\ell}} = \omega_{0}\left(2N+\left|m_{\ell}\right|+1\right)+m_{\ell}\frac{qB}{2m}, \quad \omega_0 = 2\pi n \tau_2.
\end{equation} 
In a general quantum system the heat kernel (path integral of the Euclidean action corresponding to this classical Hamiltonian) is given by
\begin{equation}
K(\textbf{r},t|\textbf{r},0) = \sum_{a}\left|\psi_{a}(\textbf{r})\right|^2e^{-E_{a}t}.
\end{equation}
Integrating the heat kernel over initial and final coordinate gives for normalized eigenfunctions
\begin{equation}
\iint dxdy K(\boldsymbol{\rho},1|\boldsymbol{\rho},0) = \sum_{a}e^{-E_{a}t} \to e^{-2\pi n \tau_2}.
\end{equation}
where we used the fact that in the large $\tau_2$ limit the values $N = m_{\ell} = 0$ dominate. \\

\noindent As a check, one can obtain the same result from the known heat kernel \cite{Jones:1971kk}. The relevant heat kernel is given by
\begin{eqnarray}
K(\boldsymbol{\rho},1|\boldsymbol{\rho},0) = \frac{m\omega_0}{2\pi \sinh(\omega_0)}\exp -\left\{\frac{m\omega_0}{\sinh(\omega_0)}(x^2+y^2)\left[\cosh(\omega_0)-\cosh\left(\frac{qB}{2m}\right)\right]\right\}.
\end{eqnarray}
As in the previous derivation, we need to integrate this over $x$ and $y$. These integrals are simple Gaussians and we obtain
\begin{equation}
\label{heeat}
\iint dxdy K(\boldsymbol{\rho},1|\boldsymbol{\rho},0)
= \frac{1}{2\left[\cosh(\omega_0)-\cosh\left(\frac{qB}{2m}\right)\right]}.
\end{equation}
Now we take the $\tau_2 \to \infty$ limit. Since $B$ is purely imaginary, the second term in the denominator has modulus bounded by one, and the first one becomes arbitrarily large. In the limit, the second term is dropped and we finally obtain
\begin{equation}
\iint dxdy K(\boldsymbol{\rho},1|\boldsymbol{\rho},0) \to \frac{1}{2\cosh(\omega_0)} \to e^{-\omega_0} = e^{-2\pi n \tau_2},
\end{equation} 
in agreement with the previous result. Also, one readily checks that (\ref{heeat}) agrees with $\sum_{N,m_{\ell}}e^{-E_{N,m_{\ell}}}$ with energies (\ref{geomsum}). \\

\noindent To get all corrections, this has to be taken to the $26^{\text{th}}$ power (the contribution of all target space fields $X^{\mu}$) and we have to take the product of all values of $n$ going from 1 to infinity.
So in all, we get
\begin{equation}
\prod_{j=1}^{26}\prod_{n=1}^{+\infty} e^{-2\pi n \tau_2} = e^{-52\pi \sum_{n=1}^{+\infty}n \tau_2} = e^{52\pi \frac{1}{12} \tau_2}
\end{equation}
where we used $\sum_{n=1}^{+\infty}n = -\frac{1}{12}$. This last step is the analogue of the zeta-regularization used in \cite{Alvarez:1985fw}\cite{Alvarez:1986sj}. \\
The only thing left to do is to include the contribution from the $bc$ ghosts. This is simply the ghost path integral on the torus worldsheet \cite{Polchinski:1998rq} and in the limit $\tau_2 \to \infty$ (after the modular transformation), this gives a factor 
\begin{equation}
e^{-8\pi\tau_2/24}.
\end{equation}
Combining everything finally gives
\begin{equation}
e^{4\pi\tau_2}
\end{equation}
which is precisely the result we got in equation (\ref{bosAdd}) when comparing with the exact result. Note that this derivation is purely from the path integral and never used the quantized string spectrum.\\
We have succeeded in determining the exact correction term for the flat spacetime bosonic string. The question immediately arises whether we can do this also for different backgrounds and for other types of strings. 

\subsection*{Non-trivial background corrections}
As an example of a non-trivial background, consider Rindler spacetime with metric
\begin{equation}
ds^2 = \left(\frac{\rho^2}{\alpha'}\right)d\tau^2 + d\rho^2 + d\mathbf{x}^2_{\perp}.
\end{equation}
In this spacetime we see that $G_{00}$ is quadratic in the field $\rho$, so the corrections are up to fourth order in the fields. Such particle path integrals are in general not exactly calculable. Also, there is a non-trivial mixing between \emph{all} Fourier modes, so we cannot integrate them out one by one. \\
We conclude that the previous tricks (most likely) only work in flat spacetime.

\subsection*{Type II Superstring corrections}

For type II superstrings, we have extra contributions from the worldsheet fermions. The fermions (and superghosts) we need, give in total the following exact contribution to the partition function\footnote{This corresponds to only one type of spin structure on the torus worldsheet. The reason for this is the thermal boundary conditions as discussed in \cite{Alvarez:1986sj}.} \cite{Alvarez:1986sj}
\begin{equation}
2^8\prod_{n=1}^{\infty}(1+q^n)^8(1+\overline{q}^n)^8e^{-4\pi\tau_2/3}
\end{equation}
where $q=\exp(2\pi i \tau)$.
Doing the modular transformation and taking $\tau_2 \to \infty$ limit yields
\begin{equation}
e^{2\pi\tau_2}e^{-\frac{8\pi\tau_2}{6}}.
\end{equation}
The bosons (and the ghosts) give the following contribution
\begin{equation}
e^{\frac{8\pi\tau_2}{6}}.
\end{equation}
We see that we end up with $e^{2\pi\tau_2}$ which is indeed the contribution we identified in section \ref{supers} by comparing with the known flat space result. Mutatis mutandis one can also see that everything works out for the heterotic string.

\section{Alternative approach: the field theory action at one loop}
\label{alternative}
So far, we have obtained a one-loop result that gives the contribution of the winding tachyon\footnote{In what follows, we will call this state (winding number $\pm1$, no discrete momentum and no oscillators) the winding tachyon, even though strictly speaking it is not tachyonic in the regime we are interested in.} to the partition function. All the previous manipulations ensured we only got this contribution and not the oscillators or other quantum numbers. There is however also another way to get the contribution for only this string state, namely the spacetime action. \\
To describe this, let us first take a worldsheet CFT point of view. Consider the one-loop partition function
\begin{equation}
Z(\tau) = \text{Tr}\left(q^{L_0-c/24}\bar{q}^{\bar{L}_0 -c/24}\right) = (q\bar{q})^{-c/24}\text{Tr}\left[e^{2\pi i \tau_1(L_0 - \bar{L}_0) -2\pi\tau_2(L_0 +\bar{L}_0)}\right].
\end{equation}
We are interested in a CFT state that dominates the above partition function as $\tau_2 \to \infty$ (where $\tau$ is living on the modular fundamental domain). This is the state with lowest $L_0 + \bar{L}_0$. To describe this in terms of field theory, we are hence interested in the `geometrization' (i.e. writing in terms of differential operators) of the string Hamiltonian $L_0 + \bar{L}_0$, along the lines of \cite{Dijkgraaf:1991ba}. In Lorentzian signature flat space for instance, this reduces to the Klein-Gordon operator with plane wave solutions $\sim e^{ipx}$. This operator describes how CFT fluctuations propagate in a background without interacting with any other fluctuations. In our case, the states appearing in the above partition function are those living on the thermal manifold and the state that dominates is the thermal scalar. 
This thermal scalar field theory action (which at lowest order in $\alpha'$ coincides with the lowest order \emph{effective} action), can be used to calculate the one-loop contribution to the free energy and should give us the critical behavior of the string gas. The path integral derivation presented in section \ref{deriv} should coincide with the field theory derivation if the large $\tau_2$ limit is correctly taken, that is by integrating out higher Fourier modes instead of setting these to zero.\footnote{As mentioned earlier, so far we have not been able to do this in general.} We will now check to what extent this story is true.\footnote{The reader might object at this point since in this section we consider winding in the modular fundamental domain whereas in the previous sections we discussed `winding' in the modular strip domain. This is however precisely the thermal scalar interpretation that was found long ago \cite{Atick:1988si}: the divergence in the modular strip is reflected in the masslessness of the winding tachyon in the fundamental domain.} \\
Since we could not find a derivation of the relevant action in the literature (and since we will utilize extensions of this derivation several times further on), we first (re)derive the winding tachyon action to lowest order in $\alpha'$ from a spacetime dimensional reduction.

\subsection{Dimensional reduction}
\label{dimred}
The bosonic closed string tachyon action to lowest order in $\alpha'$, is given by
\begin{equation}
S = \frac{1}{2}\int d^{d}x \sqrt{G}e^{-2\Phi}\left(G^{\mu\nu}\partial_{\mu}T\partial_{\nu}T + m^2T^2\right),
\end{equation}
where $T$ is a real scalar field and $\Phi$ is the dilaton field. This also holds for the closed superstring tachyon \emph{before} the GSO projection. \\
Assume now that $x^{0} \sim x^{0}+2\pi R$ while the metric does not depend on $x^{0}$ and has no components $G_{i0}$. We expand the field $T$ in Fourier modes 
\begin{equation}
T(x^{0},x^{i}) = \sum_{n \in \mathbb{Z}}T_{n}(x^{i})e^{\frac{inx^{0}}{R}},
\end{equation}
where $R = \frac{\beta}{2\pi}$.
Plugging this in the tachyon action gives for $T_{n}$
\begin{equation}
S = \pi R \int d^{d-1}x \sqrt{G_{ij}}\sqrt{G_{00}}e^{-2\Phi}\left(G^{ij}\partial_{i}T_{n}\partial_{j}T_{-n}+\frac{k^2G^{00}}{R^2}T_{n}T_{-n}+m^2T_{n}T_{-n}\right).
\end{equation}
Using the reality of $T$ $\left(T_{n} = T^{*}_{-n}\right)$, gives the action
\begin{equation}
S = \pi R \int d^{d-1}x \sqrt{G_{ij}}\sqrt{G_{00}}e^{-2\Phi}\left(G^{ij}\partial_{i}T_{n}\partial_{j}T_{n}^{*}+\frac{k^2G^{00}}{R^2}T_{n}T_{n}^{*}+m^2T_{n}T_{n}^{*}\right),
\end{equation}
and we can restrict to positive $n$. The complex field $T_{n}$ combines both $\pm n$ contributions. \\
This gives the momentum states of the tachyon field directly in the field theory action. The full action contains both winding and momentum fields. When dimensionally reducing the action, we obtain only the momentum states. But T-duality should still be present in the full field action. We thus exploit this and use a T-duality on the action with
\begin{align}
G_{00} &\to \frac{1}{G_{00}}, \\
\Phi &\to \Phi - \frac{1}{2}\ln\left(G_{00}\right), \\
T_{n} &\to T_{w}.
\end{align}
The momentum tachyon field is transformed to a winding tachyon. Also $\sqrt{G}e^{-2\Phi}$ is T-invariant. Thus we arrive at (also using $R \to \alpha'/R$)
\begin{equation}
\label{tachact}
S \sim \int d^{d-1}x \sqrt{G_{ij}}\sqrt{G_{00}}e^{-2\Phi}\left(G^{ij}\partial_{i}T_{w}\partial_{j}T_{w}^{*}+\frac{w^2R^2G_{00}}{\alpha'^2}T_{w}T_{w}^{*}+m^2T_{w}T_{w}^{*}\right).
\end{equation}
This is the tachyon action for a winding $w$ state \cite{Horowitz:1997jc}. \\
For instance for the type II superstring in polar coordinates $G^{00} = \alpha'/\rho^2$, we arrive at $m^{2} \to m^{2} + \frac{w^2\rho^2R^2}{\alpha'^3}$. With $R= \sqrt{\alpha'}$, we get the action used by the authors of \cite{Kutasov:2005rr}\cite{Giveon:2012kp} ($\alpha'=2$)
\begin{equation}
S \sim \int d^{d-1}x \sqrt{G_{ij}}\sqrt{G_{00}}e^{-2\Phi}\left(G^{ij}\partial_{i}T_{w}\partial_{j}T_{w}^{*} + \left(-1 + \frac{w^2\rho^2}{4}\right)T_{w}T_{w}^{*}\right).
\end{equation}
So in all, we start with the tachyon action for the tachyon living in the uncompactified spacetime. Dimensional reduction combined with T-duality then yields the lower-dimensional winding tachyon action. We noted that our original tachyon need not respect the GSO projection, it is enough to be there before the GSO projection to cause compactified tachyons (that \emph{do} satisfy the GSO projection) to appear. \\

\noindent This works fine for both bosonic and superstrings, but for the heterotic strings we need to be a little more clever.
The heterotic string has a left-moving tachyon in its spectrum ($m^2 = -4/\alpha'$) just as the bosonic string and a right-moving tachyon ($m^2=-2/\alpha'$) just as the superstring. The latter is projected out due to the GSO projection on the right-moving sector and the first cannot match with anything of the same mass on the right-moving side, so this state also does not exist. This is the story behind the tachyon-free heterotic string theories. \\
We know we should neglect GSO for our covering space tachyon action, but what about this left-right asymmetry? Well, it turns out to work just fine if one \emph{averages} both tachyon masses of the covering space tachyon. We can see why from the flat space spectrum (on the NS side of the right-moving sector) (see e.g. \cite{Schulgin:2011zb} for the spectrum)
\begin{eqnarray}
m^2 = -\frac{4}{\alpha'}+\frac{4N^{left}}{\alpha'}+\left(\frac{n}{R}+\frac{wR}{\alpha'}\right)^2, \\
m^2 = -\frac{2}{\alpha'}+\frac{4N^{right}_{NS}}{\alpha'}+\left(\frac{n}{R}-\frac{wR}{\alpha'}\right)^2,
\end{eqnarray}
where the constraint is $N^{right}_{NS} - N^{left} + \frac{1}{2} = nw$. The crucial difference is the term $\frac{1}{2}$, so that the non-oscillator states \emph{must} have both momentum and winding in the compact direction. Choosing no oscillators and averaging yields
\begin{equation}
m^2 = -\frac{3}{\alpha'} +\frac{n^2}{R^2}+\frac{w^2R^2}{\alpha'^2},
\end{equation}
where $nw=1/2$. Setting $ w= \pm1$ requires $n= \pm 1/2$ (with the same sign). For a general static background, we have accordingly
\begin{equation}
m^2_{total} = -\frac{3}{\alpha'} +\frac{n^2}{R^2G_{00}}+\frac{w^2R^2G_{00}}{\alpha'^2}.
\end{equation}
Specifying to the state we are interested in gives finally for the lowest order $\alpha'$ action\footnote{We dropped the subindex $w$ here for notational convenience.}
\begin{equation}
S \sim \int d^{d-1}x \sqrt{G_{ij}}\sqrt{G_{00}}e^{-2\Phi}\left(G^{ij}\partial_{i}T\partial_{j}T^{*}+\frac{1}{4R^2G_{00}}TT^{*} + \frac{R^2G_{00}}{\alpha'^2}TT^{*} - \frac{3}{\alpha'}TT^{*}\right).
\end{equation}
Hence in the above action (\ref{tachact}), we need to add also a discrete momentum contribution and choose the covering space mass correction equal to the average of the bosonic and the superstring tachyon mass. This action was also written down from a scattering amplitude perspective in \cite{Schulgin:2011zb}.\footnote{Note that our derivation of this action is not watertight in this case, but the derivation using scattering amplitudes \cite{Schulgin:2011zb} shows that this action is the correct one.}\\
In general, one can also obtain the same effective action by analyzing scattering amplitudes as we briefly discuss for the bosonic string in section \ref{appscat}. \\
For later convenience, let us assemble the different `local' mass terms for the singly wound string in a single function $m_{local}$ as follows
\begin{align}
\label{mlocall}
m_{local}^2 &= -\frac{4}{\alpha'} + \frac{R^2G_{00}}{\alpha'^2}, \quad \text{for bosonic strings}, \\
m_{local}^2 &= -\frac{2}{\alpha'} + \frac{R^2G_{00}}{\alpha'^2}, \quad \text{for type II superstrings}, \\
m_{local}^2 &= -\frac{3}{\alpha'} +\frac{1}{4R^2G_{00}}+\frac{R^2G_{00}}{\alpha'^2}, \quad \text{for heterotic strings}.
\end{align}
The extension to spacetimes with $G_{0i} \neq 0$ is straightforward (at least for bosonic and type II superstrings). We get the (discrete momentum) tachyon action
\begin{align}
S = \pi R \int d^{d-1}x \sqrt{G}e^{-2\Phi}\left(G^{ij}\partial_{i}T_{n}\partial_{j}T_{n}^{*}+\frac{n^2G^{00}}{R^2}T_{n}T_{n}^{*} \right. \nonumber \\
\left. + G^{0i}\frac{in}{R}\left(T_{n}\partial_{i}T_{n}^{*}- T_{n}^{*}\partial_{i}T_{n}\right) + m^2T_{n}T_{n}^{*}\right),
\end{align}
whereas T-duality now gives
\begin{align}
G_{00} \to \frac{1}{G_{00}}&,\quad G_{0i} \to \frac{B_{0i}}{G_{00}}=0,\quad G_{ij} \to G_{ij} - \frac{G_{0i}G_{0j}}{G_{00}}, \nonumber \\
&\Phi \to \Phi - \frac{1}{2}\ln\left(G_{00}\right),\quad T_{n} \to T_{w}.
\end{align}
The new $G_{0i}$ vanishes because $B_{0i}=0$. Also, $G'^{0i} = 0$ because $G'_{0i} = 0$. We have
\begin{equation}
\sqrt{G}e^{-2\Phi} \to \sqrt{G'}e^{-2\Phi'} = \sqrt{G'_{00}}\sqrt{G'_{ij}}e^{-2\Phi}G_{00} = \sqrt{G_{00}}e^{-2\Phi}\sqrt{G_{ij} - \frac{G_{0i}G_{0j}}{G_{00}}}.
\end{equation}
Thus we arrive at
\begin{equation}
S \sim \int d^{d-1}x \sqrt{G_{ij} - \frac{G_{0i}G_{0j}}{G_{00}}}\sqrt{G_{00}}e^{-2\Phi}\left(G'^{ij}\partial_{i}T_{w}\partial_{j}T_{w}^{*}+\frac{w^2R^2G_{00}}{\alpha'^2}T_{w}T_{w}^{*}+m^2T_{w}T_{w}^{*}\right).
\end{equation}
where $G'^{ij}$ is the matrix inverse of $G_{ij} - \frac{G_{0i}G_{0j}}{G_{00}}$. 
We thus see that the only effect of this more general case is the replacement $G_{ij} \to G_{ij} - \frac{G_{0i}G_{0j}}{G_{00}}$. In what follows we will again restrict to the case $G_{0i} = 0$, but we observe that the final result (after a simple substitution) will still hold in the more general case. This case is in nice agreement with the results in section \ref{exten}.

\subsection{The particle path integral of the field theory action}
\label{qm}
Let us compare the one-loop prediction of this action with our previous derivation from section \ref{deriv}. Since we only path integrate over the tachyon (the metric, dilaton and NS-NS fields are backgrounds), the one-loop effective action $\Gamma^{(1)}$ obtained in this way coincides with the free energy of the system (up to a factor of $\beta$). As a reminder, the first quantized stringy picture and second quantized field picture are related to the free energy as follows:
\begin{align}
F_{gas} &= -\frac{1}{\beta}Z_{part}, \\
F_{gas} &= \frac{1}{\beta}\Gamma^{(1)} = -\frac{1}{\beta}\ln\left(Z_{FT}\right),
\end{align}
where $part$ denotes the particle action derived in section \ref{deriv} and $FT$ denotes the field theory of only the winding tachyon. If the string gas indeed can be described by only the thermal scalar, both formalisms should yield the same expression for the free energy of the string gas.\\
We rewrite the one-loop result in a first-quantized way. The action (\ref{tachact}) derived in the previous section takes the following form (we have dropped the $w$ index of $T_{w=1}$ for notational convenience)
\begin{equation}
S =  \int d^{d-1}x \sqrt{G_{ij}}\sqrt{G_{00}}\left[G^{ij}\partial_{i}T\partial_{j}T^{*}+m_{local}^2TT^{*}\right]
\end{equation}
where $i$ runs over all space indices and $G^{ij}$ is a Euclidean metric on the spatial part of the manifold. We remind the reader that $m_{local}$ is a function of spacetime since it contains metric components. Our goal is to remove the $G_{00}$ contribution to the kinetic term `as much as possible' to hopefully reinterpret this as a particle on a curved background described by only the spatial part of the total manifold. We will proceed very carefully in what follows.\\
We first perform a partial integration in the action to distill the inverse propagator:
\begin{equation}
S =  \int d^{d-1}x \sqrt{G_{ij}}\sqrt{G_{00}}T^{*}\left[-\nabla^{2} - G^{ij}\frac{\partial_{j}\sqrt{G_{00}}}{\sqrt{G_{00}}}\partial_{i} + m_{local}^2\right]T.
\end{equation}
The operator $\nabla^2 = G^{ij}\nabla_{i}\partial_{j}$ denotes the covariant Laplacian on the spatial submanifold.
The operator between square brackets is readily seen to be Hermitian with respect to the inner product
\begin{equation}
\left\langle \psi_1 \right.\left|\psi_2\right\rangle = \int d^{d-1}x \sqrt{G_{ij}}\sqrt{G_{00}} \psi_1(x)^{*} \psi_2(x)
\end{equation}
so its eigenfunctions can be chosen orthonormal and its eigenvalues are real. For convenience, let us call the operator $\hat{\mathcal{O}}$. We now choose a basis of such eigenfunctions and expand
\begin{equation}
\label{eigen}
\psi(x) = \sum_{n}a_{n}\psi_{n}(x), \quad \hat{\mathcal{O}}\psi_{n} = \lambda_{n}\psi_{n}, \quad \int d^{d-1}x \sqrt{G_{ij}}\sqrt{G_{00}} \psi_{n}(x)^{*} \psi_{m}(x) = \delta_{n,m}.
\end{equation}
The one-loop action is given by the logarithm of the path integral over $T$ with the above action. In the above basis the path integral gives a product of Gaussian integrals over the $a_{n}$, resulting in\footnote{We ignore prefactors, since they will just end up as an additive contribution to the effective action (due to the logarithm) that are independent of the state in the Hilbert space. Also note that we have a \emph{complex} scalar field so we should square the contributions coming from real scalar fields.}
\begin{equation}
\prod_{n}\left(\frac{1}{\lambda_{n}}\right) = \text{det}^{-1}\hat{\mathcal{O}}.
\end{equation}
The one-loop action can be written as
\begin{equation}
\Gamma^{(1)} = -\ln \text{det}^{-1}\hat{\mathcal{O}} = \text{Tr} \ln\hat{\mathcal{O}}.
\end{equation}
Now we use the Schwinger proper time representation of the logarithm
\begin{equation}
\ln(a) = -\int_{0}^{+\infty}\frac{dT}{T}\left(e^{-aT} - e^{-T}\right).
\end{equation}
We drop the $-e^{-T}$ term\footnote{This term is proportional to the size of the Hilbert space, just as the prefactors we ignored above.} which gives
\begin{equation}
\Gamma^{(1)} = - \int_{0}^{+\infty}\frac{dT}{T} \text{Tr} e^{-T\left(-\nabla^{2}+m_{local}^2 - G^{ij}\frac{\partial_{j}\sqrt{G_{00}}}{\sqrt{G_{00}}}\partial_{i}\right)}.
\end{equation}
Note that the net effect of starting with a complex instead of a real scalar field is an overall factor $2$. This corresponds in the previous path integral derivation to the sum over the two winding states. \\

\noindent To proceed, we notice a delicate point: despite the fact that it looks like we succeeded in removing all $G_{00}$ dependence from the kinetic term, there is still a non-trivial dependence on it. The trace still contains the $\sqrt{G_{00}}$ measure as is shown in the normalization (\ref{eigen}).  
We still want to remove this factor. This can be done by a simple rescaling of the eigenfunctions. 
Let us define new basis vectors
\begin{equation}
\left|\phi_{n}\right\rangle = G_{00}^{1/4}\left|\psi_{n}\right\rangle
\end{equation}
which are by definition normalized as
\begin{equation}
\int d^{d-1}x \sqrt{G_{ij}}\phi_{n}(x)^{*} \phi_{m}(x) = \delta_{n,m}.
\end{equation}
It now readily follows that the $\left|\phi_{n}\right\rangle$ are eigenstates of the operator obtained by pulling the exponential through the $G_{00}^{-1/4}$ factor as follows
\begin{equation}
e^{-T\hat{\mathcal{O}}}G_{00}^{-1/4} = G_{00}^{-1/4}e^{-T\hat{\mathcal{D}}}.
\end{equation}
In coordinate space, the operator $\hat{\mathcal{D}}$ is simply obtained by setting
\begin{equation}
\hat{\mathcal{D}}f(x) = \frac{\hat{\mathcal{O}}\left(G_{00}^{-1/4} f(x)\right)}{G_{00}^{-1/4}},\quad \forall f(x). 
\end{equation}
We can thus rewrite the previous action as
\begin{equation}
\label{effD}
\Gamma^{(1)} = - \int_{0}^{+\infty}\frac{dT}{T} \text{Tr} e^{-T\hat{\mathcal{D}}}
\end{equation}
where now the $\sqrt{G_{00}}$ is removed from the integral measure in the trace (we have in effect changed the definition of the inner product on our Hilbert space).\\
Transforming $\hat{\mathcal{O}}$ in this fashion, we obtain the following operator in the exponential
\begin{equation}
\hat{\mathcal{D}} = -\nabla^{2} -\frac{3}{16}\frac{G^{ij}\partial_iG_{00}\partial_jG_{00}}{G_{00}^2} + \frac{\nabla^2 G_{00}}{4G_{00}}+ m_{local}^2
\end{equation}
where the terms involving only one $\partial_{i}$ have miraculously dropped out. As a check, we see that this operator $\hat{\mathcal{D}}$ is Hermitian with respect to the canonical inner product on the spatial submanifold
\begin{equation}  
\left\langle \phi_1 \right.\left|\phi_2\right\rangle = \int d^{d-1}x \sqrt{G_{ij}}\phi_1(x)^{*}\phi_2(x).
\end{equation}
The only corrections we have are two terms that behave as a potential for the particle. 
We denote these for convenience in what follows as $K(x)$:
\begin{equation}
\label{KK}
K(x) =-\frac{3}{16}\frac{G^{ij}\partial_iG_{00}\partial_jG_{00}}{G_{00}^2} + \frac{\nabla^2 G_{00}}{4G_{00}}.
\end{equation}
The exponential in (\ref{effD}) needs to be given a Lagrangian interpretation. 
So we seek a path integral description for a system with Hamiltonian
\begin{equation}
H = p_i p_j G^{ij} + m_{local}^2 + K(x).
\end{equation}
The corresponding (Euclidean) Lagrangian is given by
\begin{equation}
L_E = \frac{1}{4}\dot{x}^{i}\dot{x}^{j}G_{ij} + m_{local}^2 + K(x).
\end{equation}
The last step is then to give a path integral representation for $e^{-TH}$ which finally results in\footnote{The appearance of the $\sqrt{G_{ij}}$ in the measure is natural from coordinate invariance. It can also be explicitly derived (see \cite{Abers:1973qs} for the particle case and \cite{Weinberg:1995mt} (chapter 9) for the field theory case).}
\begin{equation}
\Gamma^{(1)} = - \int_{0}^{+\infty}\frac{dT}{T}\int_{S^{1}} \left[\mathcal{D}x\right]\sqrt{G_{ij}}e^{-\int_{0}^{T}dt\left(\frac{1}{4}G_{ij}(x)\dot{x}^{i}\dot{x}^{j} +  m_{local}^2 + K(x)\right)}.
\end{equation}
The $S^{1}$ denotes periodic boundary conditions on the path: $x(0)=x(T)$.
Performing the substitution $t \to \pi \alpha' t$ gives the path integral
\begin{equation}
\label{FTaction}
\Gamma^{(1)} = - \int_{0}^{+\infty}\frac{dT}{T}\int_{S^{1}} \left[\mathcal{D}x\right]\sqrt{G_{ij}}e^{-\frac{1}{4\pi\alpha'}\int_{0}^{T}dt\left(G_{ij}(x)\dot{x}^{i}\dot{x}^{j} + 4\pi^2\alpha'^2\left(m_{local}^2 + K(x\right)\right)}.
\end{equation}
For instance, filling in the correct value for $m_{local}$ (\ref{mlocall}) when the bosonic flat space Hagedorn temperature is to be used, gives
\begin{equation}
\label{partaction}
\Gamma^{(1)} = - \int_{0}^{+\infty}\frac{dT}{T}\int_{S^{1}} \left[\mathcal{D}x\right]\sqrt{G_{ij}}e^{-\frac{1}{4\pi\alpha'}\int_{0}^{T}dt\left(G_{ij}(x)\dot{x}^{i}\dot{x}^{j} + 4\pi^2\alpha'^2\left(-\frac{4}{\alpha'} + \frac{R^2G_{00}}{\alpha'^2} + K(x)\right)\right)}.
\end{equation}
\\
We can compare this with the result from section \ref{deriv}, when we identify $\Gamma^{(1)}$ directly with $\beta F$. For convenience, we rewrite the result given there:
\begin{equation}
\label{Firstaction} 
\beta F = -\int_0^{+\infty} \frac{d \tau_2}{\tau_2} \int_{S^{1}} \left[ \mathcal{D}X \right] \sqrt{G_{ij}} \exp -\frac{1}{4\pi \alpha'}\left[ \int_0^{\tau_2} dt \left( G_{ij} \partial_t X^i \partial_t X^j + \beta^2  G_{00} - \beta_{H,\text{flat}}^2 \right) \right]. 
\end{equation}
Translating (\ref{FTaction}) to this notation, we find
\begin{equation}
\label{pint} 
\beta F = - \int_{0}^{+\infty}\frac{d\tau_2}{\tau_2}\int_{S^{1}} \left[\mathcal{D}X\right]\sqrt{G_{ij}}\exp -S
\end{equation}
where
\begin{equation}
\label{FFTaction}
S = \frac{1}{4\pi\alpha'} \left[\int_{0}^{\tau_2} dt \left\{G_{ij}\partial_t X^{i}\partial_ tX^{j} + 4\pi^2\alpha'^2\left(m_{local}^2 + K(X)\right)\right\}\right].
\end{equation}

\noindent Now let us compare the second quantized field theory result (\ref{FFTaction}) with the first quantized result (\ref{Firstaction}).

\begin{itemize}
\item{The particle action (\ref{FFTaction}) naturally lives in one dimension less than the original problem. Here this occurs due to the dimensional reduction used to arrive at the field theory action. In (\ref{Firstaction}), this happened because we integrated out the Euclidean time coordinate explicitly.}
\item{
For bosonic and type II superstrings we have the equality\footnote{Using the appropriate value of $\beta_{H,\text{flat}}$.}
\begin{equation}
4\pi^2\alpha'^2m_{local}^2 = \beta^2G_{00} - \beta_{H,\text{flat}}^2.
\end{equation}
For the heterotic string, the story changes a bit. The mass term is now of the form
\begin{equation}
\label{massheter}
4\pi^2\alpha'^2m_{local}^2 = -\tilde{\beta}_{H}^2 + \beta^2G_{00} + \frac{\pi^2\alpha'^2G^{00}}{R^2}.
\end{equation}
Note that $\tilde{\beta}_{H}^2 = 12\pi^2\alpha'^2$. This corresponds to the averaging of the bosonic and type II tachyon masses as we argued in section \ref{dimred}. The reader should not be confused at this point: this $\tilde{\beta}_{H}$ is not equal to the Hagedorn temperature, not even in the flat space. The final term in (\ref{massheter}) is not found using the naive path integral result and should be added (just like the $\tilde{\beta}_{H}^2$-term). In flat spacetime, we found that this correction actually originated from a $\tau_1$-dependent contribution as we discussed in section \ref{heteroticz}.
}
\item{The second quantized result (\ref{FFTaction}) has an extra term (denoted as $K(x)$). 
This suggests we missed this term in the derivation in section \ref{deriv} in the `worldsheet dimensional reduction', which should be incorporated in the result in the same way as the $\beta_{H,\text{flat}}^2$ contribution. This term alters the random walk behavior discussed in \cite{Kruczenski:2005pj} and we will show in the next few chapters, based on \cite{Mertens:2013zya}\cite{Mertens:2014cia}, that it gives crucial modifications. This extra term disappears when choosing a flat metric, so in the flat case we have perfect agreement between the two approaches. The extra term cannot in general be discarded in any approximation since it is not a higher order curvature contribution (we will discuss this more extensively in what follows).}

\item{This section only focused on the case where there is only a background metric. We can extend this result to include a background NS-NS field. This is discussed in section \ref{KR}. The upshot is that we get precisely the result from section \ref{exten} but with another extra correction term.}

\end{itemize}

\subsection{General discussion of corrections to the particle action}
Let us look in general to the corrections of the particle action that we derived in section \ref{deriv}. In what follows, we define a correction term as a term that is missed in the naive worldsheet dimensional reduction for $\tau_2 \to \infty$ discussed in section \ref{deriv}. \\

\noindent Firstly, we have a correction term proportional to the mass of the most tachyonic mode in the \emph{cover} of the manifold that `unwraps' the thermal direction. Note that this only depends on the \emph{type} of string theory used and not on the manifold itself. For bosonic and type II superstrings, this equals the (flat space) Hagedorn correction to the action, but this need not be the Hagedorn temperature of the space under consideration. As an example of the latter case, consider the WZW $AdS_3$ bosonic string background (we will discuss this model extensively in chapter \ref{chwzw}). The correction is in this case the Hagedorn temperature of the flat space bosonic string, but this is not equal to the $AdS_3$ Hagedorn temperature.\footnote{The latter temperature is \emph{larger} than the flat space bosonic Hagedorn temperature.} We conclude that this term represents strings that \emph{locally} approach the Hagedorn temperature, but this need not give a \emph{global} divergence in the free energy. This is very reminiscent of the well-known fact that negative mass$^2$ particles in $AdS$ spacetimes can be stable if their mass$^2$ is not too low (the Breitenlohner-Freedman bound). \\
For heterotic strings, we found an additional correction that gives the discrete momentum contribution of the winding tachyon.\\

\noindent In flat spacetime, this is the only correction. The one-loop contribution is a simple sum over all string states without including interactions. When restricting to the winding $\pm 1$ states, this is twice the field theory vacuum bubble diagram. We saw that we have an exact matching between our path integral result and the effective action calculated from the thermal scalar field theory action.\\
 
\noindent Secondly, we found a correction term $K(x)$ (\ref{KK}) obtained from a non-trivial $G_{00}$ metric component. This term alters the random walk behavior and there is no rationale in neglecting it.\\

\noindent This need not be the end of the story however, as $\alpha'$ corrections to the spacetime action might be important. Their influence and appearance is rather subtle. The precise effect of these higher order $\alpha'$ corrections in general on the random walk behavior is something we are still further investigating. In the case of Rindler spacetime, these corrections are understood in full detail as we will show in chapter \ref{chri} \cite{Mertens:2013zya}. Other examples with and without $\alpha'$ corrections will be discussed in the remainder of this work.

\section{Summary of the random walk picture}
\label{pathderiv}
In this section, we provide a summary of the random walk picture as a reference for future chapters. \\
Following \cite{Kruczenski:2005pj}, we are interested in the torus path integral on the thermal manifold (obtained by Wick-rotating the time direction and periodically identifying with period $\beta$, the inverse temperature). After performing a modular transformation, we find the worldsheet action to be
\begin{eqnarray}
\label{action1}
S = \frac{1}{4\pi \alpha'} \left[ \left(1 + \frac{\tau_1^2}{\tau_2^2}\right) \int_0^{1/\tau_2} d\sigma \int_0^1 d\tau G_{\mu\nu} \partial_\sigma X^\mu \partial_\sigma X^\nu \right. \nonumber \\
\left.
+ 2 \frac{\tau_1}{\tau_2} \int_0^{1/\tau_2} d\sigma \int_0^1 d\tau G_{\mu\nu} \partial_\sigma X^\mu \partial_\tau X^\nu 
+ \int_0^{1/\tau_2} d\sigma \int_0^1 d\tau G_{\mu\nu} \partial_\tau X^\mu \partial_\tau X^\nu \right].
\end{eqnarray} 
To find the critical behavior, we focus on the $\tau_2 \to \infty$ contribution to the path integral and select the string state that is singly wound around the Euclidean time direction: $w = \pm 1$. We next perform a worldsheet Fourier series expansion:
\begin{align}
X^i(\sigma,\tau) & = \sum_{n=-\infty}^{+\infty} e^{i(2\pi n \tau_2) \sigma} X_n^i(\tau), \quad i=1\hdots d-1,\\
X^0(\sigma,\tau) & = \pm \beta \tau_2 \sigma +  \sum_{n=-\infty}^{+\infty} e^{i(2\pi n \tau_2) \sigma} X_n^0(\tau).
\end{align}
If we drop all higher Fourier modes and hence perform a naive worldsheet dimensional reduction, we found above that the torus path integral on the thermal manifold reduces to
\begin{equation} 
Z_p = 2\int_0^\infty \frac{d \tau_2}{2\tau_2} \int \left[ \mathcal{D}X \right] \sqrt{\prod_{t} \det G_{ij}} \exp - S_p(X) 
\end{equation}
where 
\begin{equation}
\label{act}
S_p = \frac{1}{4\pi \alpha'}\left[ \beta^2 \int_0^{\tau_2} dt G_{00} + \int_0^{\tau_2} dt G_{ij} \partial_t X^i \partial_t X^j\right].
\end{equation}
The time parameter $t$ along the worldline is related to the worldsheet coordinate $\tau$ in (\ref{action1}) as $t=\tau_2 \tau$. \\
The full string partition function has been reduced to a partition function for a non-relativistic particle moving on the purely spatial submanifold. The time evolution of the particle in its random walk is identified with the spatial form of the long highly excited string. We view this as a realization of the Wick rotation: the long string in real spacetime has a form described by the above random walk.\\
The free energy of a gas of strings can then be identified with the single string partition function as \cite{Polchinski:1985zf}
\begin{equation}
F = -\frac{1}{\beta} Z_p.
\end{equation}
We analyzed this particle path integral for several string types. The worldsheet dimensional reduction misses some correction terms that we could not determine solely from the path integral. In flat spacetime, we calculated these explicitly using known particle path integrals.

An alternative route we followed was the field theory of the thermal scalar. 
These two perspectives should yield the same outcome since both methods focus precisely on the same string state. We checked this and it works out well for flat space. For curved backgrounds however, we found a discrepancy: the spacetime action has an extra term in the action. We identified this as something we missed in the (naive) worldsheet dimensional reduction. This identifies both approaches. Both approaches (worldsheet and spacetime) have their advantages though the worldsheet path integral approach has a direct connection to the random walk picture of string thermodynamics in the microcanonical ensemble. However, this approach is computationally more challenging because the $\tau_2 \to \infty$ limit involves coupling to higher Fourier modes which we drop in a first approximation. These are however important to attain the full result in curved space.
The thermal scalar action is given by (to lowest order in $\alpha'$)
\begin{equation}
\label{lowestFT}
S \sim \int d^{d-1}x \sqrt{G_{ij}}\sqrt{G_{00}}e^{-2\Phi}\left(G^{ij}\partial_{i}T\partial_{j}T^{*} + \frac{R^2G_{00}}{\alpha'^2}TT^{*} + m^2TT^{*}\right),
\end{equation}
where $m^2$ is the tachyon mass$^2$ in flat space whose precise value will be given below. From this action, the one-loop free energy is given by
\begin{align}
\label{FT}
\beta F &= - \int_{0}^{+\infty}\frac{dT}{T} \text{Tr} e^{-T\left(-\nabla^{2}+m_{\text{local}}^2 - G^{ij}\frac{\partial_{j}\sqrt{G_{00}}}{\sqrt{G_{00}}}\partial_{i}\right)} \\
\label{randwalk}
&= - \int_{0}^{+\infty}\frac{dT}{T}\int_{S^{1}} \left[\mathcal{D}x\right]\sqrt{G}e^{-\frac{1}{4\pi\alpha'}\int_{0}^{T}dt\left(G_{ij}(x)\dot{x}^{i}\dot{x}^{j} + 4\pi^2\alpha'^2\left(m_{\text{local}}^2 + K(x\right)\right)}.
\end{align}
We denote the operator in brackets in the exponential in the first equation as $\hat{\mathcal{O}}$ in what follows. \\
We have also collected the `local' mass terms in
\begin{align}
\label{mlocal}
m_{\text{local}}^2 &= -\frac{4}{\alpha'} + \frac{R^2G_{00}}{\alpha'^2}, \quad \text{for bosonic strings}, \\
m_{\text{local}}^2 &= -\frac{2}{\alpha'} + \frac{R^2G_{00}}{\alpha'^2}, \quad \text{for type II superstrings}, \\
m_{\text{local}}^2 &= -\frac{3}{\alpha'} +\frac{1}{4R^2G_{00}}+\frac{R^2G_{00}}{\alpha'^2}, \quad \text{for heterotic strings}.
\end{align}
The function $K(x)$ denotes the following metric combination\footnote{$\nabla^2$ is the Laplacian on the spatial submanifold.}
\begin{equation}
\label{K}
K(x) =-\frac{3}{16}\frac{G^{ij}\partial_iG_{00}\partial_jG_{00}}{G_{00}^2} + \frac{\nabla^2 G_{00}}{4G_{00}}
\end{equation}
and this represents the effect of removing the $\sqrt{G_{00}}$ from the measure in the field theory action. Going from (\ref{FT}) to (\ref{randwalk}) requires some delicate manipulations that we discussed above. Equation (\ref{randwalk}) can then be identified with (\ref{act}) and hence we can see which correction terms are needed in the particle action. \\
The correction terms are of three different types.
\begin{itemize}
\item{Firstly we have a correction term coming from the mass of the flat space tachyon and this is of the following form
\begin{equation}
\Delta S = -\frac{\beta_{H,\text{flat}}^2\tau_2}{4\pi\alpha'}.
\end{equation}
For bosonic strings $\beta_{H,\text{flat}}^2 = 16\pi^2\alpha'$, for type II superstrings $\beta_{H,\text{flat}}^2 = 8\pi^2\alpha'$ and for heterotic strings $\beta_{H,\text{flat}}^2 = 12\pi^2\alpha'$. 
For bosonic and type II strings, this is the flat space Hagedorn temperature but for heterotic strings this is not the case. By abuse of notation, we will nonetheless denote this term with $\beta_{H,\text{flat}}^2$.
}
\item{Secondly we have a correction coming from the $G_{00}$ component as explained above: 
\begin{equation}
\label{corr}
\Delta S = \frac{1}{4\pi\alpha'}\int_{0}^{\tau_2}dt 4\pi^2\alpha'^2 K(x).
\end{equation}
}
\item{Finally we could have order-by-order $\alpha'$ correction terms of the field theory action (\ref{lowestFT}). These are of course not present in (\ref{randwalk}) and it is difficult to say anything specific about these for the general case. 
These will be analyzed more explicitly in the second and third parts of this work.
}
\end{itemize}
We also presented a simple extension to include a background NS-NS field. 
The string worldsheet action (\ref{action1}) has the following extra contribution
\begin{equation}
\label{KRexten}
S_{\text{extra}} = - i\frac{1}{2\pi\alpha'}\int_{0}^{1/\tau_2}d\sigma\int_{0}^{1}d\tau B_{\mu\nu}(X)\partial_{\sigma}X^{\mu}\partial_{\tau}X^{\nu}
\end{equation}
which results in the following augmentation of the particle action (\ref{act})
\begin{equation}
\label{KRexten2}
S_{\text{extra}} = \mp i\frac{\beta}{2\pi\alpha'}\int_{0}^{\tau_2}dt B_{0i}(X)\partial_{t}X^{i}.
\end{equation}
This action represents a minimal coupling of a point particle to a vector potential $A_{i} = B_{0i}$. This term breaks the symmetry between both windings because the NS-NS field breaks the orientation reversal symmetry of the string. From a point particle viewpoint, this means that the particles are oppositely charged under the electromagnetic field. \\

\noindent The generalization to stationary spacetimes was also given, resulting in
\begin{equation} 
\label{statio}
Z_p = 2\int_0^\infty \frac{d \tau_2}{2\tau_2} \int \left[ \mathcal{D} X \right] \sqrt{\prod_t \det \left(G_{ij} - \frac{G_{0i}G_{0j}}{G_{00}}\right)} \exp - S_p( X) 
\end{equation}
where 
\begin{equation}
S_p = \frac{1}{4\pi \alpha'}\left[ \beta^2 \int_0^{\tau_2} dt G_{00} +\int_0^{\tau_2} dt \left(G_{ij} - \frac{G_{0i}G_{0j}}{G_{00}}\right) \partial_t X^i \partial_t X^j\right].
\end{equation}
Like the static case described above, also this formula requires corrections, the simplest of which again being the correction for the flat space tachyon mass. \\

\noindent Several questions arise in this process: is there really a winding mode in the string spectrum, especially if the space does not topologically support winding modes? Can we get a handle on the higher correction terms? 



\section{*Extension path integral to non-static spacetimes}
\label{AppAA}
We start with the action (\ref{particleaction}) in the general case $G_{0i} \neq 0$:
\begin{equation}
S_{part} =  \frac{1}{4\pi \alpha' } \left[ \beta^2\frac{\left|\tau\right|^2}{\tau_2^2} \int_0^{\tau_2} dt G_{00} - \beta_{H,\text{flat}}^2 \tau_2 \pm 2 \frac{\tau_1}{\tau_2} \beta \int_0^{\tau_2} dt G_{0\mu} \partial_t X^\mu 
+\int_0^{\tau_2} dt G_{\mu\nu} \partial_t X^\mu \partial_t X^\nu \right].
\end{equation}
The classical equation of motion of $X^{0}$ is given by
\begin{equation}
\pm \frac{\tau_1}{\tau_2}\beta \partial_t G_{00} + \partial_t \left(G_{00}\partial_t X^{0}\right) + \partial_t \left(G_{0i}\partial_t X^{i}\right) = 0. 
\end{equation}
Integrating with respect to $t$ gives
\begin{equation}
\pm \frac{\tau_1}{\tau_2}\beta G_{00} + G_{00}\partial_t X^{0} + G_{0i}\partial_t X^{i} = C. 
\end{equation}
Again integrating and using periodicity of $X^{\mu}$ fixes 
\begin{equation}
C = \frac{1}{\left\langle G_{00}^{-1}\right\rangle}\left[\pm \tau_1 \beta + \left\langle \frac{G_{0i}\partial_t X^{i}}{G_{00}}\right\rangle\right].
\end{equation}
One readily finds for the on-shell classical action (including only the $X^{0}$-dependent contributions)
\begin{align}
4\pi\alpha'S_{on-shell} &= \mp 2\frac{\tau_1}{\tau_2}\beta \left\langle G_{0i}\partial_t X^{i}\right\rangle - \frac{\tau_1^2}{\tau_2^2}\beta^2\left\langle G_{00}\right\rangle + \frac{\tau_1^2\beta^2}{\left\langle G_{00}^{-1}\right\rangle} \pm \frac{2\tau_1\beta}{\left\langle G_{00}^{-1}\right\rangle}\left\langle \frac{G_{0i}\partial_t X^{i}}{G_{00}}\right\rangle \nonumber \\
&+ \frac{1}{\left\langle G_{00}^{-1}\right\rangle}\left\langle \frac{G_{0i}\partial_t X^{i}}{G_{00}}\right\rangle^2 - \left\langle \frac{G_{0i}G_{0j}}{G_{00}}\partial_t X^{i}\partial_t X^{j}\right\rangle.
\end{align}
Setting $X^{0} = X^{0,cl} + \tilde{X}^{0}$, one arrives at the following total action
\begin{align}
4\pi\alpha'S_{total} &=  \beta^2\int_0^{\tau_2} dt G_{00} - \beta_{H,\text{flat}}^2 \tau_2 + \frac{\tau_1^2\beta^2}{\left\langle G_{00}^{-1}\right\rangle} \pm \frac{2\tau_1\beta}{\left\langle G_{00}^{-1}\right\rangle}\left\langle \frac{G_{0i}\partial_t X^{i}}{G_{00}}\right\rangle \nonumber \\
&+ \frac{1}{\left\langle G_{00}^{-1}\right\rangle}\left\langle \frac{G_{0i}\partial_t X^{i}}{G_{00}}\right\rangle^2 + \left\langle \left(G_{ij}-\frac{G_{0i}G_{0j}}{G_{00}}\right)\partial_t X^{i}\partial_t X^{j}\right\rangle + \left\langle G_{00} \partial_t \tilde{X}^{0} \partial_t \tilde{X}^{0}\right\rangle.
\end{align}
Finally performing the $\tau_1$ integration, the first term in the second line is precisely cancelled and the $\tilde{X}^{0}$ path integral again cancels the $\left\langle G_{00}^{-1}\right\rangle$ prefactors. Since 
\begin{equation}
\sqrt{G} = \sqrt{G_{00}}\sqrt{G_{ij} - \frac{G_{0i}G_{0j}}{G_{00}}},
\end{equation}
we finally end up with
\begin{equation} 
Z_p = 2\int_0^\infty \frac{d \tau_2}{2\tau_2} \int \left[ d\vec X \right] \sqrt{\prod \det \left(G_{ij} - \frac{G_{0i}G_{0j}}{G_{00}}\right)} \exp - S_p(\vec X) 
\end{equation}
where 
\begin{equation}
S_p = \frac{1}{4\pi \alpha'}\left[ \beta^2 \int_0^{\tau_2} dt G_{00} - \beta_{H,\text{flat}}^2 \tau_2 
+\int_0^{\tau_2} dt \left(G_{ij} - \frac{G_{0i}G_{0j}}{G_{00}}\right) \partial_t X^i \partial_t X^j\right],
\end{equation}
which is the expression shown in section \ref{nonstatt}.

\section{*\boldmath Large $\tau_2$ limit of several string theories}

\subsection*{Superstring}
\label{appsuper}
The free energy for superstrings is given by the following expression \cite{Alvarez:1986sj}
\begin{equation}
F = -2 V_9\int_{0}^{+\infty}d\tau_2\int_{-1/2}^{1/2}\frac{d\tau_1}{\tau_2^{6}(2\pi^2\alpha')^5}\left[\vartheta_3\left(0,\frac{i\beta^2}{4\pi^2\alpha'\tau_2}\right)-\vartheta_4\left(0,\frac{i\beta^2}{4\pi^2\alpha'\tau_2}\right)\right]\left|\vartheta_4(0,2\tau)\right|^{-16}
\end{equation}
where 
\begin{equation}
\vartheta_3(0,\tau) = \sum_{n=-\infty}^{+\infty}q^{n^2/2}, \quad \vartheta_4(0,\tau) = \sum_{n=-\infty}^{+\infty}(-1)^n q^{n^2/2}.
\end{equation}
For the modular functions, we follow the definitions of \cite{Polchinski:1998rq}.
We do a modular transformation $\tau \to -\frac{1}{\tau}$. 
The factor $\frac{d\tau_1d\tau_2}{\tau_2^2}$ is invariant under modular transformations. Also,
\begin{equation}
\vartheta_3\left(0,\frac{i\beta^2}{4\pi^2\alpha'\tau_2}\right)-\vartheta_4\left(0,\frac{i\beta^2}{4\pi^2\alpha'\tau_2}\right) = \sum_{n=-\infty}^{+\infty}(1-(-1)^n)e^{-\frac{\beta^2n^2}{4\pi\alpha'}\frac{\tau_1'^2+\tau_2'^2}{\tau_2'}}.
\end{equation}
From now on we drop the primes.
We can take the $\tau_2 \to +\infty$ limit of the expression:
\begin{equation}
\label{integrand}
F = -2 V_9\iint_{\mathcal{A}} d\tau_2\frac{d\tau_1}{\tau_2^{2}(2\pi^2\alpha')^5}\left(\frac{\left|\tau\right|^2}{\tau_2}\right)^4\left[\sum_{n=-\infty}^{+\infty}(1-(-1)^n)e^{-\frac{\beta^2n^2}{4\pi\alpha'}\frac{\tau_1^2+\tau_2^2}{\tau_2}}\right]\left|\vartheta_4\left(0,-\frac{2}{\tau}\right)\right|^{-16}.
\end{equation}
The $\vartheta_4$ function has the modular property
\begin{equation}
\vartheta_4\left(0,-\frac{1}{\tau}\right) = (-i\tau)^\frac{1}{2}\vartheta_2(0,\tau),
\end{equation}
and the $\vartheta_2$ function has the following product expansion
\begin{equation}
\vartheta_2(0,\tau) = 2e^{\pi i \tau/4} \prod_{m=1}^{+\infty}(1-q^m)(1+q^m)^2.
\end{equation}
We notice that $\left|\vartheta_2\right|$ for $\tau_2 \to \infty$ has no contribution from the infinite product. So
\begin{equation}
\left|\vartheta_2(0,\tau)\right| \to 2e^{-\pi \tau_2 /4}
\end{equation}
irrespective of the value of $\tau_1$. We finally arrive at
\begin{equation}
\left|\vartheta_4\left(0,-\frac{2}{\tau}\right)\right|^{-16} \to \left|\tau\right|^{-8}2^{-8}e^{2\pi\tau_2}.
\end{equation}
The sum in the integrand (\ref{integrand}) is dominated by $n= \pm 1$ in the limit $\tau_2 \to +\infty$. Plugging all this into the integral finally gives
\begin{eqnarray}
F &= -2 V_9\iint_{\mathcal{A}} d\tau_2\frac{d\tau_1}{\tau_2^6(2\pi^2\alpha')^5}\left[4e^{-\frac{\beta^2}{4\pi\alpha'}\frac{\tau_1^2+\tau_2^2}{\tau_2}}\right]2^{-8}e^{2\pi\tau_2} \nonumber \\
&= -2 V_9\iint_{\mathcal{A}} \frac{d\tau_2d\tau_1}{2\tau_2}\frac{1}{(4\pi^2\alpha'\tau_2)^5}e^{-\frac{\beta^2}{4\pi\alpha'}\frac{\tau_1^2+\tau_2^2}{\tau_2}}e^{2\pi\tau_2}.
\end{eqnarray}
which is the result stated in section \ref{supers}.

\subsection*{Heterotic string}
\label{appheterotic}
The free energy for the $E_8 \times E_8$ heterotic string is given by \cite{Alvarez:1986sj}
\begin{align}
F = -2 V_9\int_{0}^{+\infty}d\tau_2\int_{-1/2}^{1/2}\frac{d\tau_1}{16\tau_2^{6}(2\pi^2\alpha')^5}\left[\vartheta_3\left(0,\frac{i\beta^2}{4\pi^2\alpha'\tau_2}\right)-\vartheta_4\left(0,\frac{i\beta^2}{4\pi^2\alpha'\tau_2}\right)\right] \nonumber \\
\times \frac{\Theta_{E_8 \oplus E_8}(-\overline{\tau})}{\vartheta_4(0,2\tau)^{8}\eta(-\overline{\tau})^{24}}.
\end{align}
We again use the modular transformation $\tau \to -1/\tau$ and the limiting behavior of the modular functions (as $\tau_2 \to \infty$)
\begin{eqnarray}
\vartheta_4\left(0,-\frac{2}{\tau}\right)^{-8} \to \frac{e^{-\pi i \tau}}{\tau^42^4}, \\
\eta\left(\frac{1}{\overline{\tau}}\right) = (i\overline{\tau})^{1/2}\eta(-\overline{\tau}) \to (i\overline{\tau})^{1/2} e^{-\frac{\pi i \overline{\tau}}{12}}.
\end{eqnarray}
Using $\Theta_{E_8 \oplus E_8} = \Theta_{E_8}^2$ and $\Theta_{E_8} = \frac{1}{2}\left(\vartheta_2^8+\vartheta_3^8+\vartheta_4^8\right)$, we see that\footnote{This result is more general: given a lattice theta function $\Theta_{\Gamma}(\tau)$, the limit for $\tau_2 \to +\infty$ is always equal to 1. This implies that this derivation also holds for the SO(32) heterotic string (based on the $\Gamma_{16}$ lattice).}
\begin{equation}
\Theta_{E_8 \oplus E_8}\left(\frac{1}{\overline{\tau}}\right) = (-i\overline{\tau})^8 \Theta_{E_8 \oplus E_8}\left(-\overline{\tau}\right) \to (-i\overline{\tau})^8
\end{equation}
where the contributions from $\vartheta_3$ and $\vartheta_4$ give a factor of 4 that cancels the denominator. We now arrive at\footnote{There is a subtlety here: we approximate $\tau_1^2 + \tau_2^2 \approx \tau_2^2$, which is only valid for $\tau_2 \gg \tau_1$. So the $\tau_2 \to \infty$ limit is taken before the integral over $\tau_1$ is performed.} 
\begin{align}
F &= -2 V_9\iint_{\mathcal{A}} \frac{d\tau_2d\tau_1 \tau_2^4}{16\tau_2^2(2\pi^2\alpha')^5}\left[4e^{-\frac{\beta^2}{4\pi\alpha'}\frac{\tau_1^2+\tau_2^2}{\tau_2}}\right]\frac{e^{-\pi i \tau}}{\tau^42^4}(i\overline{\tau})^{-12} e^{2\pi i \overline{\tau}} (-i\overline{\tau})^8 \nonumber \\
&= -2 V_9\iint_{\mathcal{A}} \frac{d\tau_2d\tau_1 }{2\tau_2(4\pi^2\alpha'\tau_2)^5}e^{-\frac{\beta^2}{4\pi\alpha'}\frac{\tau_1^2+\tau_2^2}{\tau_2}}e^{\pi i \tau_1} e^{3\pi \tau_2}.
\end{align}
which is the result stated in section \ref{heteroticz}.

\section{*Scattering amplitudes}

\label{appscat}
In this supplementary section we compute the scattering amplitude of one graviton and two winding ($w= \pm1$) tachyons for the bosonic string. We first present the amplitudes as computed from string theory and then we reproduce these amplitudes from the thermal scalar field theory action. To avoid awkwardness, we return to Lorentzian signature and consider the $25$-direction to be compactified with radius $R$. Upon Wick rotating the resulting action, we will see that the spacetime action has the form of the thermal scalar action (\ref{tachact}). For the indices, we will denote $M,N$ as 26-dimensional indices and $\mu,\nu$ as 25-dimensional indices.

\subsection*{Stringy amplitudes}
As the vertex operators we take
\begin{equation}
\frac{g_{c,25}}{\alpha'}:\left(\partial X^{M} \bar{\partial} X^{N} + \partial X^{N} \bar{\partial} X^{M}\right)e^{i k \cdot X}:
\end{equation}
for the graviton and KK scalar ($M=N=25$) and the winding tachyon vertex operators are given by
\begin{equation}
g_{c,25}:e^{ik_{L} \cdot X_{L}(z) + ik_{R} \cdot X_{R}(\overline{z})}:
\end{equation}
where we denoted $g_{c,25} = \frac{g_{c}}{\sqrt{2 \pi R}}$. \\
Following standard arguments \cite{Polchinski:1998rq}, we arrive at the following scattering amplitudes\footnote{Note that this differs slightly from the results in \cite{Polchinski:1998rq} due to a somewhat different normalization of the graviton vertex operators.} (where the two tachyons are wound with $w=1$ and $w=-1$)
\begin{eqnarray}
\label{graviton}
G^{\mu\nu} \to &-\pi i g_{c,25}(2\pi)^{25}\delta^{25}(k_1+k_2+k_3)k^{\mu}_{23}k^{\nu}_{23}, \\
G^{\mu25} \to &-\pi i g_{c,25}(2\pi)^{25}\delta^{25}(k_1+k_2+k_3)k^{\mu}_{23}(k^{25}_{L23}+k^{25}_{R23}), \\
\label{scalar}
G^{2525} \to &-\pi i g_{c,25}(2\pi)^{25}\delta^{25}(k_1+k_2+k_3)k^{25}_{L23}k^{25}_{R23}.
\end{eqnarray}
The last amplitude becomes (using $k^{25}_{L23} = \frac{2R}{\alpha'}$ and $k^{25}_{R23} = -\frac{2R}{\alpha'}$)
\begin{equation}
G^{2525} \to \pi i g_{c,25}(2\pi)^{25}\delta^{25}(k_1+k_2+k_3)\frac{4R^2}{\alpha'^2}.
\end{equation}

\subsection*{Field theory amplitudes}
The field theory action is
\begin{equation}
S = -\int d^{25}x \sqrt{-G_{\mu\nu}}\sqrt{G_{2525}}\left(G^{\mu\nu}\partial_{\mu}T\partial_{\nu}T^{*}+\frac{R^2 G_{2525}}{\alpha'^2}TT^{*}-\frac{4}{\alpha'}TT^{*}\right)
\end{equation}
where we have absorbed $\sqrt{2\pi R}$ in the tachyon field.\footnote{This action is the Lorentzian signature action for a complex tachyon with winding number $w = \pm 1$. Note that we do not include $G_{\mu25}$ dependence, although one can readily generalize the arguments given here to this more general case.} We now show that this action reproduces the graviton and Kaluza-Klein scalar amptitudes determined above. Expanding $G_{MN} = \eta_{MN} - 2\kappa_{25}e_{MN}f(x)$ results in
\begin{equation}
\sqrt{-G_{\mu\nu}}\sqrt{G_{2525}} = \sqrt{\det\left(\eta_{MN} - 2\kappa_{25}e_{MN}f(x)\right)} \approx 1 + \text{Tr}(- 2\kappa_{25}e_{MN}f(x)) =1
\end{equation}
because the polarization tensor is traceless for a graviton. We then expand the action resulting in
\begin{eqnarray}
S &= -\int d^{25}x \left(\eta^{\mu\nu}\partial_{\mu}T\partial_{\nu}T^{*} + 2\kappa_{25} e^{\mu\nu}f(x)\partial_{\mu}T\partial_{\nu}T^{*}  + \frac{R^2}{\alpha'^2}TT^{*} \right. \nonumber \\
&\left.- \frac{R^2 2\kappa_{25} e_{2525}}{\alpha'^2}f(x)TT^{*}-\frac{4}{\alpha'}TT^{*}\right).
\end{eqnarray}
We clearly see a kinetic term for the tachyon (first term) and two mass terms (third and fifth term). The two other terms describe interactions with the graviton. The second term corresponds to graviton-tachyon-tachyon scattering, while the fourth one corresponds to KK scalar-tachyon-tachyon scattering.\\
The (graviton)-(winding tachyon)-(winding tachyon) amplitude becomes
\begin{equation}
A^{\mu\nu} \propto  2i\kappa_{25}e^{\mu\nu}k^{2}_{\mu}k^{3}_{\nu}
= -i\frac{\kappa_{25}}{2}e^{\mu\nu}k^{23}_{\mu}k^{23}_{\nu}.
\end{equation}
When writing $\kappa_{25} = 2\pi g_{c,25}$ and including the kinematic factors, we get
\begin{equation}
\label{graviton2}
A^{\mu\nu} = -i\pi g_{c,25}e^{\mu\nu}k^{23}_{\mu}k^{23}_{\nu}(2\pi)^{25}\delta^{25}(k_1+k_2+k_3).
\end{equation}
The (KK scalar)-(winding tachyon)-(winding tachyon) amplitude becomes
\begin{equation}
A^{2525} \propto 2i\kappa_{25}e_{2525}\frac{R^2}{\alpha'^2} .
\end{equation}
We obtain
\begin{equation}
\label{scalar2}
A^{2525} =  4 i\pi g_{c,25}e^{2525}(2\pi)^{25}\delta^{25}(k_1+k_2+k_3)\frac{R^2}{\alpha'^2}.
\end{equation}
Note that $e^{2525} = e_{2525}$ since indices of $e_{MN}$ are raised and lowered with $\eta_{MN}$. \\
We conclude that the stringy amplitudes (\ref{graviton}) and (\ref{scalar}) agree with the field amplitudes (\ref{graviton2}) and (\ref{scalar2}) respectively.\footnote{If we choose discrete momentum states instead of winding states, this would result in the stringy amplitude with $k^{25}_{L23}k^{25}_{R23} = \frac{n^2}{R^2}$. So in all, this would differ by a minus sign and a factor of $\frac{R^4}{\alpha'^2}$, precisely the difference in the effective action: the minus sign comes from the $G^{2525}$ component (instead of $G_{2525}$) and the other factor comes from the T-duality step to obtain the winding action.} Wick rotating then immediately yields the thermal scalar action (\ref{tachact}). Obviously the above action was not entirely general, for instance the dilaton field or the $G^{\mu25}$ field couplings are not present. One can readily generalize the above to also include these contributions.

\section[*Extension to a background Kalb-Ramond field]{*Extension of the thermal scalar field theory to a background Kalb-Ramond field}
\label{KR}
We make a final extension to the field theory result and include also a non-zero NS-NS field. We know that to lowest order, the covering-space tachyon action does not couple to the NS-NS field. This however does not imply that there is no influence of the NS-NS background as one can readily check that the scattering amplitudes for a $B_{0i}$ component and two winding tachyons does not vanish. Hence we do expect a coupling. \\
The (discrete momentum) tachyon action for the complex field $T_{n}$ is given by
\begin{align}
S = \pi R \int d^{d-1}x \sqrt{G}e^{-2\Phi}\left(G^{ij}\partial_{i}T_{n}\partial_{j}T_{n}^{*} +\frac{n^2G^{00}}{R^2}T_{n}T_{n}^{*} \right.\nonumber \\
\left. + G^{0i}\frac{in}{R}\left(T_{n}\partial_{i}T_{n}^{*}- T_{n}^{*}\partial_{i}T_{n}\right)+ m^2T_{n}T_{n}^{*}\right),
\end{align}
The T-duality is given by
\begin{align}
G_{00} \to \frac{1}{G_{00}}&,\quad G_{0i} \to \frac{B_{0i}}{G_{00}},\quad G_{ij} \to G_{ij} - \frac{G_{0i}G_{0j}}{G_{00}} + \frac{B_{0i}B_{0j}}{G_{00}}, \nonumber \\
&\Phi \to \Phi - \frac{1}{2}\ln\left(G_{00}\right),\quad T_{n} \to T_{w}.
\end{align}
We have
\begin{equation}
\sqrt{G}e^{-2\Phi} \to \sqrt{G'}e^{-2\Phi'} = \sqrt{G_{00}'}\sqrt{G_{ij}'-\frac{G_{0i}'G_{0j}'}{G_{00}'}}e^{-2\Phi}G_{00} = \sqrt{G_{00}}e^{-2\Phi}\sqrt{G_{ij} - \frac{G_{0i}G_{0j}}{G_{00}}}.
\end{equation}
Thus we arrive at (also using $R \to \alpha'/R$)
\begin{align}
S \sim \int d^{d-1}x \sqrt{G_{ij} - \frac{G_{0i}G_{0j}}{G_{00}}}\sqrt{G_{00}}e^{-2\Phi} \nonumber \\
\left(G'^{ij}\partial_{i}T_{w}\partial_{j}T_{w}^{*}+\frac{w^2R^2G'^{00}}{\alpha'^2}T_{w}T_{w}^{*}+ G'^{0i}\frac{iwR}{\alpha'}\left(T_{w}\partial_{i}T_{w}^{*}- T_{w}^{*}\partial_{i}T_{w}\right)+ m^2T_{w}T_{w}^{*}\right).
\end{align}
Note that this field theory action indeed couples to the background NS-NS field. From here on we set $\Phi = constant$ (as we have done in all the other cases as well). \\
The term corresponding to $G'^{0i}$ in brackets needs to be written in the form $T^{*}\hat{\mathcal{O}}T$ to apply the manipulations as in section \ref{qm}. This gives schematically
\begin{equation}
\sqrt{G}G'^{0i}\left(T_{w}\partial_{i}T_{w}^{*}- T_{w}^{*}\partial_{i}T_{w}\right) \to -\partial_{i}\left(\sqrt{G}G'^{0i}\right)T_{w}^{*}T_{w} - 2\sqrt{G}G'^{0i}T_{w}^{*}\partial_{i}T_{w}.
\end{equation}
The first term is a spatial derivative term and is similar to $K(x)$ in section \ref{qm}. We thus include it in $K(x)$ and we ignore it in what follows. \\
Further following the logic from section \ref{qm}, we seek a path integral description of a system with Hamiltonian
\begin{equation}
H = p_i p_j g^{ij} + m_{local}^2 + K(x) + V^{i}p_{i}
\end{equation}
where $V^{i} = 2G'^{0i}wR/\alpha'$.
The Euclidean Lagrangian corresponding to this Hamiltonian equals\footnote{Expression (23.A.22) in \cite{Weinberg:1996kr} with $A_{ab} = 2g_{ab}$ and $B_{a} = V_{a}$.}
\begin{align}
L_E &= \frac{1}{4}\dot{x}^{i}\dot{x}^{j}g_{ij} + i\frac{V_{i}\dot{x}^{i}}{2} - \frac{V_{i}V^{i}}{4} + K(x)+ m_{local}^2\\
 &= \frac{1}{4}\dot{x}^{i}\dot{x}^{j}\bar{G}'_{ij} + iw\bar{G}'_{ij}G'^{0j}\frac{R}{\alpha'}\dot{x}^{i} - \bar{G}'_{ij}G'^{0i}G'^{0j}\frac{w^2R^2}{\alpha'^2} + \frac{w^2R^2G'^{00}}{\alpha'^2} + m^2 +K(x)
\end{align}
where we denoted $\bar{G}'_{ij}$ as the inverse to the purely spatial matrix $G'^{ij}$. This is \emph{not} the same as $G'_{ij}$ since the latter results from inverting the complete $G'^{\mu\nu}$ matrix and then looking at the spatial components.
Using matrix algebra, one can show the following identities
\begin{align}
\bar{G}'_{ij} &= G_{ij} - \frac{G_{0i}G_{0j}}{G_{00}}, \\
\bar{G}'_{ij}G'^{0i}G'^{0j} - G'^{00} &= - G_{00}, \\
\bar{G}'_{ij}G'^{0j} &= -B_{0i}.
\end{align}
The Euclidean Lagrangian reduces to (for $w = \pm1$)
\begin{equation}
L_E = \frac{1}{4}\dot{x}^{i}\dot{x}^{j}\left(G_{ij} - \frac{G_{0i}G_{0j}}{G_{00}}\right) \mp iB_{0i}\frac{R}{\alpha'}\dot{x}^{i} + G_{00}\frac{R^2}{\alpha'^2} + m^2 + K(x).
\end{equation}
Setting $R = \frac{\beta}{2\pi}$, we see that the $\dot{x}^{i}$ term reduces precisely to that given in section \ref{exten} and the other terms remain as before, in agreement with section \ref{nonstatt}. There is however one extra correction term that we included in $K(x)$. This term vanishes for zero NS-NS background. 

\chapter{Random Walks - Extensions and Examples}
\label{chex}

This chapter contains some mild generalizations of the formalism developed in the previous chapter, following closely \cite{Mertens:2014cia}. Despite being relatively straightforward, the computations presented here do provide some valuable lessons on random walks in specific settings.

%
%
%

\section{Introduction}

\noindent In the previous chapter \ref{chth}, we analyzed the method set forth by the authors of \cite{Kruczenski:2005pj} to analyze the near-Hagedorn thermodynamics of string theory in general curved spacetimes directly from the string path integral. The method explicitly describes the random walk picture of high-temperature string thermodynamics. We noted there that this random walk receives several corrections compared to the naive worldsheet dimensional reduction. \\ 

\noindent In this chapter, we will discuss several specific extensions and examples for which we will discuss the form of the thermal scalar action and the resulting near-Hagedorn random walk thermodynamics. These results illustrate the general discussion given in \cite{Mertens:2013pza} in the previous chapter and demonstrate the resolutions to some issues one might encounter in following the general treatment presented in section \ref{pathderiv}. Up to this point, we only analyzed flat Minkowski backgrounds explicitly, and hence we would like to see whether the above description really gives the correct answers in some explicit backgrounds. \\
For a general curved background, we have much less control on what precisely happens. One of the problems we face is the exact description of the $\alpha'$ corrections to the thermal scalar action. Without these, we can not obtain a correct random walk description.\\
To avoid having to deal with this issue in full detail, in this chapter we will focus on backgrounds that are geometrically trivial but have non-trivial topology. We will provide explicit solvable examples of the random walk picture. Each of these will highlight a different feature of the random walk (and the thermal scalar) description. \\

\noindent Firstly, in section \ref{dilatonbackground} we will discuss the linear dilaton background. This background is simple enough to be exactly solvable, yet it will teach us some valuable lessons about the interplay between dilatons in the spacetime action and their role in the string path integral. \\

\noindent In \cite{Grignani:2001ik}\cite{Grignani:2001hb}, the authors determine the Hagedorn temperature for a toroidally compactified flat spacetime geometry, when including a constant metric and Kalb-Ramond background. This background has applications in non-commutative open string theory \cite{Gubser:2000mf}. The Hagedorn temperature in this background was determined by studying the divergence in the modular integral using the summation-of-particles strategy to evaluate the partition function. In section \ref{torcompmodel} we will apply our previous results to find the same critical temperature from the winding tachyon perspective. We also generalize this to the most general flat space toroidal model and determine the critical behavior.\\

\noindent Then, in section \ref{stringsinbox}, we consider orbifolds of flat space, where the orbifolding is done along one of the Cartesian coordinates. Such orbifolds are stringy models of strings-in-a-box. This generalization of the unorbifolded case will teach us an interesting fact regarding the link between string models and the associated thermal scalar particle models, in particular with respect to possible boundary conditions. \\

\noindent Sections \ref{exti1} and \ref{exti2} contain some simple constructions of the uses of the thermal scalar beyond only the free energy of a closed string gas. Finally, section \ref{boosted} contains a discussion on how a boosted observer in flat space would describe the thermal scalar and thermodynamics in general. These sections are new and have not been incorporated in published research. \\


\noindent Several technical computations on toroidal models are given in the supplementary sections. \\

\section{An exactly solvable model: linear dilaton background}
\label{dilatonbackground}
To start with, we will analyze the linear dilaton model. This will teach us some lessons on how to treat dilaton backgrounds in the thermal scalar formalism and also on the influence of continuous quantum numbers on the critical temperature. Besides being a grateful toy model, this background is relevant since (among others) the asymptotic region of the $SL(2,\mathbb{R})/U(1)$ black hole (and the near-extremal NS5 black hole) is equal to such a background (see e.g. \cite{Sugawara:2012ag}\cite{Giveon:2013ica} and references therein). We will see several consistency checks to our general earlier results.

\subsection{Paradox concerning dilaton backgrounds}
There is seemingly a paradox when considering backgrounds with a non-trivial dilaton. In the gauge-fixed string path integral, the dilaton background is not present (because it multiplies the world-sheet Ricci-scalar), while in the field theory action it is always present (at least as a correction to the measure). This puzzle is of course well-known and while the dilaton does not appear explicitly in the string path integral, it does show up in worldsheet computations such as OPEs etc. It is interesting to see how this paradox is avoided in our case. We will restrict ourselves to the bosonic string. Consider flat space in $d<26$ dimensions ($G_{\mu\nu} = \delta_{\mu\nu}, B_{\mu\nu} = 0$). In order to have a consistent string background, we are required to include a linear dilaton, say in the $a$-direction:
\begin{equation}
\Phi = \sqrt{\frac{26-d}{6\alpha'}}X^{a}
\end{equation}
and this solution is exact in $\alpha'$. We analyze the near-Hagedorn one-loop thermodynamics from two points of view.\\

\noindent The dilaton does not appear directly in the particle path integral (\ref{act}). In the previous chapter we computed explicitly the corrections to the flat space result that one obtains by integrating out (rather than setting to zero) the higher worldsheet Fourier modes. Since we are now considering $D<26$ dimensions, the exact correction to the particle path integral is given by
\begin{equation}
\label{lindil}
e^{4\pi\tau_2}e^{\frac{\pi\tau_2}{6}(d-26)}
\end{equation}
for the linear dilaton background, the first factor being the (universally present) correction from the flat space tachyon and the second factor is an additional contribution solely present when the number of (flat) dimensions is less than $26$.\\

\noindent Next we analyze this background from the field theory point of view. The field theory can be taken off-shell so let us consider all backgrounds turned off except a non-trivial dilaton (whose form is generic for now). A non-trivial dilaton appears in the measure of the field theory action and it can be treated in the same way as the $G_{00}$ metric component discussed above. Using the substitution $G_{00} \to e^{-4\Phi}$ in (\ref{K}), this gives an extra contribution to the resulting particle action given by
\begin{align}
K(x) &= -\frac{3}{16}\frac{(\nabla e^{-4\Phi})^2}{e^{-8\Phi}} + \frac{\nabla^2 e^{-4\Phi}}{4e^{-4\Phi}} \\
&= (\nabla \Phi)^2 - \nabla^2 \Phi.
\end{align}
Up to this point, we have not imposed the background to be a consistent string background. To match this to the string path integral result, we need to go on-shell (the dilaton should be a linear dilaton). This immediately yields
\begin{equation}
K(x) = -\frac{d-26}{6\alpha'}
\end{equation}
and the action gets the correction, according to (\ref{corr})
\begin{equation}
\Delta S = \frac{1}{4\pi\alpha'}\int_{0}^{\tau_2}dt 4\pi^2\alpha'^2K(x) = -\pi\tau_2\frac{d-26}{6}.
\end{equation}
The extra contribution to the associated path integral (\ref{FT}) is then finally given by
\begin{equation}
\exp\left(\pi\tau_2\frac{d-26}{6}\right)
\end{equation}
and this precisely matches the extra second factor in the path integral result (\ref{lindil}). \\

\noindent It is curious to realize that in the path integral the number of dimensions gives the correction, while in the field theory the dilaton itself gives the correction. These are only linked when going on-shell. As discussed above, this has been known a long time, but it is reassuring to see it arise in this context as well. We conclude that there is no contradiction between the string path integral and the field theory when including dilaton backgrounds as soon as we use the equations of motion of the background fields. This actually provides an explicit example of our statements made previously: the field theory point of view allows us to go off-shell and provides a natural off-shell generalization of the worldsheet results. 

\subsection{Critical behavior}
Despite the fact that the string genus expansion in this background is not a good approximation, we can still formally determine the Hagedorn temperature both from the partition function and from the thermal spectrum. We will see that these again match as we expect. This exercise will learn us something valuable concerning continuous quantum numbers in conformal weights. 
The Hagedorn temperature of the linear dilaton background can be readily computed (both from the string path integral and from the field theory point of view) and is given by
\begin{equation}
\label{haglindil}
\beta_{H}^2 = 4\pi^2\alpha'\left(\frac{d-2}{6}\right).
\end{equation}
As usual, this can also be seen in the thermal string spectrum. We define a tachyon as a state that causes the (one-loop) free energy to diverge. We describe this in a general way. In a general bosonic string CFT, the one-loop partition function is given by
\begin{equation}
Z = \int_{F}\frac{d\tau_1 d\tau_2}{2\tau_2} \text{Tr}\left[q^{L_0-c/24}\bar{q}^{\bar{L_0}-\bar{c}/24}\right] = \int_{F}\frac{d\tau_1 d\tau_2}{2\tau_2}\left|\eta(\tau)\right|^4(q\bar{q})^{-\frac{1}{12}}\sum_{H_{\text{matter}}}{q^{h_i-1}\bar{q}^{\bar{h_i}-1}}.
\end{equation}
In the second equality, we sum over only the matter contributions (of the full $c=26$ matter CFT). We have isolated a $q\bar{q}$ combination, since this precisely compensates the ghost CFT in its asymptotic behavior, meaning
 \begin{equation}
Z \to \int_{F}\frac{d\tau_1 d\tau_2}{2\tau_2}\sum_{H_{\text{matter}}}{q^{h_i-1}\bar{q}^{\bar{h_i}-1}}
\end{equation}
as $\tau_2 \to \infty$. 
A tachyonic state in bosonic string theory is thus determined if the conformal dimension $h+\bar{h}$ in the matter sector is smaller than 2 (divergence for $\tau_2 \to \infty$ in $Z$) after integrating over continuous quantum numbers.\footnote{By this we mean that when considering the expression
\begin{equation}
\sum_{H_{\text{matter}}}{q^{h_i-1}\bar{q}^{\bar{h_i}-1}}
\end{equation}
there can be continuous states in the matter Hilbert space. The integral over these continuous states needs to be performed \emph{before} concluding whether there is really a divergence. For type II superstrings, the conformal dimension $h+\bar{h}$ needs to be less than $1$ to have a tachyonic state.} These continuous quantum numbers can give a non-vanishing contribution if they integrate into a $\tau_2$-dependent exponential. This indeed happens for the linear dilaton background as we now illustrate. \\
A general string state in the thermal linear dilaton background has conformal weight
\begin{align}
h &= \frac{\alpha'}{4}\left(\frac{2\pi n}{\beta} + \frac{w\beta}{2\pi\alpha'}\right)^2 + \frac{\alpha'k_i^2}{4} + \frac{\alpha'k_{a}}{4}\left(k_{a}-2iV\right) + N, \\
\bar{h} &= \frac{\alpha'}{4}\left(\frac{2\pi n}{\beta} - \frac{w\beta}{2\pi\alpha'}\right)^2 + \frac{\alpha'k_i^2}{4} + \frac{\alpha'k_{a}}{4}\left(k_{a}-2iV\right) + \bar{N},
\end{align}
where $d$ denotes the linear dilaton direction, $i$ represents all other flat directions and $V = \sqrt{\frac{26-d}{6\alpha'}}$.
The thermal scalar ($w=\pm1$, $n=0$) then has conformal weight
\begin{equation}
h = \bar{h} = \frac{\beta^2}{16\pi^2\alpha'} + \frac{\alpha'k_i^2}{4} + \frac{\alpha'k_{a}}{4}\left(k_{a}-2iV\right).
\end{equation}
In the partition function, we integrate over the transverse $k_i$ (like in the flat space case). These do not influence the Hagedorn temperature since these integrate to a square root prefactor. In the $a$-direction however, the integration over continuous quantum numbers is important since the integration over $k_a$ effectively gives an extra contribution of $\frac{26-d}{24}$ to the conformal weight. Since in this case $h=\bar{h}$, setting the conformal weight equal to 1 gives
\begin{equation}
\frac{\beta_H^2}{16\pi^2\alpha'} + \frac{26-d}{24} = 1
\end{equation}
which is the same expression as (\ref{haglindil}). Note that the Hagedorn temperature becomes infinitely large when $D=2$, so no tachyonic instability can set in in this spacetime dimension. This was known already a long time for the non-thermal tachyon. Note also that we focused on spacelike linear dilatons ($D<26$). Lightlike and timelike linear dilatons would correspond to mixing between the thermal compactification and the linear dilaton direction. We do not want to study this since these linear dilaton backgrounds are not static and thus it is meaningless to study thermodynamics with these.\\

\noindent For completeness, we present the relevant formulae for the type II superstring, which can be obtained analogously. The linear dilaton is now given by
\begin{equation}
\Phi = \sqrt{\frac{10-d}{4\alpha'}}X^{d}.
\end{equation}
The worldsheet correction term obtained in the large $\tau_2$ limit equals
\begin{equation}
e^{2\pi\tau_2}e^{\frac{\pi\tau_2}{4}(d-10)},
\end{equation}
which leads to the Hagedorn temperature
\begin{equation}
\beta_H^2 = 4\pi^2\alpha'\left(\frac{d-2}{4}\right).
\end{equation}

\section{Toroidally compactified background model}
\label{torcompmodel}
In chapter \ref{chth} we have discussed the result of the worldsheet dimensional reduction for topologically trivial spatial dimensions. We did not analyze what happens when spatial coordinates include topological identifications and it is this feature that we will analyze more explicitly in this section. To evade the discussions concerning the missed worldsheet contributions (as we summarized in section \ref{pathderiv}), we will take the background fields to be (almost) trivial in the following way. We consider flat Euclidean space where $X^1$ is a compact coordinate
\begin{equation}
X^1 \sim X^1 + 2\pi R_1.
\end{equation}
Moreover, we will introduce a constant (Euclidean) metric and Kalb-Ramond field of the following form \cite{Grignani:2001ik}\cite{Grignani:2001hb}
\begin{eqnarray}
\label{metriKR}
G_{\mu\nu}=\left[\begin{array}{cccc} 
1-A^2 & -iA & 0 & \hdots \\
-iA & 1 & 0 & \hdots \\
0 & 0 & 1 & \hdots \\
\hdots & \hdots & \hdots & \hdots  \end{array}\right], \quad\quad
B_{\mu\nu}=\left[\begin{array}{cccc} 
0 & -iB & 0 & \hdots \\
iB & 0 & 0 & \hdots \\
0 & 0 & 0 & \hdots \\
\hdots & \hdots & \hdots & \hdots  \end{array}\right].
\end{eqnarray}
where $A$ and $B$ are real constants. Note the factors of $i$ coming from the fact that we consider an Euclidean target space. Although utilizing the Euclidean metric for such spacetimes (with hence parts of the metric being imaginary) looks quite subtle, we will see that we nonetheless reproduce the correct results. This background will be a check on the Kalb-Ramond extension to the random walker in section \ref{pathderiv}. One cannot globally diagonalize the metric tensor without disrupting the periodicity of the coordinates (there is a topological obstruction, much like a constant Wilson loop around a compactified dimension). Such backgrounds can give quite non-trivial consequences for string thermodynamics (see e.g. \cite{Dienes:2012dc} for a recent analysis of the situation of heterotic and type I strings in backgrounds with constant Wilson loops around the thermal circle). As discussed in the introduction, such backgrounds are used in non-commutative open string theory \cite{Gubser:2000mf}\cite{Grignani:2001hb}. This background, despite being relatively simple to handle analytically, will demonstrate the usefulness of the path integral worldsheet dimensional reduction approach. \\
We will study all different types of closed string theories in this background near their Hagedorn temperature. Extra material on this model is provided in the supplementary section. In section \ref{spec} we search for a winding tachyon in the exact string spectrum. In \ref{sop} we check the correspondence between the free energy and the path integral on the thermal manifold. \\
Let us now follow the procedure outlined in section \ref{pathderiv} to obtain the critical behavior purely from the string path integral.

\subsection{Dominant Hagedorn behavior}
We will derive the dominant Hagedorn behavior from the path integral. In section \ref{pathderiv} we did not have a worldsheet instanton contribution for the $X^{i}$. Here we must include this for the $i=1$ component. The Fourier expansion we use is 
\begin{align}
\label{expann}
X^0(\sigma,\tau) & =   \pm \beta \tau_2 \sigma +  \sum_{n=-\infty}^\infty e^{i(2\pi n \tau_2) \sigma} X_n^0(\tau), \nonumber\\
X^1(\sigma,\tau) & =  w\beta_1\tau_2\sigma + n\beta_1\tau + \sum_{n=-\infty}^\infty e^{i(2\pi n \tau_2) \sigma} X_n^1(\tau), \nonumber \\
X^i(\sigma,\tau) & =  \sum_{n=-\infty}^\infty e^{i(2\pi n \tau_2) \sigma} X_n^i(\tau),\quad i=2\hdots d-1.
\end{align}
where $w$ and $n$ are new quantum numbers labeling the winding around the $X^1$ direction. We alert the reader that these quantum numbers have nothing to do with the thermal winding and momentum quantum numbers. This is a slightly more general setting than the one studied in the previous chapter. For $X^{0}$ we again only take the winding $\pm 1$ contribution, since we expect this mode to dominate (and we have explicitly checked that indeed this mode becomes massless at the Hagedorn temperature in section \ref{spec}). \\
We insert this expansion into the action (\ref{action1}) supplemented by the Kalb-Ramond extension (\ref{KRexten}).
From the worldsheet instanton contributions, we obtain the following non-oscillator contributions (coming from the terms in front of the series present in equation (\ref{expann}))
\begin{eqnarray}
S_{\text{non-osc}} = \frac{1}{4\pi\alpha'}\left[\left(1+\frac{\tau_1^2}{\tau_2^2}\right)\tau_2\beta^2(1-A^2) + \left(1+\frac{\tau_1^2}{\tau_2^2}\right)\left(\mp 2iA\beta\beta_1\tau_2w+w^2\beta_1^2\tau_2\right) \right. \nonumber \\
\left. +2\frac{\tau_1}{\tau_2}\left(\mp i A\beta\beta_1 n + w \beta_1^2 n \right)+\frac{\beta_1^2n^2}{\tau_2}\right]\mp \frac{\beta\beta_1 nB}{2\pi\alpha'}.
\end{eqnarray}
All that remains is to take the large $\tau_2$ limit of the worldsheet instanton sum
\begin{equation}
\sum_{n,w}\exp\left(-S_{\text{non-osc}}\right).
\end{equation}
Taking $\tau_2 \to \infty$ gives immediately that the $w=0$ contribution dominates. For the summation over $n$ the opposite happens: all terms contribute equally. We can get a handle on this by doing a Poisson resummation in $n$ 
\begin{eqnarray}
\sum_{n}\exp\left(-\frac{\beta_1^2n^2}{4\pi\alpha'\tau_2} \pm \frac{2i\tau_1A\beta\beta_1n}{4\pi\alpha'\tau_2} \pm \frac{B\beta\beta_1n}{2\pi\alpha'}\right) \nonumber \\
=\left(\frac{\alpha'\tau_2}{R_1^2}\right)^{1/2}\sum_{m}\exp\left(-\frac{4\pi^3\alpha'\tau_2 m^2}{\beta_1^2} \pm \frac{2\pi A\tau_1\beta m}{\beta_1} - \frac{\tau_1^2A^2\beta^2}{4\pi\alpha'\tau_2} \right. \nonumber \\
\left. \mp \frac{2\pi i B\tau_2m}{\beta_1} + \frac{\tau_2B^2\beta^2}{4\pi\alpha'} + i\frac{\tau_1AB\beta^2}{2\pi\alpha'}\right).
\end{eqnarray}
The third, fifth and sixth term in the exponential are independent of $m$ and the fourth term is an imaginary contribution. The first term dominates when $m=0$. When $m \neq 0$, the first term pushes the exponential to zero, regardless of what the second term does. So the sum is dominated by $m=0$.\footnote{These arguments can be made mathematically more precise: one first Poisson resums the series in $n$ (before setting $w=0$). The resulting double series in $w$ and $m$ is readily shown to converge uniformly in both parameters. This allows the large $\tau_2$ limit to be taken in the summand and the conclusion is the same.}
In all, we get the following particle action
\begin{eqnarray} 
S_{\text{part}} =  \frac{1}{4\pi \alpha'} \left[\frac{\tau_1^2}{\tau_2}\beta^2 + \tau_2\beta^2(1-A^2) - \tau_2B^2\beta^2 - 2i\tau_1AB\beta^2 - \beta_{H,\text{flat}}^2\tau_2  \right. \nonumber \\
\left. \pm 2 \frac{\tau_1}{\tau_2} \beta \int_0^1 d\tau G_{00} \partial_\tau X^0 \pm 2 \frac{\tau_1}{\tau_2} \beta \int_0^1 d\tau G_{0i} \partial_\tau X^i +
\frac{1}{\tau_2}\int_0^1 d\tau G_{\mu\nu} \partial_\tau X^\mu \partial_\tau X^\nu \right].
\end{eqnarray}
The first line contains all non-oscillator contributions. The last term in the first line is the result of the exact integration of the higher oscillator modes. This is the same as before since after extracting the non-oscillator contributions determined above (encoding the topological non-trivial aspects of the space), the remainder is the same as flat space. The last line contains the zero-modes of the coordinate fields. \\
The metric is constant (and the target fields are periodic), so out of the three terms in the second line, only the final term contributes.
When integrating this term, one is free to do a coordinate redefinition and change it to globally flat space. This allows one to integrate out the $X^{0}$ field, and all that remains is the $X^{i}$ fields that are non-interacting and living in flat space.\\
The $X^0$ integration produces a factor
\begin{equation}
\frac{\beta}{\sqrt{4\pi^2\alpha'\tau_2}}.
\end{equation}
Finally we integrate over $\tau_1$ using
\begin{equation}
\label{tau1}
\int_{-\infty}^{+\infty}d\tau_1\exp\left(-\frac{\beta^2}{4\pi\alpha'\tau_2}\tau_1^2 + i\frac{AB\beta^2}{2\pi\alpha'}\tau_1\right) = \sqrt{\frac{4\pi^2\alpha'\tau_2}{\beta^2}}\exp\left(-\frac{A^2B^2\beta^2}{4\pi\alpha'}\tau_2\right).
\end{equation}
Putting everything together, we arrive at
\begin{equation} 
\label{resComp}
Z_p = 2 \int_0^\infty \frac{d \tau_2}{2\tau_2} \left(\frac{\alpha'\tau_2}{R_1^2}\right)^{1/2} \int \left[ \mathcal{D}X \right]\exp - S_p(X) 
\end{equation}
where 
\begin{equation}
S_p = \frac{1}{4\pi \alpha'}\left[ - 16\pi^2\alpha' \tau_2  + \tau_2\beta^2(1-A^2-B^2+A^2B^2)
+\int_0^{\tau_2} dt \partial_t X^i \partial_t X^i\right].
\end{equation}
Convergence is achieved if
\begin{equation}
\beta^2(1-A^2)(1-B^2) > 16\pi^2\alpha',
\end{equation}
so the Hagedorn temperature is given by
\begin{equation}
\label{Hagtemp}
T_{H} = T_{H,\text{flat}}\sqrt{(1-A^2)(1-B^2)},
\end{equation}
exactly as predicted by either the string spectrum \ref{spec} or the divergence in the free energy \cite{Grignani:2001ik}. \\
We now make a few comments concerning this result. 
\begin{itemize}
\item{We get three correction terms to the Hagedorn behavior: schematically denoted as $A^2$, $B^2$ and $A^2B^2$ terms. It is curious to note that all of these have a different origin.
The $A^2$ is present simply from the $G_{00}$ component in the string worldsheet action.
The $B^2$ term only appears after a Poisson resummation in the variable $n$.
Finally the $A^2B^2$ term appears in the $\tau_1$ integration in the end. This is the second time we see that the $\tau_1$ integration can give us a crucial correction to the critical behavior: for the heterotic string in flat space, we saw that this was also the case in chapter \ref{chth}. Of course, from the perspective of second quantized thermal scalar field theory, there is no analog of $\tau_1$ and we expect that the thermal scalar action fully reproduces these results. We will not discuss the results from this point of view in this section.}
\item{Note that in expression (\ref{resComp}) there is a prefactor $\left(\frac{\alpha'\tau_2}{R_1^2}\right)^{1/2}$ present. What is the significance of this factor? \\
One easily sees that it precisely cancels the path integral contribution from the compact direction $X^1$. The interpretation is that given by \cite{Barbon:2004dd}.\footnote{Equations (3.6) and (3.7) in their paper.} The random walk now has one compact dimension and this changes the large $\tau_2$ form of a random walk precisely in the manner described by the previous prefactor. Note that this reinforces the interpretation of the Euclidean path integral as the random walk describing the long string.}
\item{In \cite{Grignani:2001ik} the critical behavior was analyzed from a partition function perspective in the strip. In this modular domain, a saddle point procedure is required to handle the $\tau_1$ integration. We have seen that here we reproduce their result in a technically easier way (i.e. without having to resort to saddle point methods). Also, when using the methods of \cite{Grignani:2001ik}, it appears a daunting task to extend the result to multiple toroidal dimensions (including constant backgrounds). We extend our method in section \ref{generalization} to the most general flat toroidal model (including arbitrary constant backgrounds $G_{mn}$ and $B_{mn}$) and we will find agreement with the thermal winding state found in the Euclidean spectrum. The Hagedorn temperature we find using the thermal scalar path integral formalism is:
\begin{equation}
T_{H} = T_{H,\text{flat}}\sqrt{G_{00} + G^{lk}B_{l0}B_{k0}}.
\end{equation}}
\item{For type II superstring theories, we expect the only difference to be that the Hagedorn correction is replaced by the superstring correction. So our prediction for the Hagedorn temperature is the same, except we use $T_{H,\text{flat}} = \frac{1}{\sqrt{8}\pi\sqrt{\alpha'}}$ now. This again agrees with the result in \cite{Grignani:2001ik}.}
\item{For the heterotic string theories, some aspects change. In the previous chapter \cite{Mertens:2013pza}, we have observed that the Hagedorn correction for heterotic strings (containing the $\beta_{H,\text{flat}}^2$ contribution to the particle action) is instead the following
\begin{equation}
e^{\pi i \tau_1}e^{3\pi \tau_2}.
\end{equation}
This $\tau_1$ contribution modifies the $\tau_1$ integration (\ref{tau1}) into
\begin{eqnarray}
\int_{-\infty}^{+\infty}d\tau_1\exp\left(-\frac{\beta^2}{4\pi\alpha'\tau_2}\tau_1^2 + i\frac{AB\beta^2}{2\pi\alpha'}\tau_1 + i \pi \tau_1\right) \nonumber \\
= \sqrt{\frac{4\pi^2\alpha'\tau_2}{\beta^2}}\exp\left(-\frac{A^2B^2\beta^2}{4\pi\alpha'}\tau_2 - AB\pi\tau_2 - \frac{\pi^3\alpha'\tau_2}{\beta^2}\right).
\end{eqnarray}

Taking all contributions, we see that for convergence we need
\begin{equation}
\beta^2(1-A^2)(1-B^2) + 4\pi^2AB\alpha'+\frac{4\pi^2\alpha'^2}{\beta^2} \geq 12\pi^2\alpha'.
\end{equation}
One recognizes this immediately as the same equation we wrote down in \ref{spec} for the critical radius when a certain string state becomes massless. The solutions to the preceding equation are hence the same as the Hagedorn temperature we wrote down in \ref{spec}.
The Hagedorn temperature is thus equal to
\begin{equation}
\beta_{H}^2 = \frac{6 - 2AB + 2\sqrt{8-6AB+A^2+B^2}}{(1-A^2)(1-B^2)}\pi^2\alpha'.
\end{equation}
}
\end{itemize}

\subsection{Interpretation of divergences}
\label{interpre}
\begin{figure}[h]
\centering
\includegraphics[width=0.2\textwidth]{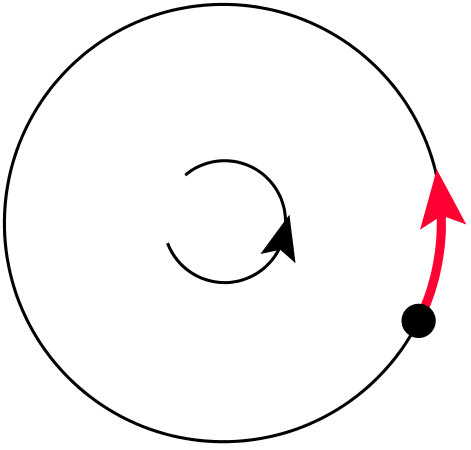}
\caption{For small $A^2$, timelike observers can remain stationary. For $A^2 \geq 1$, all observers must co-rotate with space (frame dragging effect), causing a breakdown of the field quantization along the Killing vector $\frac{\partial}{\partial t}$.}
\label{dragging}
\end{figure}

To finalize this section, let us give an interpretation of the critical values of $A$ and $B$ in this model. The toroidally compactified model has a peculiarity as $1-A^2=1$. We focus on only the $0$ and $1$ dimensions, for which the (Lorentzian signature) metric itself can be written in the standard stationary form:
\begin{equation}
ds^2 = -\alpha^2 dt^2 + (dx^1-Adt)(dx^1-Adt)
\end{equation}
with $\alpha = 1$. In the following we will use black hole terminology for convenience (despite there not being an actual black hole of course). \\
It is clear that there is no event horizon ($G_{00} - \frac{G_{01}G_{01}}{G_{11}} = -\alpha^2 \neq 0$), but there is indeed an ergoregion ($G_{00}= -(1-A^2) \stackrel{?}{=} 0$). This region is present if $A^2 \geq 1$ and is uniformly present all over space. One can interpret this as follows: space is a 1d interval with periodic identifications (a circle). With $A$ nonzero, absolute space is moving along this circle. When $A$ is sufficiently large, no timelike trajectory can remain stationary and the timelike Killing vector $\frac{\partial}{\partial t}$ that is used for quantization becomes spacelike everywhere. The situation is depicted in figure \ref{dragging}. 
In principle one can not discuss thermodynamics in terms of this observer anymore in this regime. This can also be seen in the results summarized in section \ref{pathderiv} (equation (\ref{statio})). For stationary spacetimes with Euclidean metric $G_{\mu\nu}$, one assumes both $G_{00} > 0$ and $G_{ij} - \frac{G_{0 i}G_{0 j}}{G_{00}}$ positive definite in the derivation. These conditions are related to the presence of ergoregions: $G_{00}$ vanishes at the stationary limit surface, whereas $\det\left(G_{ij} - \frac{G_{0 i}G_{0 j}}{G_{00}}\right)$ blows up there. \\ 

\noindent The $B=\pm1$ divergence on the other hand is related to the presence of a critical background field. One can relate the closed string gas in this $B$-field to an open string gas in a constant electric field around the compact dimension \cite{Grignani:2001ik}\cite{Grignani:2001hb}. For an electric field of too high strength, the string's tension cannot win from the electric field and the system becomes unstable towards string breaking. \\
We conclude by making an amusing remark. We have given intuitive explanations for both divergences. The $B$-divergence contains information on critical fields (related to the divergence of an open string gas in a constant electric field). The $A$-divergence on the other hand has to do with the spacelike behavior of the Killing vector used for describing thermodynamics. This divergence contains the information that the observer has to move faster than light when $A^2>1$ to remain stationary. It is interesting to again see that critical fields are T-dual to faster-than-light problems:\footnote{T-duality in the $X^1$-direction takes $A\leftrightarrow B$.} the DBI-action has a similar feature where the worldvolume electric fields have a critical value, T-dual to a D-brane moving faster than the speed of light. \\

\noindent We conclude that also in spaces with compact dimensions, we have a random walk behavior of the near-Hagedorn thermodynamical regime and we have derived this directly from the path integral. We have a prediction of the Hagedorn temperature in this space that agrees with both the string spectrum and with the divergence in the one-loop free energy.

\section{The thermal scalar on flat space orbifolds}
\label{stringsinbox}
A next question we ask is how boundaries affect the thermal scalar. In particular which boundary conditions should we impose at the boundary? To study this, we go one step further than the previous section and consider flat space orbifold compactifications. Because we are still geometrically in flat space, the string path integral is again exactly solvable. 

\subsection{The $S^1/\mathbb{Z}_2$ orbifold}
\begin{figure}[h]
\centering
\includegraphics[width=4cm]{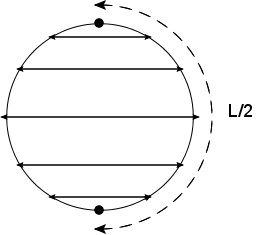}
\caption{The $S^1/\mathbb{Z}_2$ orbifold has two fixed points. It is obtained by a $\mathbb{Z}_2$ identification of a circular dimension.}
\label{orbif}
\end{figure}

Let us analyze the $S^1/\mathbb{Z}_2$ orbifold. We start with flat space (and all other background fields turned off) and make one dimension $X^1$ into a circle of circumference $L$. Then we impose the $\mathbb{Z}_2$ orbifold condition on this circle. This is a textbook model (see figure \ref{orbif}). 
First we will write down the near-Hagedorn limit of the exactly known partition function. \\
The situation outlined above corresponds to the following one-loop partition function \cite{Polchinski:1998rq}\cite{Burgess:1988qs}:
\begin{equation}
Z = Z(++) + Z(+-) + Z(-+) + Z(--),
\end{equation}
where
\begin{equation}
\label{untwisted}
Z(++) = \beta V\int_{F}\frac{d\tau_1 d\tau_2}{2\tau_2}\frac{1}{(4\pi^2\alpha'\tau_2)^{13}}\left|\eta(\tau)\right|^{-48}\sum_{p,q,n,m}\exp\left(- L^2\frac{\left|p-q\tau\right|^2}{4\pi\alpha'\tau_2} - \beta^2\frac{\left|m-n\tau\right|^2}{4\pi\alpha'\tau_2}\right)
\end{equation}
and
\begin{align}
\label{twisted}
&Z(+-) + Z(-+) + Z(--) =  \nonumber \\
&\beta A\int_{F}\frac{d\tau_1 d\tau_2}{2\tau_2}\frac{1}{(4\pi^2\alpha'\tau_2)^{25/2}}\left|\eta\right|^{-46}\left(\left|\frac{\eta}{\vartheta_2}\right| + \left|\frac{\eta}{\vartheta_3}\right| + \left|\frac{\eta}{\vartheta_4}\right| \right)\sum_{n,m}\exp\left(-\beta^2\frac{\left|m-n\tau\right|^2}{4\pi\alpha'\tau_2}\right).
\end{align}
The labels $+$ and $-$ correspond to path integral boundary conditions as we will discuss further on. In these formulas $V=A\frac{L}{2}$ where $A$ is the volume of all other (non-circular) dimensions. The quantum number $m$ and $n$ are the two wrapping numbers for the temporal dimension whereas $p$ and $q$ are the wrapping numbers for the spatial $X^1$ dimension. \\
After using the theorem of \cite{McClain:1986id}\cite{O'Brien:1987pn} to trade the fundamental domain for the strip (and dismissing the $n$-summation) and using the modular transformation $\tau \to -1/\tau$, the large $\tau_2$ limit becomes for the (+,+)-sector:
\begin{align}
\label{exact1}
&\iint_{\mathcal{A}} \frac{d\tau_1 d\tau_2}{\tau_2}\frac{\beta V}{(4\pi^2\alpha'\tau_2)^{13}}e^{4\pi\tau_2}\sqrt{\frac{4\pi^2 \alpha'\tau_2}{L^2}}e^{-\frac{\beta^2\left|\tau\right|^2}{4\pi\alpha'\tau_2}} = \iint_{\mathcal{A}} \frac{d\tau_1 d\tau_2}{2\tau_2}\frac{\beta A}{(4\pi^2\alpha'\tau_2)^{25/2}}e^{4\pi\tau_2}e^{-\frac{\beta^2\left|\tau\right|^2}{4\pi\alpha'\tau_2}},
\end{align}
and for the other sectors:
\begin{align}
\label{exact2}
\iint_{\mathcal{A}}\frac{d\tau_1 d\tau_2}{\tau_2}\frac{\beta A}{(4\pi^2\alpha'\tau_2)^{25/2}}&\left(\frac{e^{\pi\tau_2/6}}{2} + e^{-\pi\tau_2/12} + e^{-\pi\tau_2/12}\right)e^{23\pi\tau_2/6}e^{-\frac{\beta^2\left|\tau\right|^2}{4\pi\alpha'\tau_2}} \nonumber \\
&\to \iint_{\mathcal{A}}\frac{d\tau_1 d\tau_2}{2\tau_2}\frac{\beta A}{(4\pi^2\alpha'\tau_2)^{25/2}}e^{4\pi\tau_2}e^{-\frac{\beta^2\left|\tau\right|^2}{4\pi\alpha'\tau_2}},
\end{align}
upon dropping the $m=0$ temperature-independent contribution. By $\mathcal{A}$ we denoted the image of the modular transform $\tau \to -1/\tau$ on the strip region, which was shown in \cite{Kruczenski:2005pj}\cite{Mertens:2013pza}. The important part is that the $\tau_1$ integral is effectively from $-\infty$ to $+\infty$ in the large $\tau_2$ region, and one can hence readily evaluate it. Note that a factor of 2 appears due to the sum over both windings.\\

\noindent Let us now analyze this background from the string path integral point of view along the lines of chapter \ref{chth} as summarized in section \ref{pathderiv} and see whether we reproduce the above partition function. After the modular transformation $\tau \to -1/\tau$, the boundary conditions in the $X^1$ direction are
\begin{align}
X^{1}(\sigma+1/\tau_2,\tau) = \alpha X^{1}(\sigma,\tau) + p L, \\
X^{1}(\sigma,\tau+1) = \gamma X^{1}(\sigma,\tau) + q L,
\end{align}
where $\alpha$ and $\gamma$ are $\pm 1$ and represent the twisting along both torus cycles. The integers $p$ and $q$ represent the winding along both cycles. Not all of these sectors are present: in the twisted sectors with $\alpha = -1$ we have $p=q=0$. In the sector where $\alpha = 1$ and $\gamma = -1$, the $p$ and $q$ are also seen to vanish in the canonical formulation, although a priori this is not a requirement. So in all, we consider the following sectors: $(+,+,p,q)$, $(+,-,p,q)$, $(-,+,0,0)$ and $(-,-,0,0)$. We will comment on the $(+,-)$ sector further on. \\
To start, we do the worldsheet Fourier series expansion as in the previous chapter. The $\alpha = 1$ sectors are the same as before. For the $\alpha = -1$ sectors on the other hand, this is the following expansion:
\begin{align}
X^0(\sigma,\tau) & = \pm \beta \tau_2 \sigma +  \sum_{n=-\infty}^{+\infty} e^{i(2\pi n \tau_2) \sigma} X_n^0(\tau),\\
X^1(\sigma,\tau) & = \sum_{n=-\infty+1/2}^{+\infty+1/2} e^{i(2\pi n \tau_2) \sigma} X_n^i(\tau), \\
X^i(\sigma,\tau) & = \sum_{n=-\infty}^{+\infty} e^{i(2\pi n \tau_2) \sigma} X_n^i(\tau), \quad i=2\hdots d-1.
\end{align}
In the $X^{1}$ direction, half-integer modes are used. These do not have a zero-mode and are subdominant in the $\tau_2 \to \infty$ limit. These modes are the twisted sector states that are localized in the $X^{1}$ dimension. We thus drop the $\alpha = -1$ sectors.\\
Next is the sum over toroidal windings $p$ and $q$. These are dealt with just like in the toroidal model discussed in section \ref{torcompmodel}: $p=0$ from the start. \\
After this step, we have only a sum over two sectors remaining: $\gamma = \pm 1$ (and the sum over $q$). This quantum number sets the boundary conditions on the point particle path integral in the $X^1$-dimension: $X^1(\tau_2) = \gamma X^1(0) + q L$. 
Explicitly we have for the $\gamma = 1$ sector:
\begin{align}
\label{untwist}
\sum_{q\in \mathbb{Z}} \int_{X^1(0) = X^1(\tau_2) + qL} \left[\mathcal{D}X^1\right]e^{-\frac{1}{4\pi\alpha'}\int_{0}^{\tau_2}dt\left(\partial_tX^1\right)^2} &= \int_{0}^{L/2}dx \sum_{q\in \mathbb{Z}} e^{-\frac{L^2q^2}{4\pi\alpha'\tau_2}}\frac{1}{\sqrt{4\pi^2\alpha'\tau_2}} \nonumber \\
&\to \quad \frac{1}{2},
\end{align}
and for large $\tau_2$ this yields simply $1/2$.
The integration over the zero-mode is only from $0$ to $L/2$ because the particle lives in a box of length $L/2$. The $\gamma = -1$ sector gives analogously:
\begin{equation}
\label{twist}
\sum_{q\in \mathbb{Z}} \int_{X^1(0) = -X^1(\tau_2) + qL} \left[\mathcal{D}X^1\right]e^{-\frac{1}{4\pi\alpha'}\int_{0}^{\tau_2}dt\left(\partial_tX^1\right)^2} = \int_{0}^{L/2}dx \sum_{q\in \mathbb{Z}} e^{-\frac{(x+qL/2)^2}{\pi\alpha'\tau_2}}\frac{1}{\sqrt{4\pi^2\alpha'\tau_2}},
\end{equation}
which becomes\footnote{Note that in \cite{Polchinski:1998rq} only the (+,+)-sector has non-zero $p$ or $q$. The last integral looks like a sector with only zero $q$, but where the final zero-mode integral runs over the entire volume. This explicitly shows that this term is independent of $L$ as in \cite{Polchinski:1998rq}.} 
\begin{equation}
\int_{0}^{L/2}dx \sum_q e^{-\frac{(x+qL/2)^2}{\pi\alpha'\tau_2}}\frac{1}{\sqrt{4\pi^2\alpha'\tau_2}} = \int_{-\infty}^{+\infty}dx e^{-\frac{x^2}{\pi\alpha'\tau_2}}\frac{1}{\sqrt{4\pi^2\alpha'\tau_2}} = \frac{1}{2}.
\end{equation}
Finally including also $X^0$ and the other uncompactified dimensions we find agreement with the previous expressions (\ref{exact1}) and (\ref{exact2}). When we take one step back, we have for the path integral in the $X^1$ direction:
\begin{equation}
Z_{X^1} = \int_{0}^{L/2}dx\left\{\sum_q e^{-\frac{L^2q^2}{4\pi\alpha'\tau_2}}\frac{1}{\sqrt{4\pi^2\alpha'\tau_2}} + \sum_q e^{-\frac{(2x+qL)^2}{4\pi\alpha'\tau_2}}\frac{1}{\sqrt{4\pi^2\alpha'\tau_2}}\right\}.
\end{equation}
This can be interpreted as a point particle in a box corresponding to Neumann boundary conditions for a scalar field at the boundary $X^1 = 0$ and $X^1 = L/2$ \cite{Kleinert:2004ev}.

\subsubsection*{Alternative view on the boundary conditions}
One can also see that Neumann boundary conditions are the only possibility directly from the point particle perspective. Consider a non-relativistic point particle of mass $m$ in a box of length $L/2$. We are interested in the (Euclidean) propagation amplitude for a scalar spin 0 particle starting at any point in the box and returning at the same point in the limit of large propagation time $\tau_2$. Consider then Dirichlet and Neumann boundary conditions for this point particle in the box. Dirichlet boundary conditions lead to 
\begin{align}
\int \left[\mathcal{D}X\right]&e^{-\frac{m}{2}\int_{0}^{\tau_2}dt\left(\partial_tX\right)^2} 
= \int_0^{\frac{L}{2}} dx\frac{4}{L}\sum_{n=1}^{+\infty}\sin\left(\frac{2\pi n x}{L}\right)^2e^{-\frac{4\pi^2n^2}{2mL^2}\tau_2} \quad \to \, e^{-\frac{\pi^2}{2mL^2}\tau_2},
\end{align}
whereas Neumann boundary conditions give
\begin{align}
\int \left[\mathcal{D}X\right]&e^{-\frac{m}{2}\int_{0}^{\tau_2}dt\left(\partial_tX\right)^2} 
= \int_0^{\frac{L}{2}} dx\sum_{n=0}^{+\infty}\left(\frac{4}{L}-\delta_{n,0}\frac{2}{L}\right)\cos\left(\frac{\pi n x}{L}\right)^2e^{-\frac{4\pi^2n^2}{2mL^2}\tau_2} \quad \to \, 1.
\end{align}
We know that changing spacetime solely in the spatial dimensions should not change the Hagedorn temperature, an example of which we have seen in section \ref{torcompmodel}. We notice that Dirichlet boundary conditions give a Hagedorn-like correction and hence change the Hagedorn temperature. This corresponds in the point particle model to the zero-point energy of a particle in a box. Neumann boundary conditions do not have this problem and it is clear from this point of view that indeed Neumann boundary conditions are the right choice.\\
Obviously, all of the results in this section readily extend to multiple boxed dimensions when making the required modifications.

\subsection{The $\mathbb{R}/\mathbb{Z}_2$ orbifold}
\begin{figure}[h]
\centering
\includegraphics[width=4cm]{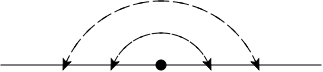}
\caption{The $\mathbb{R}/\mathbb{Z}_2$ orbifold obtained by taking $L \to \infty$ in the previous model.}
\label{orbif2}
\end{figure}

Taking $L \to \infty$, we arrive at the $\mathbb{R}/\mathbb{Z}_2$ orbifold (figure \ref{orbif2}).
Note that it is important to take $L\to\infty$ first and only then consider the large $\tau_2$ limit. For the $(+,+)$ sector we find using (\ref{untwist}):
\begin{equation}
\frac{L}{2} \frac{1}{\sqrt{4\pi^2\alpha'\tau_2}}
\end{equation}
where $L$ is the length before orbifolding (from $-\infty$ to $+\infty$) and the physical length remains $L/2$. The twisted sector computation (\ref{twist}) was exact in $\tau_2$ and remains the same (since it is independent of $L$).\footnote{We mention a technicality: taking $L \to \infty$ before rewriting the sum over $q$ as an integral, we retain the sectors $q=0$ and $q=-1$. The second term can be rewritten as an integral from $-L/2$ up to $0$ with the same integrand as the $q=0$ term. Then taking $L\to\infty$ yields the same result as before. One should be careful not to forget the $q=-1$ sector.} We notice that the $(+,+)$ sector is different than before and this has now a volume-dependence. This is consistent with the exact results, obtained by taking the large $L$ limit in (\ref{untwisted}) and (\ref{twisted}) and then performing the $\tau_2 \to \infty$ limit in the modular transformed domain:
\begin{align}
&\iint_{\mathcal{A}} \frac{d\tau_1 d\tau_2}{\tau_2}\frac{\beta V}{(4\pi^2\alpha'\tau_2)^{13}}e^{4\pi\tau_2}e^{-\frac{\beta^2\left|\tau\right|^2}{4\pi\alpha'\tau_2}}, \quad (+,+) \text{ sector}, \\
&\iint_{\mathcal{A}}\frac{d\tau_1 d\tau_2}{2\tau_2}\frac{\beta A}{(4\pi^2\alpha'\tau_2)^{25/2}}e^{4\pi\tau_2}e^{-\frac{\beta^2\left|\tau\right|^2}{4\pi\alpha'\tau_2}}, \quad \text{sum of other sectors}.
\end{align}

\subsection{Identification of the resulting particle models}
We have seen several different string approaches to making a compact model of strings in a box: toroidal models in section \ref{torcompmodel} and orbifold models in this section. We have seen that the orbifold model actually corresponds to a particle in a box with Neumann boundary conditions. Let us now make explicit the analogous particle model for the toroidally compactified string models. For the toroidal compactification discussed in the previous section (but this time \emph{without} background fields), we have the expansion
\begin{equation}
X^1(\sigma,\tau) \approx w \beta_1 \tau_2 \sigma + n \beta_1 \tau + X^{1}_{0}(\tau),
\end{equation}
where we again only retained the lowest Fourier mode which is periodic ($X^{1}_0(0) = X^{1}_0(1)$). Taking $\tau_2 \to \infty$ previously gave us that $w=0$. Instead of choosing the periodic coordinate $X^{1}_0$, we regroup the field into
\begin{equation}
X^1(\sigma,\tau) \approx X^{1}(\tau),
\end{equation}
where this last field has the property that $X^{1}(1) = nL + X^{1}(0)$ (with $L=\beta_1$). Although a trivial step, this means that there is no modification compared to the uncompactified case exhibited in section \ref{pathderiv}, except for this boundary condition. So in the end we get a particle path integral with these boundary conditions. One recognizes this immediately as a particle path integral on a circle \cite{Kleinert:2004ev} since the path integral is given by
\begin{equation}
\sum_{n=-\infty}^{+\infty}\int_{X^1(\tau_2) = X^1(0) + nL} \left[\mathcal{D}X^1\right]e^{-\frac{1}{4\pi\alpha'}\int_{0}^{\tau_2}dt\left(\partial_tX^1\right)^2} = \int_0^{L} dx\frac{1}{L}\sum_{n=-\infty}^{+\infty}\exp\left(-\frac{4\pi^3\alpha'n^2}{L^2}\tau_2\right).
\end{equation}
It asymptotes to $1$ when $\tau_2 \to \infty$ and this is the value we previously found (the Poisson dual variable was equal to zero).\footnote{See the second comment after equation (\ref{Hagtemp}).} Just like in the orbifold model, we see that as we take the $\tau_2 \to \infty$ limit in the worldsheet path integral, we can identify the intermediate expressions as point particle path integrals. In this point particle path integral, we need to take the same $\tau_2 \to \infty$ limit again, and this gives us finally the limiting Hagedorn behavior. Summarizing, we have the following correspondences between string models and particle models for the near-Hagedorn thermodynamics of the thermal scalar:
\begin{align}
\text{Toroidal model} \quad &\Leftrightarrow \quad \text{Point particle on circle}, \nonumber \\
\text{Strings in a box = } S^1/\mathbb{Z}_2 \quad &\Leftrightarrow \quad \text{Point particle in a box with Neumann BC}, \nonumber \\
\text{Strings in a halfspace = } \mathbb{R} / \mathbb{Z}_2 \quad &\Leftrightarrow \quad \text{Point particle in a halfspace with Neumann BC}. \nonumber
\end{align}
We conclude that string theory in its critical regime gives a particle theory of the thermal scalar whose boundary conditions are entirely dictated by the full string theory \cite{Burgess:1988qs}. One cannot choose other boundary conditions freely.

\section{A simple extension: free energy density}
\label{exti1}
In the previous chapter, we discussed the thermal scalar wave operator (\ref{eigen}) with spectrum 
\begin{equation}
\hat{\mathcal{O}}\psi_n = \lambda_n \psi_n.
\end{equation}
Provided the spectrum is discrete, the lowest eigenmode and -value at $n=0$ determines the Hagedorn temperature and dominant random walk behavior:
\begin{equation}
\beta F = \text{Tr}\ln(\hat{\mathcal{O}}) \approx \ln(\lambda_0).
\end{equation}
This simplification of thermodynamics can be used for several generalizations as well.
So let us make a short general detour here. One can actually find an expression for the free energy \emph{density} in a general curved background quite easily as follows.
The free energy $\beta F \approx \text{ln}(\lambda_0)$ can be just as well written in coordinate space as
\begin{equation}
\beta F = \text{Tr}\ln(\hat{\mathcal{O}}) = \int d^{d-1}x \sqrt{G} \left\langle \mathbf{x} \right|\ln(\hat{\mathcal{O}})\left|\mathbf{x}\right\rangle \approx \int d^{d-1}x \sqrt{G} \psi_0(\mathbf{x})^*\psi_0(\mathbf{x})\ln(\lambda_0),
\end{equation}
leading to an expression for the free energy density
\begin{equation}
\beta f = \psi_0(\mathbf{x})^*\psi_0(\mathbf{x})\ln(\lambda_0).
\end{equation}
To obtain the full free energy again, one should integrate this expression over the spatial slice, including the redshift factor $\sqrt{G_{00}}$. Clearly, the ground state wavefunction determines where the free energy density is localized, as we argued for before already.

\section{Another simple extension: open string random walks}
\label{exti2}
Another short computation can be performed to find the random walk picture of \emph{open} strings. \\
The one-loop free energy of an open string gas on a (Lorentzian) D-brane can be rewritten in a suggestive form. We focus on `static' branes, such that a Wick rotation of the brane one-point functions is trivial. Firstly, a closed/open duality is used to reinterpretate this as a closed string propagating between two boundary states. This can be done in general. We then can write
\begin{align}
-\beta F &= \left\langle \left\langle  p \right.\right| \frac{1}{L_0 + \bar{L}_0} \left|\left. q \right\rangle\right\rangle \\
&\approx  \left\langle \left\langle  p \right.\right| \frac{1}{\left.L_0 + \bar{L}_0\right|_{\text{th.sc.}}} \left|\left. q \right\rangle\right\rangle, \quad \beta \approx \beta_H \\
&= \iint d\mathbf{x}d\mathbf{x}' \phi_p(\mathbf{x})^* \phi_q(\mathbf{x}') \left\langle \mathbf{x}\right| \frac{1}{\left.L_0 + \bar{L}_0\right|_{\text{th.sc.}}}\left|\mathbf{x}'\right\rangle
\end{align}
where the two coordinates are possibly living in different dimensions. The states $\left|\left. q \right\rangle\right\rangle$ are D-brane states, where a string worldsheet can end. This is a random walk propagator, integrated over the two ending points with a suitable weight (the D-brane wavefunction).\\
Such formulae require a knowledge of the thermal scalar propagator in a general background (which includes its $\alpha'$-corrections). This is an issue that will take up a large portion of the remainder of this work. Assuming one knows the exact $\alpha'$-corrected thermal scalar action, one has all the ingredients to make this construction explicit. \\
For instance, if the spectrum is discrete (as will be the case in Rindler space as we will show further on), one obtains
\begin{align}
-\beta F &\approx \iint d\mathbf{x}d\mathbf{x}' \phi_p(\mathbf{x})^* \phi_q(\mathbf{x}') \sum_n \psi_n(\mathbf{x}) \frac{1}{\lambda_n} \psi_n(\mathbf{x}')^* \\
&\approx \iint d\mathbf{x}d\mathbf{x}' \phi_p(\mathbf{x})^* \phi_q(\mathbf{x}') \psi_0(\mathbf{x}) \frac{1}{\lambda_0} \psi_0(\mathbf{x}')^* \\
&= \frac{1}{\lambda_0} \left\langle \phi_p \right|\left.\psi_0\right\rangle \left\langle \psi_0 \right|\left.\phi_q\right\rangle.
\end{align}
This shows that the most dominant contribution simply equals the overlap between the thermal scalar wavefunction $\psi_0$ and the D-brane wavefunction for each end of the string.\\

\noindent Finally, the random walk picture can be made explicit by using
\begin{equation}
\frac{1}{L_0 + \bar{L}_0} = \int_{0}^{+\infty}dT e^{-(L_0 + \bar{L}_0)T}.
\end{equation}
Following standard manipulations, a Lagrangian description is readily obtained:
\begin{align}
-\beta F = \iint &d\mathbf{x}d\mathbf{x}' \phi_p(\mathbf{x})^* \phi_q(\mathbf{x}') \nonumber \\
&\times \int_{0}^{+\infty}dT \int_{\substack{X(0) = \mathbf{x}'\\ X(T) = \mathbf{x}}} \left[\mathcal{D}X\right]\sqrt{G_{ij}}e^{-\frac{1}{4\pi\alpha'}\int_{0}^{T}dt\left(G_{ij}(X)\dot{X}^{i}\dot{X}^{j} + 4\pi^2\alpha'^2\left(m_{local}^2 + K(X\right)\right)}.
\end{align}
Note that indeed for this case, one does not divide $dT$ by $T$ (in contrast to the closed random walk), since we do not need to correct for overcounting the arbitrary choice of starting point.
To fully specify the random walk, one needs to know explicitly the D-brane wavefunctions.

\section{Flat space thermal gas from boosted frame}
\label{boosted}
This section contains a discussion on how a boosted observer in flat space sees the thermal string gas and the thermal scalar. Thermodynamics in special relativity is a surprisingly confusing subject. In fact, the way temperature transforms under Lorentz boosts has been debated during the past hundred years, with no single universally accepted answer arising from these discussions. \\
In this section, we look at the partition function and use this to define the temperature. This will lead to a Lorentz-invariant temperature.\\
This section is somewhat detached from the remainder of this work and can be safely skipped.

\subsection{Boltzmann weights in boosted frame}
\begin{figure}[h]
\centering
\includegraphics[width=0.3\textwidth]{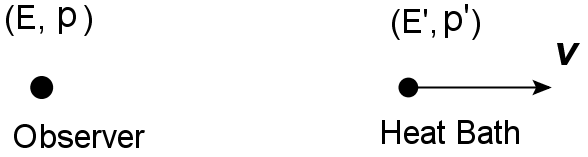}
\caption{Observer and heat bath. The heat bath is moving with relative velocity $\mathbf{v}$ w.r.t. the observer. The energy and momentum of a particle as observed by the heat bath are denoted by primed quantities.}
\label{movin}
\end{figure}

The fermion/boson density distribution function can be written in a manifestly Lorentz invariant way as
\begin{equation}
n(E', T) = \frac{1}{e^{\beta' E'}\pm 1} = \frac{1}{e^{\beta'\left|u_{\mu}p^{\mu}\right|}\pm 1},
\end{equation}
with $u^{\mu}$ the 4-velocity of the heat bath and $p^{\mu}$ the 4-momentum of the particle. $E'$ is the energy measured in the heat bath rest frame. Describing the system in the heat bath rest frame yields $u_{\mu}p^{\mu} = -E'$. Describing this in a frame where the heat bath has 3-velocity $\mathbf{v}$, one finds $u_{\mu}p^{\mu} = -\gamma(E-\mathbf{v}\cdot \mathbf{p})$, in accordance with the Lorentz boost behavior of the 4-momentum. The observers and their relative motion is illustrated in figure \ref{movin}. 

\noindent In terms of the energy $E$, one finds
\begin{equation}
\label{boltzm}
u_{\mu}p^{\mu} = -\gamma \left(E - v\cos(\theta)\sqrt{E^2-m^2}\right),
\end{equation}
where $\theta$ is the angle between $\mathbf{v}$ and $\mathbf{p}$. This expression has no Boltzmann weight interpretation. Hence from this perspective, it seems not possible to find the transformation properties of the temperature. Before continuing, let us first discuss two examples where one \emph{is} able to find a temperature transformation in this way.
\begin{itemize}
\item Massless particles: $m=0$. \\
In this case, 
\begin{equation}
u_{\mu}p^{\mu} = -\gamma \left(1 - v\cos(\theta)\right)E.
\end{equation}
The transformed temperature can then be found by
\begin{equation}
e^{-\beta' E'} = e^{-\beta' \gamma \left(1 - v\cos(\theta)\right)E},
\end{equation}
and hence
\begin{equation}
T' = \gamma\left(1-v\cos(\theta)\right) T,
\end{equation}
which relates the temperature of the observer moving with the heat bath ($T'$) to the temperature of the reference observer ($T$). 
This is the formula used in cosmology to describe the dipole anisotropy in the microwave background caused by the earth moving through the cosmic microwave background (see e.g. \cite{Weinberg:2008zzc}).

\item Suppose the spectrum of excitations is not Lorentz-invariant. Say the gas is restricted to have zero momentum: $\mathbf{p'} = \mathbf{0}$ in the heat bath rest frame. In any other frame, the 3-momentum is of the form $\mathbf{p} = \mathbf{v}E$. Then one infers that (using $E' = \gamma^{-1} E$)
\begin{equation}
e^{-\beta' E'} = e^{-\beta' \gamma^{-1} E}
\end{equation}
and hence
\begin{equation}
T = \gamma T',
\end{equation}
meaning an observer moving w.r.t. the gas measures a higher temperature ($T$) than a non-moving one ($T'$). Of course the reason is the boost non-invariance of the spectrum in this case.

\end{itemize}

\subsection{Partition function of a Lorentz-invariant system in boosted frame}
For a Lorentz-invariant system, even though the boosted Boltzmann weight itself (constructed from (\ref{boltzm})) is not of the right form, the full partition function transforms in a trivial way under Lorentz transformations and hence one can readily identify a transformed temperature directly in this way. \\
Suppose we want to determine the canonical partition function as computed by an observer, boosted w.r.t. the heat bath with velocity $-\mathbf{v}$. In the heat bath rest frame, one computes 
\begin{equation}
Z' = \text{Tr}e^{-\beta H'}
\end{equation}
where $\beta$ is the inverse temperature. The trace is over a complete set of states of the Hilbert space. The observer's definition of energy $H$ and the heat bath's $H'$ are related by the Lorentz boost formula
\begin{equation}
H' = \gamma \left(H-\mathbf{v}\cdot \mathbf{p}\right),
\end{equation}
so we find
\begin{equation}
Z' = \text{Tr}e^{-\beta \gamma \left(H-\mathbf{v}\cdot \mathbf{p}\right)}.
\end{equation}
This is actually identical to $\text{Tr}e^{-\beta H}$ where the trace is over the same Hilbert space (boosted observers have identical Hilbert spaces, the boost simply permutes all states). This is the canonical partition function defined by our observer and his Hamiltonian $H$. Hence both observers agree on the temperature and the temperature is boost-invariant. Here is the simple proof for a non-interacting massive boson. \\
The standard multi-boson arguments give for the free energy:
\begin{align}
F &= \frac{V}{\beta} \int \frac{d^{d-1}k}{(2\pi)^{d-1}}\ln\left(1-\exp(- \beta \gamma \left(E-\mathbf{v}\cdot \mathbf{k}\right))\right) \\
&= -\frac{V}{\beta} \sum_{r=1}^{+\infty}\frac{1}{r}\int \frac{d^{d-1}k}{(2\pi)^{d-1}}\exp(- r\beta \gamma\left(E-\mathbf{v}\cdot \mathbf{k}\right)).
\end{align}
To proceed, we use the identity
\begin{equation}
\frac{1}{r}\exp(-\beta \gamma r E) = \frac{\beta}{\sqrt{2\pi}}\int_{0}^{+\infty}\frac{ds}{s^{3/2}}\exp\left(-\frac{E^2s}{2}-\frac{r^2\beta^2\gamma^2}{2s}\right),
\end{equation}
and we obtain
\begin{equation}
F = -\frac{V}{\sqrt{2\pi}} \int_{0}^{+\infty}\frac{ds}{s^{3/2}} \sum_{r=1}^{+\infty} \int \frac{d^{d-1}k}{(2\pi)^{d-1}} \exp\left(-\frac{E^2 s}{2} - \frac{r^2\beta^2\gamma^2}{2s}+r\beta\gamma\mathbf{v}\cdot \mathbf{k}\right).
\end{equation}
Integrating out the momenta, one finds:
\begin{equation}
F = -V \int_{0}^{+\infty}\frac{ds}{s(2\pi s)^{d/2}} \sum_{r=1}^{+\infty}\exp\left(-\frac{m^2 s}{2} +\frac{r^2\beta^2\gamma^2 v^2}{2s}- \frac{r^2\beta^2\gamma^2}{2s}\right),
\end{equation}
which can indeed be simplified into
\begin{equation}
F = -V \int_{0}^{+\infty}\frac{ds}{s(2\pi s)^{d/2}} \sum_{r=1}^{+\infty}\exp\left(-\frac{m^2 s}{2} - \frac{r^2\beta^2}{2s}\right) = -\frac{1}{\beta}\text{ln}\text{Tr}e^{-\beta H}.
\end{equation}

\noindent Hence both partition functions are equal. This immediately implies that also the internal energies are also equal since these are related simply by a $\beta$-derivative on the partition function. Note that this leads to the somewhat counterintuitive result that the canonical internal energy is a Lorentz-invariant of the system, just like the entropy and the free energy. \\

\noindent Next let us turn to string theory and sum this free energy over the flat space string spectrum $m^2 = \frac{2}{\alpha'}\left(N+\bar{N}-2\right)$. Exactly the same arguments as usual apply in the modular strip and one finds precisely the same Hagedorn temperature for the boosted observer. \\

\noindent From the thermal manifold approach, the partition function
\begin{equation}
Z' = \text{Tr}e^{-\beta \gamma \left(H-\mathbf{v}\cdot \mathbf{p}\right)}
\end{equation}
can be interpreted as computing the vacuum amplitude on a manifold where Euclidean time and imaginary (!) space are identified simultaneously.
Standard path integral computations then lead to the factor
\begin{equation}
Z' \sim \sum_{m,w}\exp\left(-\frac{\pi\beta^2\gamma^2}{4\pi^2\alpha'\tau_2}\left|m-w\tau\right|^2 + \frac{\pi\beta^2v^2\gamma^2}{4\pi^2\alpha'\tau_2}\left|m-w\tau\right|^2\right),
\end{equation}
the first factor originating from the usual thermal wrapping, and the second factor occurs due to the simultaneous identification on the spatial circle (period $\beta v \gamma$) Wick-rotated to Lorentzian signature. For clarity, suppose the velocity $\mathbf{v}$ lies along the 1-direction. Then we note that the spacetime signature has gone from $(-+++\hdots)$ to $(+-++\hdots)$ to identify the thermal manifold. Again it is trivial that this result is the same as the conventional canonical partition function at temperature $\beta^{-1}$. \\

\noindent The thermal scalar can be interpreted as a state winding once around the thermal direction and the Wick-rotated spatial 1-direction, with weight:
\begin{equation}
h = \frac{\alpha' p^2}{4} + \frac{\alpha'}{4}\left(\frac{w \beta \gamma}{2\pi \alpha'}\right)^2 - \frac{\alpha'}{4}\left(\frac{w v\beta \gamma}{2\pi \alpha'}\right)^2 = \frac{\alpha' p^2}{4} + \frac{\alpha'}{4}\left(\frac{w \beta}{2\pi \alpha'}\right)^2,
\end{equation}
where $p^2$ denotes the remaining momentum along the other flat directions.

\section{Summary of this chapter}
In the previous chapter, we analyzed the path integral derivation of the random walk behavior of near-Hagedorn thermodynamics \cite{Kruczenski:2005pj}. In this chapter, we analyzed several examples that do not have the complications of possible curvature corrections. These will be analyzed in parts 2 and 3 of this work. \\
We started by considering the linear dilaton background. This background taught us a valuable lesson concerning the role of the dilaton field in the string path integral and the field theory action: despite treating the dilaton in a totally different way, they agree for on-shell backgrounds as it should be. Also, this background provides an explicit example where continuous quantum numbers in the thermal spectrum are important to determine the critical behavior. The toroidally compactified models showed that also in these spaces (with spatial topological identifications) we reproduce a random walk behavior. The thermal scalar approach also gives us a method to determine the Hagedorn temperature in this space. After that, we considered flat orbifold CFTs to model strings in a box. We learned that the thermal scalar path integral realizes several known topologically non-trivial particle path integrals. String theory provides us with the boundary conditions for the thermal scalar, unlike in particle theories where we are free to choose the boundary conditions. \\
We believe these examples show that the methods developed in \cite{Kruczenski:2005pj}\cite{Mertens:2013pza} lead to suggestive descriptions of the random walk picture directly from the string path integral. \\
We also looked into some small extensions (an expression for the free energy density, open string random walks and the description of a boosted observer).



\section{*Additional computations for the toroidally compactified string}
\subsection{Winding tachyon in string spectrum}
\label{spec}
We show the appearance of a state in the spectrum that becomes massless at a certain temperature in the bosonic string, the superstring and the heterotic string.
\subsubsection*{Bosonic string}
First we show that there is indeed a state in the string spectrum that becomes massless at the Hagedorn temperature. The non-linear sigma model in a general background with constant metric and Kalb-Ramond field has the following mass spectrum \cite{Polchinski:1998rq}
\begin{equation}
m^2 = \frac{1}{2\alpha'^2}G_{mn}\left(v^{m}_Lv^{n}_L+v^{m}_Rv^{n}_R\right) - \frac{4}{\alpha'}
\end{equation}
where $v_{L,R}^{m} = v^{m} \pm w^{m}R_{m}$ and $v_{m} = \frac{\alpha'n_{m}}{R_{m}}-B_{mn}w^{n}R_{n}$. Indices are raised and lowered with the $G_{mn}$ metric.
There is also a constraint
\begin{equation}
n_{m}w^{m}+N-\tilde{N} = 0.
\end{equation}
We are interested in the lowest mass state with non-zero winding in the Euclidean time direction so we set $w^{0} = 1$ and $n_{0}=n_{1}=w^{1}=0.$\footnote{Note that this satisfies the constraint with $N=\tilde{N}=0$.} With the metric and NS-NS field (\ref{metriKR}), we get the following components for the $v$'s:
\begin{align}
v^{0}_{L} = iA(-iB)R_{0}+R_{0}, \quad v^{0}_{R} = iA(-iB)R_{0}-R_{0}, \\
v^{1}_{L} = (1-A^2)(-iB)R_{0}, \quad v^{1}_{R} = (1-A^2)(-iB)R_{0}. 
\end{align}
Inserting this in the mass formula, we get
\begin{equation}
m^2 = \frac{2}{2\alpha'^2}(1-A^2)(1-B^2)\left(R_{0}\right)^2 - \frac{4}{\alpha'}.
\end{equation}
Precisely when 
\begin{equation}
\beta^2 (1-A^2)(1-B^2) = 16\pi^2\alpha'
\end{equation}
we obtain a massless state. This gives us the Hagedorn temperature
\begin{equation}
T_{H} = T_{H,\text{flat}}\sqrt{(1-A^2)(1-B^2)}
\end{equation}
where $T_{H,\text{flat}} = \frac{1}{4\pi\sqrt{\alpha'}}$ is the flat space bosonic Hagedorn temperature. 

\subsubsection*{Type II Superstring}
For type II superstrings exactly the same story holds except for the replacement 
\begin{equation}
\frac{4}{\alpha'} \to \frac{2}{\alpha'}.
\end{equation}
This only modifies the end result in that $T_{H,\text{flat}} \to \frac{1}{2\sqrt{2}\pi\sqrt{\alpha'}}$, the flat space superstring Hagedorn temperature. 

\subsubsection*{Heterotic string}
For heterotic strings, the previous result gets modified in that now the state we are looking at has $n_0 = 1/2$ to satisfy the constraint.
This slightly complicates the derivations:
\begin{align}
v^{0}_{L} = \frac{\alpha'}{2R_{0}} + iA(-iB)R_{0}+R_{0}, \quad v^{0}_{R} = \frac{\alpha'}{2R_{0}} + iA(-iB)R_{0}-R_{0}, \\
v^{1}_{L} = \frac{i\alpha'A}{2R_{0}} + (1-A^2)(-iB)R_{0}, \quad v^{1}_{R} = \frac{i\alpha'A}{2R_{0}} + (1-A^2)(-iB)R_{0}. 
\end{align}
Inserting this in the mass formula (where we now have $\frac{4}{\alpha'} \to \frac{3}{\alpha'}$), we get the following equation for a massless state
\begin{equation}
\beta^2(1-A^2)(1-B^2)+4\pi^2AB\alpha'+\frac{4\pi^4\alpha'^2}{\beta^2}=12\pi^2\alpha'.
\end{equation}
This has two solutions given by
\begin{equation}
\beta_{H\pm}^2 = \frac{6 - 2AB \pm 2\sqrt{8-6AB+A^2+B^2}}{(1-A^2)(1-B^2)}\pi^2\alpha'.
\end{equation}
Just like in the uncompactified case, the physical solution is the one with the plus sign. \\

\subsection{Path integral and summation-of-particles are equivalent for the free energy}
\label{sop}
We now only consider the bosonic string. Using the summation-of-particles strategy to calculate the free energy, the authors of \cite{Grignani:2001ik} found that the free energy is given by
\begin{align}
F = &- \sum_{n,p,l}\int_{0}^{+\infty}\frac{d\tau_2}{2\tau_2}\int_{-1/2}^{1/2}\frac{d\tau_1}{(4\pi^2\alpha'\tau_2)^{13}}\left(\frac{\alpha'\tau_2}{R_{1}^2}\right)^{1/2}\left|\eta(\tau)\right|^{-48} \nonumber\\
&\times\exp\left[-\frac{\beta^2n^2}{4\pi\alpha'\tau_2}-\pi\alpha'\tau_2\left(\frac{l^2}{R_{1}^2}+\frac{R_{1}^2p^2}{\alpha'^2}\right)-2\pi i \tau_1pl+n\beta B\frac{R_{1}p}{\alpha'} + n\beta A\frac{l}{R_{1}}\right].
\end{align}

\noindent We show that this can also be found as a torus path integral where we choose the strip as domain and we restrict attention to torus winding for $X^0$ only in the $\sigma_2$ direction. The $X^{1}$ target field includes winding in both torus cycles. So this is an explicit check that the free energy is equal to a path integral on the thermal manifold for this set-up. \\
We return to the original genus one string action (before any modular transformation)
\begin{eqnarray}
S = \frac{1}{4\pi\alpha'}\int_0^1d\sigma_1\int_0^1d\sigma_2\left[\frac{\tau_1^2+\tau_2^2}{\tau_2}\partial_1X^{\mu}\partial_1X^{\nu}G_{\mu\nu}-2\frac{\tau_1}{\tau_2}\partial_1X^{\mu}\partial_2X^{\nu}G_{\mu\nu} \right. \nonumber\\
\left. +\frac{1}{\tau_2}\partial_2X^{\mu}\partial_2X^{\nu}G_{\mu\nu}+2i\partial_1X^{\mu}\partial_2X^{\nu}B_{\mu\nu}\right].
\end{eqnarray}
with the expansions
\begin{align}
X^0(\sigma,\tau) & =  n \beta \sigma_2 +  \text{periodic}, \nonumber \\
X^1(\sigma,\tau) & =  p\beta_1\sigma_1 + m\beta_1\sigma_2 + \text{periodic}, \nonumber \\
X^i(\sigma,\tau) & =  \text{periodic}. 
\end{align}
Inserting this in the torus path integral, we get the non-oscillator contribution
\begin{eqnarray}
S_{non-osc} = \frac{1}{4\pi\alpha'}\left[\frac{\tau_1^2+\tau_2^2}{\tau_2}p^2\beta_1^2 - 2\frac{\tau_1}{\tau_2}pm\beta_1^2 + 2i\frac{\tau_1}{\tau_2}\beta\beta_1Anp + \frac{1}{\tau_2}m^2\beta_1^2 \right. \nonumber \\
\left.  + \frac{1}{\tau_2}(1-A^2)n^2\beta^2 - 2i\frac{1}{\tau_2}\beta\beta_1Anm-2B\beta\beta_1 np\right].
\end{eqnarray}
Using a Poisson resummation in $m$, we obtain precisely the previous expression (after some straightforward arithmetics and including the oscillator path integral).

\subsection{Most general flat toroidal model}
\label{generalization}
The results from section \ref{torcompmodel} can be readily generalized to the most general flat space toroidal model with an arbitrary constant metric $G_{mn}$ and arbitrary constant Kalb-Ramond field $B_{mn}$ with $m,n=0\hdots D-1$. 
Firstly, the state in the thermal spectrum is readily found as we now discuss. The general mass formula is given by \cite{Polchinski:1998rq}:
\begin{equation}
m^2 = \frac{1}{2\alpha'^2}G_{mn}\left(v^{m}_Lv^{n}_L+v^{m}_Rv^{n}_R\right) - \frac{4}{\alpha'},
\end{equation}
where $v_{L,R}^{m} = v^{m} \pm w^{m}R_{m}$ and $v_{m} = \frac{\alpha'n_{m}}{R_{m}}-B_{mn}w^{n}R_{n}$. 
For a general toroidal background, we have
\begin{align}
v_0 = 0, &\quad v_i = -B_{i0}R_0, \\
v^{0} = -G^{0i}B_{i0}R_0, &\quad v^{i} = -G^{ij}B_{j0}R_0, \\
v^{0}_{L,R} = -G^{0i}B_{i0}R_0 \pm R_0, &\quad v^{i}_{L,R} = -G^{ij}B_{j0}R_0,
\end{align}
and so
\begin{equation}
m^2 = \frac{\beta^2}{4\pi^2\alpha'^2}\left[G_{00}G^{0i}G^{0j}B_{i0}B_{j0} + G_{00} + 2G_{0i}G^{0j}G^{ik}B_{j0}B_{k0} + G_{ij}G^{il}G^{jk}B_{l0}B_{k0}\right]- \frac{4}{\alpha'},
\end{equation}
which can be simplified into
\begin{equation}
m^2 = \frac{\beta^2}{4\pi^2\alpha'^2}\left[ G_{00} + G^{lk}B_{l0}B_{k0}\right]- \frac{4}{\alpha'}.
\end{equation}
In the path integral language, the same steps as in section \ref{torcompmodel} are used. We briefly sketch the intermediate results.
\begin{itemize}
\item In the $\tau_2 \to \infty$ limit, all winding contributions around the $\sigma$-direction are dominated by $w_{i}=0$. This generalizes the case in section \ref{torcompmodel} where $w=0$ was shown to dominate.
\item The prefactors combine into $\prod_{i} \left(\frac{\alpha'\tau_2}{R_i^2}\right)^{1/2}$. After the coordinate redefinition, the flat space coordinates live in the toroidal region of size $\prod_{i}R_{i}$.
\item One needs to do $D$ Poisson resummations and then the $\tau_1$ integration. This results in 
\begin{equation}
S = \frac{\tau_2 \beta^2}{4\pi\alpha'}\left[G_{00} + \sum_{i,j}\frac{B_{i0}B_{j0}M_{0i}M_{0j}}{(\det G) M_{00}} + \sum_{i,j}\frac{B_{i0}B_{j0}M_{0i,j0}}{M_{00}}\right]
\end{equation}
where $M_{ab}$ denotes the cofactor of the matrix $G_{\mu\nu}$ and $M_{ab,cd}$ denotes the minor where rows $a$ and $c$ and columns $b$ and $d$ are deleted (with the sign $(-1)^{a+b+c+d}$).
In general the following identity is valid:\footnote{This formula is slightly more general than the so-called Desnanot-Jacobi identity, but is still a special case of the Jacobi identity \cite{Gradshteyn:2007}. }  
\begin{equation}
\frac{M_{0i}M_{0j}}{(\det G)M_{00}} + \frac{M_{0i,j0}}{M_{00}} = \frac{M_{ij}}{\det G}.
\end{equation}
This results in 
\begin{align}
S &= \frac{\tau_2 \beta^2}{4\pi\alpha'}\left[G_{00} + \sum_{i,j}\frac{M_{ij}B_{i0}B_{j0}}{\det G}\right] \\
 &=\frac{\tau_2 \beta^2}{4\pi\alpha'}\left[G_{00} + \sum_{i,j}G^{ij}B_{i0}B_{j0}\right]. 
\end{align}
This action leads to the same Hagedorn temperature as the previous calculation
\begin{equation}
T_{H} = T_{H,\text{flat}}\sqrt{G_{00} + G^{lk}B_{l0}B_{k0}}.
\end{equation}
\end{itemize}
We make several remarks concerning this result.
\begin{itemize}
\item 
Note that the mixed components $G_{0i}$ and $B_{0i}$ are both imaginary while $G_{00}$ is real and positive. The above Hagedorn temperature makes sense since $\det G$ and hence $G^{lk}$ is real so $G_{00} + G^{lk}B_{l0}B_{k0}$ is indeed a real number. It might become negative, but this simply indicates the presence of critical backgrounds, as discussed in \cite{Grignani:2001ik} and in section \ref{interpre}.
\item 
If $B_{\mu\nu} = 0$, only the metric component $G_{00}$ contributes. In particular, as long as the $i0$ components are zero, the Hagedorn temperature is unaffected (except the $G_{00}$ component). This result is more general: it holds also for smooth compactifications (with unitary CFTs) as long as the space factorizes. Note also that the Hagedorn temperature is independent of $B_{ij}$.
\item 
For type II superstrings, one again simply substitutes the correct value of $T_{H,\text{flat}}$ to find the Hagedorn temperature. 
\item
For heterotic strings, two extra terms appear in the action
\begin{align}
S_1 &= -\pi \tau_2 \sum_i \frac{B_{i0}M_{0i}}{\det G} = -\pi \tau_2 \sum_i B_{i0}G^{0i}, \\
S_2 &= \frac{\pi^3\alpha'\tau_2}{\beta^2}\frac{M_{00}}{\det G} =\frac{\pi^3\alpha'\tau_2}{\beta^2}G^{00}.
\end{align}
Explicitly working out the mass$^2$ in the thermal spectrum also gives precisely these two terms.
Finally one readily finds the Hagedorn temperature
\begin{equation}
\mbox{{\small{$\displaystyle\frac{\beta_H^2}{2\pi^2\alpha'} = \frac{3 + \sum_i B_{i0}G^{0i}+\sqrt{9 + 6\sum_i B_{i0}G^{0i} + \sum_{i,j}B_{i0}B_{j0}\left(G^{i0}G^{j0}-G^{ij}G^{00}\right)-G_{00}G^{00}}}{G_{00}+\sum_{i,j}G^{ij}B_{i0}B_{j0}}$}}}.
\end{equation} 
\end{itemize}
These results demonstrate again that a divergence in the partition function (path integral calculation) matches directly to a marginal state in the Euclidean spectrum.

\part{Application to Black Hole Horizons}

\chapter{Quantum black holes}
\label{chBH}
In this chapter, we provide the necessary background material on black holes within quantum physics, at least what is known about this and relevant for our further discussions. We will discuss briefly the Euclidean functional integral approach and the relevance of conical spaces for fluctuations around black holes. Then we briefly take a closer look at the seminal Unruh and Hawking effects. Within string theory, a thought experiment devised by Susskind plays an important role in our further story and will be treated quite elaborately. Most of the material presented in this chapter is based on the existing literature but a few new additions have been added as well.

\section{Classical black holes}
\subsection{Generalities}
Before delving into the intricacies of quantum mechanics, let us briefly recall the classical structure of black holes. \\
For simplicity, consider the Schwarzschild geometry which is of the form:
\begin{equation}
ds^2 = -\left(1-\frac{2GM}{r}\right)dt^2 + \left(1-\frac{2GM}{r}\right)^{-1}dr^2 + d\mathbf{x}^2_{\perp}.
\end{equation}
At $r=2GM=r_s$, these coordinates break down. Curvature invariants are however well-behaved and this radius, the Schwarzschild radius, is a coordinate singularity of the spacetime. At $r=0$, a genuine curvature singularity exists. The region $0<r<2GM$ is the black hole interior where every observer inevitably moves inwards towards the singularity. The structure of the classical black hole is shown below in figure \ref{clbh}.
\begin{figure}[h]
\centering
\includegraphics[width=0.3\textwidth]{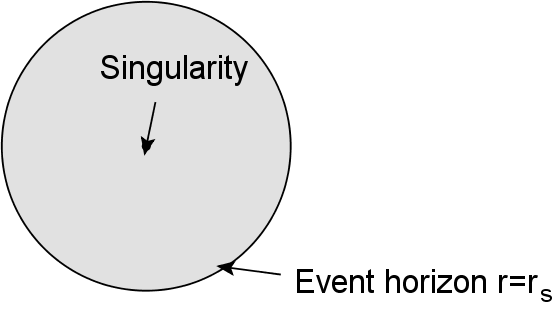}
\caption{Structure of a classical Schwarzschild black hole.}
\label{clbh}
\end{figure}
The black hole horizon $r=r_S$ is not locally special and cannot be detected by any local experiment. This is a consequence of the equivalence principle: inertial free-falling observers fall inwards and fall through the horizon only noticing problems when they approach the inner curvature singularity. That the horizon is only globally well-defined can be appreciated by noting that the horizon itself moves acausally when matter falls in: it moves outwards to anticipate matter falling in in the future. This is often translated into the slogan that we ourselves might be moving through a black hole horizon at this moment caused by matter that will fall in at a location far away in the distant future. \\
If we define a new coordinate $\rho = \sqrt{8GM(r-2GM)}$ and focus on the near horizon geometry, we reduce this metric to
\begin{equation}
\label{rinddd}
ds^2 = -\frac{\rho^2}{(4GM)^2}dt^2 + d\rho^2 + d\mathbf{x}^2_{\perp}.
\end{equation}
This is the Rindler metric and provides the near-horizon approximation of the Schwarzschild geometry. The above derivation holds as long as
\begin{equation}
\rho \ll 4GM.
\end{equation}
It is noteworthy that this is the near-horizon approximation to almost any black hole (that is non-extremal) and studying Rindler spacetime is hence quite generic for black hole physics. Hence locally there is no distinction between a uniformly accelerating observer and an observer standing still close to a black hole horizon: this is again an incarnation of the equivalence principle. \\

\noindent Observers moving parallel to $\frac{\partial}{\partial t}$ are static observers. These are called \emph{fiducial} observers (or FIDOs). The family of fiducial observers fully cover the exterior of the black hole. FIDOs close to the horizon need to apply an ever-increasing acceleration to stay put.

\subsection{Membrane Paradigm}
An interesting perspective on black holes is the so-called \emph{membrane paradigm} \cite{Thorne:1986iy}. The main idea can be conveyed quite easily. First, let us look back to classical electromagnetics, a wide-spread trick to calculate the electrostatic potential created by a point charge next to a perfectly conducting plate (assumed infinitely large) is to use the method of images: one removes the plate and inserts an (oppositely charged) image charge symmetrically on the other side of the plate. Then the potential computed for this 2-charge configuration is the same as that computed for the original set-up. The difference between the two situations is only apparent when looking beyond where the plate used to be. \\
The black hole membrane paradigm uses precisely the same philosophy for the fiducial observers: one removes the black hole interior and the horizon and replaces these by a spherical membrane hovering above the true horizon. The full effect of the intricacies of black hole near-horizon regions (where the redshift is very large) is then encoded in boundary conditions of the fields on this spherical membrane. This paradigm on black hole physics has as its major advantage that one dismisses the computationally intricate region close to the black hole (in which one is not interested in astrophysics), and thereby reduces the problem to a simpler problem that still encodes all of the physics relevant for astrophysical purposes. \\
One finds the membrane to react to outside influences as a viscous, charged fluid. The construction of the black hole theory using this perspective can be found in the textbook \cite{Thorne:1986iy} and references therein. 

\subsection{There's plenty of room near the horizon}
A feature that will be relevant in our work further on, is that infalling (non-stringy) matter tends to flatten as it approaches the horizon, which has implications on the amount of information that can be stored near the black hole horizon. We first note that the proper distance to the Schwarzschild horizon is finite, hence at first sight only a finite amount of information can be stored in any region extending to some fixed $r'$ outside the horizon. This is not true as infalling matter Lorentz contracts by an amount precisely high enough to be able to store an arbitrary amount of information close to the horizon. \\
Suppose infalling matter moving along a radial ingoing geodesic is described by a FIDO along the trajectory close to the horizon. The Lorentz contraction observed by this FIDO is then given by
\begin{equation}
L' = \frac{L}{\gamma} \sim L \rho(t),
\end{equation}
where $\rho(t) =\rho(0) e^{-\frac{t}{4GM}}$. This radial Rindler coordinate is also the proper distance to the horizon. The observed length $L'$ becomes zero as the particle approaches the horizon, its velocity approaches the speed of light as observed by the FIDO close to the horizon. We notice that the length decreases by the same factor as the distance to the horizon, allowing an arbitrary amount of data close to the horizon even though the proper distance is finite.

\section{Canonical Quantum Gravity versus String Theory}
In this section we present the main formalism to deal with quantum gravity from a Euclidean perspective, both the field theory side of the story and the string theory side. In particular, our main focus is on the differences between QFT and string theory.
\subsection{General formalism}
In the canonical approach to quantum gravity, one defines the partition function as a Euclidean path integral over matter fields and the metric as:
\begin{equation}
Z(\beta) = N\int_{\mathcal{G}} \left[\mathcal{D}g\right]\int\left[\mathcal{D}\phi\right] e^{-S\left[g,\phi\right]}.
\end{equation}
In writing this, $\mathcal{G}$ denotes the space of Euclidean metrics on the manifold $\mathcal{M}$. This space $\mathcal{G}$ is restricted by boundary conditions (such as for instance a prescribed asymptotic circumference of the time circle $\beta$). The manifolds are allowed to have different topology though (which was a long-standing puzzle in quantum gravity).\footnote{For field theory it is clear from cluster decomposition that different topological sectors are needed, but for gravity this is not clear a priori.} On this note, let us remark that in the past several semiclassical decay processes were considered for the Kaluza-Klein vacuum in \cite{Witten:1981gj} and hot flat space in \cite{Gross:1982cv}, both of which yielded results that support the presence of different topological sectors in the path integral. More conclusive evidence in favor of this was later found in string theory \cite{Aspinwall:1993nu}\cite{Witten:1993yc}, where explicit topologically changing processes were observed (which are perfectly smooth in string language but violent in geometry). \\ 
The semiclassical approach consists in expanding around a saddle point of the metric in the above (formal) integral: $ g = \bar{g} + \tilde{g}$. The partition function is then approximated by
\begin{equation}
Z(\beta) \approx N \sum_{\text{saddles }\bar{g}}\int \left[\mathcal{D}\tilde{g}\right]\left[\mathcal{D}\phi\right] e^{-S\left[\bar{g}+\tilde{g},\phi\right]},
\end{equation}
where we can expand:
\begin{equation}
S\left[\bar{g}+\tilde{g},\phi\right] \approx S\left[\bar{g},0\right] + S\left[\bar{g},\phi\right] + \int_{\mathcal{M}} \left.\frac{\delta S}{\delta g}\right|_{\bar{g}}\tilde{g} + \hdots,
\end{equation}
where additional terms corresponding to higher order graviton fluctuations have not been written. The first term is the classical term, the other terms represent the matter and graviton fluctuations in a fixed background. \\
For each saddle, the partition function factorizes in a classical part $e^{-S\left[\bar{g},0\right]}$, obtained by evaluating the on-shell action, and the quantum fluctuations around this fixed background possibly expanded in a perturbation series. \\

\noindent Within string theory, one evaluates the stringy path integral on the thermal manifold. This path integral has a standard genus expansion:
\begin{figure}[h]
\centering
\includegraphics[width=0.8\textwidth]{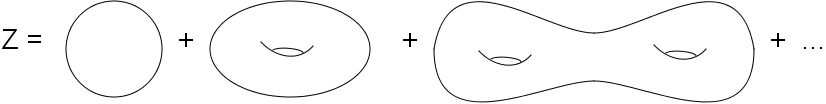}
\caption{Genus expansion of the Euclidean path integral.}
\label{genusexp2}
\end{figure}
Starting with genus 1, these can be identified with the quantum fluctuations around a fixed background (just like in QFT). The genus zero result is new: it is not present in field theory and should coincide with the ($\alpha'$-corrected) on-shell tree-level effective spacetime action \cite{Tseytlin:1988tv}\cite{Susskind:1994sm}. \\

\noindent Typically, for instance in flat space, the genus zero contribution vanishes. Likewise, the on-shell action in flat space vanishes (since all the fields are turned off). This is not the case for black hole backgrounds to which we turn next.

\subsection{Black hole thermal manifolds}
For the special case of a black hole, the Euclidean background is cigar-shaped with the horizon at the tip of the cigar. The interior of the black hole is not present on the Euclidean background. Unlike a conventional thermal manifold, in this case the Euclidean geometry has a natural value of $\beta$. This can be readily seen from the near-horizon Rindler geometry (\ref{rinddd}) continued to Euclidean signature:
\begin{equation}
ds^2 = +\frac{\rho^2}{(4GM)^2}d\tau^2 + d\rho^2 + d\mathbf{x}^2_{\perp}.
\end{equation}
This space is simply flat space written in polar coordinates! This requires $\tau \sim \tau + \beta_R$ where $\beta_R = 8\pi GM$ and any other periodicity would result in a conical singularity of the manifold. It so happens that this temperature is precisely the \emph{Hawking temperature} of the black hole $\beta_{\text{Hawking}} = 8\pi GM$ and this line of thought provides a quick way of obtaining this result. The Euclidean geometry at the Hawking temperature is shown below in figure \ref{cigarfig}.
\begin{figure}[h]
\centering
\includegraphics[width=0.3\textwidth]{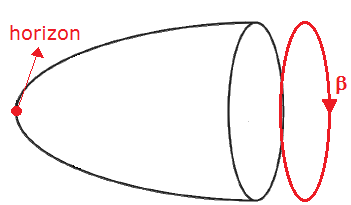}
\caption{The thermal manifold for a black hole is cigar-shaped. The asymptotic circumference shows the Hawking temperature. The tip of the cigar is the black hole horizon where the manifold smoothly caps off. The interior of the black hole is not represented on the thermal manifold.}
\label{cigarfig}
\end{figure}

\subsection{Tree level contribution to black hole entropy}
Different ideas and perspectives exist on the lowest order (tree-level) contribution. As discussed above, the tree level contribution is the tricky one to identify the microscopical degrees of freedom. The tree level Bekenstein-Hawking entropy should also follow from the effective tree level action. \\
Within the Euclidean approach to gravity, several methods to determine $S_{BH}$ exist (a concise overview of these methods can be found in \cite{Balasubramanian:2013rqa}). We focus on the simplest gravitational action:
\begin{equation}
S = \frac{1}{16\pi G}\int_{\mathcal{M}} R + \frac{1}{8\pi G}\int_{\partial \mathcal{M}} K,
\end{equation}
the Einstein-Hilbert action and the corresponding Gibbons-Hawking boundary term. The most straightforward way to get to the entropy is to compute the on-shell action, which immediately gives the free energy as $S_{\text{on-shell}} = \beta F$. The black hole entropy can then be computed in one of two ways. 
\subsubsection{First method: on-shell}
One can use
\begin{equation}
S = \beta M - \beta F,
\end{equation}
or evaluate
\begin{equation}
S = \left(\beta\partial_{\beta}-1\right)\left(\beta F\right) = \beta^2\partial_{\beta}F,
\end{equation}
where the black hole mass $M(\beta)$ is given \emph{on-shell} as a function of $\beta$. This second formula keeps fixed the on-shell relation of the black hole background. In both cases, the free energy comes from the Gibbons-Hawking boundary term only, as the Einstein-Hilbert action vanishes on a smooth Ricci-flat manifold.
\subsubsection{Second method: off-shell}
\begin{figure}[h]
\centering
\includegraphics[width=0.35\linewidth]{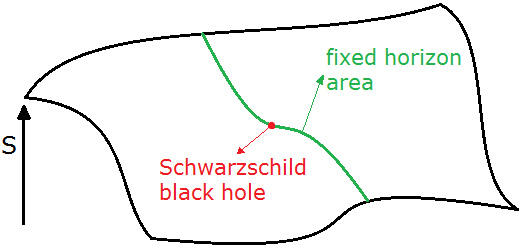}
\caption{Configuration space of solutions. The classical action is extremized at saddle points. The Schwarzschild solution is shown as a red dot. The green slice represents the restriction to solutions that have the same horizon area. Within this green slice, a 1-dimensional connected subspace of (constrained) saddle points exists passing through the Schwarzschild extremum that represents the conical spaces (not shown).}
\label{saddles}
\end{figure}

An alternative approach is to vary $\beta$ \emph{without} changing $M$. This induces a conical singularity at the tip of the cigar. \\
There are some subleties when considering Euclidean black holes with $\beta \neq \beta_{\text{Hawking}}$. Manifolds with $\beta \neq \beta_{\text{Hawking}}$ correspond to non-equilibrium Euclidean configurations, where the black hole either accretes more matter or emits radiation (using Hawking's emission process, to be explained further on) to reach its equilibrium configuration. \\
It is immediately seen that changing $\beta$ creates a conical singularity at the tip of the cigar, and the resulting space is not a solution of Einstein's equations (not a valid saddle point). There are several ways to handle this. \\
\noindent When studying conical spaces on their own, then one needs to introduce an explicit source of stress-energy at the conical singularity. \\
\noindent In thermodynamics on the other hand, one computes the entropy by varying the temperature infinitesimally away from the (non-conical) saddle point: $S = \beta^2 \partial_{\beta}F$. For this case, one does not need to introduce an explicit source at the conical tip. As was shown in \cite{Susskind:1994sm}\cite{Carlip:1993sa}, when varying the temperature in thermodynamics, we should keep fixed all other quantities in our space, including the horizon area $A$.\footnote{The thermodynamic entropy involves partial derivatives in $\beta$ after all.} Thus we vary infinitesimally (in $\beta$) to neighboring saddle points that have the same horizon area. Insisting on a fixed horizon area $A$ (the same as that for $\beta = \beta_{\text{Hawking}}$), we introduce a Lagrange multiplier into the action which effectively introduces an energy density at the tip of the cigar (a cosmic string), which causes conical spaces to be solutions to Einstein's equations and hence be valid saddle points. It is within this class of solutions that the temperature variation takes place. The Lagrange multiplier becomes linked to the conical deficit, and for any value of the Lagrange multiplier one gets a saddle point when a conical singularity is introduced at the origin in an otherwise smooth metric. The situation is sketched in the cartoon below (figure \ref{saddles}). 

\noindent This solves an initial worry one might have in that the off-shell (conical) black holes are not even stationary points of the path integral. \\
Thus the spaces needed to compute thermodynamics are the conical spaces (sourced by some energy-momentum at the tip of the cone). \\ 

\noindent In this approach, the Gibbons-Hawking boundary term is irrelevant (since it is simply proportional to $\beta$), but the Einstein-Hilbert term has a contribution from the conical tip $\left(\sqrt{g^{(2)}} R^{(2)} = 2(2\pi-\beta)\delta^{(2)}(x)\right)$:
\begin{equation}
\int_{\mathcal{M}} R = 2A (2\pi-\beta),
\end{equation}
to be found by using the Gauss-Bonnet formula on a flat cone. This yields $S_{\text{on-shell}} = -\frac{(2\pi-\beta)A}{8\pi G} = \beta F$ and using the above formula one finds
\begin{equation}
S = \frac{A}{4G},
\end{equation} 
also from this perspective. \\

\noindent Of course, these points of view are within classical field theory, and it is instructive to look at stringy and quantum extensions to this story. \\
What do we expect? \\
Firstly, stringy corrections (in $\alpha'$) provide corrections to General Relativity and it is expected that one obtains Wald's entropy instead. This should arise by evaluating the full tree-level string effective action on-shell. Secondly, loop quantum corrections are expected to contribute to the entropy in a sense that was discovered first by Susskind and Uglum in \cite{Susskind:1994sm} and which will be discussed in section \ref{flucttt} below. \\

\noindent At higher loops, the background geometry is held fixed (by definition). This means $M$ is fixed. The temperature however is (in principle) allowed to change for the thermal trace of fluctuations on top of the background, which implies studying quantum field theory on a conical space. The on-shell method discussed above seems not to be suitable for studying quantum fluctuations around a classical black hole geometry. \\

\subsection{Conical spaces in string theory}

\noindent For string theory, one starts with an on-shell background and then propagates strings on these. The tree level contribution reduces to the evaluation of the on-shell action (to all orders in $\alpha'$). Again conical spaces need to be considered to obtain the thermodynamical entropy, where these spaces should include a stress tensor at the tip of the cone again. \\
The string background equations are not the Einstein equations but are $\alpha'$-corrected. To lowest order in $\alpha'$ \emph{any} conical space can solve Einstein's equations (as long as it is sourced by a suitable energy of a cosmic string). Within the full string theory however, the only conical spaces that one can deal with are the $\mathbb{C}/\mathbb{Z}_N$ orbifolds with periodicity $\beta = 2\pi/N$ for integer $N$. In this case, string theory manages to be on-shell (with a suitable energy density inserted at the tip of the cone again). Curiously, other deficit angles do not seem to solve the full ($\alpha'$-corrected) string background equations of motion. The difficulty hence lies in the higher $\alpha'$-corrections that appear to somehow conspire to prefer these discrete values. It would be very interesting to obtain some direct evidence for this (at for instance second order in $\alpha'$), but the effective action containing degrees of freedom at the tip of the cone is not known at higher orders. Thus we are restricted to only a discrete set of cones for which we can apply the above story. \\

\noindent Note though that there is a change of language when going from field theory to string theory. In string theory, the genus zero partition function should coincide with the (on-shell) tree level action contribution to the free energy. It is in the language of the latter that the above discussion of keeping fixed the horizon area applies. The link between genus zero string theory and on-shell tree level field theory is made in \cite{Tseytlin:1988tv}\cite{Susskind:1994sm}.\footnote{Although it should be noted that the relation is more complicated than the simple equality we are envisioning here.} With the latter link in mind, the resemblence between field theory and string theory can be made more apparent.
At tree level, both reduce to the on-shell evaluation of the action (this action is vastly more complicated for string theory). Higher loop orders (quantum fluctuations) then correspond to propagating particles or strings in a fixed background (that is a solution of the tree level equations of motion). \\
In both cases, the tree level backgrounds relevant for thermodynamics include conical spaces. \\ 

\subsection{Effective string action and Bekenstein-Hawking entropy}

\noindent Going beyond Einstein-Hilbert gravity, Dabholkar \cite{Dabholkar:2001if} computed the tree-level black hole entropy within classical string theory (to all orders in $\alpha'$), using a conical manifold approach. The method differs from the conical method described above in two ways: the source of the conical singularity is explicitly taken into account and the on-shell action is renormalized by subtracting the flat space action. Consider first the effective spacetime action, to lowest order in $\alpha'$:
\begin{equation}
S = \frac{1}{16\pi G} \int_{\mathcal{M}} \sqrt{g}e^{-2\Phi}\left[R + 4 (\nabla \phi)^2 - \delta^{(2)}(x)V(T)\right] + \frac{1}{8\pi G} \int_{\partial \mathcal{M}}\sqrt{h}e^{-2\Phi} K,
\end{equation}
where the Gibbons-Hawking boundary term has been inserted to assert a valid variational principle. \\
In this action, a potential of twisted tachyons is included at the origin $x=0$ to source the conical singularity. 
We already know that the flat cone (with opening angle $2\pi/N$) must be a solution to the all-order in $\alpha'$ equations of motion, as we can apparantly consistently propagate strings on these. This geometry is accompanied by a twisted tachyon vev at the tip of the cone (supporting the conical singularity) and a constant dilaton. To lowest order in $\alpha'$, it is immediate that a constant dilaton implies a vanishing bulk on-shell action. A moment's thought shows that this property is continued to all orders in $\alpha'$. Thus the entire contribution comes from the boundary action. \\
At large distances, we have $K=\frac{1}{r}$ and the area of the boundary is $\frac{2\pi A}{N}r$, leading immediately to an on-shell action of 
\begin{equation}
S_{\text{on-shell}} = \frac{A}{4GN} - \frac{A}{4G},
\end{equation}
where the flat space action has been subtracted. 
The free energy of this solution hence becomes
\begin{equation}
F = \frac{A}{4G}\left(\frac{1-N}{2\pi}\right),
\end{equation}
leading to a thermodynamic entropy
\begin{equation}
S = \beta^2 \partial_\beta F = -2\pi \partial_N F = \frac{A}{4G},
\end{equation}
which is indeed the tree-level Bekenstein-Hawking entropy, but this time derived within string theory.\\

\noindent To conclude this paragraph, we present a puzzle and its probable resolution. The above discussion was $\alpha'$-exact, but in the end one obtains the Bekenstein-Hawking entropy and not the Wald entropy. How is this possible? This discrepancy can be traced back to the Gibbons-Hawking boundary term, which is suitable for Einstein-Hilbert gravity, but not for corrections to it. Presumably the correct boundary term would yield the Wald entropy in the end. \\
Note though that for Rindler space, the Bekenstein and Wald entropy are equal.

\subsection{Relevance of open strings on the horizon}

In string theory, Susskind and Uglum argued that exotic open strings exist which have their endpoints fixed on the black hole horizon \cite{Susskind:1994sm}. Their argument uses Euclidean time slices of the string partition functions. \\
At genus 0, a sphere that contains the origin is time-sliced as a semi-circular curve as shown in figure \ref{genus0} below. This is to be interpreted a fixed time snapshot of the computation. So it can be described as an open string that is rotated around the black hole horizon. 
\begin{figure}[h]
\centering
\includegraphics[width=0.2\textwidth]{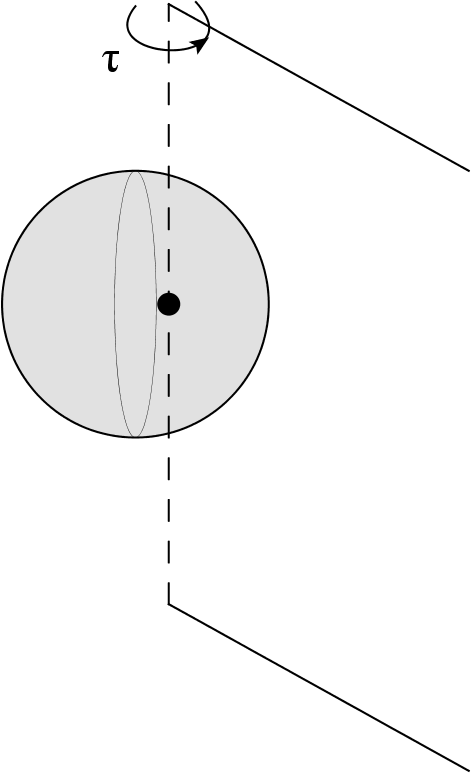}
\caption{Genus 0 sphere graph that contains the origin. Euclidean time corresponds to rotation around the dashed vertical axis. Fixed Euclidean time slices (shown as the half-plane bounded by the dashed vertical axis and the two full lines) contain a semi-circle.}
\label{genus0}
\end{figure}
This shows, at least qualitatively, that the microscopic degrees of freedom responsible for the tree-level entropy are open strings on the horizon. \\

\begin{figure}[h]
\centering
\includegraphics[width=0.3\textwidth]{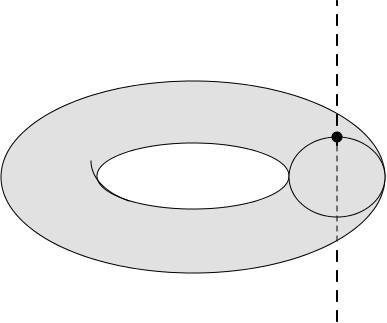}
\caption{Genus 1 torus graph that contains the origin. Fixed Euclidean time slices describe the process of emission and reabsorption of a closed string by an open string on the horizon.}
\label{genusone}
\end{figure}

\noindent At genus 1, a torus that contains the origin is time-sliced as an open string that emits and reabsorbs a closed string as shown in figure \ref{genusone}. 
This interpretation extends to arbitary genus. Susskind and Uglum argued that these strange interactions between open and closed strings cause the one-loop result to disagree with the free closed string Hamiltonian trace $\text{Tr}e^{-\beta H}$ in Rindler space (were one able to compute it). \\
Of course this argument is made in the Euclidean signature and it raises some questions on the real-time interpretation as given above. \\

\begin{figure}[h]
\centering
\includegraphics[width=0.4\textwidth]{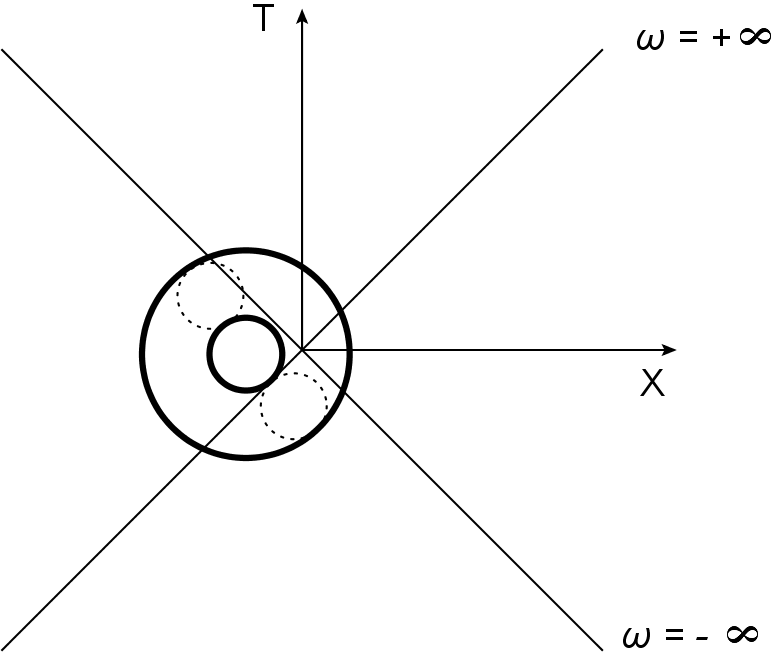}
\caption{String vacuum fluctuation that crosses the Rindler origin. This is seen by the Rindler observer as an open string with ending points fixed (and immobile) on the horizon.}
\label{topside2}
\end{figure}

\noindent To that end, it is instructive to give a very intuitive explanation on this same phenonomenon, fully from the Lorentzian point of view. Suppose we coordinatize our flat metric as
\begin{equation}
ds^2 = - dT^2 + dX^2 = -\rho^2d\omega^2 + d\rho^2.
\end{equation}
The coordinate frame is shown below in figure \ref{topsidee}. Rindler space only covers a quarter of 2d Minkowski space. Constant Rindler time slices are semi-infinite lines originating at the origin. The infinite past and infinite future (according to the Rindler observer) are the two diagonal lines drawn in the figure. \\
As we will discuss further on in this chapter, a fiducial observer in Rindler space experiences the Minkowski vacuum as being thermally populated. This effect can be understood intuitively by considering how vacuum loops are described by the accelerating observer \cite{Susskind:2005js}. The heat bath seen by the Rindler observer arises because of eternal vacuum fluctuations as shown in figure \ref{topsidee} (a). 
\begin{figure}[h]
\centering
\begin{minipage}{0.4\textwidth}
\centering
\includegraphics[width=\textwidth]{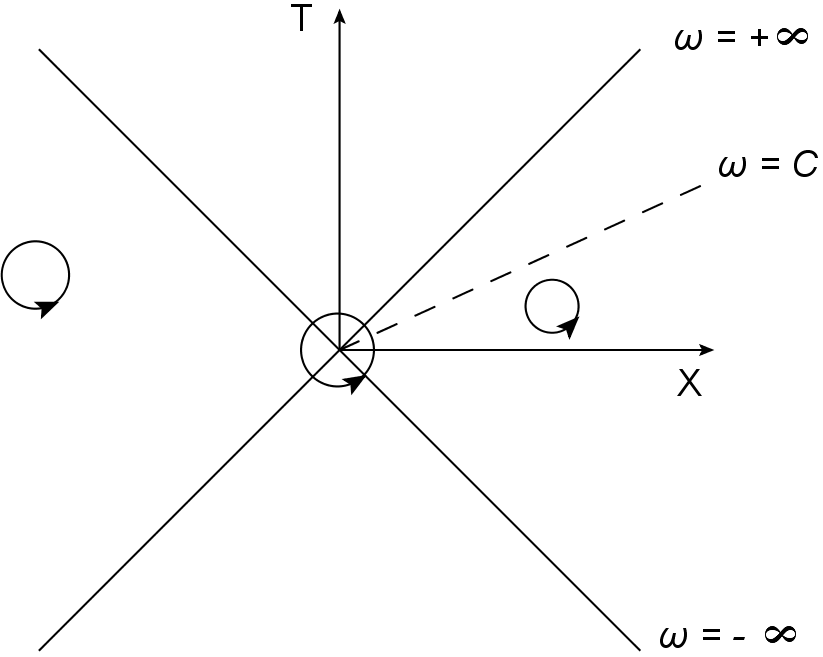}
\caption*{(a)}
\end{minipage}
\begin{minipage}{0.4\textwidth}
\centering
\includegraphics[width=\textwidth]{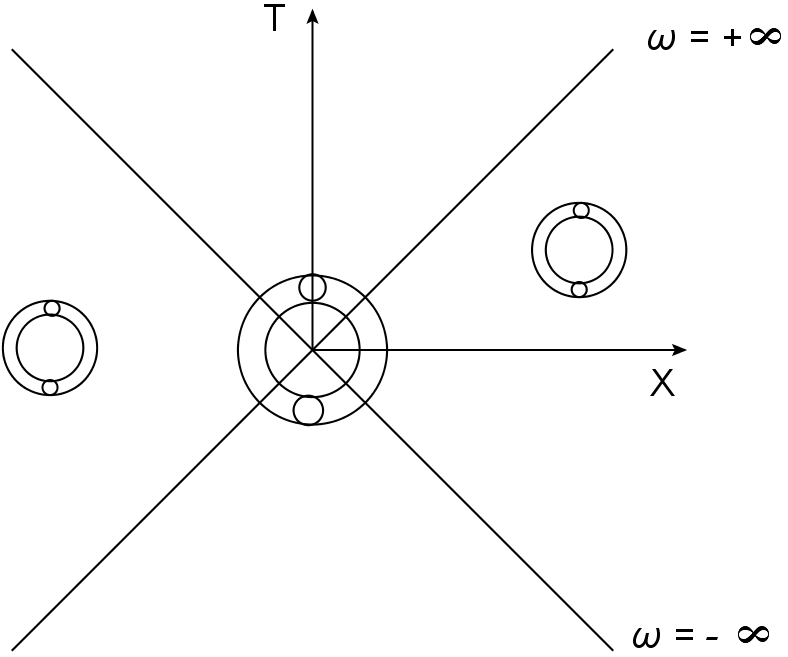}
\caption*{(b)}
\end{minipage}
\caption{(a) Vacuum fluctuations in QFT. According to the Rindler observer, the leftmost fluctuation is unseen. The rightmost fluctuation is seen as a vacuum fluctuation by the Rindler observer as well. The middle fluctuation is important: this is long-lived according to the Rindler observer. (b) The same diagrams within string theory. The interpretations are the same.}
\label{topsidee}
\end{figure}
The vacuum loop that encloses the origin is the important one: it is viewed by an accelerating observer as being eternal. The vacuum fluctuations become real close to the Rindler origin. The analogous diagrams in string theory have been drawn as well in figure \ref{topsidee} (b). \\
In string theory however, a second set of embeddings are possible, leading to an open string gas with fixed endpoints on the horizon as shown in figure \ref{topside2}.

\noindent As mentioned above, the presence of such exotic open strings was argued for from the thermal manifold \cite{Susskind:1994sm}. This intuitive Lorentzian approach to the Rindler particles directly leads to the presence of such states without having to contemplate the transition from Euclidean (thermal) manifold to the Lorentzian case. Of course, the detailed nature of these virtual particles requires a more careful study. We will come back to this strange feature several times in the remainder of this work.


\section{Fluctuations around the black hole}
\label{flucttt}
Aside from the tree-level contribution, a very important role is played by the quantum fields (or strings) around the black hole. These provide the higher loop corrections to the computed quantities and also exhibit very deep relations between thermal physics and black holes. We start first with a brief discussion on higher loop corrections.

\subsection{Renormalization of Newton's constant}
How do we expect higher loop corrections to modify the story? The answer was uncovered by Susskind and Uglum \cite{Susskind:1994sm} where they showed that upon integrating out the contribution of matter fields to the canonical partition function, these renormalize the gravitational coupling and yield $S = \frac{A}{4G_R}$. This shows how loop corrections to black hole thermodynamics affect the story: they simply renormalize Newton's constant.\\
There is a fundamental difference here between QFT and string theory. It is known that field theory has too many degrees of freedom close to the black hole horizon. When considering the entropy of quantum fields around the black hole as a quantum correction to Bekenstein's formula, one finds a divergent quantity \cite{'tHooft:1984re}\cite{Susskind:1994sm}\cite{Barbon:1994ej}\cite{Emparan:1994qa}. Physically, this arises because one can store too much information close to the black hole, infinitely much in fact. The reason is the longitudinal contraction of matter as it falls in. We will encounter a manifestation of this infinity further on. \\
Strings on the other hand are expected to be UV finite, and indeed as we will review further on, their longitudinal contraction is offset by an expansion allowing only a finite amount of information to be stored close to the black hole horizon, in corroboration with the holographic principle \cite{Susskind:1993ws}\cite{Susskind:1993ki}\cite{Susskind:1994uu}\cite{Susskind:1993aa}. Hence the quantum corrections to Newton's constant are finite in string theory.

\subsection{Unruh effect}
The two effects that are most important when considering quantum field theory around black holes, are the \emph{Unruh effect} and the \emph{Hawking effect}, both of which are derived in non-interacting field theory, although arguments can be given to extend these results to the interacting regime \cite{Gibbons:1976es}. \\
We focus here on the flat space Unruh effect in 1+1 dimensional Minkowski space. The \emph{Unruh effect} tells us that the Minkowski vacuum is described according to fiducial observers as being thermally populated. We take the flat metric as follows:
\begin{equation}
ds^2 = -dt^2 +dx^2 = -\rho^2 d\tau^2 + d\rho^2.
\end{equation} 
The thermal Unruh effect can be deduced in several ways. Historically, the first method uses the Bogoliubov transformation to link inertial with fiducial modes \cite{Fulling:1972md}\cite{Davies:1974th}\cite{Unruh:1976db}. We will here present a more recent derivation that is much more geometrical and general (it does not require a detailed knowledge about the specific field theory at hand) \cite{Callan:1994py}\cite{Susskind:2005js}. \\

\noindent For simplicity, first we look at standard QM in a 1+1 dimensional flat space. The Euclidean propagation amplitude between points $x$ and $y$ is given by
\begin{equation}
\left\langle y,T\right|\left.x,0\right\rangle \sim \int_{X(0)=x, \, X(T)=y}\left[\mathcal{D}X\right]e^{-S}.
\end{equation}
Taking $T$ to infinity, allows us to extract the ground state wavefunction as follows:
\begin{equation}
\left\langle y,T\right|\left.x,0\right\rangle \to e^{-E_0 T} \psi_0(y)\psi_0(x)^*.
\end{equation}
Integrating over the final location $y$, then yields the proportionality:
\begin{equation}
\psi_0(x)^* \sim \int_{X(0)=x}\left[\mathcal{D}X\right]e^{-S}.
\end{equation}
Hence the ground state wavefunction can be evaluated by performing a path integral over all $t>0$ with the path starting at $t=0$ at $x$. Completely analogous, one obtains the field theoretic vacuum wavefunctional:
\begin{equation}
\Psi_0[\phi]^* \sim \int_{\phi(0)=\phi} \left[\mathcal{D}\phi\right]e^{-S}.
\end{equation}
Now we split the field into a left and right contribution $\phi_L$ and $\phi_R$. The propagation amplitude can then be just as well interpreted as performing a 180 degree rotation around the origin:
\begin{equation}
\Psi_0[\phi_L,\phi_R]^* \sim \left\langle \phi_L\right|e^{-\pi H_R}\left|\phi_R\right\rangle,
\end{equation}
where $H_R$ is the boost generator, the Rindler Hamiltonian. Utilizing this operator over an imaginary value $i\pi$ performs a 180 degree rotation. The state of the system, corresponding with $\Psi_0$,  is characterized by the density matrix
\begin{equation}
\rho = \left|\psi_0\right\rangle\left\langle \psi_0\right|,
\end{equation}
where the state $\left|\psi_0\right\rangle$ is the Minkowski vacuum, invariant under the full Poincar\'e group. \\
Finally, a Rindler observer does not have access to the left degrees of freedom and he effectively integrates these out (entanglement entropy). Doing this results in the density matrix with matrix elements:
\begin{equation}
\left\langle \phi_R'\right|\rho_R\left|\phi_R\right\rangle \sim \left\langle \phi_R'\right|\text{Tr}_L(\left|\psi_0\right\rangle\left\langle \psi_0\right|)\left|\phi_R\right\rangle = \int d\phi_L \left\langle \phi_R'\right|e^{-\pi H_R}\left|\phi_L\right\rangle \left\langle \phi_L\right|e^{-\pi H_R}\left|\phi_R\right\rangle = \left\langle \phi_R'\right|e^{-2\pi H_R}\left|\phi_R\right\rangle,
\end{equation}
implying a thermal density matrix with temperature $2\pi$. This means the Rindler observer observes the Minkowski vacuum as being thermally populated, this is the Unruh effect. \\
It is noteworthy that this derivation applies fully to other spacetimes that include a bifurcate Killing horizon. \\

\noindent The Rindler Hamiltonian eigenvalues are measured by the Rindler observer: this is the observer at the location $G_{00}=1$. He measures a temperature $T=\frac{1}{2\pi}$. Other fiducial observers have a redshifted local temperature as
\begin{equation}
T_{\text{local}} = \frac{1}{2\pi\rho},
\end{equation}
which hence becomes arbitrarily large close to the Rindler origin (being the black hole horizon).

\subsection{Hawking Radiation}
One of the most profound insights in black hole physics was one by Hawking in the mid seventies. He proved that black holes formed during gravitational collapse emit a continuous flow of particles in a thermal black body spectrum. Several different derivations of this phenomenon exist \cite{Parikh:1999mf}\cite{Hartle:1976tp}. In this section we will restrict to giving a physical argument in favor of particle creation by black holes (closely following \cite{Thorne:1986iy}). \\

\begin{figure}[h]
\centering
\begin{minipage}[b]{0.45\linewidth}
\includegraphics[width=0.85\textwidth]{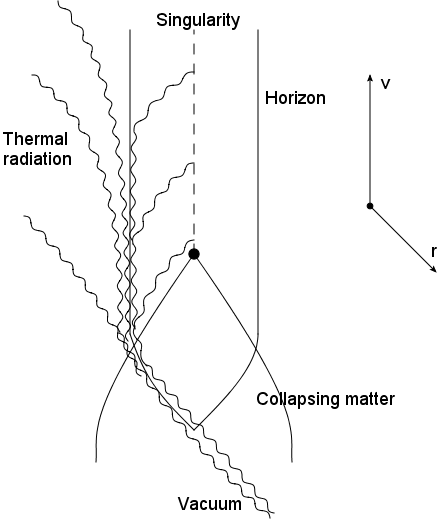}
\end{minipage}
\hfill
\begin{minipage}[b]{0.45\linewidth}
\includegraphics[width=0.85\textwidth]{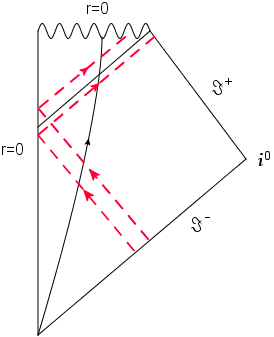}
\end{minipage}
\hfill
\begin{minipage}[t]{0.45\linewidth}
\caption{Finkelstein diagram of the gravitational collapse and Hawking radiation. The vertical axis represents (ingoing) Finkelstein time, and the other axis (which is tilted here) represents the Schwarzschild $r$-coordinate.}
\label{Hawkingrad}
\end{minipage}
\hfill
\begin{minipage}[t]{0.45\linewidth}
\caption{Carter-Penrose diagram of the same situation. The figure shows two rays in the wavepacket: one ray makes it to asymptotic infinity, whereas the other one falls into the singularity.}
\label{Hawkingrad2}
\end{minipage}
\end{figure}

\noindent Consider a wavepacket moving radially inwards and centered around a very high frequency before the collapse occurs. We assume that the initial state of the universe is such that this mode is unoccupied. This wavepacket is also chosen such that it will just cross the origin at the same time as the event horizon within the collapsing matter starts forming. The situation is displayed in figures \ref{Hawkingrad} and \ref{Hawkingrad2}. 
\noindent Part of the wavepacket is inside, and part is outside. The wavepacket is inevitably distorted as the modes that are inside the event horizon are irrevocably pulled inwards towards the singularity. The part of the packet that is exterior to the horizon escapes to future null infinity. Two aspects are crucial: firstly, since some part of the wavepacket is inaccessable to outside fiducial observers, they effectively trace over the degrees of freedom that fall into the singularity and this causes them to see the outgoing remaining wavepacket as populated (thermally populated in fact). Secondly, the outgoing part is arbitrarily close to the future horizon, implying an enormous redshift in the frequency upon emerging in the asymptotic future. Moreover, this reemerging can take an arbitrarily long amount of time (for the fiducial observers). \\
Thus outside asymptotic observers see a thermal flux of particles coming from the future horizon of a black hole formed in gravitational collapse. The thermality follows from the inaccesability of part of the information of the original modes and the ever-lasting flux follows from the time dilation close to the event horizon. 

\subsection{Short review on QFT near black holes}
In the previous paragraphs, we have already alluded to the different possible vacua one can choose in field theory near black hole horizons. It is instructive for the following to provide a more detailed overview of the different vacua and their properties. First we discuss Rindler spacetime and then we discuss Schwarzschild spacetime.
\subsubsection{Rindler space}
We consider a Rindler metric of the form
\begin{equation}
ds^2 = -\rho^2 dt^2 + d\rho^2 + d\mathbf{x}^2.
\end{equation}
There are two vacua relevant for Rindler space: 

\begin{itemize}
\item
The \textbf{Rindler vacuum} (or Fulling vacuum) $\left|R\right\rangle$ is defined by uniformly accelerating observers (positive frequency with respect to Rindler time $t$). \\
The stress tensor for a massless field of spin $s=0, 1/2, 1$ in the Rindler vacuum can be computed \cite{Sciama:1981hr} 
and is given by
\begin{equation}
\left\langle T_{\mu}^{\nu}\right\rangle_R = -\frac{h(s)}{2\pi^2\rho^4}\int_{0}^{+\infty}\frac{d\nu \nu (\nu^2 +s^2)}{e^{2\pi\nu}-(-1)^{2s}}\text{diag}\left(-1,\frac{1}{3},\frac{1}{3},\frac{1}{3}\right),
\end{equation}
where $h(s)$ is the number of helicity states for a massless field of spin $s$.
The integral can be performed explicitly and one finds for $-\left\langle T_{0}^{0}\right\rangle_R$
\begin{equation}
-\left\langle T_{0}^{0}\right\rangle_{R,s=0} = -\frac{1}{480\pi^2\rho^4}, \quad -\left\langle T_{0}^{0}\right\rangle_{R,s=1/2} = -\frac{17}{1920\pi^2\rho^4}, \quad -\left\langle T_{0}^{0}\right\rangle_{R,s=1} = -\frac{11}{240\pi^2\rho^4}.
\end{equation}
These values were written down by Dowker in \cite{Dowker:1994fi}. Note that the components of the stress tensor blow up at the Rindler horizon $\rho=0$.

\item
The \textbf{Minkowski vacuum} $\left|M\right\rangle$ is defined by inertial observers (positive frequency with respect to inertial time).\footnote{Previously, we have denoted this state as $\left|\psi_0\right\rangle$.} \\
The Minkowski vacuum is related to the Rindler vacuum by a Bogoliubov transformation that shows that the Rindler observer sees the Minkowski vacuum as a \emph{thermal} system, in the sense that
\begin{equation}
\left\langle M\right| \mathcal{O} \left|M\right\rangle = \text{Tr}_R(\hat{\rho} \mathcal{O}),
\end{equation}
where
\begin{equation}
\hat{\rho} = e^{-\beta_R H_R},
\end{equation}
with the Rindler Hamiltonian $H_R$ generating time translations in the Rindler time coordinate. $\beta_R = 2\pi$ in this case, as is required by absence of a conical singularity for the Euclidean Rindler space. \\
The stress tensor vev in the Minkowski vacuum vanishes. It can be rewritten in terms of the Rindler observer's coordinates and modes and, since the stress tensor is a tensorial quantity, it must also vanish in these coordinates. The Rindler observer explains this as due to a cancellation between the thermal contribution and the Casimir vacuum contribution $\left\langle T_{\mu}^{\nu}\right\rangle_R$. The Minkowski vacuum state hence has a stress tensor that is finite at the Rindler horizon.\\
Any thermal system can be just as well analyzed on the thermal manifold, which in this case is flat space in polar coordinates. For a thermal system with $\beta \neq \beta_R$, a conical singularity is present at the origin. In this case, the cancellation between thermal and Casimir contributions no longer occurs and one finds \cite{Dowker:1994fi}:
\begin{align}
-\left\langle T_{0}^{0}\right\rangle_{M,s=0} &= \frac{\pi^2}{30\beta^4\rho^4}-\frac{1}{480\pi^2\rho^4}, \\
-\left\langle T_{0}^{0}\right\rangle_{M,s=1/2} &= \frac{7\pi^2}{120\beta^4\rho^4} + \frac{1}{48\beta^2\rho^4} - \frac{17}{1920\pi^2\rho^4}, \\
-\left\langle T_{0}^{0}\right\rangle_{M,s=1}& = \frac{\pi^2}{15\beta^4\rho^4} + \frac{1}{6\beta^2 \rho^4} - \frac{11}{240\pi^2\rho^4}.
\end{align}
If one is interested in computing thermodynamic quantities of the quantum fields in the Minkowski vacuum, one can start from the canonical internal energy:
\begin{equation}
E = -\int d\rho \rho d\mathbf{x} \left\langle T_{0}^{0}\right\rangle_{M}
\end{equation}
and from this the free energy and entropy can be computed as
\begin{align}
E &= \partial_{\beta}(\beta F), \\
S &= \beta^2 \partial_{\beta} F.
\end{align}
One readily finds that the vacuum contribution to $\left\langle T_{0}^{0}\right\rangle_{M}$ gives a $\beta$-independent contribution to $F$ and is irrelevant when computing $S$. \\
Hence the entropy $S$ only cares for the thermal part and for $\beta = \beta_R$ and spin zero, it gives
\begin{equation}
S = \frac{A}{360\epsilon^2 \pi} + C
\end{equation}
for some finite constant $C$. The regulator $\epsilon$ is introduced at the lower end of the $\rho$-integral. Thermodynamic quantities are infinitely large due to the near-horizon region (near-origin) because field theory has much too many degrees of freedom near black hole horizons \cite{'tHooft:1984re}. These quantities scale as the area of the black hole and provide the quantum corrections to the black hole entropy, which according to the views of Susskind and Uglum should be absorbed into a renormalization of Newton's constant.
\end{itemize}

\subsubsection{Schwarzschild spacetime}
An analogous discussion can be made for Schwarzschild spacetime:
\begin{equation}
ds^2 = - \left(1-\frac{2GM}{r}\right)dt^2 + \frac{dr^2}{1-\frac{2GM}{r}} + r^2(d\theta^2 + \sin^2(\theta)d\phi^2).
\end{equation}
Instead of focusing on the spacetime relevant for gravitational collapse (Schwarzschild outside collapsing matter by Birkhoff's theorem and non-analytically continued to some internal matter metric), we focus on the eternal black hole. \\
A more suitable set of coordinates is the Kruskal coordinates $X$ and $T$, of which we can construct two null coordinates $U = T-X$ and $V=T+X$. These coordinates are the analogue of Minkowski coordinates in flat space: they do not contain a coordinate singularity and correspond to the coordinates used by freely falling observers. \\
We can discern three different vacuum states in such spacetimes.
\begin{itemize}
\item
The \textbf{Boulware vacuum} $\left|B\right\rangle$ is defined by taking modes with positive frequency with respect to the Schwarzschild time $t$. It hence has a divergent stress tensor both on the past and future event horizon. Near infinity, it shows empty space. Near $r=2GM$, it has the behavior:
\begin{equation}
\left\langle T_{\mu}^{\nu}\right\rangle_B \approx -\frac{h(s)}{2\pi^2\left(1-\frac{2GM}{r}\right)^2}\int_{0}^{+\infty}\frac{d\nu \nu (\nu^2 +s^2\kappa^2)}{e^{2\pi\nu/\kappa}-(-1)^{2s}}\text{diag}\left(-1,\frac{1}{3},\frac{1}{3},\frac{1}{3}\right)
\end{equation}
with $\kappa= 1/(4GM)$, the surface gravity of the black hole. This corresponds precisely to the vacuum polarization in Rindler space. The Boulware vacuum is the analogue of the Rindler vacuum in Rindler space. Physically it should be used to model the stress tensor of for instance a static star that does \emph{not} Hawking radiate (it is outside its Schwarzschild radius). It is found that $\left\langle T_{\mu}^{\nu}\right\rangle_B$ is a diagonal matrix. There is hence no flux of energy or momentum as we expect.

\item
The \textbf{Unruh vacuum} $\left|U\right\rangle$ is defined by taking incoming modes from infinity to be positive frequency with respect to $t$, while outgoing modes (from the past horizon) with positive frequency with respect to $U$. It has a divergent stress tensor on the past horizon. It is finite on the future event horizon. At infinity, it has an outgoing flux of energy, corresponding to Hawking radiation. It is the closest analog of the state relevant for gravitational collapse. In this case $\left\langle T_{0}^{r}\right\rangle_U$ is non-zero and corresponds to an inward flux of \emph{negative} energy near the horizon and an outward flux of positive energy near infinity.

\item
The \textbf{Hartle-Hawking vacuum} $\left|H\right\rangle$ is defined by taking modes with positive frequency with respect to $V$ (for incoming modes) and $U$ (for outgoing modes). It is finite on both past and future event horizons and at infinity it includes a thermal bath of radiation. $\left\langle T_{\mu}^{\nu}\right\rangle_B$ is found to be diagonal corresponding to a time-independent thermal equilibrium between black hole and surrounding heat bath. This vacuum state corresponds to the Minkowski vacuum state in the flat case. Just as in flat space, one can write
\begin{equation}
\left\langle H\right| \mathcal{O} \left|H\right\rangle = \text{Tr}_B(\hat{\rho} \mathcal{O})
\end{equation}
where
\begin{equation}
\hat{\rho} = e^{-\beta_B H_B},
\end{equation}
with $\beta_B = \frac{2\pi}{\kappa} = 8\pi G M$, the Hawking temperature. Just like in the Rindler case, the stress tensor vev in $\left|H\right\rangle$ is a sum of a thermal part and a Casimir part; the latter can be identified with $\left\langle T_{\mu}^{\nu}\right\rangle_B$, analogous to Rindler space. In this case however, they do \emph{not} cancel out, not even near the horizon. The Casimir part diverges at the horizon (because $\left\langle T_{\mu}^{\nu}\right\rangle_B$ does) and the thermal part is found to diverge as well on the horizon. After renormalization, a finite non-zero contribution remains. Physically it is interpreted as coming from curvature corrections, and hence the Rindler limit is incapable of seeing this \cite{Frolov:1989jh}. It is found \cite{Howard:1984qp} that the energy density has a zone of negative density close to the horizon (order $R_s$), the spatial pressures are all positive and decay as one moves away from the horizon. No mixed components are found. The lengthy analysis of Candelas \cite{Candelas:1980zt} can be approximated quite well by Page's approximation \cite{Page:1982fm}. \\
The Hartle-Hawking state corresponds to a precisely thermal state. A slight deviation in this state would cause the cancellation of divergences to be absent and the resulting stress tensor would again diverge on the horizon.

\end{itemize}

\noindent It is noteworthy that measurements done by FIDOs and ZAMOs (being the analogous observers on rotating black holes) are \emph{not} sensitive to the Casimir contribution. They hence really find a thermal atmosphere, whereas for (semi-classical) backreaction the Casimir part also contributes.

\subsection{Polar coordinates with fixed winding}
As noted above, for the stress tensor on Rindler spacetime, the Casimir contribution cancels the thermal contribution perfectly. Translating this result to Euclidean space, one finds that the zero-winding contribution is temperature-independent (Casimir) and cancels the sum of the non-zero winding contributions (which do feel the temperature). The result is hence the same, but both type of contributions have been given geometric meaning. Since the stress tensor can be obtained by applying a suitable differential operator to the Green propagator, it is instructive to know that this same property actually follows directly from the propagator itself \cite{Dowker:1977zj}\cite{Troost:1977dw}\cite{Troost:1978yk}\cite{Parentani:1989gq}. \\

\begin{figure}[h]
\centering
\includegraphics[width=0.8\textwidth]{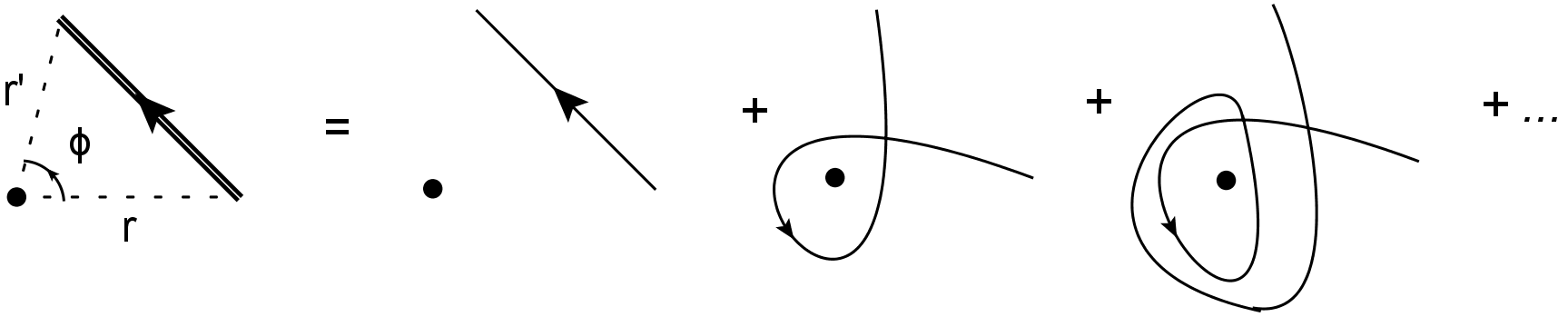}
\caption{The total Euclidean propagator between points $(r,0)$ and $(r',\phi)$ can be written as a sum over propagators with fixed winding number around the origin.}
\label{partiPolar}
\end{figure}

\noindent The Euclidean propagation amplitude can be rewritten in polar coordinates as a sum over fixed winding numbers around the origin as shown in figure \ref{partiPolar}.
In formulas:
\begin{equation}
\label{sumwind}
G(r,0;r',\phi, s) = \sum_{n\in\mathbb{Z}}G^{(n)}(r,0;r',\phi, s).
\end{equation}
The fixed winding path integral for polar coordinates was computed by \cite{Troost:1978yk}. For a Euclidean Lagrangian $ L = \frac{1}{4} \dot{x}^2$, the final expression for wrapping number $n$ (changed to the Euclidean signature) is given by
\begin{equation}
G^{(n)}(r,0;r',\phi, s) = \frac{1}{4\pi s}e^{-\frac{r^2+r'^2}{4s}}\int_{-\infty}^{+\infty}d\nu I_{\left|\nu\right|}\left(\frac{rr'}{2s}\right)e^{-2\pi i n \nu - \phi i \nu}.
\end{equation}
As a check, one can sum this expression over all $n$. Using the following formulas
\begin{align}
\sum_{n\in\mathbb{Z}}e^{-2\pi i n \nu} &= \sum_{k\in\mathbb{Z}}\delta(k-\nu), \\
\sum_{k\in\mathbb{Z}}I_k(x)t^k &= e^{\frac{x}{2}(t+1/t)},
\end{align}
and the fact that $I_k = I_{-k}$ for integer $k$, we get
\begin{equation}
G(r,0;r',\phi, s) = \frac{1}{4\pi s}e^{-\frac{r^2+r'^2-2rr'\cos\phi}{4s}},
\end{equation}
which is indeed the flat space heat kernel between these two points. \\

\noindent Can such a relation (\ref{sumwind}) hold in string theory as well? \\
Actually not. The reason for this is the same as that which we encountered before: string configurations can encircle the origin in a way that is unavailable for particle graphs. Taking the right hand side of the above formula (\ref{sumwind}) and summing over the string spectrum, one would obtain closed string tubes encircling the origin a fixed number of times. However, the sum of these do not exhaust the available string embeddings as there are diagrams that contain the origin in the interior of the worldsheet. These do not have a well-defined winding number and should be added to the right hand side in order to obtain the correct string propagator, obtained by summing the left hand side over the full string spectrum. Such a result is actually quite strange: it relies strongly on the fact that strings consist of infinitely many states (in fact what must be going on is a failure of commutativity of two infinite summations: the sum over winding numbers and the sum over string states). The left hand side of (\ref{sumwind}), when summed over the string spectrum, is precisely the string propagator (with pointlike external states as was shown in \cite{Cohen:1985sm}).\\

\noindent In any case, we will have more to say about all of this in the remainder of this work.

\subsection{Thermal structure of black holes in QFT}
As a summary, the thermal structure of a black hole in quantum field theory is shown below in figure \ref{BHstr}.
\begin{figure}[h]
\centering
\includegraphics[width=0.5\textwidth]{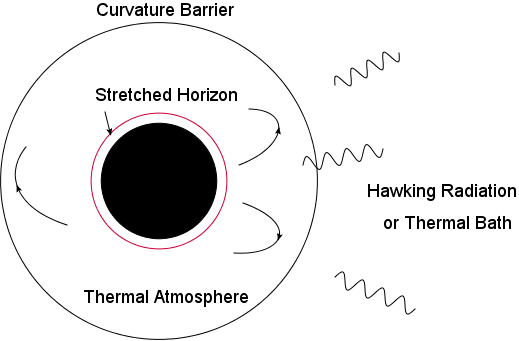}
\caption{Thermal structure of a black hole in field theory, including the near-horizon thermal atmosphere, where the Rindler approximation is used and the far region where Hawking radiation is present, evaporating the black hole.}
\label{BHstr}
\end{figure}
Depending on the asymptotic boundary conditions, either a Hawking flux emerges at infinity (causing evaporation) or a thermal heat bath is present (Hartle-Hawking state). Close to the horizon $r \lesssim 2 r_s$, the Rindler approximation is valid and a thermal heat bath (Unruh effect) is present. Physically, one can think of this heat bath as containing the quanta that do not make it to infinity because of the curvature barrier at a Schwarzschild radius from the horizon. For fiducial observers, the locally measured temperature rises as one approaches the horizon and becomes string scale at a string scale distance from the horizon. The so-called \emph{stretched horizon} is placed at this location and QFT should not be trusted closer to the black hole horizon. This stretched horizon masks all quantum gravity phenomena happening close to the horizon and demarcates where low energy field theory ceases to make sense. It is the quantum analogue of the classical black hole membrane, whose radial location is not arbitrary anymore. \\
In any case, the extremely high temperature region close to the horizon provides a zone where we expect string theory to make an important contribution to the QFT story. In the next subsection, we will already discuss a strange feature, unique to strings, that happens very close to the black hole horizon.

\subsection{Free-falling strings}
After these quantum field theoretic effects, we will discuss a special effect occuring for perturbative \emph{strings} around the black hole. The effect is typical for strings and is purely a quantum effect. \\
More concretely, in this subsection we give evidence that long strings occur near black hole horizons from a single string perspective, following closely the seminal papers by Susskind \cite{Susskind:1993ws}\cite{Susskind:1993ki}\cite{Susskind:1994uu}\cite{Susskind:1993aa}\cite{Susskind:2005js}. The experiment Susskind envisioned consisted of throwing a single string into a black hole and describing its evolution as it falls in according to a fiducial observer. \\
Due to the relevance of these ideas to our work, we will describe this process extensively. \\

\noindent Consider first a point particle in free fall close to a Schwarzschild black hole, close enough to approximate the geometry by Rindler space. The free-falling particle is described by an inertial trajectory in Minkowski space with the Rindler wedge superimposed to describe the measurements done by fiducial (stationary) observers. \\
The two-dimensional plane is described by $X^0$ and $X^1$ which are combined into the light-cone coordinates
\begin{equation}
X^{\pm} = \frac{X^0 \pm X^1}{\sqrt{2}}.
\end{equation}
These are, in turn, related to the Rindler coordinates $\rho$ and $\tau$ as
\begin{equation}
X^{\pm} = \mp \frac{\rho}{\sqrt{2}}e^{\mp \tau},
\end{equation}
where we introduced Rindler space as $ds^2 = -\rho^2 d\tau^2 + d\rho^2$. We now describe an infalling particle trajectory given by
\begin{equation}
X^- - X^+ = \sqrt{2}L.
\end{equation}
In the Minkowski coordinates $X^0, X^1$ this corresponds to a static particle. In Rindler coordinates $\rho, \tau$ this corresponds to a geodesic free-falling particle coming out of the past horizon (at $\tau=-\infty$), reaching a maximal height $\rho^*$ and then falling into the future horizon (at $\tau=+\infty$). The parameter $L$ is a measure for how high the particle reaches ($L = \rho^*$). \\
In one starts this experiment at $\tau=0$, we are describing a particle free-falling towards the future horizon with no initial velocity from a location $\rho=L$. The coordinates and trajectory are shown in figure \ref{traj} below. 
\begin{figure}[h]
\centering
\includegraphics[width=0.4\textwidth]{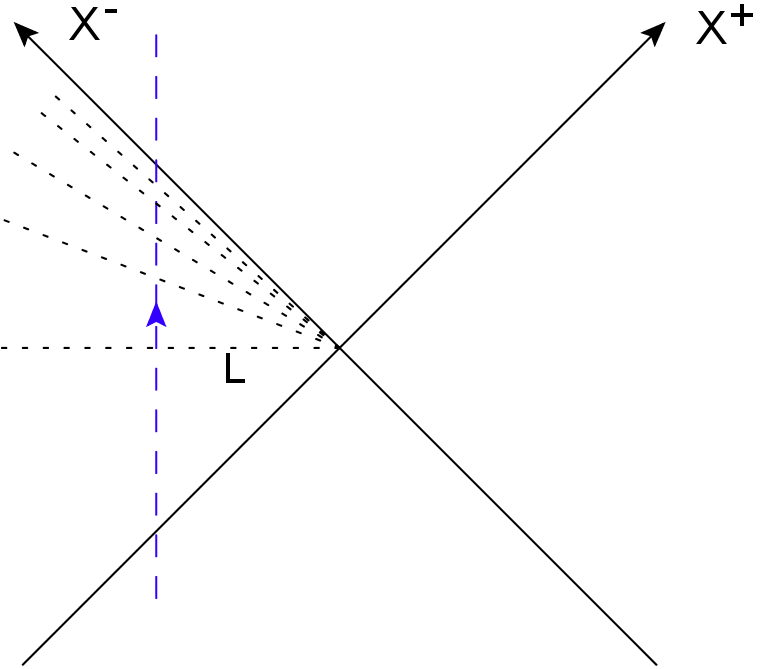}
\caption{Coordinate system and particle trajectory free-falling into the black hole horizon. As the particle is very close to the horizon, constant $\tau$ slices are nearly identical to constant $X^+$ slices.}
\label{traj}
\end{figure}
For late times $\tau \gg 1$ the trajectory is approximated by
\begin{equation}
X^+ \approx - \sqrt{2}Le^{-2\tau} , \quad X^- \approx \sqrt{2}L.
\end{equation}
Close to this trajectory, we can hence view the above equality as showing that the light-cone coordinate $X^+$ and Rindler time $\tau$ are directly related (without interference from spatial dependence). The constant $\tau$-slices and constant $X^+$-slices are approximately the same (in the left quadrant and up to the above rescaling) (figure \ref{traj}). This suggests we can relate descriptions using both of these time coordinates in a straightforward fashion. \\
The $X^+$ coordinate is hence interpreted as a time coordinate associated to the freely falling particle, whereas $\tau$ is associated to the fiducial observer.\\

\noindent In string theory, we will use the same coordinates and conventions where the trajectory described above will be reinterpreted as the center of mass trajectory of the string.\\
The strategy will then be the following: we will compute quantities associated to the string in light-cone coordinates. Using the above relationship between clock rates we will then make the transition to observations done by fiducial observers. \\

\noindent Consider now a free string in flat Minkowski space, quantized in light-cone coordinates $X^+, X^-, X^i$. Standard arguments show that the transverse coordinates have the following mode expansion
\begin{equation}
X^i(\tau_{ws},\sigma) = \underbrace{x_0^i + \alpha' p^{i} \tau_{ws}}_{X_{CM}^i(\tau_{ws})} + i \sqrt{\frac{\alpha'}{2}}\sum_{n\neq0}\frac{e^{-in\tau_{ws}}}{n}\left(\alpha_n^i e^{in\sigma} + \bar{\alpha}_n^i e^{-in\sigma}\right) 
\end{equation}
with $\left[\alpha_m^\mu , \alpha_n^{\nu}\right] = m\delta_{m,-n}\eta^{\mu\nu}$. We focus on a string in its ground state that is thrown in. Standard oscillator algebra then shows that
\begin{equation}
\label{transverseloc}
R^2 = \frac{1}{2\pi} \int_{0}^{2\pi} d\sigma \left\langle 0\right|\left|X^i - X^i_{CM}\right|^2\left|0\right\rangle = \alpha' (d-2) \sum_{n=1}^{+\infty}\frac{1}{n}
\end{equation}
and hence diverges logarithmically. The transverse spread of a free string in flat space is infinitely large. Physically, one is always restricted by the measuring apparatus which can only resolve oscillations up to a certain frequency. Frequencies higher than this cut-off are not measured. Suppose one cannot resolve $X^+$-timescales smaller than $\delta$. Note that this is light-cone time, and hence it corresponds to a time coordinate that would be measured more directly in experiments since obviously a fixed $X^0$-timeslice is never observed by any single observer. Light-cone time on the other hand is for instance the time as observed on a holographic screen capturing emitted lightrays from the source \cite{Susskind:1994vu}. One is hence instructed to take the temporal average over the coordinates as
\begin{equation}
X^i_\delta = \frac{1}{\frac{\delta}{\alpha'p^+}}\int_{0}^{\frac{\delta}{\alpha'p^+}}d\tau_{ws} X^i.
\end{equation}
The time average in $X^+$ over $\delta $ is translated through the light cone gauge condition $X^+ = \alpha' p^+ \tau_{ws}$ to a condition for a time average in $\tau_{ws}$ over $\frac{\delta}{\alpha'p^+}$.
This suggests one should replace equation (\ref{transverseloc}) by
\begin{equation}
R^2_{\delta} = \frac{1}{2\pi} \int_{0}^{2\pi} d\sigma \left\langle 0\right|\left|X^i - X^i_{CM}\right|^2\left|0\right\rangle_{\delta} = \alpha' (d-2) \sum_{n=1}^{\frac{\alpha'p^+}{\delta}}\frac{1}{n} \approx \alpha' (d-2) \ln\left(\frac{\alpha'p^+}{\delta}\right).
\end{equation}
The longitudinal size (in $X^-$) can be deduced analogously and leads to\footnote{This can be found by using the mode expansion:
\begin{equation}
X^-(\tau_{ws},\sigma) = x_0^- + \alpha'p^- \tau_{ws} + i \frac{1}{2p^+}\sum_{n\neq 0}\frac{e^{-in\tau_{ws}}}{n}\left(L_n^{\perp}e^{in\sigma} + \bar{L}_n^{\perp}e^{-in\sigma}\right),
\end{equation}
and the Virasoro algebra for the transverse modes:
\begin{equation}
\left[L_n^{\perp},L_m^{\perp}\right] = (n-m)L_{n+m}^{\perp} + \frac{d-2}{12}\left(n^3-n\right)\delta_{n,-m}
\end{equation}
for which only the first term of the central extension contributes in the high-frequency cut-off regime.} 
\begin{equation}
\left|\Delta X^-\right|^2 = \frac{1}{2\pi} \int_{0}^{2\pi} d\sigma \left\langle 0\right|\left|X^- - X^-_{CM}\right|^2\left|0\right\rangle_{\delta} \approx \left(\frac{1}{p^+}\right)^2 \sum_{n=1}^{\frac{\alpha'p^+}{\delta}}n \approx \left(\frac{\alpha'}{\delta}\right)^2.
\end{equation}
Another important quantity is the total length of the string, projected on the transverse space:
\begin{equation}
\ell = \int_{0}^{2\pi} d\sigma \sqrt{\left|\frac{\partial X^i}{\partial \sigma}\right|^2} \sim \sqrt{\alpha' (d-2) \sum_{n=1}^{+\infty}n} \approx \sqrt{\alpha'(d-2)}\left(\frac{\alpha'p^+}{\delta}\right).
\end{equation} 
To summarize, the transverse size scales as $\sim \ln(\delta)$, the longitudinal size as $\sim \frac{1}{\delta}$ and the total projected length as $\sim \frac{1}{\delta}$. \\

\begin{figure}[h]
\centering
\begin{minipage}{0.4\textwidth}
\centering
\includegraphics[width=\textwidth]{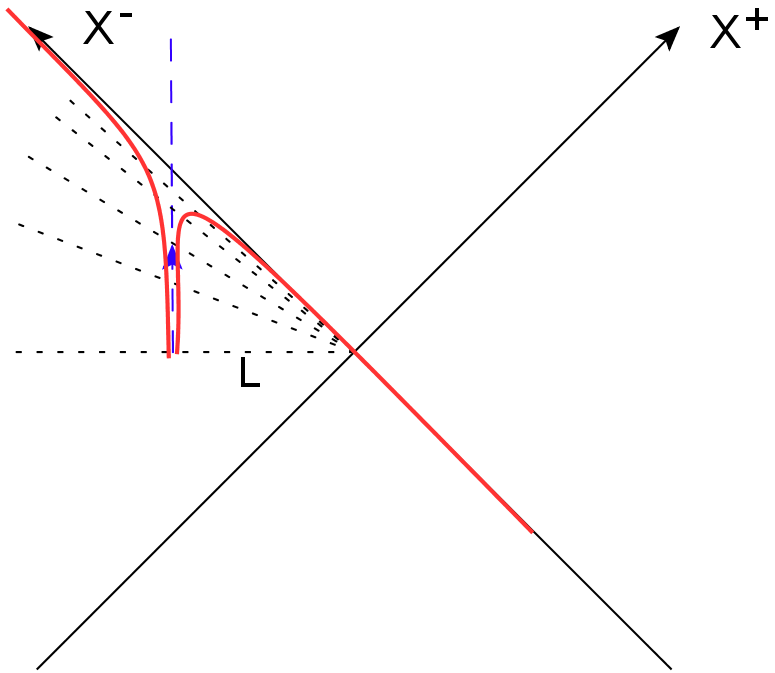}
\caption*{(a)}
\end{minipage}
\begin{minipage}{0.4\textwidth}
\centering
\includegraphics[width=\textwidth]{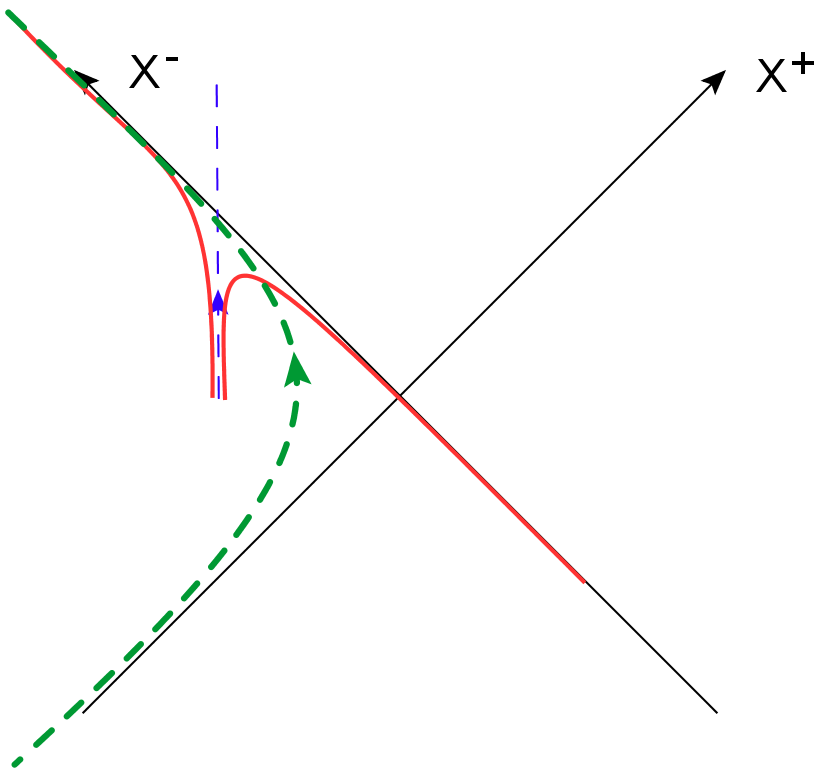}
\caption*{(b)}
\end{minipage}
\caption{(a) Free-falling string center of mass coordinate. The red lines mark the spread $\Delta X^-$ in the longitudinal direction as the string falls. The spread in reality only really starts at a string distance from the horizon and is heavily exaggerated here. (b) Fiducial observers closer than $\ell_s$ from the horizon (green dashed line) see the string fluctuations for arbitrarily long times: the Lorentz contraction offsets the stringy expansion. 
}
\label{falling}
\end{figure}

\noindent The coordinate equality $X^+ \approx -\sqrt{2}L e^{-2\tau}$ leads to 
\begin{equation}
\delta X^+ \approx 2\sqrt{2}Le^{-2\tau } \delta \tau
\end{equation}
for a small time interval. Taking $\delta \tau \sim 1$, it is immediate that $\delta X^+$ is much smaller for large $\tau$ (near the horizon). \\
The above computations were done for $\delta X^+ = \delta$. The free-falling observer using $\delta X^+ = \delta \sim \ell_s$ and sees nothing out of the ordinary. The fiducial observer on the other hand (for $\delta \tau \sim 1$) sees much smaller time scales for $\delta$ and sees the above spreading effects occuring. In particular he would obtain
\begin{align}
R^2_{\delta} &\sim 2 \alpha' \tau , \\
\left|\Delta X^-\right| &\sim \left(\frac{\alpha'}{L}\right)e^{2\tau}, \\
\ell &\sim \sqrt{\alpha'}\frac{\alpha'p^+}{L}e^{2\tau}.
\end{align}
It is clear from these results that the transverse area occupied by the infalling string increases much more slowly than the total length of the string. This implies the local density of string will become arbitrarily large and interactions, if present, will inevitably become non-negligible. \\
Secondly, one readily obtains
\begin{equation}
\Delta X^+ \Delta X^- \approx \alpha',
\end{equation}
which is a spatial uncertainty principle. This has a graphical interpretation as in figure \ref{falling}.
\noindent This means a fiducial observer at string length from the horizon would continue to see the infalling string straddling his location: the string's radial Lorentz contraction is precisely compensated by its radial extension as it falls in. Infalling information cannot Lorentz contract to below the string-scale and this eliminates the UV divergence that was present in QFT close to the horizon. \\
\noindent From this one concludes that, for large $\tau$, the infalling string has diffused over a region of thickness $\ell_s$ close to the event horizon.\\

\noindent All of this relates \emph{infalling time} to \emph{Rindler time}. One might ask what happens to other fiducial observers. Their clock rate is related to the Rindler clock rate as
\begin{equation}
d\tau_{pr} = \sqrt{G_{00}(\rho^*)}d\tau
\end{equation}
for the fiducial observer at $\rho=\rho^*$.
Hence one expects that fiducial observers closer to the horizon than the Rindler observer have a relatively cruder measuring apparatus (worse relative resolution) and would have to wait a bit longer for long strings to appear. The reverse holds for fiducial observers further from the horizon. \\
For a Schwarzschild black hole, the Rindler time is the asymptotic Schwarzschild time and all other fiducial observers would have to wait longer to see the long strings. \\

\noindent The freely falling string has taught us several lessons for physics as described by fiducial observers:
\begin{itemize}
\item{
The non-interacting string spreads without bound as it approaches the horizon, where its total length is much larger than its transverse occupied space, implying a dense packing of string pieces.}
\item{
Interactions, if present, are inevitably important close to the horizon.
}
\item{
The stringy uncertainty principle implies the long string to be localized in a string length $\ell_s$ shell surrounding the event horizon.
}
\end{itemize}
We will find analogs of all of these phenomena later on in the canonical ensemble. \\
It is instructive to make the link more direct between the effect discussed here (spreading due to a finite resolution) and the effect discussed in the previous chapters (long strings due to high excitation). The total transverse length of the string \emph{in a generic state} $\left|N,\bar{N}\right\rangle$ in flat space is given by the expression:
\begin{equation}
\ell = \left\langle N,\bar{N} \right| \int_{0}^{2\pi} d\sigma \sqrt{\left|\frac{\partial X^i}{\partial \sigma}\right|^2} \left|N,\bar{N} \right\rangle.
\end{equation} 
One can compute that the dominant contribution is given by
\begin{equation}
\label{leng}
\ell \sim \sqrt{\alpha'}\sqrt{(d-2)\sum_{n=1}^{+\infty}n + N + \bar{N}}.
\end{equation}
The first term is the vacuum contribution discussed above. We demonstrated that it should be cut off in the high-frequency limit and causes an ever-increasing stretching of the string due to zero-point fluctuations. The second contribution is the classical part $\ell \sim \sqrt{\alpha' N} \sim E \alpha'$. \\
Both contributions lead to long strings. \\

\noindent Curiously, regularizing the sum in (\ref{leng}) using zeta-regularization instead of the physical cut-off procedure, one would find for $d=26$:
\begin{equation}
\ell_{\text{zeta}} \sim \sqrt{\alpha'}\sqrt{-2 + N + \bar{N}} \sim \alpha' m.
\end{equation}
and the length is proportional to the mass of the string. This formula does not make sense for the ground state and should be viewed merely as a curiosity.

\chapter{Random Walks in Rindler Spacetime}
\label{chri}
This chapter contains the application of the random walk phenomenon to Rindler space. We utilize a combination of exact CFT methods and field theory methods. The contents of this chapter coincide with \cite{Mertens:2013zya} and parts of \cite{Mertens:2014saa}. Also a discussion on rotating black holes is provided that has not appeared elsewhere up to this point.

\section{Introduction}

In \cite{Mertens:2013pza}, we analyzed the method set forth by the authors of \cite{Kruczenski:2005pj} to analyze the near-Hagedorn thermodynamics of string theory directly from the string path integral. The method explicitly describes the random walk picture of high-temperature string thermodynamics: after heating up a gas of closed strings, the constituents coalesce into a single long, highly excited closed string.\footnote{Actually this depends on the compactness of the different dimensions \cite{Deo:1989bv}. For $D\geq3$ a single string dominates, for $D=0$ multiple-string configurations contribute, whereas for $D=1$ and $D=2$ a more detailed analysis is required \cite{Deo:1989bv}\cite{Deo:1988jj}.} This long string behaves as a random walker in a fixed background. From a Euclidean point of view, the long string is described by the thermal scalar \cite{Atick:1988si}. This is a complex scalar field that combines the winding $\pm 1$ stringy excitations around the Euclidean time circle. We noted that this random walk receives corrections compared to the naive `worldsheet dimensional reduction' to a particle theory. \\

\noindent The random walk picture also emerges when considering black hole horizons \cite{Susskind:1993ws}\cite{Susskind:2005js}. A quick way of appreciating this, is the following reasoning. As we mentioned in the previous chapter, the Euclidean black hole is cigar-shaped with the circumference (inverse temperature $\beta$) shrinking all the way to zero at the tip of the cigar (the black hole horizon). Thus the local temperature gets arbitrarily high, close to the horizon (we mentioned this in our discussion of the Unruh effect as well). But at stringy temperatures long strings are favored and Hagedorn phenomena are expected to take place. Hence we expect similar effects as in high temperature flat space to take place close to the black hole horizon (at its normal non-stringy temperature). Of course, this argument extrapolates the flat space result to hold locally in any space, which is not true in general, but for qualitative predictions it suffices. The long string(s) wraps the horizon and effectively forms the stretched horizon. This is a microscopic stringy candidate for the black hole membrane as it is called in the earlier literature.\footnote{See e.g. \cite{Thorne:1986iy} and references therein.} The stretched horizon is located at a string length outside the black hole horizon. This picture and the related correspondence principle have been studied extensively in the past by numerous authors (see e.g. \cite{Halyo:1996vi}\cite{Maldacena:1996ds}\cite{Halyo:1996xe}\cite{Halyo:2001us}\cite{Halyo:2003bt}\cite{Sen:2004dp}\cite{Giveon:2006pr}\cite{BatoniAbdalla:2007zv}\cite{Sasai:2010pz}). See also \cite{Halyo:2015ffa} and \cite{Giveon:2015cma} for some very recent developments on black hole horizons in string theory. It is our goal to analyze this picture further for a string gas close to the horizon or more specifically for a string gas in Rindler space using the explicit methods developed earlier. In particular, we want to substantiate the above local argument. Recently, in light of the so-called firewall-paradox \cite{Almheiri:2012rt},\footnote{See \cite{Braunstein:2009my} for an earlier account of this.} it has become increasingly important to better understand how string theory behaves near black hole horizons. 

\noindent The methods developed only consider the genus one worldsheet (torus) and are thus firmly rooted in perturbative string theory. We are aware that this limits the applications. In particular it has been argued \cite{Susskind:2005js} that close to the black hole horizon, one cannot ignore the higher genus contributions. 
Due to a lack of analytical methods to analyze non-perturbative string theory in this regime, we take the genus one results as a guide to what actually happens near black hole horizons. The genus one approach to Rindler string thermodynamics has been considered quite extensively in the past (see e.g. \cite{Dabholkar:1994ai}\cite{Lowe:1994ah}\cite{Parentani:1989gq}\cite{Emparan:1994bt}\cite{McGuigan:1994tg}).\\
The methods we will use are a mixture of two approaches: firstly we use a worldsheet Fourier series expansion of the string path integral. This provides an explicit random walk picture and it can be interpreted as the long string in the Lorentzian picture. Secondly, we use the field theory action of the thermal scalar. This field theory point of view will allow us to explicitly see the corrections to the particle action that the first approach misses. Also, off-shell questions are accessable using field theory. When we combine these two approaches, the difficulties of either approach are better understood and we obtain a realization of the (genus one) long string surrounding the black hole as was anticipated and argued by Susskind several years ago. \\

\noindent This chapter is organized as follows.\\
In section \ref{rindler} we discuss Rindler space thermodynamics and its link to black hole horizons. In section \ref{alphaprime} we elaborate on the higher order $\alpha'$ corrections for the specific case of Rindler space. We discuss this in general first and then we use the link to the cigar CFT as was recently proposed in \cite{Giveon:2013ica}\cite{Giveon:2012kp}\cite{Giveon:2014hfa}. Using this knowledge, we analyze the thermal scalar in Rindler space in section \ref{critical} for bosonic, type II and heterotic strings. In particular we determine the Hagedorn temperature, the location of the random walk and whether the free energy remains finite or not. We discuss the Hagedorn behavior and the density of states of the Rindler string and show that the free energy is dominated by a long string at string-scale distance from the horizon with redshifted critical temperature equal to the Rindler temperature. In section \ref{orbifold} we make the link between the critical behavior of flat space $\mathbb{C}/\mathbb{Z}_N$ orbifold models and angular orbifolds of the $SL(2,\mathbb{R})/U(1)$ model and in particular we will find that the non-standard momentum-winding duality of bosonic strings on the cigar \cite{Dijkgraaf:1991ba} has an important manifestation on the flat $\mathbb{C}/\mathbb{Z}_N$ orbifold. In section \ref{unitarity} we identify the Rindler quantum numbers with those from the cigar model. We will discuss the unitarity constraints and how they are translated to the Euclidean Rindler case. After a short discussion in section \ref{discussion}, we tie up a loose end in section \ref{heteroticc} concerning heterotic Rindler space. Section \ref{flatlimit} describes the flat limit directly at the level of the partition function. Section \ref{remarks} contains some comments on the full cigar CFT in the opposite limit (strong curvature). Rotating black holes are briefly discussed using the thermal scalar perspective in section \ref{rotat}. Some remaining open problems and our conclusions are presented in sections \ref{open} and \ref{conclusio}. Several illustrative and technical calculations are gathered in the supplementary sections. \\

\noindent In part 1 of this work, we only analyzed flat backgrounds explicitly. For a general curved background however, we have much less control on what precisely happens. Two questions need to be answered:
\begin{itemize}
\item{Is there a winding mode that becomes tachyonic when exceeding a specific temperature? We know this to be the case for flat backgrounds, but does it hold for more general (either topologically or geometrically non-trivial) backgrounds? Can we establish a regime where this thermal scalar dominates the thermodynamical quantities?}
\item{Can we get a handle on what other $\alpha'$ correction terms need to be added to the thermal scalar action for a general background?}
\end{itemize}
We will find answers to both of these questions in our study of Rindler space. While this space is geometrically quite easy, the description of string thermodynamics in terms of the Rindler observer is not straightforward. Nevertheless we will be able to explicitly solve the critical behavior of the Rindler string.
We will make contact with some previous results regarding string thermodynamics in Rindler spacetime and we will obtain a prediction of the Rindler Hagedorn temperature for all types of (conventional) string theories.

\section{Rindler thermodynamics}
\label{rindler}
We will be interested in computing the one-loop free energy of strings in the critical regime. In general one knows that a black hole is surrounded by a thin membrane called the stretched horizon. This Planck (or string) sized object is outside the reach of quantum field theory in curved spacetimes. This can be seen by e.g. the UV divergence of one-loop thermodynamical quantities \cite{'tHooft:1984re}\cite{Barbon:1994ej}\cite{Emparan:1994qa} of a gas of particles in the black hole geometry. When looking further from the horizon, a thermal zone is found that is stretched over roughly one Schwarzschild radius radially outwards. This is also the region where we can approximate the black hole background by Rindler spacetime as we demonstrated in the previous chapter \cite{Susskind:2005js}. Consider a $d$-dimensional Schwarzschild spacetime\footnote{We take the Schwarzschild black hole as an example to demonstrate our point in what follows, even though it is not an exact string background. In general, most uncharged black holes have a Rindler near-horizon limit. Rindler space on the other hand is an exact string background.}
\begin{equation}
ds^2 = -\left(1-\frac{2GM}{r}\right)dt^2 + \left(1-\frac{2GM}{r}\right)^{-1}dr^2 + d\mathbf{x}^2_{\perp}.
\end{equation}
Defining $\rho = \sqrt{8GM(r-2GM)}$ and focusing on the near horizon geometry, the metric reduces to the Rindler form:
\begin{equation}
ds^2 = -\left(\frac{\rho^2}{(4GM)^2}\right)dt^2 + d\rho^2 + d\mathbf{x}^2_{\perp}.
\end{equation}
Our goal is to analyze the near-horizon behavior of the one-loop free energy in string theory. Note that to describe thermodynamics, we need to choose a preferred time coordinate. In this case it is Rindler time (or Schwarzschild time), so we describe thermodynamics as perceived by fiducial observers in the spacetime. In particular, space ends at the horizon by a smooth capping of the cigar. In string theory, one does not expect a UV divergence in the free energy, but we should be careful about IR divergences which can (and will) occur. Considering Rindler spacetime as a string background, all background fields are turned off except the metric whose Euclidean section takes the following form
\begin{equation}
\label{Rindmetric}
ds^2 = \frac{\rho^2}{(4GM)^2}d\tau^2 + d\rho^2 + d\mathbf{x}^2_{\perp}.
\end{equation}
To avoid a conical singularity at $\rho=0$, the Euclidean time coordinate needs to have the identification $\tau \sim \tau + \beta_R$ where 
$\beta_R = 8\pi GM$. We will refer to this temperature as the canonical Rindler temperature. Euclidean Rindler space is then the same as flat space in polar coordinates. This temperature coincides with the Hawking temperature of the original black hole since we did not change the temporal coordinate. \\

\noindent We want to remark that this temperature is a global property of the space. The local temperature is equal to $\beta_{R,local} = \beta_{R}\sqrt{G_{00,Rindler}}$. This is the temperature as measured by local observers. In particular, the canonical Rindler temperature itself is measured by an observer located at $G_{00} = 1$. In this case this is at $\rho = 4GM$. \\

\noindent Since our intention is to study thermodynamics at the canonical Rindler temperature, we are actually looking at the stringy Unruh effect and hence the vacuum we are considering is the Minkowski vacuum. As is well-known, the description of this vacuum by a fiducial observer corresponds to a thermal state at the canonical Rindler temperature. \\

\noindent Calculating thermodynamical properties in such spaces presents some complications, as we mentioned in the previous chapter.
In field theory, varying $\beta$ results in conical spaces. Although one would at first sight think that a conical space represents non-equilibrium,\footnote{And thus one cannot infinitesimally vary the temperature to nearby equilibrium configurations, as one is instructed to do in thermodynamical calculations.} it can be proven that if one \emph{only} varies the temperature and hence fixes the transverse horizon area \cite{Susskind:1994sm}\cite{Carlip:1993sa}, one indeed gets these conical spaces. So it is consistent for equilibrium phenomena to infinitesimally vary $\beta$ as long as one sets it equal to $8\pi GM$ in the end. \\
In string theory the situation is more problematic: besides the previous complication, one also has the problem that for $\beta \neq \frac{8\pi GM}{N}$ with $N \in \mathbb{N}$, the worldsheet theory is not conformal and thus inconsistent.\footnote{Corresponding to the fact that string theory in its standard formulation cannot be taken off-shell.} This was in fact the rationale behind the orbifold approximations to Rindler spacetime thermodynamics by the authors of \cite{Dabholkar:1994ai}\cite{Lowe:1994ah}. We will come back to this further on. In our case, we leave $\beta$ general, and in spite of the fact that the starting point of the derivation in section \ref{pathderiv} is only valid for one value of $\beta$, we will interpret our final expression as an off-shell continuation of the conformal result.\footnote{Although such an interpretation is not really necessary for much of what is to follow.} Moreover, we have seen that one can also get the particle path integral from the field theory action and this action obviously \emph{can} be off-shell. We will return to this topic further on. \\

\noindent The free energy is a \emph{globally} defined concept and we do not draw any conclusions from the local properties of thermodynamic quantities. As discussed by \cite{Emparan:1994bt}, locality in string theory is a delicate topic and also even in field theory the thermal properties in black hole backgrounds are all global properties.\footnote{See \cite{Calcagni:2013eua} for a very recent account on string nonlocality. On the other hand, note that in chapter \ref{chex} we did obtain an expression for the local free energy density and in the final chapter we will study other local quantities. However, we only use these to study the spatial profile and not to predict convergence or divergence.} Local arguments can be a good guide in a general background, but in a cigar-shaped background they are misleading: local reasoning predicts a divergence simply because the thermal circle keeps on shrinking below the Hagedorn radius. We will see that this is wrong in general and that the thermal quantities for most string types are tachyon-free. We want to notice that this dramatic failure of local reasoning is typical for cigar-shaped backgrounds. For manifolds with topologically stable thermal circles on the other hand (like Euclidean $AdS_3$), local reasoning is qualitatively good (also see \cite{Rangamani:2007fz}).

\section{Corrections in $\alpha'$ to the thermal scalar action}
\label{alphaprime}
One of our main interests lies in getting a handle on possible higher $\alpha'$ corrections to the thermal scalar action. 
We will first discuss this in general using only general covariance as a guide. Further on we will make the link with the cigar $SL(2,\mathbb{R})/U(1)$ CFT model where we will precisely pinpoint what these corrections look like.

\subsection{General analysis}
\label{general}
\begin{figure}[h]
\centering
\includegraphics[width=11cm]{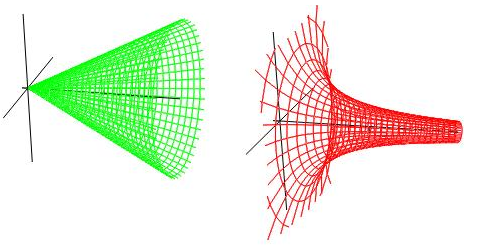}
\caption{Left figure: size of the thermal circle in Rindler space as a function of radial distance increasing towards the right. Right figure: size of the thermal circle in the T-dual of Rindler space as a function of radial distance increasing towards the right.}
\label{thermalcircles}
\end{figure}

We remind the reader that the thermal scalar action is given by the effective action for the first discrete momentum mode in the T-dual background, i.e. $n=\pm1$ and $w=0$. Corrections to the thermal scalar action can thus be analyzed by considering the discrete momentum action of the T-dual background. 
In this section we analyze the general form of possible $\alpha'$ corrections to the thermal scalar action in Rindler space and we discuss whether we can identify a regime where these, if they are present, are subdominant.
Considering Euclidean Rindler space (\ref{Rindmetric}), the T-dual Ricci tensor has components
\begin{equation}
\tilde{R}^{00} = -\frac{2}{(4GM)^2}, \quad  \tilde{R}^{\rho\rho} = -\frac{2}{\rho^2}, \quad \tilde{R}^{0\rho}=0
\end{equation}
and the T-dual Ricci scalar $\tilde{R} = -\frac{4}{\rho^2}$. Note that there is a curvature singularity at $\rho=0$ (for the string metric). The T-dual dilaton field is given by
\begin{equation}
\partial_\rho \tilde{\Phi} = -\frac{1}{\rho}.
\end{equation}
Some possible terms that could appear in the field theory action are given by
\begin{align}
m^2 TT^{*} &= -\frac{4}{\alpha'}TT^{*}, \quad \text{bosonic}\quad \text{or} \quad m^2 TT^{*} = -\frac{2}{\alpha'}TT^{*}, \quad \text{type II},\\
\tilde{G}^{\mu\nu}\partial_{\mu}T\partial_{\nu}T^{*} &=  \frac{w^2\beta^2\rho^2}{4\pi^2(4GM)^2\alpha'^2}TT^{*} + \partial_\rho T \partial_\rho T^{*} = \frac{\rho^2}{\alpha'^2}TT^{*} + \partial_\rho T \partial_\rho T^{*}, \\
\label{curv0}
\tilde{R} TT^{*} &= -\frac{4}{\rho^2}TT^{*}, \\
\label{masscorr}
\alpha'\tilde{R}^{\mu\nu}\partial_\mu T \partial_{\nu} T^{*} &= -\frac{w^2\beta^2}{2\pi^2(4GM)^2\alpha'}TT^{*} - 2\frac{\alpha'}{\rho^2}\partial_\rho T \partial_\rho T^{*} = -\frac{2}{\alpha'}TT^{*} - 2\frac{\alpha'}{\rho^2} \partial_\rho T \partial_\rho T^{*}
\end{align}
and
\begin{align}
\label{curv1}
\partial_\mu \tilde{\Phi} \partial^{\mu} \tilde{\Phi} TT^{*} &= \frac{1}{\rho^2}TT^{*}, \\
\label{curv2}
\alpha'\partial^{\mu}\tilde{\Phi} \partial^{\nu} \tilde{\Phi} \partial_{\mu}T\partial_{\nu}T^{*} &= \frac{\alpha'}{\rho^2}\partial_\rho T \partial_\rho T^{*},
\end{align}
since $\partial_0 = i\frac{\beta}{2\pi\alpha'} = i (4GM) / \alpha'$ and $\left|w\right|=1$. Note that if we do not fix the temperature to the Hawking temperature, we get $\beta^2$ corrections to the action but not other powers of $\beta$. This will be relevant further on. We see that higher order $\alpha'$ terms are not suppressed due to two reasons:
\begin{itemize}
\item[1.] The T-dual geometry has curvature that blows up at $\rho = 0$. Comparing the original Rindler space with its T-dual, the coordinate singularity of the original black hole gets transformed into a curvature singularity. If we would be interested in the far region from the horizon, we could still neglect these terms. So only for $\rho$ large, will these terms be subdominant. Some terms of this type are (\ref{curv0}), (\ref{curv1}), (\ref{curv2}) and the second term of (\ref{masscorr}).
\item[2.] Secondly, the temperature is not string scale, but is equal to the Hawking temperature. This provides extra contributions (like the first term in (\ref{masscorr})) in the action that are of the same order in $\alpha'$ as the naive lowest $\alpha'$ action whose general form was given in (\ref{lowestFT}).
\end{itemize}
This conclusion holds in general: the $\alpha'$ corrections are subdominant if the T-dual curvature radius is (much) larger than the string length and the temperature is string scale. So if the thermal circle does not deviate much from flat space, its T-dual space will also be quite well-behaved and the $\alpha'$ corrections are subdominant. An example is $AdS$ space as we will discuss in chapter \ref{chwzw}. The thermal circles for Rindler space and its T-dual partner are illustrated in figure \ref{thermalcircles}.
We conclude that in Euclidean Rindler space, there is no regime in which we can get control over the $\alpha'$ corrections to the thermal scalar field theory action if these terms are present in the first place, i.e. if their coefficients are non-zero. However, we can follow a different path and consider exact $SL(2,\mathbb{R})/U(1)$ WZW models that have a flat space limit corresponding to Euclidean Rindler space. This will be done in the next subsection.

\subsection{Exact WZW analysis}
In what follows we will rescale the time coordinate to $\tau = \frac{\sqrt{\alpha'}t}{4GM}$ to simplify several expressions and to make the link with $\mathbb{C}/\mathbb{Z}_N$ orbifolds more transparent. We have introduced the string length scale here as our reference length:
\begin{equation}
\label{snorm}
ds^2 = \left(\frac{\rho^2}{\alpha'}\right)d\tau^2 + d\rho^2 + d\mathbf{x}^2_{\perp}.
\end{equation}
To avoid a conical singularity, the Euclidean time coordinate needs to have the identification $\tau \sim \tau + \beta_R$ where $\beta_R = 2\pi \sqrt{\alpha'}$. \\
Let us now consider the $SL(2,\mathbb{R})/U(1)$ cigar CFT. This CFT was first discovered by Witten in \cite{Witten:1991yr} and has received a lot of attention since then (see e.g. \cite{Dijkgraaf:1991ba}\cite{Tseytlin:1991ht}\cite{Jack:1992mk}\cite{Giveon:1999px}\cite{Kutasov:2000jp}\cite{Kutasov:2005rr}\cite{Sugawara:2012ag}). To lowest order in $\alpha'$ the solution is given by
\begin{align}
\label{lowest}
ds^2 &= \frac{\alpha'k}{4}\left(dr^2 + 4\tanh^2\left(\frac{r}{2}\right)d\theta^2\right), \\
\label{lowest2}
\Phi &= - \ln\cosh\left(\frac{r}{2}\right),
\end{align}
where $\theta \sim \theta +2\pi$. Excitations in this background were studied extensively by \cite{Dijkgraaf:1991ba}.
The lowest lying state (tachyon) has the property
\begin{equation}
\label{ons}
(L_{0} + \bar{L}_0 - 2) \left|T\right\rangle = 0.
\end{equation}
For the cigar gauged WZW model, we can write this as a differential equation using the Laplacian on the group manifold (see \cite{Dijkgraaf:1991ba} for details):
\begin{align}
L_{0} &= -\frac{\Delta}{k-2} - \frac{1}{k}\partial_{\theta_L}^2, \\
\bar{L}_{0} &= -\frac{\Delta}{k-2} - \frac{1}{k}\partial_{\theta_R}^2,
\end{align}
where $\Delta$ denotes the Laplacian on the $SL(2,\mathbb{R})$ manifold. 
The physical state condition 
\begin{equation}
(L_0 - \bar{L}_0) \left|T\right\rangle = 0,
\end{equation} 
implies that tachyons divide in two categories for spaces where the other dimensions do not allow states with $L^{other}_{0} \neq \bar{L}^{other}_{0}$: momentum modes and winding modes. We are however interested in the one-loop free energy in which off-shell stringy states propagate in the loop. This implies we do not have to impose this physicality condition but instead we can relax it to $L_0 - \bar{L}_0 \in \mathbb{Z}$ which is required for modular invariance. For simplicity we will consider only pure winding and pure discrete momentum modes for now, though we will generalize this further on. 
The momentum modes have the following $L_0$:
\begin{equation}
L_{0} = - \frac{1}{k-2}\left[\partial_r^2 + \coth (r) \partial_r + \frac{1}{4}\left(\coth^2\left(\frac{r}{2}\right) - \frac{2}{k}\right)\partial_{\theta}^2\right].
\end{equation}
For winding modes on the other hand, the relevant operator is
\begin{equation}
L_{0} = - \frac{1}{k-2}\left[\partial_r^2 + \coth (r) \partial_r + \frac{1}{4}\left(\tanh^2\left(\frac{r}{2}\right) - \frac{2}{k}\right)\partial_{\tilde{\theta}}^2\right].
\end{equation}
For more details regarding these constructions, the reader is referred to \cite{Dijkgraaf:1991ba}.
The geometry can then be identified by writing: 
\begin{equation}
\label{metricL0}
L_0 = -\frac{\alpha'}{4e^{-2\Phi}\sqrt{G}}\partial_i e^{-2\Phi}\sqrt{G}G^{ij}\partial_j,
\end{equation}
since the on-shell relation (\ref{ons}) should contain the same content as the equations of motion for this field. This identifies immediately the effective metric and dilaton for the momentum modes as
\begin{align}
\label{momgeom}
ds^2 &= \frac{\alpha'}{4}(k-2)\left[dr^2 + \frac{4}{\coth^2\left(\frac{r}{2}\right) - \frac{2}{k}}d\theta^2\right],\\
\Phi &= -\frac{1}{2}\ln\left(\frac{\sinh(r)}{2}\sqrt{\coth^2\left(\frac{r}{2}\right) - \frac{2}{k}}\right),
\end{align}
and to lowest order in $1/k$ these agree with (\ref{lowest}) and (\ref{lowest2}). For the winding modes on the other hand, one finds
\begin{align}
\label{windgeom}
ds^2 &= \frac{\alpha'}{4}(k-2)\left[dr^2 + \frac{4}{\tanh^2\left(\frac{r}{2}\right) - \frac{2}{k}}d\tilde{\theta}^2\right],\\
\Phi &= -\frac{1}{2}\ln\left(\frac{\sinh(r)}{2}\sqrt{\tanh^2\left(\frac{r}{2}\right) - \frac{2}{k}}\right).
\end{align}
In this case the coordinate is identified as $\tilde{\theta} \sim \tilde{\theta} + \frac{2\pi}{k}$. Note that this identification is \emph{not} the same as the T-dual periodicity. The T-dual coordinate would need to have the periodicity $\theta^{Tdual} \sim \theta^{Tdual} + \frac{2\pi}{k-2}$.
This method of obtaining the background metric and dilaton proved to be very powerful for gauged WZW models \cite{Bars:1992sr}. In \cite{Dijkgraaf:1991ba} it was argued that the metric and dilaton obtained in this way are exact in $\alpha'$ and later on substantial evidence for this appeared \cite{Tseytlin:1991ht}\cite{Jack:1992mk}. This method also works for other coset models (see e.g. \cite{Bars:1992sr}).
These two backgrounds are \emph{not} related by the normal T-duality as discussed in \cite{Dijkgraaf:1991ba}. What does this mean? The momentum geometry (\ref{momgeom}) is what non-stringy excitations experience and this is the exact form: the Laplacian in this background determines the momentum modes. Performing a naive T-duality\footnote{I.e. to lowest order in $\alpha'$, using the formulas presented in section \ref{bushe}.} on the discrete momentum tachyon action, we obtain a winding tachyon action whose geometry does not correspond to the above exact dual geometry (\ref{windgeom}). This means that we can regard the mismatch as higher $\alpha'$ corrections to the T-dual tachyon action. When summing all of the $\alpha'$ corrections, we can rewrite the winding tachyon action as only a covariant kinetic term in the above exact dual background. This implies the winding tachyon action indeed gets $\alpha'$-corrected in this case.\\
For type II superstrings however, the story is different. The $L_0$ operator for momentum states is now
\begin{equation}
L_{0} = - \frac{1}{k}\left[\partial_r^2 + \coth (r) \partial_r + \frac{1}{4}\coth^2\left(\frac{r}{2}\right)\partial_{\theta}^2\right].
\end{equation}
For winding modes on the other hand, the relevant operator is
\begin{equation}
L_{0} = - \frac{1}{k}\left[\partial_r^2 + \coth (r) \partial_r + \frac{1}{4}\tanh^2\left(\frac{r}{2}\right)\partial_{\tilde{\theta}}^2\right].
\end{equation}
The overall normalization changed, but also the (innocent-looking) $\frac{2}{k}$ term in the $\theta$ derivatives disappeared. This is crucial since it is this last term that disrupted the interpretation of the dual background as just the naive T-dualized version. In this case the momentum modes perceive the metric and dilaton background as
\begin{align}
\label{metricn}
ds^2 &= \frac{\alpha' k}{4}\left[dr^2 + \frac{4}{\coth^2\left(\frac{r}{2}\right)}d\theta^2\right],\\
\label{diln}
\Phi &= -\ln\left(\cosh\left(\frac{r}{2}\right)\right),
\end{align}
and for the winding modes one finds
\begin{align}
\label{metric}
ds^2 &= \frac{\alpha' k}{4}\left[dr^2 + \frac{4}{\tanh^2\left(\frac{r}{2}\right)}d\tilde{\theta}^2\right],\\
\label{dil}
\Phi &= -\ln\left(\sinh\left(\frac{r}{2}\right)\right).
\end{align}
The coordinate periodicity is again $\theta \sim \theta +2\pi$ and $\tilde{\theta} \sim \tilde{\theta} + \frac{2\pi}{k}$ and now T-duality is manifest. This implies the winding tachyon action is obtained by simply T-dualizing the momentum tachyon action and this contains only the covariant kinetic term. The mass term finally is inserted because the on-shell conditions read\footnote{This only holds for $h = \bar{h}$ states.}
\begin{align}
(L_{0} - 1) \left|T\right\rangle = 0, \quad \text{bosonic},\\
(L_{0} - 1/2) \left|T\right\rangle = 0, \quad \text{type II}.
\end{align}
We conclude that for the type II string on the cigar, the naive lowest order $\alpha'$ effective tachyon action is exact. The dual metric (\ref{metric}) and dilaton (\ref{dil}) are in this case equal to the naive T-dual metric and dilaton of (\ref{metricn}) and (\ref{diln}) of the exact $\alpha'$ geometry. We interpret the dual metric and dilaton as a way to succinctly write down the winding tachyon action. Expanding this dual background around the naive T-dual background allows us to identify the $\alpha'$ corrections to the winding tachyon action. Since in this case, the exact dual and naive T-dual backgrounds are equal, no $\alpha'$ corrections are present.\\
So far, everything we discussed concerns the cigar CFT. Let us now consider the limit $k \to \infty$ to get to Euclidean Rindler space as discussed in \cite{Giveon:2013ica}. For type II superstrings, the authors of \cite{Giveon:2013ica} show that in this case the lowest order $\alpha'$ action is obtained by using this limit on the winding tachyon action in the cigar background. Since we saw above that this action should not get any corrections in $\alpha'$, the Rindler space winding tachyon action should also not get any corrections, at least if $k \to \infty$ is a good description of Rindler space. We conclude that the type II superstring lowest order effective winding tachyon action on Rindler space is exact.\\
Let us analyze bosonic strings more closely. As a first step, we rescale the dual angular coordinate $\tilde{\theta}$ as $\tilde{\theta}_{new} = \frac{k}{k-2}\tilde{\theta}_{old}$. The reason for doing this is that we now have a natural relation between $\theta$ and $\tilde{\theta}$ in terms of T-dual variables. Next we rescale the coordinates as $\rho = \frac{\sqrt{\alpha' (k-2)}}{2} r$, $\tau = \sqrt{\alpha'(k-2)}\theta$ and $\tilde{\tau} = \sqrt{\alpha'(k-2)}\tilde{\theta}$ in the exact bosonic cigar background and its dual. This gives
\begin{align}
ds^2 &= d\rho^2 + \frac{d\tau^2}{\coth^2\left(\frac{\rho}{\sqrt{\alpha'(k-2)}}\right) - \frac{2}{k}}, \quad \text{momentum metric},\\
\label{windmet}
ds^2 &= d\rho^2 + \left(\frac{k-2}{k}\right)^2\frac{d\tilde{\tau}^2}{\tanh^2\left(\frac{\rho}{\sqrt{\alpha'(k-2)}}\right) - \frac{2}{k}}, \quad \text{winding metric}.
\end{align}
where we now require $ \tau \sim \tau + 2\pi\sqrt{\alpha'(k-2)}$ and $\tilde{\tau} \sim \tilde{\tau} + 2\pi\sqrt{\frac{\alpha'}{k-2}}$. Note that indeed $\tau$ and $\tilde{\tau}$ have the correct T-dual periodicities. Let us call $\sqrt{\alpha'(k-2)} = L$. The winding tachyon action in the dual geometry is given by (corresponding to (\ref{ons}))
\begin{align}
&\int_{0}^{+\infty}d\rho\frac{L}{2}\sinh\left(\frac{2}{L}\rho\right)\left[\left|\partial_\rho T\right|^2 + w^2\tilde{G}^{00}\frac{\beta^2}{4\pi^2\alpha'^2}TT^{*} - \frac{4}{\alpha'}TT^{*}\right] \\
&= \int_{0}^{+\infty}d\rho\frac{L}{2}\sinh\left(\frac{2}{L}\rho\right)\left[\left|\partial_\rho T\right|^2 + w^2\frac{\beta^2}{4\pi^2\alpha'^2}\frac{k^2}{(k-2)^2}\left(\tanh^2\left(\frac{\rho}{L}\right) - \frac{2}{k}\right)TT^{*} - \frac{4}{\alpha'}TT^{*} \right]
\end{align}
and this action is exact in $\alpha'$ when considering $\beta$ equal to the inverse Hawking temperature on the cigar.\\
We follow the prescription of \cite{Giveon:2013ica} and let $k \to \infty$ in this action keeping $\rho$ fixed. We obtain
\begin{equation}
\int_{0}^{+\infty}d\rho\rho\left[\left|\partial_\rho T\right|^2 + w^2\frac{\beta^2\rho^2}{4\pi^2\alpha'^3(k-2)}TT^{*} - 2w^2\frac{\beta^2}{4\pi^2\alpha'^2k}TT^{*} - \frac{4}{\alpha'}TT^{*} \right].
\end{equation}
To further ease the identification with the string-normalized Rindler space discussed above in (\ref{snorm}), we extract a $\sqrt{k-2}$ factor from $\beta$ such that the canonical temperature becomes $\beta = 2\pi\sqrt{\alpha'}$ instead of $\beta = 2\pi\sqrt{\alpha'(k-2)}$. After doing this, we obtain for the $w=\pm1$ thermal scalar in the $k\to\infty$ limit:
\begin{equation}
\label{bosts}
\int_{0}^{+\infty}d\rho\rho\left[\left|\partial_\rho T\right|^2 + \frac{\beta^2\rho^2}{4\pi^2\alpha'^3}TT^{*} - 2\frac{\beta^2}{4\pi^2\alpha'^2}TT^{*} - \frac{4}{\alpha'}TT^{*} \right].
\end{equation}
Note that the metric component $\tilde{G}_{00}$ in equation (\ref{windmet}) has a component $\propto 1/k$ which would at first sight vanish when taking the $k\to\infty$ limit. This would leave the large $k$ action unaltered w.r.t. the lowest order (in $\alpha'$) thermal scalar action. However, $\beta^2$ is also proportional to $k$ for large $k$. Hence these factors cancel and leave a finite contribution in the limit. 
Plugging in the canonical Rindler temperature finally yields:
\begin{equation}
\int_{0}^{+\infty}d\rho\rho\left[\left|\partial_\rho T\right|^2 + \frac{\rho^2}{\alpha'^2}TT^{*} - \frac{2}{\alpha'}TT^{*} - \frac{4}{\alpha'}TT^{*} \right].
\end{equation}
We conclude that the only effect of all other higher order $\alpha'$ corrections is a mass shift. Such terms were indeed expected to correct the field theory action as we discussed in the previous subsection.\\
To summarize, for Euclidean Rindler space the bosonic string thermal scalar action receives $\alpha'$ corrections as determined above, whereas the type II superstring thermal scalar action does not get $\alpha'$-corrected.

\section{Critical behavior in Rindler space}
\label{critical}
Using the above field theory actions for the thermal scalar, we can analyze the critical behavior of the one-loop free energy using (\ref{FT}). We can then use (\ref{randwalk}) to rewrite this expression as a random walk in the purely spatial submanifold. Note that we do not yet know what `critical' means for Rindler space, since we have not yet determined the Hagedorn temperature. This will also be done using the thermal scalar action.
In this section we discuss the thermal scalar approximation to the free energy for type II, heterotic and bosonic strings, where we keep the bosonic strings for last since these present (surprisingly!) the most subtleties.

\subsection{Type II Superstrings in Rindler space}
\label{typeII}
We first consider type II superstrings since these do not present complications in the spectrum and these are also more realistic than bosonic strings. The one-loop free energy of the thermal scalar field is given by
\begin{equation}
\label{freebosonic}
\beta F = - \int_{0}^{+\infty}\frac{dT}{T} \text{Tr} e^{-T\left(-\nabla^{2} + m_{local}^2 - G^{ij}\frac{\partial_{j}\sqrt{G_{00}}}{\sqrt{G_{00}}}\partial_{i}\right)}.
\end{equation}
For Rindler spacetime with a general $\beta$ (flat conical spaces), the operator in the exponential is given by
\begin{equation}
\label{opera}
-\partial^{2}_{\rho} - \frac{1}{\rho}\partial_{\rho} - \frac{2}{\alpha'} + \frac{\beta^2\rho^2}{4\pi^2\alpha'^3}.
\end{equation}
We now search for the eigenfunctions and eigenvalues of this operator. Enforcing regularity at the origin and at infinity gives a discrete set of eigenfunctions and eigenvalues given by
\begin{align}
\label{eigenf}
\psi_{n}(\rho) &\propto \exp\left(-\frac{\beta\rho^2}{4\pi\alpha'^{3/2}}\right)L_{n}\left(\frac{\beta\rho^2}{2\pi\alpha'^{3/2}}\right), \\
\label{superspectrum}
\lambda_{n} &= \frac{\beta - 2\pi\sqrt{\alpha'}+2\beta n}{\pi \alpha'^{3/2}},
\end{align}
where $L_{n}$ is the Laguerre polynomial and $n$ is a positive (or zero) integer. The $n=0$ term has the lowest eigenvalue. Setting $\beta = 2\pi\sqrt{\alpha'}$ and $n=0$ gives
\begin{equation}
\label{gs}
\psi_0 \propto \exp\left(-\frac{\rho^2}{2\alpha'}\right), \quad \lambda_0=0.
\end{equation}
The large $T$ contribution is then given by
\begin{equation}
\label{meth22}
\text{Tr} e^{-T\left(-\nabla^{2} + m_{local}^2 - G^{ij}\frac{\partial_{j}\sqrt{G_{00}}}{\sqrt{G_{00}}}\partial_{i}\right)} \to e^{-\lambda_{0}T} = \exp\left(-\frac{\beta - 2\pi\sqrt{\alpha'}}{\pi \alpha'^{3/2}}T\right).
\end{equation}
Let us reinterpret this one-loop free energy as a random walk in the spatial submanifold as we did in going from equation (\ref{FT}) to (\ref{randwalk}).
Applying formula (\ref{randwalk}) gives
\begin{equation}
\beta F = - \int_{0}^{+\infty}\frac{dT}{T}\int_{S^{1}} \left[\mathcal{D}x\right]\sqrt{G_{sp}}e^{-S - S_{sp}}
\end{equation}
where
\begin{equation}
\label{ppp2}
S = \frac{1}{4\pi\alpha'}\int_{0}^{T}dt\left[\dot{\rho}^2+\frac{\beta^2\rho^2}{\alpha'}-\frac{\pi^2\alpha'^2}{\rho^2}-8\pi^2\alpha' \right]
\end{equation}
and $G_{sp}$ and $S_{sp}$ denote the metric and action of the other spectator dimensions needed to get a valid string background. We are not interested in these at the moment.\\
Our original goal was to get information on the corrections to the worldsheet dimensional reduction approach as discussed in \cite{Mertens:2013pza}. Let us look at the particle action (\ref{ppp2}) more closely. The first two terms are the only ones that are found in the Fourier expansion in the string path integral as discussed in \cite{Mertens:2013pza}. The third term was denoted $K(x)$ in section \ref{pathderiv} and it is caused by removing the $\sqrt{G_{00}}$ from the measure of the field theory action as we discussed above. The fourth term is the Hagedorn correction that can be found by looking at flat space models. The free energy corresponding to the above action (\ref{ppp2}) has a random walk interpretation with a modified potential. \\
The result is a particle path integral on a half-line in a harmonic oscillator and in a $1/\rho^2$ potential. From equation (\ref{meth22}) we can distill the Hagedorn temperature. To do this, of course, we need to know what the other spectator dimensions are. These can obviously influence the Hagedorn temperature. Let us choose 8 flat dimensions such that the total one-loop free energy of the thermal scalar becomes
\begin{equation}
\label{type2free}
\beta F = -V_{T}\int_{0}^{+\infty}\frac{dT}{T}\left(\frac{1}{4\pi T}\right)^{4}\exp\left(-\frac{\beta}{\pi\alpha'^{3/2}}T + \frac{2}{\alpha'}T \right).
\end{equation}
Crucially, the traced heat kernel of flat dimensions do not yield corrections to the exponential. Convergence in the large $T$ limit then determines the critical temperature as:
\begin{equation}
\label{convsuper}
\beta \geq 2\pi\sqrt{\alpha'}.
\end{equation}
We clearly find $\beta_{critical} = \beta_{R}$ so the canonical Rindler temperature (needed to avoid the conical singularity and defined in equation (\ref{snorm})) is precisely equal to the critical temperature above which the free energy would diverge in the IR. We will comment on the link to the physical black hole normalized coordinates (\ref{Rindmetric}) further on. This marginal convergence is associated with a state in the string spectrum that becomes massless precisely when $\beta = 2\pi\sqrt{\alpha'}$ and this stringy state was found by the authors of \cite{Kutasov:2000jp}\cite{Giveon:2012kp}\cite{Giveon:2013ica}. \\
For any $\beta$, the wavefunction of the lowest state ($n=0$) is given by
\begin{equation}
\label{groundstate}
\psi_0 \propto \exp\left(-\frac{\beta\rho^2}{4\pi\alpha'^{3/2}}\right).
\end{equation}
We identify the lowest eigenfunction of the associated particle heat kernel with the wavefunction of the thermal scalar stringy state. So qualitatively the type II thermal scalar in Rindler space is a massless state located close to the origin in Euclidean Rindler space. As a sidenote, we remark that expression (\ref{type2free}) is also UV divergent, although this is a remnant of considering solely the thermal scalar field theory and not the full string theory. Ultimately we should not take $0$ as the lower boundary of the $T$-integral since we assumed $T$ large. This is actually the same story as in flat space: the difference between the fundamental modular domain and the strip is the reason for the divergence.\\

\noindent Note also that even though the original string path integral was only well-defined for $\beta = \frac{2\pi\sqrt{\alpha'}}{N}$ (for $N \neq 1$ these models are flat space $\mathbb{C}/\mathbb{Z}_N$ orbifolds that have been used in the past to predict string thermodynamics \cite{Dabholkar:1994ai}\cite{Lowe:1994ah} and we will discuss these further on), the resulting particle path integral and the final result have been calculated for a general $\beta$. Thus this provides a \emph{natural} continuation of the string results to a general $\beta$. The continuation is in terms of the field theory of the states constituting the string theory, and this road to off-shell descriptions of string theory has indeed been the most fruitful one (see e.g. \cite{Taylor:2003gn} and references therein). In particular, this allows us to differentiate with respect to $\beta$ to obtain the thermodynamic entropy.\\

\noindent The authors of \cite{Kruczenski:2005pj} give a formula for the average size of the long string. They showed that it is given by the width of the ground state wavefunction of the associated particle heat kernel. Applying such a reasoning to the ground state wavefunction given in (\ref{groundstate}), we find as a measure for the width of the $n=0$ mode that
\begin{equation}
\left\langle \rho^2\right\rangle = \frac{2\pi\alpha'^{3/2}}{\beta}.
\end{equation}
Note that this assumes that values of $\beta \neq 2\pi \sqrt{\alpha'}$ are meaningful as we have discussed above.\\

\noindent To sum up, the wavefunction of the thermal scalar is identified with the lowest eigenfunction of the particle heat kernel and this determines the region where the random walk is situated. Thus the picture we arrive at is that the thermal scalar is represented by a random walk close to the Rindler origin. This interpretation is the same as in flat spacetime, the difference is that in Rindler space the walk is localized close to the origin. Since the Rindler origin is actually the black hole horizon, we conclude that the thermal scalar is localized to a string length surrounding a black hole. For clarity about the transition from this string-normalized Rindler spacetime (\ref{snorm}) to the black hole-normalized Rindler spacetime (\ref{Rindmetric}), we refer to section \ref{GM} where we translate the results from this section to the black hole case.

\subsection{Heterotic strings in Rindler space}
\label{heterotic}
We can also solve the previous model for the heterotic string in Rindler spacetime. In this case, we do not at first sight have a WZW cigar model to guide us, so we will compute the critical behavior \emph{assuming} no $\alpha'$ corrections to the thermal scalar action, just like for the type II case. We will provide arguments in favor of this further on in section \ref{heteroticc}. The operator $\hat{\mathcal{O}}$ to be considered for the heterotic string is given by
\begin{equation}
\frac{1}{2}\left(-\partial^{2}_{\rho} - \frac{1}{\rho}\partial_{\rho} - \frac{3}{\alpha'} + \frac{\beta^2\rho^2}{4\pi^2\alpha'^3} + \frac{\alpha' \pi^2}{\beta^2 \rho^2}\right),  
\end{equation}
with eigenfunctions and eigenvalues given by
\begin{align}
\psi_{n}(\rho) \propto \rho^{\frac{\pi\sqrt{\alpha'}}{\beta}}\exp\left(-\frac{\beta\rho^2}{4\pi\alpha'^{3/2}}\right)L_{n}^{\left(a\right)}\left(\frac{\beta\rho^2}{2\pi\alpha'^{3/2}}\right), \quad \lambda_{n} = \frac{\beta - 2\pi\sqrt{\alpha'}+2\beta n}{\pi \alpha'^{3/2}},
\end{align}
where in this case the generalized Laguerre polynomial is used with order $a=\frac{\pi\sqrt{\alpha'}}{\beta}$. The lowest eigenfunction is given by
\begin{equation}
\psi_0 \propto \rho^{\frac{\pi\sqrt{\alpha'}}{\beta}}\exp\left(-\frac{\beta\rho^2}{4\pi\alpha'^{3/2}}\right).
\end{equation}
The ground state has again zero eigenvalue for the canonical Rindler temperature so, like in the type II case, the convergence criterion is given by:\footnote{When we include some extra flat dimensions.}
\begin{equation}
\label{convhet}
\beta \geq 2\pi \sqrt{\alpha'}.
\end{equation}
In particular we have again that the canonical Rindler temperature equals the Rindler Hagedorn temperature. At this temperature, the zero-mode has the wavefunction
\begin{equation}
\psi_0 \propto \rho^{\frac{1}{2}}\exp\left(-\frac{\rho^2}{2\alpha'}\right), \quad \lambda_0=0.
\end{equation}
For the heterotic string, the $n=0$ width formula changes to 
\begin{equation}
\left\langle \rho^2\right\rangle = \frac{2\pi\alpha'^{3/2}}{\beta}\left(1+\frac{\pi\sqrt{\alpha'}}{\beta}\right).
\end{equation} 
In particular, for the canonical Rindler temperature, we get a size equal to $\sqrt{\frac{3}{2}}\sqrt{\alpha'}$ which is a factor of $\sqrt{\frac{3}{2}}$ \emph{larger} than the case considered above.\\
The random walk behavior has the form:
\begin{equation}
\beta F = - \int_{0}^{+\infty}\frac{dT}{T}\int_{S^{1}} \left[\mathcal{D}x\right]\sqrt{G_{sp}}e^{-S - S_{sp}}
\end{equation}
where
\begin{equation}
S = \frac{1}{4\pi\alpha'}\int_{0}^{T}dt\left[\dot{\rho}^2+\frac{\beta^2\rho^2}{\alpha'} + \frac{4\pi^4\alpha'^3}{\beta^2\rho^2}-\frac{\pi^2\alpha'^2}{\rho^2}-12\pi^2\alpha' \right].
\end{equation}
Finally note that we have lost the heterotic thermal duality \cite{O'Brien:1987pn} in this case. The duality symmetry is compromised as soon as one considers a non-trivial background.

\subsection{Bosonic strings in Rindler space}
\label{bosonic}
In this section we discuss the same story for bosonic strings. This case is more complex due to the $\alpha'$ corrections and also due to unitarity constraints that we will discuss further on in section \ref{unitarity}. \\
If we include the $\alpha'$ corrections, we need to consider the operator
\begin{equation}
-\partial^{2}_{\rho} - \frac{1}{\rho}\partial_{\rho} - 2\frac{\beta^2}{4\pi^2\alpha'^2} - \frac{4}{\alpha'} + \frac{\beta^2\rho^2}{4\pi^2\alpha'^3},
\end{equation}
where also the substitution $\frac{2}{\alpha'} \to \frac{4}{\alpha'}$ was made in comparison to the type II superstring. The eigenfunctions remain the same as in the type II case but the eigenvalues shift to
\begin{equation}
\label{bosonicspectrum}
\lambda_{n} = \frac{\beta - 4\pi\sqrt{\alpha'} - \frac{\beta^2}{2\pi\sqrt{\alpha'}}+2\beta n}{\pi \alpha'^{3/2}}.
\end{equation}
The $n=0$ term has the lowest eigenvalue. This state has the same wavefunction and hence also the same width as the type II case discussed above. A further subtlety is whether these quantum numbers are really in the string spectrum. We will discuss this further in section \ref{unitarity}, where we will conclude that actually the bosonic spectrum only starts at $n=1$. In this section we will ignore this complication because other thermodynamic quantities (like the entropy) do appear to rely on the $n=0$ mode. Setting $\beta = 2\pi\sqrt{\alpha'}$ gives a negative $n=0$ eigenvalue. 
The (naive) critical behavior is given by
\begin{equation}
\label{meth1alpha}
\text{Tr} e^{-T\left(-\nabla^{2} + m_{local}^2 - G^{ij}\frac{\partial_{j}\sqrt{G_{00}}}{\sqrt{G_{00}}}\partial_{i}\right)} \to e^{-\lambda_{0}T} = \exp\left(-\frac{\beta - 4\pi\sqrt{\alpha'}-\frac{\beta^2}{2\pi\sqrt{\alpha'}}}{\pi \alpha'^{3/2}}T\right).
\end{equation}
The random walk interpretation can again be found by applying formula (\ref{randwalk}) and including the $\alpha'$ correction term in this case gives
\begin{equation}
\beta F = - \int_{0}^{+\infty}\frac{dT}{T}\int_{S^{1}} \left[\mathcal{D}x\right]\sqrt{G_{sp}}e^{-S - S_{sp}}
\end{equation}
where
\begin{equation}
\label{pppp}
S = \frac{1}{4\pi\alpha'}\int_{0}^{T}dt\left[\dot{\rho}^2+\frac{\beta^2\rho^2}{\alpha'}-\frac{\pi^2\alpha'^2}{\rho^2}-16\pi^2\alpha'- 2\beta^2\right]
\end{equation}
with $G_{sp}$ and $S_{sp}$ the metric and action of the spectator dimensions. \\
For bosonic strings in Rindler space, we have a nice demonstration of all different types of corrections to the naive particle action (\ref{act}). 
All terms have the same origin as in the type II case, except the final term. This term combines all $\alpha'$ corrections to the field theory action for this particular background. We saw previously that in general such terms were to be expected in Rindler space and the cigar CFT approach indeed produces such a correction term. We believe that now all corrections are determined. The free energy corresponding to the above action (\ref{pppp}) has a random walk interpretation with a modified potential and with a temperature-dependent mass. \\
As a further check on these results and in particular the manipulations done to get from (\ref{FT}) to (\ref{randwalk}), we explicitly solve the particle path integral directly and get the same results as above. We present these results in section \ref{appB}, where we also discuss the modifications needed to treat superstrings and heterotic strings. \\
Again choosing 24 flat dimensions, the total one-loop free energy of the thermal scalar becomes
\begin{equation}
\label{bosfree}
\beta F = -V_{T}\int_{0}^{+\infty}\frac{dT}{T}\left(\frac{1}{4\pi T}\right)^{12}\exp\left(-\frac{\beta}{\pi\alpha'^{3/2}}T + \frac{4}{\alpha'}T + \frac{\beta^2}{2\pi^2\alpha'^2}T\right).
\end{equation}
The last term in the exponential factor is the result of the $\alpha'$ corrections and its influence is substantial. If we ignored the $\alpha'$ corrections, we would determine the convergence criterion to be
\begin{equation}
\beta \geq 4\pi\sqrt{\alpha'}.
\end{equation}
If we include this final term however, we would find a divergence in thermodynamical quantities for any value of $\beta$:\footnote{Note that, as we previously discussed, we should think of this divergence as occuring in the entropy and not in the free energy itself since this quantity only starts with the $n=1$ mode as we will discuss further on.} the Hagedorn temperature is effectively zero (there is always a negative eigenvalue). \\
If we would (naively) consider arbitrary winding modes, it is readily checked that not only the $w=\pm1$ mode, but also all higher winding modes are tachyonic at the canonical Rindler temperature. This behavior is a drastic departure from the lowest $\alpha'$ action discussed above (i.e. ignoring the last term in the exponential in (\ref{bosfree})) where $w=\pm2$ is massless and all higher winding modes would be massive. This discussion is a bit mute since we will see that all higher winding modes are actually not in the Euclidean Rindler spectrum in the first place, but it is an illustration of the importance of the $\alpha'$ corrections in a cigar-shaped background (or its flat Euclidean Rindler limit considered here). \\

\noindent To conclude the bosonic thermal scalar in Rindler space, the lowest $n=0$ state is tachyonic and is situated at a string length from the origin of Rindler space. However, we will see in section \ref{unitarity} that the string spectrum only starts at $n=1$ in this case. We will postpone any further discussion on this until then. 

\subsection{Hagedorn behavior of the Rindler string}
Above we established the critical temperatures for the string gas in Rindler space. We found that for type II and heterotic strings, the free energy is `marginally' convergent. In this section we want to elaborate on whether the free energy converges or diverges at this critical temperature. So we zoom in on the behavior of the free energy at the Hagedorn temperature. The crucial aspect is whether the spectator dimensions are compact or non-compact. Let us first take the other dimensions (tangential to the horizon) as compact dimensions. We have to do this to circumvent the Jeans instability. The effect of compact dimensions on the Hagedorn behavior is illustrated in section \ref{dom}. For type II superstrings and for heterotic strings we obtain the following free energy expression
\begin{equation}
F = -\frac{1}{\beta}\int_{0}^{+\infty}\frac{dT}{T},
\end{equation}
which is analogous to (\ref{bosfree}) by replacing the non-compact dimensions by compact ones and adjusting for the superstring or heterotic string. We clearly see a logarithmic divergence for $T \to \infty$ in this case. The free energy at the Hagedorn temperature (which is the same as the canonical Rindler temperature) diverges in the IR and this divergence is caused by the massless states. Other non-winding massless states also cause this same kind of divergence, but they are independent of the temperature and their contribution is dropped (and these do not influence the entropy). We find that the thermal scalar yields the divergent temperature-dependent contribution to thermodynamic quantities in Rindler space and it should take over the entire thermodynamics. One can compare this type of divergence with the microcanonical picture in flat space where one can never reach the Hagedorn temperature and the free energy diverges at this temperature. This does not cause a condensation process, but represents here our inability to reach this temperature. The difference with this case and the Rindler case, is that here the temperature should remain fixed at the canonical Rindler temperature and this equals the Hagedorn temperature. So in some sence, the Rindler string is held fixed at this unobtainable temperature of strings-in-a-box.
If we only take the contribution from the thermal scalar, we find a random walk behavior as was predicted by Susskind.\\

\noindent Since $F$ becomes infinite, the string density becomes arbitrarily high. It is in this way that the higher order string interactions should come in and cure this behavior. In \cite{Susskind:2005js} it was argued that interactions would cause a repulsion in such a way that the density becomes constant when nearing the horizon. Higher genus near-Hagedorn thermodynamics has not been studied much in the past.\footnote{However see \cite{Brigante:2007jv} where the Hagedorn behavior is studied using factorizations of higher genus Riemann surfaces and dual holographic matrix models.} \\

\noindent If we would take at least one non-compact dimension, the free energy density becomes finite at the Hagedorn temperature. The thermal scalar no longer dominates the free energy. However, if we differentiate with respect to $\beta$, we bring powers of the Schwinger parameter $T$ down that deteriorate the convergence properties in the IR limit $T \to \infty$. In other words, the non-analyticity of the free energy gets its dominant contribution from the thermal scalar and the number of compact dimensions determines how many times one has to differentiate to obtain a divergence. This situation is exactly as in flat space. The behavior of the free energy near the Hagedorn temperature when including $D$ non-compact dimensions is given by
\begin{equation}
\beta F \propto 
\left\{
    \begin{array}{ll}
        \left(\beta - \beta_{H}\right)^{D/2}\ln\left(\beta - \beta_{H}\right), \quad D\text{ even}, \\
        \left(\beta - \beta_{H}\right)^{D/2}, \quad D\text{ odd}.
    \end{array}\right.
\end{equation}

\noindent We will now compute the asymptotic density of single string states and provide an interpretation of the long string and the stretched horizon. For type II superstrings and heterotic strings, the single string partition function is dominated by the zero-mode as given by (\ref{type2free}): 
\begin{equation}
Z = V_{T}\int_{0}^{+\infty}\frac{dT}{T}\left(\frac{1}{4\pi T}\right)^{D/2}\exp\left(-\frac{\beta}{\pi\alpha'^{3/2}}T + \frac{2}{\alpha'}T \right),
\end{equation}
where $V_T$ denotes the volume of the transverse dimensions, i.e. the area of the black hole horizon. We have chosen $D$ non-compact spectator dimensions in this discussion. The temperature $\beta$ in Rindler space is defined as the local temperature at $\rho = 4GM$ and given in string units corresponding with the definition of the Rindler metric as in (\ref{snorm}). In physical units where the metric is given by (\ref{Rindmetric}), this becomes
\begin{equation}
\beta F = -V_{T}\int_{0}^{+\infty}\frac{dT}{T}\left(\frac{1}{4\pi T}\right)^{D/2}\exp\left(-\frac{\beta}{\pi\alpha'(4GM)}T + \frac{2}{\alpha'}T \right).
\end{equation}
Defining the energy at $\rho=4GM$ by 
\begin{equation}
E = \frac{T}{4\pi\alpha'GM},
\end{equation}
we find:
\begin{equation}
\label{star}
\beta F = -V_{T}\int_{0}^{+\infty}\frac{dE}{E^{1+D/2}}\left(\frac{1}{16\pi^2\alpha'GM}\right)^{D/2}\exp\left(-\beta E + 8\pi GM E \right).
\end{equation}
The single string density of states is given by an inverse Laplace transform of this (single string) partition function  as
\begin{equation}
-\beta F = z = \int_{0}^{+\infty}dE \omega(E) e^{-\beta E},
\end{equation}
which at high energies using (\ref{star}) yields
\begin{equation}
\omega(E) \approx V_T \left(\frac{1}{2\pi\beta_R \alpha'}\right)^{D/2}\frac{e^{\beta_R E}}{E^{1+D/2}}
\end{equation}
with $\beta_R = 8\pi GM$ the inverse Rindler temperature. The asymptotic density of states clearly displays Hagedorn behavior with critical temperature exactly equal to the Rindler temperature. Note also that the density of states is proportional to the transverse area (the horizon area). Let us now give an interpretation of the stretched horizon using these results. We can define the location of the long string (or the stretched horizon) as the distance from the horizon where the blueshifted canonical Rindler temperature becomes equal to the flat space Hagedorn temperature. Using the flat space Hagedorn expressions
\begin{align}
\beta_{H,\text{flat}} &= 2\sqrt{2}\pi\sqrt{\alpha'}, \quad \text{Type II}, \\
\beta_{H,\text{flat}} &= (2+\sqrt{2})\pi\sqrt{\alpha'}, \quad \text{Heterotic},
\end{align}
and
\begin{equation}
\beta_R \sqrt{G_{00}(\rho_{sh})} = \beta_R \frac{\rho_{sh}}{4GM} = \beta_{H,\text{flat}},
\end{equation}
where the subscript $sh$ denotes stretched horizon quantities, we can localize the stretched horizon at 
\begin{align}
\rho_{sh} &= \sqrt{2\alpha'}, \quad \text{Type II}, \\
\rho_{sh} &= \left(1 + \frac{1}{\sqrt{2}}\right) \sqrt{\alpha'}, \quad \text{Heterotic}.
\end{align}
The physical picture which arises from this, is that the thermodynamics of a gas of strings in a Rindler background is dominated by a single long string living at $\rho_{sh} \propto \sqrt{\alpha'}$ at the flat space Hagedorn temperature.\footnote{Again we should take into account the compactness of the other dimensions to determine whether one string dominates or multiple-string configurations dominate \cite{Deo:1989bv}.} The density of single string states as measured by a Rindler observer at the stretched horizon is then given by
\begin{equation}
\omega(E_{sh}) \approx V_T \left(\frac{1}{2\pi\beta_R \alpha'}\right)^{D/2}\left(\frac{\beta_R}{\beta_{H,\text{flat}}}\right)^{D/2}\frac{e^{\beta_{H,\text{flat}} E_{sh}}}{E_{sh}^{1+D/2}},
\end{equation}
where we used the redshift $\beta_R E = \beta_{H,\text{flat}} E_{sh}$ and $\omega(E) dE = \omega(E_{sh}) dE_{sh}$. For a black hole, the asymptotic (single string) density of states is the same and the Rindler temperature equals the Hawking temperature in this case, as we have discussed already before. This behavior can be interpreted as Hagedorn critical behavior which is redshifted to the Hawking temperature for the observer at infinity.

\section{Comparison to flat $\mathbb{C}/\mathbb{Z}_N$ orbifold CFTs}
\label{orbifold}
\begin{figure}[h]
\centering
\includegraphics[width=7cm]{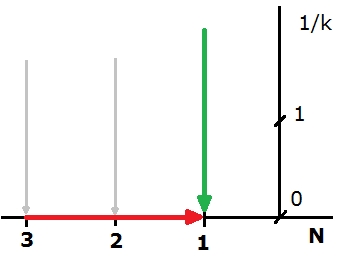}
\caption{$1/k$ versus orbifold number $N$. The horizonal red arrow represents the $\mathbb{C}/\mathbb{Z}_N$ continuation to $N=1$. The vertical green line represents the approach taken by the authors of \cite{Giveon:2013ica}, where one takes $k\to\infty$ in the cigar CFT. The vertical gray lines represent the strategy we are following in this section.}
\label{logic}
\end{figure}

In this section we will provide additional evidence for the recent claim by \cite{Giveon:2013ica} we discussed above that the $\alpha'$ corrections to the thermal scalar action can be deduced from the action on the $SL(2,\mathbb{R})/U(1)$ cigar. We will do this by comparing the resulting critical one-loop free energy with the one coming from flat space orbifolds \cite{Dabholkar:1994ai}\cite{Lowe:1994ah}. The orbifolds we have in mind are obtained by creating a conical space from a two-dimensional plane: the $\mathbb{C}/\mathbb{Z}_N$ orbifolds. We refer the reader to \cite{Dabholkar:1994ai}\cite{Lowe:1994ah} for a treatment of the spectrum and the partition function. These have been used as an approach to thermodynamics of the Rindler observer. Above we argued that we should be able to vary $\beta$ continuously to study thermodynamics. For string theory, only $\beta = \frac{2\pi\sqrt{\alpha'}}{N}$ correspond to CFTs, namely those obtained by orbifolding flat space. The spirit of \cite{Dabholkar:1994ai}\cite{Lowe:1994ah} is to take these discrete values of $\beta$ and afterwards analytically continue the resulting expression to a general $\beta$. Although the continuation and interpretation of these results to Rindler space ($N \to 1$) is not entirely airtight, for $\beta = \frac{2\pi\sqrt{\alpha'}}{N}$ the orbifold construction is well-founded in string theory and so it is an important check if we reproduce this with the cigar $k\to \infty$ approach. We will find perfect agreement. To clarify our intents, we sketch the strategy in figure \ref{logic}.
Taking $k\to\infty$ in the $SL(2,\mathbb{R})/U(1)$ model as in \cite{Giveon:2013ica}, we obtain a field theory action that \emph{supposedly} is the Rindler thermal scalar action. We do not have a verification for this result since the $\mathbb{C}/\mathbb{Z}_N$ continuation does not give a prediction for the thermal scalar action for $N=1$.\footnote{Simply setting $N=1$ in the $\mathbb{C}/\mathbb{Z}_N$ partition function of \cite{Dabholkar:1994ai}\cite{Lowe:1994ah} yields the partition function of an infinite 2d plane. This does not reflect the thermal scalar winding the Euclidean origin and is useless when we are interested in for instance the thermal entropy. This is also the reason Rindler thermodynamics is more complicated than simply stating that Rindler space is flat space.} For $N>1$, the $\mathbb{C}/\mathbb{Z}_N$ construction \emph{does} give a prediction for the thermal scalar action and we will find a precise match with this result. \\
Consider the bosonic string partition function on the $\mathbb{C}/\mathbb{Z}_N$ orbifold \cite{Dabholkar:1994ai}\cite{Lowe:1994ah}:
\begin{equation}
Z = \frac{V_{T}}{N}\int_{F}\frac{d\tau_1d\tau_2}{2\tau_2}\frac{1}{\left|\eta(\tau)\right|^{44}(4\pi^2\alpha'\tau_2)^{12}}\sum_{m,n=0}^{N-1}Z_{m,n}
\end{equation}
where 
\begin{equation}
Z_{m,n} = \left| \frac{\eta(\tau)}{\theta\left[
\begin{array}{c}
\frac{1}{2}+\frac{m}{N} \\
\frac{1}{2}+\frac{n}{N}  \end{array} 
\right](\tau)}\right|^2
\end{equation}
for $(m,n) \neq (0,0)$ and $Z_{0,0} = \frac{A}{\tau_2\left|\eta(\tau)\right|^4}$. The quantity $V_T$ denotes the volume of the other dimensions which have been chosen flat. The $m=n=0$ sector has $Z \propto \beta$ and so gives a temperature-independent contribution to the free energy and this is dropped from now on.
We study the $\tau_2 \to \infty$ limit in this case.\footnote{It is easier to take this limit in the fundamental modular domain. Below, in section \ref{flatlimit}, we will present a more elaborate comparison between both partition functions.} 
The theta function with characteristics has the asymptotic behavior
\begin{equation}
\left|\theta\left[
\begin{array}{c}
\frac{1}{2} + \frac{m}{N} \\
b  \end{array} 
\right](\tau)\right| \to e^{-\pi \left(\frac{1}{2} + \frac{m}{N}\right)^2 \tau_2}e^{2\pi\tau_2\frac{m}{N}}.
\end{equation}
So we get\footnote{To be precise, the $m=0$ sector has a different prefactor since we dropped the $m=n=0$ sector. We do not care about this, since we will drop the $m=0$ sector anyway in what follows.}
\begin{equation}
Z = V_{T}\int^{+\infty}\frac{d \tau_2}{2\tau_2}\frac{1}{(4\pi^2\alpha'\tau_2)^{12}}\sum_{m=0}^{N-1}e^{2\pi\tau_2\left(2-\frac{m}{N}+\frac{m^2}{N^2}\right)}.
\end{equation}
The $\tau_1$ integral gives a multiplicative factor of 1. The $m=1$ and $m=N-1$ sectors dominate so the critical behavior is given by
\begin{equation}
Z = V_{T}\int^{+\infty}\frac{d \tau_2}{\tau_2}\frac{1}{(4\pi^2\alpha'\tau_2)^{12}}e^{2\pi\tau_2\left(2-\frac{1}{N}+\frac{1}{N^2}\right)}.
\end{equation}
The critical behavior of the free energy becomes
\begin{equation}
\label{free}
F = -\frac{V_{T}N}{2\pi}\int^{+\infty}\frac{d \tau_2}{\tau_2}\frac{1}{(4\pi^2\alpha'\tau_2)^{12}}e^{2\pi\tau_2\left(2-\frac{1}{N}+\frac{1}{N^2}\right)}.
\end{equation}
Looking back at the bosonic thermal scalar action (\ref{bosts}), we fill in the orbifold temperatures ($\beta = \frac{2\pi\sqrt{\alpha'}}{N}$) and obtain\footnote{This state actually coincides with the most divergent state in the $SL(2,\mathbb{R})/U(1)$ conical orbifolds as it should.}
\begin{equation}
\int_{0}^{+\infty}d\rho\rho\left[\left|\partial_\rho T\right|^2 + w^2\frac{\rho^2}{N^2\alpha'^2}TT^{*} - \frac{2}{\alpha'}\frac{w^2}{N^2}TT^{*} - \frac{4}{\alpha'}TT^{*} \right].
\end{equation}
The partition function becomes
\begin{equation}
Z = V_{T}\int_{0}^{+\infty}\frac{dT}{T}\left(\frac{1}{4\pi T}\right)^{12}e^{-\frac{2}{N\alpha'} T}e^{\frac{4}{\alpha'}T}e^{\frac{2}{N^2\alpha'}T}.
\end{equation}
Performing the substitution $T = \pi\alpha'\tau_2$, we obtain
\begin{equation}
\label{orbi}
F = -\frac{V_{T}N}{2\pi}\int_{0}^{+\infty}\frac{d\tau_2}{\tau_2}\left(\frac{1}{4\pi^2\alpha' \tau_2}\right)^{12}e^{2\pi\tau_2\left(2 - \frac{1}{N} + \frac{1}{N^2}\right)},
\end{equation}
which precisely coincides with (\ref{free}).\\
Next we discuss type II and heterotic superstrings. We will suffice by comparing the criterion for divergence of the partition function, although an elaborate treatment like for the bosonic string discussed above is possible. It is shown in \cite{Lowe:1994ah} that oddly twisted sectors (with twist $w$) have tachyons with masses $M^2 = \frac{2}{\alpha'}\left(\frac{w}{N}-1\right)$. Since with our notation $w\beta = \frac{w}{N}2\pi\sqrt{\alpha'}$, we find that the convergence criterion for the free energy that we found earlier in (\ref{convsuper}) and (\ref{convhet}) can be rewritten as
\begin{equation}
\label{convcrit}
0 \leq \beta - 2\pi \sqrt{\alpha'} = 2\pi\sqrt{\alpha'}\left(\frac{1}{N} - 1\right) = \pi \alpha'^{3/2}M^2.
\end{equation}
So we find that for $N \in \mathbb{N}$, the convergence criterion is equivalent to whether the most tachyonic twisted state has $M^2 > 0$ or $M^2< 0$. The analytic continuation discussed in \cite{Dabholkar:1994ai} concerning $N$ gets translated to an analytic continuation in $\beta$. The arguments in favor of this continuation in \cite{Dabholkar:1994ai} are equivalent in our case to arguments of taking $\beta$ away from the orbifold values. In our language of field theory, the continuation in $\beta$ is quite natural. \\
We believe this is an important consistency check. To recapitulate, using the cigar WZW model and taking $k \to \infty$ we find the precise $\alpha'$ corrected action for the different types of strings. For the discrete orbifold temperatures however, we already know what the result should be \cite{Dabholkar:1994ai}\cite{Lowe:1994ah}. Since we are able to reproduce precisely the limiting free energy in the orbifold CFTs, we believe that indeed as argued above and in \cite{Giveon:2013ica}, the Rindler space thermal scalar action (including \emph{all} its $\alpha'$ corrections) can be obtained by taking $k \to \infty$ in the cigar CFT. The result is that type II superstrings do not receive $\alpha'$ corrections in their thermal scalar action, while bosonic strings do receive corrections. Without these, the orbifold result would not have been reproduced (the $1/N^2$ term in the exponent in (\ref{orbi}) would be missed). \\
Note also that we have gone full circle now: the $\tau_2 \to \infty$ limit should correspond to a state in the CFT that gives the dominant thermal behavior. From the field theory perspective we considered the thermal scalar action (with possible $\alpha'$ corrections) to give us this dominant contribution. But the link with the exact conformal description is lost. In this case we clearly see that these two descriptions match. \\
For heterotic strings, we also find precisely the same convergence criterion (\ref{convcrit}) as in the $\mathbb{C}/\mathbb{Z}_N$ orbifold models. Since the heterotic string thermal scalar action also includes a discrete momentum contribution, we again find it non-trivial to find a perfect match to the $\mathbb{C}/\mathbb{Z}_N$ orbifold. We believe that the heterotic thermal scalar action should also not receive $\alpha'$ corrections (like the type II superstring). We present a detailed argument in favor of this in section \ref{heteroticc}; the main complication is the subtleties with heterotic WZW models as discussed e.g. in \cite{Giveon:1993hm} and \cite{Sfetsos:1993bh}. \\
We note that, to treat $N\neq 1$, it suffices to simply take the correct value of $\beta = \frac{2\pi}{N}\sqrt{\alpha'}$; no additional (conical) corrections are present. If there would be conical corrections, we expect these to exist for these special cones as well. As discussed in section \ref{general}, the only place where $\beta$ enters the field theory action is in the $\partial_0$ derivatives and indeed we find here a $\beta^2$ contribution, which was the only type of $\beta$-dependent correction term we anticipated there. We believe this supports our expectation that we can safely take $\beta$ away from the CFT-values $\frac{2\pi}{N}\sqrt{\alpha'}$.\footnote{Let us elaborate on this point of view. Consider the field theory as discussed in general in section \ref{general}. Changing $\beta$ only affects the periodicity of one of the coordinates. For the T-dual background, the origin $\rho=0$ is not a fixpoint of the $U(1)$ rotation $\tau \to \tau + C$ with $C$ a real constant. So the T-dual geometry should not become `extra' singular just from the periodic identification, meaning that at $\rho=0$ we expect only the curvature singularity and this is not sensitive to the periodicity parameter $\beta$. More concretely, these arguments show that the only type of correction we can have in the thermal scalar action is 
\begin{equation}
\Delta S = f(\rho)\partial_\rho T \partial_\rho T^* + g(\rho)\beta^2 TT^{*} + h(\rho)TT^{*},
\end{equation}
with three unknown functions $f$, $g$ and $h$ that do not depend on $\beta$. Since we know the result at $\beta = \frac{2\pi}{N}\sqrt{\alpha'}$, we have $f=h=0$ and $g$ is the $\rho$-independent correction we found above. This holds now for \emph{all} values of $\beta$. In particular the $\alpha'$ corrections do not generate $\beta$-dependence, except the $\beta^2$ already discussed above. The apparent subtlety in these results is whether the thermal scalar action really can be determined only by the T-dual quantities, also off-shell. This seems to be the case and T-duality invariance is one of the biggest ideas used to construct off-shell descriptions of string theory (the so-called double field theory, see e.g. \cite{Aldazabal:2013sca} and references therein).} \\
We want to remark that there is a discrepancy in the wavefunctions: the orbifold twisted sector wavefunctions are localized at the tip of the cone, while in our case we find a string-scale spread around the tip when taking the $\beta$ continuation seriously.\footnote{Actually this might not be a real problem after all. Firstly, we notice that the link between these two wavefunctions is found by simply extracting the $\rho$-dependent part of the wavefunction. For instance for the bosonic (or type II) $w=\frac{1}{N}$ wavefunction $\psi_{n}$ and the twisted state wavefunction $F_n$ we would have (up to normalization): 
\begin{equation}
\psi_{n}(\tau,\rho,\bold{x}) = F_{n}(\tau,\bold{x})e^{-\frac{\rho^2}{4N}}L_{n}\left(\frac{\rho^2}{2N}\right).
\end{equation}
After this, we consider the resulting differential equation for $F_n(\tau,\bold{x})$, which has manifestly the same eigenvalue spectrum, as it should.\\
Secondly, our interest lies in the spread of the long string. This is written as a first-quantized path integral on the spatial submanifold. From this perspective, one could integrate out the $\rho$-field (schematically) as
\begin{equation}
\int \left[\mathcal{D}\tau\right]\left[\mathcal{D}\bold{x}\right]\left[\mathcal{D}\rho\right]e^{-S_p(\tau,\bold{x},\rho)} = \int \left[\mathcal{D}\tau\right]\left[\mathcal{D}\bold{x}\right]e^{-S_{eff}(\tau,\bold{x})}
\end{equation}
to obtain the twisted sector wavefunction point of view, though this is not what we want to do to distill the random walk picture.}
Nevertheless, we find it intriguing that we precisely reproduce the critical behavior of the partition function and we find a further explanation for the corrections to the bosonic T-duality found in \cite{Dijkgraaf:1991ba}. 

\section{Unitarity constraints in Euclidean Rindler space}
\label{unitarity}
In this section we make the link between the $SL(2,\mathbb{R})/U(1)$ cigar and Euclidean Rindler space more precise in terms of conformal field theory language. In particular we want to analyze the `induced' unitarity constraints in Euclidean Rindler space. The cigar CFT states are characterized by several quantum numbers. The winding $w$ around the cigar and discrete momentum $n$ are collected in two linear combinations
\begin{equation}
m = \frac{n+kw}{2}, \quad \bar{m} = \frac{-n+kw}{2}, \quad n,w \in \mathbb{Z}.
\end{equation}
The quantum number $j$ is a measure for the radial momentum and is given as
\begin{align}
j &= -\frac{1}{2}+is, \quad s\in\mathbb{R} , \quad \text{continuous representations}, \\
j &= M - l , \quad l=1,2,...,\quad \text{discrete representations},
\end{align}
where $M = \text{min}(m,\bar{m})$ with $m,\bar{m} > 1/2$ \cite{Aharony:2004xn}.\footnote{\label{fn}There is also the option $M = \text{min}(\left|m\right|,\left|\bar{m}\right|)$ with $m,\bar{m} < -1/2$ \cite{Aharony:2004xn}, but we will not focus on this case since it will disappear when taking the flat $k\to\infty$ limit for the cases $w>0$. This sector is though relevant for the $w=-1$ state and we will comment on this further on. Note that only states with $\left|kw\right| > \left|n\right|$ are in the spectrum, these are the so-called winding dominated states \cite{Aharony:2004xn}.} We are interested in the discrete states on the cigar, so we consider the discrete representations. In this section we are interested in pure winding states, so $m = \bar{m} = M$. For the discrete representations $j$ has the following unitarity constraints:
\begin{align}
\label{unibos}
-\frac{1}{2} &< j < \frac{k-3}{2}, \quad \text{bosonic},\\
-\frac{1}{2} &< j < \frac{k-1}{2}, \quad \text{type II}.
\end{align}
The conformal weights of these states are given by
\begin{align}
h &= \frac{m^2}{k}-\frac{j(j+1)}{k-2}, \quad \bar{h} = \frac{\bar{m}^2}{k}-\frac{j(j+1)}{k-2} , \quad \text{bosonic},\\
h &= \frac{m^2}{k}-\frac{j(j+1)}{k}, \quad \bar{h} = \frac{\bar{m}^2}{k}-\frac{j(j+1)}{k} , \quad \text{type II}.
\end{align}

\subsection{Quantum numbers in Euclidean Rindler space}
Let us first analyze how these quantum numbers are reflected in the Rindler case.
We treat the type II superstring in this section.\footnote{The bosonic string case is analogous and we present the relevant formulas in section \ref{boswind}.} The eigenvalue equation we consider for $\beta = 2\pi\sqrt{\alpha'}$ is given by\footnote{We set $\alpha'=2$ in this section to conform to the conventions of \cite{Giveon:2013ica}. }
\begin{equation}
-\frac{\partial_\rho\left(\sinh\left(\sqrt{2/k}\rho\right)\partial_{\rho}T(\rho)\right)}{\sinh\left(\sqrt{2/k}\rho\right)} + \left(-1 + w^2\frac{k}{2}\tanh^2\left(\rho/\sqrt{2k}\right)\right)T(\rho)= \lambda T(\rho).
\end{equation}
The solution that does not blow up as $\rho \to \infty$ is given by
\begin{equation}
\label{expl}
T(\rho) \propto \frac{1}{\cosh\left(\frac{\rho}{\sqrt{2k}}\right)^{1+\sqrt{\omega}}}\,\, {\mbox{$_2$F$_1$}\left(\frac{\sqrt{\omega}+1+kw}{2},\frac{\sqrt{\omega}+1-kw}{2};\,\sqrt{\omega}+1;\,\frac{1}{\cosh\left(\frac{\rho}{\sqrt{2k}}\right)^{2}}\right)}.
\end{equation}
where $\omega = 1-2k-2k\lambda +k^2w^2$.
All normalizable states should behave near $\rho \to \infty$ as
\begin{equation}
\label{asymp}
\psi \propto \exp\left(-\sqrt{2/k}\left(j+1\right)\rho\right),
\end{equation}
since the background approaches a linear dilaton background there.
For a discrete pure winding state we have $j=\frac{kw}{2}-l$ where $l=1,2,3...$.
Since $\mbox{$_2$F$_1$}(a,b;c;0) = 1$, the prefactor determines the entire asymptotic behavior. Identifying the asymptotic behavior of (\ref{expl}) with the required asymptotics of (\ref{asymp}) gives
\begin{equation}
\label{bound}
\sqrt{\omega} = kw-2l+1,
\end{equation}
leading to 
\begin{equation}
\lambda = -\frac{2l(l-1)}{k} +2wl -w-1.
\end{equation}
For $k\to\infty$ the first term drops out and we are left with
\begin{equation}
\lambda \approx 2wl -w-1,\quad l=1,2\hdots 
\end{equation}
Setting $n=l-1$, we obtain
\begin{equation}
\lambda \approx 2wn +w-1,\quad n=0,1\hdots 
\end{equation}
which coincides with the discrete spectrum (\ref{superspectrum}) in Euclidean Rindler space. 
The condition to get the discrete states also implies that the second argument of the hypergeometric function becomes a negative (or zero) integer. This causes the hypergeometric function to become a polynomial and this is well-behaved also for $\rho \to 0$.\\
Note that there are only a finite number of discrete states since $l$ is bounded from above by the requirement that the r.h.s. of (\ref{bound}) should be positive. As $k$ increases, more values of $l$ are allowed and in the limit $k\to\infty$, $l$ becomes effectively unbounded. \\
The continuous states can be found for $\omega < 0$. This corresponds to a critical eigenvalue
\begin{equation}
\lambda^* = \frac{kw^2}{2} + \frac{1}{2k}-1,
\end{equation}
where $\lambda > \lambda^*$ corresponds to the continuous spectrum. Taking $k \to \infty$ gives $\lambda^* \to +\infty$ and the continuous states disappear.\\
It is instructive to see the above identification of $n$ and $l-1$ explicitly for the eigenfunctions. 
First consider the $w=1$, $l=1$ eigenfunction as given by \cite{Giveon:2013ica}:
\begin{equation}
T(\rho) \propto \frac{1}{\cosh\left(\frac{\rho}{\sqrt{2k}}\right)^k}.
\end{equation}
Taking $\rho \to \infty$ immediately gives
\begin{equation}
T(\rho) \propto \exp\left(-\sqrt{k/2}\rho\right)
\end{equation}
as it should be. Alternatively taking $k \to \infty$ gives
\begin{equation}
T(\rho) \propto \exp\left(-\rho^2/4\right).
\end{equation}
This identifies the $l=1$ wavefunction with the $n=0$ Rindler eigenfunction.\\
As another example, consider the $w=1$, $l=2$ bound state eigenfunction:
\begin{equation}
T(\rho) \propto \frac{(k-1) - (k-2)\cosh\left(\frac{\rho}{\sqrt{2k}}\right)^2}{\cosh\left(\frac{\rho}{\sqrt{2k}}\right)^k}.
\end{equation}
The $\rho \to \infty$ limit gives
\begin{equation}
T(\rho) \propto \exp\left(-\sqrt{2/k}\left(k/2-1\right)\rho\right).
\end{equation}
while taking $k \to \infty$ yields
\begin{equation}
T(\rho) \propto L_1\left(\rho^2/2\right)\exp\left(-\rho^2/4\right),
\end{equation}
which is precisely the $n=1$ eigenfunction in Rindler space. \\
We want to remark that the $w=-1$ state can be found on the cigar with the quantum numbers $j = M - l$ where this time $M = \left|-\frac{k}{2}\right|$ as we discussed in footnote \ref{fn} earlier. This yields $j = \frac{k}{2} - l$ and the asymptotic behavior is the same as that of the $w=1$ state. The Euclidean Rindler wavefunction is also the same as that of the $w=1$ state.

\subsection{Cigar spectrum in the $k\to\infty$ limit}
In this section, we take the large $k$ limit explicitly in the cigar spectrum. Our goal is to see the Rindler states explicitly appear and to further elaborate on the effect of the unitarity bounds.
For bosonic strings, we consider the cigar spectrum with some flat spectator dimensions to identify the tachyonic character of a state in the large $k$ limit. We have
\begin{equation}
h = -\frac{\alpha'M^2}{4} + \frac{m^2}{k}-\frac{j(j+1)}{k-2} =1.
\end{equation}
Setting $j=\frac{kw}{2}-l$, $l=1$ and $m=j+l$ we obtain
\begin{equation}
-\frac{\alpha'M^2}{4} + \frac{kw^2}{4}-\frac{(\frac{kw}{2}-1)(\frac{kw}{2})}{k-2} =1,
\end{equation}
and so (taking $k \to \infty$)
\begin{equation}
-\frac{\alpha'M^2}{4} + \frac{kw^2}{4}- \frac{kw^2}{4} - \frac{w^2}{2} + \frac{w}{2} =1
\end{equation}
or 
\begin{equation}
M^2 =  \frac{2}{\alpha'}\left( -w^2 + w - 2 \right).
\end{equation}
General $l$ would yield
\begin{equation}
M^2 =  \frac{2}{\alpha'}\left( -w^2 + (2l-1)w - 2 \right).
\end{equation}
For bosonic strings on the cigar, the $l=1$ sector is prohibited by the unitarity constraints (\ref{unibos}) in the $k\to\infty$ limit.
The bosonic unitarity constraint is $-1/2 < j < \frac{k-3}{2}$, where for winding states $j= \frac{kw}{2}-l$.
Setting $l=1$ and $w=1$ gives $j= \frac{k-2}{2}$ which violates the upper bound. Setting $l=2$ and $w=1$ on the other hand gives $j= \frac{k-4}{2}$ which is a valid string state. All higher values of $l$ are also allowed. Higher winding modes always violate the upper unitarity bound when taking $k\to \infty$.\\
Let us consider $\mathbb{Z}_N$ orbifolds of the $SL(2,\mathbb{R})/U(1)$ cigar. In this case, one introduces a conical singularity at the tip of the cigar. We expect that taking $k\to\infty$ in this CFT should coincide with the flat space $\mathbb{C}/\mathbb{Z}_N$ orbifolds.
To study orbifolds of the cigar, the effect of twisting the CFT is simply the change $w \to \frac{w}{N}$. That is, we should allow fractional winding numbers \cite{Martinec:2001cf}\cite{Son:2001qm}. States that have $w \notin N\mathbb{N}$ are the twisted sectors. The others are untwisted. We elaborate on this orbifolding procedure in section \ref{orbifoldpart} using the known cigar partition function \cite{Hanany:2002ev}. The cigar orbifold CFT in the $l=1$ sector thus has masses 
\begin{equation}
M^2 =  \frac{2}{\alpha'}\left( -\frac{w^2}{N^2} + \frac{w}{N} - 2 \right).
\end{equation}
In this case, $j= \frac{kw}{2N}-l$ and for $w=0\hdots N-1$, the $l=1$ sector satisfies the upper unitarity bound. These states can be identified with the twisted sector primaries of the $\mathbb{C}/\mathbb{Z}_N$ orbifold. The $w=N$ sector only starts with $l=2$ like in the unorbifolded case discussed above. Sectors with $w>N$ are to be excluded by the unitarity bound.\\
The analogous equations for type II superstrings are
\begin{equation}
-\frac{\alpha'M^2}{4} + \frac{m^2}{k}-\frac{j(j+1)}{k} =1/2.
\end{equation}
Again we set $j=\frac{kw}{2}-l$, $l=1$ and $m=j+l$ to obtain (taking $k \to \infty$):
\begin{equation}
-\frac{\alpha'M^2}{4} + \frac{kw^2}{4}- \frac{kw^2}{4} + \frac{w}{2} =1/2
\end{equation}
or 
\begin{equation}
M^2 =  \frac{2}{\alpha'}\left(  w - 1 \right).
\end{equation}
For general $l$ we would have
\begin{equation}
M^2 =  \frac{2}{\alpha'}\left( (2l-1)w - 1 \right).
\end{equation}
The unitarity constraint for type II superstring is $-1/2 < j < \frac{k-1}{2}$. In this case, $w=1$ and $l=1$ is in the spectrum (as are higher values of $l$). Higher winding modes on the other hand are again not allowed. We treat the discrete momentum modes and mixed momentum-winding modes in section \ref{spectr} to conclude our study of the NS-NS spectrum of conformal primaries in this space.\\
To summarize, all types of strings do not include higher winding modes $\left|w\right| > 1$ in the string spectrum. The $\left|w\right| = 1$ mode differs for bosonic strings and type II strings: in the bosonic case, $l=2,3,\hdots$ are allowed while in the type II case, $l=1,2,\hdots$ are allowed. This implies that in the bosonic Rindler eigenvalues we should drop the $n=0$ term and the $n=1$ contribution becomes the lowest eigenvalue. For flat $\mathbb{C}/\mathbb{Z}_N$ orbifolds, the twisted sectors $w=1\hdots N-1$ have $l=1,2,\hdots$ so these start with $n=0$ in the Rindler case. The $w=N$ sector only starts at $l=2$ for bosonic strings and $l=1$ for superstrings. The $w>N$ sectors are not present. 

\subsection{The bosonic Rindler string revisited}
\begin{figure}[h]
\centering
\includegraphics[width=13cm]{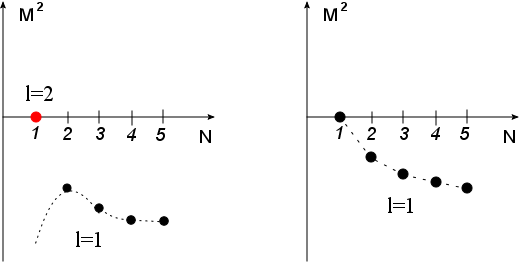}
\caption{Left figure: Most tachyonic state in the bosonic case versus the orbifold number $N$. The black dots are the $l=1$ modes. For $N=1$ the $l=1$ mode is absent and one should consider the $l=2$ mode (the red dot). The continuation in $N$ (the dashed line) suggests the entropy is divergent even though the free energy itself is not, since this only starts at $l=2$ for $N=1$. Right figure: Most tachyonic state in the type II superstring case. In this case, nothing special occurs and the $l=1$ mode is present both for the free energy and for the entropy.}
\label{cartoon}
\end{figure}

Let us reanalyze the critical behavior of the bosonic Rindler string, now incorporating these unitarity constraints. The lowest Rindler mode is the $n=1$ mode
\begin{equation}
\psi_{1}(\rho) \propto \exp\left(-\frac{\beta\rho^2}{4\pi\alpha'^{3/2}}\right)L_{1}\left(\frac{\beta\rho^2}{2\pi\alpha'^{3/2}}\right), \quad \lambda_{1} = \frac{3\beta - 4\pi\sqrt{\alpha'} - \frac{\beta^2}{2\pi\sqrt{\alpha'}}}{\pi \alpha'^{3/2}},
\end{equation}
where $L_1(x) = 1-x$ is the Laguerre polynomial of the first degree. This mode has a width
\begin{equation}
\left\langle \rho^2\right\rangle = \frac{6\pi\alpha'^{3/2}}{\beta}.
\end{equation}
The size of the wavefunction is a factor of $\sqrt{3}$ larger than the $n=0$ width. The eigenvalue is zero for $\beta = 2\pi\sqrt{\alpha'}$ and $\beta = 4\pi\sqrt{\alpha'}$. Values of $\beta$ in between these two values give a postive eigenvalue $\lambda_1$, while values of $\beta$ outside this range give rise to a negative eigenvalue and hence a divergence.
We conclude that for the canonical Rindler temperature, the free energy becomes marginally convergent, just like for the type II and heterotic strings.\\
However, as soon as one decreases $\beta$ (like for the orbifold CFTs), the $l=1$ mode reappears in the spectrum and the bosonic string free energy diverges. In particular, when computing the entropy as a derivative in $\beta$ of the free energy, we would expect a divergence. It is (presumably) only at precisely $\beta=\beta_R$ that the $l=1$ mode drops from the spectrum. For this reason, the bosonic thermal scalar is still effectively tachyonic. The situation is sketched in figure \ref{cartoon}.
Such restrictions in the Rindler quantum number $n$ are quite strange from a particle perspective. In particular the random walk picture has the caveat that we should drop the lowest mode of the associated wave operator. We see that string theory is more subtle than any limiting field theory might suggest.\\
Nonetheless, the analysis in section \ref{orbifold} concerning the $\mathbb{C}/\mathbb{Z}_N$ orbifolds remains valid (since the twisted sectors do have $l=1$ in the spectrum). The analysis concerning the thermal scalar action in section \ref{alphaprime} also remains valid: it is only the eigenmodes of this action that are subject to the unitarity constraints.\\
Perhaps even more importantly, the type II superstrings in Euclidean Rindler space do not suffer from this restriction on $l$ and ultimately we are interested in these. 

\section{Discussion}
\label{discussion}
We make several remarks concerning these results.\\

\noindent In section \ref{GM} we translate the previous results to the Schwarzschild normalization of Rindler space. We find that the state has a string width around the event horizon and the Hagedorn temperature for type II and heterotic superstrings becomes $\beta_{H} = 8\pi GM$ which is the Hawking temperature of the black hole. This temperature is much lower than the string scale unlike flat space Hagedorn temperatures. Physically, the Hawking temperature is the temperature set at $G_{00}=1$. So for Schwarzschild black holes, this temperature is the one at infinity. The size of the thermal circle goes all the way to zero at the horizon. So even choosing non-stringscale temperatures at infinity gives local stringscale temperatures near the tip of the cigar. Apparently this causes the singly wound state to be precisely massless at the Hawking temperature.\\

\noindent It is quite remarkable that for both type II superstrings and heterotic strings, the Rindler Hagedorn temperature equals the canonical Rindler temperature. It would have been surprising to obtain a divergence in these cases, since this would indicate an instability in thermodynamics and thus an instability in the theory itself. Quite reassuringly, there are no thermal tachyons in both these cases. No tachyon condensation process occurs, although an IR divergence can still occur depending on whether there are compact dimensions or not as we discussed above. Though it is peculiar that in both cases we obtain \emph{precisely} `marginal' stability. For both these string types, the winding string lives close to the event horizon. Since there is no tachyon anywhere in spacetime in these cases, this region does not correspond to a condensate but instead the marginal state is a bound state. Note that the localized state is in the perturbative spectrum of Euclidean Rindler space and its classical value is set to zero. \\ 

\noindent If we look back at the operator (\ref{opera}), we notice that the reason the thermal scalar is localized near the horizon is the $\rho^2$ potential and this is present since we study $w=\pm1$. The `normal' particles do not have this potential and can freely propagate outward (if they overcome the centrifugal barrier as discussed in e.g. \cite{Susskind:2005js}). Back on the thermal manifold, there are also $n\neq0$ states that propagate to infinity. For the type II string, the wavefunctions of these states and their relation to the cigar CFT can be found in section \ref{spectr}. This is a very important difference between these fields: non-winding states can propagate to infinity whereas the thermal scalar on the other hand is bound to the horizon. It is to be interpreted in the Lorentzian signature as a long highly excited string located at a string length from the horizon. Considering the thermal zone of a black hole, the thermal scalar is living much closer to the horizon, effectively outside the reach of low energy quantum field theory. This makes it a natural microscopic candidate for the stretched horizon (or the membrane). Note that classically, the black hole membrane is located at an arbitrary radial location that is sufficiently small \cite{Thorne:1986iy}. In our case, the stretched horizon has a fixed location at a string length outside the horizon. \\

\noindent The ground state wavefunctions determined above have a width equal to (or of the order of) $\sqrt{\alpha'}$ for $\beta = 2 \pi \sqrt{\alpha'}$, so consistency with the Rindler approximation requires this to be smaller than $4GM$. We thus arrive at the following equivalent consistency requirements 
\begin{equation}
\alpha' \ll G^2M^2 \quad \Leftrightarrow \quad \ell_{s} \ll R_{s} \quad \Leftrightarrow \quad T_{\text{Hawking}} \ll T_{H,\text{flat}}.
\end{equation}
So if the asymptotic temperature of the black hole is much smaller than the \emph{flat space} Hagedorn temperature, we are safe to use Rindler space. This is equivalent to choosing a black hole much larger than the string length. We conclude that our picture is only valid when the horizon size is much larger than the width of the dominant string state. 
The above was applied to a Schwarzschild black hole but it applies equally well to (large) AdS black holes. We are more interested in the latter since these are thermodynamically stable and we can use the $AdS$ spacetime as a `container' to mitigate the ever-present Jeans instability \cite{Barbon:2001di}\cite{Barbon:2002nw}. \\
We also point out that when we consider Rindler spacetime, we have no way of seeing the ground state width diverge as was argued in \cite{Kutasov:2005rr} to happen for sufficiently small black holes. This is obviously caused by the fact that Rindler spacetime has an ever increasing circumference of the thermal circle, while for a real black hole the circumference asymptotes to the radius corresponding to the Hawking temperature. The authors of \cite{Kutasov:2005rr} and \cite{Giveon:2005jv} interpret this diverging thermal scalar width as the black hole-string correspondence point. So we do not have the chance to study the black hole evaporation process and the correspondence principle as put forth by \cite{Horowitz:1996nw} within Rindler space. Within the full cigar CFT on the other hand, we can analyze this picture, but only for this specific 2d black hole (we will do this in section \ref{remarks}). This is in constrast with our discussion here on Rindler space, which is very generic. \\

\noindent From the Rindler example, we can learn some lessons regarding the higher winding modes. Naively, when considering a conical space (or its smooth cigar-like cousin), one can imagine that \emph{all} winding modes are tachyonic because the radius of the thermal circle shrinks all the way to zero. Thus naively we cross all Hagedorn transition temperatures for all winding modes. This reasoning is wrong. 
The $SL(2,\mathbb{R})/U(1)$ cigar CFT spectrum explicitly shows that this is not the case: it depends on the quantum numbers considered. When taking the flat limit $k\to\infty$, we instead see that all higher winding modes are simply absent from the string spectrum. This is a feature to which the field theory action is a priori insensitive. The singly wound mode is a marginal state and hence is not tachyonic. \\

\noindent This entire discussion has been for strings with all spectator dimensions geometrically flat. One could ask whether the above picture is altered when compactifying on geometrically non-trivial spaces. Let us split spacetime in a four dimensional (for string phenomenology) or five dimensional (for holographic purposes) spacetime and a compact internal manifold. 
Smooth compactifications (e.g. on a compact Calabi-Yau manifold $\mathcal{M}$) that are compact unitary CFTs do not alter our conclusions. The reason is that the conformal weights are $h \geq 0$ in this case and $h=0$ is in the spectrum (this is the unit operator and this state is automatically normalizable due to the compactness). Due to the compactness, we also do not introduce continuous quantum numbers. Thus the convergence calculation remains the same and the winding mode retains its character (marginally convergent for all string types). More explicitly, consider the partition function
\begin{equation}
Z = \int_{F}\frac{d\tau_1d\tau_2}{2\tau_2}Z_{\text{matter}}(\tau)Z_{\text{gh}}(\tau)Z_{\text{compact}}(\tau),
\end{equation}
where
\begin{align}
Z_{\text{matter}}(\tau) &= \text{Tr}\left(q^{L_0 -c_m/24}\bar{q}^{\bar{L}_0-\bar{c}_m/24}\right), \\
Z_{\text{gh}}(\tau) &= \text{Tr}\left(q^{L_0 -c_g/24}\bar{q}^{\bar{L}_0-\bar{c}_g/24}\right),\\
Z_{\text{compact}}(\tau) &= \text{Tr}\left(q^{L_0 -c_{comp}/24}\bar{q}^{\bar{L}_0-\bar{c}_{comp}/24}\right),
\end{align}
with $c_m + c_{g} + c_{comp} = 0$. We immediately cancel the $c$-dependent factors in the trace and in the limit $\tau_2 \to \infty$, upon dropping the $c$-dependent factors, we have that
\begin{equation}
\tilde{Z}_{\text{compact}}(\tau) = \text{Tr}\left(q^{L_0}\bar{q}^{\bar{L}_0}\right) \to 1.
\end{equation}
In this limit the compact part drops out of the partition function. This has an analogous manifestation in the heat kernel picture. Since the heat kernel factorizes, the ground state wavefunction becomes a product of the two ground states and the shape of the tachyon wavefunction in the $\rho$ direction remains the same. For the compact dimensions we have
\begin{equation}
K(\tau_2) = \int dV K(x,x;\tau_2) = \sum_n\int dV \psi_n(x)\psi_n(x)^{*}e^{-E_{n}\tau_2} \to 1.
\end{equation}
when we assume that we do not have negative eigenvalues of the operator that gives us this heat kernel. The constant function is an eigenfunction with eigenvalue zero, so its wavefunction is constant and equal to $\psi_0 = \frac{1}{\sqrt{V}}$ (up to a phase). The critical behavior and the Hagedorn temperature are indeed not modified and the ground state wavefunction is uniformly distributed over the compact manifold. The random walk behavior is entirely determined by the non-compact part.\\
As an explicit example, consider a $SU(2)_k$ WZW model as (part of) the compact CFT. The primaries are discrete and have conformal weights
\begin{align}
h &= \frac{1}{k+2} j(j+1), \quad \bar{h} = \frac{1}{k+2}\bar{j}(\bar{j}+1), \quad \text{bosonic string}, \\
h &= \frac{1}{k} j(j+1), \quad \bar{h} = \frac{1}{k}\bar{j}(\bar{j}+1), \quad \text{type II superstring},
\end{align}
where $0\leq j,\bar{j}\leq k/2$. Which primaries appear in the string spectrum is irrelevant for our discussion (this corresponds to choosing a specific modular invariant). We clearly see that the minimal conformal weight is indeed zero, as it should for a unitary compact CFT. \\
From the heat kernel point of view, the scalar Laplacian on $S_3$ has eigenvalues $-l(l+2)$ where $0 \leq l$ an integer. Clearly the lowest eigenvalue is zero and the Hagedorn temperature remains unchanged. \\

\noindent Now that the dust has settled, we make a comparison between the different approaches to Rindler near-Hagedorn thermodynamics. \\
As a first approach we have the random walk picture as made explicit by \cite{Kruczenski:2005pj}\cite{Mertens:2013pza} and reviewed in section \ref{pathderiv}. This realizes the picture as proposed by Susskind \cite{Susskind:2005js} that highly excited strings can be described as random walks and that near the Hagedorn transition, the string gas recombines itself into a single highly excited string.\footnote{Again with the caveats we discussed before.} The random walk of the thermal scalar in the Euclidean picture traces out the spatial form of the long string in the Lorentzian picture. The $\tau_2 \to \infty$ limit corresponds to a long walk, and thus to a long string in the Lorentzian picture. The subtle point is however that the fate of the correction terms of the worldsheet action is unclear. Also, since we started with the gauge-fixed string path integral, we cannot go off-shell.\\
The other approach is to start with the field theory action, compute the one-loop amplitude and convert this to a first-quantized form. When properly taking into account the $G_{00}$ metric component and $\alpha'$ corrections, we get a modified random walk picture and we interpret the modifications as the correction terms from above. So from this approach, we do get information about the corrections. Also, since this is a field theory, we appear to have no problem in going off-shell. However, here we have trouble in interpreting the required Wick rotation to return to the Lorentzian picture as discussed in \cite{Kutasov:2000jp}\cite{Giveon:2012kp}.\\
We see that the delicate points of one approach are explained by the other approach. So combining these two viewpoints, we have all ingredients to fully realize the Rindler string as a (modified) random walker that is confined to a string length from the event horizon. Thus this realizes (at genus one) the picture put forward by Susskind.

\section{Approach to Heterotic Euclidean Rindler space}
\label{heteroticc}
This section ties up a loose end that appeared up to this point on the heterotic string thermal scalar. Above, and in \cite{Mertens:2013zya}, we analyzed the critical one-loop string thermodynamics and found the following results for the non-interacting thermal scalar field theory. The thermal scalar action for type II superstrings is exactly given by the lowest order (in $\alpha'$) action as was shown in \cite{Giveon:2012kp}\cite{Giveon:2013ica}\cite{Giveon:2014hfa} by taking the large $k$ limit of the $SL(2,\mathbb{R})/U(1)$ black hole. The bosonic string thermal scalar on the other hand does contain $\alpha'$ corrections. We also observed that heterotic string theory on flat $\mathbb{C}/\mathbb{Z}_N$ orbifolds agrees with a thermal scalar action without any $\alpha'$ corrections (like for type II superstrings), but we did not yet give a proof of this statement. In this section we will present an argument as to why this is so. As in \cite{Giveon:2013ica}\cite{Mertens:2013zya}, we are looking for a suitable cigar CFT to take the large $k$ limit. There exist several approaches and points of view on heterotic coset models (see e.g. \cite{Johnson:1994jw}\cite{Johnson:1994kv}\cite{Johnson:2004zq}\cite{Svendsen:2005gy}\cite{Giveon:1993hm}\cite{Sfetsos:1993bh}), and also several realizations of the analog of the cigar CFT. We choose the left-right symmetric realization where the heterotic worldsheet theory actually has (1,1) supersymmetry instead of the expected (0,1) supersymmetry \cite{Giveon:1993hm}. Heterotic backgrounds can be trivially constructed from type II backgrounds by embedding the spin connection in the gauge connection. This approach was explored by \cite{Giveon:1993hm} to discuss heterotic WZW models. One of the benefits of this approach is that the techniques from type II coset models can be integrally carried over to this case, in particular the identification of the exact background fields and the resulting ($\alpha'$-exact) tachyon equation of motion. \\
For more general heterotic models it becomes less clear whether such an approach is viable. Other methods to distill the metric and dilaton exist in this case \cite{Johnson:2004zq}, but we also need to determine the tachyon equation of motion, and the approach followed in \cite{Dijkgraaf:1991ba} is ideally suited for this. \\

\noindent As a motivation to consider the left-right symmetric models, we note the following. In general, the Busher rules for heterotic strings receive $\alpha'$-corrections. Hence the thermal scalar action in heterotic string theory receives $\alpha'$-corrections, just like the bosonic string. However, if the background has an enlarged supersymmetry compared to the expected $(0,1)$ SUSY, the heterotic string effectively behaves as a type II superstring and the Busher rules do not get corrections (at least for (gauged) WZW models). The fact that for Rindler space the lowest order (in $\alpha'$) thermal scalar action appears to be $\alpha'$-exact, is evidence that in this case indeed more supersymmetry is present than expected. This shows why the left-right symmetric approach to heterotic coset models (effectively giving type II models), is the most natural place to look for realizing Euclidean Rindler space in heterotic string theory. \\

\noindent Thus to any type II background, one can associate a heterotic background by embedding the spin connection into the gauge connection. For this heterotic background, the fluctuation equations of the states is given by the same $L_0$ and $\bar{L}_0$ (written in terms of the Laplacian on the coset) as for the type II superstring \cite{Giveon:1993hm}. The only difference is in the precise on-shell conditions, which for heterotic strings are given by
\begin{align}
L_0 - 1 &=0 ,\\
\bar{L}_0 - 1/2 &= 0.
\end{align}

\begin{figure}[h]
\begin{minipage}{0.5\textwidth}
\centering
\includegraphics[width=5cm]{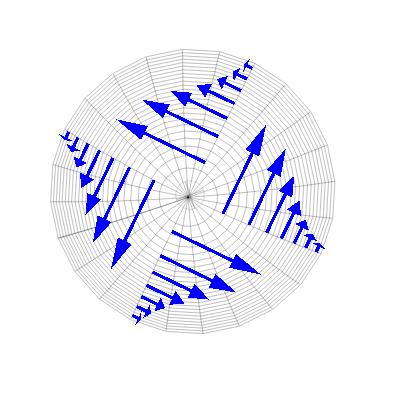}
\end{minipage}
\begin{minipage}{0.5\textwidth}
\centering
\includegraphics[width=5cm]{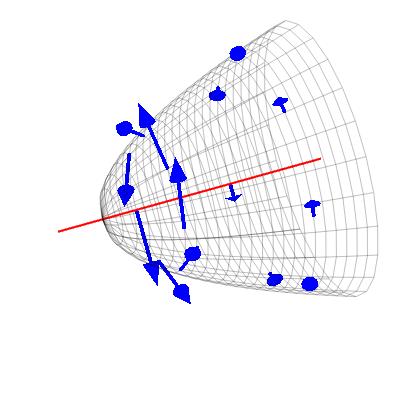}
\end{minipage}
\caption{Left figure: top view of the cigar with the background gauge field schematically shown. Right figure: side view of the cigar and the background gauge field.}
\label{topside}
\end{figure}

\noindent Within such a left-right symmetric approach, a $SL(2,\mathbb{R})/U(1)$ CFT can be found with the following background fields \cite{Giveon:1993hm}
\begin{align}
\label{bgfields1}
ds^2 &= \frac{\alpha'k}{4}\left(dr^2 + 4 \tanh\left(\frac{r}{2}\right)^2 d\theta^2\right), \\
\label{bgfields2}
\Phi &= \Phi_0 - \ln\left(\cosh\left(\frac{r}{2}\right)\right), \\
\label{bgfields3}
A_\theta &= -\frac{1}{\cosh\left(\frac{r}{2}\right)^2},
\end{align}
where the gauge connection equals the Lorentz spin connection and $A_r = 0$. The spin connection is valued in the holonomy group of the 2d space (being $U(1)$). Hence the result is an Abelian gauge field $A_\mu$ that resides in some $U(1)$ subalgebra of the full heterotic gauge algebra. The angular coordinate is identified as $\theta \sim \theta + 2\pi$. These coordinates are however singular for $r=0$, in the same way that polar coordinates are. Normally one can readily continue the solutions to include $r=0$, since no physical singularities are encountered at $r=0$. In this case however, the gauge field becomes singular at $r=0$, and this singularity has physical consequences. We therefore conclude that the above solution is only valid for $r>0$. The gauge field and its singular character are schematically depicted in figure \ref{topside}.

\noindent The dual background (corresponding to a $U(1)$ vector gauging) is given by
\begin{align}
\label{dualbg}
ds^2 &= \frac{\alpha'k}{4}\left(dr^2 + 4 \coth\left(\frac{r}{2}\right)^2 d\tilde{\theta}^2\right), \\
\tilde{\Phi} &= \tilde{\Phi_0} - \ln\left(\sinh\left(\frac{r}{2}\right)\right), \\
\tilde{A}_{\tilde{\theta}} &= \frac{1}{\sinh\left(\frac{r}{2}\right)^2},
\end{align}
where now $\tilde{\theta} \sim \tilde{\theta} + \frac{2\pi}{k}$. These background fields determine how winding modes sense the geometry. Like for the bosonic and type II case, this geometry is trumpet-shaped with a curvature singularity at the origin $r=0$. This background is $\alpha'$-exact and the tachyon equation of motion can be readily determined in this background. Like for the type II string, the winding tachyon equation of motion does not get $\alpha'$ corrections compared to the T-dual geometry of the background.\footnote{The T-dual metric, dilaton and Kalb-Ramond field are not influenced by the non-vanishing gauge field.} In this case, the dual gauge field also blows up at the origin, but this is irrelevant for the thermal scalar equation of motion since this one is only determined by the dual metric and dual dilaton. Writing down the discrete momentum tachyon equation in this background (\ref{dualbg}) gives us the winding tachyon equation of motion we are after. \\
Note that the propagation equations for the thermal scalar in this background are constructed from the same ingredients as those of the type II superstring (the $L_0$ and $\bar{L}_0$ operators have the same form) and hence the background gauge field does not couple directly to the tree-level action of the thermal scalar. This in particular seems to imply that the thermal scalar does not carry any charge corresponding to this background gauge field. \\
To see the flat limit, we substitute $\rho = \frac{\sqrt{\alpha' k}}{2} r$ and then take $k \to \infty$ keeping $\rho$ fixed. The geometry reduces to polar coordinates, the dilaton becomes constant and the gauge field also becomes constant.\\

\noindent To proceed, we first discuss the singular character of the background fields. Let us briefly look at the nature of the gauge field near the horizon. Near $r=0$, the geometry reduces to polar coordinates, and the gauge field becomes constant. So we are actually interested in a constant angular gauge field $A_\theta = C = -1$ in polar coordinates. One readily computes the Cartesian components as
\begin{equation}
\label{CS}
A_x = C\frac{d\theta}{dx}= -\frac{Cy}{x^2+y^2}, \quad A_y = C \frac{d\theta}{dy} = \frac{Cx}{x^2+y^2}
\end{equation}
and one finds that the field tensor $F_{xy}$ vanishes (almost) everywhere (or directly in polar coordinates $F_{\rho\theta} =0$).
Nevertheless, the flux does not vanish since
\begin{equation}
\int_{\text{disc}} F = \oint_{\text{circle}}d\theta A_{\theta} = 2\pi C
\end{equation}
so we find $F_{xy} = 2\pi C \delta(x,y)$ and there is a delta-source of magnetic flux present at the origin. This flux is important because charged states can acquire an Aharonov-Bohm phase upon circling the origin. From another perspective, the singularity of the gauge field at the origin is translated to the violation of the commutativity of partial derivatives of the angular coordinate at the origin:
\begin{equation}
F_{xy} = C\left(\partial_x\partial_y - \partial_y\partial_x \right)\theta \neq 0 \quad \text{at }r=0.
\end{equation}

\noindent Our strategy is to perform a (large) gauge transformation to eliminate the gauge field. The gauge transformation has two effects that need to be separately analyzed. Firstly, the other background fields might change, undoing the very thing we try to accomplish. This can be analyzed easiest by turning to the effective spacetime action: a solution of the all-order in $\alpha'$ effective field theory yields a consistent background for string propagation.\\
Secondly, the fluctuations on this background might be charged under the background gauge field. They hence can feel this gauge transformation. Both of these will be looked at now.

\subsection{Effect of gauge transformation on background}
The background gauge transformation can influence the background fields. To lowest order in $\alpha'$, it is known that the Kalb-Ramond 2-form (despite being uncharged under $A_\mu$) undergoes a simultaneous gauge transformation of the form:
\begin{align}
\delta B_2 &\propto \text{Tr}\left(\lambda dA\right), \\
\delta A &= d\lambda.
\end{align}
More generally, at higher orders in $\alpha'$, the $B_2$-form is known to transform also under gauge transformations of the Lorentz connection, though we will not need this. Note that the only (massless) field that \emph{is} charged under the background gauge field is the gaugino. This field transforms under a gauge transformation in the adjoint representation of the gauge group (homogeneously) and hence if it is turned off initially, it will not be turned on by a gauge transformation. \\
What is crucial is that this gauge transformation is dictated by the Green-Schwarz anomaly cancellation mechanism \cite{Green:1984sg}, and does \emph{not} depend on the concrete form of the spacetime effective action. Hence the above gauge transformation should hold to all orders in $\alpha'$. And indeed, the analysis of \cite{Callan:1985ia}\cite{Metsaev:1986yb}\cite{Foakes:1987bn} shows that, up to three loops on the worldsheet, it is only the gauge-independent combination 
\begin{equation}
\tilde{H}_3 = dB_2 -c \omega_{3Y} - c'\omega_{3L}
\end{equation}
that appears in the effective action. In this formula, the $\omega_3$'s are the Chern-Simons 3-forms constructed with the Yang-Mills connection $A$ or the Lorentz connection. \\
For Rindler space, we already have that $dA = 0$ for $r\neq 0$. This means $B_2$ does not become non-zero for $r\neq0$ after the gauge transformation. Note that this is very different from the original $A$. The original $A$ field had a constant angular component and hence a singularity at the origin. The new $B_2$ is zero everywhere and can be chosen zero at the origin as well: no global analysis (such as a line integral for $A$) can detect something is present at $r=0$. For all intents and purposes, the $B_2$-form is absent. This gauge transformation obviously maps solutions to solutions, and hence we can safely turn off the background $A$-field in this case: no effects on other background fields are present.

\subsection{Effect of gauge transformation on fluctuations}
\begin{figure}[h]
\centering
\includegraphics[width=5cm]{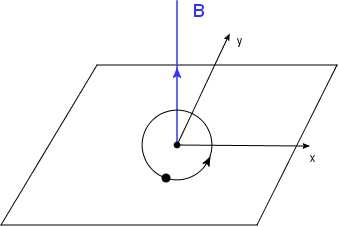}
\caption{The $(\rho, \theta)$ plane (or the $xy$ plane) and the delta-source of magnetic flux at the origin. States that circle the origin can receive an arbitrary phase factor corresponding to anyon statistics.}
\label{cart}
\end{figure}

Gauge fields of the form (\ref{CS}) are well-known from studies of matter-coupled Chern-Simons gauge theories in (2+1) dimensions and the related anyon statistics (see e.g. \cite{Dunne:1998qy} and references therein). In the $k\to\infty$ limit, the gauge field is pure gauge for $r \neq 0$ and can be eliminated. However, charged states get multiplied by an angle-dependent prefactor and their periodicity or anti-periodicity upon circling the origin gets altered to general anyon statistics. A sketch of the situation is given in figure \ref{cart}.
In somewhat more detail, we can write (for $i=1,2$, the Cartesian coordinates in the ($\rho,\theta$) plane)
\begin{equation}
A_i = C \partial_i \arg(\mathbf{x}) = -\partial_i \theta.
\end{equation}
Under a gauge transformation with $A_i \to A_i + \partial_i \chi$ where $\chi = - C \arg(\mathbf{x})$, the wavefunction of a charged state gets changed into
\begin{equation}
\psi \to e^{ie\chi} \psi = e^{ie\theta} \psi,
\end{equation}
which hence changes indeed the phase of the state upon circling the origin by $e^{2\pi i e}$. The charges of the matter depend on which $U(1)$ embedding is chosen and it is difficult to say anything concrete about these at this point.\footnote{Life would be simpler if we could choose an Abelian ideal instead, but unfortunately a semi-simple algebra does not have such ideals.} \\
Let us now return to the entire cigar. We perform a large gauge transformation that kills the singularity of the gauge field at the horizon. This introduces a non-zero Wilson loop at infinity. This kind of reasoning was performed also in \cite{Giveon:2004rw} in a different context.\footnote{See also \cite{Giveon:2005jv} for more discussions on this.} Charged string states hence `feel' the gauge transformation. In our case, we are interested in the thermal scalar. This stringy state is uncharged under the gauge field\footnote{This is because the gauge field comes from the 10 dimensional heterotic gauge field. If for example the gauge field is actually a component of the Kaluza-Klein reduced metric, the thermal scalar would be charged and the above reasoning would not hold. In fact, the thermal scalar of the heterotic string is explicitly charged under the Kaluza-Klein gauge field $G_{\mu \tau}$ and under the Kalb-Ramond gauge field $B_{\mu\tau}$, as one can for instance see very explicitly in the low energy field theory of the thermal scalar as in \cite{Schulgin:2011zb}.} (it only couples to the gauge field through $F_{\mu\nu}$ at higher orders in $\alpha'$) and hence the statistics is unchanged (i.e. it does not get premultiplied by a Wilson loop upon circling the cigar). So after performing this (large) gauge transformation, the resulting equation of motion of the thermal scalar is the same as before. \\

\noindent We summarize what we have done so far: the above background (equations (\ref{bgfields1})-(\ref{bgfields3})) is valid only for $r \neq 0$. For this background the exact fluctuation equations for the string states are known. We then perform a large gauge transformation that eliminates the gauge singularity at $r=0$. This transformation has physical consequences by introducing a non-trivial Wilson-loop that influences charged string states. The thermal scalar on the other hand is uncharged and its equation of motion is not influenced by the large gauge transformation. Finally taking $k\to\infty$ while keeping $\rho = \frac{\sqrt{k\alpha'}r}{2}$ fixed, we recover Euclidean Rindler space for which the fluctuation equations for uncharged states have remained the same during the gauge transformation. The fate of the charged string states is another question, but for the purposes of this work, we are only interested in the thermal scalar itself. For uncharged states, the gauge field has no physical impact anymore and we effectively reduce the model to Euclidean Rindler space. \\

\noindent The propagation equations for the heterotic string are
\begin{align}
L_0 - 1 &=0 ,\\
\bar{L}_0 - 1/2 &= 0.
\end{align}
Adding and subtracting gives
\begin{align}
L_0 +\bar{L}_0- 3/2 &=0 ,\\
L_0 - \bar{L}_0 - 1/2 &= 0.
\end{align}
For this left-right symmetric cigar CFT the conformal weights reduce to those of the type II superstring and so we find the conformal weights:
\begin{equation}
h = -\frac{j(j-1)}{k} + \frac{m^2}{k}, \quad \bar{h} =  -\frac{j(j-1)}{k} + \frac{\bar{m}^2}{k},
\end{equation}
so the physicality constraint becomes
\begin{equation}
h - \bar{h} = wn = 1/2,
\end{equation}
just like in flat space. The equation of motion is determined by writing $L_0 + \bar{L}_0$ in terms of the Casimir operator, and is exactly the same as for type II superstrings. The metric one obtains is again:
\begin{equation}
ds^2 = \frac{\alpha' k}{4}\left[dr^2 + \frac{4}{\coth^2\left(\frac{r}{2}\right)}d\theta^2 + \frac{4}{\tanh^2\left(\frac{r}{2}\right)}d\tilde{\theta}^2\right].
\end{equation}
The eigenvalue equation for the NS primaries, that one obtains by taking $k\to\infty$, is now:\footnote{At least for the uncharged string states as discussed above.}
\begin{equation}
-\frac{\partial_\rho\left(\rho\partial_{\rho}T(\rho)\right)}{\rho}+ \left[-\frac{3}{\alpha'} + n^2\frac{1}{\rho^2} + w^2\frac{\rho^2}{\alpha'^2} \right]T(\rho)= \lambda T(\rho).
\end{equation}
The thermal scalar equation hence becomes
\begin{equation}
\left[-\partial_\rho^2  - \frac{1}{\rho}\partial_{\rho} - \frac{3}{\alpha'} + \frac{1}{4\rho^2} + \frac{\rho^2}{\alpha'^2} \right]T(\rho)= \lambda T(\rho).
\end{equation}
From this we conclude that indeed heterotic strings also do not receive $\alpha'$ corrections to the (quadratic part of the) thermal scalar action. The physicality constraint is the same as in flat space, and the thermal scalar action combines discrete momentum and winding around the Rindler origin. We already noted \cite{Mertens:2013zya} that the critical behavior as dictated by this action (and no corrections to it) agrees with the flat space $\mathbb{C}/\mathbb{Z}_N$ orbifold thermodynamics \cite{Dabholkar:1994ai}\cite{Lowe:1994ah}. \\
When considering the entire spectrum, we note that the only difference (modulo the charged states issue) with type II strings is that the constraint is different and so different states are allowed or forbidden. The constraint is
\begin{equation}
N_L - N_{R} + nw = 1/2,
\end{equation}
even before taking the $k\to\infty$ limit. In particular, the wavefunctions of the states are the same as those for type II superstrings. One can imagine that for asymmetric constructions this constraint might be different.

\section{Flat space limit of the cigar partition function}
\label{flatlimit}
In \cite{Giveon:2014hfa} it was shown for the type II superstring that the large $k$ limit can be directly taken at the partition function level. The authors use a modular invariant regularization to deal with the internal divergences and reinterpret this regulator (together with the level $k$) as the volume divergence of flat space. The discrete states are however not found anymore in the large $k$ limit of the partition function and it was speculated that their imprint should be found as a $1/k$ effect. \\
In this section we study this limit further. We first focus on the bosonic string and its orbifolds. The benefit of studying the orbifolds is that no extra divergence arises and this allows a clean comparison between the flat $\mathbb{C}/\mathbb{Z}_N$ partition function and the cigar orbifold partition function in the large $k$ limit. This allows us to make the link between these models more precise. Above, and in \cite{Mertens:2013zya}, we only compared the dominant thermal scalar behavior of both partition functions. The results of this section can be viewed as a more elaborate comparison of the full partition functions. 

\subsection{Bosonic cigar CFT}
The partition function for the $\mathbb{Z}_N$-orbifolded cigar CFT can be written in the following form \cite{Sugawara:2012ag}\cite{Mertens:2013zya}\cite{Hanany:2002ev}:
\begin{align}
\label{startingp}
Z &= \frac{1}{N}2\sqrt{k(k-2)}\int_{\mathcal{F}}\frac{d\tau d\bar{\tau}}{\tau_2} \int_{-\infty}^{+\infty}ds_1ds_2 \nonumber \\
&\sum_{m,w=0}^{N-1}\sum_i q^{h_i}\bar{q}^{\bar{h}_i}e^{4\pi\tau_2(1-\frac{1}{4(k-2)}) -\frac{k\pi}{\tau_2}\left|(s_1 - \frac{w}{N})\tau +(s_2 - \frac{m}{N})\right|^2+2\pi\tau_2s_1^2} \nonumber \\
&\frac{1}{\left|\sin(\pi(s_1\tau + s_2))\right|^2}\left|\prod_{r=1}^{+\infty}\frac{(1-e^{2\pi i r \tau})^2}{(1-e^{2\pi i r \tau - 2\pi i (s_1\tau +s_2)})(1-e^{2\pi i r \tau + 2\pi i (s_1\tau +s_2)})}\right|^2.
\end{align}
In this formula, the $bc$-ghosts have already been included and an internal CFT with weights $h_i$ is left arbitrary. As usual, $q=\exp(2\pi i \tau)$. First we focus on the twisted sectors (with $(w, m) \neq (0,0)$), since these do not exhibit a volume divergence. The large $k$ limit implies that the $s_1$- and $s_2$-integrals are dominated by $s_1 = w/N$ and $s_2 = m/N$. The infinite product factor in the end should simply be evaluated at this point. The theta-function appearing here can be directly related to the theta function with characteristics using the following set of formulas (and setting $\nu = \frac{w}{N} \tau + \frac{m}{N}$):
\begin{align}
\vartheta_1(\nu,\tau) &= 2 e^{\pi i \tau/4}\sin(\pi\nu) \prod_{n=1}^{+\infty}(1-q^n)(1-zq^n)(1-z^{-1}q^n), \\
-\vartheta_1(\nu,\tau) &= \vartheta\left[
\begin{array}{c}
\frac{1}{2}\\
\frac{1}{2}\end{array} 
\right](\nu, \tau) = e^{\frac{\pi i \tau}{4} + \pi i (\nu + \frac{1}{2})}\vartheta\left(\nu+\frac{\tau}{2}+\frac{1}{2},\tau\right), \\
\vartheta\left[
\begin{array}{c}
\frac{1}{2}+\frac{w}{N} \\
\frac{1}{2}+\frac{m}{N}  \end{array} 
\right](\tau) &= e^{\pi i \tau \left(\frac{1}{2} +\frac{w}{N}\right)^2 + 2\pi i \left(\frac{1}{2}+\frac{w}{N}\right)\left(\frac{1}{2}+\frac{m}{N}\right)} \vartheta\left(\left(\frac{1}{2}+\frac{w}{N}\right)\tau + \left(\frac{1}{2} + \frac{m}{N}\right), \tau\right),
\end{align}
where $z=\exp(2\pi i \nu)$. After some straightforward arithmetic, we can then write for this sector:
\begin{align}
Z_{w,m} &\approx \frac{1}{N}2\sqrt{k(k-2)}\int_{F}\frac{d\tau d\bar{\tau}}{\tau_2} \int_{-\infty}^{+\infty}ds_1ds_2 \nonumber \\
&\sum_i q^{h_i}\bar{q}^{\bar{h}_i}e^{4\pi\tau_2(1-\frac{1}{4(k-2)}) -\frac{k\pi}{\tau_2}\left|(s_1 - \frac{w}{N})\tau +(s_2 - \frac{m}{N})\right|^2}
\frac{4\left|\prod_{n=1}^{+\infty}(1-q^n)\right|^6 e^{-\pi\tau_2/2}}{\left|\vartheta\left[
\begin{array}{c}
\frac{1}{2}+\frac{w}{N} \\
\frac{1}{2}+\frac{m}{N}  \end{array} 
\right](\tau)\right|^2}.
\end{align}
To evaluate the integral, we make use of polar coordinates in the form $s_1 \tau + s_2 = x_1 + ix_2$ or\footnote{To be more precise, we first shift $s_1 \to s_1 + w/N$ and $s_2 \to s_2 + m/N$.} 
\begin{align}
s_1 \tau_1 + s_2 &= \rho \cos(\phi), \\
s_1 \tau_2 &= \rho \sin(\phi).
\end{align}
The transformation from ($s_1$, $s_2$) to ($\rho$, $\phi$) has Jacobian $\rho/\tau_2$. Hence the remaining integral becomes
\begin{equation}
\frac{2\pi}{\tau_2} \int_{0}^{+\infty}d\rho \rho e^{-\frac{\pi k}{\tau_2}\rho^2} = \frac{1}{k}.
\end{equation}
In the large $k$-limit, we hence obtain
\begin{align}
Z_{w,m} \approx \frac{1}{N}2\int_{F}\frac{d\tau d\bar{\tau}}{\tau_2} 
\sum_i q^{h_i}\bar{q}^{\bar{h}_i}e^{4\pi\tau_2}\frac{4\left|\eta(\tau)\right|^6}{\left|\vartheta\left[
\begin{array}{c}
\frac{1}{2}+\frac{w}{N} \\
\frac{1}{2}+\frac{m}{N}  \end{array} 
\right](\tau)\right|^2},
\end{align}
where $\eta(\tau) = q^{1/24}\prod_{n=1}^{+\infty}(1-q^n)$ is the Dedekind eta-function. Note that the factor $e^{4\pi\tau_2} = (q\bar{q})^{-1}$ is to be interpreted as the central charge term of the internal CFT with $c=24$. Choosing this internal CFT to be flat, we obtain in the end
\begin{align}
Z_{w,m} &\approx \frac{V_T}{N}2\int_{F}\frac{d\tau d\bar{\tau}}{\tau_2}
\frac{1}{(4\pi^2\alpha'\tau_2)^{12}}\left|\eta\right|^{-48}\frac{4\left|\eta\right|^6}{\left|\vartheta\left[
\begin{array}{c}
\frac{1}{2}+\frac{w}{N} \\
\frac{1}{2}+\frac{m}{N}  \end{array} 
\right](\tau)\right|^2}.
\end{align}
which agrees with the flat ($w$, $m$) sector \cite{Dabholkar:1994ai}\cite{Lowe:1994ah}.\footnote{In fact, we are off by a factor of 32. We would like to have obtained instead
\begin{align}
Z_{w,m} &\approx \frac{V_T}{N}\int_{F}\frac{d\tau d\bar{\tau}}{4\tau_2}
\frac{1}{(4\pi^2\alpha'\tau_2)^{12}}\left|\eta\right|^{-48}\frac{\left|\eta\right|^6}{\left|\vartheta\left[
\begin{array}{c}
\frac{1}{2}+\frac{w}{N} \\
\frac{1}{2}+\frac{m}{N}  \end{array} 
\right](\tau)\right|^2}.
\end{align}
We interpret this as a factor that should be included in the result of \cite{Hanany:2002ev}. For the type II superstring, the expressions given in the literature have a different normalization, and we will not have this discrepancy anymore. In the following computation of the bosonic string, we include this factor of $1/32$.} \\ 

\noindent The sector $w=m=0$ should be dealt with separately, since the sine function in (\ref{startingp}) causes a divergence that is to be interpreted as a IR volume divergence. \\
The above analysis has the following modifications. Firstly, the saddle point is at $s_1=s_2=0$. This implies the infinite product in (\ref{startingp}) becomes equal to 1. The sine factor blows up at the origin and we regulate it by cutting out a $\epsilon$-sized circle in the $x_1, x_2$ plane (following \cite{Giveon:2014hfa}). The final saddle-point integral is given by\footnote{In writing this we used the formula for the exponential integral: 
\begin{equation}
\int dx \frac{e^{-Ax^2}}{x} = -\frac{1}{2} \text{Ei}(1,Ax^2),
\end{equation}
with the series expansion
\begin{equation}
\label{taylor}
\text{Ei}(1,Ax^2) \approx -\gamma -\ln(A) -2\ln(x) + Ax^2 - \frac{1}{4} A^2 x^4 + \mathcal{O}(x^6).
\end{equation}
The first two terms are irrelevant in a definite integral (such as the one we have here). The third term is important: it gives precisely the $\ln(\epsilon)$ dominant contribution.}
\begin{equation}
\frac{2\pi}{\tau_2}\int_{\epsilon}^{+\infty}d\rho \rho \frac{e^{-\frac{\pi k }{\tau_2}\rho^2}}{\pi^2\rho^2} = -\frac{2}{\pi \tau_2} \ln(\epsilon).
\end{equation}
The resulting expression can be compared with the $w=m=0$ sector of $\mathbb{C}/\mathbb{Z}_N$ which is given by
\begin{equation}
\label{flatbosonic}
Z = \frac{V_T \mathcal{A}}{N}\int_{F}\frac{d\tau d\bar{\tau}}{4\tau_2^2}\frac{1}{(4\pi^2\alpha'\tau_2)^{12}}\left|\eta(\tau)\right|^{-48}.
\end{equation}
This allows us to identify the transverse area with the following divergent quantities:
\begin{equation}
\mathcal{A} = -\frac{1}{2\pi}\sqrt{k(k-2)}\ln(\epsilon).
\end{equation}
In the large $k$ limit this becomes\footnote{This implies that for fixed area $\mathcal{A}$, $k$ scales as $-\frac{1}{\ln(\epsilon)}$. The higher terms in the above expansion (\ref{taylor}) are then of the form (including the $k$ prefactor present in the partition function (\ref{startingp}) itself)
\begin{equation}
k^2 \epsilon^2 \sim  \frac{\epsilon^2}{\ln(\epsilon)^2} \to 0,
\end{equation}
or for the general term
\begin{equation}
\frac{\epsilon^{2n}}{\ln(\epsilon)^{n+1}} \to 0.
\end{equation}
Hence there are no subleading corrections that survive the $k\to\infty$ limit.}
\begin{equation}
\label{area}
\boxed{
\mathcal{A} = -\frac{1}{2\pi}k\ln(\epsilon)}.
\end{equation}

\subsection{Extension to the type II superstring for odd $N$}
The above reasoning can be readily generalized to the $\mathbb{Z}_N$ orbifolds of the type II superstring on each of these spaces. The formulas are a bit long, but the logic is the same as for the bosonic string. For odd $N$, the flat conical partition function is of the form \cite{Lowe:1994ah}
\begin{align}
\label{flatpt}
Z(\tau) = \frac{1}{4N}\left(\frac{1}{\left|\eta\right|^2\sqrt{4\pi^2\alpha'\tau_2}}\right)^{6}\sum_{w,m=0}^{N-1}&\sum_{\alpha,\beta,\gamma,\delta}\omega'_{\alpha\beta}(w,m)\bar{\omega}'_{\gamma\delta}(w,m) \nonumber \\
&\times \frac{\vartheta\left[
\begin{array}{c}
\alpha \\
\beta  \end{array} 
\right]^3 \vartheta\left[
\begin{array}{c}
\alpha+\frac{w}{N} \\
\beta+\frac{m}{N}  \end{array} 
\right]\bar{\vartheta}\left[
\begin{array}{c}
\gamma \\
\delta  \end{array} 
\right]^3 \bar{\vartheta}\left[
\begin{array}{c}
\gamma+\frac{w}{N} \\
\delta+\frac{m}{N}  \end{array} 
\right]}{\left|\vartheta\left[
\begin{array}{c}
\frac{1}{2}+\frac{w}{N} \\
\frac{1}{2}+\frac{m}{N}  \end{array} 
\right]\eta^3\right|^2}.
\end{align}
The $\omega'$ prefactors are given as follows
\begin{align}
\omega'_{00}(w,m) &= 1 , \\
\omega'_{0\frac{1}{2}}(w,m) &= e^{-\frac{\pi i w}{N}}(-1)^{w+1}, \\
\omega'_{\frac{1}{2}0}(w,m) &= (-1)^{m+1}, \\
\omega'_{\frac{1}{2}\frac{1}{2}}(w,m) &= \pm e^{-\frac{\pi i w}{N}}(-1)^{w+m}.
\end{align}
The partition function on the cigar orbifold on the other hand is given by \cite{Sugawara:2012ag}\cite{Giveon:2014hfa}
\begin{align}
\label{cigarpt}
Z(\tau) = &\frac{k}{N} \sum_{\sigma_L, \sigma_R}\sum_{w,m \in \mathbb{Z}} \int_{0}^{1}ds_1ds_2 \epsilon(\sigma_L;w,m)\epsilon(\sigma_R;w,m) \nonumber \\
&\times f_{\sigma_L}(s_1\tau+s_2,\tau)f_{\sigma_R}^{*}(s_1\tau+s_2,\tau)e^{-\frac{\pi k}{\tau_2}\left|\left(s_1 - \frac{w}{N}\right)\tau + \left(s_2 - \frac{m}{N}\right)\right|^2}.
\end{align}
where 
\begin{equation}
f_{\sigma}(u,\tau) = \frac{\vartheta_{\sigma}(u,\tau)}{\vartheta_1(u,\tau)}\left(\frac{\vartheta_{\sigma}(0,\tau)}{\eta}\right)^3,
\end{equation}
where $\vartheta_{\sigma} = \vartheta_{1,2,3,4}$ for $\sigma = \tilde{R}, R , NS , \tilde{NS}$ respectively and
$\epsilon = (1, (-1)^{w+1}, (-1)^{m+1}, (-1)^{w+m})$ for ($NS$, $\tilde{NS}$, $R$, $\tilde{R}$) respectively. This partition function includes all contributions from the worldsheet fermions and their spin structure and also the superconformal ghosts. To make this into a full string partition function, only a bosonic contribution should be added.\\

\noindent To make the link between these models, we rewrite the theta-functions by linking them directly as 
\begin{align}
\vartheta\left[
\begin{array}{c}
\frac{1}{2}+\frac{w}{N} \\
\frac{1}{2}+\frac{m}{N}  \end{array} 
\right] &= - e^{\pi i \tau \frac{w^2}{N^2}+\pi i \frac{w}{N} + \frac{2\pi i w m}{N^2}} \vartheta_1\left(\frac{w}{N}\tau + \frac{m}{N},\tau\right) , \\
\vartheta\left[
\begin{array}{c}
\frac{w}{N} \\
\frac{m}{N}  \end{array} 
\right] &= e^{\pi i \tau \frac{w^2}{N^2} + \frac{2\pi i w m}{N^2}}\vartheta_3\left(\frac{w}{N}\tau + \frac{m}{N},\tau\right), \\
\vartheta\left[
\begin{array}{c}
\frac{1}{2}+\frac{w}{N} \\
\frac{m}{N}  \end{array} 
\right] &= e^{\pi i \tau \frac{w^2}{N^2} + \frac{2\pi i w m}{N^2}}\vartheta_2\left(\frac{w}{N}\tau + \frac{m}{N},\tau\right), \\
\vartheta\left[
\begin{array}{c}
\frac{w}{N} \\
\frac{1}{2}+\frac{m}{N}  \end{array} 
\right] &= e^{\pi i \tau \frac{w^2}{N^2}+\pi i \frac{w}{N} + \frac{2\pi i w m}{N^2}}\vartheta_4\left(\frac{w}{N}\tau + \frac{m}{N},\tau\right).
\end{align}
The link between the $\sigma$-index and the $(\alpha, \beta)$ couple is:
\begin{align}
(1/2,1/2) &\to \tilde{R}, \\
(1/2,0) &\to R, \\
(0,1/2) &\to \tilde{NS}, \\
(0,0) &\to NS.
\end{align}
Again the saddle point can be handled quite easily. The saddle point integral yields again $1/k$, where all other factors present in the cigar partition function (\ref{cigarpt}) are simply to be evaluated at $s_1=\frac{w}{N}$ and $s_2 = \frac{m}{N}$. The rest is simply a bookkeeping exercise.\footnote{For the reader who is interested in more details, we make the following remarks. Starting with expression (\ref{flatpt}) and focussing on the holomorphic part with $\alpha$ and $\beta$, the sectors with $\beta = 1/2$ have a $e^{-\pi i w /N}$ factor in the $\omega'$'s which cancels with the $e^{\pi i w /N}$ phase present in the above conversion formulas. After this, all sectors above have the same prefactors, thus upon including the complex conjugate expression, only their modulus contributes and this gives a global prefactor $e^{-2\pi\tau_2 \frac{w^2}{N^2}}$, which cancels with the same prefactor appearing from the denominator.} The prefactor of $\left(\frac{1}{\left|\eta\right|^2\sqrt{4\pi^2\alpha'\tau_2}}\right)^{6}$ can be generated by including 8 free bosons and the bc ghosts, giving in total the contribution of 6 free bosons indeed.\footnote{Up to a $1/\sqrt{\tau_2}$ prefactor which in the notation of \cite{Lowe:1994ah} is absorbed into the fundamental domain measure.} \\

\noindent A question that immediately arises in this process is the following. For the flat orbifold, it is known that only odd $N$ makes sense as a string theory on a cone \cite{Dabholkar:1994ai}\cite{Lowe:1994ah}\cite{Adams:2001sv}. Yet on the cigar orbifold, no mention is made of such a restriction in the literature \cite{Sugawara:2012ag}\cite{Eguchi:2010cb}\cite{Sugawara:2011vg}. Although we should remark that in most of this work, the authors were interested in constructing consistent modular invariant partition functions (which is satisfied by the above expression also for even $N$). It seems then that also for these spaces, an interpretation in terms of strings on a cone can only be given for odd $N$. We postpone a deeper investigation into this issue to possible future work.\footnote{In fact, the argument given in \cite{Adams:2001sv} can be copied exactly for the cigar orbifolds. The orbifold identification has two possible actions on the spacetime spinors:
\begin{equation}
R = e^{\frac{2\pi i J}{N}} \quad \text{or} \quad R= (-)^Fe^{\frac{2\pi i J}{N}}
\end{equation}
where $J$ is the generator of angular rotations in the $U(1)$ cigar angular direction (which becomes $J_{89}$ on the plane of \cite{Adams:2001sv} by taking $k\to\infty$). Then we have $R^N = (-)^F$ or $R^N = (-)^{(N+1)F}$. The first possibility leads to an inclusion of $(-)^F$ in the orbifold group and the absence of spacetime spinors (which is unwanted). We hence should choose the second option with odd $N$ to avoid this.} \\

\noindent The untwisted sector ($w=m=0$) was handled in \cite{Giveon:2014hfa}. In more detail, it is given by
\begin{equation}
Z(\tau) \approx -\frac{k}{N\tau_2}\ln(\epsilon)\frac{2}{\pi}\frac{1}{4} \frac{\left|\vartheta_3^4 - \vartheta_4^4 - \vartheta_2^4\right|^2}{\left|\eta\right|^{12}},
\end{equation}
which equals the flat space cosmological constant (up to the factor of $1/N$) and it vanishes again due to Jacobi's obscure identity. Including the other flat dimensions, the $bc$-ghosts and the modular integral, we obtain
\begin{equation}
\label{superF}
Z \approx -\frac{1}{N}\int_{\mathcal{F}}\frac{d\tau d\bar{\tau}}{4\tau_2^2}k\ln(\epsilon)\frac{1}{2\pi} \left(\frac{1}{\left|\eta\right|^2\sqrt{4\pi^2\alpha'\tau_2}}\right)^{6} \frac{\left|\vartheta_3^4 - \vartheta_4^4 - \vartheta_2^4\right|^2}{\left|\eta\right|^{12}}.
\end{equation}
The transverse area can again be identified in this expression as the same formula (\ref{area}). \\ 

\begin{figure}[h]
\centering
\includegraphics[width=0.8\textwidth]{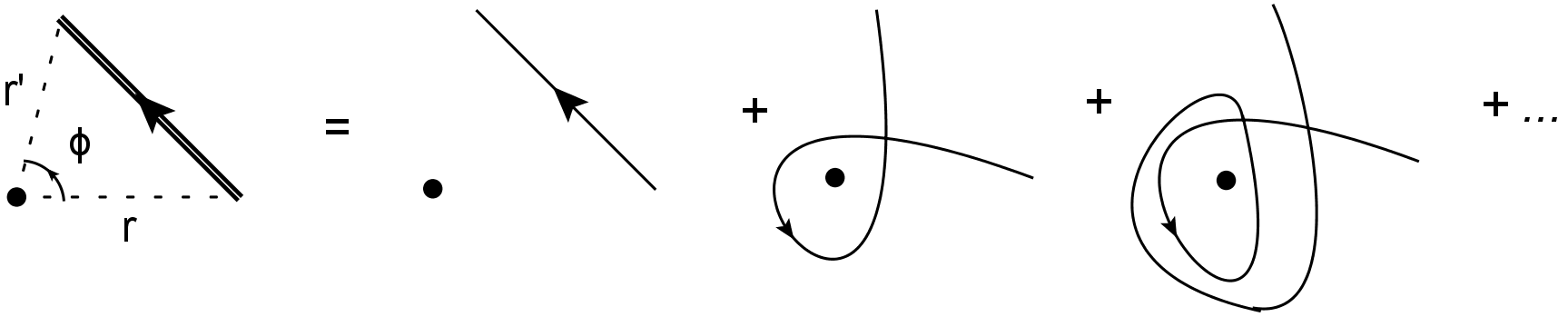}
\caption{The total Euclidean propagator in flat space between points $(r,0)$ and $(r',\phi)$ can be written as a sum over propagators with fixed winding number around the origin.}
\label{particlePolar}
\end{figure}

\noindent This seems a quite important result: even though the GSO projection assigns a special role to the angular coordinate, the partition function of Euclidean Rindler space is precisely the same as the flat space vacuum energy on an infinite 2d plane (just as it was for the bosonic string). \\
Let us end on a more speculative note here. This equality means the coordinate transformation from polar coordinates to cartesian coordinates is unhindered by the GSO projection. We remind the reader that for quantum fields in Rindler space, the stress tensor vanishes in the Minkowski vacuum, which can be rewritten in terms of the coordinates of the Rindler observer as:
\begin{equation}
\left\langle T_{\mu\nu}\right\rangle_M = \left\langle T_{\mu\nu}\right\rangle_R + \text{Tr}_R\left(T_{\mu\nu}e^{-\beta H_R}\right)_{H_R \neq 0}.
\end{equation}
The thermal bath of Rindler particles combines with the Casimir contribution to give a vanishing vev. Thus the thermal bath does not backreact on the background. The ultimate reason for this is the fact that Rindler space and Minkowski spacetime are simply related by a coordinate transformation. \\
This QFT story can be interpreted in Euclidean signature as well. It was shown in \cite{Dowker:1977zj}\cite{Troost:1977dw}\cite{Troost:1978yk}\cite{Parentani:1989gq} that the Euclidean propagator in flat space can be expanded into a sum over winding numbers around the origin:
\begin{equation}
G(r,0;r',\phi; s) = \sum_{w\in\mathbb{Z}}G^{(w)}(r,0;r',\phi; s).
\end{equation}
This relation is shown diagrammatically in figure \ref{particlePolar}.
\\
The stress tensor vev can then be obtained by applying a suitable differential operator on the Green function and taking the coincidence limit of the two points. Hence we obtain
\begin{equation}
\left\langle T_{\mu\nu}\right\rangle_E = \left\langle T_{\mu\nu}\right\rangle_{w=0} + \sum_{w'=-\infty}^{+\infty}\left\langle T_{\mu\nu}\right\rangle_{w},
\end{equation}
where the prime denotes the absence of the $w=0$ term in the sum. The $w=0$ term is the temperature-independent Casimir contribution whereas the remaining sum is the thermal contribution. In this Euclidean setting, it is apparent that the vanishing of the vev in 2d flat space implies that the sum of the Casimir and the thermal contributions vanish. Again the main reason is the coordinate equivalence between cartesian coordinates and polar coordinates. \\
Now back to string theory. The fact that the partition function of string theory in polar coordinates is the same as that in cartesian coordinates shows that the coordinate transition between both is still valid in string theory. This is a necessary condition to have a vanishing stress tensor.\footnote{It seems difficult to make this point more firmly: the full stress tensor of the string gas seems difficult to obtain; we only obtained the most dominant contribution in \cite{Mertens:2014dia} (as we will discuss in chapter \ref{chrel}). The free energy however is directly linked to the canonical internal energy which is to be interpreted as the spatial integral of $T_0^0$ and the result we obtain is then only in the weaker integrated sense.} In this sense, the tip of the cigar is not special and the GSO projection does not ruin the coordinate equivalence between polar and cartesian coordinates. This suggests there is no backreaction caused by the thermal atmosphere of the black hole (a good thing!). 

\subsection{Continuation of the flat orbifold inherited from the cigar orbifold}
Let us make a short detour here and consider a question first posed in \cite{Dabholkar:1994ai}: can we continue the partition functions on the $\mathbb{C}/\mathbb{Z}_N$ orbifolds to a non-integer $N$? An immediate response would be no, since modular invariance is no longer present for such values of $N$. Overcoming the initial shock, one might be tempted to use this approach as a possible off-shell proposal for string theory on a cone. In fact, we argued before (and in \cite{Mertens:2013zya}) that for the dominant behavior (given in field theory language) such a continuation is quite natural. Even if one believes this, the partition functions on $\mathbb{C}/\mathbb{Z}_N$ do not lend themselves towards continuation in $N$ (as was discussed by Dabholkar as well \cite{Dabholkar:1994ai}). Here, we perform this natural continuation at the level of the partition function on the cigar orbifold. Our goal is to use this continuation of the cigar CFT to tell us something about the continuation for the flat cones. \\
The partition function (\ref{startingp}) on the cigar orbifold can be rewritten in the suggestive way
\begin{align}
Z &= \frac{1}{N}2\sqrt{k(k-2)}\int_{\mathcal{F}}\frac{d\tau d\bar{\tau}}{\tau_2} \int_{0}^{1}ds_1ds_2 \nonumber \\
&\sum_{m,w=-\infty}^{+\infty}\sum_i q^{h_i}\bar{q}^{\bar{h}_i}e^{4\pi\tau_2(1-\frac{1}{4(k-2)}) -\frac{k\pi}{\tau_2}\left|(s_1 - \frac{w}{N})\tau +(s_2 - \frac{m}{N})\right|^2+2\pi\tau_2s_1^2} \nonumber \\
&\frac{1}{\left|\sin(\pi(s_1\tau + s_2))\right|^2}\left|\prod_{r=1}^{+\infty}\frac{(1-e^{2\pi i r \tau})^2}{(1-e^{2\pi i r \tau - 2\pi i (s_1\tau +s_2)})(1-e^{2\pi i r \tau + 2\pi i (s_1\tau +s_2)})}\right|^2.
\end{align}
This allows a natural continuation in $N$ as $1/N \to \frac{\beta}{\beta_{\text{Hawking}}}$, where modular invariance is lost. We investigate how this translates into a continuation of the flat cone.\\
Firstly, the $s_1$- and $s_2$-integrals are only over a unit interval here. The untwisted sector has two stationary points for each $s$-integral: $s=0$ and $s=1$.\footnote{In fact, the stationary point $s_1=1$ corresponds to $w=1$ which gets translated upon using Poisson resummation etc. into the winding $1$ mode. Higher winding modes (which would correspond to $s_1=2$ etc.) are not stationary points. This shows from this perspective as well that only singly wound modes are present in the large $k$ limit. As a reminder, the $w$ and $m$ quantum numbers are the torus cycle winding numbers. The $m$ quantum number gets Poisson resummed into the discrete momentum whereas the winding number $w$ remains the same (at least for the discrete representations) throughout the manipulations.} However, both are at the boundary of the integration interval and hence receive weight factor $1/2$. Both contributions are equal due to the periodicity of the Ray-Singer torsion. In all, one can choose one of these saddle points and neglect the weight factors. This agrees with our earlier analysis of the saddle points.\\
Upon making the replacement for $1/N$, one finds a saddle point only for those $w$ and $m$ for which
\begin{equation}
\label{inequal}
0 \leq \frac{w\beta}{\beta_{\text{Hawking}}} \leq 1 , \quad 0 \leq \frac{m\beta}{\beta_{\text{Hawking}}} \leq 1
\end{equation}
holds. The lower boundary corresponds to the untwisted sector and the upper boundary is only reached precisely for the orbifold models. For such general values of $\beta$, the points $(s_1, s_2) = (1,0),\, (0,1),\, (1,1)$ are not saddle points anymore. This implies the untwisted sector has an overall scaling of $1/4$ with respect to the orbifold points. The only difference for the twisted sectors is hence the replacement of the twisted sum by
\begin{equation}
\frac{1}{N} \sum_{m,w=0}^{N-1} \to \frac{\beta}{\beta_{\text{Hawking}}}\sum_{m,w}'
\end{equation}
where the prime indicates that $m$ and $w$ are integers restricted by (\ref{inequal}). One readily checks that for the orbifold points, one regains the earlier results. \\
In particular, this continuation implies that for $T<T_{\text{Hawking}}$ the only sector present is the $m=w=0$ sector and this is unchanged as $\beta$ is varied. This is in conflict with the free-field trace $\text{Tr}e^{-\beta H}$ which decreases monotonically as $\beta$ increases. \\
We note that the most dominant thermal state ($w=\pm1, m =0$) is present as soon as $T>T_{\text{Hawking}}$. At the Hawking temperature itself, we saw earlier that this state is in fact also present, albeit camouflaged in the flat space result. Thus one can follow this state as one lowers the temperature all the way to the Hawking temperature where this state becomes marginal. This seems to fit with our general expectations on continuing this state through a range of temperatures. \\
Also, this continuation makes it clear that this resulting off-shell proposal is \emph{non-analytic} in $N$, in contrast to the arguments made by Dabholkar. \\
The arguments presented in this section should not be viewed as a full-fledged proposal for the off-shell continuation of the flat cones, we merely link the most natural off-shell continuation of the cigar orbifolds to the inherited off-shell continuation of the flat cones. Whether these continuations make sense, is left open. \\

\noindent What is apparent is that one gets another hint that these partition functions do not appear to correspond to Hamiltonian free field traces (other hints are the discussions made by Susskind and Uglum \cite{Susskind:1994sm} on the exotic open string interactions and the absence of certain winding sectors in the thermal spectrum (i.e. the unitarity constraints of the model we discussed above), obscuring a thermal interpretation \cite{Mertens:2014cia}). We will come back to this point in chapter \ref{chgenbh}. \\


\section{The thermal scalar on the full $SL(2,\mathbb{R})/U(1)$ cigar}
\label{remarks}

In this section, we are interested in the opposite limit as that studied up to this point: we comment on the behavior of the cigar CFT as we \emph{lower} $k$ to its critical value of $k=1$ or $k=3$ for type II and bosonic strings respectively. Note that this is an on-shell change of the black hole: no conical singularity is created during the process. It is known that several phenomena occur at the critical value of $k$ \cite{Giveon:2005mi} and in this section we look at this limiting process from the thermal scalar point of view. Qualitatively, one expects the string/black hole correspondence point \cite{Horowitz:1996nw} to occur as soon as the black hole membrane \cite{Thorne:1986iy} diverges to infinity \cite{Kutasov:2005rr}\cite{Giveon:2005jv}. This corresponds to the thermal scalar becoming non-normalizable. The $SL(2,\mathbb{R})/U(1)$ black hole provides an exact setting where the corresponding string/black hole phase transition can be better studied \cite{Giveon:2005mi}\cite{Parnachev:2005qr}. Even though the genus one partition function is incapable of seeing this transition directly (the transition is driven by non-perturbative effects), several suggestive clues will arise even at the one loop level. Moreover, we will see a continuous transition between a random walk corresponding to a Hamiltonian free-field trace, and a random walk coming from the discrete thermal scalar bound to the black hole horizon, providing some explicit evidence of the final remark presented in the previous section. \\ 
The string spectrum of the cigar CFT was written down at the beginning of section \ref{unitarity}. We remind the reader that for the discrete modes, the $SL(2,\mathbb{R})$ quantum number $j = M -l$ for $l=1,2,\hdots$ and $M$ is directly related to $m$. First of all, we note that the $l=1$ state for the type II superstring (the $l=2$ state for the bosonic case) are exactly marginal for all values of $k$ (if they exist) \cite{Kutasov:2000jp}. As we lower $k$, we encounter the special value of $k=3$. For this value of $k$ several things happen:
\begin{itemize}
\item{The asymptotic linear dilaton Hagedorn temperature becomes equal to the Hawking temperature.}
\item{The wavefunctions of the discrete states all become non-normalizable as can be seen by looking at the asymptotic linear dilaton behavior.}
\item{The one-loop thermal partition function does not include any discrete states anymore, in corroboration with the non-normalizability of the wavefunctions. In this setting, the states are absent since the contour shift used in \cite{Hanany:2002ev} does not cross poles anymore.}
\item{The lowest weight state in the continuum becomes marginal.}
\end{itemize}
Proofs of these statements can be found in section \ref{proof}. Exactly the same effects happen for the superstring at $k=1$. Of course, several of these features have been known from previous work \cite{Giveon:2005mi}\cite{Nakayama:2005pk}. What we want to emphasize in this section is how the one-loop free energy (and other thermodynamical quantities) behaves when approaching the critical $k$ value. As $k$ approaches this value, the thermal scalar wavefunction spreads without bound, until it disappears from the spectrum. Immediately thereafter, a continuous state takes over its role. 
During this procedure, the dominant part of the free energy (coming from the thermal scalar) changes in extensivity. For $k>3$, this part scales as the area of the horizon. For $k=3$, the mode is sensitive to the entire volume of the space. \\

\noindent It is nice to see in the explicit formulas that this critical value indeed occurs as soon as the continuous modes show Hagedorn behavior.\footnote{In fact, we believe this continuous process sheds some light on the nature of the thermal scalar on black hole horizons. We discussed earlier that the interpretation of the genus one partition function in terms of the free Hamiltonian trace is obscured since possibly interactions are included with an open string gas on the horizon. We will make this more explicit in section \ref{modDom} by pointing out a difference in our treatment of the string path integral and the genus 1 CFT result. This is not the case for the thermal scalar in the continuous representation here since this one is asymptotically precisely the same as the linear dilaton thermal scalar, yielding long string dominance for near-Hagedorn linear dilaton temperatures: this can be made explicit by splitting the free energy in two contributions: one coming from the horizon and one from the asymptotic linear dilaton space. The latter is insensitive to the subtleties about torus embeddings (see chapter \ref{chgenbh}) and corresponds to non-interacting strings \cite{Kutasov:2000jp}\cite{Barbon:2007za}. The process of lowering $k$ results in a continuous transition between dominance by the discrete thermal scalar living close to the horizon and the asymptotic continuous thermal scalar giving long string dominance in a free-field trace of the asymptotic space. The nature of these states appears to be the same and the free-field trace is expected to be sensitive to the same thermal scalar and critical temperature as the genus one partition function (at least for this CFT), something we will argue for at the end of section \ref{modDom} further on.} \\

\noindent We can now give a full picture of thermodynamics on the $SL(2,\mathbb{R})/U(1)$ cigar (or the near-extremal NS5 brane solution) as one varies $k$. \\
It was shown by Sugawara \cite{Sugawara:2012ag} that the thermal partition function can be written as a sum of two contributions:
\begin{equation}
Z = Z_{\text{fin}} + Z_{\text{asympt}},
\end{equation}
a finite part coming from the tip of the cigar (containing both the discrete modes and part of the continuum\footnote{This continuum part comes from the phase shift contribution in the density of states (or equivalently the reflection amplitude of the gravitational potential).}) and an asymptotic part which diverges due to the infinite volume available and which is precisely equal to the linear dilaton thermal partition function. This translates into $F = F_{A} + F_{V}$, where $F_A$ scales like the transverse area of the black hole and $F_V$ scales as the volume of the entire space.\\
\noindent Let us start by taking $k$ very large. In this case, the background approaches Rindler space and the free energy ($\propto Z$) is dominated by the thermal scalar. The thermodynamical quantities have a random walk interpretation and are given by a localized contribution at string length from the horizon. They have a part proportional to the volume (subdominant) and a part proportional to the black hole area. It is the latter part that should be interpreted as intrinsic to the black hole itself. \\
Lowering $k$ invalidates the Rindler approximation. Nonetheless, roughly the same story is valid when considering the thermal scalar wavefunction on the entire cigar geometry. The random walk is spreading more and more from the tip of the cigar. Thermodynamic quantities still satisfy the same scaling properties and the dominant contribution comes from the discrete thermal scalar bound to the (Euclidean) horizon. \\
As soon as $k$ reaches 1 (type II) or 3 (bosonic), the (discrete state) thermal scalar drops out from the spectrum. When nearing this value of $k$, the random walk spreads without bound over the entire cigar and becomes non-normalizable. At the same time, the asymptotically linear dilaton thermal scalar (coming from the continuous part of $Z$) becomes massless. The random walk of the latter looks asymptotically like the one studied in \cite{Mertens:2014cia} for the linear dilaton background. Not just the thermal scalar, but in fact all bound states disappear from the spectrum at this point. The bound states are a typical feature of string theory since they only exist because of the possibility of including winding numbers along the cigar, a possibility that is absent for field theory. At $k=3$ (or $k=1$), only the continuous spectrum remains. \\

\noindent It is also known that for this value of $k$, the black hole itself becomes non-normalizable \cite{Giveon:2005mi}. Also, infalling D0-branes change their radiation pattern \cite{Nakayama:2005pk}: normally most radiation is incident on the black hole. When $k$ reaches the critical value, a significant part of the radiation propagates outwards, contrary to our intuition about black holes. 
This process of lowering $k$ is sketched in figure \ref{rw} below.
\begin{figure}[h]
\centering
\includegraphics[width=0.7\textwidth]{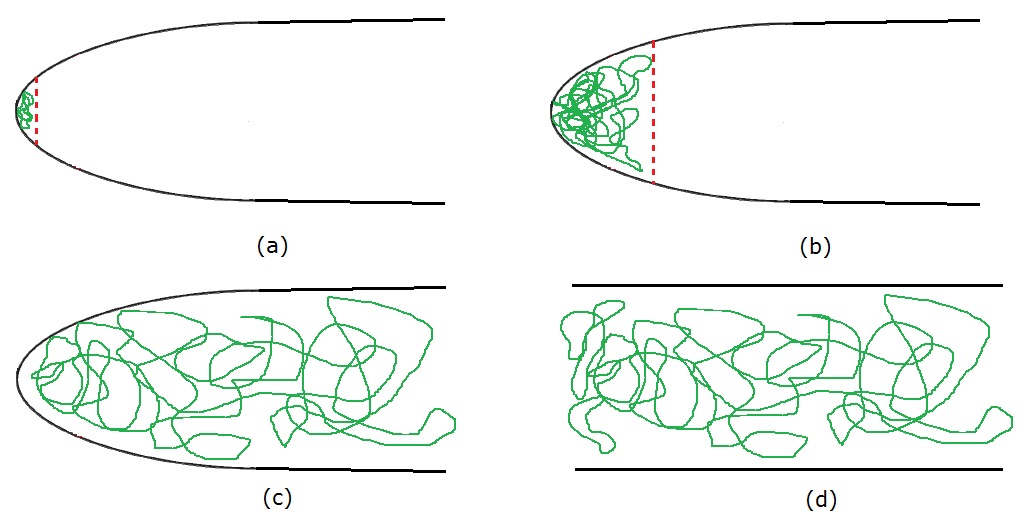}
\caption{(a) Large $k$ cigar background with random walk superimposed. Of course, in reality the random walk is on the spatial submanifold only, but for illustrative purposes we draw it like this. The (red) dashed line limits the spread of the discrete thermal scalar in the radial direction. For very large $k$ the flat Rindler approximation becomes accurate. (b) Lowering $k$ causes the discrete mode to spread radially. This invalidates the Rindler approximation. Nonetheless, the mode retains its dominant character. (c) When $k=1$, the mode spreads without bound radially. The mode becomes non-normalizable and disappears from the spectrum. Precisely at this point, the continuous thermal scalar takes over the dominant behavior. (d) The same dominant random walk on the linear dilaton background. The asymptotic parts of the random walks look the same for $k=1$.}
\label{rw}
\end{figure}

\section{Rotating black holes and the Hartle-Hawking state}
\label{rotat}
As noted before, Rindler space is the near-horizon region of a general uncharged black hole. It is also the near-horizon region of a rotating black hole (e.g. a Kerr black hole) in terms of the co-rotating ZAMO observers. Without going into details, the Kerr metric is of the form (see e.g. \cite{Thorne:1986iy})
\begin{equation}
ds^2 = - \alpha^2 dt^2 + \frac{\rho^2}{\Delta}dr^2 + \rho^2 d\theta^2 + \bar{\omega}^2\left(d\phi - \omega dt\right)^2.
\end{equation}
Close to the horizon, one can approximate $\omega \approx \Omega_H$, and define a new angular coordinate as $\bar{\phi} = \phi - \Omega_H t$. Then a redefinition of the radial coordinate directly leads to Rindler spacetime. The main point is then that in the original coordinates, the near-horizon Rindler region is moving along with the horizon angular velocity $\Omega_H$ in the $\phi$-direction. Just as before, the Kerr metric itself is not a solution of string theory but Rindler space is. \\

\noindent Let us now take a closer look at how the thermal scalar random walk formalism would work for rotating black holes. This discussion will not be fully airtight.\\
In the path integral derivation presented in chapter \ref{chth}, it is explicitly used that 
\begin{equation}
\label{condii}
G_{00} > 0, \quad \det{G_{ij} - \frac{G_{0i}G_{0j}}{G_{00}}} > 0.
\end{equation}
Let us look at a Kerr black hole and find out whether these conditions are fulfilled. \\
For a rotating black hole, the event horizon is located at $G_{00} - \frac{G_{0\phi}G_{0\phi}}{G_{\phi\phi}} = 0$. If we assume $G_{00} \neq 0$, then this is equivalent to $G_{\phi\phi} - \frac{G_{0\phi}G_{0\phi}}{G_{00}} = 0$, so restricting to the exterior (as in any Euclidean black hole), we seemingly do not run into trouble with the second condition of (\ref{condii}). \\
However, for a Kerr black hole also $G_{00} = 0$ is possible, which defines the ergoregion of the black hole and the derivation fails to hold within such regions. This region \emph{cannot} be excluded since the exterior of the ergoregion is not globally hyperbolic and therefore does not admit a (straightforward) quantum field theory construction. \\
It is therefore implicit in our derivation that we exclude spacetimes with ergoregions. This also demonstrates again that our quantization proceeded using the timelike Killing vector $\frac{\partial}{\partial t}$. This vector field turns spacelike within the ergosphere. For a rotating black hole, the horizon is a Killing horizon of the co-rotating vector field $\chi = \frac{\partial}{\partial t} + \Omega_H \frac{\partial}{\partial \phi}$ which does remain timelike inside the ergoregion. It is this vector field that is associated with the Hartle-Hawking vacuum (supposing it can be sensibly defined). This vector field, however, turns spacelike far away (unlike $\frac{\partial}{\partial t}$), corresponding to the fact that rigid rotation automatically implies faster-than-light travel for distant observers. For a hole that rotates sufficiently slow, this region is arbitrarily far away and a sensible quantization might be allowed \cite{Frolov:1989jh}. This does raise some questions regarding the level of rigor with which the Hartle-Hawking vacuum can actually be constructed. On the other hand, in AdS backgrounds, rotating black holes exist for which the co-rotating Killing vector field remains timelike throughout the full exterior of the hole \cite{Hawking:1999dp}\cite{Winstanley:2001nx}. \\
\noindent In any case, suppose we start not with a canonical ensemble but with a grand canonical ensemble of the gas around the black hole where the angular momentum of the mode contributes to the partition function as: 
\begin{equation}
\text{Tr}e^{-\beta\left(H - \Omega_H J\right)}.
\end{equation}
Then this can be viewed equivalently as a canonical ensemble using $\chi$ (Hamiltonian $H'$) to measure energies instead:
\begin{equation}
\text{Tr}e^{-\beta\left(H - \Omega_H J\right)} = \text{Tr}e^{-\beta H'}.
\end{equation}
In somewhat more detail, suppose we have a mode in the spectrum with energy $\omega$ and angular momentum $m$:
\begin{equation}
\psi_{\omega,m} \sim e^{-i\omega t}e^{im\phi}.
\end{equation}
This contributes to the grand canonical trace as $e^{-\beta\left(\omega - \Omega_H m\right)}$. Performing the coordinate redefinition $\bar{\phi} = \phi - \Omega_H t$, the mode can be written as
\begin{equation}
\psi_{\omega,m} \sim e^{-i(\omega -m\Omega_H) t}e^{im\bar{\phi}},
\end{equation}
and it now satisfies $e^{-\beta H} = e^{-\beta\left(\omega - \Omega_H m\right)}$. Hence its contribution equals that of a canonical ensemble after the coordinate redefinition. Also, one readily finds that the above coordinate transformation indeed yields
\begin{equation}
\chi = \frac{\partial}{\partial t} + \Omega_H \frac{\partial}{\partial \phi} \, \to \, \frac{\partial}{\partial t}.
\end{equation}

\noindent For a large enough black hole, or anticipating that the near-horizon region will contain the most dominant contribution, we can approximate the problem again by only looking into the near-horizon region.
We can make the transition to the near-horizon co-rotating ZAMOs and their mode spectrum, quantized by $\chi$.\footnote{$\chi$ coincides with ZAMO movement close to the horizon. When going further away, $\chi$ describes a rigid rotation whereas the ZAMO's rotation slows down (according to an asymptotic observer) as their worldlines move further out. At infinity, a ZAMO is not rotating at all anymore (w.r.t. the coordinate $\phi$).} These observers effectively see a static Rindler region. \\ 

\begin{figure}[h]
\centering
\includegraphics[width=0.5\textwidth]{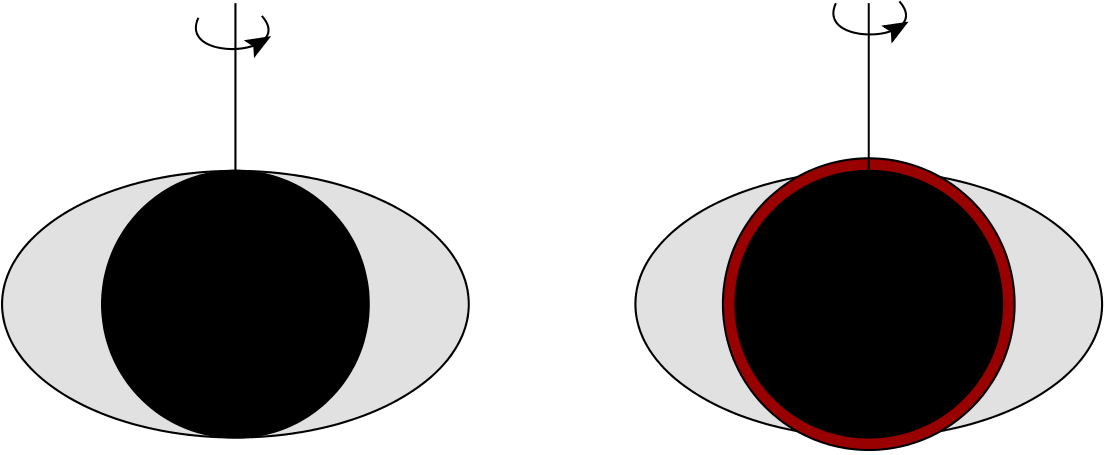}
\caption{Left figure: structure of a rotating black hole. $\frac{\partial}{\partial t}$ becomes spacelike in the ergoregion (the gray region), whereas $\chi$ remains timelike in that region. Right figure: most dominant contribution to the grand canonical trace. The Rindler thermal scalar provides the dominant contribution (colored in red) and rotates rigidly along with the hole itself.}
\label{rota}
\end{figure}

\noindent In any case, the canonical trace is dominated by the thermal scalar in the near-horizon Rindler region. Hence the full grand canonical trace, as observed at infinity by non-moving observers (or ZAMOs at infinity), is dominated by the thermal scalar mode. As this mode has no non-trivial profile tangential to the horizon, its wavefunction is unaltered as the hole rotates, and the long string again forms a region close to the horizon of string-scale thickness. Thus from infinity, the string-sized shell surrounding the black hole simply rotates rigidly along with the hole itself. The full situation is sketched in figure \ref{rota}.

\section{Open problems and speculations}
\label{open}
We take a look into some puzzles and unsolved problems that we face.

\begin{itemize}


\item{
The authors of \cite{Kutasov:2000jp} argued that on the $SL(2,\mathbb{R})/U(1)$ cigar in the type II superstring case, the marginal state should become tachyonic when considering higher loop effects: the tree level mass should get corrections that drive it tachyonic. This idea arose in order to agree with (and explain) the unstable little string thermodynamics. The one-loop free energy is \emph{only} sensitive to the tree-level spectrum, so indeed this is not divergent. The higher genus contributions to the free energy however, are expected to diverge. Whether this is the case or not is, as far as we know, not explicitly known. Note that an alternative explanation of the unstable little string thermodynamics was given in \cite{Rangamani:2001ir}\cite{Buchel:2001dg} where the instability was attributed to a gravitational zero-mode. This solution would not require the winding string zero-mode to become tachyonic at higher genera. 
}

\item{In some less recent literature (see e.g. \cite{Parentani:1989gq}, \cite{Emparan:1994bt} and \cite{McGuigan:1994tg}), it was found that the Green propagator (and the energy-momentum tensor) should diverge for the canonical Rindler temperature. Their arguments do not care about the specifics of the bosonic string theory and are equally valid for superstrings and heterotic strings. So in these papers, it is suggested that the region near the horizon causes divergences in several quantities, corresponding to a \emph{maximal acceleration} of strings near black hole horizons. In \cite{McGuigan:1994tg} it was argued that this provides a natural cut-off to the theory. It would be very interesting to know the link to our work.}

\item{One could think about expanding the random walk path integral picture to include Ramond-Ramond backgrounds. The goal would be to again identify the membrane with the long string(s). Besides the technical difficulties this would entail, we would like to point out a conceptual problem with this. For a neutral black hole, the above arguments suggest that indeed a long string dominates the thermal ensemble, exactly like in flat spacetime. This agrees with the black hole correspondence principle \cite{Horowitz:1996nw} where a neutral black hole evaporates into a long highly excited string.\footnote{In \cite{Susskind:2005js} this is viewed as the same string that constitutes the stretched membrane.} When we consider RR charged black holes however, the correspondence principle suggests a match to a gas of open strings living on a D-brane. This suggests that we should augment the closed string random walk picture with open string sectors to find the stretched membrane for such black holes. The thermal scalar appears not to be sufficient to constitute the membrane in this case. 
This behavior is not universal for all charged black holes since for electrically NS-NS charged black holes, the correspondence principle \emph{does} suggest a match to long closed strings. \\
We finally remark that our study was also limited to non-extremal black holes which have Rindler space near their horizon. Extremal black holes typically develop an infinite throat (e.g. the extremal 4d Reissner-Nordstr\"om black hole has $AdS_2 \times S^2$ near-horizon geometry).
}

\end{itemize}

\section{Summary of this chapter}
\label{conclusio}
In \cite{Mertens:2013pza}, we reviewed and extended the path integral derivation of the random walk behavior of near-Hagedorn thermodynamics \cite{Kruczenski:2005pj}. Several questions arose in the process: is there really a winding mode in the string spectrum, especially if the space does not topologically support winding modes? Can we get a handle on the higher $\alpha'$-correction terms? In this chapter, we found answers to these questions by analyzing Rindler spacetime as the near-horizon approximation of black holes. \\

\noindent We found that the random walk gets corrected but we know precisely what the correction is (due to its relation to the $SL(2,\mathbb{R})/U(1)$ cigar). Moreover the resulting particle path integral can be exactly solved and we checked that it gives the same answer as the zero-mode of the Hamiltonian operator associated with it, which was recently found in a slightly different context in \cite{Giveon:2012kp}\cite{Giveon:2013ica}. We used one-loop convergence as a criterion to obtain the Hagedorn temperature and found the critical temperatures and critical behavior for the Rindler string gas of all string types (bosonic, type II and heterotic). All string types show critical Hagedorn behavior at the canonical Rindler temperature.\footnote{Keeping in mind the complications for the bosonic string.} We interpreted the stretched horizon as a Hagedorn string at string length from the event horizon whose temperature is redshifted to the Hawking temperature at infinity. To further substantiate the claims in \cite{Giveon:2013ica} regarding the $\alpha'$ corrections, we have shown that the corrections to the bosonic string thermal scalar action are precisely such that they reproduce correctly the flat $\mathbb{C}/\mathbb{Z}_N$ limit. Disregarding the bosonic corrections would yield a different answer. The unitarity constraints from the cigar get transferred to Euclidean Rindler space, where they in particular forbid any higher winding modes. \\

\noindent Our attitude towards this approach to string thermodynamics is the following. The method reduces string theory to a particle theory of the thermal scalar. This inherently brings back the UV divergence one typically encounters in field theory. We of course know that the thermal scalar field theory should only be used in the large Schwinger parameter limit and the expressions should be cut-off effectively at small Schwinger parameter. Not the full thermal scalar field theory is hence relevant for our purposes, but it seems difficult to obtain a more quantitative prediction on how one should impose this cut-off at the field theory level. What one pays for in this aspect, we gain in another: namely because we have a field theory, we are free to go off-shell. We obtain an off-shell description of the dominant behavior of string thermodynamics near black holes where we are free to vary $\beta$ \emph{without} having to worry about violating (super)conformal invariance. We can see that black holes when equilibrated with a gas of strings are surrounded by a thin stringsize shell of the thermal scalar, that we identify with the black hole membrane. This membrane differs for the different types of strings: for bosonic strings, we have the complications we addressed in section \ref{unitarity}: the lowest mode is not in the string spectrum and the free energy does not have a tachyonic divergence. Flat orbifolds on the other hand always include tachyonic divergences and this suggests quantities like the entropy also are tachyonically divergent. For both superstrings and heterotic strings, the situation is simpler and we have a marginal convergence: the membrane is composed of a zero-mode. The heterotic membrane is also slightly larger than the superstring membrane.\\

\noindent We finally compared the two approaches to Rindler thermodynamics (string path integral Fourier expansion versus field theory) and noted that, when combined, they describe the picture by Susskind of a single long string surrounding the event horizon, at least for the genus one worldsheet.



\section{*Schwarzschild normalization of Rindler space}
\label{GM}
We translate the results from sections \ref{typeII}, \ref{heterotic} and \ref{bosonic} to the Schwarzschild-normalized Rindler space:
\begin{equation}
ds^2 = -\left(\frac{\rho^2}{(4GM)^2}\right)dt^2 + d\rho^2 + d\mathbf{x}^2_{\perp}.
\end{equation}
For type II superstrings, the eigenfunctions and eigenvalues in these coordinates are
\begin{equation}
\psi_{n}(\rho) \propto \exp\left(-\frac{\beta\rho^2}{4\pi\alpha'(4GM)}\right)L_{n}\left(\frac{\beta\rho^2}{2\pi\alpha'(4GM)}\right), \quad \lambda_{n} = \frac{\beta - 2\pi(4GM)+2\beta n}{\pi \alpha'(4GM)}.
\end{equation}
The lowest eigenmode has the form
\begin{equation}
\psi_0 \propto \exp\left(-\frac{\beta\rho^2}{16\pi\alpha'GM}\right).
\end{equation}
For $\beta = \beta_R = 8\pi GM$, we find
\begin{equation}
\psi_0 \propto \exp\left(-\frac{\rho^2}{2\alpha'}\right)
\end{equation}
and we clearly see that the mode has stringscale width. \\
For heterotic strings, the eigenfunctions and eigenvalues become
\begin{align}
\psi_{n}(\rho) &\propto \rho^{\frac{\pi(4GM)}{\beta}}\exp\left(-\frac{\beta\rho^2}{4\pi\alpha'(4GM)}\right)L_{n}^{(a)}\left(\frac{\beta\rho^2}{2\pi\alpha'(4GM)}\right), \\
\quad \lambda_{n} &= \frac{\beta - 2\pi(4GM)+2\beta n}{\pi \alpha'(4GM)},
\end{align}
where $a=\frac{\pi(4GM)}{\beta}$.\\
Both for type II superstrings and heterotic strings, the Hagedorn temperature is equal to the canonical Rindler temperature which is (with this normalization) also equal to the Hawking temperature:
\begin{equation}
\beta_{Hawking} = 8\pi GM.
\end{equation}
For bosonic strings, the eigenfunctions are the same as those of the type II superstring, and the eigenvalues are given by
\begin{equation}
\lambda_{n} = \frac{\beta +2\beta n}{\pi \alpha'(4GM)} - \frac{4}{\alpha'},
\end{equation}
whereas the $\alpha'$-corrected eigenvalues are
\begin{equation}
\lambda_{n} = \frac{\beta  +2\beta n}{\pi \alpha'(4GM)} - \frac{4}{\alpha'} - \frac{\beta^2}{2\pi^2\alpha'^2}.
\end{equation}

\section{*Explicit solution of the Rindler random walk path integral}
\label{appB}
In this section we explicity solve the random walk particle path integrals and show that they match with the field theory computations as they should. We start with bosonic strings. The thermal scalar action in Rindler space is governed by the particle action\footnote{Substitute $t \to t/(2\pi\alpha')$ in expression (\ref{act}).}
\begin{equation}
\label{pppi}
S = \frac{1}{2}\int_{0}^{T}dt\left[\dot{\rho}^2+\frac{\beta^2\rho^2}{4\pi^2\alpha'^3}-\frac{1}{4\rho^2}-\frac{4}{\alpha'}- 2\frac{\beta^2}{4\pi^2\alpha'^2}\right].
\end{equation}
This particle path integral lives on a half-line in a harmonic oscillator and in a $1/\rho^2$ potential. This model can be exactly solved \cite{Khandekar:1975}. The non-trivial part of the particle action is given by
\begin{equation}
S = \int_{0}^{T}dt\left(\frac{1}{2}m\dot{\rho}^2 + \frac{1}{2}m\omega^2\rho^2 + \frac{g}{\rho^2}\right).
\end{equation}
The resulting heat kernel is given by
\begin{equation}
K(\rho,T|\rho,0) = \rho\left[\frac{m\omega}{\sinh(\omega T)}\right]\exp\left(-m\omega \rho^2\coth(\omega T)\right)I_{\kappa}\left(\frac{\rho^2m\omega}{\sinh(\omega T)}\right)
\end{equation}
where
\begin{equation}
m=1,\quad \omega^2 = \frac{\beta^2}{4\pi^2\alpha'^3},\quad g= -\frac{1}{8}, \quad \kappa^2 = 2mg + 1/4 = 0.
\end{equation}
Since $g < 0$ the inverse potential is attractive. Note that since $\kappa = 0$, we have precisely saturated the stability limit of this problem: any stronger attraction would result in the `fall to the center' as discribed in \cite{Khandekar:1975}.
This propagator (heat kernel) was derived in \cite{Khandekar:1975} using an explicit path integration. It was noted there that due to the singular perturbation $\propto \frac{1}{\rho^2}$, there is no `contact' between the states living at $\rho>0$ and those at $\rho<0$. The integration range of the intermediate integrations in the path integral is from $\rho=0$ to $\rho \to \infty$. So this explicitly demonstrates that our space is restricted to a half-line. If we neglect the singular perturbation, the origin $\rho=0$ would be regular, and we would need to think about which boundary conditions to impose at the origin. In our case here, the nature of the potential solves this problem. Notice that this singular perturbation term is precisely the correction term $K(x)$ that we discussed previously, so discarding it would fundamentally alter the problem.\\
One easily checks that the integrals converge for $\rho \to 0$ and for $\rho \to \infty$.\footnote{The modified Bessel function has the following behavior
\begin{align}
I_{\alpha}(x) &= \frac{1}{\Gamma(\alpha+1)}\left(\frac{x}{2}\right)^{\alpha},\quad x \approx 0, \\
I_{\alpha}(x) &= \frac{e^{x}}{\sqrt{2\pi x}},\quad x \gg 1.
\end{align}} The $T \to 0$ limit diverges, corresponding to the field theory UV divergence.
The integral over $\rho$ can be done and results in\footnote{$\int_{0}^{+\infty}dx \exp\left(-\omega x \coth(\omega T)\right)I_{A}\left[\omega x / \sinh(\omega T)\right] = \frac{1}{\omega}\exp(-A\omega T)$.} 
\begin{equation}
\int_{0}^{+\infty}d\rho K(\rho,T|\rho,0) = \frac{1}{2\omega}\frac{\omega}{\sinh(\omega T)}.
\end{equation}
When examining the interesting $T \to \infty$ limit, the traced heat kernel becomes\footnote{Here we reincluded the $\rho$-independent contributions to the action (\ref{pppi}).}
\begin{equation}
\sim e^{-\omega T}e^{\frac{2}{\alpha'}T}e^{\frac{\beta^2}{4\pi^2\alpha'^2}T}.
\end{equation}
which indeed is the same as equation (\ref{meth1alpha}) as soon as one substitutes $T \to T/2$ in the $T$-integral in (\ref{freebosonic}).\\
Also the Schr\"odinger equation corresponding to the particle path integral (\ref{pppi}) can be directly solved \cite{Khandekar:1975} and yields the following eigenfunctions and eigenvalues:
\begin{equation}
\phi_{n}(\rho) \propto \rho^{\kappa + 1/2}L^{(\kappa)}_{n}(m\omega \rho^2)e^{-\frac{m\omega}{2}\rho^2}, \quad \lambda_{n} = (2n+1+\kappa)\omega
\end{equation}
where $n=0,1,2,\hdots$ We observe that these eigenfunctions differ from those in equation (\ref{eigenf}) by a factor of $G_{00}^{1/4} \propto \sqrt{\rho}$ which is a consequence of the different inner product on the Hilbert space in extracting the $\sqrt{G_{00}}$ from the measure in the field theory action as we discussed in \cite{Mertens:2013pza}.\\
For type II superstrings, the changes are trivial: one only needs to substitute $\frac{4}{\alpha'} \to \frac{2}{\alpha'}$ for the covering space mass term in the particle path integral.\\
For heterotic strings, we can also compute the particle path integral. Starting with the particle path integral (\ref{pppi}), we again first perform the substitution $t \to t/(2\pi\alpha')$.
The extra discrete momentum term corresponds to a shift in the parameter $g$ of the particle path integral model
\begin{equation}
g \to g + \frac{\pi^2\alpha'}{2\beta^2}
\end{equation}
and this leads to
\begin{equation}
\kappa^2 = \frac{\pi^2\alpha'}{\beta^2}.
\end{equation}
One readily finds that the resulting traced heat kernel and eigenvectors and eigenvalues of the associated Schr\"odinger operator are consistent with those found in section \ref{heterotic}.

\section{*Thermal scalar dominance}
\label{dom}
We will clarify in what sense the thermal scalar dominates the free energy in the canonical ensemble. The main question one should ask is whether the target space is compact or not. This distinction can be understood on quite general terms from a random walk perspective \cite{Barbon:2004dd} (see also \cite{Deo:1991mp} for more extensive discussions). In this section we illustrate these claims with the simple flat space example and we explicitly demonstrate the significance of the compactness of the target space. For simplicity we only consider bosonic strings.\\
All divergences in the free energy expression come from the $\tau_2 \to \infty$ limit in the fundamental modular domain.
A flat non-compact dimension gives a factor
\begin{equation}
Z_X(\tau) = \frac{\left|\eta\right|^{-2}}{\sqrt{4\pi^2\alpha'\tau_2}} \to \frac{e^{\pi\tau_2/6}}{\sqrt{4\pi^2\alpha'\tau_2}},
\end{equation}
while a flat compact dimension gives
\begin{equation}
2\pi RZ_X(\tau) \sum_{n,m}\exp\left(-\pi R^2\frac{\left|m-n\tau\right|^2}{\alpha'\tau_2}\right) \to e^{\pi\tau_2/6},
\end{equation}
where in both cases we have taken the $\tau_2 \to \infty$ limit.
For the temperature-dependent part we have dropped the $n=0$ contribution.
If all dimensions are non-compact, the free energy in flat space can be written as
\begin{equation}
F = -V\int_{F}\frac{d\tau d\bar{\tau}}{2\tau_2}\frac{1}{(4\pi^2\alpha'\tau_2)^{25/2}}\left|\eta(\tau)\right|^{-46}\frac{1}{(4\pi^2\alpha'\tau_2)^{1/2}}\left|\eta(\tau)\right|^{-2}\sum_{n,m}\exp\left(-\beta^2\frac{\left|m-n\tau\right|^2}{4\pi\alpha'\tau_2}\right)
\end{equation}
and yields in the critical limit
\begin{equation}
F \to -\frac{V}{\beta}\int_{F}\frac{d\tau d\bar{\tau}}{\tau_2}\frac{1}{(4\pi^2\alpha'\tau_2)^{25/2}}e^{4\pi\tau_2}e^{-\frac{\beta^2\left|\tau\right|^2}{4\pi\alpha'\tau_2}}.
\end{equation}
We see that this expression converges at $\beta = \beta_H$ due to the suppressing $\tau_2$ factors from the non-compact dimensions.
If on the other hand we choose only compact dimensions, we get
\begin{equation}
\label{comp}
F = -V\int_{F}\frac{d\tau d\bar{\tau}}{2\tau_2}\frac{1}{(4\pi^2\alpha'\tau_2)^{13}}\left|\eta(\tau)\right|^{-48}\prod_{i}\sum_{n_i,m_i}\exp\left(-\pi R_i^2\frac{\left|m_i-n_i\tau\right|^2}{\alpha'\tau_2}\right),
\end{equation}
and this gives
\begin{equation}
F \to -\frac{1}{\beta }\int_{F}\frac{d\tau d\bar{\tau}}{\tau_2}e^{4\pi\tau_2}e^{-\frac{\beta^2\left|\tau\right|^2}{4\pi\alpha'\tau_2}}.
\end{equation}
For $\tau_2 \to \infty$, this diverges logarithmically at $\beta = \beta_H$. Choosing at least one non-compact dimension gives convergence. We also remark that the two limits $\tau_2 \to \infty$ and $R_{i} \to \infty$ do not commute. \\
Note that the `normal' massless non-winding strings contribute to the free energy (\ref{comp}) as
\begin{equation}
F \to -\frac{1}{\beta }\int_{F}\frac{d\tau d\bar{\tau}}{2\tau_2}.
\end{equation}
This also diverges logarithmically, but gives a temperature-independent contribution to $\beta F$ and we are not interested in this. \\
So for fully compact spaces, when the temperature is close to the Hagedorn temperature (but still smaller), the free energy contribution from the thermal scalar dominates the full free energy as
\begin{equation}
\beta F \propto \ln(\beta-\beta_H) + \text{infinite but independent of $\beta$} + \text{finite}.
\end{equation}
More generally, for fully compact spaces (not necessarily flat) we expect a behavior $\beta F \propto \ln(\beta - \beta_H)$ that diverges at $\beta = \beta_{H}$. Noncompact spaces do not yield such a divergence.\\
If at least one dimension is non-compact the free energy does not diverge and the thermal scalar does not dominate the free energy (since it remains finite): it gives the leading non-analytic behavior of the thermodynamic quantities. If on the other hand, all dimensions are compact the thermal scalar gives an infinitely large contribution. \\
In order to circumvent the Jeans instability, which is always present for sufficiently large thermal systems if we include gravity, we should consider only compact dimensions and the thermal scalar really represents the dominant contribution.

\section{*Cigar orbifold partition function}
\label{orbifoldpart}
In this section we compute the partition function of $\mathbb{Z}_N$ orbifolds of the $SL(2,\mathbb{R})/U(1)$ model. We follow the computation of \cite{Hanany:2002ev} and indicate where differences occur. In \cite{Son:2001qm} the orbifolds of $AdS_3$ were considered which are closely related to the ones we study in this section. We are interested in these cigar orbifolds since we want to make sure that indeed orbifolding corresponds to including fractional winding numbers with the \emph{same} unitarity bounds as the ones obtained for the unorbifolded case. Also, we want to investigate whether some extra states occur like those in the (Lorentzian signature) $AdS_3$ orbifold model found in \cite{Son:2001qm}. For discussions concerning the superstring case, we refer to \cite{Eguchi:2010cb}. \\
The strategy is to write down an expression for the partition function by explicitly integrating the string path integral. Then this expression needs to be rewritten in terms of the $\widehat{SL(2,\mathbb{R})}$ characters such that it has the form 
\begin{equation}
Z(\tau) = \sum_i N_{i\tilde{i}}\chi_{i}(\tau) \chi_{\tilde{i}}^{*}(\tau)
\end{equation}
and we can identify the different states that occur in the string spectrum. Following \cite{Hanany:2002ev}, we start with a gauged WZW model with coordinates $\theta$, $\tilde{\theta}$ and $r$, where $\theta = \frac{1}{2}(\theta_L - \theta_R)$ and $\tilde{\theta} = \frac{1}{2}(\theta_L + \theta_R)$.
The gauge field degrees of freedom are translated into scalars $\rho$ and $\tilde{\rho}$ and we define two new coordinates
$\kappa = \theta + \rho$ and $\tilde{\kappa} = \tilde{\theta} - \tilde{\rho}$.
Next we do a coordinate transformation
\begin{align}
v &= \sinh(r/2) e^{i\kappa}, \\
\bar{v} &= \sinh(r/2) e^{-i\kappa}, \\
\phi &= i\tilde{\kappa} - \log\cosh(r/2).
\end{align}
The $\theta$ coordinate is identified with period $2\pi$ and represents the angular coordinate on the cigar. This periodicity is passed over to $\kappa$. If we want to consider $\mathbb{Z}_N$ orbifolds of the cigar, we should change the periodicity to $2\pi/N$. This means that we consider the sectors
\begin{equation}
v(z+2\pi) = H(v)(z)e^{i2\pi a/N}, \quad v(z+2\pi \tau) = H(v)(z)e^{i2\pi b/N},
\end{equation}
where $H(v)$ denotes the `normal' effect of the gauge holonomies on $v$ and $a,b = 0...N-1$ label the different sectors. Extracting the non-periodic parts from $v$, we obtain
\small
\begin{align}
v(z) &= \hat{v}\exp\left[-\frac{1}{4\tau_2}((u_1 \bar{\tau}-u_2)z - (u_1 \tau -u_2) \bar{z}) - \frac{1}{2\tau_2}\left(\left(\frac{a}{N}\bar{\tau}-\frac{b}{N}\right)z - \left(\frac{a}{N} \tau - \frac{b}{N}\right) \bar{z}\right)\right], \\
\bar{v}(z) &= \hat{\bar{v}}\exp\left[\frac{1}{4\tau_2}((u_1 \bar{\tau}-u_2)z - (u_1 \tau -u_2) \bar{z}) + \frac{1}{2\tau_2}\left(\left(\frac{a}{N}\bar{\tau}-\frac{b}{N}\right)z - \left(\frac{a}{N} \tau - \frac{b}{N}\right) \bar{z}\right)\right], \\
\rho(z) &= \hat{\rho} + \frac{1}{4\tau_2}((u_1 \bar{\tau}-u_2)z + (u_1 \tau -u_2) \bar{z}), \\
\phi(z) &= \hat{\phi} + \frac{1}{4\tau_2}((u_1 \bar{\tau}-u_2)z + (u_1 \tau -u_2) \bar{z}).
\end{align}
\normalsize
where $\hat{v}, \hat{\bar{v}}, \hat{\phi}, \hat{\rho}$ denote periodic fields on the torus.\\
When including all the contributions to the path integral \cite{Hanany:2002ev}, we arrive at
\begin{align}
Z &= \frac{1}{N}\sum_{a,b=0}^{N-1}2\sqrt{k(k-2)}\int_{F}\frac{d\tau d\bar{\tau}}{\tau_2} \int_{-\infty}^{+\infty}du_1du_2 \nonumber \\
&\sum_i q^{h_i}\bar{q}^{\bar{h}_i}e^{4\pi\tau_2(1-\frac{1}{4(k-2)}) -\frac{k\pi}{\tau_2}\left|u_1\tau -u_2\right|^2+2\pi\tau_2\tilde{u}_1^2} \nonumber \\
&\frac{1}{\left|\sin(\pi(\tilde{u}_1\tau - \tilde{u}_2))\right|^2}\left|\prod_{r=1}^{+\infty}\frac{(1-e^{2\pi i r \tau})^2}{(1-e^{2\pi i r \tau - 2\pi i (\tilde{u}_1\tau -\tilde{u}_2)})(1-e^{2\pi i r \tau + 2\pi i (\tilde{u}_1\tau -\tilde{u}_2)})}\right|^2,
\end{align}
where we denoted $\tilde{u}_1 = u_1 + \frac{a}{N}$ and $\tilde{u}_2 = u_2 + \frac{b}{N}$.
Firstly we shift the integration variables as $u_1 \to u_1 - \frac{a}{N}$, $u_2 \to u_2 - \frac{b}{N}$.
Now we split the holonomy integrals in an integer and fractional part
\begin{equation}
\int_{-\infty}^{+\infty}du_i \to \int_{0}^{1}ds_i \sum_{n_i=-\infty}^{+\infty}
\end{equation}
and we combine the summation over $n_i$ and $a$ (or $b$) in a single sum:
\begin{equation}
\sum_{a=0}^{N-1} \sum_{n_1=-\infty}^{+\infty} \to \sum_{w/N, w=-\infty}^{+\infty}, \quad \sum_{b=0}^{N-1} \sum_{n_2=-\infty}^{+\infty} \to \sum_{m/N, m=-\infty}^{+\infty}.
\end{equation}
In all, we arrive at
\begin{align}
Z &= \frac{1}{N}2\sqrt{k(k-2)}\int_{F}\frac{d\tau d\bar{\tau}}{\tau_2} \int_{0}^{1}ds_1ds_2 \nonumber \\
&\sum_{m,w=-\infty}^{+\infty}\sum_i q^{h_i}\bar{q}^{\bar{h}_i}e^{4\pi\tau_2(1-\frac{1}{4(k-2)}) -\frac{k\pi}{\tau_2}\left|(s_1 + \frac{w}{N})\tau -(s_2 + \frac{m}{N})\right|^2+2\pi\tau_2s_1^2} \nonumber \\
&\frac{1}{\left|\sin(\pi(s_1\tau - s_2))\right|^2}\left|\prod_{r=1}^{+\infty}\frac{(1-e^{2\pi i r \tau})^2}{(1-e^{2\pi i r \tau - 2\pi i (s_1\tau -s_2)})(1-e^{2\pi i r \tau + 2\pi i (s_1\tau -s_2)})}\right|^2.
\end{align}
The next step is the Poisson resummation to extract the discrete momentum quantum number:
\begin{equation}
\sum_{m=-\infty}^{+\infty}e^{-\frac{k\pi}{\tau_2}\left[\frac{m^2}{N^2}-2\frac{m}{N}\left(\left(s_1 + \frac{w}{N}\right)\tau_1 -s_2\right)\right]} = N \sqrt{\tau_2}{k}\sum_{n=-\infty}^{+\infty}e^{-\frac{\pi\tau_2}{k}\left[Nn + \frac{ik}{\tau_2}\left(\left(s_1+\frac{w}{N}\right)\tau_1 - s_2\right)\right]^2}.
\end{equation}
In all, the net effect in comparison to the $N=1$ result is simply
\begin{equation}
n \to N n , \quad w \to w/N.
\end{equation}
This leads to the constraints
\begin{align}
&q-\bar{q} = Nn, \\
&q + \bar{q} + 2j = -k\frac{w}{N}.
\end{align}
The first constraint leads to the conclusion that only discrete momenta $\in N \mathbb{N}$ are allowed. The second constraint combined with a contour-shift argument as given in \cite{Hanany:2002ev}, gives the same unitarity constraints for all sectors as the $N=1$ case:
\begin{equation}
\frac{1}{2} < j < \frac{k-1}{2}.
\end{equation}
Changing the conventions of the $SL(2,\mathbb{R})$ quantum numbers to those we used in section \ref{unitarity}, this condition is the same as (\ref{unibos}).
The remainder of the analysis is the same as the $N=1$ case.
We conclude that the only effect of orbifolding the cigar CFT is including all fractional winding numbers and constraining the discrete momentum. No extra sectors or special cases occur in contrast to the Lorentzian $AdS_3/\mathbb{Z}_N$ case \cite{Son:2001qm}.

\section{*Quantum numbers for the bosonic Rindler string}
\label{boswind}
In this section we briefly describe the modifications for the bosonic string. For an early treatment of the $SL(2,\mathbb{R})/U(1)$ spectrum, we refer to \cite{Jatkar:1992np}. Here we focus again on the transition from the cigar to flat Euclidean Rindler space. The eigenvalue equation is given by:\footnote{We set $\alpha'=2$ in this section.}
\begin{align}
&-\frac{\partial_\rho\left(\sinh\left(\sqrt{2/(k-2)}\rho\right)\partial_{\rho}T(\rho)\right)}{\sinh\left(\sqrt{2/(k-2)}\rho\right)} \nonumber \\
&\quad \quad \quad  + \left(-2 + w^2\frac{k^2}{2(k-2)}\left(\tanh^2\left(\rho/\sqrt{2(k-2)}\right)-\frac{2}{k}\right)\right)T(\rho)= \lambda T(\rho).
\end{align}
The solution that does not blow up as $\rho \to \infty$ is given by
\begin{equation}
T(\rho) \propto \frac{1}{\cosh\left(\frac{\rho}{L}\right)^{1+\sqrt{\omega}}}\,\, {\mbox{$_2$F$_1$}\left(\frac{\sqrt{\omega}+1+kw}{2},\frac{\sqrt{\omega}+1-kw}{2};\,\sqrt{\omega}+1;\,\frac{1}{\cosh\left(\frac{\rho}{L}\right)^{2}}\right)}.
\end{equation}
where $L = \sqrt{2(k-2)}$ and $\omega = 1-4(k-2)-2(k-2)\lambda +k(k-2)w^2$.
The bound states are again determined when the hypergeometric function reduces to a polynomial. This entails
\begin{equation}
\sqrt{\omega} = kw -2l+1.
\end{equation}
This is again the same condition as the identification with the asymptotic linear dilaton primaries. As an example, the lowest state has $l=1$ and
\begin{equation}
T(\rho) \propto \frac{1}{\cosh\left(\frac{\rho}{\sqrt{2(k-2)}}\right)^{kw}}.
\end{equation}
For large $k$ this gives us again the behavior $e^{-\rho^2/4}$. The eigenvalues are given by
\begin{equation}
\lambda = \frac{-2l(l-1) +2wlk -kw -kw^2 -2k + 4}{k-2},
\end{equation}
whose large $k$ limit gives (upon setting $l=n+1$)
\begin{align}
\lambda &\approx 2wl -w -w^2 -2 \nonumber \\
&= 2wn +w -w^2 -2,
\end{align}
which is indeed the correct result (\ref{bosonicspectrum}). We conclude that also for the bosonic string, $l=n+1$, where $n$ denotes again the quantum number labeling Rindler eigenmodes as in section \ref{bosonic}.

\section{*Type II Euclidean Rindler spectrum}
\label{spectr}
In this section we discuss the discrete momentum modes and mixed modes (containing both momentum and winding) for type II superstrings in Euclidean Rindler space. We then collect the results in the resulting string spectrum. Our main goal is to identify the cigar quantum numbers with the Rindler quantum numbers. We set $\alpha'=2$ again to lighten the notation.
\subsection*{Discrete momentum states}
The discrete momentum states have wavefunctions obeying the following eigenvalue equation
\begin{equation}
-\frac{\partial_\rho\left(\sinh\left(\sqrt{2/k}\rho\right)\partial_{\rho}T(\rho)\right)}{\sinh\left(\sqrt{2/k}\rho\right)} + \left(-1 + n^2\frac{1}{2k}\coth^2\left(\rho/\sqrt{2k}\right)\right)T(\rho)= \lambda T(\rho)
\end{equation}
where $n$ labels the discrete momentum. The resulting eigenfunctions are almost the same as the winding eigenfunctions (\ref{expl}) except for the replacements $\omega = 1- 2k -2k\lambda + n^2$ and $\cosh \to \sinh$. This is crucial since $\frac{1}{\sinh}$ blows up when its argument goes to zero. This immediately implies that there are no discrete states. The continuous states are determined by a critical eigenvalue
\begin{equation}
\lambda^* = \frac{n^2}{2k} + \frac{1}{2k} -1.
\end{equation}
Taking $k\to\infty$ implies $ \lambda^* \to \frac{n^2}{2k} - 1$ and for $\lambda > \lambda^*$ one finds the continuum. The eigenfunctions one finds in the $k\to \infty$ limit are given by
\begin{equation}
\psi(\rho) \propto J_n\left(\sqrt{1+\lambda}\rho\right) 
\end{equation}
and $1+\lambda$ is indeed positive for the continuous states. The unitarity bounds are not present for the continuous states and all $n \in \mathbb{Z}$ are allowed. Note that even though $J_n$ is damped for large $\rho$, the physical probability density equals $\sqrt{\rho}J_n$ and this does not damp. One can rewrite the wavefunction as
\begin{equation}
\psi \propto J_n\left(\sqrt{t+\frac{n^2}{2k}}\rho\right) 
\end{equation}
with $t \in \mathbb{R}^{+}$. Using the asymptotic behavior, one can identify 
\begin{equation}
t = \frac{2s^2}{k}.
\end{equation} 
In the limit $k \to \infty$, we see that $n$ represents the order of the Bessel function. We see that we should consider only $n < \mathcal{O}(k)$ to have a finite order.\footnote{States that have larger $n$ sense the asymptotic linear dilaton geometry. These states are important for the cigar, but are scaled out for the flat limit we focus on here.} This implies the wavefunctions simplify to
\begin{equation}
\psi \propto J_n\left(\sqrt{t}\rho\right) 
\end{equation}
where still $t = \frac{2s^2}{k}$. This implies $s^2$ \emph{can} be of order $k$. These states represent propagating states. One can also see these conclusions in the cigar spectrum:
\begin{equation}
-\frac{\alpha'M^2}{4} + \frac{n^2}{4k} + \frac{s^2+1/4}{k} = 1/2.
\end{equation}
The second term disappears in the large $k$ limit but the third term remains. We find
\begin{equation}
M^2 = \frac{2}{\alpha'}\left(\frac{2s^2}{k} - 1\right) = \frac{2}{\alpha'}\left(t - 1\right),
\end{equation}
exactly analogous to the expressions we found before.\\
We have considered here only the CFT primaries and no oscillators are present here. In fact, the GSO projection does not retain any primaries that have $w=0$, so in fact we should add $1/2$ to $h$ (one oscillator). Secondary states have the same wavefunctions as their primaries but shifted eigenvalues: this will be discussed further in a few pages. This shifts the weight to $h=t/2+1/2$ which is non-tachyonic. In fact, arguments on the closely related $SL(2,\mathbb{R})$ theory in subsection \ref{dos} will show that the integral over $t$ cannot correct the criterion for relevant or irrelevant states and the singly excited state will hence be massless (marginal). Despite being massless, the eigenvalue is temperature-independent, implying that these states are not immediately relevant for thermodynamics (as the massless contribution discussed in section \ref{dom}), just like the $n=w=0$ states (with one oscillator activated).

\subsection*{Winding and momentum states}
If we relax the condition $L_0^{Rindler} = \bar{L}_0^{Rindler}$, we can find other states. Since we are interested in the one-loop path integral, string states can be off-shell and hence this condition need not be applied. We only require the less strict condition $L_0 - \bar{L}_0 \in \mathbb{Z}$ to preserve modular invariance.\footnote{However, the dominant contribution to the free energy involves also the $\tau_1$ integral. For large $\tau_2$, this integral effectively projects onto states satisfying $L_0 = \bar{L}_0$. Nevertheless, even with this restriction on states, one only requires the total Virasoro operators to satify this property, and a mismatch in the Rindler sector could be compensated by another sector.} The field theory equation of motion is found by writing the Virasoro zero mode as
\begin{align}
L_{0} &= - \frac{1}{k}\left[\partial_r^2 + \coth (r) \partial_r + \frac{1}{4}\coth^2\left(\frac{r}{2}\right)\partial_{\theta}^2 + \frac{1}{4}\tanh^2\left(\frac{r}{2}\right)\partial_{\tilde{\theta}}^2 + \frac{1}{2} \partial_\theta \partial_{\tilde{\theta}}\right], \\
\bar{L}_{0} &= - \frac{1}{k}\left[\partial_r^2 + \coth (r) \partial_r + \frac{1}{4}\coth^2\left(\frac{r}{2}\right)\partial_{\theta}^2 + \frac{1}{4}\tanh^2\left(\frac{r}{2}\right)\partial_{\tilde{\theta}}^2 - \frac{1}{2} \partial_\theta \partial_{\tilde{\theta}}\right].
\end{align}
The third term represents the discrete momentum, the fourth term the winding and the final term mixes these two contributions. If we use a variant of equation (\ref{metricL0}), we get the following effective metric
\begin{equation}
ds^2 = \frac{\alpha' k}{4}\left[dr^2 + \frac{4}{\coth^2\left(\frac{r}{2}\right)}d\theta^2 + \frac{4}{\tanh^2\left(\frac{r}{2}\right)}d\tilde{\theta}^2\right].
\end{equation}
In this metric, coordinates conjugate to the momentum and winding are both present simultaneously. Such a description is very much in the spirit of double field theory \cite{Aldazabal:2013sca}.
The resulting eigenvalue equation one finds is
\begin{align}
&-\frac{\partial_\rho\left(\sinh\left(\sqrt{2/k}\rho\right)\partial_{\rho}T(\rho)\right)}{\sinh\left(\sqrt{2/k}\rho\right)} \nonumber \\
&\quad\quad\quad+ \left[-1 + n^2\frac{1}{2k}\coth^2\left(\rho/\sqrt{2k}\right) + w^2\frac{k}{2}\tanh^2\left(\rho/\sqrt{2k}\right) \right]T(\rho)= \lambda T(\rho),
\end{align}
whose well-behaved solutions are products of $\cosh$, $\sinh$ and hypergeometric functions $\mbox{$_2$F$_1$}$. Again requiring the hypergeometric functions to reduce to polynomials identifies the discrete states by the condition\footnote{We focus on $n \geq 0$.}
\begin{equation}
\sqrt{1-2k-2k\lambda+k^2w^2+n^2} = 1 - kw + n + 2q
\end{equation}
where $q=0,1,2,...$. For instance, when $q=0$, the eigenfunction is given by
\begin{equation}
T(\rho) \propto \frac{\sinh(\rho/\sqrt{2k})^n}{\cosh(\rho/\sqrt{2k})^{kw}},
\end{equation}
whereas for $q=1$ we find
\begin{equation}
T(\rho) \propto \frac{\sinh(\rho/\sqrt{2k})^{n+2}}{\cosh(\rho/\sqrt{2k})^{kw}}\left(1-\coth(\rho/\sqrt{2k})^2\frac{(n+1)}{kw-1}\right).
\end{equation}
The asymptotic behavior of these functions is given by
\begin{equation}
\psi \propto \exp\left(-\sqrt{\frac{2}{k}}\left(\frac{kw}{2} - \frac{n}{2} - q\right)\rho\right),
\end{equation}
and this identifies $q=l-1$. Taking $k$ large\footnote{Either in the differential equation or in the cigar eigenfunctions.}, one finds the corresponding Euclidean Rindler states with
\begin{equation}
T(\rho) \propto \frac{\text{WhittakerM}\left(\frac{n}{2}+\frac{1}{2} + q, \frac{n}{2},w\frac{\rho^2}{2}\right)}{\rho} \propto e^{-w\rho^2/4}\rho^{n+1}L_{q}^{(n)}(w\rho^2/2)
\end{equation}
where $q=0,1,2,...$ and the generalized Laguerre polynomials appear again. These states are localized close to the origin. The eigenvalues are given by
\begin{equation}
\label{eigspectr}
\lambda = wn + 2qw + w -1.
\end{equation}
These states indeed coincide with the large $k$ string spectrum. We should be a little careful in this case. The on-shell condition is
\begin{equation}
L_0 + \bar{L}_0 = 1,
\end{equation}
which reduces to
\begin{equation}
-\frac{\alpha'M^2}{2} + \frac{\left(\frac{kw}{2}+\frac{n}{2}\right)^2}{k} + \frac{\left(\frac{kw}{2}-\frac{n}{2}\right)^2}{k} - 2\frac{\left(\frac{kw}{2}-\frac{n}{2}-l\right)\left(\frac{kw}{2}-\frac{n}{2}-l+1\right)}{k} = 1.
\end{equation}
Taking $k$ large yields
\begin{equation}
-\frac{\alpha'M^2}{2} + \frac{kw^2}{2}  - \frac{kw^2}{2} + wn + 2wl - w = 1
\end{equation}
or
\begin{equation}
M^2 = \frac{2}{\alpha'}\left(wn + 2wl - w - 1\right).
\end{equation}
Setting $l=q+1$, this gives the same spectrum as in (\ref{eigspectr}). To satisfy the unitarity constraints we should again take $w=1$. The quantum numbers are also constrained as $2l > -n+1$, which is trivial. To deal with the $n<0$ states, the reader can readily check that we should simply replace $n \to -n$, such that all formulas work for any $n$ if we would write $n \to \left|n\right|$. \\
Finally the case $w=-1$ has $j = M -l$ with $M = \left|-\frac{k}{2} + \frac{\left|n\right|}{2}\right| = \frac{k}{2} - \frac{\left|n\right|}{2}$.\footnote{Actually, $M$ can be written explicitly as $M = \frac{k\left|w\right|}{2} - \frac{\left|n\right|}{2}$. This expression and the winding-dominance inequality $\left|kw\right| > \left|n\right|$ together are the same restrictions as those displayed in \cite{Dijkgraaf:1991ba}.} The Euclidean Rindler wavefunctions are the same as those of the $w=+1$ case. \\
To sum up, the primaries (in the NS-NS sector) in the type II string spectrum consist of:
\begin{itemize}
\item{Discrete states with $w = \pm 1$, $j= \frac{k}{2}-l$ and $l=1,2,3,\hdots$. 
These states are localized to the origin and have wavefunctions
\begin{equation}
\psi \propto L_{l-1}\left(\rho^2/2\right)e^{-\rho^2/4}.
\end{equation}}
\item{Continuous states with $n \in \mathbb{Z}$ and $j=-1/2 + is$ with $s \in \mathbb{R}$. The wavefunctions are of the propagating form:
\begin{equation}
\psi \propto J_n\left(\sqrt{2/k}\left|s\right|\rho\right). 
\end{equation}}
\item{Discrete states with $w=\pm1$, and $n \neq 0$. Such states are only allowed on-shell if the physical constraint $L_0 - \bar{L}_0 = 0$ can be satisfied by compensating the Rindler quantum numbers by suitable spectator quantum numbers (such as a flat toroidal dimension). 
These states are localized to the origin and are of the form
\begin{equation}
\psi \propto e^{-\rho^2/4}\rho^{\left|n\right|+1}L_{l-1}^{(\left|n\right|)}(\rho^2/2).
\end{equation}
} 
\end{itemize}
All of the $w\neq0$ states satisfy the GSO projection inherited from the cigar as discussed in \cite{Giveon:2013ica}. The pure momentum states on the other hand require one oscillator to satisfy the GSO projection.

\subsection*{Higher oscillator modes}
Higher oscillator modes are given by the same equation of motion with a shifted mass. This can be seen as follows. All conformal secondaries (these are the oscillator modes) can be obtained by applying the raising operators of the affine Lie algebra to the primary states, e.g.
\begin{equation}
J_{-p}^{b}\left|\psi\right\rangle
\end{equation}
where $p$ is a positive integer and $b$ is a group index. Since we have
\begin{equation}
\left[L_0 , J_{-p}^{b}\right] = p J_{-p}^{b},
\end{equation}
the conformal weights of such states is simply shifted by an amount of $p$ with respect to the primaries. This implies the same equation of motion with a shifted mass term. In particular, all winding modes are localized to the Rindler origin. They are however not massless anymore and so are not really relevant for low energy physics. The equation of motion for such (superstring) oscillator states is effectively given by
\begin{equation}
\left(\tilde{L}_0 + \tilde{\bar{L}}_0 + p - 1\right) \left|\phi\right\rangle = 0,
\end{equation}
where we denoted by $\tilde{L}_0 + \tilde{\bar{L}}_0$ the operator used before that can be written in terms of the Laplacian on the group manifold.

\subsection*{Scaling of thermodynamical quantities of the Euclidean Rindler states}
\begin{figure}[h]
\centering
\includegraphics[width=5cm]{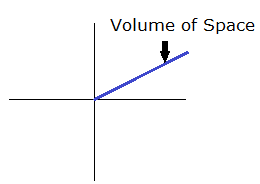}
\caption{The volume of space in Euclidean Rindler space is a half-line starting at the origin: this is the fixed time submanifold.}
\label{volumefig}
\end{figure}

We now ask the question whether the modes we found give contributions to thermodynamical quantities that scale as the volume of the space or as the area transverse to the $\rho$-direction. We previously found that the $w=\pm1$ modes yield a free energy that scales as the transverse area. What about (pure) discrete momentum modes? For this it is convenient to reconsider the heat kernel we explicitly constructed in section \ref{appB}. The particle action is given by
\begin{equation}
S = \int_{0}^{T}dt\left(\frac{1}{2}m\dot{\rho}^2 + \frac{1}{2}m\omega^2\rho^2 + \frac{g}{\rho^2}\right),
\end{equation}
with the resulting heat kernel:
\begin{equation}
K(\rho,T|\rho,0) = \rho\left[\frac{m\omega}{\sinh(\omega T)}\right]\exp\left(-m\omega \rho^2\coth(\omega T)\right)I_{\kappa}\left(\frac{\rho^2m\omega}{\sinh(\omega T)}\right)
\end{equation}
The momentum modes are characterized by a positive $g$ and $\omega = 0$. Taking $\omega \to 0$ in the heat kernel yields
\begin{equation}
K(\rho,T|\rho,0) = \rho\left[\frac{m}{T}\right]\exp\left(-\frac{m \rho^2}{T}\right)I_{\kappa}\left(\frac{\rho^2m}{T}\right)
\end{equation}
where $\kappa$ is a strictly positive real number for the momentum modes. This traced heat kernel is sensitive to the size of the space (unlike for the winding modes). This can be seen by evaluating
\begin{align}
\int_{0}^{R} d\rho K(\rho,T|\rho,0) &= \int_{0}^{R} d\rho \rho\left[\frac{m}{T}\right]\exp\left(-\frac{m \rho^2}{T}\right)I_{\kappa}\left(\frac{\rho^2m}{T}\right)
 &= \int_{0}^{\frac{R^2m}{T}} \frac{du}{2} \exp\left(-u\right)I_{\kappa}\left(u\right)
\end{align}
where we regularized the integral by a cut-off at $\rho = R$. One readily checks that the resulting integral is divergent and scales as $R$. The volume of the space is a half-line in the two-dimensional space with a fixed angle since the angular coordinate is a time coordinate. This is depicted in figure \ref{volumefig}.
So we find that the thermodynamical quantities of the discrete momentum modes scale like the volume of the space instead of only the transverse area (like the winding modes).\footnote{Note that this concerns each discrete momentum mode (labeled by $n\in\mathbb{Z}$) separately: we did \emph{not} consider the scaling behavior when summing all of these.} To summarize, the Euclidean Rindler modes give contributions to $F$ and $S$ that scale like
\begin{itemize}
\item{$w=0$. These modes scale like the volume. This can also be seen from their wavefunctions: these modes oscillate and can reach $\rho \to \infty$. Hence they should be sensitive to the entire volume as the above calculation shows.}
\item{$w=\pm1$. These modes scale like the transverse area and these modes are bound to the Euclidean Rindler origin.}
\item{$\left|w\right|>1$. These modes are absent.}
\item{All secondaries (oscillator modes) scale in the same way as the primaries from which they originated (i.e. in the same Verma module).}
\end{itemize}
When considering a cigar geometry instead of flat Euclidean Rindler space, one also has winding modes that scale like the volume (a continuum of states) \cite{Sugawara:2012ag}. The reason is that, while in Euclidean Rindler space the size of the thermal circle keeps increasing, for a cigar it asymptotes to a finite value. This causes a continuum of eigenstates to appear and these scale as the volume of space. These can be interpreted as the asymptotically flat contribution to thermodynamics: the free energy is a sum of two parts, an area-scaling part that gives the contribution of bound states, and a volume-scaling part that gives the usual contribution of a flat space continuum of states \cite{Sugawara:2012ag}.

\section{*The critical value of $k$}
\label{proof}
In this section we prove the claims made in section \ref{remarks} on what happens as $k\to3$ for the bosonic cigar CFT (or $k\to1$ for type II superstrings). \\
Let us take the cigar metric and dilaton (for bosonic strings) as \cite{Dijkgraaf:1991ba}\cite{Giveon:2013ica}\cite{Mertens:2013zya}
\begin{align}
ds^2 &= \frac{\alpha'}{4}(k-2)\left[dr^2 + \frac{4}{\coth^2\left(\frac{r}{2}\right) - \frac{2}{k}}d\theta^2\right],\\
\Phi &= -\frac{1}{2}\ln\left(\frac{\sinh(r)}{2}\sqrt{\coth^2\left(\frac{r}{2}\right) - \frac{2}{k}}\right).
\end{align}
We rescale the coordinates as $\rho = \frac{\sqrt{\alpha'(k-2)}}{2}r$ and $\theta_{new} = \sqrt{\alpha'(k-2)}\theta_{old}$, which gives
\begin{align}
ds^2 &= \left[dr^2 + \frac{1}{\coth^2\left(\frac{\rho}{\sqrt{\alpha'(k-2)}}\right) - \frac{2}{k}}d\theta^2\right],\\
\Phi &= -\frac{1}{2}\ln\left(\frac{\sinh\left(\frac{2}{\sqrt{\alpha'(k-2)}}\rho\right)}{2}\sqrt{\coth^2\left(\frac{\rho}{\sqrt{\alpha'(k-2)}}\right) - \frac{2}{k}}\right).
\end{align}
Taking the coordinates very large $\rho \gg \sqrt{\alpha'(k-2)}$, we arrive at a linear dilaton background (as expected), but with an unconventional coordinate scaling:
\begin{align}
ds^2 &= \left[d\rho^2 + \frac{1}{1 - \frac{2}{k}}d\theta^2\right],\\
\Phi &= \Phi_0 -\frac{1}{\sqrt{\alpha'(k-2)}}\rho.
\end{align}
We see now that a further rescaling as $\theta_{new} = \sqrt{\frac{k}{k-2}}\theta_{old}$ yields the standard linear dilaton background.
The Hawking temperature in the new $\theta$ coordinate is $\beta_{\text{Hawking}} = 2\pi\sqrt{\alpha'k}$. The linear dilaton background has a Hagedorn temperature equal to
\begin{equation}
\beta_{H}^2 = 4\pi^2\alpha'\left(4-\frac{1}{k-2}\right).
\end{equation}
The value of $k$ where the asymptotic Hagedorn temperature (being the linear dilaton case) and the Hawking temperature are equal is at $k=3$. Analogously (but slightly easier) one can perform the computation for the type II superstring and one finds $k=1$ as the critical value.\\

\noindent Let us look at the cigar spectrum (written down in section \ref{unitarity}) and determine when $w=1$ states drop from the spectrum. 
We focus on $n\geq 0$ states (the other case is actually the same). For a discrete state we have that
\begin{equation}
j = \frac{n+kw}{2} - l , \quad l = 1,2,\hdots
\end{equation}
The bosonic unitarity constraints reduce to (for $w=1$)
\begin{align}
l &> \frac{n+3}{2}, \\
l &< \frac{n+k+1}{2}.
\end{align}
Clearly the value of $n$ is irrelevant for the number of allowed $l$-values and only for $k > 3$, we can have states satisfying these constraints.
Moreover, at precisely $k=3$, the constraints 
\begin{align}
l &> \frac{n+3w}{2}, \\
l &< \frac{n+3w+1}{2},
\end{align}
cannot be satisfied for \emph{any} $n$ and $w$, implying that there are no discrete states at all. For lower values of $k$ some discrete states may reappear into the spectrum. One can also see this in the thermal partition function \cite{Hanany:2002ev}, where these unitarity constraints are represented as a contour-shift that either encircles poles or not.
For the type II superstring, the first constraint is replaced by $l > \frac{n+1}{2}$ leading to $k=1$ as the critical value.\\

\noindent One can also find this critical value of $k$ by looking at the asymptotics (large $\rho$) of the wavefunctions (imposing normalizability). Setting $\alpha'=2$, the measure factor $\sqrt{G}e^{-2\Phi}$ is proportional to 
\begin{equation}
e^{\sqrt{2/(k-2)}\rho}
\end{equation}
and the states have asymptotics 
\begin{equation}
\label{stasympt}
\psi \propto e^{-\sqrt{2/(k-2)}(j+1)\rho},
\end{equation}
which leads to the same value $k=3$ (or $k=1$ for type II strings). \\

\noindent A final characterization of this value is found by looking at the continuum. Assuming that the continuous quantum number $s$ does not give a contributing $\tau_2$-dependent exponential correction (as we have seen happens in fact for the linear dilaton background discussed elsewhere \cite{Mertens:2014cia}), we find that the continuum state becomes marginal when\footnote{In principle one should integrate over the $s$-quantum number. The density of states of this model and that of the $AdS_3$ WZW model are given by the same expression, and we have shown in \cite{Mertens:2014nca} that the integration over $s$ can not yield a correction to the critical temperature. This will be discussed in chapter \ref{chwzw}.}
\begin{equation}
\frac{1}{4(k-2)} + \frac{k}{4} = 1 \Leftrightarrow k=3
\end{equation}
and analogously for the type II superstring.\\
We remark that for both $k$ larger and smaller than this critical value, this state is non-tachyonic, it can only become marginal when this critical value of $k$ is reached.\\
Note also that this state is the thermal scalar of the asymptotic linear dilaton background. Indeed, allowing $\beta$ to vary, we would write\footnote{From the cigar perspective, this off-shell generalization can be found by first considering conical orbifold models. This would cause the replacement $\frac{kw^2}{^4} \to \frac{kw^2}{4N^2}$ in the conformal weights. Then we reinterpret $\frac{1}{N}$ as $\frac{\beta}{\beta_{\text{Hawking}}}$ to arrive at the correct result.}
\begin{equation}
\frac{1}{4(k-2)} + \frac{k}{4}\frac{\beta^2}{4\pi^2k\alpha'} = 1 \,\, \Leftrightarrow \,\, \beta_H = 2\pi\sqrt{\left(4-\frac{1}{k-2}\right)\alpha'},
\end{equation}
or for type II strings:
\begin{equation}
\frac{1}{4k} + \frac{k}{4}\frac{\beta^2}{4\pi^2k\alpha'} = \frac{1}{2} \,\, \Leftrightarrow \,\, \beta_H = 2\pi\sqrt{\left(2-\frac{1}{k}\right)\alpha'},
\end{equation}
which are indeed the correct Hagedorn temperatures of the linear dilaton background. In particular for only flat extra dimensions, we can relate this to the non-critical flat spaces upon setting $Q^2 = \frac{4}{(k-2)\alpha'} = 4\frac{26-D}{6\alpha'}$,\footnote{For type II superstrings, we have instead $Q^2 = \frac{4}{k\alpha'} = \frac{10-D}{\alpha'}.$} where the parameter $Q$ is conventionally defined as $\Phi = \frac{Q}{2}X^d$ with $d$ the linear dilaton direction.

\part{Application to 3d WZW models}

\chapter{Introduction to WZW conformal field theories}
\label{chinwzw}
This chapter contains a concise introduction to the more salient features of WZW models. We will omit almost all proofs as they can be found in nearly any textbook on conformal field theory. The first section explores the construction of these models in general. The second section provides the detailed application to the non-compact $SL(2,\mathbb{R})$ group manifold as will be needed in our study of $AdS_3$ in the next chapter.

\section{General introduction}
We follow the conventions of \cite{Martinec:2001cf}\cite{Rangamani:2007fz}. The Wess-Zumino-Witten (WZW) model for a compact group $G$ is given by
\begin{equation}
S = \frac{k}{8\pi}\int d^2\sigma \text{Tr}\left(g^{-1}\partial_{\mu}g g^{-1} \partial^{\mu}g\right) + k \Gamma_{WZ},
\end{equation}
where 
\begin{equation}
\Gamma_{WZ} = \frac{i}{12\pi}\int_{M^3}\text{Tr}\left(\omega \wedge \omega \wedge \omega\right).
\end{equation}
In these formulas, $g$ is a 2d field evaluated in the group $G$ under study. Typically compact groups are studied although in the next section we will extend the formulas to the non-compact case. The number $k$ is an integer for compact groups, but can be real for non-compact groups. The Maurer-Cartan 1-form is denoted by $\omega = g^{-1}dg$. The first term of the WZW action is actually the traced square of the Maurer-Cartan 1-form and is as such the natural metric to put on a group manifold (originating from the Cartan-Killing metric on the algebra).\footnote{This holds for semi-simple groups. For non-semi-simple groups, different bilinear symmetric forms exist other than the Cartan-Killing form. The Cartan-Killing form itself is degenerate in this case and is not a valid starting point to construct a metric on the group manifold. See \cite{Nappi:1993ie}.} From this we conclude that the string model has a metric background equal to the Cartan-Killing metric. The second Wess-Zumino term is necessary to ensure conformal invariance at the quantum level and can be interpreted as a background Kalb-Ramond background for the string model. \\

\noindent The main reason why these models are used in string theory is that they provide a conformal field theory that can be naturally interpreted as a non-linear sigma model on the string worldsheet. Let us sketch how this procedure works (we will be more explicit in our construction in the next section). \\

\noindent One starts by identifying a set of conserved currents of this action on the 2d space, valued in the Lie algebra of the group $G$. In complex coordinates, these can be split into a set of holomorphic currents $J(z)$ and a set of antiholomorphic currents $\bar{J}(\bar{z})$ given by the following expressions:
\begin{equation}
J(z) \propto \partial g g^{-1}, \quad \overline{J}(\bar{z}) \propto g^{-1} \bar{\partial} g.
\end{equation} 
After expanding in an algebra basis: $J(z) = \sum_{a} J^{a}(z)t^a$, these currents can be Laurent-decomposed as 
\begin{align}
J^a(z) = \sum_{n\in\mathbb{Z}}\frac{J^a_n}{z^{n+1}}
\end{align}
The currents can be shown to satisfy a current algebra, which gets translated to the following commutator relations:
\begin{equation}
\left[J^a_n, J^b_m\right] = if_{abc}J^{c}_{n+m} + k n \delta_{ab}\delta_{n+m,0}.
\end{equation}
This is a \emph{Kac-Moody} algebra, a graded centrally-extended Lie algebra. \\
The crucial part of this algebra is that it directly leads to a Virasoro algebra, signaling a worldsheet CFT. It can be shown that the following combination
\begin{equation}
T(z)= \frac{1}{2(k + g)}\sum_{a}(J^{a}J^{a})(z)
\end{equation}
defines a suitable worldsheet stress tensor of a CFT. This is called the \emph{Sugawara construction}. In this formula, $g$ denotes the dual Coxeter number of the algebra and use is made of the standard Killing form on a compact Lie algebra. \\
The resulting CFT has central charge
\begin{equation}
c = \frac{k \text{dim}\mathfrak{g}}{k+g}.
\end{equation}
The only thing left to do to get a viable string theory is to combine such building blocks to have $c=0$. \\
The beauty of this construction is that it entirely circumvents the beta-functionals and the $\alpha'$-expansion: one obtains a string background to all orders in $\alpha'$. The identification of the precise background is performed by comparing with the non-linear sigma model. \\
It is interesting to note that the flat boson CFT is a special case of this construction by defining the currents as
\begin{equation}
J(z) \sim \partial X,
\end{equation}
and hence forms a $\widehat{U(1)}$ Kac-Moody algebra. The modes $\alpha^{\mu}_n$ are obtained in the Laurent expansion of $\partial X^\mu$ and are the analogs of the $J^a_{n}$. \\
The construction is typically performed for compact Lie algebras (for which the Killing form is positive definite). In the next section, we will apply this construction concretely to the non-compact $SL(2,\mathbb{R})$ group. \\
One can extend the procedure to coset manifolds $G/H$ (which we needed in describing the 2d cigar CFT in previous chapters). The algebraic construction goes under the name of Goddard-Kent-Olive (GKO) construction which describes the worldsheet CFT with the difference of the $L_0$ operators (and central charges) of $G$ and $H$ figuring as the Virasoro algebra of the coset. A path integral realization of this is given by the so-called \emph{gauged WZW models}, in which a worldsheet gauge field is introduced. We will not go into detail on these matters. \\
The additional structure of the affine Lie algebra allows one to further break down the string spectrum into irreducible representations of the algebra. However, to fully obtain the string spectrum one should know which representations are present (and their degeneracy), a question that group theory alone cannot answer. 

\section{*\boldmath $SL(2,\mathbb{R})$ and $SL(2,\mathbb{C})/SU(2)$ WZW models}
\label{WZWapp}

In this section we provide some background material concerning the two WZW models relevant for the $AdS_3$ background. We also establish our conventions and provide several formulas for later reference.

\subsection{$SL(2,\mathbb{R})$ model}
In our case $g$ is a $SL(2,\mathbb{R})$ matrix. We choose the following basis of generators for the $\mathfrak{sl}(2,\mathbb{R})$ Lie algebra
\begin{equation}
\tau^{1} = \frac{i}{2}\sigma^{3}, \quad \tau^{2} = \frac{i}{2}\sigma^{1}, \quad \tau^{3} = \frac{1}{2}\sigma^{2}.
\end{equation}
where the $\sigma^{i}$ are the Pauli matrices:
\begin{equation}
\sigma^{1} = \left[ 
\begin{array}{cc}
0 & 1 \\
1 & 0  \end{array} 
\right],
\quad
\sigma^{2} = \left[ 
\begin{array}{cc}
0 & -i \\
i & 0  \end{array} 
\right],
\quad
\sigma^{3} = \left[ 
\begin{array}{cc}
1 & 0 \\
0 & -1  \end{array} 
\right].
\end{equation}
These generators satisfy the Lie algebra
\begin{equation}
\left[\tau^{a},\tau^{b}\right]=i{\epsilon^{ab}}_{c}\tau^{c}
\end{equation}
where $\epsilon^{123} = 1$ and indices are raised and lowered with the metric $\eta^{ab} = $diag$(+1,+1,-1)$.
The Cartan-Killing metric is not proportional to the unit matrix in this case:
\begin{equation}
\text{Tr}(\tau^{a}\tau^{b})= -\frac{1}{2}\eta^{ab}.
\end{equation}
It is not negative definite due to the non-compactness of the $SL(2,\mathbb{R})$ manifold. The $SL(2,\mathbb{R})$ group element can be written in general as\footnote{For convenience, we now rescale the coordinate fields by $\sqrt{\alpha'}$ to make them dimensionless.}
\begin{eqnarray}
&g = e^{i\frac{t+\phi}{2}\sigma_2}e^{\rho\sigma_3}e^{i\frac{t-\phi}{2}\sigma_2} \\
&= \left[
\begin{array}{cc}
\cos(t)\cosh(\rho)+\cos(\phi)\sinh(\rho) & \sin(t)\cosh(\rho)-\sin(\phi)\sinh(\rho) \\
-\sin(t)\cosh(\rho)-\sin(\phi)\sinh(\rho) & \cos(t)\cosh(\rho)-\cos(\phi)\sinh(\rho) \end{array} 
\right].
\end{eqnarray}
This is a parameterization of the group manifold in coordinates ($t$, $\rho$, $\phi$). Notice that the WZW model is written in a manifestly coordinate invariant way (intrinsic on the group manifold). Coordinate transformations simply correspond to choosing a different parameterization of the element $g$.

\subsubsection{String background field from WZW action}
Let us first compute the background fields by identifying with the non-linear sigma model.
With the above parametrization of $g$, we can evaluate the WZW action here explicitly. We need to read off the background metric and NS field by comparing with the standard non-linear sigma model
\begin{equation}
S = \frac{1}{4\pi\alpha'}\int d^{2}\sigma \left(\delta^{ab}G_{\mu\nu} + i \epsilon^{ab} B_{\mu\nu}\right) \partial_a X^{\mu}\partial_b X^{\nu}
\end{equation}
where $\epsilon^{12} = 1$ and a flat worldsheet metric was chosen. \\
The first term in the WZW action corresponds indeed to the background metric since 
\begin{equation}
\text{Tr}\left(g^{-1} \partial^{a} g g^{-1} \partial_a g\right) = \text{Tr}\left(g^{-1} \frac{\partial g}{\partial X^{\mu}} g^{-1} \frac{\partial g}{\partial X^{\nu}}\right) \frac{\partial X^{\mu}}{\partial \sigma_a}\frac{\partial X^{\nu}}{\partial \sigma_a}.
\end{equation}
Using the explicit parameterization of the $SL(2,\mathbb{R})$ group element given above, we can read off the metric as
\begin{equation}
ds^2 = \alpha'k \left(-\cosh^2(\rho) dt^2 + d\rho^2 + \sinh^2(\rho) d\phi^2\right).
\end{equation}
We next focus on the WZ term. First we compute
\begin{align}
\text{Tr}\left(\omega^3\right) &= \text{Tr} \left(g^{-1}\frac{\partial g}{\partial X^{\mu}}g^{-1}\frac{\partial g}{\partial X^{\nu}}g^{-1}\frac{\partial g}{\partial X^{\sigma}}\right)\frac{\partial X^{\mu}}{\partial \sigma_i}\frac{\partial X^{\nu}}{\partial \sigma_j}\frac{\partial X^{\sigma}}{\partial \sigma_k} d\sigma^{i} \wedge d\sigma^{j} \wedge d\sigma^{k} \\
&= d\left(6\sinh^2(\rho)\frac{\partial \phi}{\partial \sigma_j}\frac{\partial t}{\partial \sigma_k} d\sigma^{j} \wedge d\sigma^{k}\right)
\end{align}
from which we can read off the Wess-Zumino term. The background Kalb-Ramond field is given by
\begin{equation}
B_{t \phi} = - \alpha' k \sinh^2(\rho) \quad \text{or} \quad B = - \alpha' k \sinh^2(\rho) dt \wedge d\phi.
\end{equation}
The corresponding $H$-field is then
\begin{equation}
H = dB = - \alpha' k \sinh(2\rho) d\rho \wedge dt \wedge d\phi.
\end{equation}

\subsubsection{Currents, Ward identities and OPEs}
From now on we go to complex worldsheet coordinates ($\sigma_1$,$\sigma_2$) $\to$ ($z$,$\bar{z}$) as usual. The general WZW model is invariant under 
\begin{equation}
g(z,\bar{z}) \to g'(z,\bar{z}) = \Omega(z)g(z,\bar{z})\overline{\Omega}(\bar{z})^{-1}
\end{equation}
where in this case $\Omega$ and $\overline{\Omega}$ are two (independent) $SL(2,\mathbb{R})$ matrices. This corresponds to an isometry of the metric (and Kalb-Ramond background) since it states that $g'$ parametrized by the transformed coordinates ($t'$, $\rho'$, $\phi'$) gives the same metric as $g$ parametrized by ($t$, $\rho$, $\phi$). The isometry group is thus $SL(2,\mathbb{R}) \times SL(2,\mathbb{R})$.
Infinitesimal transformations give $g \to \omega g - g \overline{\omega}$ where $\omega(z)$ is traceless and real. The symmetry currents corresponding to these symmetries are proportional to 
\begin{equation}
J(z) \propto \partial g g^{-1}, \quad \overline{J}(\bar{z}) \propto g^{-1} \bar{\partial} g.
\end{equation}
Following \cite{DiFrancesco:1997nk}, we choose them as
\begin{equation}
J(z) = -\frac{k}{2} \partial g g^{-1}, \quad \overline{J}(\bar{z}) = \frac{k}{2} g^{-1} \bar{\partial} g.
\end{equation}
The sign of the antiholomorphic currents is chosen differently than in \cite{Martinec:2001cf} and these currents give hence an extra minus sign in the flat space $k \to \infty$ limit compared to those in \cite{Martinec:2001cf}. This symmetry entails a corresponding Ward identity for a general field $A$ given by 
\begin{equation}
\label{ward}
\delta A = -\oint_{w} \frac{dz}{2\pi i}\omega_{a}J^{a}(z)A(w,\bar{w}) + \oint_{w} \frac{d\bar{z}}{2\pi i}\overline{\omega_{a}}\overline{J^{a}}(z)A(w,\bar{w})
\end{equation}
where we have expanded the functions in the Lie algebra basis
\begin{equation}
J(z) = J_{a}\tau^{a}, \quad \overline{J}(\bar{z}) = \overline{J_{a}}\tau^{a}, \quad \omega(z) = \omega_{a}\tau^{a}, \quad \overline{\omega}(\bar{z}) = \overline{\omega_{a}}\tau^{a}
\end{equation}
and we should be careful with indices since upper and lower indices are not equal. Note that $\omega_{a}$ is an imaginary number. Multiplying the $J(z)$ expansion by $\tau^{b}$ and tracing gives
\begin{align}
\label{currents}
J^{a} = k \text{Tr}\left(\tau^{a}\partial g g^{-1}\right), \\
\label{currents2}
\overline{J}^{a} = -k \text{Tr}\left(\tau^{a} g^{-1} \bar{\partial} g\right).
\end{align}
If we now take a WZW primary field for which $\delta A = \omega(z)A - A \overline{\omega}(\overline{z})$, we can match this with the Ward identity and read off the following OPEs
\begin{equation}
J^{a}(z)A(w,\bar{w}) \sim -\frac{\tau^{a}A(w,\bar{w})}{z-w}, \quad\quad \overline{J}^{a}(\bar{z})A(w,\bar{w}) \sim \frac{A(w,\bar{w})\tau^{a}}{\bar{z}-\bar{w}},
\end{equation}
which immediately lead to the commutation relations of the zero mode of the current with the field $A$ 
\begin{equation}
\left[J^{3}_{0}, A\right] = -\tau^{3} A, \quad\quad \left[\overline{J}^{3}_{0}, A\right] = A\tau^{3}.
\end{equation}
The currents we have constructed satisfy the following Kac-Moody algebra relations:
\begin{align}
J^{a}(z)J^{b}(w) \sim \frac{k\eta^{ab}}{2(z-w)^2} + \frac{i{f^{ab}}_cJ^{c}(w)}{z-w}, \\
\overline{J}^{a}(\bar{z})\overline{J}^{b}(\bar{w}) \sim \frac{k\eta^{ab}}{2(\bar{z}-\bar{w})^2} + \frac{i{f^{ab}}_c\overline{J}^{c}(\bar{w})}{\bar{z}-\bar{w}}.
\end{align}
The reason we work with this sign-convention for the antiholomorphic currents is that in this case we can identify them directly with the $\overline{\Omega}$ transformations (and not their inverses). \\
Next, we identify the generator of spacetime time translations. This generator is defined by
\begin{equation}
\delta_{t} A = -i\delta t \left[Q_{t}, A\right].
\end{equation}
Since in the general parameterization it holds that
\begin{eqnarray}
\delta_{t} A &= \frac{i \delta t}{2} \sigma^{2} A +\frac{i \delta t}{2} A \sigma^{2} \nonumber\\
&= i \delta t \tau^{3} A + i \delta t A \tau^{3} \nonumber\\
&= -i \delta t \left[J^{3}_{0} - \overline{J}^{3}_{0}, A \right]
\end{eqnarray}
we see that $Q_{t} = J^{3}_{0} - \overline{J}^{3}_{0}$. Analogously one shows that $Q_{\phi} = J^{3}_{0} + \overline{J}^{3}_{0}$.
For later reference, we state the currents in terms of the global coordinates:
\begin{align}
J^{3} &= ik\left(\cosh(\rho)^2\partial t - \sinh(\rho)^2 \partial \phi\right)\\
J^{1} &= ik\left(\sin(\phi+t)\cosh(\rho)\sinh(\rho)\partial t -\sin(\phi+t)\cosh(\rho)\sinh(\rho)\partial \phi +\cos(t+\phi)\partial \rho\right) \nonumber \\
J^{2} &= ik\left(\cos(\phi+t)\cosh(\rho)\sinh(\rho)\partial t -\cos(\phi+t)\cosh(\rho)\sinh(\rho)\partial \phi -\sin(t+\phi)\partial \rho\right) \nonumber \\
\overline{J}^{3} &= -ik\left(\cosh(\rho)^2\bar{\partial} t + \sinh(\rho)^2 \bar{\partial} \phi\right)\\
\overline{J}^{1} &= -ik\left(\sin(t-\phi)\cosh(\rho)\sinh(\rho)\bar{\partial} t +\sin(t-\phi)\cosh(\rho)\sinh(\rho)\bar{\partial} \phi +\cos(t-\phi)\bar{\partial} \rho\right) \nonumber \\
\overline{J}^{2} &= ik\left(\cos(t-\phi)\cosh(\rho)\sinh(\rho)\bar{\partial} t +\cos(t-\phi)\cosh(\rho)\sinh(\rho)\bar{\partial} \phi -\sin(t-\phi)\bar{\partial} \rho\right). \nonumber
\end{align}

\subsection{$SL(2,\mathbb{C})/SU(2)$ model}
We describe the analytic continuation of this model and its relation to the $SL(2,\mathbb{C})/SU(2)$ model. Analytically continuing $t \to i\tau$ immediately gives
\begin{eqnarray}
&g = e^{i\frac{i\tau+\phi}{2}\sigma_2}e^{\rho\sigma_3}e^{i\frac{i\tau-\phi}{2}\sigma_2} \\
&=\left[
\begin{array}{cc}
 \cosh(\tau)\cosh(\rho)+\cos(\phi)\sinh(\rho) & i\sinh(\tau)\cosh(\rho)-\sin(\phi)\sinh(\rho) \\
 -i\sinh(\tau)\cosh(\rho)-\sin(\phi)\sinh(\rho) & \cosh(\tau)\cosh(\rho)-\cos(\phi)\sinh(\rho)\end{array} 
\right].
\end{eqnarray}
This group element obviously still has unit determinant, but is no longer real. It is however Hermitian. This identifies the continuation in global coordinates as the $SL(2,\mathbb{C})/SU(2)$ model. It has previously been noted that analytic continuation in the Poincar\'e patch time coordinate of the $AdS_3$ manifold corresponds to going from the $SL(2,\mathbb{R})$ to the $SL(2,\mathbb{C})/SU(2)$ WZW model \cite{Hemming:2002kd}. \\
This WZW model is invariant under
\begin{equation}
g(z,\bar{z}) \to g'(z,\bar{z}) = \Omega(z)g(z,\bar{z})\overline{\Omega}(\bar{z})^{-1}
\end{equation}
where $\Omega$ and $\overline{\Omega}$ are $SL(2,\mathbb{C})$ matrices such that $g'$ is a Hermitian matrix. This immediately implies $\Omega = \left(\overline{\Omega}^{\dagger}\right)^{-1}$. On an infinitesimal level, this means that $\omega(z) = - \overline{\omega}(\bar{z})^{\dagger}$. So the symmetry group is $SL(2,\mathbb{C})$. \\
In general $\delta_\tau g = -i\delta \tau \left[Q_{\tau},g\right]$. An infinitesimal Euclidean time translation corresponds in our parameterization to 
\begin{equation}
g \to g -\frac{\delta\tau}{2} \sigma_2 g - \frac{\delta\tau}{2} g \sigma_2,
\end{equation}
which we can rewrite as
\begin{equation}
g \to g + \delta\tau  \left[J^{3}_0-\overline{J}^{3}_0,g\right].
\end{equation}
This identifies the Euclidean time translation operator as
\begin{equation}
Q_{\tau} = i(J^{3}_0 - \overline{J}^{3}_0).
\end{equation}
Let us now take a closer look at the link between the $SL(2,\mathbb{R})$ and the $SL(2,\mathbb{C})/SU(2)$ models in terms of the currents. To see the link, we first take a step back and consider the $SL(2,\mathbb{C})$ WZW model. As a Lie algebra basis we choose the same three generators as for the $SL(2,\mathbb{R})$ model and allow for complex expansion parameters. The infinitesimal transformation $\omega(z)$ is a complex traceless matrix and the $\omega_{a}$ are complex numbers. The currents are still given by (\ref{currents}) and (\ref{currents}). So for arbitrary infinitesimal transformations the Ward identity still reads as in equation (\ref{ward}), but now with complex $\omega_{a}$ and with the current $J^{a}$ calculated with the Wick-rotated $g$ matrix. This last step is simply the analytic continuation in the currents directly. \\
The Euclidean model is however, not the general $SL(2,\mathbb{C})$ model but a right coset of this. So the left and right moving infinitesimal transformation are linked according to $\omega(z) = - \overline{\omega}(\bar{z})^{\dagger}$. This implies for the $\omega_{a}$
\begin{equation}
\omega_{1} = \overline{\omega}_{1}, \quad \omega_{2} = \overline{\omega}_{2}, \quad \omega_{3} = -\overline{\omega}_{3}.
\end{equation}
So in all, we double the number of effective currents by going to fully complex expansion parameters, but we then retain only half of these due to the left-right identification.

\subsection{WZW currents as differential operators}
In this subsection we discuss how to associate differential operators to the action of the Lie algebra currents.
Vertex operators of the WZW model are functions of the field $g$ which in turn is parametrized by the group manifold coordinates. The zero-mode symmetry operators have an action on functions as
\begin{equation}
J^{a}_0 (f(g)) = i\left.\frac{\partial}{\partial t} f \left(e^{it \tau^a} g \right)\right|_{t = 0}
\end{equation}
and
\begin{equation}
\overline{J}^{a}_0 (f(g)) = i\left.\frac{\partial}{\partial \bar{t}} f \left( g e^{-i\bar{t} \tau^{a}}\right)\right|_{\bar{t} = 0}.
\end{equation}
This is in fact simply the action of vector fields as differential operators on functions. We will denote the corresponding operators as $\hat{D}^a$ and $\hat{\overline{D}}^a$. Their action is defined through the infinitesimal group transformations. This operator has to satisfy
\begin{equation}
\hat{D}^{a} f(g) = \frac{df}{d g}(g) \left(\hat{D}^{a} g\right) = \frac{df}{d g}(g) \left( - \tau^{a} g\right),
\end{equation}
or 
\begin{align}
\hat{D}^{a} g &=  - \tau^{a} g, \\
\hat{\overline{D}}^{a} g &= g \tau^{a}.
\end{align}
One should compare this with the OPEs we derived earlier for the currents:
\begin{equation}
J^{a}(z)g(w,\bar{w}) \sim -\frac{\tau^{a}g(w,\bar{w})}{z-w}, \quad\quad \overline{J}^{a}(\bar{z})g(w,\bar{w}) \sim \frac{g(w,\bar{w})\tau^{a}}{\bar{z}-\bar{w}},
\end{equation}
and we conclude that indeed the normalization of the currents is precisely such that they are the algebra generators in the sense of the operator equations above. These formulas identify the differential operators as the dual vectors of the right (respectively left) invariant Maurer-Cartan 1-forms, with an extra minus sign for the right-invariant vector. \\
This suggests a first method to compute these differential operators: find the Maurer-Cartan forms and then dualize these into vector fields. We will however follow a more pedestrian path and simply compute the operators from the above conditions using some Pauli matrix algebra. From the above formula, one can see that these differential operators satisfy the zero-mode Lie-algebra since
\begin{align}
\left[\hat{D}^a, \hat{D}^b\right] g &= - \left[\tau^a, \tau^b\right] g = -i{f^{ab}}_c \tau^{c} g = i{f^{ab}}_c (\hat{D}^c g), \\
\left[\hat{\overline{D}}^a, \hat{\overline{D}}^b\right] g &= g \left[\tau^a, \tau^b\right] = i{f^{ab}}_c g \tau^{c} = i{f^{ab}}_c (\hat{\overline{D}}^c g).
\end{align}
The same algebra is satisfied when applying these operators on arbitrary functions on the group manifold.
The on-shell equation for a string state is simply $L_0 + \bar{L}_0 = 2$ and when one rewrites this using the Sugawara construction, we have the full stringy wave equation for the state. \\
Let us discuss this in full detail for the holomorphic part of the algebra. The general $SL(2,\mathbb{C})/SU(2)$ element was parametrized as 
\begin{equation}
g = e^{(-\tau+i\phi)\tau^3}e^{-2i\rho\tau^1}e^{(-\tau-i\phi)\tau^3}.
\end{equation}
This leads to
\begin{align}
\partial_\tau g &= -\tau^3 g - g\tau^3, \\
\partial_\phi g &= i\tau^3 g - i g\tau^3, \\
\partial_\rho g &= -2i e^{(-\tau+i\phi)\tau^3} \tau_1 e^{-2i\rho\tau^1} e^{(-\tau-i\phi)\tau^3}
\end{align}
Our goal now is to rewrite this in a form where all algebra generators are in front of the group element. This requires some rearranging of the Pauli-matrices using the following three lemmas:
\begin{equation}
e^{A\tau^3}\tau^1 = \left(\cosh(A) \tau^1 + i\sinh(A) \tau^2 \right) e^{A\tau^3},
\end{equation}
\begin{equation}
e^{B\tau^1}\tau^3 = \left(\cos(B) \tau^3 - i\sin(B) \tau^2 \right) e^{B\tau^1},
\end{equation}
\begin{equation}
e^{C\tau^3}\tau^2 = \left(\cosh(C) \tau^2 - i\sinh(C) \tau^1 \right) e^{C\tau^3}.
\end{equation}
Using these, we obtain
\begin{align}
\partial_\tau g &= -\tau^3 g - \cos(2i\rho)\tau^3 g - i \sin(2i\rho)\cosh(\tau - i \phi) \tau^2 g + \sin(2i\rho)\sinh(\tau - i \phi) \tau^1 g , \\
\partial_\phi g &= i\tau^3 g - i\cos(2i\rho)\tau^3 g + \sin(2i\rho)\cosh(\tau - i \phi) \tau^2 g + i\sin(2i\rho)\sinh(\tau - i \phi) \tau^1 g, \\
\partial_\rho g &= -2i\cosh(\tau - i \phi) \tau^1 g - 2 \sinh(\tau - i \phi) \tau^2 g. 
\end{align}
After solving the equalities $\hat{D}^{a} g = -\tau^{a} g$ for a general first-order differential operator, we obtain the unique solution for the differential operators:
\begin{align}
\hat{D}^3 &= -\frac{1}{2i}\partial_\phi + \frac{1}{2}\partial_\tau, \\
\hat{D}^{1} &= \frac{1}{2}\left[i \sinh(\tau-i\phi)\tanh(\rho)\partial_\tau - \sinh(\tau-i\phi)\coth(\rho)\partial_\phi - i\cosh(\tau-i\phi)\partial_\rho\right] ,\\
\hat{D}^{2} &= \frac{1}{2}\left[\cosh(\tau-i\phi)\tanh(\rho)\partial_\tau + i\cosh(\tau-i\phi)\coth(\rho)\partial_\phi - \sinh(\tau-i\phi)\partial_\rho\right],
\end{align}
and for the $+$ and $-$ operators, defined as $ \hat{D}^{\pm} = \hat{D}^{1} \pm i \hat{D}^{2}$, we obtain
\begin{align}
\hat{D}^{+} &= i\frac{e^{\tau-i\phi}}{2}\left[-\partial_\rho + i\coth(\rho) \partial_\phi + \tanh(\rho)\partial_\tau \right], \\
\hat{D}^{-} &= -i\frac{e^{-\tau+i\phi}}{2}\left[\partial_\rho + i\coth(\rho) \partial_\phi + \tanh(\rho)\partial_\tau \right].
\end{align}
One can explicity check using the above parametrization of $g$ that indeed
\begin{equation}
\hat{D}^{3} g = -\tau^3 g, \quad \hat{D}^{+} g = - (\tau^{1} + i\tau^2) g, \quad \hat{D}^{-} g = - (\tau^1 - i \tau^2 ) g.
\end{equation}
One can also check that
\begin{align}
\hat{\overline{D}}^3 &= -\frac{1}{2i}\partial_\phi - \frac{1}{2}\partial_\tau, \\
\hat{\overline{D}}^{+} &= i\frac{e^{-\tau-i\phi}}{2}\left[\partial_\rho - i\coth(\rho) \partial_\phi + \tanh(\rho)\partial_\tau \right], \\
\hat{\overline{D}}^{-} &= -i\frac{e^{\tau+i\phi}}{2}\left[-\partial_\rho - i\coth(\rho) \partial_\phi + \tanh(\rho)\partial_\tau \right],
\end{align}
satisfy
\begin{equation}
\hat{\overline{D}}^3 g = g \tau_3 ,\quad \hat{\overline{D}}^{+} g =  g (\tau_1 +i\tau_2), \quad \hat{\overline{D}}^{-} g =  g (\tau_1 - i\tau_2).
\end{equation}
We already saw above from the explicit construction that there is a unique solution to these conditions.

%
\chapter{The Thermal Spectrum in $AdS_3$ and $BTZ$}
\label{chwzw}

In this chapter we study the thermal spectrum on two WZW models in 3 spacetime dimensions based on the $SL(2,\mathbb{R})$ group manifold: $AdS_3$ and $BTZ$ black holes, based on \cite{Mertens:2014nca}. The construction requires some delicate worldsheet manipulations. On a technical level, the main computation is contained in section \ref{lengthy} which, due to its technicality, is considered supplementary to the story developed here. The reader is encouraged though to at least skim this section to get a feeling for the required computations.

\section{Introduction}
\label{pathderivv}
String theory models on WZW $AdS_3$ and BTZ have been fruitful toy models to study string dynamics on non-trivial target spaces. There has been a great deal of work on this topic: see e.g. \cite{Balog:1988jb}\cite{Petropoulos:1989fc}\cite{Hwang:1990aq}\cite{Hwang:1991ana}\cite{Bars:1995mf}\cite{Bars:1999ik}\cite{deBoer:1998pp}\cite{Giveon:1998ns}\cite{Kutasov:1999xu} for early treatments. Progress on the topic was hampered by confusion about unitarity in these models. A no-ghost theorem was proven by \cite{Evans:1998qu} but the issues with the model were only fully resolved in the work of \cite{Maldacena:2000hw}. The holographic interpretation of these theories combined with their solvability have shown to be explicit tests for the AdS/CFT correspondence. It is our interest to study string thermodynamics in these backgrounds and in particular learn about the behavior of thermodynamical quantities in geometrically non-trivial spaces. Previous studies on the thermodynamics in these spaces include \cite{Berkooz:2007fe} and \cite{Lin:2007gi}. \\
Due to the exact CFT description of Wess-Zumino-Witten models, it seems worthwhile to study the Hagedorn phenomenon and the random walk picture also in $AdS_3$ and BTZ spacetimes. The $AdS_3$ spacetime is a mild generalization of flat space, since the thermal time circle still is topologically stable for winding strings. The BTZ black hole on the other hand presents a temporal cigar geometry where strings can simply slip off. We have already analyzed similar geometries in chapter \ref{chri} where we studied the string gas in Rindler spacetime. The Hagedorn transition in $AdS$ spacetime has also been related to the confinement/deconfinement phase transition in the dual gauge theory (see for instance \cite{Sundborg:1999ue}\cite{Aharony:2003sx}).\\

\begin{figure}[h]
\centering
\includegraphics[width=10cm]{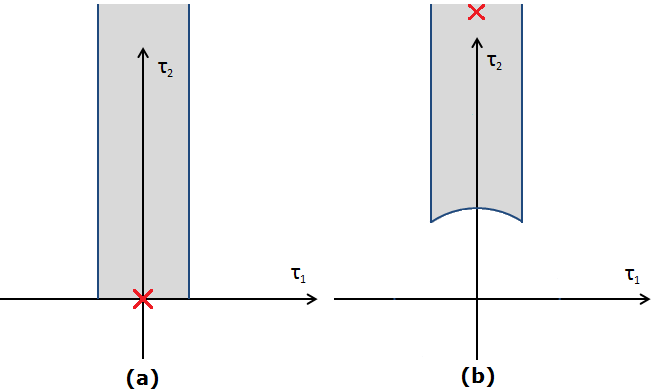}
\caption{(a) Critical thermodynamics from the strip domain. The cross depicts the critical limit. This is the approach used in \cite{Berkooz:2007fe} and \cite{Lin:2007gi}. (b) Critical thermodynamics from the fundamental domain. In this picture, the critical regime is determined by a perturbative state in the thermal spectrum: the thermal scalar. One finds this regime by taking $\tau_2 \to \infty$.}
\label{approach}
\end{figure}

\noindent Our primary goal in this chapter is to study the above picture for the specific case of the $AdS_3$ and BTZ WZW models. Our study focuses on the approach to the critical thermodynamics through the thermal manifold. The different canonical approaches are depicted in figure \ref{approach}.

\noindent Our objectives can be summarized as follows.
\begin{itemize}
\item[I.]{Determine the thermal spectrum on the $AdS_3$ and $BTZ$ model.}
\item[II.]{Analyze the critical near-Hagedorn behavior of thermodynamical quantities.}
\item[III.]{Find out to what extent field theory results reproduce this.}
\item[IV.]{Determine the complete random walk picture of a highly excited string gas in these spacetimes.}
\end{itemize}
The chapter is organized as follows. In section \ref{WZWmodel} we present an exact computation of the thermal string spectrum on $AdS_3$. We utilize twisting techniques on the worldsheet to determine the states. We then search for the thermal scalar in the resulting spectrum. In section \ref{BTZsection} we make the transition to the WZW BTZ thermal black hole and study the same questions. After that, in section \ref{conical} we take a look at conical orbifolds, due to their relevance for thermodynamics. Section \ref{chemical} treats a slight generalization of the results of section \ref{WZWmodel} in which we include an angular chemical potential for the string gas. Finally in section \ref{fieldtheory}, we look at the naive lowest order $\alpha'$ thermal scalar action and study to what extent it captures the critical behavior of the thermal scalar. We end with some conclusions in section \ref{conclusion}. We present several more elaborate calculations in the supplementary sections.

\section{Exact $AdS_3$ WZW model}
\label{WZWmodel}
\subsection{Random walks in $AdS_3$}
\label{randwalksads}
We consider $AdS_3$ spacetime with a non-zero Kalb-Ramond background field. The metric is given in the global coordinates of the $AdS$ spacetime as:
\begin{equation}
ds^2 = \alpha'k\left(-\cosh(\rho)^2dt^2+d\rho^2+\sinh(\rho)^2d\phi^2\right)
\end{equation}
where $\phi \sim \phi + 2\pi$ and the space includes a NS-NS two-form:
\begin{equation}
B = -\alpha'k\sinh(\rho)^2dt \wedge d\phi.
\end{equation}
The crucial aspect of this string background is that it is an exact (up to all orders in $\alpha'$) CFT because it can be written as a Wess-Zumino-Witten (WZW) model: it is the $SL(2,\mathbb{R})$ WZW model. This causes the string spectrum to be composed of irreducible representations of the affine Lie algebra underlying the WZW model. The string spectrum in this background was determined in \cite{Maldacena:2000hw}. Performing a Wick rotation on this model, yields another WZW model: the $SL(2,\mathbb{C})/SU(2)$ model. We identify the Euclidean time coordinate as $\tau \sim \tau + \beta$. Note that this time coordinate is dimensionless in these conventions. For more information regarding these WZW models and their link through analytic continuation, we refer the reader to the previous chapter.
Moreover, the full string path integral on the thermal $AdS_3$ manifold can be exactly computed and is given by \cite{Maldacena:2000kv}
\begin{align}
Z = \frac{\beta\sqrt{k-2}}{8\pi}\int_{E}\frac{d\tau_1d\tau_2}{\tau_2^{\frac{3}{2}}}&e^{4\pi\tau_2\left(1-\frac{1}{4(k-2)}\right)}\sum_{h,\overline{h}}D(h,\overline{h})e^{2\pi i \tau(h+\overline{h})} \nonumber \\
&\times \sum_{m=1}^{+\infty}e^{-\frac{(k-2)m^2\beta^2}{4\pi\tau_2}}\frac{\left|\eta(\tau)\right|^4}{\left|\vartheta_{1}\left(-\frac{im\beta}{2\pi},\tau\right)\right|^2},
\end{align}
where $E$ denotes the modular strip region and the sum over $h$ and $\bar{h}$ corresponds to the internal CFT, required to make the space a valid string background. The Hagedorn temperature in this background was determined directly from this partition function in \cite{Berkooz:2007fe}\cite{Lin:2007gi} and was found to be
\begin{equation}
\label{hag}
\beta_{H}^{2} = \frac{4\pi^2}{k}\left(4-\frac{1}{k-2}\right).
\end{equation}
The results from section \ref{pathderiv} predict that the critical behavior of the free energy of the string gas is determined by 
\begin{equation}
F = -\frac{1}{\beta}\sum_{w=\pm 1}\int_{0}^{+\infty}\frac{d\tau_2}{2\tau_2}\int\left[\mathcal{D}X\right]\exp\left(-S_p\right)
\end{equation}
where $S_p$ is given by
\begin{align}
\label{randwalkk}
S_p = \frac{k}{4\pi}\int_{0}^{\tau_2}dt\left[(\partial_t \rho)^2 + (\beta^2\cosh(\rho)^2-\beta_{H,\text{flat}}^2)+\sinh(\rho)^2(\partial_t\phi)^2 + 2w \frac{\beta}{2\pi\alpha'}\sinh(\rho)^2\partial_t\phi \right.\nonumber \\
\left. + \frac{4\pi^2}{k^2}\left\{\frac{3}{4} + \frac{1}{4\cosh(\rho)^2}\right\}\right].
\end{align}
This represents a particle moving in a two-dimensional curved space in a potential determined by $\cosh(\rho)^2$ and interacting with a specific electromagnetic field. The first two corrections to the random walk discussed in the previous section have already been included. Firstly, the tachyon mass correction was included, e.g. for the bosonic string $\alpha'k\beta_{H,\text{flat}}^2 = 16\pi^2\alpha'$. Secondly, the extra final term in the action comes from the $G_{00}$ metric component and represents a mild potential that slightly damps paths that come close to the origin $\rho=0$.\footnote{In section \ref{KR} in chapter \ref{chth} we discussed that another term should be incorporated in the particle action when considering non-zero NS-NS flux. However, one readily checks that this term vanishes in this case using the explicit form of the metric and NS-NS field.}  We remark that we have not considered possible $\alpha'$ corrections to the thermal scalar action so there might be more contributions to the particle action (\ref{randwalkk}) that have been neglected. We will turn to this problem next. 
 
\subsection{General analysis of $\alpha'$ corrections}
\begin{figure}[h!!!!]
\centering
\includegraphics[width=11cm]{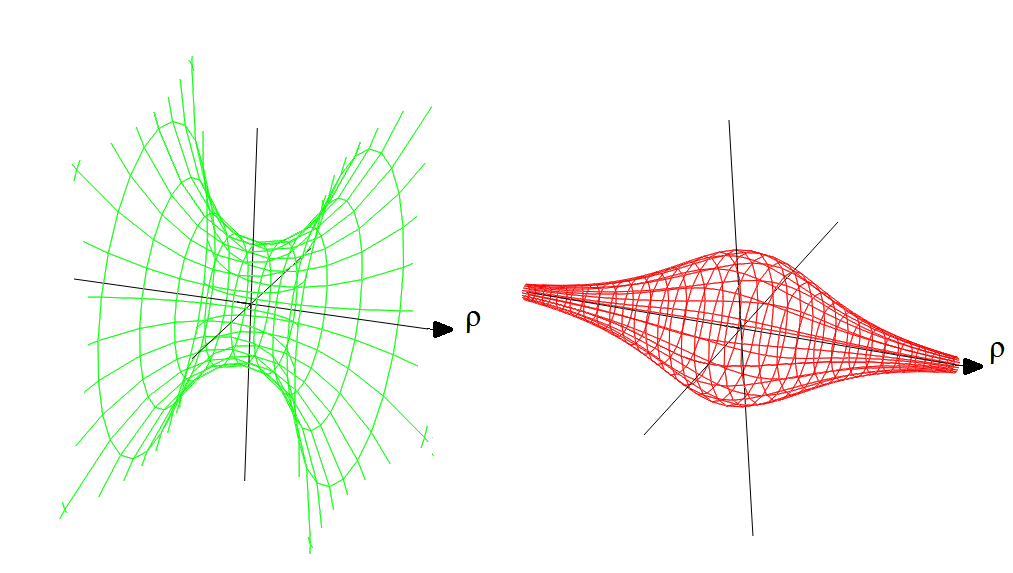}
\caption{Left figure: size of the thermal circle in $AdS$ space as a function of radial distance. The center of the figure is at $\rho=0$. Right figure: size of the thermal circle in the T-dual of $AdS$ space as a function of radial distance.}
\label{thermcircle}
\end{figure}

Now we ask whether the above random walk action is $\alpha'$-exact. Let us therefore first analyze possible corrections in general and see whether there is at least a regime in which they can be neglected. We know from previous work \cite{Mertens:2013zya} that this is not the case for the Euclidean Rindler string. In this section only, we rescale the coordinates such that they are not dimensionless.\footnote{$k\alpha'\tau^2 \to \tau^2$, $k\alpha'\rho^2 \to \rho^2$ and $k\alpha'\phi^2 \to \phi^2$. The $AdS$ length has been introduced as $l^2=k\alpha'$. The Euclidean time coordinate is obtained as $t \to i\tau$.} The Euclidean metric and Kalb-Ramond field are
\begin{equation}
ds^2 = \cosh(\rho/l)^2d\tau^2+d\rho^2+\sinh(\rho/l)^2d\phi^2
\end{equation}
and
\begin{equation}
B = -i\sinh(\rho/l)^2d\tau \wedge d\phi.
\end{equation}
We identify $\tau \sim \tau + \beta$ and note that this differs from the temperature we have introduced earlier by a factor of $l$, the $AdS$ length. The thermal scalar action consists of diffeomorphism invariants constructed with T-dual quantities. 
The T-dual Ricci tensor components are given by
\begin{equation}
\tilde{R}^{00} = 0, \quad  \tilde{R}^{\rho\rho} = \frac{2}{l^2\cosh(\rho/l)^2}, \quad \tilde{R}^{\phi\phi} = \frac{2}{l^2\sinh(\rho/l)^2},
\end{equation}
and all components with mixed indices vanish. The T-dual Ricci scalar is given by
\begin{equation}
\tilde{R} = \frac{4}{l^2\cosh(\rho/l)^2},
\end{equation}
and the T-dual dilaton has the expression:
\begin{equation}
\partial_{\rho}\tilde{\Phi} = - \frac{1}{l} \tanh(\rho/l).
\end{equation}
A peculiarity of this background is that $\tilde{B}_{\mu\nu} = 0$. Some possible terms that could appear in the thermal scalar action are given by:
\begin{align}
m^2 TT^{*} &= -\frac{4}{\alpha'}TT^{*}, \quad \text{bosonic}\quad \text{or} \quad m^2 TT^{*} = -\frac{2}{\alpha'}TT^{*}, \quad \text{type II},\\
\tilde{G}^{\mu\nu}\partial_{\mu}T\partial_{\nu}T^{*} &= \frac{\beta^2}{4\pi^2\alpha'^2}TT^{*} + \partial_\rho T \partial_\rho T^{*} - \frac{\beta}{2\pi\alpha'}(T\partial_\phi T^{*}- T^{*}\partial_\phi T) \nonumber \\
&\quad + \frac{1}{\sinh(\rho/l)^2}\partial_\phi T \partial_\phi T^{*},\\
\tilde{R} TT^{*} &= \frac{4}{\cosh(\rho/l)^2}\frac{1}{l^2} TT^{*},\\
\label{higher1}
\alpha'\tilde{R}^{\mu\nu}\partial_\mu T \partial_\nu T^{*} &=  \frac{2}{\cosh(\rho/l)^2}\frac{\alpha'}{l^2}\partial_\rho T \partial_\rho T^{*} +\frac{2}{\sinh(\rho/l)^2}\frac{\alpha'}{l^2}\partial_\phi T \partial_\phi T^{*},
\end{align}
and
\begin{align}
\partial_\mu \tilde{\Phi} \partial^{\mu} \tilde{\Phi} TT^{*} &= \frac{1}{l^2}\tanh(\rho/l)^2 TT^{*}, \\
\label{higher2}
\alpha'\partial^{\mu}\tilde{\Phi} \partial^{\nu} \tilde{\Phi} \partial_{\mu}T\partial_{\nu}T^{*} &= \frac{\alpha'}{l^2}\tanh(\rho/l)^2\partial_\rho T \partial_\rho T^{*}. 
\end{align}
In this case, all terms originating from higher order corrections such as (\ref{higher1}) and (\ref{higher2}) are suppressed as $\alpha'/l^2$. This ratio is suppressed since the T-dual geometry is only slowly varying with $\rho$. This is in sharp contrast to the black hole case, where a curvature singularity in the T-dual spaces sets in at the event horizon \cite{Mertens:2013zya}. The thermal circles for $AdS_3$ and its T-dual are sketched in figure \ref{thermcircle}. 
In \cite{Mertens:2013zya} we argued that in general two conditions need to be met in order to suppress higher $\alpha'$ terms. The first condition is the one discussed above. The second requirement is that the temperature is of order the string scale. This condition is not met for black holes where the temperature equals the Hawking temperature. In our case, this condition is not required but this is again a peculiarity of this specific model. Considering for instance the same background but with the Kalb-Ramond field turned off, one readily finds that $\beta^2/l^2 \ll 1$ is also necessary to suppress all higher $\alpha'$ corrections. \\
From these generalities we conclude, that we can find a regime (large $l$ in string units and (although not necessary here) string-scale temperatures) where we can neglect all possible $\alpha'$ corrections, if they are present in the first place. We will come back to this issue further on.

\subsection{Thermal string spectrum from a $SL(2,\mathbb{R})$ point of view}
\label{spectrumsection}
Let us now try to answer a different question. In order for the starting point of our story to be valid, we need to find a winding tachyon in the string spectrum on the thermal manifold that becomes massless at the Hagedorn temperature. The spectrum in both the Lorentzian and the Euclidean model are known, so all that is left to do is to compactify the imaginary time direction and see how the spectrum changes. In \cite{Berkooz:2007fe} a mini-superspace analysis is used to find the thermal tachyon. The authors of \cite{Argurio:2000tb}\cite{Rangamani:2007fz}\cite{Martinec:2001cf} developed methods to find the spectrum in general orbifold CFTs by introducing twist operators \cite{Dixon:1986qv}. We will show that this approach can be succesfully applied in this context. \\
We first cite the results for the string spectrum on the Lorentzian signature $SL(2,\mathbb{R})$ model \cite{Maldacena:2000hw}. \\
The spectrum in (Lorentzian) $AdS_3$ is built on two types of $SL(2,\mathbb{R})$ representations. In what follows, the quantum number $m$ is the eigenvalue of the $J^{3}_{0}$ operator in the zero-mode Lie algebra. 
\begin{itemize}
\item{$\mathcal{D}_{j}$ where $\frac{1}{2} < j < \frac{k-1}{2}$. These are the so-called principal discrete representations. These can be further classified in lowest weight principal discrete representations given by
\begin{equation}
\mathcal{D}_{j}^{+} = \left\{\left|j,m\right\rangle, m = j, j+1, j+2, ...\right\}, 
\end{equation}
and highest weight principal discrete representations
\begin{equation}
\mathcal{D}_{j}^{-} = \left\{\left|j,m\right\rangle, m = -j, -j-1, -j-2, ...\right\}.
\end{equation}
}
\item{$\mathcal{C}_{j,\alpha}$ where $j = \frac{1}{2} + is$ ($s \in \mathbb{R}$) and $0 \leq \alpha < 1$. These are the continuous representations. In this case the representations are given by
\begin{equation}
\mathcal{C}_{j,\alpha} = \left\{\left|j,\alpha,m\right\rangle , m = \alpha, \alpha \pm 1, \alpha \pm 2, ...\right\}.
\end{equation}
}
\end{itemize}
CFT primaries are labeled by these quantum numbers. Descendants can then be constructed by applying the affine algebra raising operators. The primary and its descendants in one $SL(2,\mathbb{R})$ represenation of the zero-mode algebra together with all their affine algebra descendants form a $\widehat{SL(2,\mathbb{R})}$ representation. The full affine algebra has an automorphism called \emph{spectral flow} given by:
\begin{equation}
\tilde{J}^{3}_{n} = J^{3}_{n} - \frac{k}{2}w\delta_{n,0}, \quad \tilde{J}^{\pm}_n = J^{+}_{n\pm w},
\end{equation}
which preserves the commutation relations and maps one representation into another. The amount of spectral flow is labeled by an integer $w \in \mathbb{Z}$. The Virasoro operators associated to the $\tilde{J}^{a}_n$ are given by the Sugawara construction as
\begin{equation}
\tilde{L}_n = L_n + wJ^{3}_n - \frac{k}{4}w^2 \delta_{n,0}.
\end{equation}
For compact groups this does not lead to new representations whereas for non-compact groups (such as the one we have here) this results in new representations that should be incorporated. For more details, we refer the reader to \cite{Maldacena:2000hw}. The strategy is to start with a representation of the $\tilde{J}^{a}_n$ algebra whose primary satisfies
\begin{align}
\tilde{J}^{\pm}_n \left|\tilde{j},\tilde{m}\right\rangle &= 0, \quad \tilde{J}^{3}_n\left|\tilde{j},\tilde{m}\right\rangle = 0, \quad n \geq 1, \\
&\tilde{J}^{3}_0 \left|\tilde{j},\tilde{m}\right\rangle = \tilde{m}\left|\tilde{j},\tilde{m}\right\rangle.
\end{align}
One then finds the conformal weight of this state by applying $L_0$. The conformal dimensions of the primaries of Lorentzian $AdS_3$ are given by
\begin{equation}
h^{w}_{jm\overline{m}} = -\frac{\tilde{j}(\tilde{j}-1)}{k-2}- \tilde{m}w - \frac{kw^2}{4} + h_{int},
\end{equation}
where $w$ denotes the spectral flow used to generate all primary states and $h_{int}$ is the conformal weight of an internal CFT needed to obtain the right central charge. An analogous expression holds for the antiholomorphic components. \\
The Lorentzian $AdS_3$ metric is given in global coordinates by
\begin{equation}
ds^2 = \alpha'k\left(-\cosh(\rho)^2dt^2+d\rho^2+\sinh(\rho)^2d\phi^2\right)
\end{equation}
with the periodic identification $\phi \sim \phi + 2\pi$. To obtain the thermal manifold we should also impose periodicity in imaginary time: $t \sim t + i\beta$. We will use the Lorentzian signature vertex operators and impose periodicity in imaginary time on these. The reader might feel a bit uneasy about this, but we will nonetheless obtain the expected result. Moreover, in the next section we will rederive this from a fully Euclidean point of view as well. The Lorentzian spectrum consists of normal affine $\widehat{SL(2,\mathbb{R})}$ representations and the spectral flowed ones. The latter can also be obtained by twisting the CFT by the twist operator associated with the $\phi$ identification \cite{Argurio:2000tb}. So we now utilize this method and twist the CFT in \emph{both} the $\phi$ and the $t$ direction. \\
The operators generating (spacetime) time translations and angular rotations are given by (as proven in section \ref{WZWapp})
\begin{eqnarray}
Q_{t} = J^{3}_{0} - \overline{J}^{3}_{0}, \\
Q_{\phi} = J^{3}_{0} + \overline{J}^{3}_{0}.
\end{eqnarray}
It is important to note that both are generated by the same set of operators.\\
We demand that states respect the periodicity of spacetime, so translation by $2\pi$ in the angular direction should reproduce the same state. In other words
\begin{equation}
\exp\left(i2\pi Q_{\phi}\right) = 1.
\end{equation}
This implies that 
\begin{equation}
\label{con1}
m + \overline{m} \in \mathbb{Z}.
\end{equation}
Analogously for the imaginary time periodicity we get
\begin{equation}
\label{con2}
\frac{i\beta}{2\pi}(m-\overline{m}) \in \mathbb{Z}.
\end{equation}
String states need to respect these conditions. However, we know that a consistent string theory should also include twisted states, so we are not finished yet. We can do this by constructing local operators that implement the above restrictions and hence twist the preceding (still inconsistent!) CFT \cite{Argurio:2000tb}. We remark that we determined the above restrictions for untwisted sectors only. The twisted sector states could have different restrictions imposed on their quantum numbers. We will come back to this later.\\ 
To proceed, we use a parafermionic representation of the current algebra by diagonalizing the $J^3$ operator: 
\begin{equation}
J^{3} = -\sqrt{\frac{k}{2}}\partial{X}, \quad J^{\pm} = \psi^{\pm} e^{\pm\sqrt{2/k}X}
\end{equation}
where the $X$ and $\psi^{\pm}$ satisfy
\begin{equation}
X(z)X(w) \sim -\ln(z-w), \quad \psi^{+}(z)\psi^{-}(w) \sim \frac{k}{(z-w)^{2+2/k}}, \quad \psi^{\pm}(z)\psi^{\mp}(w) \sim 0,
\end{equation}
and with analogous relations for the antiholomorphic copy of the current algebra. The (untwisted) primaries are then represented as
\begin{equation}
\Phi_{jm\overline{m}} = \Psi_{jm\overline{m}}e^{\sqrt{\frac{2}{k}}(mX + \overline{m}\overline{X})},
\end{equation}
where the $\Psi_{jm\overline{m}}$ are uncharged under $J^3$ and $\overline{J}^{3}$.\\
Now we construct the twist operators and demand mutual locality of the OPEs (this will correspond to projecting onto invariant states) and we demand closure of the OPE (this corresponds to inclusion of twisted sectors). In analogy with \cite{Argurio:2000tb}\cite{Rangamani:2007fz}\cite{Martinec:2001cf}, the twist operators are given by
\begin{align}
t^{\phi}_{w} &= e^{\sqrt{\frac{k}{2}}w(X-\overline{X})}, \\
t^{t}_{p} &= e^{\sqrt{\frac{k}{2}}\frac{i\beta p}{2\pi}(X+\overline{X})},
\end{align}
where $w$ denotes the twisting in the $\phi$ direction (this is the same as the spectral flow parameter $w$ used in \cite{Maldacena:2000hw}) and $p$ denotes the twisting in the imaginary time direction. Let us consider the OPE of a twist operator and an untwisted primary. For a boson field satisfying $X(z)X(w) \sim -\ln(z-w)$ the following OPE holds
\begin{equation}
e^{\alpha X(z)}e^{\beta X(w)} \sim (z-w)^{-\alpha\beta}e^{(\alpha+\beta)X(w)} + (z-w)^{-\alpha\beta+1}\alpha :\partial X (w) e^{(\alpha+\beta)X(w)}: + \hdots
\end{equation}
and higher powers of $z-w$ as dictated by the Taylor series expansion of $e^{\alpha X(z)}$ around $z=w$. This OPE holds for general complex values of both $\alpha$ and $\beta$. Obviously, depending on these values, the number of singular terms varies. Important to note is that all powers of $z-w$ are integrally shifted from $\alpha\beta$ and so it is this combination that provides restrictions on the quantum numbers as we now show. The OPEs of the twist operators with the untwisted primaries are given by:
\begin{align}
\label{eq1}
t^{\phi}_{w}(z,\bar{z})\Phi_{jm\overline{m}}(w,\bar{w})&\sim (z-w)^{-wm}(\bar{z}-\bar{w})^{w\overline{m}}\Psi_{jm\overline{m}}e^{\sqrt{\frac{2}{k}}\left[(m+w\frac{k}{2})X + (\overline{m}-w\frac{k}{2})\overline{X}\right]} + \hdots, \\
\label{eq2}
t^{t}_{p}(z,\bar{z})\Phi_{jm\overline{m}}(w,\bar{w})&\sim (z-w)^{-\frac{i\beta p}{2\pi}m}(\bar{z}-\bar{w})^{-\frac{i\beta p}{2\pi}\overline{m}}\Psi_{jm\overline{m}}e^{\sqrt{\frac{2}{k}}\left[(m+\frac{i\beta p}{2\pi}\frac{k}{2})X + (\overline{m}+\frac{i\beta p}{2\pi}\frac{k}{2})\overline{X}\right]} + \hdots.
\end{align}
The OPE of a twist operator with a primary generates new operators that must be included in the operator spectrum to close the OPE. These are the twisted primaries and these can be written as
\begin{align}
:t^{\phi}_{w}(z,\bar{z})t^{t}_{p}(z,\bar{z})\Phi_{jm\overline{m}}(z,\bar{z}):
\end{align}
for general $w$ and $p$. The most general primary vertex operator is then
\begin{equation}
\Phi^{wp}_{jm\overline{m}} = \Psi_{jm\overline{m}}e^{\sqrt{\frac{2}{k}}\left[(m+\frac{k}{2}w+\frac{k}{2}\frac{i\beta}{2\pi}p)X + (\overline{m}-\frac{k}{2}w+\frac{k}{2}\frac{i\beta}{2\pi}p)\overline{X}\right]}
\end{equation}
with conformal weight
\begin{equation}
h^{wp}_{jm\overline{m}} = -\frac{j(j-1)}{k-2} + \frac{m^2}{k} - \frac{\left(m+\frac{k}{2}w + \frac{kp}{2}\frac{i\beta}{2\pi}\right)^2}{k} + h_{int}.
\end{equation}
Next we need to determine the range of the quantum numbers $m$ and $\overline{m}$ and it is at this point that an important subtlety sets in. The conserved charges determined above are in fact in general not correct for the twisted sector states. This is related to an ambiguity of the Noether current: adding a divergence of an antisymmetric tensor gives the same conservation equation, although the conserved charge is different in topologically non-trivial sectors. 
To avoid branch cuts (mutual locality), the $m$ and $\overline{m}$ quantum numbers of the untwisted primaries are restricted from (\ref{eq1}) and (\ref{eq2}) by the following two conditions:
\begin{eqnarray}
m+\overline{m} \in \mathbb{Z}, \\
\frac{i\beta}{2\pi}\left(m-\overline{m}\right) \in \mathbb{Z},
\end{eqnarray}
which are indeed the conditions required for projecting on invariant states which we wrote down in equations (\ref{con1}) and (\ref{con2}). One can also (too naively) apply the twist operators to already twisted vertex operators. This would give us then\footnote{We denoted the $J^{3}_0$ eigenvalue as $m_J$ to distinguish it with the $m$ quantum number. These are not equal for twisted sectors. Analogous comments hold for the antiholomorphic sector.}
\begin{eqnarray}
\label{naive}
m_J+\overline{m}_J \in \mathbb{Z}, \\
\label{naive2}
\frac{i\beta}{2\pi}\left(m_J-\overline{m}_J\right) \in \mathbb{Z},
\end{eqnarray}
with 
\begin{align}
\label{relation}
m_J = m + \frac{kw}{2} + \frac{i\beta k p}{4\pi}, \\
\label{relation2}
\overline{m}_J = \overline{m} - \frac{kw}{2} + \frac{i\beta k p}{4\pi}.
\end{align}
This is however wrong: it is known that the conserved charges can be different in the twisted sectors. We will demonstrate that this is indeed the case here by using two different arguments. This failure of the twist operator construction was also previously observed in a different context in \cite{Parsons:2009si} where extremal Lorentzian BTZ black holes were considered. \\
As a first argument, let us consider the level-matching condition: $L_0 - \bar{L}_0 \in \mathbb{Z}$. It is known from studies in the past concerning heterotic 4d black holes \cite{Giddings:1993wn} and rotating WZW BTZ black holes \cite{Natsuume:1996ij} that this condition provides us with the correct projection operator \cite{Hemming:2002kd}. The general primaries have weights given by:
\begin{align}
\label{weig1}
h^{wp}_{jm\overline{m}} = -\frac{j(j-1)}{k-2} + \frac{m^2}{k} - \frac{\left(m+\frac{k}{2}w + \frac{kp}{2}\frac{i\beta}{2\pi}\right)^2}{k} + h_{int}, \\
\label{weig2}
\bar{h}^{wp}_{jm\overline{m}} = -\frac{j(j-1)}{k-2} + \frac{\overline{m}^2}{k} - \frac{\left(\overline{m}-\frac{k}{2}w + \frac{kp}{2}\frac{i\beta}{2\pi}\right)^2}{k} + \bar{h}_{int},
\end{align}
and we assume that the internal CFT is on its own level-matched: $h_{int} - \bar{h}_{int} \in \mathbb{Z}$. This gives for the level-matching condition:
\begin{align}
h - \bar{h} &= -w(m+\overline{m}) -\frac{pi\beta}{2\pi}(m-\overline{m}) - \frac{ki\beta}{2\pi}pw \in \mathbb{Z} \\
\label{levelmatch}
&= -w(m_{J}+\overline{m}_{J}) -\frac{pi\beta}{2\pi}(m_{J}-\overline{m}_{J}) + \frac{ki\beta}{2\pi}pw \in \mathbb{Z}
\end{align}
and it is clear that this is in contradiction with (\ref{naive}) and (\ref{naive2}) unless $pw=0$ which is impossible to satisfy in general since interactions of states having $pw=0$ could in principle create states that have $pw\neq0$.
Noether ambiguities can spoil the projection condition by additional terms only present in twisted sectors. Let us keep an open mind and consider the general deformation (in a non-technical sense) of the conserved charges:
\begin{align}
Q_{\phi} &= J^3_0 + \overline{J}^{3}_0 + f(w,p), \\
Q_{t} &= J^3_0 - \overline{J}^{3}_0 + g(w,p),
\end{align}
with $f$ and $g$ functions of the twists with the property that $f(0,0) = g(0,0)=0$. Now we will determine these functions using what we already know. 
Consider first the sector $p=0$ which coincides, up to the projection onto invariant states, with the non-thermal Lorentzian $AdS_3$. It was shown in \cite{Maldacena:2000kv}\cite{Maldacena:2000hw} that the energy and angular momentum really are measured by $J^3_0 \mp \overline{J}^3_0$. Applying this result here, we obtain $f(w,p) = f(p)$ and $g(w,p)=g(p)$.\footnote{Note that we neglect the possibility that $f$ (or $g$) include `mixing' terms such as $pw$. Such terms are unnatural when considering the resulting conformal weights and we will find agreement with other arguments further on.} When considering equation (\ref{levelmatch}), it is clear that choosing $g=0$ and $f=-\frac{ki\beta}{2\pi}p$ satisfies the level-matching condition which reduces to
\begin{equation}
-w\left(m_{J}+\overline{m}_{J}-\frac{ki\beta}{2\pi}p\right) \in \mathbb{Z},
\end{equation}
and which is different than the requirement (\ref{naive}). Note that this line of thought is not a rigorous derivation, but merely demonstrates that the above solution is the most natural one to choose.\\
One can also use a different argument to demonstrate this result by taking the flat space $k \to \infty$ limit defined by keeping $k\rho^2$ fixed. This argument was used by the authors of \cite{Rangamani:2007fz} to obtain the correct spacetime energy of twisted states on the Lorentzian BTZ manifold. The generator of $\phi$-translations was naively identified as 
\begin{equation}
Q_{\phi} = J^{3}_{0} + \overline{J}^{3}_{0}.
\end{equation}
The Lorentzian currents are determined in section \ref{WZWapp} and the relevant components are given by
\begin{align}
J^{3}(z) &= i k\cosh(\rho)^2 \partial t - ik\sinh(\rho)^2 \partial \phi, \\
\overline{J}^{3}(\bar{z}) &=  -ik\cosh(\rho)^2 \bar{\partial} t - ik\sinh(\rho)^2 \bar{\partial} \phi.
\end{align}
In the large $k$ limit (keeping $k\rho^2$ fixed) these currents become
\begin{align}
J^{3}(z) &= i k\partial t - ik\rho^2 \partial \phi, \\
\overline{J}^{3}(\bar{z}) &=  -ik\bar{\partial} t - ik\rho^2 \bar{\partial} \phi.
\end{align}
We can thus rewrite the conserved $\phi$ charge as
\begin{equation}
J^{3}_0 + \overline{J}^{3}_0 = \oint dz J^{3}(z) - \oint d\bar{z}\overline{J}^{3}(\bar{z}) = ik\oint dz \partial t + ik\oint d\bar{z} \bar{\partial} t + ik \oint dz \rho^2 \partial \phi - ik\oint d\bar{z} \rho^2\bar{\partial}\phi,
\end{equation}
where $1/(2\pi i)$ factors are left implicit in the contour integrals. The final two terms are giving us the angular rotation we seek. The first two terms however are not what we want, but these are dominant in the large $k$ limit. These can be explicitly written as
\begin{equation}
\frac{ik\beta p}{2\pi},
\end{equation}
which corresponds to a winding contribution. The generator of angular rotations can then be obtained by subtracting this part as
\begin{equation}
Q_{\phi} = J^{3}_0 + \overline{J}^{3}_0 - \frac{ik\beta p}{2\pi}.
\end{equation}
Note that for $Q_t$ on the other hand, the winding contribution is subleading in the large $k$ limit and can be neglected \cite{Rangamani:2007fz}. This agrees with the expression we determined above using the level-matching argument.\\
We conclude that the projection conditions are 
\begin{align}
\label{cond1}
m_J+\overline{m}_J - \frac{ik\beta p}{2\pi} &\in \mathbb{Z}, \\
\label{cond2}
\frac{i\beta}{2\pi}\left(m_J-\overline{m}_J\right) &\in \mathbb{Z}.
\end{align}
The two conditions (\ref{cond1}) and (\ref{cond2}) can be solved and this gives together with (\ref{relation}) and (\ref{relation2}):
\begin{align}
m &= \frac{q}{2} + i\frac{\pi n}{\beta} -\frac{kw}{2}, \\
\overline{m} &= \frac{q}{2} - i\frac{\pi n}{\beta}  +\frac{kw}{2},
\end{align}
where $q,n \in \mathbb{Z}$. Substituting these in (\ref{weig1}) and (\ref{weig2}), we finally obtain the conformal weights of the primaries:
\begin{align}
\label{ads3spectrum1prelim}
h^{wp}_{jqn} &= -\frac{j(j-1)}{k-2} -\frac{qw}{2} -\frac{i\pi nw}{\beta} + \frac{kw^2}{4}- i\frac{qp\beta}{4\pi} + \frac{ p n}{2} + \frac{kp^2\beta^2}{4(2\pi)^2} + h_{int}, \\
\label{ads3spectrum2prelim}
\bar{h}^{wp}_{jqn} &= -\frac{j(j-1)}{k-2} + \frac{qw}{2} -\frac{i\pi nw}{\beta} + \frac{kw^2}{4} - i\frac{qp\beta}{4\pi} - \frac{ p n}{2} + \frac{kp^2\beta^2}{4(2\pi)^2} + \bar{h}_{int}.
\end{align}
Note that indeed $h - \bar{h} \in \mathbb{Z}$, provided the internal CFT is on its own level-matched.

\subsection{Comments on the Euclidean $SL(2,\mathbb{C})/SU(2)$ point of view}
We now reanalyze this result from a purely Euclidean point of view. Euclidean $AdS_3$ is the hyperbolic 3-plane and can be seen as a coset $SL(2,\mathbb{C})/SU(2)$. The isometry group is given by $SL(2,\mathbb{C})$. We are interested in $AdS_3$ (and its continuation) in the \emph{global} coordinates and the continuation in that particular time coordinate. In section \ref{WZWapp} we discussed this continuation in somewhat more detail.
Expanding the symmetry current in the algebra generators (which we choose to be the same as those of $\mathfrak{sl}(2,\mathbb{R})$; the only difference is that the expansion coefficients $J^{a}(z)$ are now allowed to be arbitrary complex numbers), we can identify which symmetry current is responsible for Euclidean time translations. This was again done in section \ref{WZWapp} and the result is
\begin{equation}
Q_{\tau} = i(J^{3}_0 - \overline{J}^{3}_0).
\end{equation}
This differs a factor of $i$ compared to the earlier result for $Q_t$. However, the thermal identification is with parameter $\beta$ now, so in all nothing changes and the derivation of the previous subsection still holds. However, we start with the principal representations of the $\mathfrak{sl}(2,\mathbb{C})$ algebra, which do not contain discrete representations nor spectral flowed representations. This sets $w=0$ from the start and $j=1/2+is$ with real $s$. This mismatch is caused by the fact that after Wick rotating all the other representations (discrete and the spectral flowed), these do not correspond to states in the Euclidean string spectrum \cite{Maldacena:2001km}, so we have computed `too much' in our first derivation.\footnote{We want to remark one subtlety: one could set $w=0$ from the start in the entire previous derivation, and consider only one twist operator. The level-matching condition is satisfied without any deformation to the energy. To determine the spacetime angular momentum of such states (find the correct form of $Q_{\phi}$), we should then resort to the large $k$ limit as explained in the previous section. The level-matching condition is of no help here. So in this case, the large $k$ limit becomes a necessary part of the procedure (and not just an alternative).} To summarize, we give the conformal weights of all primaries:
\begin{align}
\label{ads3spectrum1}
h^{p}_{jqn} &= \frac{s^2 +1/4}{k-2} - i\frac{qp\beta}{4\pi} + \frac{ p n}{2} + \frac{kp^2\beta^2}{4(2\pi)^2} + h_{int}, \\
\label{ads3spectrum2}
\bar{h}^{p}_{jqn} &= \frac{s^2 +1/4}{k-2} - i\frac{qp\beta}{4\pi} - \frac{ p n}{2} + \frac{kp^2\beta^2}{4(2\pi)^2} + \bar{h}_{int},
\end{align}
where $q,n \in \mathbb{Z}$ and $p \in \mathbb{Z}$ denotes the winding around the Euclidean time dimension. For the antiholomorphic part, one simply changes the sign of both $p$ and $q$. From here on, we assume the internal CFT to be unitary and compact such that to analyze possible tachyons, we can restrict ourselves to $h_{int} = \bar{h}_{int} = 0$.

\subsection{Atick-Witten tachyon}
\label{AWtach}
It is beneficial to now clearly state how we will identify a tachyonic state in the string spectrum. In a general bosonic string CFT, the one-loop partition function is given by
\begin{equation}
Z = \int_{F}\frac{d\tau_1 d\tau_2}{2\tau_2} \text{Tr}\left[q^{L_0-c/24}\bar{q}^{\bar{L_0}-\bar{c}/24}\right] = \int_{F}\frac{d\tau_1 d\tau_2}{2\tau_2}\left|\eta(\tau)\right|^4(q\bar{q})^{-\frac{1}{12}}\sum_{H_{matter}}{q^{h_i-1}\bar{q}^{\bar{h_i}-1}}
\end{equation}
In the second equality, we sum over only the matter contributions (of the full $c=26$ matter CFT). We have isolated a $q\bar{q}$ combination, since this precisely compensates the ghost CFT in its asymptotic behavior, meaning
\begin{equation}
Z \to \int_{F}\frac{d\tau_1 d\tau_2}{2\tau_2}\sum_{H_{matter}}{q^{h_i-1}\bar{q}^{\bar{h_i}-1}}
\end{equation}
as $\tau_2 \to \infty$. 
A tachyonic state in bosonic string theory is thus determined if the conformal dimension $h+\bar{h}$ in the matter sector is smaller than 2 (divergence for $\tau_2 \to \infty$ in $Z$) after integrating over continuous quantum numbers.\footnote{For type II superstrings, the only modification in this definition is that the conformal dimension $h+\bar{h}$ needs to be smaller than $1$ to have a tachyonic state.} Continuous quantum numbers can give a non-vanishing contribution if they integrate into a $\tau_2$-dependent exponential. Let us make some comments regarding the above conformal weights (\ref{ads3spectrum1}) and (\ref{ads3spectrum2}). Firstly note that the conformal weights have an imaginary part when both $p$ and $q$ are non-zero. This is due to the non-unitarity of the $SL(2,\mathbb{C})/SU(2)$ model \cite{Teschner:1997ft}. Complex conformal weights are not uncommon on these Euclidean signature manifolds \cite{Hemming:2002kd} and we will see them appear again below when we study BTZ black holes. Imaginary parts of conformal weights are harmless when considering divergence properties. When considering the partition function in the fundamental modular domain, each string state makes a contribution proportional to
\begin{equation}
\label{imag}
q^{h}\bar{q}^{\bar{h}} = e^{2\pi i \tau_1 (h - \bar{h})}e^{-2\pi \tau_2 (h + \bar{h})}.
\end{equation}
Since $\tau_1$ ranges from $-1/2$ to $+1/2$, the first factor is not capable of causing divergences and thus an imaginary contribution of a conformal weight can not cause divergences.
Secondly, the term $\frac{ p n}{2}$ might appear alarming, since this could be arbitrarily large and negative apparently causing tachyonic divergences. However, one should note that the antiholomorphic part has the opposite sign and the contribution to $h+\bar{h}$ hence vanishes. As a summary, only $\Re(h + \bar{h})$ matters for determining instabilities. \\
Let us make one final remark on these conformal weights. The primaries we determined above satisfy the physicality constraint $L_0 - \bar{L}_0 \in \mathbb{Z}$ by construction. The on-shell condition $L_0 + \bar{L}_0 = 2$ is in general not satisfied for all these states (and neither is $L_0 = \bar{L}_0$). This means for instance that only a subset of these can be used as vertex operators in scattering amplitudes. However, we are interested in the one-loop vacuum amplitude and the states that circle the loop are clearly off-shell. Thus we will not apply the on-shell restriction to the conformal weights. Our definition of tachyon is rooted in the one-loop amplitude and a tachyon state can hence be an off-shell state. Note that in \cite{Rangamani:2007fz} tachyons are identified only as on-shell physical states. Their definition of tachyon is hence not completely the same as ours.\\

\noindent To find the thermal tachyon, we simply set $s$ to zero and we will find the right state where we expect it. This will be an a posteriori verification that the integration over $s$ (with the correct density of states) does not yield a $\tau_2$-dependent exponential contribution, unlike for instance the linear dilaton background that we discussed before in chapter \ref{chex}. We find for the $p=\pm1$ state:
\begin{equation}
\frac{1}{4(k-2)} + \frac{k\beta^2}{4(2\pi)^2} = 1,
\end{equation}
which determines indeed the Hagedorn temperature (\ref{hag}) given in section \ref{randwalksads}. The left hand side is equal to 1 if $p = \pm 1$, so this represents a state that becomes `marginally convergent' at the Hagedorn temperature. The state is in the twisted sector and can be interpreted as a winding 1 state. So we have found a state in the thermal spectrum that becomes tachyonic at temperatures higher than the Hagedorn temperature: this is the Atick-Witten tachyon \cite{Atick:1988si}\cite{Berkooz:2007fe}. \\
An important aspect of the above analysis is that not only the $q=n=0$ state is marginal but in fact the states with arbitrary $q$ and $n$ are all marginal simultaneously at the Hagedorn temperature. This implies that the critical limit of $Z(\tau)$ includes all of these states since they are equally dominant. Note though that we are not interested in the thermal partition function for fixed $\tau$, but instead integrate over the fundamental modular domain. Tachyonic divergences are located at large $\tau_2$ and in that region, the $\tau_1$ integral is simply from $-1/2$ to $+1/2$ and acts as a projector onto states satisfying $L_0 = \bar{L}_0$ \cite{Kutasov:1990sv}. Thus when considering the critical regime of the free energy, the $n \neq 0$ states are irrelevant, but the sum over $q$ remains. The quantum number $q$ is to be interpreted as discrete momentum along the spatial cigar. This is a priori very surprising since one generally expects states that include discrete momentum quantum numbers to be more massive than those without discrete momentum. This observation will lead to some important ramifications further on when we look at the random walk behavior of the critical free energy. Of course, for this to be valid, the integration over $s$ should not alter the critical behavior: it must not give a $\tau_2$-dependent exponential for each $q$. If this is not the case, several of these states can actually be subdominant. This however does not occur. We present quantitative arguments in favor of this in supplementary section \ref{dos}. In subsection \ref{numads} we will further discuss, using numerical methods, that indeed all $q \in \mathbb{Z}$ states are needed to produce the critical behavior.

\subsection{Type II Superstring in $AdS_3$ space}
\label{Hagsup}
The modification to obtain the Hagedorn temperature for type II superstrings in $AdS_3$ is the following:
\begin{equation}
\frac{\frac{1}{4}+s^2}{k}+\frac{k}{4}\frac{\beta_{H}^{2}}{4\pi^2} = \frac{1}{2}.
\end{equation}
Two things have changed: the denominator of the $SL(2,\mathbb{R})$ term is now $k$ and the r.h.s. is $1/2$ which is the condition needed to ensure convergence of the thermal partition function. This immediately leads to 
\begin{equation}
\beta_{H}^{2} = \frac{4\pi^2}{k}\left(2-\frac{1}{k}\right),
\end{equation}
which satisfies the correct flat space limit when taking $k \to \infty$. This provides confidence in our method to determine the winding tachyon. \\
Our analysis of the type II superstring is not rigorous, but looks very plausible. It is known though, that in a wide variety of models based on the $SL(2,\mathbb{R})$ WZW model \cite{Giveon:1999px}\cite{Aharony:2004xn}\cite{Israel:2003ry}\cite{Israel:2004ir}\cite{Rangamani:2007fz}, the only difference with the bosonic string is the replacement $k-2 \to k$ in the first term of the conformal weights and a modification of the unitarity constraints for discrete representations. We will assume that also in this case, these are the only modifications.\\
We want to remark that the tachyon state we found (both for the bosonic string and for the type II string) is in the continuous representation as in \cite{Berkooz:2007fe} and the quantum number $s$ can be interpreted as a measure for the radial momentum. The state is delocalized over the $AdS$ space due to the repulsion from the Kalb-Ramond background just as is the case for the long strings in (Lorentzian signature) $AdS_3$ \cite{Maldacena:2000hw}.

\subsection{Summary}
Using CFT twist techniques, we have found the thermal string spectrum on the WZW $AdS_3$ background. This required introducing two twist operators and a subtle discussion on the Noether ambiguities in the projection conditions on invariant states. The thermal scalar state is clearly visible in the spectrum and it predicts the correct Hagedorn temperature. In the following we will come back to these results several times, providing more clarifications as we go along.

\section{Euclidean BTZ model}
\label{BTZsection}
Next we look at the Euclidean BTZ black hole. This is the same background as the Euclidean $AdS_3$ background, so the results of the previous sections apply. We first briefly review the link between the thermal ensembles of both backgrounds \cite{Maldacena:1998bw}. \\
Consider the following Euclidean metric
\begin{equation}
ds^2 = l^2\left(\cosh(\rho)^2d\tau^2+d\rho^2+\sinh(\rho)^2d\phi^2\right)
\end{equation}
where $l$ is the $AdS$ length. For WZW models this is related to the string length as $l^2 = k \alpha'$. We identify $\tau \sim \tau + \beta$ and $\phi \sim \phi + 2\pi$ (fixed to avoid a conical singularity at $\rho=0$) to obtain the thermal $AdS_3$ metric.
To obtain Euclidean BTZ black holes, we set $r=r_+\cosh(\rho)$, $\varphi = \frac{l}{r_+}\tau$ and $t = \frac{l}{r_+}\phi$ to obtain
\begin{equation}
ds^2 = (r^2-r_+^2)dt^2 + \frac{l^2dr^2}{r^2-r_+^2}+r^2d\varphi^2.
\end{equation}
We then identify $t \sim t + \beta_{BTZ}$ (to avoid a conical singularity at $r=r_+$) and $\varphi \sim \varphi + 2\pi$ where
\begin{equation}
\beta_{BTZ} = \frac{2\pi l }{r_+}.
\end{equation}
To connect these, note that using the BTZ manifold coordinates we have
\begin{equation}
\varphi \sim \varphi + 2 \pi \quad \Rightarrow \quad \tau \sim \tau + \frac{2\pi r_+}{l},
\end{equation}
where the $\tau$-periodicity is equal to the inverse temperature $\beta$ from the $AdS_3$ manifold.
So we have that
\begin{equation}
\beta_{BTZ} = \frac{2\pi l }{r_+} = \frac{4\pi^2}{\beta}.
\end{equation}
In all, we obtain
\begin{equation}
Z_{BTZ}\left(\beta\right) = Z_{AdS_3}\left(\frac{4\pi^2}{\beta}\right)
\end{equation}
and this is the precise link between thermodynamics in both backgrounds.\\
The generators of Euclidean time translations and angular rotations are given by
\begin{align}
Q_{t} &= \frac{2\pi}{\beta_{BTZ}}\left(J^{3}_{0} + \overline{J}^{3}_{0}\right), \\
Q_{\varphi} &= i\frac{2\pi}{\beta_{BTZ}}\left(J^{3}_{0} - \overline{J}^{3}_{0}\right),
\end{align}
where $t \sim t + \beta_{BTZ}$ and $\varphi \sim \varphi + 2\pi$. With the same caveats and modifications as in the $AdS_3$ background, this leads to the restrictions (for untwisted primaries):
\begin{align}
m+\overline{m} &\in \mathbb{Z}, \\
\frac{i 2\pi}{\beta_{BTZ}}\left(m-\overline{m}\right) &\in \mathbb{Z}.
\end{align}
This is indeed simply the substitution $\beta \to \frac{4\pi^2}{\beta}$ as discussed above. The Euclidean BTZ spectrum is thus obtained simply by evaluating the thermal $AdS_3$ spectrum at a temperature $\frac{4\pi^2}{\beta}$:
\begin{equation}
h^{w}_{jm\overline{m}} = \frac{s^2+1/4}{k-2} + \frac{m^2}{k} - \frac{\left(m + \frac{kw}{2}\frac{ir_+}{l}\right)^2}{k} + h_{int}.
\end{equation}
where the winding $w$ is around the $\varphi$ direction which is the $\tau$-direction in the $AdS_3$ language. The $m$ and $\overline{m}$ quantum numbers are similarly given by
\begin{align}
m &= \frac{q}{2} + i\frac{\beta n}{4\pi}, \\
\overline{m} &= \frac{q}{2} - i\frac{\beta n}{4\pi},
\end{align}
where $q,n \in \mathbb{Z}$. 
To sum up, the conformal weights of the primaries are given by
\begin{align}
h^{w}_{jqn} &= \frac{s^2+1/4}{k-2} - qwi\frac{r_+}{2l} + \frac{wn}{2} + \frac{kw^2r_{+}^2}{4l^2} + h_{int}, \\
\bar{h}^{w}_{jqn} &= \frac{s^2+1/4}{k-2} - qwi\frac{r_+}{2l} - \frac{wn}{2} + \frac{kw^2r_{+}^2}{4l^2} + h_{int},
\end{align}
with $q,n \in \mathbb{Z}$ and $w \in \mathbb{Z}$ denotes the winding around the angular $\varphi$ direction. These conformal weights were obtained in \cite{Hemming:2002kd}, but there the projection operation on invariant states (which determines the $m$ and $\overline{m}$ quantum numbers) was not considered. \\
Let us now discuss some peculiarities of this spectrum.

\subsection{Thermal tachyons}
\begin{figure}[h]
\centering
\includegraphics[width=15cm]{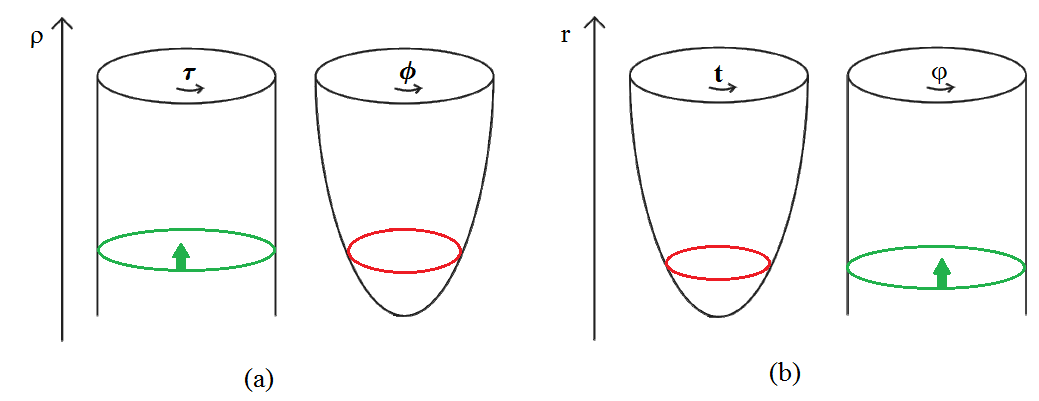}
\caption{(a) Thermal circle and spatial cigar for thermal $AdS_3$. The arrows denote the Kalb-Ramond repulsion. String states can wind the cilinder where the gravitational attraction to the center precisely compensates the Kalb-Ramond repulsion such that the winding strings are in the continuous representations. States that would wind the cigar are not allowed: these states are simply absent from the spectrum. (b) The same situation for the thermal BTZ black hole. Strings cannot wind the temporal cigar in this case.}
\label{adsbtzfigure}
\end{figure}

Firstly (and most importantly for our purposes) there are no string states that wind the $t$ direction for the same reason that the Euclidean $AdS_3$ manifold does not have cigar-winding states. Thus the thermal scalar is not even in the spectrum!\footnote{In \cite{Lin:2007gi} it was argued that this BTZ partition function might be incomplete but we believe this not to be the case as we discuss further on.} Even though such states are normalizable (they can be determined numerically from a field theory point of view as discussed in section \ref{fieldtheory}), they are simply absent from the spectrum. There are also no discrete states bound to the black hole horizon. This is a consequence of the repulsive NS-NS background field. The situation is sketched in figure \ref{adsbtzfigure}.
These results show that the WZW BTZ model is markedly different from the standard behavior of strings near uncharged black holes where we expect a zero-mode localized at string length from the horizon \cite{Mertens:2013zya}. 
We can be more explicit about this. One can consider the $AdS_3$ partition function (as for instance given in equation (\ref{partfunct})) and rewrite this in terms of the characters of $\widehat{SL(2,\mathbb{R})}$. This has been done in terms of twisted characters in \cite{Gawedzki:1991yu}\cite{Gawedzki:1988nj}. In subsections \ref{toy} and \ref{Ham1} we go one step further and rewrite these in terms of normal characters where the conformal weights of the primaries are those we determined above. The result is that no spectral flowed states (in the angular $\phi$ dimension) are present: the partition function is entirely reproduced by the primaries (both untwisted and twisted in the Euclidean time dimension) and their descendants. For the BTZ partition function (obtained by simply setting $\beta \to 4\pi^2/\beta_{BTZ}$ in the $AdS_3$ partition function (\ref{partfunct})), one obtains the same result upon switching the interpretation of time and angular: thermal winding states are not present in the BTZ partition function.

\subsection{Cylinder-winding tachyons}
Secondly, there is the possibility of a winding tachyon in the $\varphi$ direction. The winding states in the $\varphi$ direction depend on $r_+$, the horizon location. 
Let us look at a $q=n=0$ state:
\begin{equation}
h^{w}_{j00} = \frac{s^2+1/4}{k-2}  + \frac{kw^2\left(\frac{r_+}{l}\right)^2}{4} + h_{int}.
\end{equation}
To have convergence for the BTZ partition function, we need
\begin{equation}
\frac{1}{4(k-2)}  + \frac{kw^2\left(\frac{r_+}{l}\right)^2}{4} \geq 1.
\end{equation}
Decreasing the BTZ temperature, decreases $r_+$. So there is a critical BTZ temperature below which a $\varphi$-winding tachyon appears.
The location of the horizon at this critical temperature is given by
\begin{equation}
r_+^2 = \alpha'\left(4-\frac{1}{k-2}\right),
\end{equation}
and the size of the black hole is clearly string size ranging from $r_+ = \sqrt{3}\sqrt{\alpha'}$ (for the limiting $k=3$ case) to $r_+ = 2 \sqrt{\alpha'}$ (as $k \to \infty$). So this tachyon is a consequence of shrinking the horizon to string scale and has nothing to do with the thermal scalar that we are interested in. The change in $\beta_{BTZ}$ that we have considered here in this paragraph is an on-shell change (changing the temperature changes the black hole size), whereas for the thermal scalar we are interested in an off-shell change of the temperature and the ensuing introduction of conical singularities. The $\varphi$-winding state causes the $AdS_3/BTZ$ black hole condensation process as was discussed in \cite{Berkooz:2007fe}. \\
For the type II superstring, the results of this and the preceding subsection do not change qualitatively.

\subsection{Summary}
Let us summarize the BTZ WZW black hole. There is no zero-mode surrounding the black hole. This is in sharp contrast with the generic uncharged black hole where such a zero-mode is present. It is only for small black holes that a stringy state becomes marginal, but this state is a Lorentzian state that wraps the cylindrical $\varphi$ dimension. It signals the $AdS_3/BTZ$ transition and it is not the high temperature thermodynamical state that we seek.

\section{$AdS_3$ orbifolds: conical spaces}
\label{conical}
In the previous section, we saw that the Euclidean BTZ WZW model does not contain cigar-winding string states in the thermal spectrum. For thermodynamical purposes, conical orbifolds of the cigar-shaped subspace are also important since these correspond to the string gas at a temperature different than the Hawking temperature. An intriguing possibility would be that the winding state is not present for the black hole itself, but when considering conical spaces the state might reappear. This could then possibly still give an important effect. In this section we analyze the thermal spectrum on such conical spaces, and in particular consider the question whether the thermal scalar is present or not. Our primary focus is again on the lowest weight state.
\subsection{Thermal spectrum}
Let us take a closer look at the orbifolds obtained by creating conical deficits with opening angle $\frac{2\pi}{N}$ at the tip of the cigar. First we remark that the prodedure of section \ref{spectrumsection} can be applied to this case and leads simply to $w \to \frac{a}{N}$ with $a\in\mathbb{Z}$. Hence one considers \emph{fractional} winding numbers \cite{Martinec:2001cf}. We have however no control on which of these states actually appear in the spectrum. To analyze this, a Hamiltonian analysis of the exact partition function is needed, to which we turn now. The thermal partition function of these orbifolds (for $AdS_3$ or BTZ) has the schematic form \cite{Son:2001qm}
\begin{equation}
Z = \frac{1}{N}\sum_{a,b}Z_{ab},
\end{equation}
where $a$ runs over the twisted states and the sum over $b$ realizes the projection on invariant states. The sum over $b$ ranges from $0$ to $N-1$. The individual partition functions that are summed over are given by
\begin{align}
\label{numer}
Z_{ab}(\tau) = \frac{\beta\sqrt{k-2}}{8\pi\sqrt{\tau_2}}\sum_{l,p}\frac{e^{-k\beta^2\left|l-p\tau\right|^2/4\pi\tau_2 + 2\pi \Im(U_{lp})^2/\tau_2}e^{\frac{\pi\tau_2}{2}}}
{\left|\sin(\pi U_{lp})\right|^2\left|\prod_{r=1}^{+\infty}(1-q^{r})(1-q^re^{2\pi i U_{lp}})(1-q^{r}e^{-2\pi i U_{lp}})\right|^2},
\end{align}
with
\begin{equation}
U_{lp} = \frac{b}{N} + \frac{a}{N}\tau_1 - i\frac{\beta}{2\pi}(p\tau_1-l) + i\frac{a}{N}\tau_2 + \frac{p\beta}{2\pi}\tau_2.
\end{equation}
In subsections \ref{Ham2}, \ref{Poisson} and \ref{Ham3} we rewrite this orbifolded partition function in a Hamiltonian way and we identify which states actually occur in the spectrum. The upshot is that in this case there are string states that wind the cigar, but only for $w=a/N$ with $\left|w\right| < \frac{1}{2}$ and integer $a$. The reason we restrict to this interval for $w$ will be explained in the next subsection.\footnote{A priori \emph{any} range of length $N$ is allowed for $a$. This follows from the fact that $Z_{a,b} = Z_{a+N,b}$ and $Z_{a,b} = Z_{a,b+N}$ which can be seen explicitly in equation (\ref{numer}). This is related to the periodicity of the Ray-Singer torsion.} In particular, one can choose the range as follows:
\begin{align}
a&= -\frac{N-1}{2} \to \frac{N-1}{2}, \quad N \text{ odd}, \\
a&= - \frac{N-2}{2} \to \frac{N}{2}, \quad N \text{ even}.
\end{align}
A crucial observation made in subsection \ref{Ham3} is that the sectors with $w \neq 0$ can include discrete states. This depends on whether $k\left|w\right|$ is larger or smaller than 1. Roughly, these discrete modes appear as follows. To rewrite the partition function in a Hamiltonian manner, one needs to employ the general Poisson summation formula. For $w \neq 0$ however, one actually needs a proper analytic continuation of Poisson's summation formula which is presented in subsection \ref{Poisson}. The naive substitution of complex arguments in the normal summation formula gives the continuous representations. This is not enough however: one needs to include extra terms in Poisson's summation formula corresponding to simple poles of the complex function. These precisely correspond to the discrete states that are otherwise completely missed.\\
For BTZ, it is clear that now the thermal spectrum does contain twisted strings that wind the temporal cigar. The situation is similar to Euclidean Rindler space and its orbifold cousins.\\
In terms of $AdS_3$ parameters, the thermal spectrum includes continuous states with conformal weights
\begin{align}
h^{wp}_{jqn} &= \frac{s^2+1/4}{k-2} +\frac{qw}{2} +\frac{i\pi nw}{\beta} + \frac{kw^2}{4}- i\frac{qp\beta}{4\pi} + \frac{ p n}{2} + \frac{kp^2\beta^2}{4(2\pi)^2}, \\
\bar{h}^{wp}_{jqn} &= \frac{s^2+1/4}{k-2} - \frac{qw}{2} +\frac{i\pi nw}{\beta} + \frac{kw^2}{4} - i\frac{qp\beta}{4\pi} - \frac{ p n}{2} + \frac{kp^2\beta^2}{4(2\pi)^2},
\end{align}
and discrete states with weights
\begin{align}
\label{disc1}
h &= -\frac{\tilde{j}(\tilde{j}-1)}{k-2} + \frac{qw}{2} - \frac{\pi i w n }{\beta} + \frac{kw^2}{4} - \frac{i \beta  pq}{4\pi} - \frac{pn}{2} + \frac{kp^2\beta^2}{4(2\pi)^2}, \\
\label{disc2}
\bar{h} &= -\frac{\tilde{j}(\tilde{j}-1)}{k-2} - \frac{qw}{2} - \frac{\pi i w n }{\beta} + \frac{kw^2}{4}  - \frac{i \beta  pq}{4\pi} + \frac{pn}{2} + \frac{kp^2\beta^2}{4(2\pi)^2},
\end{align}
where $\tilde{j} = M - l= \frac{k\left|w\right|}{2} - \frac{\left|q\right|}{2} \pm \frac{i\pi n }{\beta} - l$ and $l=0,1,2,\hdots$. We also have $q\in N\mathbb{Z}$, $n\in\mathbb{Z}$, $p\in\mathbb{Z}$ and $w = \frac{a}{N}$ with $a$ in the range determined above. The discrete states are present for $\Re(\tilde{j}) > \frac{1}{2}$. In particular for $k\left|w\right| < 1$, no discrete states are present at all. This is important in what follows. We remark further that for the thermal manifold, the discrete states really are discrete, unlike the discrete representations utilized in the Lorentzian $AdS_3$ spacetime \cite{Maldacena:2000hw} which are actually continuous because $\tilde{j}$ and $\tilde{m}$ are continuous quantum numbers there due to the fact that one considers the universal cover of the $SL(2,\mathbb{R})$ manifold.\\
Note that the discrepancy between $M$ and $\tilde{m}$ or $\tilde{\bar{m}}$ is expected and the same sort of situation occurs for the $SL(2,\mathbb{R})/U(1)$ black hole in which case it is well understood \cite{Dijkgraaf:1991ba}\cite{Aharony:2004xn}. Actually, the resemblance of the discrete weights with those of the $SL(2,\mathbb{R})/U(1)$ black hole is remarkable: in that case one finds for $M$:
\begin{equation}
M = \frac{k\left|w\right|}{2} - \frac{\left|q\right|}{2},
\end{equation}
which is of precisely the same form as above for the quantum numbers associated to the cigar-submanifold.

\subsection{Numerical analysis}
\label{numerical}
As a first method to analyze the critical large $\tau_2$ regime, we present the results of a numerical analysis of the expression (\ref{numer}). To start, we drop the infinite product present in the denominator. As discussed in subsection \ref{Ham1}, we expect this product to arise from oscillator states. Further on we will confirm numerically that this product does not influence the critical behavior. So we analyze numerically the expression
\begin{equation}
\label{EE}
E = \lim_{\tau_2 \to \infty} \sum_{l\in \mathbb{Z}}\frac{e^{(2-k)\frac{\pi l^2}{\tau_2}\frac{\beta^2}{4\pi^2} + 4\pi l w\frac{\beta}{2\pi} + 2 \pi w^2\tau_2}}{\left|\sin(\pi\left(iw\tau_2 + il\frac{\beta}{2\pi}\right))\right|^2}.
\end{equation}
This expression gives the asymptotic limit of part of the partition function (since $w$ is fixed) for the case $\tau_1=0$, $b=0$ and $p=0$. We will comment on the more general cases we are interested in below.\\
Firstly (crucially!), notice that we can not bring the large $\tau_2$ limit into the summation over $l$. \\
Numerically we can analyze the expression $E$ by truncating the series and then taking $\tau_2$ sufficiently large but still sufficiently small compared to the truncation index.\footnote{By `sufficiently' we mean that we varied the values of these parameters up to the point where the numerical result is neglegibly influenced by any further variation.} Using numerics as a guidance, we found the following asymptotic behavior.
\begin{itemize}
\item{
If $k\left|w\right| < 1$, one finds the following asymptotic behavior\footnote{Obtained by taking the logarithm of the numerical computation and then determing the slope of the resulting line.}
\begin{equation}
\label{expone}
E \propto e^{-k\pi w^2 \tau_2}.
\end{equation}
This is the expected behavior corresponding to a continuous state. Note that indeed one finds here that the density of states does not correct the critical behavior.
The prefactor of (\ref{expone}) has a periodicity in $\tau_2$ which equals\footnote{This was determined numerically and is not manifest in expression (\ref{EE}).}
\begin{equation}
\tau_2 \to \tau_2 + \frac{N\beta}{2\pi w}, \quad N \in \mathbb{Z}
\end{equation}
and indeed, this is the symmetry one expects from the CFT point of view of the critical behavior:
\begin{equation}
\sum_{n \in \mathbb{Z}} \rho(n) e^{\frac{4\pi^2 i n w \tau_2}{\beta}},
\end{equation}
with $\rho$ the density of string states.}

\item{
If $k\left|w\right| > 1$, we find
\begin{equation}
E \propto e^{-k\pi w^2 \tau_2}e^{\frac{\pi(k\left|w\right|-1)^2}{k-2}\tau_2}.
\end{equation}
More precisely, the prefactors can be determined as
\begin{equation}
E \to \frac{8\pi}{\beta}\sqrt{\frac{\tau_2}{k-2}} e^{-k\pi w^2 \tau_2}e^{\frac{\pi(k\left|w\right|-1)^2}{k-2}\tau_2}.
\end{equation}
We note that this behavior of the prefactor is also correct even when $\tau_1 \neq 0$ or $p\neq 0$ or $b\neq0$. One can readily see that this asymptotic behavior corresponds exactly to a discrete state including the prefactors.}
\end{itemize}
Let us now briefly discuss how the numerics change when we consider the more general case. Firstly, let us set $b \neq 0$. Numerically we checked that $b\neq0$ gives precisely the same asymptotics.\\
The case with $p\neq0$ is also easily analyzed. We consider now
\begin{equation}
\lim_{\tau_2 \to \infty} \sum_{l\in \mathbb{Z}}\frac{e^{(2-k)\frac{\pi l^2}{\tau_2}\frac{\beta^2}{4\pi^2} + 4\pi l w\frac{\beta}{2\pi} + 2 \pi w^2\tau_2-k\pi p^2\tau_2\frac{\beta^2}{4\pi^2}}}{\left|\sin(\pi\left(p\tau_2\frac{\beta}{2\pi} + iw\tau_2 + il\frac{\beta}{2\pi}\right))\right|^2}.
\end{equation}
The result gives the expected extra correction $\sim e^{-k\pi p^2\tau_2}$ but no mixing between $w$ and $p$ is generated by this, as indeed our analytical results also predict.\\
The above expressions do not fully coincide with what we are interested in: from the point of view of the spectra written down in the previous subsection, we want to set $q=0$, which is enforced by the $\tau_1$ integral. Simply setting $\tau_1$ equal to zero still gives us the sum over $q$ 
with the $q$-dependent density of states. We can extend the above numerical analysis to include $\tau_1$ dependence by studying instead
\begin{equation}
\lim_{\tau_2 \to \infty} \sum_{l\in \mathbb{Z}}\frac{e^{(2-k)\frac{\pi l^2}{\tau_2}\frac{\beta^2}{4\pi^2} + 4\pi l w\frac{\beta}{2\pi} + 2 \pi w^2\tau_2}}{\left|\sin(\pi\left(w\tau_1 + iw\tau_2 + il\frac{\beta}{2\pi}\right))\right|^2}.
\end{equation}
This expression was written down with $b=0$ and $p=0$. Numerical analysis yields the same asymptotic behavior as before, irrespective of the value of $\tau_1$ (it does however influence the prefactors). Integrating $\tau_1$ from $-1/2$ to $+1/2$ hence does not alter the asymptotic form.\footnote{For instance, consider the mean value theorem from elementary integral calculus: the $\tau_1$-integral equals the length of the interval (=1) times the function value at some intermediate point. But all intermediate points display the same asymptotic form. Thus the asymptotic form cannot be influenced by the integral over $\tau_1$.} Note that the sum over $b$ becomes irrelevant since the $\tau_1$ integral enforces $q=0$ and the restriction of $q \in N\mathbb{Z}$ achieved by summing over $b$ does not influence the final result.

\subsubsection*{Infinite oscillator product}
Up to this point, we did not analyze the infinite product present in equation (\ref{numer}). Its treatment is tricky. We dismissed it somewhat carelessly in subsection \ref{Ham1} when considering the Hamiltonian picture. A delicate point is that the Taylor expansion we should use for each of the factors of the infinite product depends on the precise value of $l$. Therefore we should split the sum over $l$ in different pieces. But we needed the entire sum over $l$ to perform the Poisson resummation formula. This is related to the fact that one cannot take the large $\tau_2$ limit through the summation over $l$. This is an issue that requires more thought and we will not discuss this further here.\\
Numerically however, we can analyze this infinite product (albeit in a truncated way of course).
One finds the following. Firstly periodicity $w \to w+1$ is recovered. This symmetry is exactly present in (\ref{numer}) due to the periodicity properties of the Ray-Singer torsion, but is compromised upon dropping the infinite product. It is nice to find numerical evidence that when including a truncated version of this infinite product, the symmetry becomes more and more restored.\footnote{The symmetry restoration is apparent only for values of $\left|w\right|$ not too large. Roughly speaking, if one wants the periodicity domain to increase by an integer, one should include one more factor in the numerical treatment of the infinite product.} Secondly, for values of $w$ that satisfy $\left|w\right|<\frac{1}{2}$, the infinite product does not contribute to the large $\tau_2$ limit and one can hence trust the above expressions to yield the dominant behavior. If $w$ is outside this interval, the infinite product makes a contribution that survives the large $\tau_2$ limit, precisely to restore the periodicity $w \to w+1$. Hence we restrict $w$ to $\left|w\right|<\frac{1}{2}$ and drop the infinite product. All other intervals of $w$ should be found by periodicity. As noted before, this numerical treatment is actually the only indication we have on how the infinite product affects the critical behavior.

\subsection{A brief look back at the $AdS_3$ string gas}
\label{numads}
Before we proceed, let us briefly return to the normal $AdS_3$ space. The numerical methods discussed in the previous subsection can also be used for this (easier) case. We hence analyze
\begin{equation}
E = \lim_{\tau_2 \to \infty} \sum_{l\in \mathbb{Z}}\frac{e^{(2-k)\frac{\pi l^2}{\tau_2}\frac{\beta^2}{4\pi^2} - k\pi p^2\frac{\beta^2}{4\pi^2}\tau_2}}{\left|\sin(\pi\left(\frac{p\beta}{2\pi}\tau_2 + i\frac{\beta}{2\pi}l\right))\right|^2}.
\end{equation}
One finds
\begin{equation}
E \sim e^{-k\pi p^2\frac{\beta^2}{4\pi^2}\tau_2}
\end{equation}
with a prefactor periodic in $\tau_2$ as
\begin{equation}
\tau_2 \to \tau_2 + \frac{2\pi N}{\beta p}, \quad N \in \mathbb{Z},
\end{equation}
which is in accord with what we argued for in subsection \ref{AWtach}. There we discussed the fact that all $q\in\mathbb{Z}$ stringy states contribute to the critical regime. And indeed, their sum respects this symmetry since it gives (schematically):
\begin{equation}
\sum_{q \in \mathbb{Z}} \rho(q) e^{i q p \beta \tau_2}.
\end{equation}
It is hard to imagine how this periodicity would be generated if the $q\in\mathbb{Z}$ states would be subdominant.\footnote{For instance setting $q=0$ would not give a critical behavior that respects this symmetry.} A periodic prefactor of the critical behavior was also explicitly determined in \cite{Lin:2007gi}. \\
Note that in this case no change of dominant behavior occurs. This case is hence for all values of the parameters similar to the $k\left|w\right| < 1$ case of the orbifolds discussed above. 

\subsection{Dominant state}
\begin{figure}[h]
\centering
\begin{minipage}{.4\textwidth}
  \centering
  \includegraphics[width=\linewidth]{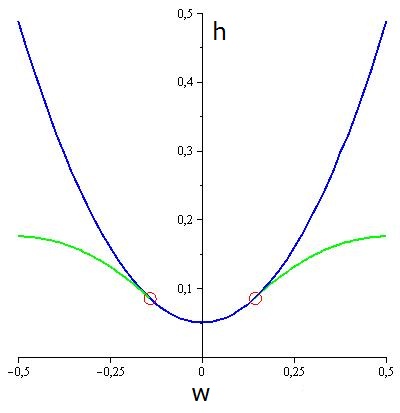}
  \caption*{(a)}
\end{minipage}%
\begin{minipage}{.4\textwidth}
  \centering
  \includegraphics[width=\linewidth]{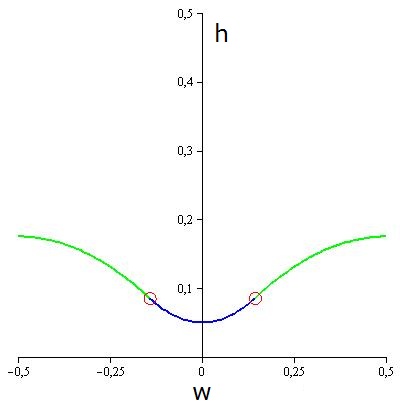}
  \caption*{(b)}
\end{minipage}
\caption{(a) Conformal weight $h$ of the most tachyonic mode of both types of representations (continuous and discrete) as a function of $w$. The blue curve originates from the continuous representations. The green curve represents the lowest mode of the discrete representations. The latter only appears when $\left|w\right| > \frac{1}{k}$, represented by the red circles. (b) Conformal weight of the overall lowest weight state relevant for stability issues. As soon as the discrete representations are present, they dominate the continuous modes.}
\label{weight}
\end{figure}

Let us now study the large $\tau_2$ limit analytically. Which state is the dominant one? This is easy to analyze. First let us take a look at the continuous states. The situation is the same as that analyzed in section \ref{AWtach} and one can repeat the entire discussion given there, including the arguments presented in section \ref{dos} as to why the density of states does not influence the critical weight.\\
Next, we focus on the discrete states. Consider (part of) the conformal weight $h = - \frac{\tilde{j}(\tilde{j}-1)}{k-2}$. The most negative conformal weight dominates. Since $\tilde{j}>1/2$ is required, the dominant state has the largest value of $\tilde{j}$ allowed by the constraints. This is then obviously a state which has $q=l=0$. The $n$ quantum number has an $n^2$ contribution in the conformal weight as given before. It is hence dominated by $n=0$.\\
Note that this is a major qualitative difference between the continuous states and the discrete states. For continuous states in $AdS_3$ we noted before in subsection (\ref{AWtach}) that the dominant state has arbitrary $q\in\mathbb{Z}$ and one should sum these to get the critical behavior. The same story applies here for the continuous states: arbitrary $n$ is allowed and we should sum these for the critical behavior (and in order to satisfy the periodicity in $\tau_2$ of the prefactors as discussed in the previous subsections). For discrete states on these orbifolds, $n$ and $q$ are both set to zero. Hence no summation is required to obtain the critical behavior. The reason for this peculiar behavior of the continuous states is the absence of $n^2$ and $q^2$ terms in the expressions for the conformal weights (\ref{ads3spectrum1prelim}) and (\ref{ads3spectrum2prelim}). We will discuss this further in section \ref{thscaction}. \\
For any value of $N$, one can then determine whether the system is stable or not by considering the mode with lowest conformal weight. The conformal weight $h$ as a function of $w = \frac{a}{N}$ is given in figure \ref{weight}. Even though this figure is drawn for $w$ a continuous variable, one should keep in mind that in principle we only determined this for $w= \frac{a}{N}$. From the numerical approach, we learned that we should restrict to $\left|w\right| < 1/2$. The lowest mode for a fixed value of $N$, is given by taking $w = \frac{1}{N}$. In particular, the dominant state is discrete for $N < k$ and is continuous otherwise.
We remark that the curve of the lowest conformal weight (as displayed in figure \ref{weight}), is smooth across the `joints' $\left|w\right| = \frac{1}{k}$. This analysis is in precise agreement with the numerical study presented in the previous subsection.\\
Note further that since $k > 2$, these orbifold models \emph{always} include twisted discrete states in their spectrum. \\
If we analyze solely the large $\tau_2$ behavior with $w$ arbitrary and not summed over, i.e. one of the expressions we studied numerically in the previous subsection, then the lowest conformal weight displayed in figure \ref{weight} should be periodically continued with period 1 both to the left and to the right.

\subsection{Hagedorn temperature}
\begin{figure}[h]
\centering
\includegraphics[width=7cm]{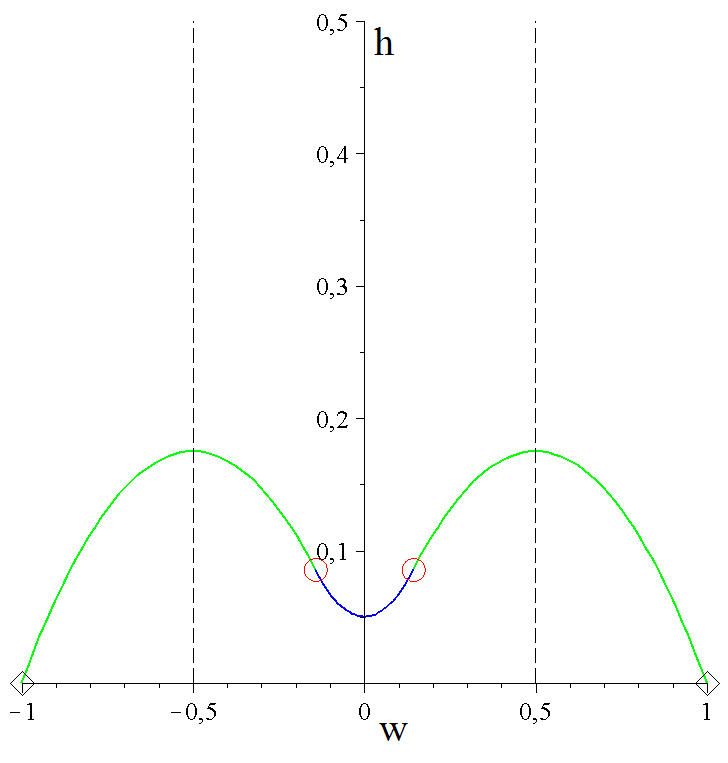}
\caption{Conformal weight of lowest weight state when continuing the expressions up to $\left|w\right| =1$. The vertical dashed lines denote the boundary $\left|w\right| = \frac{1}{2}$. The two black diamonds depict the weight of the state at $\left|w\right| =1$, where it becomes zero.}
\label{weight3}
\end{figure}

\begin{figure}[h]
\centering
\begin{minipage}{.33\textwidth}
  \centering
  \includegraphics[width=\linewidth]{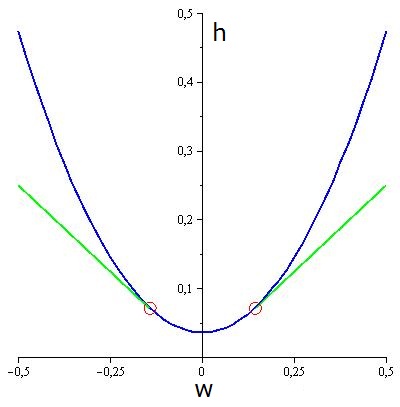}
  \caption*{(a)}
\end{minipage}%
\begin{minipage}{.33\textwidth}
  \centering
  \includegraphics[width=\linewidth]{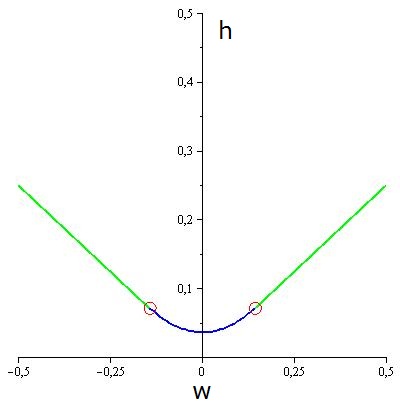}
  \caption*{(b)}
\end{minipage}
\begin{minipage}{.33\textwidth}
  \centering
  \includegraphics[width=\linewidth]{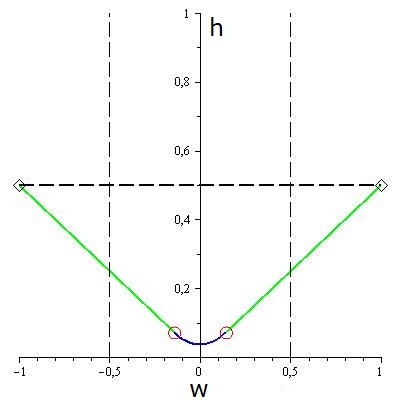}
  \caption*{(b)}
\end{minipage}
\caption{(a) and (b) represent the analogous figures for the type II superstring as figure \ref{weight} does for the bosonic string. (c) The curves are drawn all the way to $\left|w\right|=1$, where the green curve is found to precisely equal $\frac{1}{2}$. This point is depicted with black diamonds. The vertical dashed lines denote the boundary $\left|w\right| = \frac{1}{2}$.}
\label{IIweight}
\end{figure}

Can we utilize these observations to determine the Hagedorn temperature of the BTZ black hole? Analogous to the treatment of conical orbifolds of the flat plane \cite{Dabholkar:1994ai}\cite{Lowe:1994ah}\cite{Mertens:2013zya}, the $\frac{1}{N}$ parameter should be interpreted as $\frac{\beta}{\beta_{BTZ}}$, the ratio of the actual temperature and the Hawking temperature. The lowest state for each $N$ is given by choosing $\left|w\right|=\frac{1}{N}$, thus setting $a=\pm1$. This corresponds to taking $w = \frac{\beta}{\beta_{BTZ}}$. To determine whether the BTZ black hole itself is stable, we are hence interested in taking $N\to 1$. Thus we would like to continue the above expressions to $N = 1$. The discussion that follows is hence necessarily more speculative than the previous results. Unlike for the flat $\mathbb{C}/\mathbb{Z}_N$ cones, in this case this continuation in $N$ is more ambiguous: we have a piecewise definition of the lowest conformal weight and it is a priori unclear which is the correct way to proceed. A naive application of the above periodicity results would suggest then that the lowest weight state for $w=1$ is the same as that for $w=0$ and the BTZ black hole would be unstable. However, we believe this is not correct for the following reasons. Firstly, from a Lorentzian point of view, the canonical partition function is given by\footnote{Let us remark that the equality of the partition function on the thermal manifold at one loop and the Lorentzian thermal free field trace is not a settled issue for black hole spacetimes \cite{Susskind:1994sm}. Different arguments are presented below.}
\begin{equation}
\label{Lorpart}
Z = \sum_{n\in\mathcal{H}}e^{-\beta E_n}.
\end{equation}
It is clear from this formula that periodic behavior of the system as $w \to w + 1$ or $\beta \to \beta + \beta_{BTZ}$ is impossible. The reason for this periodicity symmetry can be traced back to the following. As we discuss in subsection \ref{Ham1}, the first step in evaluating the partition function using path integral methods consists of performing a coordinate transformation that makes manifest that $\phi$ is an angular coordinate. However, this coordinate transformation is 1:1 only when the range of $\phi$ is less than $2\pi$. Thus the periodicity $w \to w+1$ is an artifact of the new coordinates and is \emph{not} a symmetry of the original space. \\
A better approach, which we believe to be the correct one, is to ignore this periodicity. The presence of discrete states causes the partition function to diverge for $w \approx 1$. In figure \ref{weight3} we draw the lowest conformal weight when we continue the discrete state all the way to $\left|w\right| = 1$. Curiously, the conformal weight becomes zero and the conclusion seems to be that the bosonic BTZ WZW model is divergent at the Hawking temperature. In fact, it appears to be divergent for any value of $\beta$. The computation of the critical weight for the discrete states is actually completely the same as that for the discrete modes in Euclidean Rindler space \cite{Mertens:2013zya}. In both cases, it is very ambiguous how to correctly interpret the continuation $N\to1$ in terms of convergence or divergence of thermodynamic quantities.\footnote{Note further that even if we did take the $\beta \to \beta + \beta_{BTZ}$ periodicity seriously, our conclusion would remain unaltered: the singly wound state yields a divergence for $\left|w\right| \to 1$.}
Despite the speculative nature of the above discussion, for \emph{conical} BTZ spaces the dominant behavior is well-defined. The thermal scalar on these BTZ conical spaces for $N < k$ is characterized by $\left|w\right|=\frac{1}{N}$,\footnote{Winding $\pm 1$ on the cone is equivalent to winding $\pm \frac{1}{N}$ on the `covering' space of the cone, i.e. the unorbifolded space.} $q=0$, $l=0$, $p=0$ and $n=0$. Unlike the thermal scalar on the $AdS_3$ space which included a summation over $q$ to distill the critical random walk behavior, in this case no summation is required.\\
Let us look at the analogous formulas for the type II superstring. We did not determine these ab initio, but it seems obvious how to modify the resulting conformal weights: one simply replaces $k-2 \to k$ in the denominator of the first term of the conformal weights.\footnote{We have already argued for this before in section \ref{Hagsup}.} After integrating out the continuous quantum number $s$, the most tachyonic continuous state hence has
\begin{equation}
h = \bar{h} = \frac{1}{4k} + \frac{kw^2}{4},
\end{equation}
whereas the most tachyonic discrete state (for $k\left|w\right| > 1$) has
\begin{equation}
h = \bar{h} = - \frac{\frac{kw}{2}\left(\frac{k\left|w\right|}{2}-1\right)}{k} + \frac{kw^2}{4} = \frac{\left|w\right|}{2}.
\end{equation}
Crucially, unlike the bosonic string, the term quadratic in $w$ cancels out and we are left with a linear dependence on $w$. Moreover, and this is the intriguing part of this analysis, if we demand that $h = \frac{1}{2}$, we find precisely $\left|w\right|=1$, i.e. $\beta = \beta_{BTZ}$. The situation is drawn in figure \ref{IIweight}.
A further field-theoretic argument in favor of this continuation in $N$ will be given in section \ref{cigarwind}. We are inclined to believe these results since they are very reminiscent on the results for $SL(2,\mathbb{R})/U(1)$ black holes and its Euclidean Rindler limit. In fact, qualitatively the situation is almost the same. Bosonic strings have (save for the non-thermal closed string tachyon) a convergent free energy, but as soon as the temperature is varied, thermodynamic quantities (such as the thermal entropy) diverge. Type II superstrings precisely have $T_H = T_{Hawking}$, meaning a marginal convergence is achieved for thermodynamic quantities. The major difference is that the thermal scalar state is absent for the BTZ black hole itself.

\subsection{Summary}
In this section we analyzed conical orbifolds obtained by identifying points on the cigar after a certain rotation. The partition function contains states that wind the cigar now. A surprising result is that the spectrum includes also a set of discrete states in the twisted sectors. These states appear only for $k\left|w\right| >1$. Mathematically, the appearance of the discrete states can be traced back to a correct analytic continuation of Poisson's summation formula. These discrete states are crucial, since if they are present, they dominate the continuous states. We also presented a numerical analysis of the partition function in the large $\tau_2$ limit. The results agree with the predictions in terms of the spectrum. The two major lessons we learned from the numerical analysis are the following. Firstly, the summation over $l$ and the large $\tau_2$ limit cannot be interchanged. Secondly, the infinite product gives a subdominant contribution as long as $\left|w\right| < \frac{1}{2}$. Else, the infinite product simply restores the periodicity $w \to w + 1$ that is not present in the partition function if one simply drops the infinite product. We then discussed how the Hagedorn temperature emerges for the BTZ black holes. We continued the conformal weight of the discrete representations all the way to $\left|w\right|=1$, and found a divergence for the bosonic string BTZ black hole. For type II superstrings on the BTZ black holes however, we find a marginal convergence: $\beta_H = \beta_{BTZ}$.

\section{The inclusion of a chemical potential for the $AdS_3$ string gas}
\label{chemical}
\subsection{Thermal spectrum}
A simple generalization of the $AdS$ WZW model at finite temperature is by substituting $\beta \to \beta(1+i\mu)$ in $U_{lp}$ as defined in equation (\ref{Ulp}) in the exact path integral treatment. This corresponds physically to the introduction of a chemical potential for the angular momentum of the string gas around the cigar-shaped angular submanifold. On the Euclidean manifold this is realized as the simultaneous identification $\tau \sim \tau + \beta$ and $\phi \sim \phi + \mu\beta$. As a small reminder, this can be seen by writing down the grand-canonical partition function:
\begin{equation}
\mathcal{Z} = \text{Tr}e^{-\beta(H+i\mu Q)}
\end{equation}
with $Q$ the conserved charge. This is the generator of angular rotations in our case. Again as usual anti-periodic boundary conditions for the fermions are required. These identifications lead to skew tori as fundamental domains in the ($\phi$, $\tau$) plane as shown in figure \ref{chemi} below.
\begin{figure}[h]
\centering
\includegraphics[width=10cm]{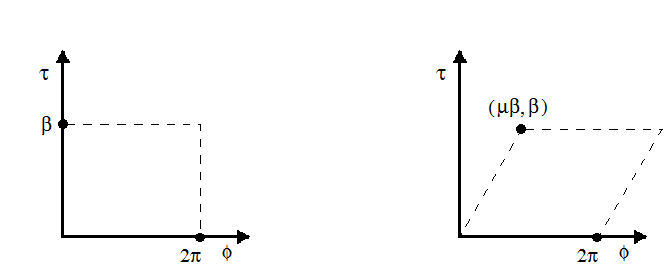}
\caption{Left figure: The identifications of the coordinates $\tau \sim \tau + \beta$ and $\phi \sim \phi + 2\pi$ define a rectangular torus as fundamental domain. Right figure: When including $\mu$, the two identifications are more involved. The first is the simultaneous identification of $\tau \sim \tau + \beta$ and $\phi \sim \phi + \mu\beta$. The second identification is again $\phi \sim \phi + 2\pi$. These define a more general skew torus as fundamental domain.}
\label{chemi}
\end{figure}
The path integral and Hamiltonian interpretation techniques that we analyzed in subsection \ref{Ham1} provide the fastest way to get the string spectrum in this case. In fact, our results on angular orbifolds in subsections \ref{Ham2}, \ref{Poisson} and \ref{Ham3} can be almost exactly copied to study this case. We present the computational details in subsection \ref{chemapp}. One finds a continuous spectrum of states with conformal weights
\begin{align}
\label{chemw1}
h^{p}_{sqn} &= \frac{s^2 +1/4}{k-2} + i\frac{\mu n p}{2} - i\frac{qp\beta}{4\pi} + \frac{ p n}{2} + \frac{kp^2\beta^2}{4(2\pi)^2}(1+\mu^2) - i \frac{\mu^2\beta qp}{4\pi}+ h_{int}, \\
\label{chemw2}
\bar{h}^{p}_{sqn} &= \frac{s^2 +1/4}{k-2} + i\frac{\mu n p}{2} - i\frac{qp\beta}{4\pi} - \frac{ p n}{2} + \frac{kp^2\beta^2}{4(2\pi)^2}(1+\mu^2) - i \frac{\mu^2\beta qp}{4\pi} + \bar{h}_{int},
\end{align}
and a set of discrete states with weights
\begin{align}
\label{chemw3}
h^{p}_{jqn} &= -\frac{\tilde{j}(\tilde{j}-1)}{k-2} - i\frac{\mu n p}{2} - i\frac{qp\beta}{4\pi} - \frac{ p n}{2} + \frac{kp^2\beta^2}{4(2\pi)^2}(1+\mu^2) - i \frac{\mu^2\beta qp}{4\pi}+ h_{int}, \\
\label{chemw4}
\bar{h}^{p}_{jqn} &= -\frac{\tilde{j}(\tilde{j}-1)}{k-2} - i\frac{\mu n p}{2} - i\frac{qp\beta}{4\pi} + \frac{ p n}{2} + \frac{kp^2\beta^2}{4(2\pi)^2}(1+\mu^2) - i \frac{\mu^2\beta qp}{4\pi} + \bar{h}_{int},
\end{align}
where now $ \tilde{j} = \frac{k\left|\mu p\right|\beta}{4\pi} - \frac{\left|q\right|}{2} - \frac{i\mu q}{2} \pm \frac{in\beta}{\pi} - l$. The quantum numbers take values as follows: $q \in \mathbb{Z}$, $n\in\mathbb{Z}$, $p\in\mathbb{Z}$ and $l = 0,1,2,\hdots$. The discrete states include all states that satisfy $\Re(\tilde{j}) > 1/2$. The unitarity constraint $\Re(\tilde{j}) < \frac{k-1}{2}$ is trivially satisfied for all such states, provided $k>2$.

\subsection{Dominant state and critical Hagedorn thermodynamics}
\begin{figure}[h]
\centering
\includegraphics[width=12cm]{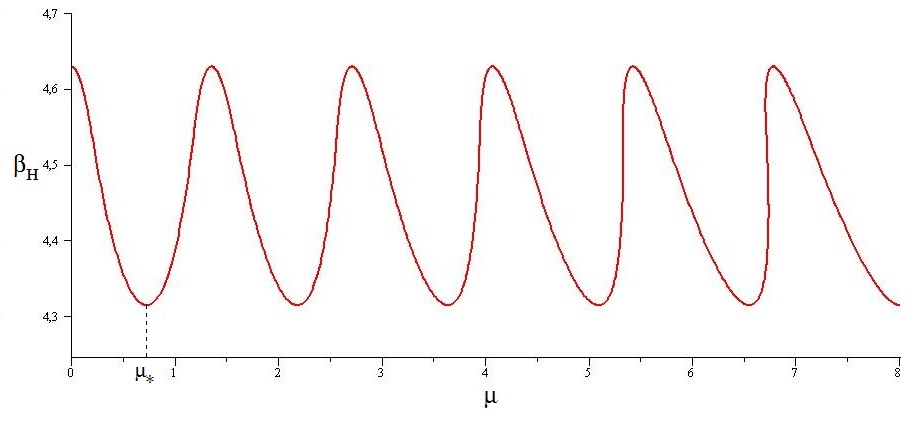}
\caption{$\beta_H$ as a function of chemical potential $\mu$ for the case $k=7$. Above the displayed curve, the system is stable. Below the curve, the system is unstable. Note that for increasing chemical potential, the curve becomes deformed.}
\label{Hagechem}
\end{figure}

\begin{figure}[h]
\centering
\includegraphics[width=6cm]{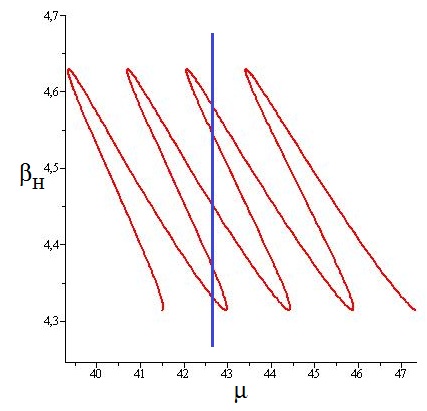}
\caption{$\beta_H$ as a function of chemical potential $\mu$ for sufficiently large $\mu$ for $k=7$. The blue line denotes a thermodynamic path one could follow in heating (or cooling) the system at fixed chemical potential. It is now apparent that multiple crossing with the critical curve are possible, indicating an alternating convergent and divergent system.}
\label{Hagechem2}
\end{figure}

\begin{figure}[h]
\centering
\includegraphics[width=7cm]{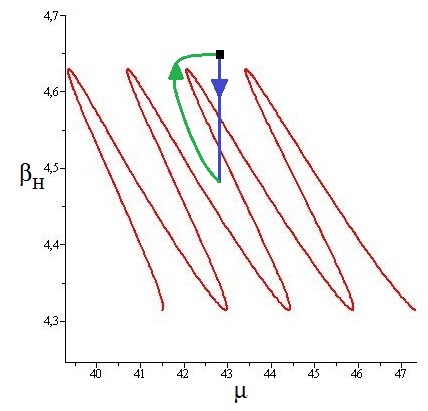}
\caption{Thermodynamical path through the ($\mu$, $\beta$) plane. We start at the black square. First the system is heated following the blue path. Then the green path is followed to finally end up at the same point from which we started.}
\label{TD}
\end{figure}

\begin{figure}[h]
\centering
\includegraphics[width=7cm]{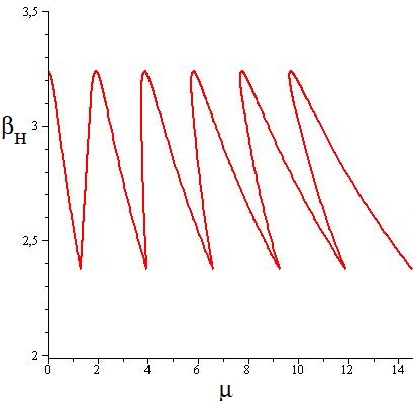}
\caption{$\beta_H$ as a function of chemical potential $\mu$ for the case $k=7$ for the type II superstring. Above the displayed curve, the system is stable. Below the curve, the system is unstable. Again the curve deforms for increasing $\mu$.}
\label{IIchem}
\end{figure}

As for the $AdS_3$ orbifolds, the dominant state can be either continuous or discrete depending on the value of $k\left|w\right|$.\\
For $k\frac{\left|\mu\right|\beta}{2\pi}<1$, the dominant state is continuous and characterized by $p=\pm1$, $n=0$ but arbitrary $q$.\\
The dominant discrete state for $k\frac{\left|\mu\right|\beta}{2\pi}>1$ is again given by considering $q=l=n=0$ and $p = \pm 1$. It is the same state as the one in the conical $AdS_3$ space characterized by $p=\pm1$, $w= \frac{\mu\beta}{2\pi} p$. \\
The numerical analysis done in the previous section can be readily extended to include this case as well.\footnote{More precisely, one should analyze
\begin{equation}
\lim_{\tau_2 \to \infty} \sum_{l\in \mathbb{Z}}\frac{e^{(2-k)\frac{\pi l^2}{\tau_2}\frac{\beta^2}{4\pi^2} + 4\pi l w\frac{\beta}{2\pi} + 2 \pi w^2\tau_2-k\pi p^2\frac{\beta^2}{4\pi^2}\tau_2}}{\left|\sin(\pi\left(\frac{p\beta}{2\pi}\tau_2 + iw\tau_2 + i\frac{\beta}{2\pi}l - \frac{\beta}{2\pi}l\mu\right))\right|^2}, 
\end{equation}
which yields results in agreement with the analytical predictions. Actually, an additional numerical consistency check can be performed. In the regime where the continuous state dominates, the prefactor of the leading behavior is expected to exhibit periodicity in $\tau_2$. According to the precise form of the weights (\ref{chemw1}) and (\ref{chemw2}), the periodicity is 
\begin{equation}
\text{lcm}\left(\frac{1}{\mu},\frac{2\pi}{\beta(1+\mu^2)}\right), 
\end{equation}
which as a sidenote exists only when the ratio of these two numbers is rational. Such a periodicity is indeed what is observed numerically.} \\
After including the spectator dimensions, the condition to find the Hagedorn temperature is given by
\begin{equation}
\frac{1}{4(k-2)} + k \frac{\beta^2}{16\pi^2}(1+\mu^2) = 1
\end{equation}
when $\frac{\left|\mu\right|\beta}{2\pi} < \frac{1}{k}$ or by
\begin{equation}
\label{discrHag}
\frac{1}{4(k-2)}-\frac{(\frac{k\left|\mu\right|\beta}{2\pi}-1)^2}{4(k-2)} + k \frac{\beta^2}{16\pi^2}(1+\mu^2) = 1
\end{equation}
when $\frac{1}{k} < \frac{\left|\mu\right|\beta}{2\pi} < \frac{1}{2}$. For even larger values of the chemical potential, periodicity of the system under $w \to w+1$ should be used. This periodicity is obvious from a Lorentzian point of view as well, since the grand-canonical partition function is given by \cite{Maldacena:2000hw}\cite{Maloney:2007ud}
\begin{equation}
\label{grandcan}
\mathcal{Z} = \sum_{n\in\mathcal{H}}e^{-\beta E_n}e^{il_n \mu\beta}
\end{equation}
for integer angular momentum $l_n$, which is manifestly periodic under
\begin{equation}
\label{periodic}
\mu \to \mu + \frac{2\pi N}{\beta}, \quad N \in\mathbb{Z}.
\end{equation}
If $\frac{\left|\mu\right|\beta}{2\pi} < \frac{1}{k}$, the Hagedorn temperature is readily found and is given by
\begin{equation}
\beta_H^2 = \frac{4\pi^2}{k(1+\mu^2)}\left(4-\frac{1}{k-2}\right),
\end{equation}
which is the expression written down in \cite{Lin:2007gi}. However, for other values of the chemical potential, the Hagedorn temperature disagrees with the above formula and should instead be determined by equation (\ref{discrHag}). For completeness, the Hagedorn temperature in this regime is given by the not-so-transparant expression:
\begin{equation}
\beta_H = -2\pi\frac{k\left|\mu\right|-\sqrt{(-7k^2\mu^2+16k\mu^2-16k^2+4k^3+16k)}}{k(-2\mu^2+k-2)}.
\end{equation}
For even larger values of the chemical potential, one needs to use the periodicity (\ref{periodic}). However, the periodic parameter is $\mu\beta$ and not $\mu$ itself. This then leads to a distortion of the $\beta_H(\mu)$ curve for larger values of $\mu$. Note that the Hagedorn temperature only depends on $\left|\mu\right|$, thus the critical curve is symmetric under $\mu \to - \mu$. The resulting critical curve is depicted in figure \ref{Hagechem}.
Note that indeed the critical curve is smooth at the points where a different formula for the Hagedorn temperature should be used, as we noticed before.
For very large values of $\mu$, the curve can even `fall over' such that for a single value of $\mu$, multiple critical temperatures exist. The interpretation is then as follows. Starting from a low-temperature gas at fixed chemical potential, we turn up the temperature. At the beginning, the system is thermodynamically stable. Then at some temperature (which is at least as high as the $\mu=0$ Hagedorn temperature), the system becomes unstable. However, when bearing through this region, one again encounters a regime where the system is stable. This stability is soon after again compromised and one re-enters the divergent Hagedorn phase. For very large temperatures, being larger than the temperature associated to $\mu_{*} = \sqrt{\frac{k(k-2)}{14k-32}}$, the system always becomes unstable. Thus there is an interval of temperatures, where depending on the chemical potentials, the system can alternate between convergent and divergent behavior. This strange zone is bounded as follows:
\begin{equation}
\frac{\sqrt{k}}{2\pi}\sqrt{\frac{k-2}{4k-9}} < T < \frac{\sqrt{k}(k-2)}{\sqrt{2}\pi\sqrt{7k^2-30k+32}},
\end{equation}
where the upper temperature is computed as $T(\mu = \mu_{*})$. This feature is illustrated in figure \ref{Hagechem2}.
Let us finally remark that the Hagedorn temperature is always at least as high as the $\mu=0$ Hagedorn temperature. This is intuitively obvious \cite{Lin:2007gi}, since we expect string states with prescribed average angular momentum to be less numerous than the general string states. Thus a higher temperature is necessary to get a sufficient number of string states to yield Hagedorn behavior.\\
The alternating divergence behavior is quite strange. In fact, in the past a similar situation arose when computing the one-loop free energy of heterotic strings in flat space \cite{O'Brien:1987pn}.\footnote{The flat space heterotic string has two critical temperatures and divergences occur only in between these two temperatures. This suggests at first sight that the heterotic string is again stable at high temperature.} It was shown \cite{Atick:1988si} that this behavior is unphysical and incompatible with the monotonicity of a canonical partition function in $\beta$. The genus zero condensate of the thermal heterotic string in flat space cannot disappear at higher temperatures. In this case however, we are using the grand-canonical partition function (\ref{grandcan}) which need not be monotonic in $\beta$. Hence the above behavior is in principle allowed. Even more so, we now present an argument that the persistence of the genus zero condensate as in \cite{Atick:1988si} cannot occur here. The crucial ingredient is the fact that we have an extra parameter $\mu$ to play with. Consider a thermodynamical path as given in figure \ref{TD}.
First we follow the blue curve by heating the system. Then we follow a suitable trajectory in the ($\mu$, $\beta$) plane (the green curve) that does not cross the critical curve. We end up at the same point. This point could have been chosen arbitrarily low in temperature and hence we expect the system to converge there in the low-temperature phase, both initially and finally after following the thermodynamical process. However, a genus zero condensate forms as soon as one crosses the critical curve the first time. If this condensate persists after crossing the critical curve the second time, then this path shows that it is possible to cool down the system without any more crossings, which implies that the low temperature system we obtain after this entire process would still contain a genus zero condensate. This is impossible.\\
Generalizing this argument, if the convergent region in the plane of parameters (here $\mu$ and $\beta$) is connected then the entire critical curve is important and the genus zero condensate can disappear again at higher temperatures. If it is not connected, one cannot a priori know what will happen at higher temperature as is the case for the heterotic string in flat space. In that case, the canonical partition function itself shows that one cannot return to a convergent region with the same degrees of freedom as the low-temperature phase.\\
For completeness, we present the analogous picture for the type II superstring (assuming the obvious replacements) in figure \ref{IIchem}.
Peculiar to note is that the smooth gluing of the piecewise defined function for the bosonic string is not present here: the critical curve exhibits points where it is non-differentiable.

\subsection{BTZ with $\mu \neq 0$}
In the previous section, we noted that the BTZ thermal spectrum does not include the thermal scalar state. We then considered conical orbifolds and found that the thermal scalar reappears. A different question one could ask is whether introducing a chemical potential for the BTZ black hole can cause the thermal scalar to appear. Let us briefly look into this.
One readily finds the generalization of the relation between $AdS_3$ and BTZ for non-vanishing chemical potential (in terms of $AdS_3$ parameters):
\begin{align}
\label{firstid}
\tau &\sim \tau + \frac{2\pi r_+ }{l}, \\
\phi &\sim \phi + 2\pi, \, \tau \sim \tau + 2\pi \mu_{BTZ}.
\end{align}
This is a situation we have not yet analyzed, both identications include the temporal dimension. The $\phi$ identification is trivial and we can drop it. Since the identification is purely in the $AdS$ time direction, we do not encounter the problems with the $w \to w + 1$ periodicity shift. The result is an $AdS_3$ where the thermal direction includes two independent identifications: a double toroidal model. More elaborate computations for this set-up will not be treated here. We just remark that cigar-winding states make a reappearance here due to the non-trivial identification $\tau \sim \tau + 2\pi \mu_{BTZ}$.
We hope to come back to this model in the future.

\subsection{$SL(2,\mathbb{Z})$ family of black holes}
The treatment above for the chemical potential in $AdS$ actually immediately allows us to determine the thermal spectrum on the entire $SL(2,\mathbb{Z})$ family of black holes in $AdS_3$ \cite{Maldacena:1998bw}. This is the black hole family treated in the so-called Farey Tail story \cite{Dijkgraaf:2000fq}\cite{deBoer:2006vg}. A difference however is that we include a background Kalb-Ramond field to obtain a valid string background. We will not keep track of this field in this section. This section is a generalization of the treatment of the $BTZ$ black hole in section \ref{BTZsection} and was not included in \cite{Mertens:2014nca}. \\
Consider the thermal $AdS_3$ metric:\footnote{We have set $l=1$ here for ease of notation.}
\begin{equation}
ds^2 = \cosh(\rho)^2d\tau_E^2 + d\rho^2 + \sinh(\rho)^2d\phi^2,
\end{equation}
with the identifications $\phi \sim \phi + 2\pi$ and $\tau_E \sim \tau_E + \beta$. This can be generalized to an entire family of classical saddle points as follows. Define $ \tau = i\frac{\beta}{2\pi}$. Then the identifications can be rewritten as
\begin{equation}
u = \phi + i \tau_E \sim u + 2\pi \sim u + 2\pi \tau.
\end{equation}
Performing a $PSL(2,\mathbb{Z})$ transformation on the modulus $\tau$, we obtain an entire family of solutions, with identifications as
\begin{equation}
\label{identi}
\phi \sim \phi + 2 \pi \quad \text{and} \quad \tau_E \sim \tau_E +2 \pi \Im \tau, \quad \phi \sim \phi + 2\pi \Re \tau.
\end{equation}
Some of these are actually the same solution, such as changing $\tau \to \tau + n$ for the above $AdS_3$ metric does not change anything. \\

\noindent It can be proven that the set of independent $\tau$'s are of the form
\begin{equation}
\tau = \frac{A\tau_0 + B}{C\tau_0 + D},
\end{equation}
for some reference $\tau_0$ with relatively prime $C$ and $D$ and an arbitrary choice of $A$ and $B$ associated to this (satisfying $AD-BC=1$). We choose $\tau_0 = i\frac{\beta}{2\pi}$ to be the $AdS_3$ case. To prove this, one uses the following two simple lemmas (whose proofs are omitted). \\

\noindent \textbf{Lemma 1}: $C$ and $D$ are relatively prime. \\

\noindent \textbf{Lemma 2}: Given $C$ and $D$, then $A$ and $B$ are uniquely determined up to shifts $(A,B) \to (A,B) + N(C,D)$ for $N \in \mathbb{Z}$.\\

\noindent The identifications above (\ref{identi}) (in fact the invariance under $\tau \to \tau + n$) however show that the transformation $(A,B) \to (A,B) + N(C,D)$ for $N \in \mathbb{Z}$, is a trivial modification. Hence we conclude that the different sectors are only those with relatively prime $C$ and $D$ and one associated choice of $A$ and $B$ (chosen at will). Mathematically, the resulting set of geometries is labeled by the left coset $\Gamma_{\infty}\backslash \Gamma$ where $\Gamma_{\infty}$ is the parabolic subgroup of translations of $\Gamma = PSL(2,\mathbb{Z})$ generated by 
\begin{equation}
\left[ 
\begin{array}{cc}
1 & 1 \\
0 & 1  \end{array} 
\right].
\end{equation}

\noindent The above identifications (\ref{identi}) have already been analyzed when considering the thermal gas in $AdS_3$ with a non-zero chemical potential. The results given there for the primaries can hence be integrally carried over, upon setting
\begin{equation}
\beta \to \frac{\beta}{\frac{C^2\beta^2}{4\pi^2}+D^2}, \quad \mu \to \frac{\frac{AC\beta^2}{2\pi} + 2\pi BD}{\beta}.
\end{equation}
Hence upon substituting these values into equations (\ref{chemw1}-\ref{chemw4}), one finds the weights of the primaries on the $SL(2,\mathbb{Z})$ family of black holes. \\

\noindent For completeness, we provide the coordinate transformation needed to go to the conventional form of the black hole family.
To get there, we perform the following coordinate transformation
\begin{equation}
\phi + i \tau_E = \frac{1}{C\tau_0 + D} (\phi' + i\tau_E')
\end{equation}
for new coordinates $\phi'$ and $\tau_E'$. The identifications get transformed into
\begin{align}
\label{wzfirs}
\phi \sim \phi + 2 \pi \quad &\to \quad \phi' \sim \phi' + 2\pi D ,\quad \tau_E' \sim \tau_E' - 2\pi i C \tau_0, \\
\label{wzsec}
\phi + i \tau_E \sim \phi + i \tau_E + 2\pi\tau \quad &\to \quad \phi' \sim \phi' +2\pi B,\quad \tau_E' \sim \tau_E' - 2\pi i A\tau_0.
\end{align}
One readily shows that these are actually equivalent to\footnote{
First, define two integers $n$ and $m$ such that $nA-mC=0$, which is always possible, and take $n$ and $m$ to be relatively prime. Let us call $X = nB-mD$. Then we have that
$AX = AnB - AmD = AnB - m(1+BC) = B(nA-mC) -m = -m$. So $X= -m/A$, which implies $m$ is a multiple of $A$, say $m=Ap$. But then $n = pC$, so $p$ is a common factor of both $n$ and $m$ (which we supposed was not present) and hence $m=A$, $n=C$ and $X=-1$. \\
Thus taking $m$ times (\ref{wzfirs}) minus $n$ times (\ref{wzsec}), we obtain that $\phi' \sim \phi' + 2\pi$. We can hence rewrite the identifications as three identifications
\begin{align}
\phi' \sim \phi' +2 \pi , \\
\tau_E' \sim \tau_E' - 2 \pi i A \tau_0 , \\
\tau_E' \sim \tau_E' - 2\pi i C \tau_0.
\end{align}
These final identifications can be combined by asking whether there exist two integers $N$ and $M$ such that $AN - CM =1$. Obviously this is indeed the case by taking $N=D$ and $M=B$ by definition of $SL(2,\mathbb{Z})$.}
\begin{align}
\phi' &\sim \phi' + 2 \pi, \\
\tau_E' &\sim \tau_E' + \beta.
\end{align}
Having done this, we perform a second coordinate transformation on the radial coordinate
\begin{equation}
r^2 = \frac{D^2\sinh^2(\rho) - C^2\tau_0^2\cosh^2(\rho)}{\left|C\tau_0 + D\right|^4}.
\end{equation}
After some straightforward algebra, one finds that the metric takes the form
\begin{equation}
ds^2 = H(r)d\tau_E'^2 + \frac{1}{H(r)}dr^2 + r^2\left(d\phi' + F(r)d\tau_E'\right)^2,
\end{equation}
where
\begin{align}
H(r) &= \frac{\left(r^2-\frac{C^2\beta^2}{4\pi^2\left|C\tau_0 + D\right|^4}\right)\left(r^2+\frac{D^2}{\left|C\tau_0 + D\right|^4}\right)}{r^2}, \\
F(r) &= \frac{-CD\beta}{2\pi\left|C\tau_0 + D\right|^4r^2}.
\end{align}
This describes a Euclidean rotating BTZ black hole with 
\begin{equation}
r_+ =  \frac{C\beta}{2\pi\left|C\tau_0 + D\right|^2}, \quad r_- = \frac{D}{\left|C\tau_0 + D\right|^2}.
\end{equation}
This metric is real, to obtain the associated Lorentzian metric one sets $t=i\tau_E'$ and also continues $r_-$ to imaginary values.\\
Note that the case with $D=0$, implies the inner horizon is at the origin and the hole is non-rotating. This is indeed the BTZ black hole found under $\tau \to -1/\tau$ on the boundary.
Note also that setting $C=0$, yields $A=D=\pm1$ with arbitrary $B$. The space reduces to thermal $AdS_3$ and its $\tau \to \tau + n$ symmetries.\\

\noindent The horizon angular velocity of this black hole is given by $-F(r_+)$, so
\begin{equation}
\Omega_H = \frac{iD}{C\tau_0} = \frac{2\pi D}{ C \beta }.
\end{equation}
The Hawking temperature of the rotating BTZ black hole is given by
\begin{equation}
\beta_{\text{Hawking}} = \frac{2\pi r_+}{r_+^2 + r_-^2} = -2\pi i C \tau_0 = C \beta.
\end{equation}
This value matches with the expected periodicity $(\phi' \sim \phi' + \Omega_H \beta_{\text{Hawking}} \, , \, \tau_E' \sim \tau_E' + \beta_{\text{Hawking}})$ of the grand canonical trace $\text{Tr}e^{-\beta_{\text{Hawking}}(H + i\Omega_H J)}$ and one can convince oneself that this periodicity is on its own sufficient to determine $\beta_{\text{Hawking}}$ independently.

\subsection{Summary}
In this section, we presented results on the treatment of the string gas in $AdS_3$ spacetimes (and its orbifolds) when including a chemical potential for the angular momentum. The techniques used are very similar to those we employed in the previous section when analyzing conical orbifolds. We found the thermal spectrum and the thermal scalar for $AdS_3$ in this case. The periodicity of $\mu$ then allowed us to construct the entire critical curve in the ($\mu,\beta)$ plane. This extends the work of \cite{Lin:2007gi}. A curious feature we discovered was that for large enough chemical potentials, the system can change between stable and unstable several times over a certain range of temperatures. This is, as far as we know, an effect that has not been noted before. The results given here have immediate applications in the thermal spectrum on the entire $SL(2,\mathbb{Z})$ family of black holes used in the Farey Tail construction.

\section{The $AdS_3$ field theory point of view}
\label{fieldtheory}
After the exact analysis of the primaries in the previous sections, we now take a closer look at the thermal scalar field equation and its possible $\alpha'$-corrections. Armed with the exact thermal spectrum on $AdS_3$ in (\ref{ads3spectrum1}) and (\ref{ads3spectrum2}), we look at the lowest order (in $\alpha'$) action and we analyze to what extent the eigenvalues reproduce the exact spectrum. 

\subsection{Winding states from the field theory action}
Firstly, we look into winding states. In general, the lowest order (in $\alpha'$) action for winding modes (with winding $p$) including a NS-NS background is given by
\begin{align}
S \sim \int d^{D-1}x \sqrt{G}e^{-2\Phi} \nonumber \\
\times\left(\tilde{G}^{ij}\partial_{i}T_{p}\partial_{j}T_{p}^{*}+\frac{p^2R^2\tilde{G}^{00}}{\alpha'^2}T_{p}T_{p}^{*}+ \tilde{G}^{0i}\frac{ipR}{\alpha'}\left(T_{p}\partial_{i}T_{p}^{*}- T_{p}^{*}\partial_{i}T_{p}\right)+ m^2T_{p}T_{p}^{*}\right),
\end{align}
where $\tilde{G}^{\mu\nu}$ denotes the T-dual metric. In our case, the T-dual metric and its inverse are given in matrix notation with ordering ($\tau$, $\rho$, $\phi$) by
\begin{equation}
\tilde{G}_{\mu\nu}=\left[\begin{array}{ccc} 
\frac{1}{l^2\cosh(\rho)^2} & 0 & -i\tanh(\rho)^2 \\
0 & l^2 & 0 \\
-i\tanh(\rho)^2 & 0 & l^2\sinh(\rho)^2 - l^2\sinh(\rho)^2\tanh(\rho)^2 \\
\end{array}\right]
, \quad \tilde{G}^{\mu\nu}=\left[\begin{array}{ccc} 
l^2 & 0 & i\\
0 & 1/l^2 & 0 \\
i & 0 & \frac{1}{l^2\sinh(\rho)^2} \\
  \end{array}\right],
\end{equation}
where $l^2=k\alpha'$. Note that the $\tilde{G}^{00}$ metric component is constant: the background NS-NS field has cancelled the confining $G_{00}$ potential $\propto \cosh^2(\rho)$.
So we obtain
\begin{align}
S \propto \int d^{D}x \sinh(\rho)\cosh(\rho)&\left[\frac{1}{l^2}\partial_\rho T \partial_\rho T^* + \frac{1}{l^2\sinh(\rho)^2}\partial_\phi T \partial_\phi T^* \right. \nonumber \\
&\left.- \frac{pR}{\alpha'}\left(T\partial_{\phi}T^{*}- T^{*}\partial_{\phi}T\right) + \frac{R^2p^2l^2}{\alpha'^2}TT^*+ m^2 TT^*\right].
\end{align}
Since $\phi \sim \phi + 2\pi$, we can expand in Fourier modes $\propto e^{iq\phi}$ for $q\in\mathbb{Z}$. The $q^{th}$ mode corresponds to the following eigenvalue equation:
\begin{equation}
\frac{1}{l^2}\left(-\partial_{\rho}\partial_{\rho} T - 2\coth(2\rho)\partial_{\rho} T + \frac{q^2}{\sinh(\rho)^2}T\right) + \frac{2iqpR}{\alpha'}T + \frac{R^2p^2l^2}{\alpha'^2}T - \frac{4}{\alpha'}T = \lambda T.
\end{equation}
When looking for $\delta$-normalizable eigenmodes, the set of eigenfunctions need to decay faster than $e^{-\rho}$ to compensate for the growing measure factor $\sim e^{2\rho}$ as $\rho\to\infty$. 
One can show that the eigenvalues of the first two terms are given by
\begin{equation}
\lambda = \frac{4s^2 + 1}{l^2},
\end{equation}
where $s$ is a real number. The restriction to $\lambda > 1/l^2$ arises precisely due to this restriction on normalizability of the wavefunctions. Using numerical methods, we checked that the same spectrum is obtained when including the third term containing the $q$ quantum number. The eigenvalues $\lambda$ of the above operator are hence
\begin{equation}
\label{sspp1}
\lambda = \frac{4s^2 + 1}{l^2} + \frac{iqp\beta}{\pi\alpha'} + \frac{\beta^2p^2l^2}{(2\pi)^2\alpha'^2} - \frac{4}{\alpha'}.
\end{equation}
Multiplying by $\alpha'/4$, we can clearly see aspects of the exact conformal weight spectrum (\ref{ads3spectrum1}) and (\ref{ads3spectrum2}) appear:
\begin{equation}
h = \frac{s^2 + 1/4}{k} + \frac{iqp\beta}{4\pi} + \frac{k\beta^2p^2}{4(2\pi)^2} - 1.
\end{equation}
The criterion for non-negative eigenvalues is equivalent to the criterion for conformal weights larger than one. The only discrepancy is the $k \to k-2$ appearing in the denominator of the Laplacian, to which this field theory analysis is a priori insensitive.

\subsection{Exact WZW analysis for non-winding states}
Now we look into the non-winding states, but instead of using the lowest order (in $\alpha'$) spacetime action, we utilize the geometrization of the (inverse) string propagator $L_0 + \bar{L}_0$. Following the treatment of \cite{Dijkgraaf:1991ba} for the gauged WZW case, we will use this geometrization to deduce the form of the spacetime action of the non-winding modes. Using formulas given in section \ref{WZWapp}, one can write the zero-modes of the currents as differential operators when acting on vertex operators. We find the following form for the holomorphic currents:
\begin{align}
\hat{D}^3 &= -\frac{1}{2i}\partial_\phi + \frac{1}{2}\partial_\tau, \\
\hat{D}^{+} &= i\frac{e^{\tau-i\phi}}{2}\left[-\partial_\rho + i\coth(\rho) \partial_\phi + \tanh(\rho)\partial_\tau \right], \\
\hat{D}^{-} &= -i\frac{e^{-\tau+i\phi}}{2}\left[\partial_\rho + i\coth(\rho) \partial_\phi + \tanh(\rho)\partial_\tau \right].
\end{align}
Analogous formulas hold for the antiholomorphic sector as given in section \ref{WZWapp}. 
One can readily check that indeed these operators satisfy the $\mathfrak{sl}(2,\mathbb{R})$ algebra:
\begin{align}
\left[\hat{D}^3, \hat{D}^{\pm}\right] = \pm \hat{D}^{\pm}, \\
\left[\hat{D}^+, \hat{D}^-\right] = -2 \hat{D}^{3}.
\end{align}
After some more algebra, the $L_{0}$ and $\bar{L}_0$ operators can be found using the Sugawara construction. These are equal for this case and given by
\begin{equation}
L_0 = \bar{L}_0 = \frac{1}{4(k-2)}\left(-\partial_{\rho}\partial_{\rho} T - 2\coth(2\rho)\partial_{\rho} T - \frac{1}{\sinh(\rho)^2}\partial_{\phi}^2T - \frac{1}{\cosh(\rho)^2}\partial_{\tau}^2T\right). 
\end{equation}
This expression coincides with the scalar Laplacian on the group manifold. We arrive at the eigenvalue equation:
\begin{equation}
\frac{1}{4(k-2)}\left(-\partial_{\rho}\partial_{\rho} T - 2\coth(2\rho)\partial_{\rho} T - \frac{1}{\sinh(\rho)^2}\partial_{\phi}^2T - \frac{1}{\cosh(\rho)^2}\partial_{\tau}^2T\right) - T = \lambda T. 
\end{equation}
Like before, modes with $\phi$ dependence do not give a modification of the spectrum. Similarly, modes with $\tau$ dependence also do not modify the spectrum (obviously both of these do modify the eigenfunctions). The resulting eigenvalues are then 
\begin{equation}
\lambda = \frac{s^2+1/4}{k-2}.
\end{equation}
This is exactly as expected: discrete momentum modes in the $\phi$ and/or the $\tau$ directions do not alter the conformal weights.

\subsection{Thermal scalar action}
\label{thscaction}
Combining the results from the previous subsections and the exactly known spectrum, we propose the following exact form of the eigenvalue equation for primary operators on $AdS_3$:
\begin{equation}
\label{tsequation}
\frac{1}{4(k-2)}\left(-\partial_{\rho}^2 T - 2\coth(2\rho)\partial_{\rho} T + \frac{q^2}{\sinh(\rho)^2}T + \frac{n^2 4\pi^2}{\beta^2\cosh(\rho)^2}T\right) - \frac{iqp\beta}{4\pi}T + \frac{k\beta^2p^2}{4(2\pi)^2}T = T.
\end{equation}
In particular, the thermal scalar action (with $q=0$, $n=0$ and $p=\pm1$) becomes
\begin{equation}
\frac{1}{4(k-2)}\left(-\partial_{\rho}^2 T - 2\coth(2\rho)\partial_{\rho} T\right)  + \frac{k\beta^2}{4(2\pi)^2}T - T = 0.
\end{equation}
Compared to the lowest order (in $\alpha'$) thermal scalar action, the only difference is the shift $k \to k-2$ for the Laplacian term. For the type II superstring, this shift does not occur, and just like for the WZW cigar discussed in \cite{Dijkgraaf:1991ba}, the lowest order effective action is exact (for describing the on-shell conditions). \\
We see that the field theory action reproduces the string spectrum (up to $k\to k-2$ for the bosonic string). This also clearly shows the physical interpretation of the different quantum numbers we introduced in section \ref{spectrumsection}: $q$ represents the discrete momentum around the angular $\phi$-cigar, $w$ is the winding around this cigar, $n$ denotes the discrete momentum around the thermal circle whereas $p$ is the winding around the thermal circle.

\subsection{Flat space limit}
In \cite{Giveon:2013ica}\cite{Giveon:2014hfa}, the authors utilize the flat limit of the $SL(2,\mathbb{R})/U(1)$ to get information on string theory in polar coordinates (Euclidean Rindler space). Also for the thermal $AdS_3$ manifold, one has the opportunity to look at the flat limit as $k\to\infty$. What happens in this case? \\
The temporal part of the background becomes a flat toroidal dimension. However, looking back at the spectrum (\ref{ads3spectrum1}) and (\ref{ads3spectrum2}), we see that there is no $n^2$ contribution. One can see how this comes about by looking at the spacetime equation of motion (\ref{tsequation}). Discrete momentum states clearly do not influence the spectrum for any finite value of $k$, which one can see by looking at the large $\rho$ asymptotics. However, if $k\rho^2$ is held fixed, $\cosh \to 1$, and the discrete momentum term no longer vanishes in the asymptotic large $\rho$ region. Thus this term represents an additive contribution to the eigenvalue which becomes the addition $\frac{n^2\pi^2\alpha'}{\beta^2}$ to the conformal weight. This is precisely the missing term. The same story happens for the $\phi$ part. The geometry asymptotes to flat space in polar coordinates, but the discrete momentum part $q^2$ is missing. Although geometrically the space turns to flat space in polar coordinates, the potential term is still cancelled by the Kalb-Ramond background, a feature which is not present in purely flat space. Despite the resemblances with the $SL(2,\mathbb{R})/U(1)$ cigar (as discussed in section \ref{conical}), this shows that taking the large $k$ limit of this space is not a good starting point to analyze string theory in polar coordinates, unlike the $SL(2,\mathbb{R})/U(1)$ cigar CFT itself as studied by \cite{Giveon:2012kp}\cite{Giveon:2013ica}\cite{Giveon:2014hfa}. \\
It is precisely the absence of the $n^2$ and $q^2$ terms in the conformal weights that causes the simultaneous marginality of these states at the Hagedorn temperature. Unlike the majority of the pathological properties of this model, the physical reason for this is \emph{not} the Kalb-Ramond field. The ever-increasing circumference of the angular cigar and the temporal cylinder, i.e. the asymptotic geometry, is the culprit here. Naively, when a compact dimension has a large circumference, discrete momentum modes are contributing little energy. This is the analogous effect of winding modes becoming light when the compact dimension becomes very small. \\
Note that also in Euclidean Rindler space the thermal circle becomes infinitely large at infinity. However, inspection of the conformal weights \cite{Mertens:2013zya} shows that in that case the degeneracy of marginal states does not occur. The above feature is hence not generic.

\subsection{Random walk behavior in $AdS_3$}
Equiped with the field theory action corresponding to (\ref{tsequation}) for the primaries, the random walk picture can be completed now. If the only critical state present were the state with $q=n=0$ (the thermal scalar), then the random walk displayed in (\ref{randwalkk}) would not be modified (except for the bosonic $k\to k-2$ shift). However, we saw in section \ref{spectrumsection} that the states with arbitrary $q$ (and $n$, but as discussed above this is irrelevant for thermodynamical quantities) all become marginal at the Hagedorn temperature. Hence all of these contribute to the critical behavior. The random walk should hence contain a sum over these quantum numbers: 
\begin{equation}
Z_p = \sum_{w=\pm 1}\sum_{q\in\mathbb{Z}}\int_{0}^{+\infty}\frac{d\tau_2}{2\tau_2}\int\left[\mathcal{D}X\right]\exp\left(-S_p\right).
\end{equation}
For definiteness, we focus on the type II superstring. Utilizing the explicit action for the primaries, the particle action becomes
\begin{align}
S_p = \frac{k}{4\pi}\int_{0}^{\tau_2}dt\left[(\partial_t \rho)^2 + (\beta^2\cosh(\rho)^2-\beta_{H,flat}^2)+\sinh(\rho)^2(\partial_t\phi)^2 + 2w \frac{\beta}{2\pi\alpha'}\sinh(\rho)^2\partial_t\phi \right.\nonumber \\
\left. + \frac{4\pi^2}{k^2}\left\{\frac{3}{4} + \frac{1}{4\cosh(\rho)^2}\right\} - \frac{i4\pi\beta qw}{k} + \frac{4\pi^2 q^2}{k^2\sinh(\rho)^2} \right].
\end{align}

\subsection{Cigar-winding states}
\label{cigarwind}
Previously we stated that the cigar-winding string states are absent for $AdS_3$ and BTZ. However, we did (formally) obtain these in section \ref{spectrumsection}. Moreover, we see in subsection \ref{Ham2} that conical spaces reintroduce these states. It thus seems worthwile to look at the field theory equation for such states. One can obtain these simply by T-dualizing along the $\phi$ direction. The T-dual (in the $\phi$-direction) metric components are given by
\begin{equation}
\tilde{G}_{\phi\phi} = \frac{1}{l^2\sinh(\rho)^2}, \quad \tilde{G}_{\phi \tau} = \frac{B_{\phi\tau}}{G_{\phi\phi}} = i, \quad \tilde{G}_{\tau\tau} = G_{\tau\tau} + \frac{B_{\phi\tau}B_{\phi\tau}}{G_{\phi\phi}} = l^2.
\end{equation}
The T-dual metric and its inverse are hence given by (the ordering of the coordinates in the matrices is now ($\phi$, $\rho$, $\tau$)):
\begin{equation}
\tilde{G}_{\mu\nu}=\left[\begin{array}{ccc} 
\frac{1}{l^2\sinh(\rho)^2} & 0 & i \\
0 & l^2 & 0 \\
i & 0 & l^2 \\
\end{array}\right]
, \quad \tilde{G}^{\mu\nu}=\left[\begin{array}{ccc} 
l^2\tanh(\rho)^2 & 0 & -i\tanh(\rho)^2\\
0 & 1/l^2 & 0 \\
-i\tanh(\rho)^2 & 0 & \frac{1}{l^2\cosh(\rho)^2} \\
  \end{array}\right]
\end{equation}
This leads to the following eigenvalue equation for pure winding states:
\begin{equation}
\label{phiwinding}
\frac{1}{l^2}\left(-\partial_{\rho}\partial_{\rho} T - 2\coth(2\rho)\partial_{\rho} T \right) + \frac{w^2l^2}{\alpha'^2}\tanh(\rho)^2 T - \frac{4}{\alpha'}T = \lambda T.
\end{equation}
In general, the spectrum of this eigenvalue equation contains both a discrete part and a continuous part. The continuous eigenvalues are:
\begin{equation}
\lambda = \frac{4s^2+1}{k} + \frac{w^2l^2}{\alpha'^2} - \frac{4}{\alpha'}
\end{equation}
and the second term is clearly the pure cigar-winding contribution $\frac{kw^2}{4}$ to the conformal weights. \\
The field theory of point view also exhibits the discrete states with the correct eigenvalues, again modulo the substitution $k \to k-2$ in the Laplacian operator. Let us briefly analyze this in more detail. We study the eigenvalue equation
\begin{equation}
\left(-\partial_{\rho}\partial_{\rho} T - 2\coth(2\rho)\partial_{\rho} T \right) + (kw)^2\tanh(\rho)^2 T  = \lambda T
\end{equation}
which is obtained from equation (\ref{phiwinding}) by setting $\lambda_{new} = l^2\lambda_{old} + \frac{4}{\alpha'}l^2$. \\
Numerically, one finds discrete bound states when 
\begin{align}
\lambda &= 2nk\left|w\right| - (n^2-1), \quad n = 1,3,5, \hdots  \\
&= -4\left(\frac{k\left|w\right|}{2}-l\right)\left(\frac{k\left|w\right|}{2}-1-l\right) + (kw)^2, \quad l = 0,1,2, \hdots
\end{align}
and indeed, precisely for these values of $\lambda$ one encounters the discrete states of expressions (\ref{disc1}) and (\ref{disc2}).\footnote{Again, with the remark that one should take $k\to k-2$ in the term associated to the Laplacian.} The continuum starts at the eigenvalue
\begin{equation}
\lambda^{*} = 1 + (kw)^2
\end{equation}
and discrete states have eigenvalues lower than this bound since
\begin{equation}
2nk\left|w\right| - (n^2-1) < 1 + (kw)^2 \quad \Leftrightarrow \quad (k\left|w\right|-n)^2 > 0
\end{equation}
which is automatically satisfied obviously. As an illustration, we draw the wavefunctions for the case $kw=6$ in the following figures \ref{bs} and \ref{cs}. Three bound state wavefunctions are found, and indeed the inequality
\begin{equation}
\frac{k\left|w\right|}{2}-l > \frac{1}{2}, \quad l=0,1,2,\hdots
\end{equation}
has three solutions. The unitarity bound $\tilde{j} < \frac{k-1}{2}$ is also satisfied for $\left|w\right|<\frac{1}{2}$ as we have shown earlier.\\
\begin{figure}[h]
\centering
\begin{minipage}{.3\textwidth}
  \centering
  \includegraphics[width=\linewidth]{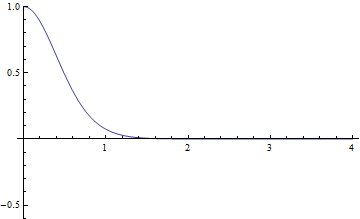}
  \caption*{(a)}
\end{minipage}
\begin{minipage}{.3\textwidth}
  \centering
  \includegraphics[width=\linewidth]{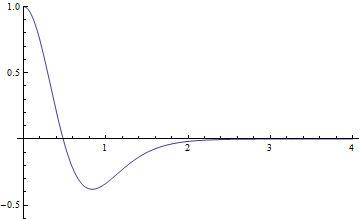}
  \caption*{(b)}
\end{minipage}
\begin{minipage}{.3\textwidth}
  \centering
  \includegraphics[width=\linewidth]{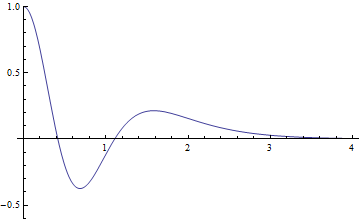}
  \caption*{(c)}
\end{minipage}
\caption{Bound state solutions for $kw=6$ as a function of radial distance $\rho$ with $T(0)=1$ chosen as normalization. (a) Lowest bound state with $\lambda=12$. (b) Second bound state with $\lambda = 28$. (c) Final bound state with $\lambda = 36$.}
\label{bs}
\end{figure}
\begin{figure}[h]
\centering
  \includegraphics[width=0.4\linewidth]{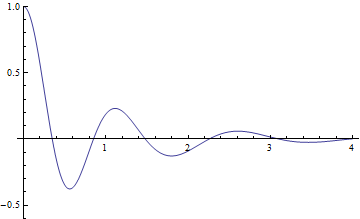}
  \caption{Example of a continuous wavefunction for $kw=6$ with $\lambda=50$ as a function of radial distance $\rho$ with $T(0)=1$ chosen as normalization.. Such states have eigenvalues larger than those of the discrete spectrum.}
  \label{cs}
\end{figure}
\noindent The equation of motion (\ref{phiwinding}) provides another argument in favor of the fact that one should not use the $w \to w+1$ periodicity of the partition function to obtain the Hagedorn divergence for the BTZ black hole. Equation (\ref{phiwinding}) determines the dominant state for $w=\frac{1}{N}$ for the $\mathbb{Z}_N$ orbifold of the BTZ black hole. We expect that one need only continue this wave equation to $N=1$ to obtain the dominant behavior for the BTZ black hole itself. This wave equation includes both continuous and discrete states (if present) and the lowest eigenvalue determines the critical behavior of the string gas. It would be very strange indeed if this equation would need to be drastically altered as soon as we are interested in $\left|w\right| > \frac{1}{2}$.

\subsection{Summary}
In this section we have taken a field theory point of view. This allowed us to clearly observe the physical meaning of the quantum numbers $q$ and $n$ that we introduced in section \ref{spectrumsection}. Since we have seen that the thermal scalar action is not modified (save for the $k\to k-2$ for bosonic strings) from the lowest $\alpha'$ effective thermal scalar action, the random walk picture exhibited in (\ref{randwalkk}) is also at first sight not modified. However, there are infinitely many string states becoming marginal at the Hagedorn temperature. Hence the random walk should contain a sum over discrete momenta around the angular cigar. This random walk is not localized to the $AdS$ origin, since the Kalb-Ramond background field precisely compensates the gravitational potential. These results solve the random walk problem in the $AdS_3$ WZW target space. Cigar-winding modes exhibit explicitly a discrete part in their spectrum, precisely corresponding to the expected conformal weights.

\section{Summary of this chapter}
\label{conclusion}
We have discussed thermal properties of the $AdS_3$ and BTZ WZW models from the thermal spectrum. Firstly we briefly analyzed the form of $\alpha'$ corrections to the thermal scalar action. Then we used CFT twisting techniques to fully determine the string spectrum on the thermal manifolds. We explicitly found the thermal scalar in the string spectrum with the correct mass. The thermal scalar is not localized to the $AdS$ origin but instead fluctuates all over space (it is in a continuous representation of the symmetry group). The reason is the Kalb-Ramond flux whose repulsion precisely compensates the gravitational attraction. For the BTZ black hole, the thermal scalar is not present in the string spectrum. The state however reappears when considering sufficiently conical spaces. We discovered that the twisted sectors on the cone also exhibit discrete modes. From a mathematical perspective, these are found by properly performing the analytic continuation of Poisson's summation formula. We discovered that these discrete states, if they are present, dominate the continuous states. Hence they are important for the critical thermodynamics. After that, using techniques we employed to analyze the temporal BTZ orbifolds, we have analyzed the generalization of the $AdS_3$ string gas by including a chemical potential corresponding to the angular momentum on the cigar. We found the critical curve in the ($\mu,\beta$) plane for the $AdS_3$ string gas and discovered a peculiar effect where the string gas alternated between stable and unstable in a certain temperature interval. Also, the application to the $SL(2,\mathbb{Z})$ family of black holes was explored. In the final section, we looked at the field theory point of view. The lowest order $\alpha'$ thermal scalar action reproduces the correct conformal weights for the type II superstring and we expect it to be exact. Bosonic string actions are nearly exact, the only difference is the shift $k\to k-2$ in the kinetic term. With all these techniques, we have written down the random walk picture of the string on $AdS_3$ spacetime. A strange feature is that arbitrary $q$ quantum numbers are all simultaneously dominant. \\
Throughout this chapter, we have presented four different methods to analyze the thermal spectrum on $AdS_3$ space and its orbifolds. Let us compare these.
\begin{itemize}
\item{
The first method we presented in section \ref{spectrumsection} utilized worldsheet CFT twisted operator methods to twist the non-thermal Euclidean CFT into the thermal one. The approach is elegant and computationally easy to carry out, though it has some subtle points in that the derivation is not airtight: we needed to take a plausible guess when considering the conserved charges in the twisted sectors. In spite of this, this approach leads to a solid prediction of the thermal spectrum whose quantum numbers can be directly related to physical quantities. However, this approach is not sensitive to some constraints on the quantum numbers. We view it as a blueprint of the exact result: the precise thermal spectrum needs to be taylored to the form predicted using this vertex operator method.
}
\item{
A second approach we followed utilized the (lowest order in $\alpha'$) field theory equations of motion in the curved background (using the $SL(2,\mathbb{R})$ Laplacian). This approach is a priori not sensitive to possible $\alpha'$ corrections, and it misses the $k \to k-2$ shift in the denominator of the kinetic Laplacian term. For type II superstrings, this approach does not have this problem. Modulo this difficulty, one correctly predicts all possible string states. This method clearly demonstrated the possibility of discrete states when $w\neq0$. The downside is that, just like the vertex operator method, one is not sensitive to constraints, although normalizability $\tilde{j}>\frac{1}{2}$ can be directly checked and this constraint is most transparent when using this method. 
}
\item{
Thirdly, we looked at the exact partition function using numerical methods (i.e. truncating the series and taking the large $\tau_2$ limit cautiously). This approach, when utilized with care, is the most failsafe option we have. Of course, one is limited to certain questions only, for instance the large $\tau_2$ limit is accessible but moderate values of $\tau_2$ are hard to analyze. We detected the presence of discrete states with this and this was our only way to analyze what the infinite product contributes to the final result. It is of course difficult to analyze what one finds if we do not know where to look in advance: the physical meaning of the results are obscured.
}
\item{
Finally, we turned to the Hamiltonian formulation of the one-loop partition function. This approach is in principle the best one, since one can distill all states and the unitarity constraints should be visible. However, this approach is the most cumbersome to use. Moreover, in order to rewrite this in the desired way, one needs to know in advance what the conformal weights of the states might look like. It is at this point that the vertex operator method is ideally suited to provide insight.
}
\end{itemize}
In this chapter, we used a combination of all four of these methods to analyze the thermal spectrum and the critical Hagedorn behavior. \\
Some open avenues that can be further explored are for instance a more elaborate treatment of the type II superstring on these spaces. We have merely displayed the expected changes, though a more thorough analysis would be welcome. \\
Another possibility is to try to get further insight in the Hagedorn temperature for the BTZ black hole. The continuation in the orbifold integer $N$ is somewhat dubious in this case, more so than for the Euclidean Rindler case or the $SL(2,\mathbb{R})/U(1)$ cigar CFT.\\
A further front on which progress can be made is the treatment of the infinite product of oscillators in (\ref{numer}). Out of the four methods discussed above, the only approach we have successfully applied to treat this product is the numerical approach. The difficulty is that the factors in the infinite product must in principle be series-expanded in different series depending on the value of $l$. But the sum over $l$ needs to be treated all at once to be able to apply the Poisson summation techniques to it. We leave a better treatment of this as an open issue.\\
Another path that can be explored, is the treatment of the $AdS_3$ string gas with chemical potential using saddle point methods along the lines of \cite{Lin:2007gi} and see whether the critical Hagedorn curve can be fully explained.\\
We have also merely provided a starting point for the treatment of BTZ black holes when including chemical potentials. Also here progress can be made.\\
It is also known for quite some time now that the WZW $AdS_3$ model is related to other CFTs by exactly marginal perturbations on the worldsheet \cite{Forste:1994wp}\cite{Giveon:1993ph}\cite{Israel:2003ry}. It would be interesting to learn how the thermal spectrum and the critical Hagedorn thermodynamics of the string gas behaves along this marginal line of CFTs.\\
To conclude, we have illustrated the general results of \cite{Mertens:2013pza} (discussed in chapter \ref{chth}) in a concrete non-trivial example. The apparently marginal behavior of all $q \in \mathbb{Z}$ states and the absence of the thermal scalar for BTZ black holes studied here, urges us to be careful when considering non-trivial geometries, especially spaces with topologically trivial thermal circles. 

\section{*The density of states does not influence the critical behavior}
\label{dos}
In this section we argue that the density of states can not alter our conclusions about the critical temperature.\\
So far in the main text we have not written down an explicit expression for the density of states. The correct expression is given in equation (\ref{dosorig}) \cite{Maldacena:2000kv}. \\ 
A first argument as to why the $s$-integration does not make a state more tachyonic is the following. In general, the integration over $s$ is of the following form:
\begin{equation}
\int_{0}^{+\infty}ds \rho(s) e^{-\tau_2 s^2} \, \to \, 0 \quad \text{for} \quad \tau_2 \to \infty
\end{equation}
so this integral cannot yield a contribution that behaves as $e^{A\tau_2}$ for positive $A$ as long as $\rho(s)$ is well-behaved near $s=0$ which it is.\footnote{This follows largely from equation (\ref{dosorig}) with $n=0$: for any fixed $q$ the limit $s \to 0$ is zero.} \\
A more general argument goes as follows. Firstly, as remarked before, the $\tau_1$-integral forces $np = 0$, which implies $n=0$ for our purposes. Important to note is that the contribution from $q$ is an imaginary exponential. Consider the following general expression\footnote{The $q$ quantum number and its prefactors are rescaled into a new number $q$.}:
\begin{equation}
\int_{0}^{+\infty}ds \sum_{q\in\mathbb{Z}}\rho(s,q)e^{iq\tau_2}e^{-s^2\tau_2}
\end{equation}
where $\rho(s,q)$ denotes the density of states with $n=0$.
Performing the sum first, we can rewrite this as
\begin{equation}
\int_{0}^{+\infty}ds F(s,\tau_2)e^{-s^2\tau_2}
\end{equation}
where $F$ has Fourier coefficients $\rho$ and is periodic in $\tau_2$. We are only interested in whether the integral is capable of producing a $\tau_2$-dependent exponential after integration (like $e^{\pm C \tau_2}$ for some constant $C$). Laplace's method gives
\begin{equation}
\int_{0}^{+\infty}ds F(s,\tau_2)e^{-s^2\tau_2} \approx \sqrt{\frac{\pi}{\tau_2}}\frac{F(0,\tau_2)}{2} + \hdots
\end{equation}
since the periodic function does not correct the leading behavior. One simple way to see this is to take a discrete limit with steps precisely equal to the periodicity of $F$ in $\tau_2$. In this case $F$ becomes effectively independent of $\tau_2$ and one can use the textbook Laplace method to get the above result. The $\tau_2$-dependence of the final result may indicate that the limit itself is ill-defined but regardless it cannot influence the critical behavior and we only care about this.
Note that the assumption that the integration over $s$ does not influence the tachyonic nature of a state, was made implicitly by the authors of \cite{Maldacena:2000hw} and \cite{Rangamani:2007fz}.

\section{*Hamiltonian description of thermal $AdS_3$ and its orbifolds}
\label{lengthy}
In this lengthy section, we describe in detail how the Hamiltonian description of the spectrum is obtained. For clarity, we first use a trivial toy model to illustrate the strategy that we will employ.

\subsection{Flat space toy model to illustrate the strategy}
\label{toy}
Consider the 1d flat space Laplacian. The operator commutes with $\hat{J} = i\partial_x$. We choose eigenfunctions
\begin{equation}
\label{eigf}
\psi_k(x) \propto e^{ikx},
\end{equation}
satisfying
\begin{align}
\label{rel1}
\Delta \psi &= -k^2 \psi, \\
\label{rel2}
\hat{J} \psi = i \partial_x \psi &= -k \psi.
\end{align}
We then evaluate
\begin{equation}
\text{Tr}\left[e^{t\Delta}e^{2\pi i U \hat{J}}\right]
\end{equation}
for some fixed number $U$. Firstly, in configuration space this equals 
\begin{equation}
\int dx \left\langle x\right|e^{t\Delta}\left|x + 2\pi U\right\rangle = V\frac{1}{2\sqrt{\pi t}}e^{-\frac{\pi^2U^2}{t}},
\end{equation}
where we used $ e^{i a \hat{J}}\left|x\right\rangle = \left|x + a\right\rangle$ and the flat space heat kernel. Secondly, we can evaluate it using the eigenfunctions (\ref{eigf}) and the relations (\ref{rel1}) and (\ref{rel2}) as:
\begin{equation}
\int dk \delta(0) e^{-tk^2}e^{-2\pi i U k} = \frac{V}{2\pi}\sqrt{\pi/t}e^{-\frac{\pi^2U^2}{t}}.
\end{equation}
We see that both expressions are manifestly the same. Note that the density of states is present in the form of $\delta(0)$. It is the second description that we are after, since then the trace over the quantum numbers (in this case only $k$) is apparent. In the following subsections, we will apply this same idea to the much more complicated $AdS_3$ space and its orbifolds.

\subsection{$AdS_3$ Thermal partition function}
\label{Ham1}
In this subsection we give a Hamiltonian description of the thermal $AdS_3$ partition function. With a suitable substitution of parameters, this is also the thermal BTZ partition function as discussed in section \ref{BTZsection}. The partition function is given by
\begin{align}
\label{partfunct}
Z(\tau) = \frac{\beta\sqrt{k-2}}{8\pi\sqrt{\tau_2}}\sum_{l,p}\frac{e^{-k\beta^2\left|l-p\tau\right|^2/4\pi\tau_2 + 2\pi \Im(U_{lp})^2/\tau_2}e^{\frac{\pi\tau_2}{2}}}
{\left|\sin(\pi U_{lp})\right|^2\left|\prod_{r=1}^{+\infty}(1-q^{r})(1-q^re^{2\pi i U_{lp}})(1-q^{r}e^{-2\pi i U_{lp}})\right|^2}
\end{align}
where $q=e^{2\pi i \tau}$ and\footnote{This differs in two ways from the expression written down in \cite{Maldacena:2000kv}: firstly the $(q\bar{q})^{-3/24}$ was inserted in the above expression. Secondly, we utilize the complex conjugate of the $U_{lp}$ as defined in \cite{Maldacena:2000kv}. This is in accord with the Ray-Singer torsion \cite{Ray:1973sb} and this was noticed in \cite{Son:2001qm} by one of the authors themselves.} 
\begin{equation}
\label{Ulp}
U_{lp} = -\frac{i\beta}{2\pi}(p\tau-l).
\end{equation}
The quantum number $p$ will correspond to thermal winding, whereas the quantum number $l$ will be Poisson-resummed into the discrete momentum. Our goal is to rewrite this in the form
\begin{equation}
\label{gen}
Z(\tau) = \sum_i N_{i\tilde{i}}\chi_{i}(\tau) \chi_{\tilde{i}}^{*}(\tau) = \text{Tr} q^{L_0 - c/24}\bar{q}^{\bar{L}_0 - c/24},
\end{equation}
with the characters
\begin{equation}
\chi_i(\tau) = \text{Tr}_i q^{L_0 - c/24}.
\end{equation}
The second equality in (\ref{gen}) traces over all primaries and their secondaries in the string spectrum. In these expressions the conformal weights $h$ and $\bar{h}$ are possibly a subset of those we determined in section \ref{spectrumsection} using CFT arguments. \\
First of all, we have that
\begin{equation}
\sum_{N=1}^{+\infty}\sum_{\mathcal{P} \in P(N)}q^{N}e^{2\pi i U_{lp} O(\mathcal{P})} = \prod_{r=1}^{+\infty}\frac{1}{1 - q^r e^{2\pi i U_{lp}}}
\end{equation}
where $P(N)$ denotes the different partitions of the integer $N$ and $O(\mathcal{P})$ denotes the size of the partition $\mathcal{P}$. It is clear that the infinite product in (\ref{partfunct}) corresponds to the different oscillator states and we will not be interested in this. Note though that this is a bit naive since the Taylor expansion we should use depends on the precise value of $U_{lp}$. Nonetheless, as a first step towards understanding the partition function (\ref{partfunct}) we choose to drop the infinite product. Comments on this are provided in the main text. \\
The method to proceed was developed by \cite{Gawedzki:1991yu}\cite{Gawedzki:1988nj} and we adapt it for our purposes.\footnote{Note that in \cite{Gawedzki:1991yu}, the author only considers twisting in one torus direction. Here we consider both.} We first evaluate the following trace (for fixed $l$ and $p$):
\begin{equation}
\label{trace}
\text{Tr}\left[\exp\left(-\tau_2\frac{\beta^2 k p^2}{4\pi}\right)\exp\left(4\pi\tau_2\frac{\Delta}{k-2}\right)\exp(2\pi i (U_{lp} J^{3}_0 + \bar{U}_{lp} \overline{J}^{3}_0))\right]
\end{equation}
where we trace over a basis of all normalizable functions $\psi_a(g)$ on $H_3^+$ and $J^{3}_0$ and $\overline{J}^{3}_0$ are differential operators acting on these functions. As a basis, we choose eigenfunctions of $\Delta$, $J^{3}_0$ and $\overline{J}^{3}_0$. The explicit form of these eigenfunctions will not be needed; the interested reader can take a closer look at appendix A of \cite{Teschner:1997fv} to find elaborate expressions. More explicitly, consider the three operators $\Delta$, $J_0^3 + \overline{J}_0^3 = i\partial_{\phi}$ and $J_0^3 - \overline{J}_0^3 = - \partial_\tau$ in coordinates $\tau,\phi,\rho$. These mutually commute so let us choose functions $\psi_{s,m,q}$ such that
\begin{align}
\label{eigf1}
\Delta \psi = (-s^2 - 1/4)\psi, \\
\label{eigf2}
(J_0^3 + \overline{J}_0^3)\psi = q \psi, \\
\label{eigf3}
-i(J_0^3 - \overline{J}_0^3)\psi = m \psi.
\end{align} 
It can be shown that when $m$ is real and $q$ is an integer, $\Delta$ is a Hermitian operator w.r.t. the standard inner product on $L_2(H_3^+)$. Therefore the $\psi_{s,m,q}$ form a basis when restricting to these quantum numbers. The starting point is then the evaluation of
\begin{equation}
\sum_{s,m,q}\left\langle \psi_{s,m,q}\right|\hat{\mathcal{O}}\left|\psi_{s,m,q}\right\rangle
\end{equation}
with $s$ a real positive number, $m$ a real number and $q$ an integer. So explicitly
\begin{equation}
\label{summations}
\sum_{s,m,q} = \int_{\mathbb{R}^{+}}ds \int_{\mathbb{R}}dm \sum_{q\in\mathbb{Z}}.
\end{equation}
In our case, the operator $\hat{\mathcal{O}}$ is given as:
\begin{equation}
\hat{\mathcal{O}} = e^{-\tau_2\frac{\beta^2 k p^2}{4\pi}}e^{4\pi\tau_2 \frac{\Delta+\frac{1}{4}}{k-2}}e^{2\pi i (U_{lp}J_0^3 + \bar{U}_{lp}\overline{J}_0^3)}.
\end{equation}
We rewrite this as 
\begin{equation}
\int dg \sum_{s,m,q}\psi_{s,m,q}(g)^{*}\psi_{s,m,q}(g) \lambda_{s,m,q} = \sum_{s,m,q}\delta(0) \lambda_{s,m,q},
\end{equation}
where $g$ denotes a group element of the $H_3^+ = SL(2,\mathbb{C})/SU(2)$ group manifold and $\lambda_{s,m,q}$ is the eigenvalue of $\hat{\mathcal{O}}$. Here $\delta(0) = \rho(s,m,q)$ is the density of states and depends on $s,m$ and $q$. The operator $\hat{\mathcal{O}}$ is labeled by $p,l$ quantum numbers and the entire expression is then summed over $l$ and $p$. 
The expression for the density of states on such spaces was written down in \cite{Maldacena:2000kv}. We use a slight modification of this expression:
\begin{equation}
\label{dosh3}
\mbox{{\small{$\displaystyle\rho(s,m,q) = 2\frac{L}{2\pi}\left[\frac{1}{2\pi}2\log(\epsilon) + \frac{1}{2\pi i}\frac{d}{2ds}\log\left(\frac{\Gamma(\frac{1}{2} - is - q/2 - im/2)\Gamma(\frac{1}{2} - is -q/2 + im/2)}{\Gamma(\frac{1}{2} + is -q/2- im/2)\Gamma(\frac{1}{2} + is -q/2+ im/2)}\right)\right]$}}}.
\end{equation}
The expression in square brackets is the usual one, used in \cite{Maldacena:2000kv}. The parameter $\epsilon$ is an IR regulator ($\epsilon \to 0$) corresponding to the radial direction. We here multiply this expression by a further factor of $\frac{L}{2\pi}$ where $L$ is the IR regulator ($L \to \infty$) of the Euclidean time direction. The reason that we include it here, is that we are considering the Euclidean space, for which the density of states also includes the temporal direction on the same foot as the other space directions. A proper definition of $L$ follows shortly. We also include an extra factor of $2$.\footnote{The reason is that we consider the entire prefactor of the $s$-integral and this includes an extra factor of 2. See e.g. \cite{Hanany:2002ev}\cite{Maldacena:2000kv} where this factor is written explicitly.} \\
We need to sum the trace (\ref{trace}) over $l$ and $p$ and divide by the range of this summation. In this case, $l$ and $p$ run over $\mathbb{Z}$ so the range is infinite (let us call it $P$), although this infinity will cancel with $L$ further on to yield a finite contribution. \\
As in the previous subsection \ref{toy}, we will evaluate this operator trace in two different ways. Let us first focus on the configuration space evaluation. We choose the free-field coordinates ($\Phi$, $v$, $\bar{v}$) introduced in \cite{Gawedzki:1991yu} to parametrize the group element:
\begin{equation}
g = \left[ 
\begin{array}{cc}
e^{\Phi}(1+v\bar{v})^{1/2} & v \\
\bar{v} & e^{-\Phi}(1+v\bar{v})^{1/2} \end{array} 
\right].
\end{equation}
for real $\Phi$ and complex $v$ (and $\bar{v}$). In these coordinates it was shown in \cite{Gawedzki:1991yu} that the zero-modes of the currents have the following form:
\begin{align}
J^{3}_0 - \overline{J}^{3}_0 &= -v\partial_v + \bar{v}\partial_{\bar{v}} = i \partial_{\phi}, \\
J^{3}_0 + \overline{J}^{3}_0 &= - \partial_{\Phi} = -\partial_{\tau},
\end{align}
where the first equality is in terms of the coordinates above, and the second equality is in terms of the global coordinates.\footnote{A few technicalities are in order here. The free-field coordinates ($\Phi$, $v$, $\bar{v}$) are related to the global $AdS$ coordinates as
\begin{align}
\label{cotrans}
\begin{cases}
v &= \sinh(\rho)e^{i\phi} \\
\bar{v} &= \sinh(\rho)e^{-i\phi} \\
\Phi &= t - 2\text{log}\cosh(\rho)
\end{cases} 
\end{align}
The $\Phi$ coordinate used in \cite{Gawedzki:1991yu} is related to that used in \cite{Maldacena:2000kv} (denoted $\tilde{\Phi}$ here) by $\Phi = \tilde{\Phi} + \frac{1}{2}\text{log}(1+\left|v\right|^2)$, which explains the factor of $2$ appearing in the final line of (\ref{cotrans}).} To conform to our conventions, we change the sign of the antiholomorphic sector, such that:
\begin{align}
J^{3}_0 + \overline{J}^{3}_0 &= -v\partial_v + \bar{v}\partial_{\bar{v}} = i \partial_{\phi}, \\
-i(J^{3}_0 - \overline{J}^{3}_0) &= i \partial_{\Phi} = i\partial_{\tau}.
\end{align}
Then these generators are indeed the angular and Euclidean time generators defined before. \\
From the form of the generator in terms of differential operators, we have that
\begin{align}
\exp(2\pi i U_{lp} J^{3}_0) g &= \exp(-\pi i U_{lp} \sigma^{3}) g, \\
\exp(2\pi i \bar{U}_{lp} \overline{J}^{3}_0) g &=  g \exp(\pi i \bar{U}_{lp} \sigma^{3}).
\end{align}
We can rewrite the trace as
\begin{align}
\sum_{a}&\left\langle \psi_a\right|\exp\left(-\tau_2\frac{\beta^2 k p^2}{4\pi}\right)\exp\left(4\pi\tau_2\frac{\Delta}{k-2}\right)\exp\left(2\pi i (U_{lp} J^{3}_0 + \bar{U}_{lp} \overline{J}^{3}_0)\right)\left|\psi_a\right\rangle \\
&= \int d g \left\langle g\right|\exp\left(-\tau_2\frac{\beta^2 k p^2}{4\pi}\right)\exp\left(4\pi\tau_2\frac{\Delta}{k-2}\right)\left|\exp(-\pi i U_{lp} \sigma^{3})g\exp(\pi i \bar{U}_{lp} \sigma^{3})\right\rangle.
\end{align}
In the last line, we integrate the heat kernel over the group manifold, but with twisted boundary conditions. Next we explicitly perform the integration on the group manifold, i.e. over the $v$, $\bar{v}$ and $\Phi$ coordinates. The measure is given by $dg = d\Phi dv d\bar{v}$. The group metric (to which the Laplacian above is associated) is given by
\begin{equation}
ds^2 = d\tilde{\Phi}^2 + (dv + v d\tilde{\Phi})(d\bar{v} + \bar{v}d\tilde{\Phi})
\end{equation} 
and is independent of $\tilde{\Phi}$ (or $\Phi$ itself). Thus integrating over $\Phi$ can be done by using this fact: since the heat kernel is a sum over paths between two points, it is independent of the `center of mass' $\Phi$ coordinate of both points.\footnote{Note also that both $J^{3}_0$ and $\overline{J}^{3}_0$ are independent of the $\Phi$ coordinate, which is a necessary condition for this statement.} We define $\int d\Phi = L$ and combining it with the $\frac{1}{P}$ present in the partition sum, it produces $\beta$; just like it does in the flat toroidal case. \\
The heat kernel on $H_3^+$ is given by (see e.g. \cite{David:2009xg})
\begin{equation}
e^{t\Delta}(g_1,g_2) = (\pi t)^{-3/2}\frac{d}{\sinh d}e^{-t/4-d^2/t},
\end{equation}
where $d$ is the geodesic (= hyperbolic) distance between the 2 points. In particular, between $g$ and $\exp(-\pi i U_{np} \sigma^{3})g\exp(\pi i \bar{U}_{np} \sigma^{3})$ with $\Phi = 0$, this is given by:
\begin{equation}
\cosh d = \left(1+\left|v\right|^2\right)\cosh(2\pi U_2) - \left|v\right|^2\cos(2\pi U_1).
\end{equation}
The integration can then be done by some simple substitutions and leads to
\begin{align}
&\int d g \left\langle g\right|\exp\left(4\pi\tau_2\frac{\Delta}{k-2}\right)\left|\exp(-\pi i U_{lp} \sigma^{3})g\exp(\pi i \bar{U}_{lp} \sigma^{3})\right\rangle \\
&= L \frac{\sqrt{k-2}}{8\pi\sqrt{\tau_2}}e^{-\pi \tau_2/(k-2)}e^{-\pi (k-2)\Im(U_{lp})^2/\tau_2}\left|\sin(\pi U_{lp})\right|^{-2}.
\end{align}
When we put everything together, we obtain
\begin{align}
\label{fulltrace}
\frac{1}{P}\text{Tr}&\left[\exp\left(-\tau_2\frac{\beta^2 k p^2}{4\pi}\right)\exp\left(4\pi\tau_2\frac{\Delta+1/4}{k-2}\right)\exp(2\pi i (U_{lp} J^{3}_0 + \bar{U}_{lp} \overline{J}^{3}_0))\right] \\
 &= \frac{\beta\sqrt{k-2}}{8\pi\sqrt{\tau_2}}\exp\left(-\tau_2\frac{\beta^2 k p^2}{4\pi}\right)e^{-\pi (k-2)\beta^2 (p\tau_1-l)^2/4\pi^2\tau_2}\left|\sin(\pi U_{lp})\right|^{-2} \\
 \label{fulltrace2}
 &=\frac{\beta\sqrt{k-2}}{8\pi\sqrt{\tau_2}}e^{- k\beta^2\left|l-p\tau\right|^2/4\pi\tau_2 + 2\pi \Im(U_{lp})^2/\tau_2}\left|\sin(\pi U_{lp})\right|^{-2}.
\end{align}
To fully agree with (\ref{partfunct}) we should multiply the trace by $e^{\frac{\pi\tau_2}{2}}$ and sum this expression over $l$ and $p$. This concludes the first computation: the partition function has been rewritten in terms of a trace of some operator $\hat{\mathcal{O}}$. \\ 
Next we would like to rewrite it fully in terms of the quantum numbers of the states we wrote down in section \ref{spectrumsection}. So we start afresh with the operator trace (\ref{fulltrace}) and we evaluate the trace by using the basis of eigenfunctions that we discussed around equations (\ref{eigf1}), (\ref{eigf2}) and (\ref{eigf3}). To proceed, we focus on the integral over $m$ in equation (\ref{summations}) and the summation over $l$. The part of the trace that depends on $l$ gives us
\begin{align}
\label{projexpo}
\frac{1}{P}\int_{\mathbb{R}}dm\sum_{l\in\mathbb{Z}}\exp\left(2\pi i \left(\frac{\beta l}{2\pi}m\right)\right) &= \frac{1}{P}\frac{2\pi}{\beta}\int_{\mathbb{R}}dm\sum_{n\in\mathbb{Z}}\delta\left(m-\frac{2\pi n }{\beta}\right) \\
 &= \frac{2\pi}{L}\int_{\mathbb{R}}dm\sum_{n\in\mathbb{Z}}\delta\left(m-\frac{2\pi n }{\beta}\right).
\end{align}
Note the appearance of a prefactor $\frac{2\pi}{L}$. Hence the integral over $m$ only has contributions for $m = \frac{2\pi n }{\beta}$ for integer $n$.
With this value of $m$ and the fact that
\begin{equation}
J^{3}_0 + \overline{J}^{3}_0  = q
\end{equation} 
on the eigenfunctions $\psi$, we obtain using (\ref{Ulp}) for the remaining parts of $U_{lp}$:
\begin{align}
\beta p \tau\left(\frac{q}{2} + \frac{i\pi n }{\beta} \right) &\subset 2\pi i U_{lp} J^{3}_0, \\
-\beta p \bar{\tau}\left(\frac{q}{2} - \frac{i\pi n }{\beta} \right) &\subset 2\pi i \bar{U}_{lp} \overline{J}^{3}_0.
\end{align}
This determines all different factors of the operator trace (\ref{fulltrace}) in terms of the quantum numbers we are interested in.\\
The factor of $e^{\pi\tau_2/2}$ that we manually added after equation (\ref{fulltrace2}) can be written as
\begin{equation}
e^{\pi\tau_2/2} = (q\bar{q})^{-3/24}.
\end{equation}
Also multiplying (and dividing) the trace by $e^{\frac{-\pi\tau_2}{k-2}}$, we have precisely rewritten the partition function as
\begin{equation}
\text{Tr}q^{L_0 - c/24}\bar{q}^{\bar{L}_0-c/24},
\end{equation}
where
\begin{align}
h &= \frac{s^2 +1/4}{k-2} - i\frac{qp\beta}{4\pi} + \frac{ p n}{2} + \frac{kp^2\beta^2}{4(2\pi)^2}, \\
\bar{h} &= \frac{s^2 +1/4}{k-2} - i\frac{qp\beta}{4\pi} - \frac{ p n}{2} + \frac{kp^2\beta^2}{4(2\pi)^2},
\end{align}
and with $c = 3 + 6/(k-2)$, the central charge of the $SL(2,\mathbb{R})$ model. We trace only over the continuous states with the density of states:
\begin{equation}
\label{dosorig}
\rho(s,n,q) = 2\left[\frac{1}{2\pi}2\log(\epsilon) + \frac{1}{2\pi i}\frac{d}{2ds}\log\left(\frac{\Gamma(\frac{1}{2} - is - q/2 - i\frac{\pi n }{\beta})\Gamma(\frac{1}{2} - is -q/2 + i\frac{\pi n }{\beta})}{\Gamma(\frac{1}{2} + is -q/2- i\frac{\pi n }{\beta})\Gamma(\frac{1}{2} + is -q/2+ i\frac{\pi n }{\beta})}\right)\right].
\end{equation}
The factor $\frac{L}{2\pi}$ has dropped out in this expression: this makes sense, since this direction has become compact due to the thermal identification and compact dimensions do not give volume-scaling prefactors when considering the Hamiltonian formulation of the partition function (see for instance any textbook on string theory).\\
No discrete states are present and there are also no states that wind the angular cigar. 

\subsection{Angular orbifolds}
\label{Ham2}
We consider orbifolds obtained by identifying $\phi \sim \phi + \frac{2\pi}{N}$. These angular orbifolds were extensively studied in \cite{Son:2001qm} and \cite{Martinec:2001cf}. The thermal partition function was computed in \cite{Son:2001qm}. It was shown there that the thermal partition function on such orbifolds has the form
\begin{equation}
\label{pforbi}
Z = \frac{1}{N}\sum_{a,b}Z_{ab},
\end{equation}
where each $Z_{ab}$ is obtained from the untwisted partition function (\ref{partfunct}) by the simple substitution
\begin{equation}
U_{lp} \to U_{lp} + \frac{a}{N}\tau + \frac{b}{N}.
\end{equation}
These parameters are hence given by
\begin{equation}
U_{lp} = \frac{b}{N} + \frac{a}{N}\tau_1 - i\frac{\beta}{2\pi}(p\tau_1-l) + i\frac{a}{N}\tau_2 + \frac{p\beta}{2\pi}\tau_2,
\end{equation}
whose imaginary part equals
\begin{equation}
\Im(U_{lp}) = -\frac{\beta}{2\pi}(p\tau_1-l) + \frac{a}{N}\tau_2.
\end{equation}
Performing the above heat kernel computation again, we obtain
\begin{align}
\text{Tr}&\left[\exp\left(-\tau_2\frac{\beta^2 k p^2}{4\pi}\right)\exp\left(4\pi\tau_2\frac{\Delta+1/4}{k-2}\right)\exp(2\pi i (U_{lp} J^{3}_0 + \bar{U}_{lp} \overline{J}^{3}_0))\right] \\
 &=\frac{\beta\sqrt{k-2}}{8\pi\sqrt{\tau_2}}e^{- k\beta^2\left|l-p\tau\right|^2/4\pi\tau_2 + 2\pi \Im(U_{lp})^2/\tau_2 -\frac{\pi k}{\tau_2}\left(-(p\tau_1-l)\frac{a}{N}\frac{\beta}{\pi}\tau_2 + \frac{a^2}{N^2}\tau_2^2\right)}\left|\sin(\pi U_{lp})\right|^{-2}.
\end{align}
By a slight rearrangement of this expression, we get 
\begin{align}
\label{expree}
\text{Tr}&\left[\exp\left(-(p\tau_1-l) k\frac{a}{N}\beta+ \pi k\frac{a^2}{N^2}\tau_2\right)\exp\left(-\tau_2\frac{\beta^2 k p^2}{4\pi}\right)\right. \nonumber \\
&\quad\quad\quad\quad \left.\times\exp\left(4\pi\tau_2\frac{\Delta+1/4}{k-2}\right)\exp(2\pi i (U_{lp} J^{3}_0 + \bar{U}_{lp} \overline{J}^{3}_0))\right] \\
 &=\frac{\beta\sqrt{k-2}}{8\pi\sqrt{\tau_2}}e^{- k\beta^2\left|l-p\tau\right|^2/4\pi\tau_2 + 2\pi \Im(U_{lp})^2/\tau_2 }\left|\sin(\pi U_{lp})\right|^{-2}.
\end{align}
Let us again interpret this from a CFT point of view. Firstly we determine and solve the analogous conditions as (\ref{projexpo}) for this case. These will be called the projection conditions in what follows. These are given by
\begin{align}
\sum_{b}\exp\left(2\pi i \frac{b}{N}(J^{3}_0 + \overline{J}^{3}_0)\right), \\
\sum_{l}\exp\left(2\pi i \left(\frac{i\beta}{2\pi} l J^{3}_0 - \frac{i\beta}{2\pi} l\overline{J}^{3}_0\right) + lkw\beta\right),
\end{align}
where $w= \frac{a}{N}$. We already have that $m_{J} + \overline{m}_J \in \mathbb{Z}$ from the covering space. Then the sum over $b$ gives us
\begin{equation}
\frac{1}{N} \sum_{b=0}^{N-1}e^{2\pi i \frac{b}{N} (m_{J} + \overline{m}_J)} = 1 \quad \text{iff} \quad m_{J} + \overline{m}_J \in N \mathbb{Z}
\end{equation}
and it vanishes in the other cases. The sum over $l$ is more problematic: the $e^{lkw\beta}$ contribution is real. We will nevertheless utilize formally the same strategy as in the previous subsection. The treatment we present here is not rigorous. We will come back to this in the next few sections, but for now let us continue this line of thought.
The above conditions project the values of $m_J$ and $\overline{m}_J$ on a discrete set given by
\begin{align}
m_{J} &= \frac{q}{2} + \frac{i\pi n }{\beta} + \frac{kw}{2}, \\
\overline{m}_{J} &= \frac{q}{2} - \frac{i\pi n }{\beta} - \frac{kw}{2},
\end{align}
for $n, w \in \mathbb{Z}$ and $q \in N\mathbb{Z}$. The prefactor of the Poisson summation in $l$ again precisely cancels the $\frac{L}{2\pi}$ present in the density of states. We remark that for the twisted sectors, the above $J^3_0$ operators are not the same as the $J^3_0$ operators we used to determine the spectrum in section \ref{spectrumsection}. Thus the above operators are not the actual $J$ operators and should be better denoted by $J'$ but we refrain from doing this. With the above form for $U_{lp}$ and these eigenvalues of the $J^3_0$ and $\overline{J}^3_0$ operators, one finds
\begin{align}
\label{ads3spectrum3}
h^{wp}_{jqn} &= \frac{s^2+1/4}{k-2} +\frac{qw}{2} +\frac{i\pi nw}{\beta} + \frac{kw^2}{4}- i\frac{qp\beta}{4\pi} + \frac{ p n}{2} + \frac{kp^2\beta^2}{4(2\pi)^2}, \\
\label{ads3spectrum4}
\bar{h}^{wp}_{jqn} &= \frac{s^2+1/4}{k-2} - \frac{qw}{2} +\frac{i\pi nw}{\beta} + \frac{kw^2}{4} - i\frac{qp\beta}{4\pi} - \frac{ p n}{2} + \frac{kp^2\beta^2}{4(2\pi)^2},
\end{align}
which, upon setting $q \to -q$, $n \to -n$ and $p \to -p$, coincides with equations (\ref{ads3spectrum1prelim}) and (\ref{ads3spectrum2prelim}) 
where $w = \frac{a}{N}$. Which values of $a$ should we sum over? The partition function itself (\ref{pforbi}) is periodic under $a \to a +N$. This symmetry is absent in our analysis since we dropped the infinite product. This issue is settled in subsection \ref{numerical} where we take a numerical approach to analyze the infinite product. The result is that one should restrict to $\left|w\right| < \frac{1}{2}$, which is indeed an interval of length 1. Only in this interval does the infinite product not yield a contribution that corrects the conformal weights of the primaries. This implies the following range for $a$:
\begin{align}
a&= -\frac{N-1}{2} \to \frac{N-1}{2}, \quad N \text{ odd}, \\
a&= - \frac{N-2}{2} \to \frac{N}{2}, \quad N \text{ even}.
\end{align}
Strings that are wound more times than this are not in the spectrum. Discrete momentum on the cigar on the other hand is present for all $q \in N\mathbb{Z}$. As a consistency check, note that the resulting spectrum satisfies $h - \bar{h} \in \mathbb{Z}$. To arrive at the Euclidean BTZ orbifold string spectrum, one should simply replace $\beta \to \frac{4\pi^2}{\beta_{BTZ}}$ as discussed in section \ref{BTZsection}.\\
Is this the end of the story? Not quite, our analysis of the projection exponential was not complete. In principle, the summation over $l$ gives a divergent result on its own. Afterwards we integrate $m$ over the real axis which is only sensitive to dirac-poles at real values. \\
These two operations, while separately nonsense, are given meaning by Poisson's summation formula, in which we naively substitute complex arguments instead of real ones. The answer then turns out to be related to the \emph{proper} analytic continuation of Poisson's summation formula to which we now turn.

\subsection{Interlude: analytic continuation of Poisson's summation formula}
\label{Poisson}
\begin{figure}[h]
\centering
\includegraphics[width=7cm]{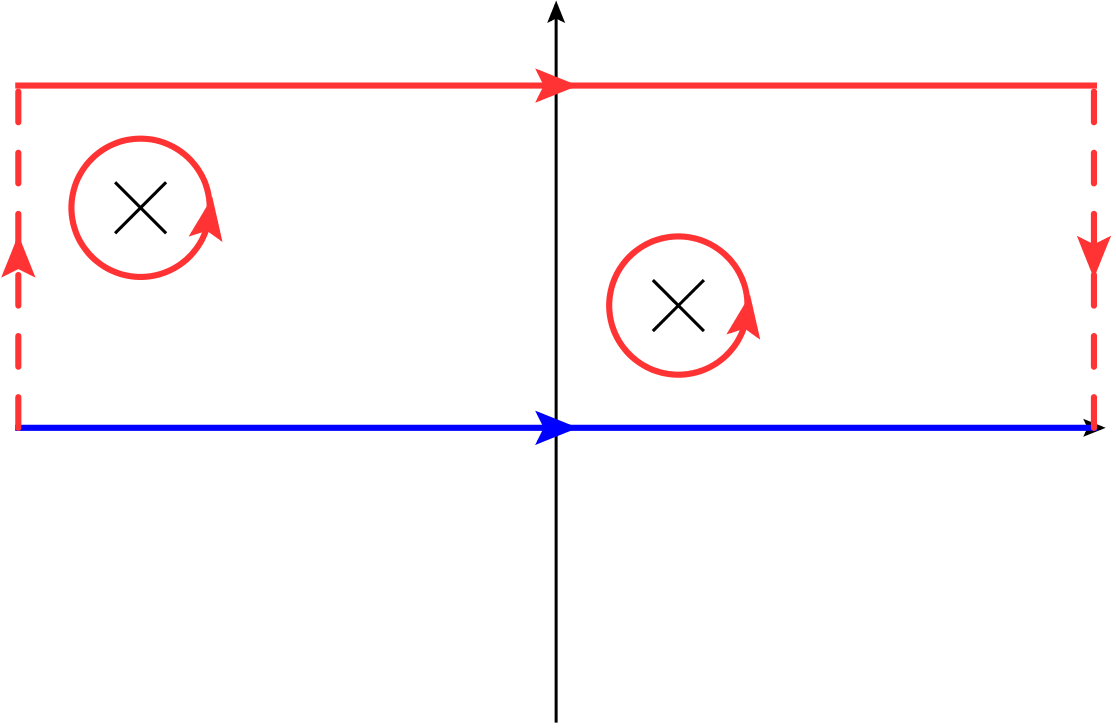}
\caption{Integration contour. The blue contour is the original one. The red contour is the result of the upwards shift. Possible poles of $f$ need to be accounted for and the two vertical segments need to be handled properly as well.}
\label{poisson}
\end{figure}

The Poisson summation formula on the real axis reads
\begin{equation}
\label{ps}
2\pi \sum_{k\in\mathbb{Z}}f(x+2\pi k) = \sum_{n\in\mathbb{Z}}e^{inx}\hat{f}(n).
\end{equation}
We are interested here in deriving the analogous formula for arbitrary \emph{complex} $x$. 
Let us evaluate
\begin{equation}
I = \sum_{n\in\mathbb{Z}}e^{i n z }\int_{\mathbb{R}}dx e^{-inx} f(x)
\end{equation}
where $z$ is an arbitrary complex number and $f$ is a complex function evaluated on the real axis.
Firstly, this equals
\begin{equation}
I = \sum_{n\in\mathbb{Z}}e^{i n z}\hat{f}(n),
\end{equation}
where $\hat{f}$ respresents the Fourier transform of $f$. Secondly, we shift the integration contour up to $+i\Im(z)$ as shown in figure \ref{poisson}.
Assuming no contributions from the short strips at infinity, then up to possible pole contributions, the resulting integral equals
\begin{equation}
I = \sum_{n\in\mathbb{Z}}e^{i n \Re(z) }\int_{\mathbb{R}}dx e^{-inx} f(x+i\Im(z)).
\end{equation}
Now, the sum over $n$ can be readily evaluated as
\begin{equation}
\sum_{n\in\mathbb{Z}}e^{i n \Re(z) }e^{-inx} = 2\pi \sum_{k\in\mathbb{Z}}\delta(-x+\Re(z) + 2\pi k).
\end{equation}
We obtain
\begin{equation}
I = 2\pi \sum_{k\in\mathbb{Z}}  f(\Re(z)+i\Im(z) + 2\pi k) = 2\pi \sum_{k\in\mathbb{Z}}  f(z + 2\pi k).
\end{equation}
Denote the collection of simple poles of $f$ whose imaginary part lies between 0 and $\Im(z)$ as $\mathcal{P}$, then the upwards contour shift also produces 
\begin{equation}
2\pi i \sum_{p_{i} \in \mathcal{P}}\text{Res}_{p_i}(f).
\end{equation}
Putting the pieces together, we finally obtain
\begin{equation}
\sum_{n\in\mathbb{Z}}e^{i n z}\hat{f}(n) = 2\pi \sum_{k\in\mathbb{Z}}\left[  f(z + 2\pi k) + i \sum_{p_{i} \in \mathcal{P}}\text{Res}_{p_i}(f)e^{ik(z-p_i)}\right].
\end{equation}
One sees that Poisson's summation formula still holds, up to the series of the second term.\\
In our case, it is precisely this series that we have neglected. 

\subsubsection*{Example}
Let us discuss a small example that will help us rewrite the above formula. We choose $f(z) = \delta(z)$. Obviously the two vertical segments do not contribute. The function also has no simple poles. Hence, the above reasoning yields
\begin{equation}
\sum_{n\in\mathbb{Z}}e^{inz} = 2\pi \sum_{k\in\mathbb{Z}}\delta(z+2\pi k),
\end{equation}
which is an analytic continuation of Dirac's comb function. We can utilize this formula to rewrite the general Poisson summation formula as
\begin{equation}
\label{ps2}
\boxed{
\sum_{n\in\mathbb{Z}}e^{i n z}\hat{f}(n) = 2\pi \sum_{k\in\mathbb{Z}}\left[  f(z + 2\pi k) + 2\pi i \sum_{p_{i} \in \mathcal{P}}\text{Res}_{p_i}(f)\delta(z-p_i + 2\pi k)\right]}.
\end{equation}
This formula is actually a quite nice example of the identity theorem from elementary complex analysis. Using this theorem, one immediately infers that Poisson's summation formula (\ref{ps}) should hold on the entire complex plane except on the poles and across branch cuts of the complex function $f$ (and their $2\pi k$ shifts). The above formula (\ref{ps2}) describes, in a distributional sense, what the correct formula is when incorporating the poles of the function $f$. Note that if branch points are present in $f$, the above formula does not hold, though it is clear what one should do to obtain the correct formula.

\subsection{Elaborate treatment}
\label{Ham3}
Equiped with this knowledge on the correct analytic continuation of Poisson's summation formula, let us now re-analyze the expression (\ref{expree}) in a more rigorous way. For clarity, let us lump together all $l$- and $b$-independent exponentials into a function $F$ and let us call $G=F\rho$ where $\rho$ is the density of states on $H_3^+$ as given by equation (\ref{dosh3}). For fixed $w$ and $p$, we are interested in:
\begin{align}
&\frac{1}{N}\sum_{l}\sum_{b}\sum_{q} \int_{\mathbb{R}^{+}}ds \int_{\mathbb{R}}dm e^{-4\pi\tau_2 s^2/(k-2)} e^{2\pi i \frac{b}{N}q}e^{-i\beta l m + lkw\beta}F(w,p,m,q)\rho(s,m,q) \\
\label{C77}
&= \sum_{l}\sum_{q\in N\mathbb{Z}} \int_{\mathbb{R}^{+}}ds \int_{\mathbb{R}}dm e^{-4\pi\tau_2 s^2/(k-2)} e^{-i\beta l m + lkw\beta}G(w,p,m,q,s) \\
&= \frac{1}{\beta}\sum_{l}\sum_{q\in N\mathbb{Z}} \int_{\mathbb{R}^{+}}ds e^{-4\pi\tau_2 s^2/(k-2)} e^{lkw\beta}\hat{G}(w,p,l,q,s),
\end{align}
where $\hat{G}$ is the Fourier transform of $G(w,p,\frac{m}{\beta},q,s)$. Using naive Poisson summation in $l$, we would get
\begin{align}
\frac{2\pi }{\beta}\sum_{n}\sum_{q\in N\mathbb{Z}} \int_{\mathbb{R}^{+}}ds e^{-4\pi\tau_2 s^2/(k-2)} F\left(w,p,-ikw+\frac{2\pi n }{\beta},q\right)\rho\left(s,-ikw+\frac{2\pi n }{\beta},q\right).
\end{align}
We see from this that we should substitute $m \to -ikw+\frac{2\pi n }{\beta}$ in both $F$ (representing the remaining $l$- and $b$-independent exponentials) and in $\rho$, given by expression (\ref{dosh3}). We hence obtain for the continuous states the following density of states
\begin{equation}
\label{doscone}
\mbox{{\small{$\displaystyle\rho(s,m,q) = 2\left[\frac{1}{2\pi}2\log(\epsilon) + \frac{1}{2\pi i}\frac{d}{2ds}\log\left(\frac{\Gamma(\frac{1}{2} - is - \frac{q}{2} - \frac{i\pi n}{\beta} -\frac{kw}{2})\Gamma(\frac{1}{2} - is -\frac{q}{2} + \frac{i\pi n}{\beta} +\frac{kw}{2})}{\Gamma(\frac{1}{2} + is -\frac{q}{2}- \frac{i\pi n}{\beta} -\frac{kw}{2})\Gamma(\frac{1}{2} + is -\frac{q}{2}+ \frac{i\pi n}{\beta} + \frac{kw}{2})}\right)\right]$}}}.
\end{equation}
This is again the result of our previous naive treatment in \ref{Ham2}. We now know that this is not entirely correct as possible poles might be present. It is known from earlier work on related models that discrete modes typically arise by crossing poles \cite{Maldacena:2000kv}\cite{Hanany:2002ev}\cite{Israel:2003ry}. In full generality, the computations that follow are quite tedious. We will hence first study the simple case where $\tau_1 = q = p = 0$ and $w>0$ to demonstrate the procedure and then slowly `turn up the heat' to work towards the general case.
\subsubsection*{Simplest case as a warm-up}
In this paragraph only we set $\tau_1 = q = p = 0$ and $w>0$. Right before the Poisson resummation, we have the expression (\ref{C77}):\footnote{An overall factor of $e^{\pi k w^2 \tau_2}$ was not written down here. We will reincorporate this factor in the end.}
\begin{align}
\label{C81}
\sum_l \int_{\mathbb{R}^{+}}ds \int_{\mathbb{R}}dm e^{-4\pi\tau_2 s^2/(k-2)} e^{- i \beta l  m + lkw\beta} \rho(s,m)e^{-2\pi i w m \tau_2}.
\end{align}
We now analyze this step by step. The integration over $m$ is on the real axis. Just like in the proof of the analytic continuation of Poisson's summation formula, we wish to shift the contour to imaginary value $-ikw$. The horizontal piece of the resulting contour and its analysis are what we have done above: they generate the continuous spectrum of states.\footnote{This is indeed simply obtained by substituting complex arguments in the real Poisson summation formula and this is what we did in the previous sections.} What we are interested in in this section, is the possibility of a pole in the complex $m$ plane. Can this occur? Obviously the exponentials in expression (\ref{C81}) have no poles. The density of states is given by
\begin{equation}
\label{dosresid}
\rho(s,m) = 2\frac{L}{2\pi}\left[\frac{1}{2\pi}2\log(\epsilon) + \frac{1}{2\pi i}\frac{d}{2ds}\log\left(\frac{\Gamma(\frac{1}{2} - is - im/2)\Gamma(\frac{1}{2} - is + im/2)}{\Gamma(\frac{1}{2} + is - im/2)\Gamma(\frac{1}{2} + is + im/2)}\right)\right].
\end{equation}
The first part is divergent, but has no poles as a function of $m$. This part hence entirely translates to the contour-shifted contribution. The second part however does allow poles. Firstly we split the logarithm in four parts, then we perform the derivative. The result is four terms of the form of a Digamma function, schematically:
\begin{equation}
\frac{\Gamma^{'}}{\Gamma} = \Psi.
\end{equation}
The Gamma function has no zeros. It has simple poles at all negative integers (including zero). The above combination hence has only simple poles when the imaginary part of $m$ equals $\pm 1$, $\pm 3$, $\pm 5$, $\hdots$. The poles then occur for
\begin{equation}
m = \pm 2s + (2n+1)i, \quad n \in \mathbb{Z}.
\end{equation}
The contour and some of the poles are illustrated in figure \ref{contour} below.\footnote{The two vertical contour segments are more problematic. If fact, the original complex function evaluated on the real axis which we started with, has an ill-defined limit for large real values of $m$. It is of the form $ \lim_{m\to\pm\infty}e^{imx}\ln(\left|Cm\right|)$ where the density of states has a logarithmic form for large values of $m$ (and we only care for the functional form of this equation) and $C$ is some constant. This same asymptotic form holds also on the two vertical contour segments. In a distributional sense, such limits are finite though and are equal to zero. In fact, it holds that $\lim_{n\to\infty} e^{inx}f(n) = 0$ as long as $f(n)$ is of order $\mathcal{O}(n^p)$ for large enough $n$ and for some $p\in\mathbb{R}$. 
For completeness, let us provide a small proof of the statement we need. To have $ \lim_{m\to\pm\infty}e^{imx}\ln(\left|Cm\right|) = 0$ for continuous $m$, it needs to hold for every subsequence so we focus on a subsequence $m_n$ for $n\in\mathbb{N}$ which satisfies $\text{lim}_{n\to\infty} m_n = \pm \infty$. For any testfunction $\varphi$ of compact support (say $L$), we then have
\begin{equation}
\left|\int dx e^{im_n x}\ln(\left|Cm_n\right|)\varphi(x)\right| = \left|\int dx e^{im_n x}\frac{\ln(\left|Cm_n\right|)}{m_n}\varphi'(x)\right| \leq \text{max}_{L}(\left|\varphi'\right|)L\left|\frac{\ln(\left|Cm_n\right|)}{m_n}\right|,
\end{equation}
which goes to zero as $n$ goes to infinity.}
\begin{figure}[h!!!!]
\centering
\includegraphics[width=8cm]{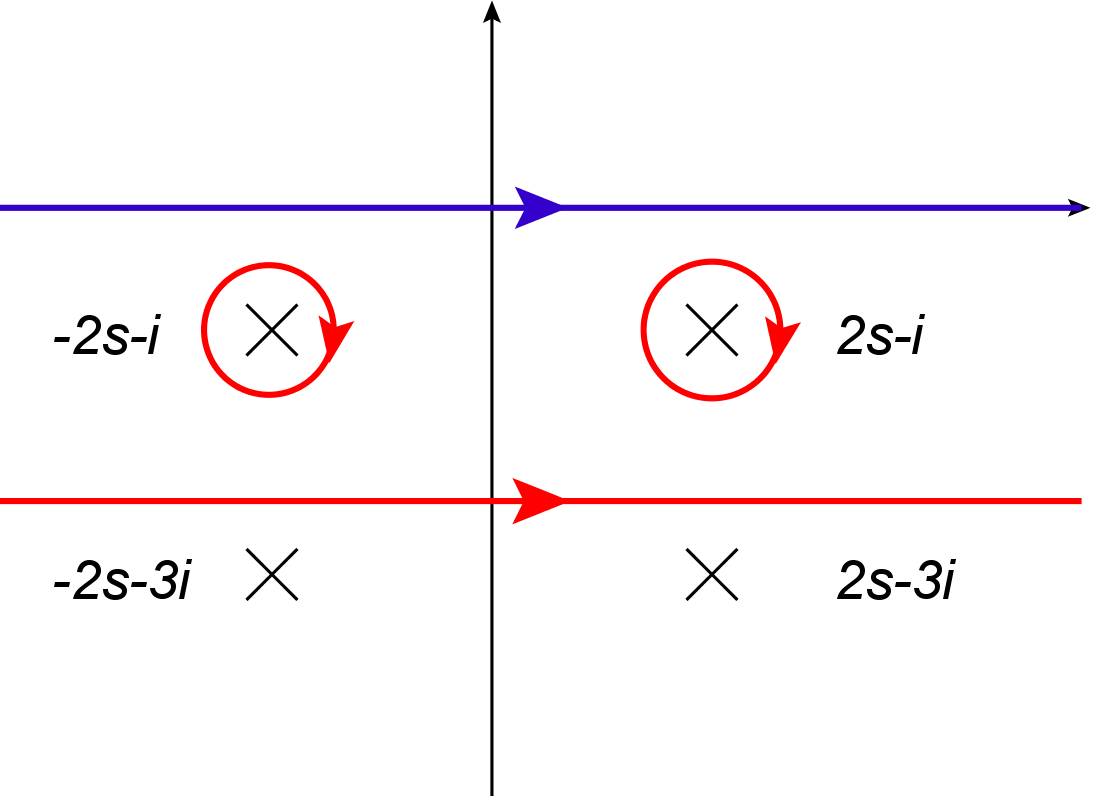}
\caption{The complex $m$ plane with the original integration contour in blue. The shifted contour is drawn in red. Two poles are crossed when $1< kw<3$ as is illustrated.}
\label{contour}
\end{figure}
We hence see that if $kw<1$, no poles are crossed and the analysis presented in \ref{Ham2} remains valid. However, when $kw>1$, at least two poles are present in the region and discrete states appear. Both poles have the same residue and one readily finds that the part of the residue coming from the density of states (\ref{dosresid}) equals\footnote{We again set $\frac{L}{P} = \beta$, with $P$ introduced in subsection \ref{Ham1}.} 
\begin{equation}
-2\pi i \text{Res}\rho(s,m) = -2\pi i 2\frac{\beta}{2\pi} \left(\frac{-2}{4\pi i}\right) = \frac{\beta}{\pi}.
\end{equation}
In the remainder of this paragraph, we focus on the case where exactly two poles are crossed. We generalize this in the next paragraphs. Next we sum the residues of both poles. These differ only in the sign preceding $s$, so schematically we can write
\begin{equation}
\int_{0}^{+\infty}ds e^{As}e^{-Bs^2} + \int_{0}^{+\infty}ds e^{-As}e^{-Bs^2} = \int_{-\infty}^{+\infty}ds e^{As}e^{-Bs^2},
\end{equation}
and both poles are taken care of simultaneously by simply integrating $s$ over the entire real axis. We obtain
\begin{equation}
\frac{\beta}{\pi}\sum_l \int_{\mathbb{R}}ds e^{-4\pi\tau_2 s^2/(k-2)} e^{- i \beta l  (2s-i) + lkw\beta} e^{-2\pi i w (2s-i) \tau_2}.
\end{equation}
The integral over $s$ is a simple Gaussian, yielding
\begin{equation}
\frac{\beta}{2\pi}\sqrt{\frac{k-2}{\tau_2}}\sum_l e^{\beta l(kw-1)} e^{-2\pi w \tau_2}e^{-\frac{(\beta l+2\pi w \tau_2)^2(k-2)}{4\pi\tau_2}}.
\end{equation}
A last Poisson resummation will yield the desired result. For the reader's comfort, we write down the Poisson resummation formula:
\begin{equation}
\label{PRformula}
\sum_{l\in\mathbb{Z}}\exp\left[-\pi a l^2 + 2\pi i b l\right] = a^{-1/2}\sum_{n\in\mathbb{Z}}\exp\left[-\pi\frac{(n-b)^2}{a}\right].
\end{equation}
One then finds
\begin{equation}
\sum_{n\in\mathbb{Z}} e^{-\frac{4\pi^3 n^2}{\beta^2(k-2)}\tau_2}e^{-\frac{4\pi^2 i (kw-1)n}{\beta (k-2)}\tau_2}e^{\frac{4\pi^2 i w n}{\beta}\tau_2}e^{\pi \frac{(kw-1)^2}{k-2}\tau_2}e^{-2\pi kw^2\tau_2}.
\end{equation}
From this, one needs to distill a factor $e^{\frac{\pi\tau_2}{k-2}}$ to serve as (part of) the central charge factor in the partition function. Extracting this and including the extra piece $e^{\pi k w^2 \tau_2}$ we inserted in the operator trace (\ref{expree}), one can see that the following conformal weights can be read off:
\begin{equation}
h = \bar{h} = -\frac{(kw-1)^2}{4(k-2)} + \frac{1}{4(k-2)} + \frac{\pi^2n^2}{\beta^2 (k-2)} + \frac{\pi i (kw-1) n }{\beta (k-2)} - \frac{\pi i w n }{\beta} + \frac{kw^2}{4}.
\end{equation}
We will rewrite this in a more clear way after we incorporate the other quantum numbers.

\subsubsection*{The general case for $w>0$}
\begin{figure}[h]
\includegraphics[width=15cm]{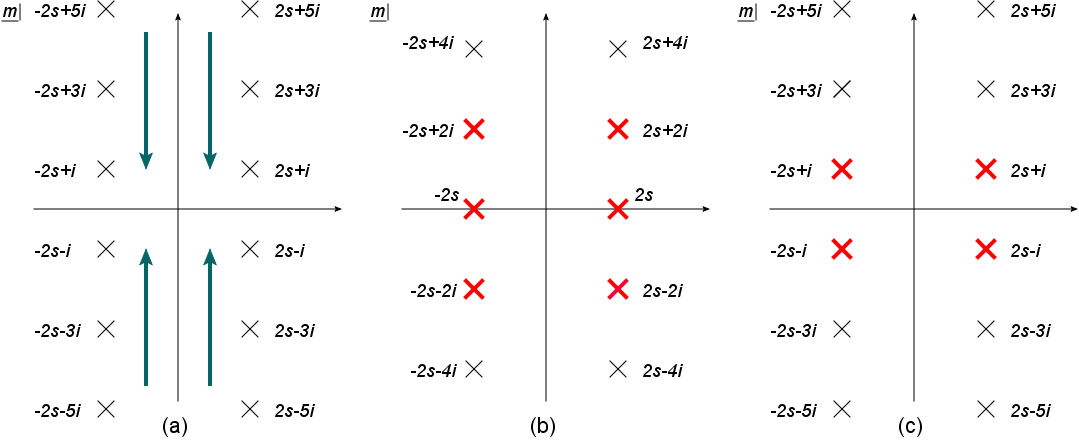}
\caption{(a) Poles in the complex $m$ plane for $q=0$. The big green arrows depict how the poles shift when $q>0$. (b) Poles when $q=3$. The bold red poles are doubly degenerate. The residues cancel out and the pole is effectively absent alltogether. General odd values of $q$ produce the same pattern of poles but with more (or less) poles doubly degenerate and hence absent. (c) Poles when $q=2$. All even values of $q$ produce the same pole pattern with again more poles doubly degenerate for increasing $q$.}
\label{polen}
\end{figure}

We now turn to the general case. We need to incorporate non-zero $\tau_1$, $p$ and $q$ quantum numbers, though we still focus on $w>0$. From now on, we also allow a general number of crossed poles. Let us first take a look at non-zero $q$ quantum numbers since these present the most elaborate modifications. The effect of $q$ is to shift the location of the poles of the Gamma function. The density of states is given by
\begin{equation}
\mbox{{\small{$\displaystyle\rho(s,m,q) = 2\frac{L}{2\pi}\left[\frac{1}{2\pi}2\log(\epsilon) + \frac{1}{2\pi i}\frac{d}{2ds}\log\left(\frac{\Gamma(\frac{1}{2} - is -q/2 - im/2)\Gamma(\frac{1}{2} - is -q/2+ im/2)}{\Gamma(\frac{1}{2} + is -q/2- im/2)\Gamma(\frac{1}{2} + is -q/2+ im/2)}\right)\right]$}}}.
\end{equation}
Poles can be found whenever 
\begin{align}
m &= \pm 2s + (-(2n+1)+q)i, \quad n \in \mathbb{N}, \\
m &= \pm 2s + ((2n+1)-q)i, \quad n \in \mathbb{N},
\end{align}
where $n$ equals $\left\{0, 1, 2 , \hdots\right\}$. The poles hence shift as shown in the figure \ref{polen}(a).\\
Poles originally in the upper half plane shift downwards (for positive $q$) and poles originally in the lower half plane shift upwards.\\
Since $q$ is an integer, one of two situations can occur. 
\begin{itemize}
\item{$q$ is odd. The poles go halfway in between where they are at $q=0$. However, depending on the value of $q$, several poles closest to the real axis become `degenerate'. Computing the residue at these double poles shows that they cancel. The situation is illustrated in figure \ref{polen}(b).}
\item{$q$ is even. The resulting set of poles is exactly equal to those with $q=0$. Likewise, multiply degenerate poles can occur and if they do, the residue becomes zero. The situation is illustrated in figure \ref{polen}(c).}
\end{itemize}
For $q$ negative, the poles in the upper and lower half plane move away from each other. The resulting picture of poles is actually the same as that for $-q$ (which is positive). Hence we can combine both case of $q$ and only consider negative $q$ and the resulting set of poles is then labeled by a positive (or zero) integer $l$, the quantum number associated to the $SL(2,\mathbb{R})$ discrete representation.
The net effect is thus to replace $+q/2 \to -\left|q\right|/2$. \\
The computation is not that hard, though it is important to keep track of all different contributions. Therefore let us first write down a list of every contribution we have. Firstly, we have explicit factors inserted in the operator trace (\ref{expree}). These are
\begin{align}
\label{remaining}
e^{\pi k w^2 \tau_2}e^{-p\tau_1 kw \beta}e^{-\frac{\beta^2kp^2}{4\pi}\tau_2}e^{lkw\beta}.
\end{align}
Secondly, the contributions from the $e^{2\pi i UJ}$ factors give
\begin{align}
\label{remaining2}
e^{2\pi i qw \tau_1}e^{-4\pi\tau_2\left(\frac{-ipq\beta}{4\pi}\right)}
\end{align}
and
\begin{align}
e^{-i\beta l m}e^{-2\pi i w m \tau_2}e^{p\beta m i \tau_1}.
\end{align}
The poles we cross are of the form $m = \pm2s - (odd)i - \left|q\right|i$ with $odd$ an odd integer $ = 1,3,5,\hdots$. We start with
\begin{align}
\sum_l \int_{\mathbb{R}}ds \int_{\mathbb{R}}dm e^{-4\pi\tau_2 s^2/(k-2)} e^{- i \beta l  m + lkw\beta} \rho(s,m)e^{-2\pi i w m \tau_2}e^{p\beta i m \tau_1}.
\end{align}
The first step is extracting the residues of this expression for the poles of the $m$-integral. After this, we perform the Gaussian integral. The result of this step is
\begin{align}
\frac{\beta}{2\pi}\sum_l e^{- \beta l  (odd + \left|q\right|) + lkw\beta} e^{-2\pi w (odd + \left|q\right|) \tau_2}e^{p\beta (odd + \left|q\right|) \tau_1}e^{-\frac{(k-2)}{4\pi\tau_2}(\beta l + 2\pi w \tau_2 - \beta p \tau_1)^2}.
\end{align}
Finally, we Poisson resum this expression. The parameters to be used in formula (\ref{PRformula}) are
\begin{align}
a &= \frac{\beta^2 (k-2)}{4\pi^2\tau_2}, \\
b &= \frac{\beta (kw- odd - \left|q\right|)}{2\pi i} - \frac{(k-2)\beta}{4\pi^2 i \tau_2}\left[2\pi w \tau_2 - \beta p \tau_1\right].
\end{align}
A careful, but straightforward analysis of the resulting factors in combination with the remaining prefactors written in (\ref{remaining}) and (\ref{remaining2}) shows that
\begin{align}
h &= -\frac{\tilde{j}(\tilde{j}-1)}{k-2} + \frac{qw}{2} - \frac{\pi i w n }{\beta} + \frac{kw^2}{4} - \frac{i \beta  pq}{4\pi} - \frac{pn}{2} + \frac{kp^2\beta^2}{4(2\pi)^2}, \\
\bar{h} &= -\frac{\tilde{j}(\tilde{j}-1)}{k-2} - \frac{qw}{2} - \frac{\pi i w n }{\beta} + \frac{kw^2}{4}  - \frac{i \beta  pq}{4\pi} + \frac{pn}{2} + \frac{kp^2\beta^2}{4(2\pi)^2},
\end{align}
where $\tilde{j} = \tilde{m} -l = \frac{kw}{2} - \frac{\left|q\right|}{2} - \frac{i\pi n }{\beta} - l$ where $l=0,1,2,\hdots$. The relation between the parameter $odd$, labeling the poles, and the $SL(2,\mathbb{R})$ parameter $l$ is $ l = \frac{odd -1}{2}$.

\subsubsection*{Negative $w$}
Now we briefly mention the differences for the case $w<0$. In this case, the contour needs to be shifted to the upper half plane. The pole also has its $q$-contribution reversed.\\
A first feature is the overall sign: the contour surrounding the poles is oriented in the opposite direction as before, but also the residue itself has the opposite sign. In all, no overall sign is present.\\
The changes with respect to the case $w>0$ are that for the pole contributions: $kw-odd \to kw+odd$ and $q\to -q$.\footnote{This swap of sign of $q$ is only present obviously for the factors originating from an $m$ quantum number, and not for the factors coming from $e^{2\pi i U J}$ contributions.}

\subsubsection*{Everything combined}
In all, the conformal weights in the most general case are given by
\begin{align}
h &= -\frac{\tilde{j}(\tilde{j}-1)}{k-2} + \frac{qw}{2} - \frac{\pi i w n }{\beta} + \frac{kw^2}{4} - \frac{i \beta  pq}{4\pi} - \frac{pn}{2} + \frac{kp^2\beta^2}{4(2\pi)^2}, \\
\bar{h} &= -\frac{\tilde{j}(\tilde{j}-1)}{k-2} - \frac{qw}{2} - \frac{\pi i w n }{\beta} + \frac{kw^2}{4}  - \frac{i \beta  pq}{4\pi} + \frac{pn}{2} + \frac{kp^2\beta^2}{4(2\pi)^2},
\end{align}
where $\tilde{j} = M - l= \frac{k\left|w\right|}{2} - \frac{\left|q\right|}{2} \pm \frac{i\pi n }{\beta} - l$ where $l=0,1,2,\hdots$ and the $\pm$ symbol equals $+$ if $w<0$ and $-$ if $w>0$. \\
A simple substitution $n \to -n$ allows us to compare these expressions to the continuous weights of equations (\ref{ads3spectrum3}) and (\ref{ads3spectrum4}): all terms are the same except the first one.\\
Since we have a set of discrete states, let us see whether they satisfy the `improved' unitarity constraints \cite{Maldacena:2000hw}: $ \frac{1}{2} < \tilde{j} < \frac{k-1}{2}$ with $\tilde{j}=\frac{k\left|w\right|}{2} - \frac{\left|q\right|}{2} \pm \frac{i\pi n }{\beta} - l$. Since $\tilde{j}$ is a complex quantity, we consider instead $\Re(\tilde{j})$ and it is this number that obeys the inequality in our case as we now show.\\
The first inequality $\frac{1}{2} < \Re(\tilde{j})$ corresponds precisely to the pole-crossing properties of the contour and is hence indeed satisfied in our case. \\
The second inequality $\Re(\tilde{j}) < \frac{k-1}{2}$ requires some input from the infinite product. We have previously argued that (due to brute force numerical computations) $\left|w\right| < 1/2$. It is clear that the worst case scenario for this inequality occurs when $\left|w\right| = 1/2$ and $q=l=0$. But then we have
\begin{equation}
\frac{k}{4} < \frac{k-1}{2} \quad \Leftrightarrow \quad k>2,
\end{equation}
which is obviously satisfied.\footnote{Note that the less strict upper bound $j < k/2$ which follows from the no-ghost theorem would yield an inequality in our case that is satisfied as long as $\left|w\right| < 1$.} Thus every discrete state we constructed as a pole that was crossed by the contour shift, satisfies indeed the unitarity constraints.\\
When taking a larger perspective on this derivation, we find it quite remarkable to find discrete representations, since our original starting point used only the complete set of continuous representations on $H_3^+$. Somehow, these wavefunctions `know' in advance what the discrete representations should look like.

\subsection{Chemical potential}
\label{chemapp}
For the sake of brevity, we will only discuss here the additional steps required compared to the derivations presented in the above analysis. \\
Firstly, we have
\begin{align}
U_{lp} &= -\frac{i\beta}{2\pi}(p\tau - l)(1+i\mu), \\
\Im(U_{lp}) &= - \frac{\beta}{2\pi}(p\tau_1-l) + \frac{\mu p \beta}{2\pi} \tau_2.
\end{align}
Two aspects should be taken care of to evaluate such partition functions. Firstly, one can relate the partition function to that of the conical spaces discussed in the previous subsections by taking $w \to \frac{\mu\beta}{2\pi}p$ in that analysis. The $q$ quantum number however runs over $\mathbb{Z}$ and not $N\mathbb{Z}$ in this case. Secondly, one should add an extra $l$-dependent exponential
\begin{equation}
\label{extra}
e^{-i\beta \mu ql},
\end{equation}
compared to the above analysis. The trace we should evaluate to agree with the path integral derivation is then
\begin{align}
\text{Tr}&\left[\exp\left(-(p\tau_1-l)k\frac{\mu p \beta}{2\pi}\beta+ \pi k\left(\frac{\mu p \beta}{2\pi}\right)^2\tau_2\right)\exp\left(-\tau_2\frac{\beta^2 k p^2}{4\pi}\right)\right. \nonumber \\
&\quad\quad\quad\quad \left.\times\exp\left(4\pi\tau_2\frac{\Delta+1/4}{k-2}\right)\exp(2\pi i (U_{lp} J^{3}_0 + \bar{U}_{lp} \overline{J}^{3}_0))\right] \\
 &=\frac{\beta\sqrt{k-2}}{8\pi\sqrt{\tau_2}}e^{- k\beta^2\left|l-p\tau\right|^2/4\pi\tau_2 + 2\pi \Im(U_{lp})^2/\tau_2 }\left|\sin(\pi U_{lp})\right|^{-2}.
\end{align}
Again we first analyze the continuous representations, obtained by the naive analytic continuation of Poisson's summation formula. The projection conditions now require
\begin{align}
&J_0^3 + \overline{J}_0^3 \in \mathbb{Z}, \\
&\frac{i\beta}{2\pi}(J_0^3 - \overline{J}_0^3) - \frac{\beta \mu}{2\pi}(J_0^3 + \overline{J}_0^3) + \frac{k\beta}{2\pi i}\frac{\mu p \beta}{2\pi} \in \mathbb{Z},
\end{align}
which is solved by
\begin{align}
m_{J} &= \frac{q}{2}(1-i\mu) + \frac{i\pi n }{\beta} + \frac{k\mu p \beta}{4\pi}, \\
\overline{m}_{J} &= \frac{q}{2}(1+i\mu) - \frac{i\pi n }{\beta} - \frac{k\mu p \beta}{4\pi},
\end{align}
for $q,n \in \mathbb{Z}$. Again these states are associated with the horizontal part of the shifted contour. The additional factor (\ref{extra}), although $l$-dependent, is of modulus one and hence it does not affect the location of the shifted contour. For the continuous states, this is relevant since the location of the shifted contour also dictates the substitution one needs to do in the density of states. Hence we see that here we can simply take the density of states (\ref{doscone}) with the replacement $w \to \frac{\mu\beta}{2\pi}p$.
The analysis of the resulting conformal weights is identical to the analysis for the angular orbifolds presented in \ref{Ham2}, except one extra term corresponding to the $\mp i \mu \frac{q}{2}$ in the above relations. This final term gives the corrections to the conformal weights
\begin{align}
h_{extra} &= -\frac{\beta \mu qp}{4\pi} - \frac{i\mu^2 qp\beta}{4\pi}, \\
\bar{h}_{extra} &= \frac{\beta \mu qp}{4\pi} - \frac{i\mu^2 qp\beta}{4\pi},
\end{align}
finally yielding
\begin{align}
h^{p}_{jqn} &= \frac{s^2 +1/4}{k-2} + i\frac{\mu n p}{2} - i\frac{qp\beta}{4\pi} + \frac{ p n}{2} + \frac{kp^2\beta^2}{4(2\pi)^2}(1+\mu^2) - i \frac{\mu^2\beta qp}{4\pi}+ h_{int}, \\
\bar{h}^{p}_{jqn} &= \frac{s^2 +1/4}{k-2} + i\frac{\mu n p}{2} - i\frac{qp\beta}{4\pi} - \frac{ p n}{2} + \frac{kp^2\beta^2}{4(2\pi)^2}(1+\mu^2) - i \frac{\mu^2\beta qp}{4\pi} + \bar{h}_{int}.
\end{align}
As a check, we see that $h - \bar{h} \in \mathbb{Z}$, a necessary condition for modular invariance. \\

\noindent Discrete states can be found by the same strategy as the one used before. Again, we only need to take a closer look at one contribution (\ref{extra}) while the remaining terms can be readily found from the conical spaces by taking $w \to \frac{\mu\beta}{2\pi}p$. 
One can then go through exactly the same computations as before to handle the discrete states. The extra exponential (\ref{extra}) simply goes along for the ride during the computations and only makes its appearance when the Poisson resummation needs to be applied. The new Poisson resummation parameters to be used in (\ref{PRformula}) are now
\begin{equation}
a = \frac{\beta^2(k-2)}{4\pi^2\tau_2}, \quad b = \frac{\beta(kw-odd - \left|q\right|- i\mu q))}{2\pi i } - \frac{(k-2)\beta}{4 \pi^2 i \tau_2}\left[2\pi w \tau_2 - \beta p \tau_1\right].
\end{equation}
The only difference is hence the replacement $\left|q\right| \to \left|q\right| + i\mu q$. Without going into details, we report the final result:
\begin{align}
h^{p}_{jqn} &= -\frac{\tilde{j}(\tilde{j}-1)}{k-2} - i\frac{\mu n p}{2} - i\frac{qp\beta}{4\pi} - \frac{ p n}{2} + \frac{kp^2\beta^2}{4(2\pi)^2}(1+\mu^2) - i \frac{\mu^2\beta qp}{4\pi}+ h_{int}, \\
\bar{h}^{p}_{jqn} &= -\frac{\tilde{j}(\tilde{j}-1)}{k-2} - i\frac{\mu n p}{2} - i\frac{qp\beta}{4\pi} + \frac{ p n}{2} + \frac{kp^2\beta^2}{4(2\pi)^2}(1+\mu^2) - i \frac{\mu^2\beta qp}{4\pi} + \bar{h}_{int},
\end{align}
where now $\tilde{j} = \frac{k\left|\mu p\right|\beta}{4\pi} - \frac{\left|q\right|}{2} - \frac{i\mu q}{2} \pm \frac{in\beta}{\pi} - l$. Again the imaginary part of $\tilde{j}$ is irrelevant for satisfying the unitarity constraints. We also note that the $+ i\mu q/2$ term does not contain an absolute value.

\part{String Thermodynamics revisited}

\chapter{General Lessons on Thermodynamics}
\label{chgentd}
Armed with the two specific examples of Rindler space and $AdS_3$, we are ready to draw some general conclusions here concerning thermodynamics in string theory. \\
This chapter is based on parts of \cite{Mertens:2014cia}. Additional unpublished results and ideas are also presented in this chapter. \\ 

\noindent This chapter is organized as follows. \\
Section \ref{higherwind} contains some general results on the different perspectives on string thermodynamics (the strip versus the fundamental domain). In section \ref{generalform} we discuss, using the two explicit examples discussed previously, what the form of the thermal scalar action would be in general. 

\section{Higher winding modes and multistring states}
\label{higherwind}
String thermodynamics in general can be rewritten in different ways: on the modular strip or on the fundamental domain. An interesting question is how the associated quantum numbers are related to corrections to thermodynamical quantities. For instance, higher winding modes and discrete momentum modes along the thermal circle on the fundamental domain do not represent physical (Lorentzian) states. Instead they represent corrections to thermodynamics. The precise sense in which this happens will be explored in this section, both for flat space and for a specific curved space example: the WZW $AdS_3$ model. It is known that the strip and fundamental domain both are viable routes to string thermodynamics. The main question we would like to examine here is how precisely the different modes of the fundamental domain encode thermodynamical properties. Partial results on this story are known in the literature \cite{Kruczenski:2005pj}\cite{Bowick:1989us}\cite{Deo:1989bv}\cite{Deo:1988jj}\cite{Spiegelglas:1988hr}, yet the full story and to what extent it can be generalized to curved space has not been studied before. \\
First we look at the strip domain and what its single quantum number $r$ (only winding around one torus cycle) tells us about thermodynamics. This quantum number is the one exhibited in the worldsheet dimensional reduction approach to critical string thermodynamics in section \ref{pathderiv}.\\
We start with the general expression of the partition function of a multiparticle bosonic system:
\begin{equation}
Z_{\text{mult}} = \prod_{k}\frac{1}{1-e^{-\beta E_k}}.
\end{equation}
The free energy of the gas is hence
\begin{equation}
\label{expans}
-\beta F = \sum_{r'=1}^{+\infty}\frac{\sum_k e^{-r'\beta E_k}}{r'}.
\end{equation}
If instead one would assume Maxwell-Boltzmann statistics, we would have $Z_{\text{mult}} = \exp{z}$, which would imply
\begin{equation}
-\beta F = \sum_k e^{-\beta E_k}.
\end{equation}
Comparing the above equations, we see that the $r'=1$ term corresponds to the Maxwell-Boltzmann classical approximation. All higher terms in the Taylor expansion result in Bose-Einstein statistics. 

\subsection{Multistring Instabilities}
Let us look more closely into the structure of the (non-interacting) multistring partition function. For simplicity we only discuss a bosonic system (bosonic string theory) explicitly. The free energy of the string gas can be written as
\begin{equation}
\label{loga}
F= - \frac{1}{\beta}\ln\left(1+z^{\beta}_1 + z^{\beta}_2 + \hdots\right),
\end{equation}
where $z^{\beta}_i$ denotes the partition function for $i$ strings at temperature $\beta^{-1}$:\footnote{So in our notation: $z^{\beta}_1 = z$.}
\begin{equation}
z^{\beta}_i = \text{Tr}_{i\text{ strings}}e^{-\beta H}.
\end{equation}
This is also equal to 
\begin{equation}
F= - \frac{1}{\beta}\left(Z^{\beta}_{\left|r\right|=1} + Z^{\beta}_{\left|r\right|=2} + \hdots\right)
\end{equation}
where we consider the single string path integrals on the thermal manifold for higher winding numbers (strip quantum number $r$). In writing this, we have used the notation $Z^{\beta}_{\left|r\right|=i} = Z^{\beta}_{r=i} + Z^{\beta}_{r=-i}$. The idea is then to series-expand the logarithm in (\ref{loga}) while using relations between the multi-boson partition functions. Let us show this in a bit more detail. Assuming  $z^{\beta}_1 = Z^{\beta}_{\left|r\right|=1}$ (this is the single-string equivalent of Polchinski's result \cite{Polchinski:1985zf}), the authors of \cite{Kruczenski:2005pj} prove that the next term in the series matches for flat space provided:\footnote{The second order term in the expansion of the logarithm is of the form
\begin{equation}
z^{\beta}_2 - \frac{1}{2}\left(z^{\beta}_1\right)^2 = \frac{1}{2}z^{2\beta}_1,
\end{equation}
where properties of the two-boson partition function were used.}
\begin{equation}
Z^{\beta}_{\left|r\right|=2} = \frac{1}{2}z^{2\beta}_1,
\end{equation}
which is indeed satisfied for flat space.
As an explicit illustration, we here go one step further and prove that the third term also matches. The partition function for three bosonic strings can in general be written as
\begin{align}
z^{\beta}_3 &= \sum_{k>l>m}e^{-\beta(E_k+E_l+E_m)} + \sum_{k,l}e^{-\beta(2E_k + E_l)} \\
 &= \frac{1}{6}\left[\sum_{k,l,m}e^{-\beta(E_k+E_l+E_m)} -3\sum_{k,l}e^{-\beta(2E_k+E_l)}+2\sum_ke^{-3\beta E_k} \right] + \sum_{k,l}e^{-\beta(2E_k + E_l)} \\
 &= -\frac{1}{3}\left(z^{\beta}_{1}\right)^3 + z^{\beta}_2z^{\beta}_1 + \frac{1}{3}z^{3\beta}_1,
\end{align} 
where we used $z^{2\beta}_1 = 2z^{\beta}_2 - \left(z^{\beta}_1\right)^2$. We can rewrite this as
\begin{equation}
\frac{1}{3}z^{3\beta}_1 = \frac{1}{3}\left(z^{\beta}_{1}\right)^3 - z^{\beta}_2z^{\beta}_1 + z^{\beta}_3.
\end{equation}
When expanding the logarithm (\ref{loga}), the third order terms are given by
\begin{equation}
z^{\beta}_3 - z^{\beta}_2z^{\beta}_1 + \frac{1}{3}\left(z^{\beta}_{1}\right)^3.
\end{equation}
The result is hence proven if we can prove that
\begin{equation}
\label{above}
Z^{\beta}_{\left|r\right|=3} = \frac{1}{3}z^{3\beta}_1.
\end{equation}
This feature is manifest in the explicit flat space free energy expression \cite{Alvarez:1986sj}.\footnote{It is also present in the WZW $AdS_3$ thermal partition function, see for instance equation (31) in \cite{Maldacena:2000kv}. We will discuss more aspects of this model further on.} It also holds in the string path integral derivation sketched in section \ref{pathderiv}.\footnote{When following the derivation in chapter \ref{chth}, the following changes are made. Changing the winding number causes $\beta \to k \beta$ in the string action. The result of the $\tau_1$-integral hence also has the same substitution. However, the integration over the quantum fluctuation $\tilde{X}_0$ gives the same result as before: the periodicity $\beta$ itself is not changed. In all, we get a prefactor $1/\left|k\right|$ to the single string partition function.} Thus in this case we have $Z^{k\beta}_{r=1} = \left|k\right| Z^{\beta}_{r=k}$ for $k\in\mathbb{Z}$ leading to
\begin{equation}
Z^{k\beta}_{\left|r\right|=1} = k Z^{\beta}_{\left|r\right|=k}, \quad k \in \mathbb{N}
\end{equation}
and the above relation (\ref{above}) is a special case of this. The derivation holds now also for curved spaces: the only step that remained to be proven was the above relation and using the string path integral approach of section \ref{pathderiv}, we see that it should hold also for curved spaces. As a first conclusion, it seems that the single quantum number $r$ in the strip should be identified with the $r'$ quantum number in the expansion of the non-interacting bosonic string partition function (\ref{expans}).\footnote{Actually it is the contribution of $+r$ and $-r$ that, upon adding, equals the $r'$ term in the bosonic partition function expansion.} \\

\noindent The benefit of the above discussion is that now we can see explicitly how the multistring partition functions are determined in terms of those containing fewer strings. Take for instance the two-string partition function $z_2^{\beta}$, which satisfies
\begin{equation}
\frac{1}{2}z^{2\beta}_{1} = z^{\beta}_2 - \frac{1}{2}\left(z^{\beta}_1\right)^2.
\end{equation}
We can read off where this partition function becomes singular: at $T > T_H$, the single particle partition function $z_1^{\beta}$ diverges. Since $z^{2\beta}_{1}$ does not diverge, it is clear that the two-string partition function $z_2^{\beta}$ also diverges. For $T > 2T_H$, if the expression for the partition function can be trusted, an extra singularity develops since now also $z^{2\beta}_{1}$ diverges. Analogously one can treat the other multiparticle partition sums. What one learns is that at $T=T_H$, not only single-string partition functions diverge, but also all multiparticle sums. In \cite{Spiegelglas:1988hr} it was argued that at $T= lT_H$, the gas becomes unstable for $l$-string states whereas $(l+1)$-string states are still stable. This is interpreted as determining how many long strings are present in the near-Hagedorn gas. Here however we see that multiparticle partition functions all diverge simultaneously at $T=T_H$ whereas formally additional instabilities develop at higher temperatures that are only present in the multistring sector.

\noindent This discussion shows that the number of long strings is not fixed by the temperature (opposed to \cite{Spiegelglas:1988hr}). Thus multi-long-string configurations are also possible and near $T_H$, the gas reconfigures to a system of (possibly multiple) long strings.

\subsection{Interpretation of Hamiltonian quantum numbers in the strip}
\begin{figure}[h]
\centering
\includegraphics[width=12cm]{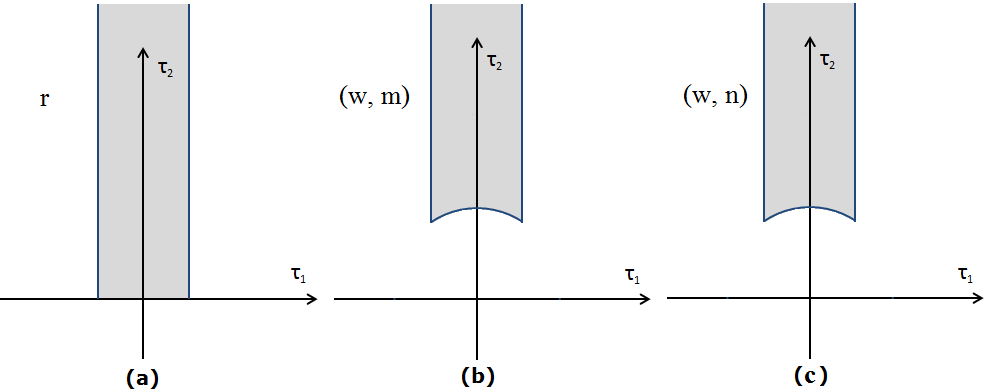}
\caption{(a) The single quantum number $r$ in the strip. (b) The two quantum numbers $w$ and $m$ in the fundamental domain. These appear when performing the torus worldsheet path integral. The link between (a) and (b) is made by utilizing the theorem of \cite{McClain:1986id}\cite{O'Brien:1987pn}. (c) The quantum number $m$ is exchanged for the number $n$ by performing a Poisson resummation. The quantum numbers ($w$, $n$) have meaning in terms of winding and discrete momentum around the compactified time direction and are hence apparent in a Hamiltonian formulation of the theory.}
\label{qn}
\end{figure}

A question that really interests us here is, given a sector ($w$, $n$) in the Hamiltonian (field theory) approach, what does it correspond to in the free energy? In particular, to what part of the strip-sum does it contribute? The answer requires a simple (albeit confusing) number-theoretic reasoning.\\
The translation of the single strip quantum number $r$ and the discrete momentum and winding quantum numbers ($w$, $n$) proceeds in two steps. We shall illustrate these in flat space\footnote{We will come back to curved space further on.} where special attention is paid to analyzing what the contribution is of a specific Hamiltonian sector ($w$, $n$) to any $r$ in the strip. For clarity, the different quantum numbers and their links are shown in figure \ref{qn}.
We will make the link in two steps. In the language of figure \ref{qn}, we first make the link between (a) and (b) and then between (b) and (c).

\subsubsection*{Step 1: Going from (a) to (b)} 
The starting point is the single quantum number $r$ in the strip domain. We start by considering the $\left|r\right|= 1$ contribution. A well-known startegy, is to restrict the modular integral to the fundamental domain while introducing an extra quantum number: we arrive at the double ($w$, $m$) quantum numbers where $m$ and $w$ are coprime \cite{Kruczenski:2005pj} to get the strip $\left|r\right|= 1$ result. These quantum numbers are obtained when one computes the partition function through path integral methods on the fundamental modular domain.
Now we generalize this: we focus on the strip $\left|r\right|= q$ contribution with $q$ an arbitrary positive integer. The coset expansion results in the set ($w$, $m$) with $\frac{m}{q}$ and $\frac{w}{q}$ relatively prime integers. In particular, $m$ and $w$ are multiples of $q$. The different classes exhibited above, can be written as ($w$, $wk + N$) for $N=0\hdots w-1$, for fixed $w$ and $N$, and $k$ runs over $\mathbb{Z}$. In detail, we have for the lowest lying quantum numbers:
\begin{align}
\begin{array}{c|ccccccc}
\hline
F & (w, m) \text{ sectors} \\
\hline
\left|r\right|= 1 & (\pm1,k) & (\pm2, 2k+1)& (\pm3, 3k+1) & (\pm3, 3k+2) & (\pm4, 4k+1) & (\pm4, 4k+3) &\hdots\\
\left|r\right|= 2 & (\pm2, 2k) &  (\pm4, 4k+2) & (\pm6, 6k+2) & (\pm6, 6k+4) &\hdots \\
\left|r\right|= 3 & (\pm3, 3k) & (\pm6, 6k+3) & \hdots \\
\left|r\right|= 4 & (\pm4, 4k) &  \hdots 
\end{array}
\end{align}

\noindent Up to this point, we have merely used the theorem of \cite{McClain:1986id}\cite{O'Brien:1987pn}, but we have paid extra attention to which sectors ($w$, $m$) correspond to a single $r$ quantum number. 

\subsubsection*{Step 2: Going from (b) to (c)}
The second step requires a Poisson resummation on the quantum number $m$. Setting $R_0 = \frac{\beta}{2\pi}$, we remind the reader of the standard textbook result:
\begin{equation}
\label{origi}
\sum_{m\in \mathbb{Z}}\text{exp}\left(-\frac{\pi R_0^2}{\alpha'\tau_2}\left|m-w\tau\right|^2\right) = \sqrt{\frac{\alpha'\tau_2}{R_0^2}}\sum_{n\in\mathbb{Z}}\text{exp}\left(-\pi\tau_2\left(\frac{\alpha' n^2}{R_0^2} + \frac{w^2R_0^2}{\alpha'}\right)+2\pi i \tau_1 nw\right).
\end{equation}
Here we wish to analyze this expression in more detail, tracking the influence of specific subsectors (like we did in the previous subsection). \\
We have discussed above that the transition from one quantum number to two quantum numbers splits these in some classes of the form ($w$, $m=wk + N$) where $N=0\hdots w-1$. Some of these classes will correspond to the same $r$, but for the moment we do not care about this. For one such sector (fix $w$ and $N$), we have
\begin{equation}
\sum_{m = wk+N, \, k\in\mathbb{Z}}\exp\left(-\frac{\pi R_0^2}{\alpha'\tau_2}\left|m-w\tau\right|^2\right) = \sum_{k=-\infty}^{+\infty}\exp\left(-\frac{\pi R_0^2}{\alpha'\tau_2}\left|(wk+N)-w\tau\right|^2\right),
\end{equation}
which can be Poisson resummed into
\begin{equation}
\label{vg1}
\sqrt{\frac{\alpha'\tau_2}{R_0^2}}\frac{1}{w}\sum_{\tilde{n}=-\infty}^{+\infty}e^{2\pi i \frac{N}{w}\tilde{n}}\exp\left(-\frac{\pi \alpha'\tau_2}{R_0^2w^2}\tilde{n}^2 + 2\pi i \tau_1 \tilde{n} - \frac{\pi R_0^2w^2}{\alpha'\tau_2}\tau_2^2\right).
\end{equation}
An at first sight troubling aspect is the appearance of $\frac{\tilde{n}^2}{w^2}$ in the exponential, suggesting some sort of fractional discrete momentum. Note though that summing the above expression over $N$, restricts $\tilde{n} = w \mathbb{Z}$, such that one obtains (with $\tilde{n} = w n$ and including the sum over $N$) 
\begin{equation}
\sqrt{\frac{\alpha'\tau_2}{R_0^2}}\sum_{n=-\infty}^{+\infty}\exp\left(-\frac{\pi \alpha'\tau_2}{R_0^2}n^2 + 2\pi i \tau_1 n w- \frac{\pi R_0^2w^2}{\alpha'\tau_2}\tau_2^2\right),
\end{equation}
which is indeed the full original result (\ref{origi}). To conclude, the fractional discrete momenta disappear when summing the different sectors and in the end we only care about the non-fractional ones. With this in mind, equation (\ref{vg1}) can be rewritten as
\begin{equation}
\sqrt{\frac{\alpha'\tau_2}{R_0^2}}\frac{1}{w}\sum_{n=-\infty}^{+\infty}\underbrace{e^{2\pi i N n}}_{1}\exp\left(-\frac{\pi \alpha'\tau_2}{R_0^2}n^2 + 2\pi i \tau_1 n w- \frac{\pi R_0^2w^2}{\alpha'\tau_2}\tau_2^2\right).
\end{equation}
What we have found, is that a single sector ($w$, $m=wk + N$) (for fixed $w$ and $N$) gets contributions from the Hamitonian sectors ($w$, $n$) for all $n$ but with weight $1/w$.

\noindent Turning this conclusion around, the above shows that a Hamiltonian sector ($w$, $n$) contributes equally to the different sectors ($w$, $m=wk + N$), $N=0\hdots w-1$. For instance, in each sector the $n=0$ mode dominates and contributes equally $\sim 1/w$. Note that all discrete momentum states and winding states correspond to thermal corrections to the free energy and these are \emph{not} the physical particles appearing in the Lorentzian theory. 

\subsubsection*{Final result}
Having found how a Hamiltonian sector ($w$, $n$) contributes to a class of sectors ($w$, $m=wk+N$), we now ask how many of the latter classes of sectors are included in a single $r$ sector.
We have already written down the tabel including the few lowest quantum numbers above. For instance, the $w=\pm4$ states are highlighted below:
\begin{align}
\begin{array}{c|ccccccc}
\hline
F & (w, m) \text{ sectors} \\
\hline
\left|r\right|= 1 & (\pm1,k) & (\pm2, 2k+1)& (\pm3, 3k+1) & (\pm3, 3k+2) & \textcolor{blue}{(\pm4, 4k+1)} & \textcolor{blue}{(\pm4, 4k+3)} &\hdots\\
\left|r\right|= 2 & (\pm2, 2k) &  \textcolor{blue}{(\pm4, 4k+2)} & (\pm6, 6k+2) & (\pm6, 6k+4) &\hdots \\
\left|r\right|= 3 & (\pm3, 3k) & (\pm6, 6k+3) & \hdots \\
\left|r\right|= 4 & \textcolor{blue}{(\pm4, 4k)} &  \hdots 
\end{array}
\end{align}
It is hence clear that a state ($4, n$) for any fixed $n$ contributes with half weight to the $\left|r\right|=1$ sector and with quarter weight to both the $\left|r\right|=2$ and $\left|r\right|=4$ sectors. This procedure can be readily generalized into the following algorithm for finding the contribution of a ($w$, $n$) Hamiltonian state to the different $r$-sectors. 
\begin{itemize}
\item[$(i)$] Find all divisors of $w$.
\item[$(ii)$] For each divisor $i$, find all multiples of $i$ (smaller or equal to $w$) that are \emph{not} multiples of any other larger divisor of $w$ (except 1).
\item[$(iii)$] The number of such multiples for every $i$ is the relative weight to the $\left|r\right|=i$ sector.
\end{itemize}
Note that the same strategy is used for every $n$ and the sum of the weights is $w$. A few examples make this clear. \\
Let us look at the ($w=11$, $n$) state. The above algorithm has the following outcome:
\begin{align}
\begin{array}{|c|c|c|}
\hline
$r$\text{-sector} & \text{multiples} & \text{relative weight} \\
\hline
\left|r\right|=1 & (1,2,3,4,5,6,7,8,9,10) & 10/11 \\
\left|r\right|=11 & (11) & 1/11 \\
\hline
\end{array}
\end{align}
The reason for this simplicity is obviously that 11 is a prime number. A more involved example is the case of a ($w=12$, $n$) state. The above algorithm has the following outcome:
\begin{align}
\begin{array}{|c|c|c|}
\hline
$r$\text{-sector} & \text{multiples} & \text{relative weight} \\
\hline
\left|r\right|=1 & (1,5,7,11) & 4/12 \\
\left|r\right|=2 & (2,10) & 2/12 \\
\left|r\right|=3 & (3,9) & 2/12 \\
\left|r\right|=4 & (4,8) & 2/12 \\
\left|r\right|=6 & (6) & 1/12 \\
\left|r\right|=12 & (12) & 1/12 \\
\hline
\end{array}
\end{align}
Note that in particular, the $w=\pm 1$ state only contributes to the $\left|r\right|=1$ sector with full weight. Hence for these sectors, roughly speaking, the winding number $w$ in the fundamental domain coincides with the number $r$ in the strip, explaining the success we had with equating the strip path integral derivation and the thermal scalar field theory. 

\subsubsection*{Extension to $AdS_3$}
It is interesting to note that the above story applies fully to the $AdS_3$ WZW model as well.
Firstly, the mapping from strip $r$ to ($w$, $m$) fundamental domain sectors appears generally true: it is explicitly discussed for $AdS_3$ in \cite{Maldacena:2000kv} and we will return to it in general in section \ref{Mcr}. Secondly, we have discussed this model in detail in chapter \ref{chwzw} and in particular section \ref{Ham1} there allows us to discuss the second step as well. In the notation given there where $l$ takes the role of our $m$, an intermediate step in the computation involves the following Poisson resummation:
\begin{align}
\frac{1}{P}\int_{\mathbb{R}}dm\sum_{l\in\mathbb{Z}}\exp\left(2\pi i \left(\frac{\beta l }{2\pi}m\right)\right) &= \frac{1}{P}\frac{2\pi }{\beta}\int_{\mathbb{R}}dm\sum_{n\in\mathbb{Z}}\delta\left(m-\frac{2\pi n }{\beta}\right).
\end{align}
Focussing now on sectors with $l\to wk + N$ for fixed $N=0\hdots w-1$ and $w$, yields instead for the Poisson resummation step\footnote{The reader is referred to chapter \ref{chwzw} for details on where this computation comes from.}
\begin{align}
\frac{1}{P}\int_{\mathbb{R}}dm&\sum_{k\in\mathbb{Z}}\exp\left(2\pi i \frac{\beta m }{2\pi}N\right)\exp\left(2\pi i \left(\frac{\beta kw }{2\pi}m\right)\right) \nonumber \\
&= \frac{1}{P}\frac{2\pi }{w\beta}\int_{\mathbb{R}}dm \exp\left(2\pi i \frac{\beta m }{2\pi}N\right) \sum_{\tilde{n}\in\mathbb{Z}}\delta\left(m-\frac{2\pi \tilde{n} }{w\beta}\right) \nonumber \\
&= \frac{1}{P}\frac{2\pi }{w\beta}\int_{\mathbb{R}}dm \sum_{\tilde{n}\in\mathbb{Z}} \exp\left(2\pi i \frac{\tilde{n} }{w}N\right) \delta\left(m-\frac{2\pi \tilde{n} }{w\beta}\right).
\end{align}
As a check, summing over $N$ gives us a global factor of $w$, and restricts to $\tilde{n} = w \mathbb{Z}$, where we call this new integer $n$ again which will correspond to discrete momentum around the cylinder. \\
Again looking at a fixed sector, we see that if $\tilde{n} \neq w \mathbb{Z}$, the contribution will disappear upon summing. When $\tilde{n} = w \mathbb{Z}$, the phase factor disappears and all sectors contribute equally. The situation is hence exactly as in flat space. It is tempting to speculate that this story actually applies in general to spaces where the thermal circle is topologically stable.

\subsubsection*{Some concluding remarks}
Let us note that the Hamiltonian field theory approach is much better suited to describe subleading corrections to thermodynamical quantities in the $\beta \to \beta_H$ limit, since these are identified with the next-to-lowest mass states. Such corrections are `dispersed' over the different $r$ sectors from the thermodynamical strip point of view.\footnote{For instance in flat space, the subleading corrections require a series-expansion of the Dedekind $\eta$-function that needs to be combined with the series of exponentials in $r$.} \\

\noindent The detailed picture exhibited in the two preceding subsections appears more complicated for black hole spacetimes (where the thermal circle pinches off at some point). In chapters \ref{chri} and \ref{chwzw} we discussed that for such spaces not every $w$ mode is present in the thermal spectrum \cite{Mertens:2013zya}\cite{Mertens:2014nca}. Therefore the above story cannot hold for such spaces and one needs to look more carefully. Note though that, provided the one-loop thermal path integral corresponds to the free field state counting, the genus one thermal partition function should include all $r$ sectors to satisfy general Bose-Einstein statistics.

\section{General form of the thermal scalar action}
\label{generalform}
Equiped with our knowledge on the $\alpha'$ corrections in two example backgrounds, we will here try to draw some general conclusions. First we look at the restrictions obtained on the general thermal scalar action and after that we will present an argument in favor of the absence of $\alpha'$ corrections for type II superstrings.

\subsection{Constraint on the form of the thermal scalar action}
We can obtain a partial constraint on terms that could appear in the thermal scalar action by using what we know from the $AdS$ and Euclidean Rindler actions. Our focus is on bosonic string theory for which $\alpha'$-corrections are definitely present. Obviously, we can only determine the action up to terms that cancel by the background equations of motion. In fact, we will find that the derivative parts of the lowest order correction in $\alpha'$ to the thermal scalar action is determined uniquely up to this freedom. We will see that the Rindler correction term is fully given by this lowest order correction (a priori it could come from arbitrary order in $\alpha'$). Let us write down some of the possible terms that could appear in the thermal scalar action for both $AdS_3$ and Rindler space. \\
$\mathbf{AdS_3}$:
\begin{align}
m^2 TT^{*} &= -\frac{4}{\alpha'}TT^{*}, \quad \text{bosonic}\quad \text{or} \quad m^2 TT^{*} = -\frac{2}{\alpha'}TT^{*}, \quad \text{type II},\\
\tilde{G}^{\mu\nu}\partial_{\mu}T\partial_{\nu}T^{*} &= \frac{\beta^2}{4\pi^2\alpha'^2}TT^{*} + \partial_\rho T \partial_\rho T^{*} - \frac{\beta}{2\pi\alpha'}(T\partial_\phi T^{*}- T^{*}\partial_\phi T) \nonumber \\
&\quad + \frac{1}{\sinh(\rho/l)^2}\partial_\phi T \partial_\phi T^{*},\\
\tilde{R} TT^{*} &= \frac{4}{\cosh(\rho/l)^2}\frac{1}{l^2} TT^{*},\\
\alpha'\tilde{R}^{\mu\nu}\partial_\mu T \partial_\nu T^{*} &=  \frac{2}{\cosh(\rho/l)^2}\frac{\alpha'}{l^2}\partial_\rho T \partial_\rho T^{*} +\frac{2}{\sinh(\rho/l)^2}\frac{\alpha'}{l^2}\partial_\phi T \partial_\phi T^{*},
\end{align}
and
\begin{align}
\partial_\mu \tilde{\Phi} \partial^{\mu} \tilde{\Phi} TT^{*} &= \frac{1}{l^2}\tanh(\rho/l)^2 TT^{*}, \\
\alpha'\partial^{\mu}\tilde{\Phi} \partial^{\nu} \tilde{\Phi} \partial_{\mu}T\partial_{\nu}T^{*} &= \frac{\alpha'}{l^2}\tanh(\rho/l)^2\partial_\rho T \partial_\rho T^{*}. \\ 
\alpha'\nabla_{\mu}\nabla_{\nu}\tilde{\Phi}\partial_{\mu}T\partial_{\nu}T^{*} &= -\frac{\alpha'}{l^2}\left(\frac{1}{\cosh(\rho/l)^2}\partial_\rho T \partial_\rho T^{*} + \frac{1}{\sinh(\rho/l)^2}\partial_\phi T \partial_\phi T^{*}\right).
\end{align}
\textbf{Rindler}:
\begin{align}
m^2 TT^{*} &= -\frac{4}{\alpha'}TT^{*}, \quad \text{bosonic}\quad \text{or} \quad m^2 TT^{*} = -\frac{2}{\alpha'}TT^{*}, \quad \text{type II},\\
\tilde{G}^{\mu\nu}\partial_{\mu}T\partial_{\nu}T^{*} &= \frac{\rho^2}{\alpha'^2}TT^{*} + \partial_\rho T \partial_\rho T^{*}, \\
\tilde{R} TT^{*} &= -\frac{4}{\rho^2}TT^{*}, \\
\alpha'\tilde{R}^{\mu\nu}\partial_\mu T \partial_{\nu} T^{*} &= -\frac{2}{\alpha'}TT^{*} - 2\frac{\alpha'}{\rho^2} \partial_\rho T \partial_\rho T^{*},
\end{align}
and
\begin{align}
\partial_\mu \tilde{\Phi} \partial^{\mu} \tilde{\Phi} TT^{*} &= \frac{1}{\rho^2}TT^{*}, \\
\alpha'\partial^{\mu}\tilde{\Phi} \partial^{\nu} \tilde{\Phi} \partial_{\mu}T\partial_{\nu}T^{*} &= \frac{\alpha'}{\rho^2}\partial_\rho T \partial_\rho T^{*}, \\
\alpha'\nabla_{\mu}\nabla_{\nu}\tilde{\Phi}\partial_{\mu}T\partial_{\nu}T^{*} &=  \frac{1}{\alpha'}TT^{*} + \frac{\alpha'}{\rho^2} \partial_\rho T \partial_\rho T^{*}.
\end{align}
We remind the reader that for $AdS_3$, $\tilde{B}_{\mu\nu}=0$. The fact that for both backgrounds $\tilde{R}^{\mu\nu}+2\nabla_{\mu}\nabla_{\nu}\tilde{\Phi}=0$, is the lowest order $\alpha'$ Einstein equation for a vanishing Kalb-Ramond field. \\
For $AdS$ we have
\begin{equation}
\frac{1}{k-2} = \frac{1}{k} + \frac{2}{k^2}+ \hdots
\end{equation}
We know from the exact action that this expansion should premultiply $\partial_\rho T \partial_\rho T^{*}$ and $\partial_\phi T \partial_\phi T^{*}$. Thus at order $\alpha'$ we require the corrections\footnote{Only the relative prefactor (w.r.t. the $1/k$ leading term) was written down here.}
\begin{equation}
\frac{2}{k}\left(\partial_\rho T \partial_\rho T^{*} + \frac{1}{\sinh(\rho/l)^2} \partial_\phi T \partial_\phi T^{*}\right).
\end{equation}
This can only be obtained by the term
\begin{equation}
\left(\left[A\alpha'\tilde{R}^{\mu\nu} - 2 \alpha'(1-A) \nabla_{\mu}\nabla_{\nu}\tilde{\Phi} \right] + 2\alpha'\partial^{\mu}\tilde{\Phi}\partial^{\nu}\tilde{\Phi}\right)\partial_{\mu}T\partial_{\nu}T^{*},
\end{equation}
where $A$ is an arbitrary constant, that precisely corresponds to the freedom in the on-shell action. \\
Using also this term for Euclidean Rindler space yields:
\begin{equation}
\left(\left[A\alpha'\tilde{R}^{\mu\nu} - 2 \alpha'(1-A) \nabla_{\mu}\nabla_{\nu}\tilde{\Phi} \right]+2\alpha'\partial^{\mu}\tilde{\Phi}\partial^{\nu}\tilde{\Phi}\right)\partial_{\mu}T\partial_{\nu}T^{*} = -\frac{2}{\alpha'}TT^{*},
\end{equation}
precisely the (entire!) correction term, thus for Rindler space this is actually the only correction: all possible terms from higher $\alpha'$ corrections cancel out. The fact that this term is the entire Rindler correction term was not an a priori necessity. What \emph{is} necessary, is that the remaining terms that scale like $\propto \frac{\alpha'}{\rho^2}$ cancel since these do represent an $\alpha'$ perturbation series and hence they should cancel order by order. This is indeed the case. \\
Note that for on-shell backgrounds, the above $\alpha'$ correction is unambiguous. \\
Secondly, for type II superstrings, the above reasoning implies that the lowest $\alpha'$ derivative correction in that case is at most a term that vanishes by the equations of motion. For on-shell backgrounds, we have no correction. \\
Should this coincide with the $\alpha'$-correction of the tachyon effective action? Actually not. The reason is that we are envisioning the non-interacting action of the thermal scalar in this work. \\
Thus to conclude, we expect the above corrections to differ from the ones used in the effective action, since for the latter the vertices are effective and incorporate integrating out the massive string fluctuations. In our case, these are simply set to zero.

\subsection{Heuristic argument for the absence of $\alpha'$-corrections for the type II thermal scalar}
\label{argu}
We use the following two conjectures on CFTs:
\begin{itemize}
\item{Any rational CFT can be realized as a coset model.}
\item{Any CFT can be arbitrarily well approximated by a rational CFT. A textbook example is the compact scalar which for rational $R^2/\alpha'$ can be rewritten in terms of a larger chiral algebra (and forms a rational CFT).}
\end{itemize}
Thus any SCFT can be approximately written as a coset model. Since the non-winding states have a propagation equation that is simply the Laplacian on the coset manifold, and since both ungauged and gauged WZW models do not receive $\alpha'$ corrections for the type II superstring, all non-winding states simply propagate using the lowest order background fields (Klein-Gordon in curved background).\footnote{Such an argument is not true for bosonic strings: it is known that for instance in WZW $AdS_3$ spacetime (which is an ungauged WZW model), the bosonic string propagation equations have the shift $k\to k-2$ in the Laplacian.} Since this is now for \emph{all} SCFTs, the type II propagation equations in a general background are not $\alpha'$ corrected. Now taking the T-dual equations of motion, also these do not suffer from $\alpha'$ corrections and we conclude that type II superstrings have a thermal scalar equation of motion that coincides with the lowest order in $\alpha'$ effective thermal scalar field equation. \\
We remind the reader that we are only considering terms quadratic in the fields in the action: self-interactions of the thermal scalar will be present in general (even for type II superstrings) but at the non-interacting level we are focusing on in this work, they are not needed. In particular, the Hagedorn temperature is defined at the one-loop level and hence no self-interactions of the thermal scalar field should be considered. \\
This argument shows that the (non-self-interacting) thermal scalar action for type II superstrings does not get $\alpha'$-corrected. For the derivation presented in section \ref{HP} however, we will require more: we need to be able to vary with respect to the background metric. The result is an off-shell background. For this we do not have the CFT methods available. Nonetheless, it seems plausible that a suitable off-shell generalization of string theory (e.g. using string field theory) should not alter this result.

\subsection{Compact thermodynamics and the prefactor: a technical intermezzo}
Assuming the action for the type II thermal scalar is not $\alpha'$-corrected, we can tie up a loose end on the lowest thermal scalar eigenvalue (that we will utilize in chapter \ref{chrel}). \\
For a purely compact space, the most dominant part of the free energy is of the form
\begin{equation}
\label{nopref}
\beta F \approx \ln(\lambda_0),
\end{equation}
with $\lambda_0$ the lowest eigenvalue of the thermal scalar wave equation $\hat{\mathcal{O}}\psi_n = \lambda_n \psi_n$. For flat space this eigenvalue is proportional to the difference in inverse temperature close to $T_H$: $\lambda_0 \sim \beta-\beta_H$. Inserting this in the above form of the free energy, one immediately finds
\begin{equation}
\beta F \approx \ln(\beta-\beta_H)
\end{equation}
with no prefactor. As a consequence of the absence of $\alpha'$ corrections to the (non-interacting part of the) thermal scalar action in type II superstring theory, this property can be directly generalized as we now show. Upon normalization of the eigenmodes of $\mathcal{\hat{O}}$, one can write
\begin{equation}
\left\langle \psi_n\right|\hat{\mathcal{O}}\left|\psi_n\right\rangle = \lambda_n.
\end{equation}
The operator depends on $\beta$ only through the additive term $C\beta^2 G_{\tau\tau}$ for some (positive) constant $C$. Hence taking the derivative of the above equation with respect to $\beta$, one finds (using the Hellmann-Feynman theorem) that 
\begin{equation}
\left\langle \psi_n\right|2C\beta G_{\tau\tau}\left|\psi_n\right\rangle = \partial_\beta\lambda_n.
\end{equation}
Since $G_{\tau\tau}$ is strictly positive, the left hand side is strictly positive as well, implying $\partial_\beta\lambda_n \neq 0$. \\
For the lowest eigenvalue $\lambda_0$ (defined by the fact that it vanishes as $\beta=\beta_H$) with Taylor expansion around $\beta_H$:
\begin{equation}
\lambda_0 \approx A(\beta-\beta_H) + B(\beta-\beta_H)^2 + \hdots
\end{equation}
this implies that $A \neq 0$ (else the derivative w.r.t. $\beta$ vanishes at $\beta = \beta_H$) and this is all we need to establish $\lambda_0 \sim \beta-\beta_H$ showing in general that no prefactor can be included in the most dominant part of the free energy (\ref{nopref}).\\

\noindent We finally remark that $\partial_{\beta}\lambda_n >0$, meaning the eigenvalues decrease monotonically as the temperature increases. This was indeed the case for the Rindler solution (\ref{superspectrum}) and the WZW background (\ref{sspp1}).

\chapter{General Lessons on String Theory near Black Holes}
\label{chgenbh}
Let us next look at general lessons we can draw for string theory around black hole horizons. \\
This chapter is based on parts of \cite{Mertens:2014saa} and some unpublished results and ideas. \\ 

\noindent This chapter is structured as follows. \\
First, in section \ref{relevance}, we discuss why neglecting higher mass modes is an incorrect approximation for the fields around the black hole. After that, we discuss the different modular domains in section \ref{modDom} for black holes and point out the possibility of different one-loop partition functions in a cigar-like geometry. This will be used on a more speculative level in section \ref{firewall} where we attempt to make a link with the firewall paradox. Finally, section \ref{higher} contains a detailed discussion on the higher genus corrections to the one-loop result for cigar-shaped geometries. \\
Some more detailed material is contained in the supplementary sections.

\section{Relevance of long strings for black hole thermodynamics}
\label{relevance}
The next few sections try to draw some general conclusions on string thermodynamics in black hole spactimes. \\
\noindent First let us ask a general question: can one ignore the massive string modes when computing loop corrections to thermodynamical quantities in black hole spacetimes? Hence we wish to contemplate whether approximating the one loop free energy by only the free energy of the massless modes (in the Lorentzian spectrum) (i.e. gravitons, photons etc.) is a good approximation. \\
String loop corrections to black hole thermodynamics have a long history riddled with controversy, see e.g. \cite{Susskind:1994sm}\cite{Susskind:1993ws}\cite{Dabholkar:1994ai}\cite{Lowe:1994ah}\cite{Susskind:2005js}\cite{Parentani:1989gq}\cite{Emparan:1994bt}\cite{McGuigan:1994tg}. In general one expects higher worldsheet corrections to be neglible when considering the exterior of black holes. The argument is well-known: higher worldsheet corrections (or massive string modes) manifest themselves in the low energy effective action as corrections of higher order in $\alpha'/R^2$ with $R$ some curvature radius. The curvature outside a (large) black hole horizon is much smaller than the inverse string length. Hence these corrections are very small and can be neglected. \\
The situation is completely different however when considering one-loop thermodynamical quantities. This can be appreciated from different perspectives.\\
A first primitive argument is as follows. We noted previously in \cite{Mertens:2013zya} that for Rindler space plus a fully compact remainder, the free energy itself diverges as $\beta F \approx \ln(\beta-\beta_H)$ where $\beta_H = \beta_{\text{Hawking}}$ and one should set $\beta$ equal to the Hawking temperature as well in the end. This implies $F$ diverges on the nose. This is obviously not achieved by only considering the massless fields around the black hole.\footnote{A divergence sets in in this case as well, though it is temperature-independent.} \\
For a different argument, consider the thermal partition function. The radius of the thermal circle is an extra curvature parameter. We have explicitly demonstrated elsewhere \cite{Mertens:2013zya} that higher order $\alpha'$ corrections constructed with the inverse temperature $\beta$ are \emph{not} subdominant for thermal winding modes. \\

\noindent A more physical point of view can be given on the Lorentzian signature manifold.\footnote{Although we are somewhat reluctant to have too much faith in it due to the comments in the next footnote.} To that effect, let us first look at the formulas for the flat space string. The free energy of a (bosonic) field of mass $m$ in $d$ dimensions is given by
\begin{equation}
\beta F = V\int\frac{d^{d-1}k}{(2\pi)^{d-1}} \ln\left(1-e^{-\beta\sqrt{k^2+m^2}}\right).
\end{equation}
Clearly a higher mass field has a lower free energy. In the large mass limit, we can approximate
\begin{equation}
\ln(1-x) \approx -x,
\end{equation}
which makes the integrand proportional to $\propto e^{-\beta E}$. When considering string theory, we should multiply this by the degeneracy of states $\propto e^{\beta_H E}$ (for large mass). We conclude that for $T \ll T_H$, the lowest $m^2$ modes give the largest contribution to the free energy: the degeneracy of high $m^2$ states cannot compete with the lower mass. For $T \lesssim T_H$, the higher $m^2$ modes are not subdominant but give important contributions to the free energy: the full string theory is relevant. \\
For black holes, the above computation goes through almost identically.\footnote{Note though that this has been questioned in \cite{Susskind:1994sm} in the following way. It was suggested that the genus one result on the thermal manifold does \emph{not} correspond to the free-field trace, but instead includes some interactions with open strings whose endpoints are fixed on the horizon. Despite being an explicit proposal on the stringy microscopic degrees of freedom, little success has been booked in using and/or proving aspects of this proposal since then. We will nonetheless assume that the free-field Hamiltonian trace has the same critical Hagedorn temperature as predicted by genus one string thermodynamics. The reason is that we believe we have given evidence that the critical temperature of the thermal scalar is tightly linked to the random walk phenomenon close to black hole horizons and this long string picture is precisely what is expected near the horizon on general arguments \cite{Susskind:1993ws}\cite{Susskind:1993ki}\cite{Susskind:1994uu}\cite{Susskind:1993aa}, strongly suggesting the equality of $\beta_H$ and $\beta_{\text{Hawking}}$ also for the free-field trace. We will in fact further investigate this issue in the next section \ref{modDom}.} The free energy of a non-interacting Bose (and/or Fermi) gas of strings is given by
\begin{equation}
\beta F =  \pm \sum_{\text{species}}\ooalign{$\displaystyle\sum$\cr\hidewidth$\displaystyle\int$\hidewidth\cr}_{E_i} \ln\left(1 \mp e^{-\beta E_i}\right),
\end{equation}
where we sum over all Lorentzian string states in the spectrum. The high energy states again provide a factor of $e^{-\beta E}$, with $\beta$ the inverse Hawking temperature. Since this precisely coincides with the Hagedorn temperature (determining the degeneracy of high energy string states), the highly excited modes are very relevant and it is incorrect to approximate the one-loop free energy of strings by that given solely by the massless modes. \\
In the physical picture we have, the massless modes alone do not give the random walker surrounding the horizon; this is only obtained by considering the highly excited strings. This is to be contrasted with several holographic computations (e.g. \cite{Denef:2009yy} where the authors compute the one-loop free energy in a holographic black hole background using only one class of charged matter).\footnote{Of course we do not claim in any way that these authors are wrong, we simply point out some tension between the gravity-plus-matter approach and the full string picture at one loop. Moreover, these authors consider charged black holes while in our case uncharged black holes are studied.}

\section{Modular domains for contractible thermal circles}
\label{modDom}
The discussions made in the previous chapters allow us to discuss more deeply the role of our starting point: does one define string thermodynamics in the fundamental domain or in the strip? Are these descriptions always identical? We shall first answer this question in the affirmative in a general background, after which we will uncover a puzzle with the precise partition functions. The discrepancy will be explained by a more detailed discussion on the different torus embeddings that are actually path integrated over.

\subsection{Generalization of the McClain-Roth-O'Brien-Tan theorem}
\label{Mcr}
The general torus path integral on the fundamental domain for a general background\footnote{For simplicity we write down only a metric background here, but the result is more general.}
\begin{equation}
\label{funda} 
Z_{T_2} = \int_{\mathcal{F}} \frac{d\tau_2}{2\tau_2} d\tau_1 \Delta_{FP} \int \left[\mathcal{D}X\right]\sqrt{G}
\exp -\frac{1}{4\pi\alpha'} \int d^2\sigma \sqrt h h^{\alpha \beta} \partial_\alpha X^\mu \partial_\beta X^\nu G_{\mu\nu}(X),
\end{equation}
where $\Delta_{FP}$ is the Faddeev-Popov determinant $\left|\eta\right|^4$ and with the torus boundary conditions (for some periodic field $X$)
\begin{align}
X(\sigma^1+2\pi,\sigma^2) & =  X(\sigma^1,\sigma^2) + 2\pi w R, \\
X(\sigma^1+2\pi\tau_1,\sigma^2+2\pi\tau_2) & =  X(\sigma^1,\sigma^2) + 2\pi m R,
\end{align}
can be rewritten in the strip domain as
\begin{equation}
\label{stri} 
Z_{T_2} = \int_0^\infty \frac{d\tau_2}{2\tau_2} \int_{-1/2}^{1/2} d\tau_1 \Delta_{FP} \int \left[\mathcal{D}X\right]\sqrt{G}
\exp -\frac{1}{4\pi\alpha'} \int d^2\sigma \sqrt h h^{\alpha \beta} \partial_\alpha X^\mu \partial_\beta X^\nu G_{\mu\nu}(X),
\end{equation}
with torus boundary conditions
\begin{align}
X(\sigma^1+2\pi,\sigma^2) & =  X(\sigma^1,\sigma^2), \\
X(\sigma^1+2\pi\tau_1,\sigma^2+2\pi\tau_2) & =  X(\sigma^1,\sigma^2) + 2\pi r R. 
\end{align}
In flat space, this equality was established by \cite{McClain:1986id}\cite{O'Brien:1987pn} some time ago. In their proof, the authors make explicit use of the flat space worldsheet action. It turns out (almost trivially) that one can make the argument independent of the flat space action and hence generalize it to an arbitrary conformal worldsheet model. \\
Starting in the fundamental domain, the proof uses that the effect of a modular transformation can be undone by a redefinition of the wrapping numbers:
\begin{align}
T&: m \to m+ w, \\
S&: m \to -w, \quad w \to m.
\end{align}
One does not need the precise action for this. What is required is that the worldsheet theory is conformally invariant. The proof then follows exactly the same strategy as for flat space: one can map each $(m, w)$ sector into $(r, 0)$ by a modular transformation, precisely building up the strip modular domain in the process. These steps are made with much more care in section \ref{MROB}. \\

\noindent Note that no use is made of the non-contractibility of the $X$-cycle: in fact one can apply this to a contractible circle as well (like the angular coordinate in polar coordinates). This shows in a very general way that the fundamental domain and the strip domain give equal results in any spacetime.

\subsection{What do these path integrals represent for Euclidean Rindler space?}

\begin{figure}[h]
\begin{minipage}{0.32\textwidth}
\centering
\includegraphics[width=0.9\textwidth]{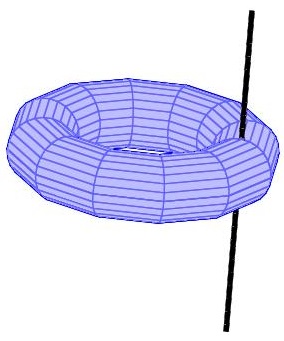}
\caption*{(a)}
\end{minipage}
\begin{minipage}{0.32\textwidth}
\centering
\includegraphics[width=0.9\textwidth]{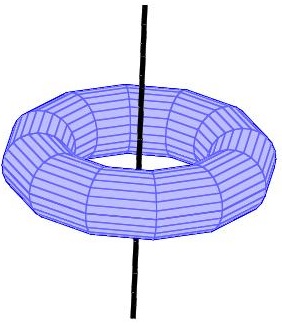}
\caption*{(b)}
\end{minipage}
\begin{minipage}{0.32\textwidth}
\centering
\includegraphics[width=0.99\textwidth]{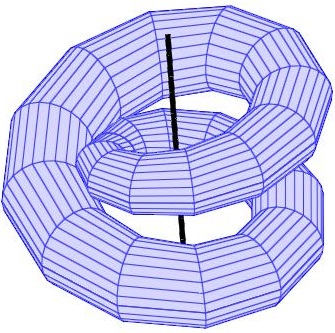}
\caption*{(c)}
\end{minipage}
\caption{(a) Singly wound torus that intersects the axis (perpendicular to the Rindler plane) through the Euclidean Rindler origin. Such configurations are not path integrated over in (\ref{stri}). (b) Singly wound torus that does not intersect the axis through the Euclidean Rindler origin. Such configurations are path integrated over in (\ref{stri}). (c) Twice wound torus that again does not intersect the axis through the origin. Such a configuration corresponds to a free closed string thermal trace.}
\label{torusPolar1}
\end{figure}

The above configurations (\ref{funda}) and (\ref{stri}) do not represent the most general embedding of the worldsheet torus into the target space, when this space has a contractible $X$-circle. It is most transparent to discuss this in the strip domain. The string path integral with winding only along one of the torus cycles (\ref{stri}) represents a restricted set of tori embeddings in the target space when the thermal circle is contractible. Upon shifting the dependence on the moduli from the boundary conditions to the worldsheet metric, the torus wrapping (along only the temporal worldsheet direction) is imposed as
\begin{equation}
X^0(\sigma^1, \sigma^2+1) = X^0(\sigma^1,\sigma^2) + r \beta,
\end{equation}
where $\sigma^1$ is the spatial coordinate on the worldsheet and $\sigma^2$ is the timelike coordinate. The interpretation of this boundary condition is that \emph{all} points along a fixed $\sigma^2$-slice rotate along the (Euclidean) time dimension to form a 2-torus. As an example, let us take a closer look at Rindler space. For Euclidean Rindler space, this means that configurations such as \ref{torusPolar1}(a) are not integrated over: points in the ``inner'' path do not rotate around the Rindler origin. Configurations displayed in figure \ref{torusPolar1}(b) on the other hand are the ones that we take into account.

\noindent It is interesting to point out that the configurations shown in figure \ref{torusPolar1}(a) are associated in \cite{Susskind:1994sm} to open-closed interactions in the Lorentzian picture and these are entirely missed in the above path integral approach to string thermodynamics, suggesting that the non-interacting closed string trace only is contained in the above path integral. The higher winding numbers are associated to tori that wind around the Rindler origin multiple times, such as that displayed in the right most figure of figure \ref{torusPolar1}.

\noindent Above we argued that the strip and the single wrapping number (\ref{stri}) miss out on a specific set of torus embeddings. This is also the case for the fundamental domain (\ref{funda}) with 2 wrapping numbers. Crucial in this argument is that the above torus boundary conditions do not allow mixed wrapping numbers (a property that could be allowed in a contractible target space), such as the torus embedding in the left figure of figure \ref{torusPolar1}. This means the Susskind-Uglum interactions are completely missed in the above path integrals, suggesting that they give only non-interacting (free-field) closed strings.\footnote{A point of critique is to be mentioned here: this path integral represents one string loop on the thermal manifold. Even in field theory, it is unclear whether the free field trace and the one-loop result on the thermal manifold agree for gauge fields (and presumably for higher spin fields as well) \cite{Kabat:1995eq}\cite{Kabat:1995jq}\cite{Kabat:2012ns}\cite{Donnelly:2012st}\cite{Donnelly:2014fua}. For a very interesting recent development, see \cite{He:2014gva}. 
Presumably, this string path integral should hence be compared with the Hamiltonian trace over Lorentzian fields.} \\

\subsection{Explicit CFT result}
The other approach to Euclidean Rindler space (starting from the cigar CFT and then taking the small curvature limit), leads to apparently different results.
The bosonic partition function \cite{Hanany:2002ev} is given by
\begin{align}
\label{bossonic}
Z &= 2\sqrt{k(k-2)}\int_{\mathcal{F}}\frac{d\tau d\bar{\tau}}{\tau_2} \int_{0}^{1}ds_1ds_2 \nonumber \\
&\sum_{m,w=-\infty}^{+\infty}\sum_i q^{h_i}\bar{q}^{\bar{h}_i}e^{4\pi\tau_2(1-\frac{1}{4(k-2)}) -\frac{k\pi}{\tau_2}\left|(s_1 - w)\tau -(s_2 - m)\right|^2+2\pi\tau_2s_1^2} \nonumber \\
&\frac{1}{\left|\sin(\pi(s_1\tau - s_2))\right|^2}\left|\prod_{r=1}^{+\infty}\frac{(1-e^{2\pi i r \tau})^2}{(1-e^{2\pi i r \tau - 2\pi i (s_1\tau -s_2)})(1-e^{2\pi i r \tau + 2\pi i (s_1\tau -s_2)})}\right|^2.
\end{align}
As explicitly demonstrated in chapter \ref{chri} in equation (\ref{flatbosonic}), the large $k$ limit yields the partition function of an infinite 2d plane. \\
In \cite{Sugawara:2012ag}, Sugawara wrote down the partition function obtained upon setting $w=0$ and simultaneously extending the modular integration to the entire strip. This application of the McClain-Roth-O'Brien-Tan theorem is (in our opinion) a bit naive as we show here. The candidate partition function equals
\begin{align}
Z_{\text{candidate}} &= Z_{m=w=0} + 2\sqrt{k(k-2)}\int_{E}\frac{d\tau d\bar{\tau}}{\tau_2} \int_{0}^{1}ds_1ds_2 \nonumber \\
&\sum_{m=-\infty}^{'+\infty}\sum_i q^{h_i}\bar{q}^{\bar{h}_i}e^{4\pi\tau_2(1-\frac{1}{4(k-2)}) -\frac{k\pi}{\tau_2}\left|s_1\tau -(s_2 - m)\right|^2+2\pi\tau_2s_1^2} \nonumber \\
&\frac{1}{\left|\sin(\pi(s_1\tau - s_2))\right|^2}\left|\prod_{r=1}^{+\infty}\frac{(1-e^{2\pi i r \tau})^2}{(1-e^{2\pi i r \tau - 2\pi i (s_1\tau -s_2)})(1-e^{2\pi i r \tau + 2\pi i (s_1\tau -s_2)})}\right|^2,
\end{align}
where the prime on the summation denotes that we do not include the $m=0$ term. Taking the large $k$ limit, one finds (upon dropping the $(m,w) = (0,0)$ sector) almost the same expression as (\ref{flatbosonic}):
\begin{equation}
Z_{\text{candidate}} \sim \frac{V_T \mathcal{A}}{N}\int_{E}\frac{d\tau d\bar{\tau}}{4\tau_2^2}\frac{1}{{\tau_2}^{12}}\left|\eta(\tau)\right|^{-48},
\end{equation}
the only difference being the change in modular integration domain. What one can say about this, is that the simple replacement of the strip $E$ with $\mathcal{F}$ and the inclusion of a second sum (over $w$) does \emph{not} give equal quantities. If it did, then the large $k$ limits should be the same as well.\footnote{It is instructive to follow the arguments in either of these papers \cite{McClain:1986id}\cite{O'Brien:1987pn} as far as possible. It turns out that the holonomy integrals over $s_1$ and $s_2$ in the end are causing the mismatch: these parameters transform as a doublet under modular transformations (just as $m$ and $w$), the problem then finally is the region of integration of these variables, which does not allow a clean extraction of the sum over different modular images to build up the strip domain.} \\
For type II superstrings, one would obtain instead (again dropping the $(m,w) = (0,0)$ sector)
\begin{equation}
\label{typeIIstri}
Z_{\text{candidate}} \sim \frac{V_T\mathcal{A}}{N}\int_{E}\frac{d\tau_1 d\tau_2}{2\tau_2^2}\frac{1}{\tau_2}\left(\frac{1}{\left|\eta\right|^2\sqrt{\tau_2}}\right)^{6} \frac{\left|\vartheta_3^4 - \vartheta_4^4 - \vartheta_2^4\right|^2}{\left|\eta\right|^{12}} = 0.
\end{equation}\\

\noindent We want to emphasize two points on these observations. Firstly, the fundamental domain and strip results of this partition function are \emph{not} equal, unlike the arguments given in the previous subsection \ref{Mcr}. Secondly, for type II superstrings in the strip, the partition functions (\ref{superF}) and (\ref{typeIIstri}) vanish, a property which is (nearly) impossible for a free-field thermal trace. \\

\noindent How then is this consistent with our discussion in the two previous subsections? To that effect, let us take a closer look at the path integral boundary conditions used to obtain equation (\ref{bossonic}). In \cite{Hanany:2002ev}, coordinate transformations were made obscuring the wrapping numbers around the cigar. Indeed, the temporal (i.e. angular) coordinate $\phi$ ($\phi \sim \phi + 2\pi$) was transformed into a single-valued coordinate $v$ as\footnote{In this formula, $r$ is the radial coordinate and $\rho$ is related to the gauge field of the gauged WZW model. These coordinates are not relevant for our discussion here.}
\begin{equation}
v = \sinh(r/2) e^{i\phi}e^{i\rho}
\end{equation}
and the remainder of the derivation focused on this coordinate. However, no fixed wrapping numbers along $\phi$ were specified in advance, and in the end both tori with fixed wrapping numbers along both cycles (i.e. those that do \emph{not} intersect the origin) and those that are partially wrapped are all in principle considered: they all get mapped into the same $v$ coordinate. Only in the end one again (re)identifies the winding numbers. \\
Moreover, we have shown previously (section \ref{flatlimit}) that the flat limit indeed reproduces the \emph{entire} 2d plane partition function, meaning these partially wrapped torus configurations are indeed taken into account. These facts strongly suggest that indeed the partially wrapped tori are considered as well for these partition functions.
 
\subsection{Discussion}
We conclude that the partition function result in \cite{Hanany:2002ev} actually is the genus-1 result on the thermal manifold and includes torus embeddings with non-definite winding number. The free-field trace on the other hand corresponds to the path integrals (\ref{stri}) or (\ref{funda}) in which the temporal coordinate has a fixed winding number and these expressions are not the same. \\

\noindent To proceed then, imagine we focus on manifolds where all winding modes are present on the thermal manifold (topologically stable thermal circles). In previous work \cite{Mertens:2013pza}, as discussed in chapter \ref{chth}, we obtained the most dominant (random walk) contribution directly from the string path integral (\ref{stri}) with torus boundary conditions (\ref{stri}), thereby reducing the string theory on the modular strip to a particle theory of the thermal scalar. Alternatively, from the field theory (and CFT) point of view, the most dominant mode (the thermal scalar) on the thermal manifold (on the modular fundamental domain) can be used to arrive at the same dominant behavior. Both of these approaches should match for spaces with topologically stable thermal circles and we used this in chapter \ref{chth} to identify possible corrections to the random walk picture and to obtain the Hagedorn temperature on such a manifold. \\
 
\noindent This determines the random walk corrections fully for \emph{any} manifold: it seems impossible to imagine what sort of local corrections could be added to the (non-interacting) thermal scalar action that vanish on all manifolds with topologically stable thermal circles but are in general non-zero on the others. \\

\noindent From this perspective, when the manifold is topologically unstable in the thermal direction (such as for black holes or Rindler space), one could follow the worldsheet (i.e. string path integral) derivation of the thermal scalar starting from equation (\ref{stri}) and observe that the thermal scalar still determines the critical behavior, \emph{irrespective of whether it is present in the thermal spectrum (on the fundamental domain) or not}.\footnote{However, see \cite{Mertens:2015adr} for more recent remarks that appear to lead to the opposite conclusion.} \\
\noindent In particular, all winding numbers are present in the path integral result of section \ref{Mcr} and we can use the thermal spectrum simply as a tool to extract the form of the thermal scalar action. After that, the path integral of section \ref{Mcr} with its angular wrapping number stands on its own and represents the contribution from non-interacting closed strings. \\
\noindent Such a scenario would also solve our understanding of the BTZ WZW models where thermal winding modes are simply absent \cite{Mertens:2014nca}. The free string path integral (\ref{stri}) on the other hand has no problem with non-zero thermal windings. This discrepancy hence seems to be again related to the special torus embeddings that in the end cause the absence of thermal winding modes. \\
\noindent We finally remark that this would imply that the bosonic non-interacting free energy for Rindler space diverges (due to \emph{all} windings since all of them are tachyonic (these are simply not present in the genus-1 result)). All winding (non-oscillator) modes are tachyonic and localized at a string length from the horizon (higher winding modes are localized even more closely to the event horizon). This seems to be in agreement with the maximal acceleration phenomenon of \cite{Parentani:1989gq}. \\

\noindent With this understanding of the two different ways of studying string thermodynamics, one can ask which one is the most natural. The path integral approach leads to non-interacting strings in a fixed background, whereas the thermal manifold (CFT) approach leads to the full genus 1 result, which includes interactions with open strings stuck on the horizon. These interactions are quite exotic, since for instance dialing down the string coupling $g_s$ does \emph{not} decouple in any way these interactions: they are inherent to the torus path integral on the thermal manifold. In summing over genera to obtain the full (perturbative) thermodynamics, they should be included as well. In the next sections this will be our state of mind. \\
Then what does the non-interacting trace do for us? It represents the sum over non-interacting closed strings and should correspond to the free-field trace, an object that can be constructed in principle as soon as the Lorentzian (non-interacting) string spectrum is known. The results of section \ref{Mcr} show that this quantity is modular invariant as well and hence has a well-defined meaning in string theory (as coming from a torus worldsheet). The difference between these two approaches can be illustrated by realizing that in fact the path integral approach of section \ref{Mcr} actually considers the perforated space where the fixpoint is removed from the space: one can arrive at this setting by for instance taking the pinching limit of a topologically supported thermal circle. String worldsheets are hence not allowed to cross this fixpoint. Reincluding this point in the geometry leads to the full genus one result, to be interpreted as both free strings and exotic open-closed interactions. \\

\noindent Finally let us remark that the fact that for the cigar CFT (and its flat limit) the singly wound string state is present in the thermal spectrum, implies that the dominant regime of both of these different partition functions (\ref{stri}) and (\ref{bossonic}) is actually the same: both are dominated by the thermal scalar. This implies the open-closed interactions are a subdominant effect near the Hagedorn temperature for Rindler space.

\section{The Firewall}
\label{firewall}
We cannot resist the temptation to add some discussion on the firewall paradox from the above perspective.\footnote{Some important references on the firewall are e.g. \cite{Almheiri:2012rt}\cite{Verlinde:2012cy}\cite{Harlow:2013tf}\cite{Almheiri:2013hfa}\cite{Bousso:2013wia}.} This section is largely speculative, but is strongly based on the ideas presented in the previous section. \\
In the firewall debate (and in the information paradox), one assumes the Hilbert space has a clean splitting between degrees of freedom internal and external of the black hole. For strings, due to their spatial extent, this property is not obvious. For instance, defining entanglement entropy in a half-space for strings is exceedingly difficult. It is related to defining it for gauge theories where one has Wilson loops extending through the boundary. The total Hilbert space does not factorize in left and right degrees of freedom, but includes mixed boundary states that pierce the boundary surface. \\

\noindent In string theory, we will see, in a relatively concrete fashion, that if one excludes these boundary-piercing configurations (and hence treats it as a particle theory), one finds a violation of the equivalence principle very close to the horizon. Such a treatment effectively assumes the Hilbert space to be a direct product. Upon including the horizon-piercing strings, one finds a restoration of the equivalence principle. \\

\noindent Up to this point, we have only been interested in the description of physics according to the fiducial observer. This priviledged the polar coordinate description and our interest in thermodynamics assigned a special role to the origin (as the location of the conical singularity for non-equilibrium). It was found that the singly wound state $\left|w\right|=1$ plays the main role in this story and describes the long string dominant configuration close to the black hole horizon. The firewall paradox is a problem concerning the infalling observer. We would hence like to consider coordinate covariant objects such as the stress tensor vev in different vacuum states. Note that at this point, the singly wound state is not special anymore compared to the other massless modes since the temperature interpretation only holds for fiducial observers. If the stress tensor vev has a non-trivial contribution at the black hole horizon, then \emph{any} observer (including the infalling one) would conclude that there is structure at the horizon and this is then the cause of drama at the horizon. In the usual firewall argument, one concludes that if the inner and outer degrees of freedom are not precisely entangled such as to give the Minkowski near-horizon vacuum, then a wall of infinite stress-energy is present at the horizon. Within a quantum theory of gravity, the infinities of the stress tensor (associated with the infinite redshift at the horizon) should be cured, but still one can use the criterion that a finite contribution of stress-energy near the horizon (i.e. present in the Rindler approximation) violates the equivalence principle. The authors of \cite{Giveon:2013ica}\cite{Giveon:2012kp} argue that the GSO projection is the ultimate reason for this structure, but afterwards it is found in \cite{Giveon:2014hfa} that the partition function does not contain this (observer-independent!) structure. In this section we will argue that the story might be more intricate.\\

\noindent There are basically two different questions that should be examined when contemplating a violation of the equivalence principle. The first is whether the near-horizon regime of a black hole really is Rindler spacetime in the quantum theory. We will assume this to be true and we will focus on Rindler spacetime from here on. A second question, which is specific to string theory, is whether the free-falling observer (Minkowski space) and the fiducial observer (Rindler space) really are related only by a coordinate transformation. Any discrepancy here is also enough cause for a firewall. In the spirit of \cite{Giveon:2013ica}\cite{Giveon:2012kp}, it is the latter question that we will examine here. Let us already remark that we are not studying old black holes but instead properties that are present for eternal black holes as well and in particular in Rindler space.\\

\noindent Before we start, it is instructive to emphasize that (Euclidean) Rindler space is a solution to the $\alpha'$-exact string equations of motion and hence provides a consistent background of string propagation. Different modular invariants then correspond to different string spectra and vacuum states or ensembles one can construct within this background. For instance in flat space on the thermal manifold, one immediately finds two modular invariants: the invariant sector $w=m=0$, which corresponds to the usual vacuum $\beta\to\infty$ and the full thermal modular invariant, corresponding to the thermal trace on the Lorentzian manifold. \\

\noindent To study the paradox, we focus only on the non-interacting theory (the original paradox is in fact formulated in the non-self-interacting theory). We consider type II superstrings on a near-horizon black hole geometry. In doing this, we approximate the black hole as a fixed tree-level geometry and consider the quantum vacuum in string theory as the matter sector on top of the black hole. This is in line with the non-linear sigma model description: perturbative excitations propagate on top of a geometry and the absence of any excitations defines the (perturbative) vacuum. For instance, the full thermal vacuum corresponds to the Lorentzian Hartle-Hawking (or Minkowski) vacuum. Looking into only the $w=m=0$ sector, one instead finds the analog of the Boulware vacuum. Near the horizon, these states become the Minkowski and Rindler vacuum respectively. \\
In the previous section, we found that (at least) two different vacua are possible, defined through boundary conditions in the path integral on the thermal manifold. The first is the analogue of the Minkowski vacuum, with partially wrapped torus configurations. The second is the vacuum corresponding to a non-interacting closed string theory. \\
First, we look at the problem as a sum over field theories of all the different string states. For each individual field on the Lorentzian manifold, the Minkowski state is the unique state that has a non-divergent stress tensor at both the future and past horizon. This implies the equivalence principle for a free-falling observer who will not detect much (except a small curvature correction) at the future horizon upon falling in. Taking the direct product of this field theory vacuum gives the candidate string vacuum:
\begin{equation}
\label{cand}
\left|M\right\rangle \stackrel{?}{=} \bigotimes_{p} \left|M\right\rangle_{p}.
\end{equation}
Each Minkowski-state can be rewritten in terms of Rindler time $t$ as a thermal ensemble of Rindler excited states. The Rindler vacuum corresponds to the vacuum defined by Rindler time $t$ and corresponds to no particles according to the fiducial observer. Thus this candidate superstring vacuum (\ref{cand}) corresponds to a thermal ensemble of non-interacting closed strings, on the thermal manifold defined through the closed string boundary conditions with definite wrapping numbers. But we have seen before that this manifestly non-interacting construction is \emph{not} equal to the full 2d flat space construction upon coordinate redefinition (we actually saw this for the Euclidean polar coordinates and Cartesian coordinates). Even though, for each individual string state, the Minkowski vacuum is precisely such that the equivalence principle remains intact, this is not the case when summing these, as the resulting partition function is not simply a coordinate transformation of flat space. \\

\noindent Next we look into the second vacuum, defined as the path integral with the partially wrapped tori. It suffices here to only consider the Euclidean manifold so we focus solely on this here. As argued before, the stress tensor in the Euclidean (stringy!) Minkowski vacuum (chosen to vanish) can be rewritten in terms of the fiducial observer time $t$ to give a thermal trace that includes exotic open strings fixed at the horizon. When doing this, it is found that higher winding modes ($\left|w\right|>1$) are not present in the resulting partition function. Because of this equality (simply a coordinate transformation), this partition function (and the stress tensor) does not have any structure at the horizon and the equivalence principle holds: no firewalls are present. This is what was found in \cite{Giveon:2014hfa}: for the partition function one recovers the flat partition function. \\

\noindent The non-interacting closed string partition function (and the stress tensor) on the other hand do not exhibit the equivalence principle. To what extent? The difference between both of these partition functions is present only in the discrete modes.\footnote{This makes sense since the partially wrapped tori correspond to winding configurations which sense a $G_{00}$ potential and are hence bound to the origin. From another perspective, this is a purely stringy phenomenon, and hence it should come from winding configurations and not only discrete momentum modes.} In particular the higher wound sectors $\left|w\right|>1$ are the main difference. All of these states are localized at a string length from the black hole horizon. Hence, upon subtracting the flat stress tensor vev, one finds the stress tensor expectation value in this non-interacting closed string vacuum to be supported only very close to the horizon; and one can identify this with the firewall. \\
For bosonic strings on the other hand, the higher winding sectors are tachyonic. This implies the non-interacting vacuum actually is unstable towards tachyon condensation (black hole growth). \\

\noindent In this sense, the firewall originates from the naive extrapolation of a string theory as a sum over field theories. Upon including the additional open strings stuck at the horizon, the firewall disappears and the equivalence principle again holds. Like Silverstein's resolution to the firewall debate \cite{Silverstein:2014yza}, it appears here as well that non-adiabaticity in string theory might come to the rescue. \\

\noindent In terms of the thermal manifold, the singly wound string state is Susskind's long string according to the fiducial observer and the multiply wound string states (if present) are the firewall. The total picture is summarized in the table below.
\begin{figure}[h]
\centering
\includegraphics[width =\linewidth]{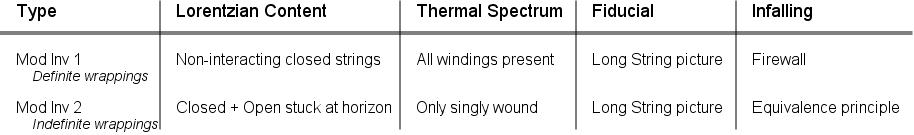}
\end{figure}
Both invariants realize Susskind's long string picture. The first modular invariant realizes the firewall whereas the second modular invariant contains the equivalence principle.\footnote{A more detailed analysis of the first modular invariant appeared in \cite{Mertens:2015adr}.} \\

%

\noindent What do we conclude about the firewall discussion? The firewall argument concludes that a single quantum violates the equivalence principle near black holes and it is hard to mitigate the argument since the regimes of validity seem satisfied. We have argued that string theory agrees with this conclusion and non-interacting closed string theory also has a firewall. However, upon including the necessary open strings at the horizon, one finds that the stress-energy at the horizon is cancelled and the equivalence principle is restored. \\
To conclude, string theory agrees with the firewall argument but at the same time shows that additional degrees of freedom (purely stringy and absent in effective field theory) are neglected and including these destroys the firewall and reinstates the equivalence principle. In some sense, the structure at the horizon is cancelled by additional structure coming from the open strings. \\

\noindent Comparing to AMPS, it seems that EFT fails in this case. Our argument is in the spirit of \cite{Larjo:2012jt} which critizes the original argument in that the stretched horizon and its high energy contents are entirely neglected. The open string degrees of freedom very close to the horizon are important and naively missed. If excluded, the equivalence principle fails; if included, EFT fails. \\

\noindent Some reservations are in order. Firstly, this entire discussion uses only the exterior region of the black hole, and we have looked into structure there. This by no means implies that there is nothing out of the ordinary inside the black hole. Secondly, the discussion is performed at genus 1 only (non-interacting). This is the same setting as the typical firewall discussion itself, but it does leave open modifications at higher genera, especially since the genus expansion in black hole backgrounds seems to break down (as we will discuss in the next section). A further point of critique to be made is that it is not clear whether the firewall discussed in this section is the same as the one in the original arguments, especially since we are discussing Rindler space and it would apply to an accelerating observer as well. Nonetheless, for non-interacting closed strings, we did see a version of a firewall-like puzzle and its resolution through non-adiabatic effects in string theory.

\section{Higher genus partition functions and the thermal scalar}
\label{higher}
In this section we consider the influence of the higher genera on the critical behavior. In \cite{Brigante:2007jv} arguments were given using dual field theories and worldsheet factorization that the Hagedorn temperature in general theories is non-limiting and that string perturbation theory (in $g_s$) actually breaks down near this temperature. It is of course interesting to try to apply some of these arguments to the case of black hole horizons. Similarly to \cite{Brigante:2007jv}, we will only focus on purely compact spaces for which the spectrum of the operator $\hat{\mathcal{O}}$ (as defined in (\ref{eigen})) is entirely discrete. In the first subsection we will analyze the influence of a fixed higher genus contribution and in the second subsection we will consider the effect of summing over the higher genus corrections. We follow \cite{Brigante:2007jv} quite closely.
\subsection{Higher genus partition functions}
In this subsection we focus on the thermal partition function at fixed genus. Consider the $SL(2,\mathbb{R})/U(1)$ CFT (and its flat limit $k\to\infty$). We follow the decomposition of higher genus Riemann surfaces into propagators and 3-punctured spheres \cite{Polchinski:1988jq}. In general, higher genus Riemann surfaces (can) have divergences when various cycles pinch. The moduli of the propagators are sufficient to give the contribution from the boundary of moduli space, which is what we need. In case of topologically stable thermal circles, it was shown \cite{Brigante:2007jv} that winding conservation prohibits all propagators from containing the thermal scalar. For cigar-shaped thermal circles however, winding conservation is violated in scattering amplitudes \cite{Giveon:1999px}\cite{Giveon:1999tq}. \\
In \cite{Giveon:1999px}\cite{Giveon:1999tq} it was shown that $n$-point amplitudes can violate winding conservation by up to $n-2$ units, thus an amplitude such as $\left\langle V_{w=+1} V_{w=+1}V_{w=+1}\right\rangle$ or $\left\langle V_{w=-1} V_{w=-1}V_{w=-1}\right\rangle$ vanishes whereas the amplitudes $\left\langle V_{w=+1} V_{w=+1}V_{w=-1}\right\rangle$ and $\left\langle V_{w=+1} V_{w=-1}V_{w=-1}\right\rangle$ do not vanish and provide the relevant 3-point amplitudes; a property which persists in the large $k$ limit. Hence it \emph{is} possible in this case that all propagators contain the thermal scalar, simplifying the analysis.\footnote{One readily shows that this is possible without having to resort to giving any 3-point vertex a winding number violation by more than one unit.}
An explicit computation of this 3-string scattering amplitude will not be considered here, but we make some further remarks on these in section \ref{nnversusn}. In what follows we will describe this decomposition in much more detail.\\

\begin{figure}[h]
\centering
\includegraphics[width=0.8\textwidth]{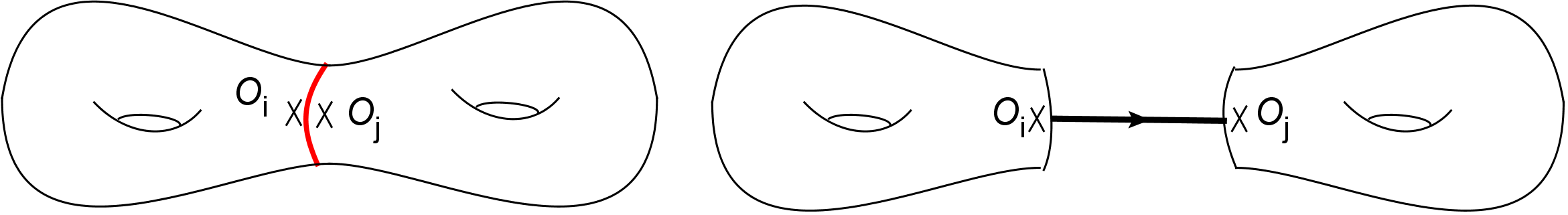}
\caption{Cutting open of a genus two worldsheet. For the cutting displayed here, this reduces it to two tori with single punctures and an intermediate propagator.}
\label{toruscutting}
\end{figure}

\noindent Following Polchinski \cite{Polchinski:1988jq}, we cut open the worldsheet (path integral) along a cycle and insert a complete set of states of local operators. 
Such a set of local operators have the normalization on the 2-sphere:
\begin{equation}
\left\langle O_i(\infty) O_j(0)\right\rangle_{S_2} \propto \delta_{h_i h_j}.
\end{equation}
We do not keep track of the overall normalization present in the Zamolodchikov metric written above.\\ 
We then have ($g=g_1+g_2$)\footnote{The index $j$ is restricted by $h_j = h_i$.}
\begin{equation}
\left\langle 1 \right\rangle_{g} = \int_{\left|q\right|<1} \frac{d^2q}{q\bar{q}}\sum_{i,j}q^{h_i}\bar{q}^{\bar{h}_i}\left\langle O_i(z_1) \right\rangle_{g_1}\left\langle O_j(z_2)\right\rangle_{g_2},
\end{equation}
where a new modulus $q$ is introduced (the sewing parameter), which should be integrated over a unit disk.\footnote{The alternative way of cutting the CFT reduces the genus by 1 as:
\begin{equation}
\left\langle 1 \right\rangle_{g} = \int_{\left|q\right|<1} \frac{d^2q}{q\bar{q}}\sum_{i,j}q^{h_i}\bar{q}^{\bar{h}_i}\left\langle O_i(z_1) O_j(z_2)\right\rangle_{g-1}.
\end{equation}
This way of cutting the worldsheet should be used as well in reducing the general genus $g$ amplitude and one readily adapts the formulas to incorporate also this procedure.} The intermediate set of states (labeled by $i$ (and $j$)) come in a propagator contribution, as can be seen by explicitly integrating over the modulus $q$. The cutting of the surface is demonstrated in figure \ref{toruscutting}.
The limit $q\to0$ corresponds to the pinching limit. In this limit, the lowest conformal weight dominates in the same way as that the lowest states dominate the one-loop partition function in the $\tau_2 \to \infty$ limit. We are interested in the most dominant contribution that is temperature-dependent, implying that the thermal scalar only is present in the intermediate vertex operators. The non-thermal massless operators are not temperature-dependent even though their conformal weight becomes equal to that of the thermal scalar at $\beta=\beta_H$. In the present section, we suppose that the wave operator $\hat{\mathcal{O}}$ of the thermal scalar (defined in equation (\ref{operatorO})) has a discrete spectrum $\psi_n$. Integrating over $q$ and keeping only the dominant contribution, this yields:
\begin{equation}
\left\langle 1 \right\rangle_{g} \approx \frac{1}{h_0+\bar{h}_0}\left\langle T_0(z_1) \right\rangle_{g_1}\left\langle T_0(z_2)\right\rangle_{g_2}
\end{equation}
where $0$ labels the state with lowest conformal weight of the thermal scalar $T$; this can be done cleanly due to the discreteness of the spectrum of $\hat{\mathcal{O}}$. We can rewrite it as
\begin{equation}
\left\langle 1 \right\rangle_{g} \approx \left\langle T_0 \right|\frac{1}{L_0+\bar{L}_0} \left| T_0 \right\rangle \left\langle T_0(z_1) \right\rangle_{g_1}\left\langle T_0(z_2)\right\rangle_{g_2}
\end{equation}
for the thermal scalar state $T_0$. Cutting along several cycles finally reduces the higher genus amplitude to a set of propagators and spheres with 3 punctures. One of the possible outcomes for the genus 2 worldsheet is shown in figure \ref{cuttingfull} below. 
\begin{figure}[h]
\centering
\includegraphics[width=0.4\textwidth]{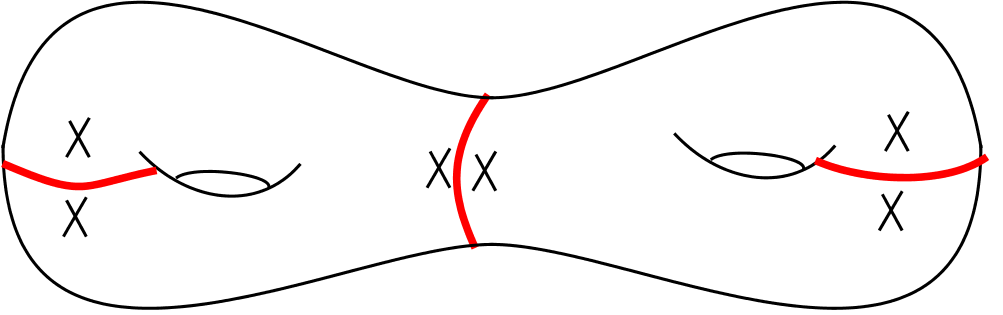}
\caption{Example of a full cutting of the genus two worldsheet into three propagators and two 3-punctured spheres.}
\label{cuttingfull}
\end{figure}
The 3-punctured spheres we need are the amplitudes of tree-level 3-thermal scalar scattering, which can in principle be computed.\\

\begin{figure}[h]
\centering
\includegraphics[width=5cm]{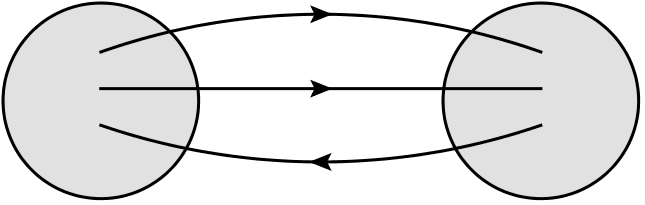}
\caption{Degenerate limit of the genus two surface in which the thermal scalar propagates along each internal line.}
\label{genustwo}
\end{figure}

\noindent Consider as an example the degenerate limit of a genus-2 amplitude shown in figure \ref{genustwo}.
Each propagator is dominated by the thermal scalar and behaves as $\sim 1/\lambda_0$. In string language we obtain for this vacuum amplitude $\mathcal{A}$:
\begin{equation}
\label{vacugen2}
\mathcal{A} = \frac{g_s^2}{\lambda_0^3}\left\langle T_0 T_0 T_0^* \right\rangle_{S^2}\left\langle T_0 T_0^* T_0^* \right\rangle_{S^2}.
\end{equation}
Next we would like to obtain a spacetime interpretation of these 3-vertex interactions. \\

\begin{figure}[h]
\centering
\includegraphics[width=3cm]{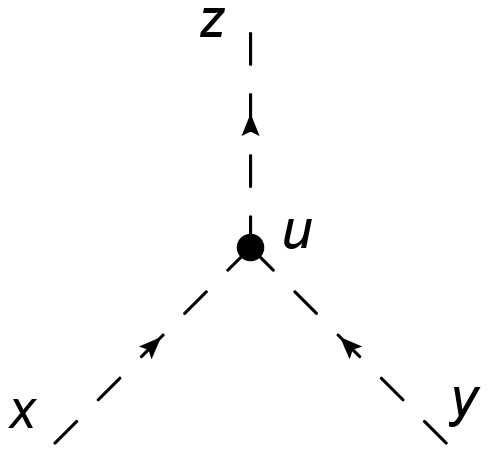}
\caption{Three-point amplitude in field theory.}
\label{threevertex}
\end{figure}

\noindent To that effect, let us consider the following off-shell 3-point function (figure \ref{threevertex}) in the field theory of the thermal scalar in coordinate space. For simplicity we are assuming that the three-vertex has amplitude $\lambda$ and does not include derivative interactions. Comments on these issues are provided further on.
With $\Delta(\mathbf{u},\mathbf{x})$ denoting the coordinate space scalar propagator, this amplitude equals
\begin{align}
\mathcal{A}_{\mathbf{x}\mathbf{y}\mathbf{z}} &= g_s\lambda \int d\mathbf{u} \Delta(\mathbf{u},\mathbf{x}) \Delta(\mathbf{u},\mathbf{y}) \Delta(\mathbf{z},\mathbf{u}) \nonumber \\
&= g_s\lambda \int d\mathbf{u} \sum_{m,n,o} \frac{\psi_m(\mathbf{u}) \psi_m(\mathbf{x})^*}{\lambda_m}\frac{\psi_n(\mathbf{u}) \psi_n(\mathbf{y})^*}{\lambda_n} \frac{\psi_o(\mathbf{z}) \psi_o(\mathbf{u})^*}{\lambda_o}.
\end{align}
It is clear from this expression that the amplitude gets its main support from that region in space where the wavefunctions live. \\
Instead focussing on external states with fixed quantum numbers ($l$, $r$, $s$), we would multiply this expression by the suitable wavefunctions:
\begin{equation}
\mathcal{A}_{lrs} = \int d\mathbf{x} \psi_l(\mathbf{x}) \int d\mathbf{y} \psi_r(\mathbf{y}) \int d\mathbf{z} \psi_s(\mathbf{z})^* \mathcal{A}_{\mathbf{x}\mathbf{y}\mathbf{z}}.
\end{equation}
This is the analog of Fourier transforming to momentum space in a translationally invariant spacetime. We obtain
\begin{align}
\mathcal{A}_{lrs} = g_s\lambda \int d\mathbf{u} \frac{\psi_l(\mathbf{u})}{\lambda_l}\frac{\psi_r(\mathbf{u})}{\lambda_r} \frac{\psi_s(\mathbf{u})^*}{\lambda_s}.
\end{align}
The propagator denominators are amputated to obtain the S-matrix elements and the $\mathbf{u}$-integral is the analog of $\delta(\sum k)$ in a translationally invariant space. From this expression, one sees that the amplitude $\mathcal{A}_{lrs}$ is mainly supported at the locations where the wavefunctions at the interaction location are relatively large. \\
In string theory, the three-vertex interaction (on the 2-sphere) can then be written in terms of the above field theoretical formulas as
\begin{equation}
\left\langle T_0 T_0 T^*_0 \right\rangle_{S^2} = \lambda \int d\mathbf{u} \psi_0(\mathbf{u})\psi_0(\mathbf{u}) \psi_0(\mathbf{u})^*.
\end{equation}
Thus the stringy amplitude determines $\lambda$ by stripping away the $\mathbf{u}$-integral given above.\\
Having gained insight into the space dependence of the 3-point interactions, we return to the genus 2 example given in equation (\ref{vacugen2}) and we obtain:
\begin{equation}
\label{3ptampl}
\mathcal{A} = \frac{g_s^2}{\lambda_0^3} \lambda^2 \int d\mathbf{u} \psi_0(\mathbf{u})\psi_0(\mathbf{u})\psi_0(\mathbf{u})^* \int d\mathbf{v} \psi_0(\mathbf{v})\psi_0(\mathbf{v})^*\psi_0(\mathbf{v})^*,
\end{equation}
We should remark that in principle the field theoretic coupling constant $\lambda$ will depend on the quantum numbers of the field factors present in the interaction vertex in the Lagrangian, but this does not cause any real difficulties.\footnote{In flat space, dependence of $\lambda$ on the momentum quantum number is interpreted as a derivative interaction. In general, such a derivative is not diagonal in the basis $\psi_l$. One should presumably sum an entire array of such derivative terms to obtain a diagonal quantity. The Lagrangian description in coordinate space is hence not particularly useful. Writing the Lagrangian in coordinates diagonal in $\psi_l$ is much better, but obscures the spacetime interpretation. 
Let us be more explicit about this point. A general 3-point interaction in field theory contains three field factors with some differential operators $\hat{D}_1$, $\hat{D}_2$ and $\hat{D}_3$ acting on them. Expanding the fields in a complete set of eigenmodes of $\hat{\mathcal{O}}$, one gets:
\begin{equation}
\mathcal{L} \supset g_s\lambda \int d\mathbf{x} \hat{D}_1 T(\mathbf{x}) \hat{D}_2 T(\mathbf{x}) \hat{D}_3 T^*(x) = g_s\lambda \sum_{l,m,n} a_l a_m a^*_n \int d\mathbf{x} \hat{D}_1 \psi_l(\mathbf{x}) \hat{D}_2 \psi_m(\mathbf{x}) \hat{D}_3 \psi_n^*(\mathbf{x}),
\end{equation}
where the c-numbers $a_n$ represent the modes that should be integrated in the path integral. This basis is diagonal in the relevant quantum numbers. The identification with string theory then proceeds by identifying field theory and string amplitudes as
\begin{equation}
\lambda \int d\mathbf{x} \hat{D}_1 \psi_l(\mathbf{x}) \hat{D}_2 \psi_m(\mathbf{x}) \hat{D}_3 \psi_n^*(\mathbf{x}) = \left\langle T_l T_m T^*_n \right\rangle_{S^2},
\end{equation}
which allows an identification of the differential operators and of $\lambda$ (which is now independent of the quantum numbers $l$, $m$ and $n$). This allows us to rewrite the interaction term in the field theory Lagrangian as
\begin{equation}
\mathcal{L} \supset g_s\lambda \int d\mathbf{x} \hat{D}_1 T(\mathbf{x}) \hat{D}_2 T(\mathbf{x}) \hat{D}_3 T^*(x) = g_s\sum_{l,m,n} \left\langle T_l T_m T^*_n \right\rangle_{S^2} a_l a_m a^*_n.
\end{equation}
Finally, one should modify equation (\ref{3ptampl}) into
\begin{equation}
\mathcal{A} = \frac{g_s^2}{\lambda_0^3} \lambda^2 \int d\mathbf{u} \hat{D}_1\psi_0(\mathbf{u})\hat{D}_2\psi_0(\mathbf{u})\hat{D}_3\psi_0(\mathbf{u})^* \int d\mathbf{v} \hat{D}_1^*\psi_0(\mathbf{v})^*\hat{D}_2^*\psi_0(\mathbf{v})^*\hat{D}_3^*\psi_0(\mathbf{v}).
\end{equation}
} \\

\begin{figure}[h]
\begin{minipage}{0.5\textwidth}
\centering
\includegraphics[width=0.75\textwidth]{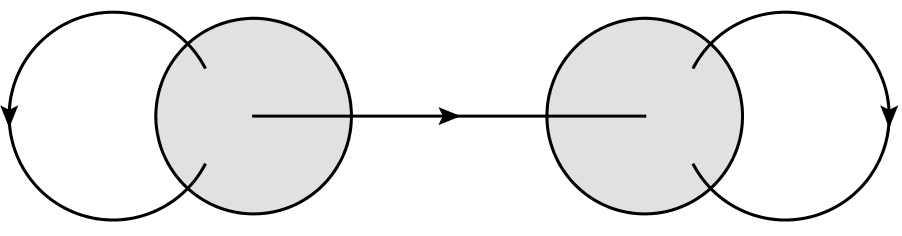}
\caption*{(a)}
\end{minipage}
\begin{minipage}{0.5\textwidth}
\centering
\includegraphics[width=0.75\textwidth]{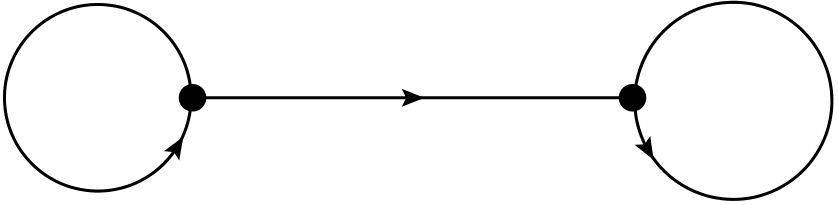}
\caption*{(b)}
\end{minipage}
\caption{(a) Other degenerate limit of the genus two surface. (b) Point particle graph of the alternate degenerate limit of the genus two surface.}
\label{alt1}
\end{figure}

\noindent This amplitude can now be rewritten as
\begin{align}
\mathcal{A} &= g_s^2 \lambda^2 \iint d\mathbf{u} d\mathbf{v} \frac{\psi_0(\mathbf{u})\psi_0(\mathbf{v})^*}{\lambda_0}\frac{\psi_0(\mathbf{u})\psi_0(\mathbf{v})^*}{\lambda_0}\frac{\psi_0(\mathbf{u})^*\psi_0(\mathbf{v})}{\lambda_0} \\
\label{secondl}
&\approx g_s^2 \lambda^2 \iint d\mathbf{u} d\mathbf{v} \left\langle \mathbf{u}\right| \frac{1}{L_0 + \bar{L}_0}\left| \mathbf{v}\right\rangle \left\langle \mathbf{u}\right| \frac{1}{L_0 + \bar{L}_0}\left| \mathbf{v}\right\rangle \left\langle \mathbf{v}\right| \frac{1}{L_0 + \bar{L}_0}\left| \mathbf{u}\right\rangle.
\end{align}
where in the second line, one should focus on the most dominant mode of the thermal scalar (hence the approximation symbol).\footnote{Again suitable insertions of the differential operators $\hat{D}_1$, $\hat{D}_2$, $\hat{D}_3$ should be made whenever necessary.} \\
Both of these equalities teach us something about the higher genus amplitudes. \\
The first line makes it clear that the amplitude gets its contribution from the near-horizon region (since that is where the thermal scalar wavefunction is supported). It shows that higher genus amplitudes have in their dominating contribution only support close to the tip of the cigar. This implies that even higher loop corrections cannot modify the fact that the random walk has a string-scale spread from the horizon.\footnote{Note that for instance in flat space, higher genus contributions would oscillate all over space, much like the genus one contribution. In a microcanonical scene, one expects these corrections to contract the long string into a stringy ball.} This random walk will have self-intersections (according to the above 3-point vertex), but it will still remain close to the horizon. We are led to the conclusion that Susskind's picture even holds when including higher loop interactions. \\
The second line (\ref{secondl}) in the above formula is in the first-quantized particle language, where the amplitude has the meaning of the following figure \ref{loop}.
\begin{figure}[h]
\centering
\includegraphics[width=4cm]{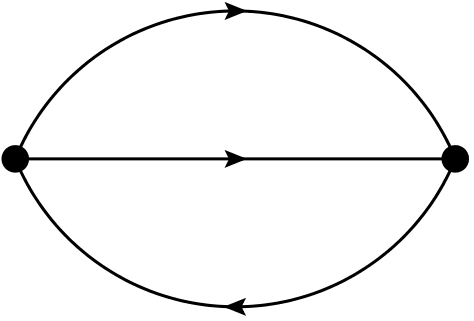}
\caption{First quantized Feynman graph corresponding to the above degenerate limit of the genus two surface.}
\label{loop}
\end{figure}
The spatial shape of the first-quantized thermal scalar should be interpreted as the shape of the long string. Note that the other possible degenerate limit of the genus-2 surface yields the same most dominant contribution (\ref{3ptampl}) as shown in figure \ref{alt1}.

\noindent The random walk picture is immediately distilled from this in a very explicit manner by using the Schwinger trick on the propagators: 
\begin{equation}
\frac{1}{L_0 + \bar{L}_0} = \int_{0}^{+\infty}dT e^{-(L_0 + \bar{L}_0)T}
\end{equation}
and then giving a Lagrangian interpretation to this Hamiltonian picture amplitude. In this picture, each propagator contains a proper time parameter (Schwinger parameter). What one finds is that there is a set of open random walks with 3-point intersections. The locations of these interactions are also integrated over the entire space. The amplitude receives the largest contribution from the near-horizon region. Note that the genus $g$ diagram gives rise to a $g$-loop random walk. \\

\begin{figure}[h]
\begin{minipage}{0.5\textwidth}
\centering
\includegraphics[width=0.4\textwidth]{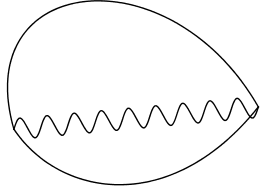}
\caption*{(a)}
\end{minipage}
\begin{minipage}{0.5\textwidth}
\centering
\includegraphics[width=0.3\textwidth]{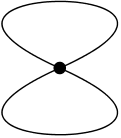}
\caption*{(b)}
\end{minipage}
\caption{(a) Graviton exchange by the long string. (b) Effective vertex of graviton exchange by integrating out virtual particles.}
\label{grav1}
\end{figure}

\noindent Also note that for instance graviton exchange by the long string is a subdominant effect (compared to other amplitudes at the same genus) for topologically trivial thermal circles: it has two separate 3-point interactions with a virtual graviton in between. This gets translated to an effective 4-point vertex, which dominates if winding number is conserved in interactions, which it is not in our case. This is shown in figure \ref{grav1}.

\noindent For topologically non-trivial thermal circles, the lowest interaction vertices are 4-point vertices, implying a different random walk intersection behavior. Such a 4-vertex arises by summing over all possible internal lines between two 3-vertices. 3-vertices on the other hand can only be associated to a pointlike interaction between three fields. \\

\noindent Of course, this reduction to particle path integrals, reintroduces UV divergences (at higher genera manifested as point-vertices), although we know this is merely an artifact of our approach. \\

\noindent For type II superstrings, we expect this picture to remain intact: the 3-punctured spheres would have different values, the Hagedorn temperature would be different, but the strategy and result would be the same. \\

\noindent As a summary, we conclude that at any \emph{fixed} genus, the partition function is dominated by the thermal scalar. The partition function behaves as a random walk with 3-point interactions and is localized close to the black hole event horizon. Note though that the propagator denominators actually force the partition function to diverge at $\beta = \beta_{\text{Hawking}}$ (as for the one-loop amplitude). \\
Of course, in string theory the different genera partition functions should be summed over and we now turn to this question. 

\subsection{The perturbative genus expansion and its limitations}
The single string partition function for topologically trivial thermal circles has the leading form (as determined above):
\begin{equation}
\label{expan}
Z_1 = - \ln(\beta-\beta_H) + \sum_{n=1}^{+\infty}C_n\frac{g_s^{2n}}{(\beta-\beta_H)^{3n}}.
\end{equation}
In this case, for each genus only the leading divergence is kept. This result can be found by using a double scaling limit as in \cite{Brigante:2007jv} where $g_s \to0$ and $\beta \to \beta_H$ while keeping $g_s^2 \propto (\beta-\beta_H)^3$. This limit ensures all other divergences at each genus scale out. \\
This expression can be seen to come directly from a field theory action. Consider the thermal scalar field theory action:\footnote{The $\hat{D}$ operators are differential operators that have been defined in a footnote earlier. $cc$ denotes the complex conjugate term of the second term in this expression.}
\begin{equation}
S = \int dV\sqrt{G}e^{-2\Phi}\left[T^* \hat{\mathcal{O}} T  + g_s \lambda \hat{D}_1 T\hat{D}_2 T \hat{D}_3 T^* +(cc)\right].
\end{equation}
In expanding the field $T$ in a complete set of eigenfunctions of $\hat{\mathcal{O}}$ with canonical normalization, one only retains the lowest mode in the critical limit:
\begin{equation}
T(\mathbf{x}) = \sum_n a_n \psi_n(\mathbf{x}) \approx a_0 \psi_0(\mathbf{x}).
\end{equation}
The action then reduces to
\begin{equation}
S \approx \lambda_0 a_0 a_0^*  + g_s \left\langle T_0 T_0 T_0^* \right\rangle_{S^2} a_0 a_0 a_0^* + (cc).
\end{equation}
The resulting critical diagrams then originate from the following theory:
\begin{equation}
Z_1 = \log \int d\phi d\phi^* e^{-\lambda_0 \phi \phi^* - g_s\phi\phi^*\left(\tilde{\lambda} \phi + \tilde{\lambda}^*\phi^*\right)},
\end{equation}
where $\phi = a_0$ is a complex number and the coupling $\tilde{\lambda} = \left\langle T_0 T_0 T_0^* \right\rangle_{S^2}$. The lowest eigenvalue $\lambda_0 \sim \beta-\beta_H$. With these definitions, one readily finds explicitly the expansion
\begin{equation}
Z_1 = -\ln\lambda_0 + \frac{6g_s^2\left|\lambda\right|^2}{\lambda_0^3} + \frac{162g_s^4\left|\lambda\right|^4}{\lambda_0^6} + \hdots
\end{equation}
agreeing with the previous expansion (\ref{expan}) and concretely giving values for the coefficients. \\

\noindent Clearly taking $\lambda_0 \to 0$ results in an infinite partition function: higher order terms are needed to determine the full (interacting) thermal scalar action. \\

\noindent If this scaling limit is not followed, additional contributions should be added to the functional integral (corresponding to the subleading corrections at each genus). These introduce 4-point interactions (and higher). As it stands, the above integral is non-perturbatively defined (although infinite in the absence of higher order corrections). Just like in \cite{Brigante:2007jv}, the thermal scalar potential can in principle be determined in this way.\\

\noindent The punch line is that the string perturbation series (the genus expansion) is not good in this case. In \cite{Brigante:2007jv} this was argued for for topologically stable thermal circles, where the perturbation series was seen to break down near the Hagedorn temperature. In this case however, since $T_H = T_{\text{Hawking}}$, we conclude that \emph{string perturbation theory on the thermal manifold breaks down for a general uncharged black hole at its Hawking temperature}. Despite several less rigorous steps in the above derivation\footnote{We did not explicitly construct the Zamolodchikov metric nor the three-point functions. The only thing we need from these however is that they are finite (and non-zero), which is something we did discuss here.}, we believe this conclusion is difficult to avoid. In any case, the one-loop result already shows that problems arise and this feature is apparently \emph{not} solved by summing higher genus contributions.\\

\noindent Other relevant work concerning the Hagedorn transition at finite string coupling (but using holographic methods instead) can be found in \cite{Liu:2004vy}\cite{AlvarezGaume:2005fv}\cite{Aharony:2003sx}. \\

\begin{figure}[h]
\centering
\includegraphics[width=0.5\linewidth]{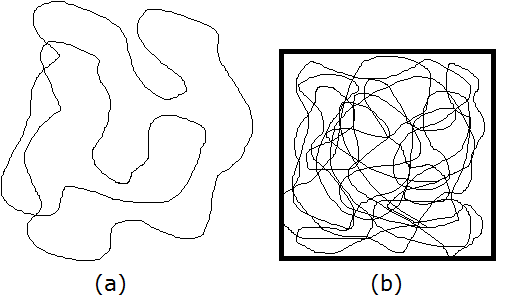}
\caption{(a) Random walk in an uncompact space. The random walk is not densily packed as it can safely avoid an unconstrained number of self-crossings. (b) In a compact space, the random walk is forced to pass through the same point over and over. This causes the density of string to diverge which manifests itself in an infinite free energy density.}
\label{cont}
\end{figure}

\noindent If $g_s$ is strictly zero however, all higher genus amplitudes vanish and Susskind's (non-interacting) long string picture is valid. As soon as any form of interaction is allowed ($g_s$ not strictly zero, but arbitrarily small), the higher genus amplitudes do not vanish and a resummation is necessary. This is in agreement with Susskind's prediction that interactions, if present, will \emph{always} become important close to the event horizon. The fact that setting $g_s$ to zero gives an infinite free energy density does not seem so strange: the string becomes arbitrarily long and intersects itself numerous times effectively giving an infinite density of string everywhere. This is intuitively obvious: non-compact dimensions are always larger than the string itself thus reducing the `chance' of a piece of string returning to the same location. For compact spaces, the string feels the boundaries and is forced to pass through the same location over and over again \cite{Barbon:2004dd}. A cartoon is given in figure \ref{cont}.
A related point was also made by Susskind: he argues that higher order interactions should become important as $g_s^2 \rho \ell_s^{d-2}$ becomes of order 1, with $\rho$ the number of string crossings per unit horizon area. This argument was made in a single-string (microcanonical) picture. In our case, the condition (for the most dominating contribution; this is the worst case scenario) is instead $g_s^2\frac{1}{(\beta-\beta_H)^3} \ell_s^{3}$ of order 1. This suggests some proportionality:
\begin{equation}
\rho \propto \frac{\ell_s^{5-d}}{(\beta-\beta_H)^{3}}
\end{equation}
and equating the temperature to the Hawking temperature hence immediately leads to an infinite number of string crossings $\rho$ per unit horizon area. As long as $g_s \neq 0$, higher order terms in the perturbation series are not negligible.\\

\noindent We conclude that as soon as $g_s$ is non-zero, one should not use the perturbation series anymore.

\section{*Path integral proof of the McClain-Roth-O'Brien-Tan theorem}
\label{MROB}
In this supplementary section, we will be more explicit in the general proof of this theorem \cite{McClain:1986id}\cite{O'Brien:1987pn}. The proof splits in two parts: first the modular transformation properties of the two wrapping numbers is distilled. Then in the second part it is shown that these are sufficient to perform the extension to the strip domain while dropping one wrapping number. 
\subsection{Part 1: Modular transformations of the wrapping numbers}
Assume the general path integral on the fundamental domain has a periodic coordinate $X \sim X + 2\pi R$. Then this path integral is considered with the set of boundary conditions
\begin{align}
X(\sigma^1 + 2\pi , \sigma^2) &= X(\sigma^1,\sigma^2) + 2\pi w R, \\
X(\sigma^1 + 2\pi\tau_1 , \sigma^2+2\pi\tau_2) &= X(\sigma^1,\sigma^2) + 2\pi m R,
\end{align}
with a flat worldsheet metric in the action. This form is particularly suited to our task since all dependence on $\tau$ and on the wrapping numbers is fully extracted from the action into the boundary conditions. \\
The entire path integral is modular invariant and the modular group $PSL(2,\mathbb{Z})$ is generated by the $T$ and $S$ transformations. Consider first the $T$ transformation $\tau \to \tau +1$. The second boundary condition then gets translated into
\begin{align}
X(\sigma^1 + 2\pi(\tau_1 +1), \sigma^2+2\pi\tau_2) = X(\sigma^1 + 2\pi\tau_1 , \sigma^2+2\pi\tau_2) + 2\pi w R = X(\sigma^1,\sigma^2) + 2\pi m R,
\end{align}
yielding
\begin{align}
X(\sigma^1 + 2\pi\tau_1, \sigma^2+2\pi\tau_2) =  X(\sigma^1,\sigma^2) + 2\pi (m-w) R.
\end{align}
Hence upon making the replacement $m \to m+w$ (or $m'=m-w$), the invariance is established. \\

\begin{figure}[h]
\centering
\includegraphics[width=0.25\textwidth]{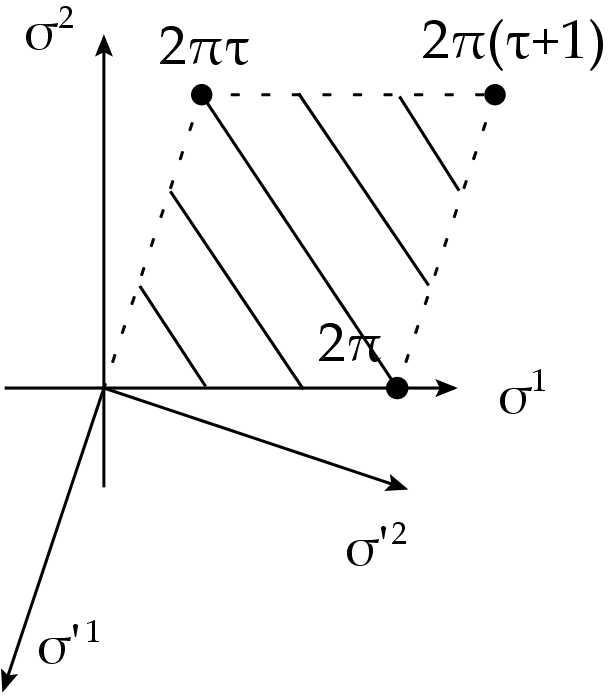}
\caption{Worldsheet coordinates and torus with modulus $\tau$. The new (primed) coordinates are also shown.}
\label{plane}
\end{figure}

\noindent Secondly we consider the $S$-transformation $\tau \to -1/\tau$. To start with, we first make a scaling coordinate transformation on the worldsheet as
\begin{equation}
z' = -\frac{z}{\tau}, \quad z = \sigma^1 + i \sigma^2, \quad z' = \sigma'^1 + i \sigma'^2.
\end{equation}
As long as the non-linear sigma model satisfies the Einstein equations, it is conformal and the above scaling symmetry has no effect on the action itself (it gets transformed into the same action with primed coordinates). The boundary conditions however are not the same anymore (corresponding to the fact that these transformations are not in the CKG of the torus). Explicitly, we obtain
\begin{equation}
\sigma'^1 = - \frac{\sigma^1\tau_1 + \sigma^2\tau_2}{\left|\tau\right|^2}, \quad \sigma'^2 = - \frac{-\sigma^1\tau_2 + \sigma^2\tau_1}{\left|\tau\right|^2}.
\end{equation}
These new coordinates are shown in figure \ref{plane}.
In these new worldsheet coordinates, the two torus cycles are described as
\begin{align}
(\sigma^1 , \sigma^2) \sim (\sigma^1 + 2\pi, \sigma_2) &\to (\sigma'^1, \sigma'^2) \sim \left(\sigma'^1 - \frac{2\pi \tau_1}{\left|\tau\right|^2}, \sigma'^2 + \frac{2\pi \tau_2}{\left|\tau\right|^2}\right), \\
(\sigma^1 , \sigma^2) \sim (\sigma^1 + 2\pi \tau_1, \sigma_2 + 2\pi\tau_2) &\to (\sigma'^1, \sigma'^2) \sim (\sigma'^1 - 2\pi , \sigma'^2).
\end{align}
Up to this point, only passive transformations were used. Next, we do the $S$-transformation, which yields the final transformed boundary conditions:
\begin{align}
X(\sigma'^1 + 2\pi \tau_1, \sigma'^2 + 2\pi\tau_2) &= X(\sigma'^1,\sigma'^2) + 2\pi w R, \\
X(\sigma'^1 - 2\pi, \sigma'^2) &= X(\sigma'^1,\sigma'^2) + 2\pi m R.
\end{align}
The substitution
\begin{equation}
w \to m ,\quad m\to -w,
\end{equation}
or ($w'=-m$ and $m'=w$) then shows that the $S$-transformation is a symmetry. \\
To summarize, the following modular transformation properties are found:
\begin{align}
T &: \quad m \to m+w , \\
S &: \quad m \to -w, \quad w \to m.
\end{align}

\subsection{Part 2: $SL(2,\mathbb{Z})$ manipulations}
The second phase of the proof consists of showing that the above modular transformation properties of $m$ and $w$ are enough to transform the fundamental domain into the strip. This part actually goes identically the same as for flat space \cite{McClain:1986id}\cite{O'Brien:1987pn}. Because of this reason, we will go rather quickly through the necessary steps. The goal is clear: find out what modular transformation on $\tau$ needs to be done to ensure that the ($m$, $w$) doublet gets transformed into ($m$, $0$) and at the same time build up the remaining regions of the strip. \\
Firstly, it is known that the set of modular transformations on $\tau$ (with composition of transformations) is isomorphic (as a group) to the set of $PSL(2,\mathbb{Z})$ matrices, equiped with matrix multiplication. A modular transformation and a $SL(2,\mathbb{Z})$ matrix are identified as
\begin{equation}
\frac{a\tau + b}{c\tau +d} \quad \Leftrightarrow \quad 
\left[\begin{array}{cc}
a & b  \\
c & d \end{array}\right],
\end{equation}
and the $T$ and $S$ generators of this group are given by
\begin{equation}
T = \left[\begin{array}{cc}
1 & 1  \\
0 & 1 \end{array}\right], \quad S = \left[\begin{array}{cc}
0 & -1  \\
1 & 0 \end{array}\right].
\end{equation}
As is well-known, $c$ and $d$ should be relatively prime by the determinantal condition $ad-bc=1$. The quantum numbers $m$ and $w$ then form a doublet of this group action, where the column vector $\left[\begin{array}{c}
m   \\
w  \end{array}\right]$ transforms under $PSL(2,\mathbb{Z})$ by left multiplication. The $SL(2,\mathbb{Z})$ element that we seek is hence of the form
\begin{equation}
\left[\begin{array}{cc}
a & b  \\
c & d \end{array}\right]\left[\begin{array}{c}
m   \\
w  \end{array}\right] = \left[\begin{array}{c}
X   \\
0  \end{array}\right],
\end{equation}
where the two bottom elements $c$ and $d$ are determined by $cm + dw = 0$ or $cm = -dw$. Let the gcd of $m$ and $w$ be $r$. Then the only possibility for relatively prime $c$ and $d$ is the solution $c=\frac{w}{r}$ and $d=-\frac{m}{r}$. \\
For fixed $c$ and $d$, the top elements $a$ and $b$ are restricted by the determinantal condition and one readily shows that the only freedom left is $(a,\, b) \to (a+\lambda c, \, c+\lambda d)$ for $\lambda \in \mathbb{Z}$. These elements can all be obtained from one such element $M$ by left multiplication by $T$ as $T^\lambda M$. Hence it is clear that only 1 value of $\lambda$ exists for which the resulting modulus is located in the strip region. Moreover, any two matrices that do not have the same $c$ and $d$ do not have any overlap in the strip region. It is also clear that the strip is built up entirely since $PSL(2,\mathbb{Z})$ fully generates the upper half plane from the fundamental domain. The $SL(2,\mathbb{Z})$ element is fully fixed by this and the transformed modulus is related to the initial one as
\begin{equation}
\tau' = \frac{a\tau+b}{c\tau+d}.
\end{equation}
The value $X$ equals $-r$ (due to the determinantal condition). \\

\noindent As a summary of these steps, in a schematic fashion, one manipulates the expression as
\begin{equation}
\label{scheme}
\sum_{m,w} \int_{\mathcal{F}(m,w)} \to \sum_{r=1}^{+\infty} \sum_{\left[c,d\right]=1}\int_{\mathcal{F}(m,w)} \to 2 \sum_{r=1}^{+\infty} \int_{E(-r,0)}
\end{equation}
and $r$ takes over the role of the single wrapping number in the strip. \\
Note further that the worldsheet transformation $z \to -z$ is equivalent to replacing $m\to-m$ and $w\to-w$. Doing this, one finds that $X=+r$ instead, showing the symmetry $r\to-r$ in the resulting path integral. We can hence double the range of $r$ which destroys an extra factor of 2 we created earlier (\ref{scheme}) by having both ($c$, $d$) and ($-c$, $-d$) map to the same $r$-number. \\

\noindent In all of these manipulations, the ($m=0,\, w=0$) sector transforms as a singlet under the modular group and this is hence not altered: no build-up of the strip domain is present for this state and in the end it is still integrated over the modular fundamental domain.

\section{*Non-normalizable versus normalizable vertex operators}
\label{nnversusn}
The computations given in \cite{Giveon:1999px}\cite{Giveon:1999tq} and \cite{Aharony:2004xn} focus on non-normalizable observables and these give a series of poles for the $n$-point functions. One should distill the residue of these poles to obtain the S-matrix elements. These then correspond to $n$-point functions of the associated normalizable operators and these should be finite. This solves an initial worry one might have in that the Zamolodchikov metric might diverge or the 3-point functions might vanish. As an extra reassurance, the four-point amplitude for non-normalizable observables for the $AdS_3$ model was computed in \cite{Maldacena:2001km} where an intermediate virtual discrete state was found indeed, suggesting the inverse Zamolodchikov metric for such states is non-zero. These correlation functions in $AdS_3$ are directly related to those of the cigar CFT and again one should zoom in on the LSZ residue instead to obtain the required stringy amplitudes obtained by normalizable vertex operators. \\
The link between the normalizable and non-normalizable vertex operators is
\begin{equation}
\mathcal{O}_{\text{non-norm}} = \frac{1}{\text{pole}}\mathcal{O}_{\text{norm}}.
\end{equation}
In more detail, for large $\phi$ the vertex operators behave as \cite{Aharony:2003vk}:\footnote{In this expression, $Y$ denotes a free boson and $\phi$ is a coordinate coming from the $AdS_3$ ancestor of this vertex operator. In the notation given in this work, $\phi=\rho$, the rescaled radial coordinate introduced in chapter \ref{chri}.}
\begin{equation}
V_{j,m,\bar{m}} \sim e^{iQmY}\left(e^{Qj\phi} + \frac{1}{\text{pole}} e^{-Q(j+1)\phi} + \hdots\right).
\end{equation}
Then multiplying by the pole contribution distills purely the normalizable part of the vertex operator. Indeed, it is this asymptotic behavior that we used in constructing the normalizable string fluctuations for Euclidean Rindler space in chapter \ref{chri} \cite{Mertens:2013zya}. The reader should compare this to equation (\ref{stasympt}). Moreover, the pole spectrum of the non-normalizable 2-point functions, originally constructed to be the spectrum of states in the dual LST \cite{Aharony:2004xn}, is also the discrete spectrum of the cigar CFT.\footnote{Note that in \cite{Hanany:2002ev}, the cigar partition function was rewritten in CFT language and the spectrum was distilled. The precise discrete spectrum shown at the beginning of section \ref{unitarity} was not found though it seems obvious that a more careful study of their argument would indeed reproduce the spectrum. Also, interestingly, we saw in chapter \ref{chwzw} \cite{Mertens:2014nca} that the spectrum of discrete states in thermal $AdS_3$ orbifolds actually also shows precisely this same structure.} The discrete part of both Hilbert spaces is then indeed the same. \\
Since the 2-point function of non-normalizable vertex operators has only a single pole, we immediately deduce that the 2-point function of normalizable operators (the scattering amplitude) vanishes, just like in flat space. Schematically,
\begin{equation}
\left\langle \mathcal{O}_{\text{norm}} \mathcal{O}_{\text{norm}} \right\rangle = \text{pole}^2\left\langle \mathcal{O}_{\text{non-norm}} \mathcal{O}_{\text{non-norm}} \right\rangle = 0. 
\end{equation}


\chapter{The Relevance of the Thermal Scalar}
\label{chrel}
In this chapter we study whether the thermal scalar encodes also anything else besides the canonical partition function. In particular we will find that the near-Hagedorn stress tensor and string charge can also be written down in terms of the thermal scalar field. This chapter contains the results of \cite{Mertens:2014dia} and \cite{Mertens:2015hia} near the end.

%
%
%

\section{Introduction}
The relevance and interpretation of the thermal scalar for string thermodynamics has been a source of confusion in the past (see e.g. \cite{Barbon:2004dd} and references therein). It has been known for quite some time that the critical behavior of a Hagedorn system can be reproduced by a random walk of the thermal scalar. This picture was made explicit in curved spacetimes in \cite{Kruczenski:2005pj}\cite{Mertens:2013pza} in chapter \ref{chth}. These results show that thermodynamical quantities such as the free energy and the entropy can be rewritten in terms of the thermal scalar. In this chapter we investigate whether other properties of the near-Hagedorn string gas (the energy-momentum tensor and the string charge) can be written in terms of the thermal scalar. This has been largely done by Horowitz and Polchinski in \cite{Horowitz:1997jc} for the high energy behavior of the energy-momentum tensor. Here we extend their analysis.\\
On a larger scale, we are interested in finding a suitable description to handle the long highly excited strings that critical string thermodynamics predicts. The thermal scalar is intimately related to this phase of matter though it is, in general, unclear precisely how. Here we report on some modest progress in this direction.\\ 
This chapter is organized as follows. \\
In section \ref{HP} we rederive the final result of Horowitz and Polchinski in detail. We will provide details and clarifications as we go along. Section \ref{nonstat} contains the extension to spacetimes with non-zero $G_{\tau i}$ metric components, necessary for computing components of the stress tensor with mixed (time-space) components and for treating stationary spacetimes. In section \ref{thermaver} we will extend the result to the canonical ensemble. Then in section \ref{stringcharge} we will provide the analogous result for the string charge. Section \ref{correl} contains some results on a specific class of correlators. In section \ref{second} we provide an interesting alternative derivation of the same results using a second quantized (Green function) formalism which illustrates some issues from a different perspective. We demonstrate the formulas in section \ref{examples} on several examples: flat space, the $AdS_3$ WZW model and Rindler spacetime. The Rindler case allows us to provide a derivation of Bekenstein's formula in section \ref{beke} for the black hole entropy in terms of long strings by building up the black hole gradually. A summary of the results is presented in section \ref{summary} and some supplementary material is given in the end.

\section{Derivation of the Horowitz-Polchinski result}
\label{HP}
In \cite{Horowitz:1997jc}, the authors consider long strings with self-interactions. This study was further analyzed in \cite{Damour:1999aw}. As a byproduct in their calculations, they obtain an expression for the energy-averaged stress tensor in terms of the thermal scalar. It is this result that we will focus on in this chapter. We already emphasize that we will only be interested in the non-self-interacting part of the stress tensor: interactions with a background are taken into account but no self-interactions between the stringy fluctuations. \\

\noindent The spacetime energy-momentum tensor of a single string in a general background is given by\footnote{The subscript $l$ denotes the Lorentzian signature tensor, whereas a subscript $e$ denotes the Euclidean signature tensor. Also, to avoid any confusion, in this chapter we will write $0$ for the Lorentzian time index and $\tau$ for the Euclidean time index.}
\begin{equation}
\label{firsteq}
T^{\mu\nu}_{l}(\mathbf{x},t) = - \frac{2}{\sqrt{-G}}\frac{\delta S_l}{\delta G_{\mu\nu}(\mathbf{x},t)} = \frac{1}{2\pi\alpha'}\int_{\Sigma} d^{2}\sigma \sqrt{h} h^{ab}\partial_aX^{\mu}\partial_bX^{\nu}\frac{\delta^{(D)}((\mathbf{x},t) - X^{\rho}(\sigma,\tau))}{\sqrt{-G(\mathbf{x},t)}}.
\end{equation}
Note that from a worldsheet point of view this definition is problematic, since varying the metric does not guarantee a conformal background. So although for classical string theory this definition is useful, as soon as one considers the quantum theory we encounter conceptual problems. This is why we shall continue fully from the spacetime (field theory) point of view to rederive the result of \cite{Horowitz:1997jc}. \\
The backgrounds we have in mind have no temporal dependence ($\partial_0 G^{\mu\nu}=0$). To proceed, let us consider the Euclidean energy-momentum tensor (sourcing the Euclidean Einstein equations):
\begin{equation}
\label{euclstress}
T^{\mu\nu}_{e}(\mathbf{x},\tau) = - \frac{2}{\sqrt{G}}\frac{\delta S_e}{\delta G_{\mu\nu}(\mathbf{x},\tau)}. 
\end{equation}
Instead of using the worldsheet non-linear sigma model (as in equation (\ref{firsteq})), we focus on the spacetime action of all string modes: the string field theory action. Then we restrict this action to the sum of the non-interacting parts of the different string states, let us call the resulting action $S_e$.

\subsection{The stress tensor as a derivative of the Hamiltonian}
This free action $S_e$\footnote{Or better, its Lorentzian signature counterpart.} can be used to write down a corresponding (Lorentzian) Hamiltonian $H_l$. 
First we wish to establish a relationship between the above Euclidean stress tensor and the (non-interacting) string field theory Hamiltonian as
\begin{equation}
\label{question}
T^{\mu\nu}_{e}(\mathbf{x}) \stackrel{?}{=} - \frac{2}{\sqrt{G}}\frac{\delta H_l}{\delta G_{\mu\nu}(\mathbf{x})}. 
\end{equation}
This statement does \emph{not} hold as an operator identity in the full non-interacting string field theory. Luckily, we only need it in suitable quantum expectation values. \\

\noindent We start by focusing on the thermal ensemble at temperature $\beta$. \\
The Euclidean stress tensor, averaged over $\beta$ in time equals
\begin{equation}
\label{betaav}
T^{\mu\nu}_{e,\beta} = \frac{1}{\beta} \int_{0}^{\beta} d\tau T^{\mu\nu}_{e}(\mathbf{x},\tau) = \frac{1}{\beta} \int_{0}^{\beta} d\tau \frac{-2}{\sqrt{G}}\frac{\delta S_{e}}{\delta G_{\mu\nu}(\mathbf{x},\tau)}.
\end{equation}
This averaging over $\beta$ looks a bit funny, but it will turn out to be quite helpful.\footnote{For black hole spacetimes on the other hand, the Euclidean manifold and the thermal manifold are naturally identified and hence this integral over $\beta$ is not that strange in that case.} Note though that the spacetimes that we consider are static (or stationary), implying a time-independent stress tensor expectation value. So the averaging procedure is trivial in the end. \\
We are interested in computing
\begin{equation}
\text{Tr}\left(T^{\mu\nu}_{e,\beta}e^{-\beta H_l}\right),
\end{equation}
where we sum over the Fock space of free-string states. The Lorentzian Hamiltonian $H_l$ is a sum over the different field Hamiltonians present in the string spectrum. Due to the integral over $\tau$, we immediately obtain
\begin{equation}
\label{previous}
\int_{0}^{\beta} d\tau T^{\mu\nu}_{e}(\mathbf{x},\tau) = \frac{-2}{\sqrt{G}}\frac{\delta S_{\text{thermal}}}{\delta G_{\mu\nu}(\mathbf{x})},
\end{equation}
with $S_{\text{thermal}} = \int_{0}^{\beta}d\tau L_e$, the string field theory action on the thermal manifold, restricted to the quadratic parts (non-interacting). This action differs by $S_e$ only in the range of the temporal coordinate. Important to note is that this action does not contain thermal winding modes, since we only Wick-rotated the Lorentzian fields. \\
\noindent Note that in equation (\ref{previous}), we only vary $S_{\text{thermal}}$ with respect to time-independent metric variations. This result is elementary, consider for instance a general functional in two space dimensions:
\begin{equation}
F = \int dx \int dy \rho(x,y) L(x,y).
\end{equation}
Varying only with respect to $y$-independent configurations $\rho(x)$, we obtain
\begin{equation}
\frac{\delta F}{\delta \rho(x)} = \int dy L(x,y).
\end{equation}

\noindent In general we can write for the (non-interacting) multi-string partition function:
\begin{equation}
\label{feyn}
Z_{\text{mult}} = \text{Tr}\left(e^{-\beta H_l}\right) = \int \left[\mathcal{D}\phi^a\right]e^{-S_{\text{thermal}}}.
\end{equation}
This relation is obtained by writing down this equality for each string field separately. In this formula $\phi^a$ labels all the different string fields with suitable (anti)periodic boundary conditions around the thermal circle.\footnote{Again we emphasize that no winding modes around the thermal circle are included at this stage: only the string fields obtained by a Wick rotation of the theory.} As is well known, this identity also holds for operator insertions. After some juggling with the above formulas, we immediately obtain
\begin{align}
\text{Tr}\left(T^{\mu\nu}_{e,\beta}e^{-\beta H_l}\right) &= \int \left[\mathcal{D}\phi^a\right] T^{\mu\nu}_{e,\beta}e^{-S_{\text{thermal}}} \\
\label{second2}
&= \frac{2}{\beta \sqrt{G}}\frac{\delta }{\delta G_{\mu\nu}} \int \left[\mathcal{D}\phi^a\right]  e^{-S_{\text{thermal}}} \\
&= \text{Tr}\left(\frac{-2}{\sqrt{G}}\frac{\delta H_l}{\delta G_{\mu\nu}}e^{-\beta H_l}\right).
\end{align}

\noindent As we noted before, the temporal average is actually irrelevant and we obtain
\begin{equation}
\label{box}
\boxed{
\text{Tr}\left( T^{\mu\nu}_{e} e^{-\beta H_l}\right) = \text{Tr}\left(\frac{-2}{\sqrt{G}}\frac{\delta H_l}{\delta G_{\mu\nu}}e^{-\beta H_l}\right).}
\end{equation}
This equality holds for any positive value of $\beta$ (larger than $\beta_H$). We can hence inverse Laplace transform both sides of (\ref{box}) and we obtain
\begin{equation}
\boxed{
\text{Tr}\left( T^{\mu\nu}_e \delta(H_l - E)\right) = \text{Tr}\left(\frac{-2}{\sqrt{G}}\frac{\delta H_l}{\delta G_{\mu\nu}}\delta(H_l - E)\right),}
\end{equation}
and this provides the microcanonical equivalent to (\ref{box}). It is in this sense that we understand the equality (\ref{question}) between stress tensor and Hamiltonian functional derivative to the metric. \\

\noindent To make the above more concrete, we pause here and follow the same logic for a massless complex scalar field. Suppose the Lorentzian spectrum of string fluctuations on some manifold contains a massless complex scalar field. First we Wick-rotate the theory to write down the Euclidean stress tensor that we wish to consider. Explicitly:
\begin{align}
S_e &\propto \int_{-\infty}^{+\infty} d\tau \int dV \sqrt{G} G^{\mu\nu} \partial_{\mu}\phi \partial_{\nu} \phi^*, \\
T^{\mu\nu}_e &\propto \left(\partial^{\mu}\phi \partial^{\nu}\phi^* + (\mu \leftrightarrow \nu)\right) - \frac{1}{2}G^{\mu\nu}\partial^{\rho}\phi \partial_{\rho}\phi^*.
\end{align}
This stress tensor, as an operator in the canonical formalism, is averaged over Euclidean time $\beta$ and then inserted in the thermal trace as above:
\begin{align}
\label{result}
\text{Tr}\left(T^{\mu\nu}_{e,\beta}e^{-\beta H_l}\right) = \int_{\phi(\mathbf{x},\tau) = \phi(\mathbf{x},\tau+\beta)} \left[\mathcal{D}\phi\right]\left[\mathcal{D}\phi^*\right]T^{\mu\nu}_{e,\beta}e^{-\int_{0}^{\beta} d\tau \int dV \sqrt{G} G^{\mu\nu} \partial_{\mu}\phi \partial_{\nu} \phi^*}.
\end{align}
The thermal action written down in the exponent is equal to $S_e$, up to the different temporal integration. Then one applies the manipulations as above to extract the stress tensor in terms of a metric derivative of the Hamiltonian and the result follows. \\

\noindent The extensions of this to other higher spin fields is straightforward and we hope the concrete treatment of the scalar field provided some insight in this procedure: one does not need the concrete form of the action nor the stress tensor: we only need the fact that the action and stress tensor both are quadratic in the fields at the non-interacting level (allowing a full decoupling of all fields) and a time-independent background. \\
In string theory we should finally sum this quantity over all the string fields present in the Lorentzian spectrum. 

\subsection{Microcanonical stress tensor}
\label{microc}
Let us now compute this energy-momentum tensor when \emph{averaging} over all (Lorentzian) string states with fixed energy $E$, for very large $E$.\footnote{The concrete criterion for `very large' will follow shortly.} 
\begin{equation}
\label{micro}
\left\langle T^{\mu\nu}_{e}(\mathbf{x})\right\rangle_E = \frac{\text{Tr}\left[T^{\mu\nu}_{e}(\mathbf{x})\delta(H_l-E)\right]}{\text{Tr}\delta(H_l-E)},
\end{equation}
where the $E$ outside the expectation value denotes the averaging over states with energy $E$.
We can rewrite this as
\begin{align}
\left\langle T^{\mu\nu}_{e}(\mathbf{x})\right\rangle_E &= \frac{2}{\sqrt{G(\mathbf{x})}\text{Tr}\delta(H_l-E)}\text{Tr}\frac{\delta}{\delta G_{\mu\nu}(\mathbf{x})}\theta (E - H_l) \\
 &= \frac{2}{\sqrt{G(\mathbf{x})}\text{Tr}\delta(H_l-E)}\frac{\delta}{\delta G_{\mu\nu}(\mathbf{x})}\text{Tr}\theta (E - H_l),
\end{align}
where in the second line we used the Hellmann-Feynman theorem to extract the functional derivative out of the expectation value. 
Now we evaluate the traces in the high energy regime. Let us take as the density of string states
\begin{equation}
\label{doss}
\rho (E) \propto \frac{e^{\beta_H E}}{E^{D/2+1}},
\end{equation}
with $D$ the number of non-compact dimensions. Such a density of states is quite general and encompasses a wide class of Hagedorn systems \cite{Horowitz:1997jc}. The integral can be done\footnote{Only the large $E$ part is taken into consideration. The lower boundary of the integral is left arbitrary.}
\begin{equation}
\int^E d\tilde{E}\frac{e^{\beta_H \tilde{E}}}{\tilde{E}^{D/2+1}} = -(-\beta_H)^{D/2} \Gamma(-D/2,-\beta_H E)
\end{equation}
in terms of the incomplete Gamma function and for large $E$ it behaves as
\begin{equation}
\Gamma(-D/2,-\beta_H E) \propto (-\beta_H E)^{-D/2-1}e^{\beta_H E}.
\end{equation}
This term carries the large $E$ behavior of the integral and we obtain:
\begin{equation}
\int^E d\tilde{E}\frac{e^{\beta_H \tilde{E}}}{\tilde{E}^{D/2+1}} \approx \frac{e^{\beta_H E}}{\beta_H E^{D/2+1}}.
\end{equation}
Note that one still has a choice here: one can choose either one string state (thus considering the single-string energy-momentum tensor as was done in \cite{Horowitz:1997jc}) or the entire string gas. The only difference is in what we use for the density of states at energy $E$. The dominant (at large $E$) part of both of these is actually the same (up to irrelevant prefactors or factors of $E$ in the denominator for $D=0$). The interested reader is advised to take a closer look at section 3 of \cite{Bowick:1989us}, where the link between the single-string density of states and the multi-string density of states is made explicit. Moreover the argument they presented in making the link between $\rho_{\text{single}}$ and $\rho_{\text{multi}}$ is independent of the background spacetime and applies equally well to a general curved (stationary) spacetime. Further discussions (in flat space) are given in \cite{Barbon:2004dd}\cite{Deo:1989bv}. From the above calculation we see that both yield the same result:\footnote{Note that the precise proportionality constant in (\ref{doss}) depends on the background fields. But taking the functional derivative with respect to the metric of this term is always subdominant in the large $E$ limit. Hence, in the dominant contribution, the proportionality constant cancels immediately between numerator and denominator of (\ref{micro}).} 
\begin{equation}
\label{cancell}
\left\langle T^{\mu\nu}_{e}\right\rangle_E \approx \frac{2}{\sqrt{G}}e^{-\beta_H E}\frac{\delta}{\delta G_{\mu\nu}}\left(\frac{e^{\beta_H E}}{\beta_H}\right).
\end{equation}
Note that this is irrespective of the value of $D$, the number of non-compact dimensions. 
We rewrite this as
\begin{equation}
\label{equ1}
\left\langle T^{\mu\nu}_{e}\right\rangle_E \approx \frac{1}{\beta_H^{2}\sqrt{G}}\frac{\delta \beta_H^2}{\delta G_{\mu\nu}}\left( E - \frac{1}{\beta_H}\right) \approx \frac{E}{\beta_H^{2}\sqrt{G}}\frac{\delta \beta_H^2}{\delta G_{\mu\nu}}
\end{equation}
where we consider energies $E \gg \frac{1}{\beta_H}$, which is the large $E$ criterion alluded to at the start of this subsection. \\

\begin{figure}[h]
\centering
\includegraphics[width=0.5\textwidth]{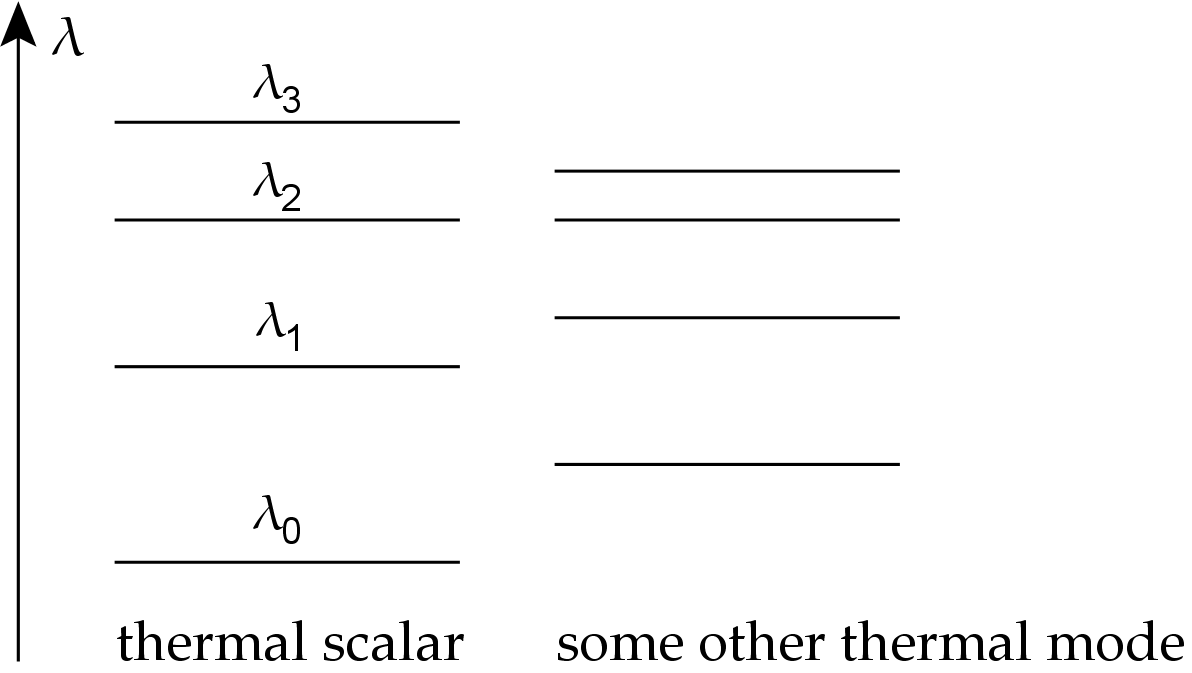}
\caption{Eigenvalue spectrum of the thermal scalar whose lowest eigenvalue $\lambda_0$ gives the dominant Hagedorn behavior of the thermal gas of strings. Higher eigenvalues are subdominant, even to some lower eigenvalues of other thermal modes (such as for instance the twice wound string state).}
\label{spectrum}
\end{figure}

\noindent At this point, it is instructive to recall some of the salient properties of the thermal scalar. This state is a singly wound state (around the thermal direction) on the thermal manifold that becomes massless precisely at $\beta = \beta_H$. At higher temperatures it becomes tachyonic. Just as above, we only consider the non-interacting gas so we only need the propagator part of the thermal scalar and no interactions with either itself or with other fluctuations. It is this propagator part that fully determines the Hagedorn temperature, by considering the eigenvalue problem associated to it. More in detail, suppose we are considering the thermal scalar partition function:
\begin{equation}
Z_{\text{th.sc.}} = \int \left[\mathcal{D}\phi\right]e^{-S_{\text{th.sc.}}},
\end{equation}
where $S_{\text{th.sc.}}$ contains only the quadratic parts of the field theory action of the thermal scalar. This action can be rewritten (after integration by parts) in the form:
\begin{equation}
\label{operatorO}
S_{\text{th.sc.}} \sim \int dV e^{-2\Phi}\sqrt{G}T^* \hat{\mathcal{O}}T
\end{equation}
for some quadratic differential operator $\hat{\mathcal{O}}$. Then construct a complete set of eigenfunctions of $\hat{\mathcal{O}}$ as $\hat{\mathcal{O}}T_n = \lambda_n T_n$ (normalized in the canonical way). Then it is immediate that 
\begin{equation}
Z_{\text{th.sc.}} = \text{det}^{-1}\hat{\mathcal{O}},
\end{equation}
or in terms of the free energy $\beta F \approx - \text{ln} Z_{\text{th.sc.}}$:
\begin{equation}
\beta F \approx \text{Tr}\text{ln}\hat{\mathcal{O}}.
\end{equation}
The approximation arises in the above formulas because we are near the Hagedorn temperature where the full thermal ensemble can be approximated by only the thermal scalar. \\
For instance for a discrete spectrum of $\hat{\mathcal{O}}$, we obtain
\begin{equation}
\beta F \approx \sum_n\text{ln}\lambda_n
\end{equation}
and it is clear that if an eigenvalue approaches zero from above, it will dominate the free energy. This is the mode of the thermal scalar that is the important one for the dominant near-critical thermodynamics. This point is worth emphasizing in a slightly different way: string theory reduces near the Hagedorn temperature to the most dominant mode of the thermal scalar. Higher modes of the thermal scalar field theory are subdominant and can be just as subdominant as modes from thermal fields other than the thermal scalar. A cartoon of this (for a discrete spectrum) is shown in figure \ref{spectrum}.
For the remainder of this chapter, our focus will be only on this lowest eigenmode and the dominant thermodynamic behavior it entails. \\

\noindent Let us remark that this formula can be rewritten using the Schwinger proper time trick as
\begin{equation}
\label{heatk}
\beta F = -\int_0^{+\infty}\frac{ds}{s}\text{Tr}e^{-s\hat{\mathcal{O}}}
\end{equation}
and it is this formula that is directly related to the worldsheet evaluation of the free energy by the identification $s=\tau_2$, the imaginary part of the torus modulus, as we made explicit in chapter \ref{chth}. \\

\noindent The thermal scalar action determines the critical temperature $\beta_H$ by the condition that for this value of the temperature, the lowest eigenvalue $\lambda_0$ of the associated operator $\hat{\mathcal{O}}$ is precisely zero. The eigenmodes and -values of this operator are denoted as $T_n$ and $\lambda_n$.\footnote{At this point, we include the additional assumption that the spectrum of the thermal scalar wave equation is discrete. If this is not the case, one should integrate out the continuous quantum numbers and consider the resulting operator. The meaning of this statement can be given in the heat kernel language of equation (\ref{heatk}). Suppose we are interested in some operator $\hat{\mathcal{O}}$ that includes a continuous quantum number $k$. Then we can write schematically
\begin{equation}
\text{Tr}e^{-s \hat{\mathcal{O}}} = \sum_n \int dk \rho_n(k) e^{-s\lambda_n(k)} = \sum_n f(n,s) e^{-s \tilde{\lambda_n}}
\end{equation}
where in the final line we integrated over $k$, the continuous quantum number and we extracted the dominant large $s$ exponential factor. The reasoning then applies if the lowest eigenvalue is left unchanged: $\lambda_0 = \tilde{\lambda_0}$, and this for all metrics infinitesimally displaced from the metric of interest.}
A concrete form of $\hat{\mathcal{O}}$ will be given further on for the type II superstring, but for now we remain more general. \\
So we have $\lambda_0(G_{\mu\nu}, \Phi, \beta_H) = 0$, where this equality holds for all backgrounds by definition. $\beta_H$ is on its own still a function of the background fields. Taking the total functional derivative of 
\begin{equation}
\lambda_0(G_{\mu\nu}, \Phi, \beta_H(G_{\mu\nu}, \Phi)) = 0
\end{equation}
with respect to the metric, we obtain
\begin{equation}
\left.\frac{\delta \lambda_0}{\delta G_{\mu\nu}(\mathbf{x})}\right|_{\text{Total}} = 0 = \frac{\delta \lambda_0}{\delta G_{\mu\nu}(\mathbf{x})} + \left.\frac{\partial \lambda_0}{\partial \beta^2}\right|_{\beta = \beta_H}\frac{\delta \beta_H^2}{\delta G_{\mu\nu}(\mathbf{x})},
\end{equation}
which can be rewritten as
\begin{equation}
\label{chain}
\frac{\delta \beta_H^2}{\delta G_{\mu\nu}(\mathbf{x})} = - \frac{\frac{\delta \lambda_0}{\delta G_{\mu\nu}(\mathbf{x})}}{\left.\frac{\partial \lambda_0}{\partial \beta^2}\right|_{\beta = \beta_H}}.
\end{equation}
The variations of the eigenvalues are given by definition as:
\begin{equation}
\frac{\delta \lambda_n}{\delta G_{\mu\nu}(\mathbf{y})} = \frac{\delta}{\delta G_{\mu\nu}(\mathbf{y})}\int d\mathbf{x}e^{-2\Phi}\sqrt{G}T_n^* \hat{\mathcal{O}} T_n = \frac{\delta}{\delta G_{\mu\nu}(\mathbf{y})} S_{\text{th.sc.}}\left[T_n\right]
\end{equation}
where the $n^{th}$ eigenmode $T_n$ (of the thermal scalar eigenvalue equation) is used.\footnote{Note that this functional derivative is only with respect to the metric, the parameter $\beta_H$ is left unchanged. In writing these expressions, we have also assumed the eigenfunctions to be normalized as 
\begin{equation}
\int d\mathbf{x}e^{-2\Phi}\sqrt{G}T_n^* T_n =1,
\end{equation}
although this is not strictly required.} In the final line we used a partial integration to relate the operator $\hat{\mathcal{O}}$ (being the inverse propagator) to the action $S_{\text{th.sc.}}$ evaluated on the $n^{th}$ eigenfunction. \\ 
Since the denominator in (\ref{chain}) is not $\mathbf{x}$-dependent, after using these results in equation (\ref{equ1}) we arrive at the conclusion that
\begin{equation}
\left\langle T^{\mu\nu}_{e}\right\rangle_E \propto \left.T^{\mu\nu}_{\text{th.sc.}}\right|_{\text{on-shell}},
\end{equation}
where the meaning of the subscript \emph{on-shell} is that the stress tensor of the thermal scalar should be evaluated on the zero-mode wavefunction solution. This zero-mode wavefunction coincides with the classical field satisfying the Euler-Lagrange equations. Hence the \emph{classical} thermal scalar energy-momentum tensor determines the time-averaged high energy stress-momentum tensor of Lorentzian string states.

\subsection{Explicit form for type II superstrings}
One can be more explicit if we know the precise form of (the non-interacting part of) the thermal scalar action (including all its $\alpha'$-corrections). For the bosonic string, we discussed in several specific examples \cite{Mertens:2013zya}\cite{Mertens:2014nca} that this action gets corrections and we do not know for a general background what these look like. For type II superstrings however, the corrections appear not to be present. We presented a heuristic argument in favor of this in subsection \ref{argu}. We hence focus on type II superstrings and, assuming there are no corrections for any background, we can compute the above energy-momentum tensor explicitly.\\
In particular the thermal scalar action for type II superstrings is given by
\begin{equation}
\label{thaction}
S_{\text{th.sc.}} = \int d^{d-1}x e^{-2\Phi}\sqrt{G}\left[G^{ij}\partial_i T \partial_j T^* + \frac{\beta^2G_{\tau\tau}-\beta_{H,\text{flat}}^2}{4\pi^2\alpha'^2}TT^*\right],
\end{equation}
where $G$ denotes the determinant of the metric (including $G_{\tau\tau}$) and $d-1$ the total number of spatial dimensions. Note that an overall factor of $\beta$ arising from the integration over $\tau$ has been dropped, and in fact we define the thermal scalar action with the above (canonical) normalization. This will prove to be the correct approach. Using the above action, one obtains
\begin{align}
\label{thenergy}
\frac{e^{2\Phi}}{\sqrt{G}}\frac{\delta \lambda_0}{\delta G_{\tau\tau}} &= \frac{\beta^2 TT^*}{4\pi^2\alpha'^2} + \frac{1}{2}G^{\tau\tau}\left[G^{ij}\partial_i T \partial_j T^* + \frac{\beta^2G_{\tau\tau}-\beta_{H,\text{flat}}^2}{4\pi^2\alpha'^2}TT^*\right], \\
\frac{e^{2\Phi}}{\sqrt{G}}\frac{\delta \lambda_0}{\delta G_{ij}} &= -\frac{\nabla^{i}T^* \nabla^{j} T + \nabla^{j}T^* \nabla^{i} T}{2} + \frac{1}{2}G^{ij}\left[G^{kl}\partial_k T \partial_l T^* + \frac{\beta^2G_{\tau\tau}-\beta_{H,\text{flat}}^2}{4\pi^2\alpha'^2}TT^*\right], \\
\frac{\partial \lambda_0}{\partial \beta^2} &= \int d^{d}x \sqrt{G} e^{-2\Phi} G_{\tau\tau}\frac{TT^*}{4\pi^2\alpha'^2}. 
\end{align}
The final line above is a fixed number, let us call it $N$. Integrating the time component of this energy-momentum tensor yields\footnote{
From here on, we focus on constant dilaton backgrounds, so $\Phi = \Phi_0$. In writing these expressions, we used the fact that the thermal scalar is a zero-mode: its on-shell action vanishes.} 
\begin{align}
-\int \left\langle T_{\tau,e}^{\tau}\right\rangle_E \sqrt{G} dV &= -\int G_{\tau\tau}\left\langle T^{\tau\tau}_{e}\right\rangle_E \sqrt{G} dV=  \frac{E}{\beta_H^{2}} \int \frac{\frac{G_{\tau\tau}}{\sqrt{G}}\frac{\delta \lambda_0}{\delta G_{\tau\tau}}}{\left.\frac{\partial \lambda_0}{\partial \beta^2}\right|_{\beta = \beta_H}} \sqrt{G} dV \\ 
&= \frac{E}{\beta_H^{2}} \frac{\int dV \sqrt{G} G_{\tau\tau}\frac{\beta_H^2 TT^*}{4\pi^2\alpha'^2}}{\left.\frac{\partial \lambda_0}{\partial \beta^2}\right|_{\beta = \beta_H}} = E, 
\end{align}
which means $-\left\langle T_{\tau,e}^{\tau}\right\rangle$ can be interpreted as the (Lorentzian signature) energy density. The precise equality between the quantum high-energy stress tensor and the classical thermal scalar stress tensor then becomes
\begin{equation}
\label{mmicroo}
\left\langle T^{\mu\nu}_{e}\right\rangle_E \approx \left.\frac{E}{2N \beta_H^2} T^{\mu\nu}_{\text{th.sc.}}\right|_{\text{on-shell}},
\end{equation}
where the classical on-shell thermal scalar energy-momentum tensor is computed using the thermal scalar wavefunction with an arbitrary normalization in principle,\footnote{The normalization cancels between the numerator $\left.T^{\mu\nu}_{\text{th.sc.}}\right|_{\text{on-shell}}$ and the denominator $N$.} although we shall assume it is normalized in the standard fashion.

\section{Extension to spacetimes with $G_{\tau i} \neq 0$}
\label{nonstat}
In the previous section we focused on $G_{\tau i} = 0$. There are two cases when one wants to extend this assumption. Firstly, if one wishes to know $T^{\tau i}$, we need to compute $\frac{\delta S_{\text{th.sc.}}}{\delta G_{\tau i}}$. Secondly, a stationary (non-static) spacetime (such as a string-corrected Kerr black hole) contains $G_{\tau i}$ metric components. Note that even the metric of static spacetimes may be rewritten in the reference frame of a moving observer which then possibly includes non-zero $G_{\tau i}$ components (such as flat space as seen by a rotating observer). \\
The entire derivation presented in the previous section does not explicitly use the fact that $G_{\tau i} = 0$, except in writing down the form of the thermal scalar action given in (\ref{thaction}). In a more general spacetime, the action to be used is:
\begin{equation}
S_{\text{th.sc.}} = \int d^{d-1}x \sqrt{G}e^{-2\Phi}\left(\bar{G}^{ij}\partial_i T \partial_j T^{*} + \frac{\beta^2 G_{\tau\tau} - \beta_{H,\text{flat}}^2}{4\pi^2\alpha'^2}TT^*\right),
\end{equation}
where $\bar{G}^{ij}$ is the matrix inverse of $G_{ij} - \frac{G_{\tau i}G_{\tau j}}{G_{\tau\tau}}$ as we derived in detail in chapter \ref{chth}. 
In fact, a simple calculation shows that $\bar{G}^{ij} = G^{ij}$. It is interesting to note that the above effective spatial geometry (encoded in $\bar{G}^{ij}$) is precisely the spatial metric one would obtain when operationally defining distance in a stationary (non-static) gravitational field.\footnote{See for instance section \S 84 in the second volume of Landau and Lifshitz (Classical Field Theory).} The random walking particle (as described by the thermal scalar) is sensitive to this effective metric. It appears that T-duality in the (Euclidean) time direction encodes this information. \\
The only thing left to do is to compute $\frac{\delta \lambda_0}{\delta G_{\tau i}}$. For this, we need to determine $\frac{\delta \bar{G}^{ij}}{\delta G_{\tau k}}$. 
A simple calculation shows that\footnote{To compute this, one needs to be careful about varying with respect to a symmetric tensor: the factor of $2$ in the denominator is easily missed.}
\begin{equation}
\label{c1}
\frac{\delta \bar{G}^{ij}}{\delta G_{\tau k}} = - \frac{G^{ik} G^{\tau j} + (i \leftrightarrow j)}{2}.
\end{equation}
Using this result we obtain
\begin{align}
\frac{e^{2\Phi}}{\sqrt{G}}\frac{\delta \lambda_0}{\delta G_{\tau i}} &= -\left( G^{ki} G^{\tau l} + (k \leftrightarrow l)\right)\left(\frac{\partial_k T^* \partial_{l} T + \partial_{l}T^* \partial_{k} T}{4}\right) \nonumber \\
&+ \frac{1}{2}G^{\tau i}\left[G^{kl}\partial_k T \partial_l T^* + \frac{\beta^2G_{\tau\tau}-\beta_{H,\text{flat}}^2}{4\pi^2\alpha'^2}TT^*\right].
\end{align}
Importantly, if $G_{\tau i}$ vanishes for \emph{every} $i$, both terms are zero and $T^{\tau i}_{\text{th.sc.}} = 0$. This implies the highly excited string(s) does not carry any spatial momentum in a static spacetime, as we expect. \\

\noindent For completeness, we analogously obtain
\begin{equation}
\label{c2}
\frac{\delta \bar{G}^{ij}}{\delta G_{\tau \tau}} = - G^{i\tau} G^{j \tau},
\end{equation}
which should be used when computing $T^{\tau\tau}$ in a non-static, stationary spacetime (such as a Kerr black hole).\footnote{We remark that for Kerr black holes, one runs into trouble with the ergoregion where the asymptotically timelike Killing vector used to define thermodynamics becomes spacelike. We will not pursue a concrete example of stationary spacetimes in this chapter.} \\

\noindent A consistency check can be performed by once more computing the total energy as a volume integral of a suitable stress tensor component. The computations are a bit more involved and we present the calculations as supplementary material in section \ref{stationary}. Suffice it to say that the results are consistent.

\section{Thermal average}
\label{thermaver}
As a slight modification, we study the thermal average of the time-averaged Euclidean energy-momentum tensor of a string gas:
\begin{equation}
\label{starting}
\left\langle T^{\mu\nu}_{e}\right\rangle_{\text{thermal}} = \frac{\text{Tr}\left(e^{-\beta H_l}T^{\mu\nu}_{e}\right)}{\text{Tr}\left(e^{-\beta H_l}\right)}
\end{equation}
in the near-Hagedorn regime $\beta \approx \beta_H$. We want to remark that it has been observed in the past that at high temperature, the canonical ensemble suffers from large fluctuations and one should resort to the microcanonical picture instead \cite{Mitchell:1987hr}\cite{Mitchell:1987th}\cite{Brandenberger:1988aj}\cite{Bowick:1989us}\cite{Deo:1989bv}\cite{Kutasov:2000jp}. Keeping in mind this issue, it is nevertheless instructive to proceed because we will see how the stress tensor of the thermal scalar contains information on the above expectation value. Moreover, at least naively, for theories with holographic duals one would expect the dual field theory to have a well-defined canonical ensemble, or alternatively one could be interested in more deeply analyzing how the discrepancy between both ensembles is realized both in the bulk and on the holographic boundary \cite{Berkooz:2000mz}. Hence we will proceed in the canonical ensemble.
Using (\ref{second2}), we can write
\begin{equation}
\label{starting2}
\frac{\text{Tr}\left(e^{-\beta H_l}T^{\mu\nu}_{e}\right)}{\text{Tr}\left(e^{-\beta H_l}\right)} = \frac{2}{\beta \sqrt{G} Z_{\text{mult}}}\frac{\delta }{\delta G_{\mu\nu}} Z_{\text{mult}},
\end{equation}
where $Z_{\text{mult}}$ is the multi-string partition function. Using $Z_{\text{mult}} = e^{-\beta F}$, we can also write
\begin{equation}
\label{expre}
\left\langle T^{\mu\nu}_{e}\right\rangle_{\text{thermal}} = - \frac{2}{\sqrt{G}}\frac{\delta F}{\delta G_{\mu\nu}}.
\end{equation}
In QFT, the expectation value of the energy-momentum tensor is badly divergent in curved spacetimes. The above formula relates this divergence to that of the free energy. But string theory does not contain the UV divergence of field theories. Therefore the stress tensor does not diverge in the UV.\\
In the near-Hagedorn limit, the thermal scalar provides the dominant contribution to the partition function in the sense that:
\begin{equation}
Z_{\text{mult}} \approx Z_{\text{th.sc.}} = \int \left[\mathcal{D}\phi\right]e^{-\beta S_{\text{th.sc.}}},
\end{equation}
where the thermal scalar field theory is used. One then obtains 
\begin{align}
\label{ththerm}
\left\langle T^{\mu\nu}_{e}\right\rangle_{\text{thermal}} &\approx \frac{1}{\beta Z_{\text{th.sc.}}} \frac{2}{\sqrt{G}}\frac{\delta }{\delta G_{\mu\nu}}\int \left[\mathcal{D}\phi\right]e^{-\beta S_{\text{th.sc.}}} \\
&= \frac{1}{ Z_{\text{th.sc.}}}\int \left[\mathcal{D}\phi\right]\left(- \frac{2}{\sqrt{G}}\frac{\delta S_{\text{th.sc.}}}{\delta G_{\mu\nu}}\right)e^{-\beta S_{\text{th.sc.}}},
\end{align}
and indeed we see here that the normalization of the thermal scalar action we chose before (extracting a factor of $\beta$ from the action) is consistent. \\
Thus the energy-momentum tensor of the near-Hagedorn string gas (\ref{starting}) can be computed by looking at the one-loop expectation value of the thermal scalar energy-momentum tensor (\ref{ththerm}). 
From this procedure we see explicitly what the stress tensor of the thermal scalar encodes: it gives information on the time-averaged stress tensor of the near-Hagedorn string gas. 
\noindent We can proceed as follows. Let us start with (\ref{expre}). Assuming a density of high-energy single-string states as:
\begin{equation}
\omega(E) \propto \frac{e^{\beta_H E}}{E^{D/2+1}}
\end{equation}
we have
\begin{equation} 
\label{asyformF}
Z_1 = -\beta F = \text{Tr}_{\text{single}}\left(e^{-\beta H_l}\right) = \int^{+\infty}dE \omega(E)e^{-\beta E} \approx -C(\beta - \beta_H)^{D/2}\ln(\beta-\beta_H),
\end{equation}
where we trace over the single-string Hilbert space only. In writing this, we have used the Maxwell-Boltzmann approximation relating the single-string partition function $Z_1$ to the multi-string partition function $Z_{\text{mult}}$ through exponentiation. This equality is valid as soon as the singly wound string dominates on the thermal manifold, meaning for $\beta$ sufficiently close to $\beta_H$. The temperature-independent constant $C$ is present only for $D>0$ and contains the factors originating from integrating out continuous quantum numbers in $Z_1$. This expression holds for $D$ even. If $D$ is odd, the final logarithm should be deleted. The most singular term then yields\footnote{The constant $C$ does depend on the geometry, but the term $\frac{\delta C}{\delta G_{\mu\nu}} (\beta - \beta_H)^{D/2}\ln(\beta-\beta_H)$ is subdominant since $C$ does not depend on $\beta$.} 
\begin{equation}
\frac{\delta }{\delta G_{\mu\nu}} C(\beta - \beta_H)^{D/2}\ln(\beta-\beta_H) = -C \frac{D}{2}(\beta - \beta_H)^{D/2-1}\ln(\beta-\beta_H)\frac{\delta \beta_H}{\delta G_{\mu\nu}}.
\end{equation}
Collecting the results, we obtain
\begin{equation}
\left\langle T^{\mu\nu}_{e}\right\rangle_{\text{thermal}} \approx -\frac{C}{N}\frac{2}{\sqrt{G}}\left.\frac{\delta S_{\text{th.sc.}}}{\delta G_{\mu\nu}}\right|_{\text{on-shell}}\frac{(\beta - \beta_H)^{D/2-1}\ln(\beta-\beta_H)}{2\beta\beta_H}\frac{D}{2},
\end{equation}
which can be rewritten as
\begin{equation}
\label{mainresult}
\boxed{
\left\langle T^{\mu\nu}_{e}\right\rangle_{\text{thermal}} \approx \left. \frac{C}{N} T^{\mu\nu}_{\text{th.sc.}}\right|_{\text{on-shell}}\frac{(\beta - \beta_H)^{D/2-1}\ln(\beta-\beta_H)}{2\beta_H^2}\frac{D}{2}}
\end{equation}
and we see that we obtain the \emph{classical} thermal scalar energy-momentum tensor evaluated at $\beta = \beta_H$, multiplied by temperature-dependent factors that diverge (or are non-analytic) at $\beta = \beta_H$. The degree of divergence is related to that of the free energy by one $\beta$ derivative. Again the logarithm should be dropped when $D$ is odd.\footnote{For $D=0$ an exception occurs. Then the logarithm should be dropped as well and the $\frac{D}{2}$ factor should be deleted.}\\
As already noted above, this formula is derived in the canonical ensemble near the Hagedorn temperature. The large energy fluctuations of the canonical ensemble invalidate the equality between the microcanonical and canonical ensembles. This can be seen explicitly here by comparing the above result with formulas given in \cite{Brandenberger:2006xi}\cite{Brandenberger:2006vv} where a microcanonical approach is followed to determine the energy.\\
An interesting consistency check can be performed by integrating $G_{\mu \tau}\left\langle T^{\mu \tau}_{e}\right\rangle_{\text{thermal}}$ over the entire space. This yields\footnote{This check explicitly demonstrates that the result is consistent in the canonical ensemble, whereas a microcanonical approach would yield a different result. Note that
\begin{equation}
\int dV \sqrt{G} \left.{T_\tau^\tau}_{\text{th.sc.}}\right|_{\text{on-shell}} = -2 \beta_H^2 N.
\end{equation}
}
\begin{equation}
-\int \left\langle T^{\tau}_{\tau}\right\rangle_{\text{thermal}} \sqrt{G} dV = C\frac{D}{2}(\beta - \beta_H)^{D/2-1}\ln(\beta-\beta_H)
\end{equation}
which equals the internal energy $E$ of the thermodynamic system. This energy can also be determined as
\begin{equation}
\label{cannen}
E = \partial_{\beta}(\beta F) = \partial_{\beta} \left(C(\beta - \beta_H)^{D/2}\ln(\beta-\beta_H)\right) \approx C\frac{D}{2}(\beta - \beta_H)^{D/2-1}\ln(\beta-\beta_H)
\end{equation}
as it should be.\footnote{Note that in principle the above energy is formally negative below the Hagedorn temperature (except for $D=0$ \cite{Brandenberger:1988aj}), which is impossible. This is a typical feature of high-temperature string thermodynamics and points towards the fact that only including the most dominant contribution is not enough: one should include also subdominant contributions \cite{Deo:1988jj}\cite{Kutasov:2000jp}. Moreover, one should resort to the microcanonical ensemble instead.} \\
We also remark that substituting the canonical energy-temperature relation (\ref{cannen}) in equation (\ref{mainresult}) reproduces the microcanonical expression (\ref{mmicroo}). \\
We conclude that, much like the general expression for the free energy (\ref{asyformF}), the energy-momentum tensor can also be written down in a general background near its Hagedorn temperature (\ref{mainresult}). \\

\noindent The results of this section utilize the canonical ensemble, whereas the results in section \ref{HP} use the microcanonical ensemble. Qualitatively, both methods agree on the spatial distribution of the stress-energy.

\section{Extension to the String Charge}
\label{stringcharge}
We are interested in writing down an analogous formula for the string charge:
\begin{equation}
J^{\mu\nu}_{e} = - \frac{2}{\sqrt{G}}\frac{\delta S_e}{\delta B_{\mu\nu}}.
\end{equation}
Let us consider the high energy-averaged string charge and again focus on time-independent backgrounds: $\partial_\tau B_{\mu\nu} =0$, although we do allow temporal indices for $B$ here. \\
A closer look at the arguments presented in section \ref{HP} shows that the entire derivation can be copied to handle this case as well.
Hence upon replacing $G_{\mu\nu}$ with $B_{\mu\nu}$, we obtain an expression for the high-energy averaged string charge tensor as
\begin{equation}
\left\langle J^{\mu\nu}_{e}\right\rangle_E \approx  -\frac{E}{N\beta_H^2}\frac{1}{\sqrt{G}}\left.\frac{\delta S_{\text{th.sc.}}}{\delta B_{\mu\nu}}\right|_{\text{on-shell}}.
\end{equation} 
The expectation value of the string charge when averaged over high energy states is given by the classical charge tensor of the thermal scalar. \\
The thermal scalar itself, being a complex scalar field, has an additional $U(1)$ symmetry. It turns out that this $U(1)$ charge symmetry and the string charge of the thermal scalar are actually the same as we now show.\footnote{In fact this is readily shown as follows. Winding strings are charged under $B_{\tau i}$, hence it can be coupled to it through a conserved current. It is then no real surprise to find
\begin{equation}
\frac{\delta S_{\text{th.sc.}}}{\delta B_{\tau i}} \sim J^i
\end{equation}
with $J^i$ the conserved $U(1)$ current of the complex thermal scalar. It is interesting however to see it more explicitly.} The thermal scalar action is in general given by \cite{Mertens:2013pza}:
\begin{align}
\label{wind}
&S_{\text{th.sc.}} \sim \int d^{d-1}x \sqrt{G}e^{-2\Phi} \nonumber \\
&\times\left(G'^{ij}\partial_{i}T\partial_{j}T^{*}+\frac{w^2\beta^2G'^{\tau\tau}}{4\pi^2\alpha'^2}TT^{*}+ G'^{\tau i}\frac{iw\beta}{2\pi\alpha'}\left(T\partial_{i}T^{*}- T^{*}\partial_{i}T\right)+ m^2TT^{*}\right),
\end{align}
where primes denote T-dual quantities. This action was written down in general, but we focus here on trivial dilatons: $\Phi = \Phi_0$. The $U(1)$ symmetry of this action leads to the following Noether current:
\begin{equation}
\label{noether}
J^{k} \propto G^{'ik}\left(T^{*}\partial_i T - T \partial_i T^{*}\right) + \frac{iw \beta}{\pi\alpha'}G^{'\tau k}TT^{*}.
\end{equation}
Explicitly, one can show that\footnote{These formulas hold also when $G_{\tau i} \neq 0$.}
\begin{align}
\frac{\partial G^{'\tau\tau}}{\partial B_{\tau k}} &= -2G^{'\tau k}, \\
\frac{\partial G^{'\tau i}}{\partial B_{\tau k}} &= -G^{'ik}, \\
\frac{\partial G^{'ij}}{\partial B_{\tau k}} &= 0.
\end{align}
Using these formulas to evaluate the expression 
\begin{equation}
\left.J^{\mu\nu}_{\text{th.sc.}}\right|_{\text{on-shell}} = -\frac{2}{\sqrt{G}}\left.\frac{\delta S_{\text{th.sc.}}}{\delta B_{\mu\nu}}\right|_{\text{on-shell}},
\end{equation}
we obtain that $\left.J^{ij}_{\text{th.sc.}}\right|_{\text{on-shell}} = 0$ and $\left.J^{\tau k}_{\text{th.sc.}}\right|_{\text{on-shell}} \propto \left.J^{k}\right|_{\text{on-shell}}$, which shows the equality between the $U(1)$ Noether current and the string charge of the thermal scalar. In terms of the highly excited string gas we are interested in, this leads to
\begin{align}
\left\langle J^{ij}_{e}\right\rangle_E &= 0 ,\\
\left\langle J^{\tau k}_{e}\right\rangle_E &\propto \left.J^{k}\right|_{\text{on-shell}}.
\end{align}
The Noether current of the thermal scalar determines the time-averaged, high-energy-averaged, string charge of a long string in a non-trivial background.\\

\noindent The surprising part of this analysis is that the high energy string average is sensitive to $B_{\mu\nu}$ even though each individual state (except for spatially wound states) is not (at the non-interacting level we consider here). \\

\noindent Finally we note that one can readily study this string charge also in the canonical ensemble (much like what was done in the previous section). One readily finds completely analogous expressions.

\section{Some Correlators}
\label{correl}
In this section we will be interested in studying a specific class of stress tensor correlators. The computations are more natural to perform in the canonical ensemble and we focus on this approach in this section. The canonical picture will allow us to make some specific clarifications on the factorization of the correlators as will be discussed further on.

\subsection{Correlators in the canonical ensemble}
\label{cancor}
We extend the analysis of section \ref{thermaver} to an energy-momentum correlator in the canonical ensemble:\footnote{Notice that a subscript $\beta$ has been written, denoting the averaging over Euclidean time as was considered earlier in equation (\ref{betaav}). Also, in principle this product of operators should be time-ordered, where the time-dependence of both of these operators is inside the temporal average. To avoid any more cluttering of the equations, we will not write this.}
\begin{align}
\left\langle T^{\mu\nu}_{e,\beta}(\mathbf{x}) T^{\rho\sigma}_{e,\beta}(\mathbf{y})\right\rangle_{\text{thermal}}.
\end{align}
Both energy-momentum tensors are individually time-averaged and evaluated at a spatial point $\mathbf{x}$ (or $\mathbf{y}$). Of course, such a correlator is not what we are interested in in the end, but it is at first sight the only type of correlator the methods described above can handle. This leads to the result that this correlator in the canonical ensemble is given by the same correlator of only the thermal scalar field theory ($Z_{\text{mult}} \approx Z_{\text{th.sc.}}$):
\begin{align}
\left\langle T^{\mu\nu}_{e,\beta}(\mathbf{x}) T^{\rho\sigma}_{e,\beta}(\mathbf{y})\right\rangle_{\text{thermal}} &\approx \frac{1}{\beta^2 Z_{\text{th.sc.}}} \frac{4}{\sqrt{G(\mathbf{x})}\sqrt{G(\mathbf{y})}}\frac{\delta }{\delta G_{\mu\nu}(\mathbf{x})}\frac{\delta }{\delta G_{\rho\sigma}(\mathbf{y})}\int \left[\mathcal{D}\phi\right]e^{-\beta S_{\text{th.sc.}}}. 
\end{align}
Throughout this section, we neglect contact terms. It seems hard to try to generalize this computation to more general types of correlators, since from the thermal scalar point of view, we are using the only way we can think of to compute its energy-momentum correlators. \\
To proceed, the same strategy as utilized before in section \ref{thermaver} can be followed. We write
\begin{equation}
\label{eq}
\frac{\frac{\delta^2 Z_{\text{th.sc.}}}{\delta G_{\mu\nu}\delta G_{\rho \sigma}}}{Z_{\text{th.sc.}}} = \frac{\delta^2 Z_1 }{\delta G_{\mu\nu} \delta G_{\rho\sigma}} + \frac{\delta Z_1}{\delta G_{\mu\nu}} \frac{\delta Z_1}{\delta G_{\rho\sigma}}.
\end{equation}
The second term represents the disconnected part of the correlator where the two stress tensors factorize. With the limiting behavior (\ref{asyformF}) in mind, we can compare the behaviors of both of these terms. We will focus here on fully compact spactimes (since these are the relevant ones for thermodynamics). For fully compact spaces, one has $Z_1 = - \ln(\beta-\beta_H)$ and one readily shows that both terms in (\ref{eq}) are equal. This then leads to 
\begin{equation}
\label{correla}
\left\langle T^{\mu\nu}_{e,\beta}(\mathbf{x}) T^{\rho\sigma}_{e,\beta}(\mathbf{y})\right\rangle \approx \frac{1}{2N^2\beta_H^{4}}\frac{1}{(\beta-\beta_H)^2}\left.T^{\mu\nu}_{\text{th.sc.}}(\mathbf{x})\right|_{\text{on-shell}}\left.T^{\rho\sigma}_{\text{th.sc.}}(\mathbf{y})\right|_{\text{on-shell}}.
\end{equation}
Such equalities can be generalized to more than two stress tensors and one finds in the near-Hagedorn limit that in fact all terms are proportional such that
\begin{equation}
\frac{\frac{\delta^n Z_{\text{th.sc.}}}{\delta G_{\mu\nu}\delta G_{\rho \sigma}\hdots}}{Z_{\text{th.sc.}}} = n! \frac{\delta Z_1}{\delta G_{\mu\nu}}\frac{\delta Z_1}{\delta G_{\rho\sigma}} \hdots.
\end{equation}
The prefactors of the analog of (\ref{eq}) in the expansion correspond precisely to the number of diagrams one can draw with a fixed number of disconnected components (each of which connecting a fixed number of points), as we will analyze more deeply in what follows. The main message here is that for a fully compact space, the correlator looks factorized because both connected and disconnected contributions become equal in the Hagedorn limit. \\
As a byproduct, the expectation value of $n$ stress tensors is given by:
\begin{equation}
\left\langle T^{\mu\nu}_{e,\beta} T^{\rho\sigma}_{e,\beta} \hdots \right\rangle \approx \frac{n!}{N^n}\frac{1}{2^n\beta_H^{2n}}\frac{1}{(\beta-\beta_H)^n}\left.T^{\mu\nu}_{\text{th.sc.}}\right|_{\text{on-shell}}\left.T^{\rho\sigma}_{\text{th.sc.}}\right|_{\text{on-shell}} \hdots 
\end{equation}
The analogous formulas for non-compact spacetimes will not be presented here.

\subsection{A puzzle on factorization}
The result that the stress tensor correlator (\ref{correla}) looks factorized may sound strange at first sight. After all, even in flat space one would expect (for the connected part) a behavior in terms of the distance between the two points. It is quite instructive to look into this matter in more detail. Actually, we do not need the added complication of the composite operators in the stress tensor to illustrate this, and so we concentrate on the free field propagator. We focus on flat space here and hope that the lesson we learn here will convince the reader that also in curved space, the near-Hagedorn limit taken above is consistent. \\
In flat $D$-dimensional Euclidean space, the Klein-Gordon propagator for a massive (complex) scalar field can be written down explicitly as
\begin{align}
\left\langle \phi(\mathbf{x}) \phi^*(\mathbf{0})\right\rangle = \frac{1}{(2\pi)^D}\int d^D \mathbf{q} \frac{e^{i \mathbf{q} \cdot \mathbf{x}}}{\mathbf{q}^2 + m^2} =  \left(\frac{1}{2\pi}\right)^{D/2}\left(\frac{m}{\left|\mathbf{x}\right|}\right)^{\frac{D-2}{2}}K_{\frac{D-2}{2}}(m\left|\mathbf{x}\right|).
\end{align}
In our case, we have $m \sim (\beta-\beta_H)^{1/2}$. We are interested in the massless limit and we only want the most dominant (non-analytic) contribution. Taylor expanding the modified Bessel function, one finds the series\footnote{Exactly the same caveats as before (below equation (\ref{asyformF})) apply here: the logarithm is only present for even $D$, except when $D=0$.}
\begin{align}
\label{seri}
&\left\langle \phi(\mathbf{x}) \phi^*(\mathbf{0})\right\rangle \approx \frac{C}{\left|\mathbf{x}\right|^{D-2}} + \sum_{i=0}^{+\infty} \frac{C_i m^{2i}}{\left|\mathbf{x}\right|^{D-2-2i}}+ C'm^{D-2}\ln m + \hdots
\end{align}
The first term is mass-independent (it becomes the UV cut-off in the coincident limit for a massless field), the second sum is regular as $m\to0$ and does not display non-analytic behavior. The dots denote terms of higher power in $m$ that result from multiplication of the third term by positive powers of $m$\footnote{And also a series in $m$ multiplied by $\text{ln}(\left|\mathbf{x}\right|)$.} and hence are less non-analytic than the third term. This term is hence the one we are interested in here and it does \emph{not} depend on the distance between the two points. It is straightforward to generalize this argument to the stress tensor correlator by including additional $\phi$ operator insertions and suitable derivatives. Also for the stress tensor itself, this propagator can be used (we will be more explicit in the next section). But one can already see that, upon setting $m \sim (\beta-\beta_H)^{1/2}$, the third term in this Taylor expansion has precisely the same temperature dependence as equation (\ref{mainresult}).\\

\noindent As as side remark, we note that in \cite{Horowitz:1997jc}, the authors write down the following correlator
\begin{equation}
\left\langle \phi\phi^*(\mathbf{x}) \phi\phi^*(\mathbf{0})\right\rangle \sim e^{-2m\left|\mathbf{x}\right|},
\end{equation}
which is also different from the above results: it is obtained by studying the asymptotic (large $\left|\mathbf{x}\right|$) behavior of the modified Bessel function (in the connected correlator). This equation is hence written down in the large $\left|\mathbf{x}\right|$ limit and only after that the $m\to0$ limit is taken. Our interest here lies in keeping $\left|\mathbf{x}\right|$ fixed while letting $m\to0$ (or $\beta \to \beta_H$): we focus on the regime $m\left|\mathbf{x}\right| \ll 1$. It is this change in limits that causes us to have a different final result. We further remark that this is the cause of the apparent violation of the cluster decomposition principle of the connected correlator: we are not taking the long distance limit, we are keeping the distance fixed and focus on the dominant (or non-analytic) parts of the correlator as the temperature reaches the Hagedorn temperature. To state this more explicitly, setting $m=0$ in (\ref{seri}), we obviously find
\begin{align}
\left\langle \phi(\mathbf{x}) \phi^*(\mathbf{0})\right\rangle \approx \frac{C}{\left|\mathbf{x}\right|^{D-2}},
\end{align}
where all other terms vanish. This is \emph{not} the procedure we are interested in here: we are discussing the process of letting $m$ go to zero and looking at the non-analytic behavior in $m$ during this procedure. All of this is related to the fact that, as was discussed in subsection \ref{microc}, we only focus on the lowest eigenmode of the thermal scalar and \emph{not} on the full scalar field theory to obtain the most dominant contribution to string thermodynamics in the Hagedorn regime. \\

\noindent A further way of appreciating these arguments, is to write the 2-point function (for a fully compact space) as 
\begin{equation}
\label{greenf}
\left\langle \phi(\mathbf{x}) \phi^*(\mathbf{\mathbf{y}})\right\rangle = \sum_n \frac{\psi_n(\mathbf{x})\psi_n^*(\mathbf{y})}{\lambda_n}
\end{equation}
and on taking the limit where $\lambda_0 \to 0$, this is approximated by
\begin{equation}
\left\langle \phi(\mathbf{x}) \phi^*(\mathbf{\mathbf{y}})\right\rangle \approx \frac{\psi_0(\mathbf{x})\psi_0^*(\mathbf{y})}{\lambda_0}.
\end{equation}
This makes it clear that a factorization is present. This type of argument (in terms of the Green function) will be made more explicit in the next section for the stress tensor and its correlators.
 
\section{Green function analysis}
\label{second}
In previous sections, we have found that the dominant contribution of the stress tensor in the canonical ensemble can be written in terms of the \emph{classical} stress tensor of the thermal scalar. It is instructive to study this from a Green function perspective as well. The stress tensor can be defined by a point-splitting procedure of the composite operator and related to a Green function. In this section we will analyze the stress tensor expectation value in the canonical ensemble again (using this point of view) and then continue our study of the correlators by elaborating on the comment made above around equation (\ref{greenf}). 

\subsection{Discrete spectrum}
For clarity, we focus first on purely compact spaces where the solution of $\hat{\mathcal{O}}\psi_n = \lambda_n \psi_n$ yields a discrete spectrum. Second quantization gives for the energy-momentum tensor of the thermal scalar (using point-splitting methods):
\begin{equation}
\label{operatorgf}
\left\langle \hat{T}^{ij}_{\text{th.sc.}}(\mathbf{x})\right\rangle = \lim_{\substack{\mathbf{x_1}\to \mathbf{x}\\ \mathbf{x_2}\to \mathbf{x}}} \left(\nabla_{1}^{i}\nabla_{2}^{j} + \nabla_{1}^{j}\nabla_{2}^{i} - G^{ij}\left[G^{kl}\nabla_{k}^{1}\nabla_{l}^{2} + \frac{\beta^2G_{\tau\tau}-\beta_{H,\text{flat}}^2}{4\pi^2\alpha'^2}\right]\right) G(\left.\mathbf{x_1}\right|\mathbf{x_2})
\end{equation}
where $G(\left.\mathbf{x_1}\right|\mathbf{x_2})$ is the scalar propagator of the thermal scalar given by
\begin{equation}
G(\left.\mathbf{x_1}\right|\mathbf{x_2}) = \sum_{n,m}\left\langle \mathbf{x_1}\right|\left.\psi_n\right\rangle\frac{\delta_{nm}}{\lambda_n}\left\langle \psi_{m}\right.\left|\mathbf{x_2}\right\rangle
\end{equation}
where the eigenfunctions and eigenvalues are found by solving:
\begin{equation}
\hat{\mathcal{O}}\psi_n(\mathbf{x}) = \left(-\nabla^2 - G^{ij}\frac{\partial_j G_{\tau\tau}}{G_{\tau\tau}}\partial_i + \frac{\beta^2G_{\tau\tau}-\beta_{H,\text{flat}}^2}{4\pi^2\alpha'^2} \right) \psi_n(\mathbf{x}) = \lambda_n \psi_n(\mathbf{x})
\end{equation}
and where $\nabla^2$ is the Laplacian on only the spatial submanifold. Likewise, the temporal component yields
\begin{equation}
\left\langle \hat{T}^{\tau\tau}_{\text{th.sc.}}(\mathbf{x})\right\rangle = \lim_{\substack{\mathbf{x_1}\to \mathbf{x}\\ \mathbf{x_2}\to \mathbf{x}}}\left(-\frac{\beta^2}{2\pi^2\alpha'^2} - G^{\tau\tau}\left[G^{kl}\nabla_{k}^{1}\nabla_{l}^{2} + \frac{\beta^2G_{\tau\tau}-\beta_{H,\text{flat}}^2}{4\pi^2\alpha'^2}\right]\right) G(\left.\mathbf{x_1}\right|\mathbf{x_2}).
\end{equation}
For a discrete spectrum and near the Hagedorn temperature, the propagator becomes:
\begin{equation}
G(\left.\mathbf{x_1}\right|\mathbf{x_2}) \approx \psi_0(\mathbf{x_1})\frac{1}{\lambda_0}\psi_{0}^*(\mathbf{x_2}).
\end{equation}
Since $\lambda_0 \sim \beta - \beta_H$, we obtain finally
\begin{align}
\left\langle \hat{T}^{ij}_{\text{th.sc.}}(\mathbf{x})\right\rangle &\approx \lim_{\substack{\mathbf{x_1}\to \mathbf{x}\\ \mathbf{x_2}\to \mathbf{x}}} \frac{\tilde{C}}{\beta-\beta_H}\left(\nabla_{1}^{i}\psi_0(\mathbf{x_1})\nabla_{2}^{j}\psi_{0}^*(\mathbf{x_2}) + \nabla_{1}^{j}\psi_0(\mathbf{x_1})\nabla_{2}^{i}\psi_{0}^*(\mathbf{x_2})\right. \\ \nonumber
&\left. \quad\quad\quad - G^{ij}\left[G^{kl}\nabla_{k}^{1}\psi_0(\mathbf{x_1})\nabla_{l}^{2}\psi_{0}^*(\mathbf{x_2}) + \frac{\beta^2G_{\tau\tau}-\beta_{H,\text{flat}}^2}{4\pi^2\alpha'^2}\psi_0(\mathbf{x_1})\psi_{0}^*(\mathbf{x_2})\right]\right) \\
&= \left.T^{ij}_{\text{th.sc.}}(\mathbf{x})\right|_{\text{on-shell}}\frac{\tilde{C}}{\beta-\beta_H}
\end{align}
where we obtain the \emph{classical} energy-momentum tensor of the thermal scalar. Analogously, we obtain
\begin{align}
\left\langle \hat{T}^{\tau\tau}_{\text{th.sc.}}(\mathbf{x})\right\rangle &\approx \lim_{\substack{\mathbf{x_1}\to \mathbf{x}\\ \mathbf{x_2}\to \mathbf{x}}} \frac{\tilde{C}}{\beta-\beta_H}\left( -\frac{\beta^2}{2\pi^2\alpha'^2}\psi_0(\mathbf{x_1})\psi_{0}^*(\mathbf{x_2})\right. \\ \nonumber 
&\left. \quad\quad\quad- G^{\tau\tau}\left[G^{kl}\nabla_{k}^{1}\psi_0(\mathbf{x_1})\nabla_{l}^{2}\psi_{0}^*(\mathbf{x_2}) + \frac{\beta^2G_{\tau\tau}-\beta_{H,\text{flat}}^2}{4\pi^2\alpha'^2}\psi_0(\mathbf{x_1})\psi_{0}^*(\mathbf{x_2})\right]\right) \\
&= \left.T^{\tau\tau}_{\text{th.sc.}}(\mathbf{x})\right|_{\text{on-shell}}\frac{\tilde{C}}{\beta-\beta_H}.
\end{align}
We have introduced here a constant $\tilde{C}$. These expression agree with those written down in equation (\ref{mainresult}).\footnote{Recall that for $D=0$, the logarithm and the $D/2$ factor should be dropped. The constant $\tilde{C}$ can then be fixed by comparing more closely.}\\

\begin{figure}[h]
\centering
\includegraphics[width=0.5\textwidth]{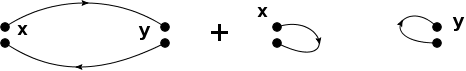}
\caption{Two possible Feynman diagrams for the (non-interacting) stress tensor correlator.}
\label{diagr}
\end{figure}

\noindent For correlators, we can schematically write\footnote{We have set $m_{local}^2 = \frac{\beta^2G_{\tau\tau}-\beta_{H,\text{flat}}^2}{4\pi^2\alpha'^2}$ representing the local mass term.} 
\begin{align}
\left\langle \hat{T}^{ij}_{\text{th.sc.}}(\mathbf{x}) \hat{T}^{kl}_{\text{th.sc.}}(\mathbf{y})\right\rangle &= \lim_{\substack{\mathbf{x_1}\to \mathbf{x}\\ \mathbf{x_2}\to \mathbf{x}}}\lim_{\substack{\mathbf{x_3}\to \mathbf{y}\\ \mathbf{x_4}\to \mathbf{y}}} \left\langle \left(\nabla^i \phi(\mathbf{x_1})\nabla^j \phi^*(\mathbf{x_2}) + \nabla^j \phi(\mathbf{x_1})\nabla^i \phi^*(\mathbf{x_2})\right.\right. \nonumber \\
&\quad\quad\quad\quad \left.- G^{ij}\left[G^{ab}\nabla_a \phi(\mathbf{x_1}) \nabla_b \phi^*(\mathbf{x_2}) + m_{local}^2 \phi(\mathbf{x_1})\phi^*(\mathbf{x_2})\right]\right) \nonumber \\
& \times\left(\nabla^k \phi(\mathbf{x_3})\nabla^l \phi^*(\mathbf{x_4}) + \nabla^l \phi(\mathbf{x_3})\nabla^k \phi^*(\mathbf{x_4}) \right. \nonumber \\ 
&\quad\quad\quad\quad \left.\left. - G^{kl}\left[G^{cd}\nabla_c \phi(\mathbf{x_3}) \nabla_d \phi^*(\mathbf{x_4}) + m_{local}^2 \phi(\mathbf{x_3})\phi^*(\mathbf{x_4})\right]\right)\right\rangle \\
&= \lim_{\substack{\mathbf{x_1}\to \mathbf{x}\\ \mathbf{x_2}\to \mathbf{x}}}\lim_{\substack{\mathbf{x_3}\to \mathbf{y}\\ \mathbf{x_4}\to \mathbf{y}}}\hat{\mathcal{D}}^{ijkl} \left\langle \phi(\mathbf{x_1})\phi^*(\mathbf{x_2})\phi(\mathbf{x_3})\phi^*(\mathbf{x_4})\right\rangle
\end{align}
with $\hat{\mathcal{D}}^{ijkl}$ a non-local differential operator. Next, we should contract this expectation value, which can only be done in two ways resulting in a connected diagram or a diagram consisting of two disconnected components since $\left\langle \phi \phi \right\rangle = \left\langle \phi^* \phi^* \right\rangle=0$ (figure \ref{diagr}).

\noindent Here again 
\begin{equation}
\left\langle \phi(\mathbf{x}) \phi^*(\mathbf{y}) \right\rangle \approx \frac{\psi_0 (\mathbf{x})\psi_0^*(\mathbf{y})}{\lambda_0}
\end{equation}
so we can just simply replace $\phi \to \psi_0$ in the correlator. The resulting expression is then twice\footnote{From the sum over all contractions of two $\phi$ and two $\phi^*$ vertices.} the product of the thermal scalar stress tensor:
\begin{equation}
\left\langle \hat{T}^{ij}_{\text{th.sc.}}(\mathbf{x}) \hat{T}^{kl}_{\text{th.sc.}}(\mathbf{y})\right\rangle \approx 2\left.T^{ij}_{\text{th.sc.}}(\mathbf{x})\right|_{\text{on-shell}} \left.T^{kl}_{\text{th.sc.}}(\mathbf{y})\right|_{\text{on-shell}} \frac{\tilde{C}^2}{(\beta-\beta_H)^2}.
\end{equation}
These results shed new light on the fact that we saw in subsection \ref{cancor} that for fully compact spaces the correlator is seen to factorize in the critical limit: as was briefly discussed there, all diagrams (both connected and disconnected) are equal in the Hagedorn limit. The combinatorics for multiple stress tensors yields a factor of $n!$ in agreement with our analysis in section \ref{cancor}. \\

\noindent It should not come as a big surprise to the reader that one can readily extend these formulas to the string charge and obtain very analogous results. 

\subsection{Continuous spectrum}
For a continuous spectrum, an analogous discussion can be made. Let us consider a continuous spectrum of states for the thermal scalar spectrum with the property that at $\beta = \beta_H$, the lowest eigenvalue of the associated eigenvalue problem vanishes.\footnote{As noted before, for a continuous spectrum this property is not trivial: it is possible that an integral over continuous quantum numbers shifts the critical eigenvalue to a non-zero value, as for instance happens in a linear dilaton background as is discussed elsewhere \cite{Mertens:2014cia}.} We write down
\begin{align}
\left\langle \hat{T}^{\mu\nu}_{\text{th.sc.}}\right\rangle = \lim_{\substack{\mathbf{x_1}\to \mathbf{x}\\ \mathbf{x_2}\to \mathbf{x}}} \hat{\mathcal{D}}^{\mu\nu}(\mathbf{x_1},\mathbf{x_2}) \int d k \rho(k)\frac{\psi_k(\mathbf{x_1})^{*}\psi_k(\mathbf{x_2})}{\lambda(k)}.
\end{align}
In this equation, $\hat{\mathcal{D}}^{\mu\nu}$ denotes the differential operator acting on the Green function to obtain the stress tensor (as written down in equation (\ref{operatorgf})) and $k$ labels the continuous quantum numbers with density of states $\rho(k)$. For a mixed continuous and discrete spectrum, one first focuses on the lowest discrete mode and then applies the method explained here. We are interested in the dominant limit as $\beta\to\beta_H$ or $\lambda(0) \to 0$. The dominant contribution to this limit can be found in the region of integration where $k \approx 0$. Hence we immediately extract the non-singular contributions:
\begin{align}
\left\langle \hat{T}^{\mu\nu}_{\text{th.sc.}}\right\rangle \approx \underbrace{\lim_{\substack{\mathbf{x_1}\to \mathbf{x}\\ \mathbf{x_2}\to \mathbf{x}}} \hat{\mathcal{D}}^{\mu\nu}(\mathbf{x_1},\mathbf{x_2}) \psi_0(\mathbf{x_1})^{*}\psi_0(\mathbf{x_2})}_{\left.T^{\mu\nu}_{\text{th.sc.}}\right|_{\text{on-shell}}} \int d k\frac{\rho(k)}{\lambda(k)}.
\end{align}
To derive the dominant contribution to this final integral in general, it is convenient to recall the form of the free energy in the critical regime in terms of the eigenvalues of the thermal scalar wave equation:
\begin{equation}
\label{todiff}
Z_1 = - \beta F \approx - \int d k \rho(k) \ln(\lambda(k)) \approx -C(\beta-\beta_H)^{D/2}\ln(\beta-\beta_H),
\end{equation}
where the same eigenvalues $\lambda(k)$ are used. This eigenvalue is generically such that $\lambda(0) \sim \beta-\beta_H$ and hence we can Taylor-expand:
\begin{equation}
\lambda(k) \approx B(\beta-\beta_H) + f(\beta,\beta_H) k + \mathcal{O}(k^2),
\end{equation}
for some proportionality constant $B$.
Differentiating (\ref{todiff}) with respect to $\beta$, we obtain\footnote{The contributions where $\rho(\mathbf{n})$ is differentiated is subdominant in the $\beta \approx \beta_H$ limit.}
\begin{equation}
- \int d k\frac{\rho(k)}{\lambda(k)}\left(B + \frac{\partial f}{\partial \beta} k + \mathcal{O}(k^2)\right) \approx
- \int d k\frac{B \rho(k)}{\lambda(k)} \approx -C\frac{D}{2}(\beta-\beta_H)^{D/2-1}\ln(\beta-\beta_H).
\end{equation}
Hence in the end we obtain
\begin{align}
\left\langle \hat{T}^{\mu\nu}_{\text{th.sc.}}\right\rangle \approx \left.T^{\mu\nu}_{\text{th.sc.}}\right|_{\text{on-shell}} \tilde{C}\frac{D}{2}(\beta-\beta_H)^{D/2-1}\ln(\beta-\beta_H),
\end{align}
for some constant $\tilde{C}$ and this is precisely the form we found in section \ref{thermaver}.\footnote{This perspective shows that, at least in the canonical ensemble, as soon as the continuous quantum numbers do not affect the critical temperature of the space we are interested in, the formulas discovered previously should be valid. In particular this implies that the rather awkward assumption made in section \ref{HP} about continuous spectra (integrating out continuous quantum numbers does not affect $\lambda_0$, also for variations on the background) seems in this case easier to handle.}

\section{Examples}
\label{examples}
The general strategy to proceed is then as follows. 
\begin{itemize}
\item[I.] Construct the lowest eigenmode of the thermal scalar wave equation.
\item[II.] Compute the classical stress tensor of this mode. This then gives the dominant contribution to the stress tensor of the highly excited or near-Hagedorn string gas. Analogously for the string charge tensor.
\end{itemize}
Let us apply these results to some specific examples.
\subsection{Flat space}
Flat space in principle has a continuous spectrum of the thermal scalar eigenvalue problem. Instead of working with this continuum, we here take a more pragmatic road and simply introduce a finite-volume regularization (with periodic boundary conditions). This leads to the (normalized) thermal scalar wavefunction:
\begin{equation}
T_0 = \frac{1}{\sqrt{V}},
\end{equation}
which gives for the time-averaged, high-energy-averaged stress tensor:
\begin{align}
\left\langle T^{\tau\tau}_{e}(\mathbf{x})\right\rangle_E &\approx -\frac{E}{V}, \\
\left\langle T^{ij}_{e}(\mathbf{x})\right\rangle_E &\approx 0,
\end{align}
and indeed one expects the spacetime energy density to be $E/V$. The pressure vanishes in the high energy regime: thus high energy string(s) behave as a pressureless fluid (or as dust). Of course, what this means is that one should look at the next most dominant contribution. Nonetheless, the pressure is subleading with respect to other thermodynamic quantities at high energies.\\
Incorporating again the extra temperature factors to obtain the canonical average of the energy-momentum tensor, it is clear that the pressure still vanishes at $\beta \approx \beta_H$. It is interesting to learn that the Hagedorn string gas in flat space has vanishing pressure, very reminiscent of the final configuration of D-brane decay (see e.g. \cite{Sen:2004nf} and references therein). This fact was observed in the microcanonical approach of string gas cosmology as well \cite{Bassett:2003ck}\cite{Takamizu:2006sy}: the Hagedorn phase in flat space behaves as a pressureless fluid. The fact that our computation yields the same result provides confidence to our approach.\\

\noindent We also remark that choosing an \emph{arbitrary} compact spatial manifold (such as a compact Calabi-Yau) also leads to a pressureless state of matter. The reason is that such spatial manifolds have as their lowest thermal scalar eigenfunction the constant mode. A constant thermal scalar immediately leads to a vanishing $T^{ij}_{\text{th.sc.}}(\mathbf{x})$.

\subsection{$AdS_3$ space}
Let us look at the time-averaged, high energy string charge tensor in the WZW $AdS_3$ spacetime \cite{Gawedzki:1991yu}\cite{Maldacena:2000hw}\cite{Berkooz:2007fe}\cite{Lin:2007gi}. We will not give a self-contained treatment of this model here and the reader is refered elsewhere for details concerning the thermal spectrum.\\
A first subtlety here is again the fact that the thermal scalar is part of a band of continuous states. Just like in the flat space case, no corrections are generated by integrating the heat kernel over continuous quantum numbers \cite{Mertens:2014nca}. Assuming this property also holds for small variations in the $AdS_3$ metric and Kalb-Ramond background (or alternatively, working in the canonical ensemble where such a technicality does not present itself), we can proceed.\\
A second subtlety is that the density of high energy states includes a periodic part in $E$ \cite{Lin:2007gi}. A moment's thought shows that such periodic parts do not alter any of the conclusions.\footnote{More in detail, the density of single string states takes the schematic form
\begin{equation}
\omega(E) \sim E^p \frac{e^{\beta_H E}}{\left|\sin(a E)\right|}
\end{equation}
for some numbers $p$ and $a$ that we do not want to specify. In principle, an integral such as 
\begin{equation}
\int^{E}d\tilde{E} \tilde{E}^p \frac{e^{\beta_H \tilde{E}}}{\left|\sin(a \tilde{E})\right|}
\end{equation}
is infinite due to the extra poles caused by the sine factor. It is known that these are caused by the infinite $AdS_3$ volume felt by the so-called long strings. The difference with flat space is that this infinity does not cleanly factorize. Nonetheless, we are not interested in these divergences, but in the Hagedorn divergence. We hence obtain
\begin{equation}
\int^{E}d\tilde{E} \tilde{E}^p \frac{e^{\beta_H \tilde{E}}}{\left|\sin(a \tilde{E})\right|} \approx E^p \frac{e^{\beta_H E}}{\beta_H\left|\sin(a E)\right|}.
\end{equation}
The extra sine factor generated eventually cancels in the ratio of the thermal expectation values in (\ref{micro}), much like the $E^{D/2+1}$ factor did earlier on in equation (\ref{cancell}).} \\ 
A third subtlety is that the dominant thermal state is not just a single state, but is in fact all states labeled by an integer $q$ with $w=\pm1$ \cite{Mertens:2014nca}.\footnote{This is actually the reason for the appearance of the periodic factor in the high-energy density of states.} The $q$ quantum number corresponds to discrete momentum around the spatial cigar subspace. We choose $q=0$ as the thermal scalar state that has the information on the Hagedorn temperature.\footnote{We expect in general that the $q\neq0$ states will not contain the information on the Hagedorn temperature for infinitesimal metric variations. Note though, that we are only interested in how the critical temperature varies with the metric, for which the $q=0$ state is sufficient without caring about the $q\neq0$ states. We therefore pick $q=0$ and the situation is reduced to that discussed above. This issue is related to the fact that we only use $w=1$ and not both $w=\pm1$: one can choose between these for the critical temperature, and the answer is the same. In any case, the $AdS_3$ WZW model is quite pathological but we hope our treatment here will show how to handle other (more well-behaved) models.} \\

\noindent The background Kalb-Ramond field causes the string charge not to vanish. This can be seen from equation (\ref{noether}), where a nonvanishing $G'^{\tau k}$ is caused by turning on a Kalb-Ramond background. For $AdS_3$, we have $G'^{\tau\phi} = i$. Hence this yields a contribution to the string charge as
\begin{equation}
\left\langle J^{\tau \phi}_{e}(\mathbf{x})\right\rangle_E \propto \left.J^{\phi}(\mathbf{x})\right|_{\text{on-shell}} \propto \frac{\beta_H}{\alpha' \pi}T_0(\mathbf{x}) T^{*}_0(\mathbf{x}),
\end{equation}
where the wavefunction $T_0$ belongs to a continuous representation of the $SL(2,\mathbb{R})$ symmetry group and hence is not confined to the $AdS$ origin. Its precise form will not be written down here. What we emphasize in this context is that the charge does not vanish. The charge is directed alongside the angular cigar direction.

\subsection{Rindler space}
Let us finally apply these methods to Rindler space. We consider highly excited strings according to the Rindler observer. \\
The thermal metric is 
\begin{equation}
ds^2 = \frac{\rho^2}{\alpha'}d\tau^2 + d\rho^2 + \hdots
\end{equation}
where $\tau \sim \tau + 2 \pi \sqrt{\alpha'}$, the Rindler temperature. As it stands, we have used string-normalization for the metric. \\
We have shown in \cite{Mertens:2013zya} (see also \cite{Giveon:2012kp}\cite{Sugawara:2012ag}\cite{Giveon:2013ica}\cite{Giveon:2014hfa}) that the Hagedorn temperature in this space equals the Rindler temperature: $\beta_H = 2 \pi \sqrt{\alpha'}$ for type II superstrings. The thermal scalar was found to have a wavefunction of the form:
\begin{equation}
T_0(\rho) \sim e^{-\frac{\rho^2}{2\alpha'}}
\end{equation}
localized to the Rindler origin, or in terms of the black hole of which this is the near-horizon limit, localized to the black hole horizon.\\
In field theory, it is known that the Rindler observer (accelerated observer or fiducial observer) constructs his Fock space using his definition of positive-frequency modes. Moreover the vacuum constructed by an inertial observer (the \emph{Minkowski} vacuum) is seen by the Rindler observer as a thermal ensemble in terms of his coordinates and Fock space. The \emph{Rindler} vacuum is the vacuum the Rindler observer himself defines. Here we first focus on a high energy string(s) constructed on top of the Rindler vacuum.\\
In the end we will make a few remarks concerning the thermal ensemble itself (which upon including the Casimir contribution should coincide with the Minkowski stress tensor):
\begin{equation}
\left\langle T^{\mu\nu}\right\rangle_{M} = \text{Tr}_{R}\left(T^{\mu\nu}e^{-\beta H}\right).
\end{equation}
We are interested in the temperature-dependent part and hence the Casimir part does not interest us that much. In fact, for observations made by accelerated observers, the Casimir contribution is not detectable: it is only relevant when looking at the backreaction of the string(s) in the semi-classical Einstein equations. \\

\begin{figure}[h]
\centering
\begin{minipage}{.3\textwidth}
  \centering
  \includegraphics[width=\linewidth]{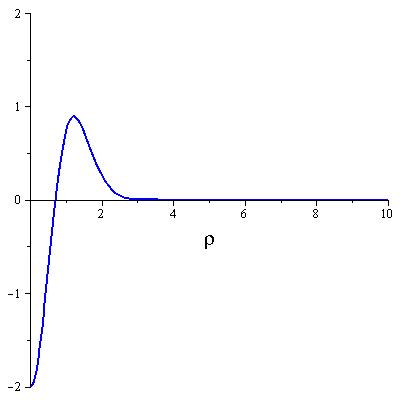}
  \caption*{(a)}
\end{minipage}
\begin{minipage}{.3\textwidth}
  \centering
  \includegraphics[width=\linewidth]{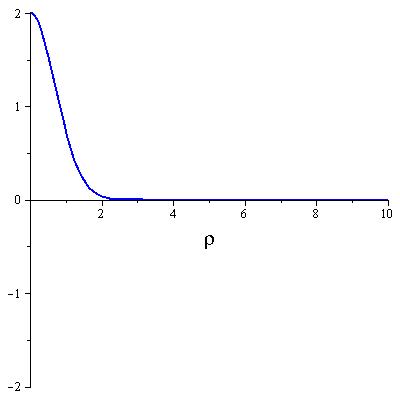}
  \caption*{(b)}
\end{minipage}
\begin{minipage}{.3\textwidth}
  \centering
  \includegraphics[width=\linewidth]{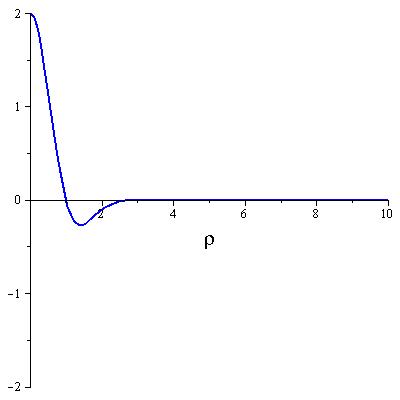}
  \caption*{(c)}
\end{minipage}
\caption{(a) Energy density $-\left\langle T^{\tau}_{\tau}(\mathbf{x})\right\rangle$ as a function of radial distance $\rho$ in units where $\alpha'=1$. (b) Radial pressure $\left\langle T^{\rho}_{\rho}(\mathbf{x})\right\rangle$. (c) Transverse pressure $\left\langle T^{i}_{i}(\mathbf{x})\right\rangle$.}
\label{form}
\end{figure}

\noindent These highly excited long strongs surrounding the event horizon are candidates for a microscopic description of the black hole membrane \cite{Susskind:1993ws}\cite{Kutasov:2005rr} thus it seems worthwhile to look into only this piece. We study the type II superstring and find the following quantities:
\begin{align}
\frac{e^{2\Phi}}{\sqrt{G}}\frac{\delta \lambda_0}{\delta G_{\tau\tau}} &= \frac{\beta_H^2}{4\pi^2\alpha'^2}e^{-\rho^2/\alpha'} + \frac{\alpha'}{2\rho^2}\left[\frac{\rho^2}{\alpha'^2} + \frac{\beta_H^2 \frac{\rho^2}{\alpha'}-\beta_{H,\text{flat}}^2}{4\pi^2\alpha'^2}\right]e^{-\rho^2/\alpha'}, \\
\frac{\partial \lambda_0}{\partial \beta^2} &= \frac{1}{4\pi^2\alpha'^2}\int_{0}^{+\infty}d\rho \left(\frac{\rho}{\sqrt{\alpha'}}\right)^3e^{-\rho^2/\alpha'} = \frac{1}{4\pi^2\alpha'^2}\frac{\sqrt{\alpha'}}{2}.
\end{align}
From this, one finds that
\begin{align}
\left\langle T^{\tau\tau}_{e}(\mathbf{x})\right\rangle_E &= -\frac{2E}{\sqrt{\alpha'}}\left(2-\frac{\alpha'}{\rho^2}\right)e^{-\rho^2/\alpha'},
\end{align}
and we readily check explicitly that indeed
\begin{equation}
-\int \left\langle T^{\tau}_{\tau,e}\right\rangle_E \sqrt{G} d^{D-1}x = \frac{2E}{\sqrt{\alpha'}}\int_{0}^{+\infty}d\rho \left(\frac{\rho}{\sqrt{\alpha'}}\right)^3\left(2-\frac{\alpha'}{\rho^2}\right)e^{-\rho^2/\alpha'} = E.
\end{equation}
Thus $-\left\langle T^{\tau}_{\tau,e}\right\rangle_E$ can be interpreted as the energy density profile whose form is given by figure \ref{form}(a).
One finds that the energy density turns negative for $\rho < \sqrt{\frac{\alpha'}{2}}$. This implies that the highly excited string matter violates the weak-energy condition. Such behavior was observed in the past for quantum fields near black holes as well \cite{Candelas:1980zt}\cite{Page:1982fm}\cite{Howard:1984qp}\cite{Frolov:1986ut}\cite{Visser:1996ix}\cite{Carlson:2003ub}.\footnote{Note though that there are differences between the results obtained in these papers and our result. In the cited papers, the authors sum the thermal stress tensor and the Casimir stress tensor (both of which are infinite). The sum is finite and has a negative-energy zone for Schwarzschild black holes. For Rindler space on the other hand, the sum is strictly zero \cite{Dowker:1994fi} which equals the vev of the stress tensor in the Minkowski vacuum. In the field theory case, there is a discrepancy between the Rindler result and the Schwarzschild result close to the horizon. It is known that this is related to curvature corrections to the stress tensor vev \cite{Frolov:1989jh}\cite{Zurek:1985gd}. Inspection of our result (the thermal scalar classical stress tensor) for the full black hole (for instance the $SL(2,\mathbb{R})/U(1)$ cigar that we used in \cite{Mertens:2013zya} to arrive at Rindler space) and Rindler space itself shows that for our purposes the results agree: no curvature corrections are needed and Rindler space captures the near-horizon region of the black hole. This is presumably because we only care for the most dominant contribution. For some more recent accounts of the stress tensor near black hole horizons in light of the holographic correspondence, see \cite{Haddad:2013tha}\cite{Figueras:2013jja}.} An obvious difference with field theory is that for string theory the stringy matter becomes of negative energy density at a string scale distance from the black hole (instead of the Schwarzschild scale).\footnote{One can check this explicitly by transforming the string-normalized Rindler space to the Schwarzschild normalization as is done for instance in \cite{Mertens:2013zya}.} With this stringy localization in mind, let us emphasize a point which we skimmed over in \cite{Mertens:2013zya}. In a black hole-normalized geometry
\begin{equation}
ds^2 = \frac{\rho^2}{(4GM)^2}d\tau^2 + d\rho^2 + \hdots,
\end{equation}
the Klein-Gordon type equations for non-winding modes give eigenfunctions that do not depend on the string length $l_s$.\footnote{Their eigenvalues do depend on the string length for massive string states.} The peculiarity of the winding modes is that T-duality explicitly introduces the string length into the eigenvalue problem. This causes the thermal scalar in Rindler space to be localized at string length from the horizon; pure discrete momentum modes on the other hand oscillate in space with an oscillation length of the order of the black hole scale $GM$ (\emph{not} the string scale). \\

\noindent Analogously one computes the spatial components of the stress tensor and one finds
\begin{align}
\left\langle T^{\rho\rho}_{e}(\mathbf{x})\right\rangle_E &= \frac{2E}{\sqrt{\alpha'}}e^{-\rho^2/\alpha'}, \\
\left\langle T^{ij}_{e}(\mathbf{x})\right\rangle_E &= \delta_{ij}\frac{2E}{\sqrt{\alpha'}}\left[1- \frac{\rho^2}{\alpha'}\right]e^{-\rho^2/\alpha'},
\end{align}
The radial pressure $\left\langle T^{\rho}_{\rho}(\mathbf{x})\right\rangle$ is given in figure \ref{form}(b) and is found to be positive and localized to the horizon. Curiously, the transverse pressure, depicted in figure \ref{form}(c), changes sign at $\rho=\sqrt{\alpha'}$. An obvious feature is that, since the stress tensor is quadratic in the fields, it decays at twice the rate of $T_0$. Note also that, as discussed above, the highly excited string(s) does not have non-zero net impulse in any direction: $\left\langle T^{\tau \rho}_{e}(\mathbf{x})\right\rangle_E = \left\langle T^{\tau i}_{e}(\mathbf{x})\right\rangle_E = 0$. This agrees with for instance the field theory results for the Hartle-Hawking vacuum of Schwarzschild black holes. The Unruh vacuum for a Schwarzschild black hole on the other hand would include non-zero $T^{\tau \rho}$ corresponding to the flux of particles emitted by a black hole formed in gravitational collapse (Hawking radiation). \\

\noindent Let us discuss some generalities on occurence of negative energy density. Firstly, it is not possible to have $-\left\langle T^{\tau}_{\tau}\right\rangle < 0$ everywhere. This follows immediately from the fact that the total energy is $E$. Secondly, since the spatial kinetic part $G^{ij}\partial_i T \partial_j T^*$ in (\ref{thenergy}) is positive semi-definite, a necessary condition for negative energy density is that 
\begin{equation}
\label{local}
\beta_H^2 G_{\tau\tau} < \beta_{H,\text{flat}}^2.
\end{equation}
This is the condition that the local thermal circle becomes smaller that the flat space Hagedorn circle. For instance for Rindler space, we have that $\rho < \sqrt{2\alpha'}$, and indeed the negative energy density region lies inside this domain. Is satisfying the condition (\ref{local}) uncommon? Actually, all backgrounds must satisfy this condition for some points. The reason is that the thermal scalar is a zero-mode at the Hagedorn temperature (by definition). If this condition is nowhere satisfied, the on-shell action for the thermal scalar (\ref{thaction}) would be positive definite which is impossible.\footnote{An exception occurs if the thermal scalar is space-independent: only then is the spatial kinetic term zero and there is no a priori requirement to have a negative energy density anywhere. Hence this occurs only for spaces with a non-varying thermal circle (meaning constant $G_{\tau\tau}$) since in that case, one can find a constant mode as a solution to the thermal scalar equation. An immediate example of this is flat space. Also for instance 4d flat space combined with a 6d compact unitary CFT (e.g. a compact Calabi-Yau 3-fold) falls in this category.}\\

\noindent For the string charge in Rindler space, we obtain $\left\langle J^{\tau k}_{e}(\mathbf{x})\right\rangle_E = 0$ for all $k$. \\ 

\noindent From a canonical point of view, the situation is analogous: the dominant part of the stress tensor of the near-Hagedorn string gas is given by the classical thermal scalar stress tensor, suitably multiplied by a temperature-dependent factor. Note though that it was previously remarked \cite{Mertens:2013zya}\cite{Sugawara:2012ag} that the Hagedorn temperature equals the Hawking temperature in this case, meaning that in principle the stress tensor becomes infinitely large. The significance of this infinity is the same as that of the free energy; higher loop corrections order-by-order are not really suited to cancel it (as we discussed in the previous chapter).\footnote{In the case $g_s=0$, the one-loop contribution will be exact and infinitely large (for only compact dimensions). Using $E = \frac{1}{\beta-\beta_H}$ in the canonical ensemble, the long string gas has infinite energy and entropy. This is not a priori a problem since the black hole itself needs to have $M\to+\infty$ to have a finite Schwarzschild radius when $G \sim g_s^2 \to 0$. Also the Bekenstein-Hawking entropy diverges in this limit.} \\ 

\noindent Up to this point, we only analyzed type II superstrings in detail. Bosonic strings were dismissed since their thermal scalar action receives corrections in $\alpha'$. Heterotic strings on the other hand do not receive corrections in Rindler space. Can we then use this action to find the stress tensor? Actually, this seems not to be the case. Using the uncorrected thermal scalar action to compute the stress tensor is unjustified as nearby backgrounds will definitely break the worldsheet supersymmetry down to the expected $(0,1)$ where the thermal scalar action \emph{will} receive corrections.
A summary of the different string types and whether their thermal scalar action and near-Hagedorn stress tensor gets $\alpha'$-corrected is given below.
\begin{figure}[h]
\centering
\includegraphics[width=0.65\textwidth]{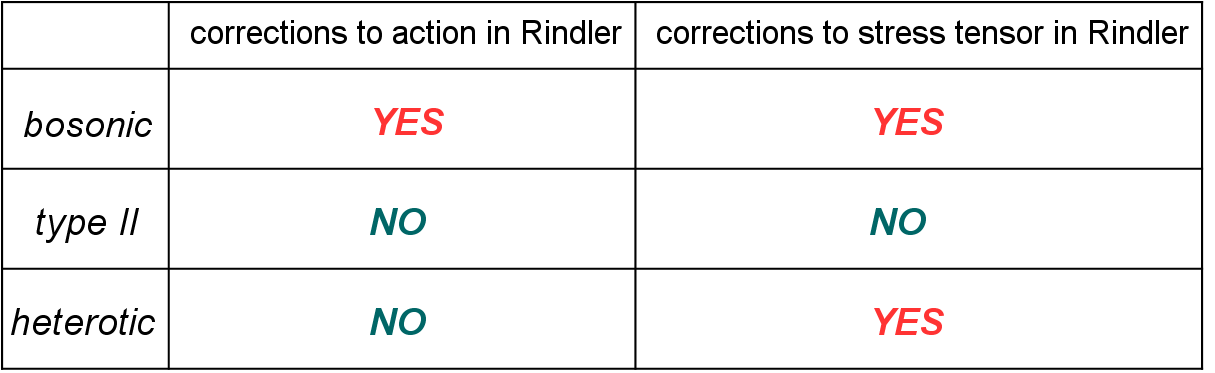}
\end{figure}

\section{Bekenstein-Hawking entropy}
\label{beke}
The results obtained here for the high energy stress tensor are actually the first results we obtained in this work that have to do with the microcanonical ensemble. We will put these to good use here to demonstrate that $T_H = T_{\text{Hawking}}$ is actually a \emph{necessary} condition for consistency by showing how these highly excited strings can account for $S_{BH}$. \\
This computation will be the final result we obtain in this work. \\
Before going there, we will make a brief detour to discuss why there is in fact an equilibrium to begin with for black holes. 

\subsection{Existence of equilibrium}
The basic conceptual question we discuss here is: how can an equilibrium for matter around a black hole even exist? Is not everything perpetually falling inwards, never truly settling down to an equilibrium configuration? \\

\noindent For the canonical ensemble, this has been understood even for QFT: an equilibrium sets in between the matter infalling from the heat bath and radiation emitted by the black hole. These two fluxes cancel each other macroscopically and a thermal (Unruh) heat bath is created close to the horizon. \\

\noindent The microcanonical ensemble is trickier and in QFT no meaningful equilibrium exists. Let us first ignore the thermal atmosphere of the hole. Fixing the energy in the box, how can the matter content be in equilibrium? It should simply fall in. E.g. if a star starts falling into a black hole, no equilibrium configuration is reached at all: the matter flattens out as it approaches the horizon.\footnote{Note that, upon including the thermal atmosphere, equilibration is expected to occur by diffusion of the infalling matter in the Unruh heat bath and stretched horizon, but this process requires interactions to be turned on, something we do not do here.}  \\
This is indeed so for field theory and is the cause of the entropy divergence: field modes Lorentz contract close to the horizon and allow an infinite amount of information to be stored close to the horizon. Not so for string theory. \\
Even though each individual field mode in the Lorentzian string spectrum has this same problem, this is not the case for the summed spectrum. Indeed, the string entropy is finite as a consequence of the better UV behavior. At high energies, the thermal scalar wavefunction determines an energy density profile bound to the horizon (figure \ref{form}(a)), a feature impossible in QFT. The high-energy gas is in equilibrium (macroscopically no stringy matter falls in), because strings fail to Lorentz contract close to the horizon. The strings do not fall in further and build up a radial pressure profile (figure \ref{form}(b)), compensating for the gravitational infall and allowing an equilibrium configuration. \\
Hence the bound states (wound in Euclidean signature) encode this general non-contraction feature of strings, a feature previously predicted by Susskind. \\

\noindent Now suppose we throw some matter of fixed total energy (microcanonical gas) into a black hole that also includes its thermal atmosphere.
The presence of this thermal gas close to the horizon presents an added complication. For the most dominant part (and at the non-interacting level), we can distinguish between this thermal gas (which has some canonical total energy) and the matter thrown in. These will not interact and their spatial profiles for energy and pressure will simply add: we indeed saw above that the most dominant part of the stress tensor is simply proportional to the total energy. Hence adding some energy and equilibrating again will result in the sum of the two stress tensors, provided both stress tensors are evaluated in the high-energy regime. The added matter hence needs to have sufficiently high energy as well for this argument to work. \\ 

\noindent All of this suggests that, within non-interacting string theory, it is meaningful to throw in some energy into a black hole and wait for it to equilibrate (it might take an arbitrarily long time, but a sensible equilibrium configuration does exist).

\subsection{Bekenstein-Hawking entropy from long strings}
Consider a large Schwarzschild black hole equiped with its thermal atmosphere and also near-horizon long string stretched horizon. We want to show that the black hole entropy is carried by long strings. To that effect, on top of this thermal ensemble, we put a fixed high-energy microcanonical string gas (obtained from throwing in some fixed energy matter and waiting for equilibration). It is known that the high-energy string gas is dominated by long string(s). What we basically have in mind here is adding some energy to the black hole and counting whether the added matter correctly accounts for the added black hole entropy. \\
In QFT language, the (initial) state we consider is\footnote{The state $\left|0\right\rangle$ is the Rindler vacuum that we called $\left|R\right\rangle$ in chapter \ref{chBH}.} 
\begin{equation}
a^{R\dagger}_{E}\left|M\right\rangle = \sum_n  \frac{1}{\sqrt{Z}}e^{-\pi E_n}\left|E_n\right\rangle_{L}\otimes a^{R\dagger}_{E}\left|E_n\right\rangle_R \sim \sum_n  \frac{1}{\sqrt{Z}}e^{-\pi E_n}\left|E_n\right\rangle_{L}\otimes \left|E_n+E\right\rangle_R,
\end{equation}
where a quantum of energy $E$ is added to the thermal ensemble, with $a^{R}_{E}\left|0\right\rangle = \left|0\right\rangle_L \otimes a^{R}_{E}\left|0\right\rangle_R = 0$. After this, one traces over the left Rindler wedge. As discussed above, the added microcanonical gas settles down to the near-horizon configuration determined earlier at a string length from the horizon. \\
The energy of the microcanical gas satisfies $E \gg \beta_H^{-1}$ in the high-energy regime and with the Hagedorn temperature equaling the Hawking temperature, we need
\begin{equation}
E \gg \frac{1}{GM}.
\end{equation}
Note that this is in fact a relatively small energy. But no matter how small the energy is measured at infinity, the blueshift will make the locally measured energy arbitrarily high and string-scale. \\
Also, we need this extra high-energy gas to be subdominant compared to the original black hole itself (small backreaction). This implies $M\gg E$. There exists an energy range for $E$ where both inequalities can be satisfied together if
\begin{equation}
GM^2 \gg 1,
\end{equation}
meaning a large black hole entropy. Our argument hence applies to large black holes where the near-horizon Rindler approximation makes sense.\\

\noindent With the extra microcanonical gas (in the high-energy regime), we have added an entropy to the black hole system of
\begin{equation}
\label{longs}
\delta S = \beta_H \delta E.
\end{equation}
Note that this equality assumes a density of states of the form $\rho(E) \sim E^\alpha e^{\beta_H E}$. The numerical and polynomial prefactors are subdominant in the large energy limit ($\beta_H E \gg 1$).
Likewise, from a gravitational point of view, we have added an entropy of
\begin{equation}
\label{longss}
\delta S = \beta_{\text{Hawking}} \delta M,
\end{equation}
where $\delta M$ is the increase of the black hole mass due to the extra string gas. Because $\beta_H = \beta_{\text{Hawking}}$, these equalities are manifestly the same. Or, said differently, we can integrate equation (\ref{longs}):
\begin{equation}
\delta S = 8\pi GM \delta M \quad \Rightarrow \quad  S= \frac{A}{4G},
\end{equation}
which is Bekenstein's black hole entropy. This demonstrates that the long string is capable of yielding the correct number of states to account for the microstructure of the black hole. \\

\noindent Some comments are in order at this place. 
\begin{itemize}
\item In contrast to everything else in this work, this derivation is at tree-level (not one-loop). We concluded that the tree-level black hole entropy can be generated by long closed strings. What about interactions? Are they not important? \\
Just like in Susskind's and Uglum's work \cite{Susskind:1994sm}, we expect these will renormalize Newton's constant, but we make no additions at this point.  

\item For which type of string theory does this argument hold? \\
The only information we used is that $\beta_H = \beta_{\text{Hawking}}$. Since we are using a microcanonical argument (i.e. free closed strings), this seems to imply bosonic strings always have a thermal divergence, since we discussed in the previous chapter that for the non-interacting partition function one needs to include higher winding modes as well, which are tachyonic in bosonic string theory. So the argument does not hold for bosonic strings. For type II and heterotic strings, the argument holds. These arguments show that $\beta_H = \beta_{\text{Hawking}}$ is necessary for consistency of our story. If this were not the case, the high-energy strings that we throw in would have too much or too little entropy. 

\item How does this relate to Susskind's redshift argument \cite{Susskind:1993ws} and Horowitz-Polchinski's correspondence principle \cite{Horowitz:1996nw}? \\
Because of the linearization at the infinitesimal level, we do not need a redshift argument nor a specific correspondence point. The downside on the other hand, is that we need a large black hole and the argument fails for small (string-scale) black holes. We are also not describing the black hole itself as a single long string, instead we are describing the infalling matter entropy as a long string (which in the end should be the same thing). An important point is that, unlike both Susskind's argument and Horowitz-Polchinski's argument, all prefactors come out correctly and no proportionality constant is missed.

\item Another interesting point is that, because we rely only on the result in Rindler space, this applies to all uncharged black holes that have Rindler space as their near-horizon approximation. One only needs $\beta_H = \beta_{\text{Hawking}}$.

\item The Wald entropy is a generalization of the Bekenstein entropy to a general diffeomorphism invariant theory. The expression has corrections compared to the famous formula of the horizon area in Planck units. However, the thermodynamic first law still holds:
\begin{equation}
\delta S_{\text{Wald}} = \beta \delta M,
\end{equation}
and it is precisely this infinitesimal relation that was satisfied for the long string as it builds up the black hole (\ref{longss}). Hence, we are automatically in agreement with the curvature corrections to GR and to Bekenstein's entropy.\footnote{We thank K. Van Acoleyen for suggesting taking a closer look at the Wald entropy.}
\end{itemize}

\noindent To conclude this work, we would like to point out that our approach sheds some new light on the old puzzle that black holes appear to be described by $c=6$ CFTs \cite{Maldacena:1996ya}\cite{Halyo:1996vi}\cite{Tseytlin:1996qg}. It is interesting to note that the shift in the Hagedorn temperature from its flat space value $\beta_H = 2\sqrt{2}\pi\sqrt{\alpha'}$ to its value $\beta_H = 2\pi\sqrt{\alpha'}$ in Rindler space, as measured at the stretched horizon, precisely corresponds to the shift in central charge from $c=12$ to $c=6$. \\
\noindent Suppose in flat space, we inject a small amount of energy $\delta E \gg T_H$. This added gas is on its own a long string gas and hence adds an entropy of
\begin{equation}
\label{added1}
\delta S = 2\sqrt{2}\pi\sqrt{\alpha'}\delta E = 2\pi \sqrt{\frac{c}{6}}2\delta \sqrt{N},
\end{equation}
where the second equality is Cardy's formula. Equating then gives with $E = \frac{2}{\ell_s}\sqrt{N}$ that $c=12$, indeed the flat space type II central charge. \\

\noindent Next, suppose we add a small amount of string gas to a black hole, with energy $\delta E \gg T_{\text{Hawking}}$. Then the added entropy is of the form
\begin{equation}
\delta S = \beta_{\text{Hawking}} \delta E = 2\pi\sqrt{\alpha'}\delta E_{\text{sh}} = 2\pi \delta E_{R},
\end{equation}
where $E$ is the energy measured at infinity, $E_{\text{sh}}$ is the energy measured at the stretched horizon and $E_R$ is the dimensionless energy measured at $\rho=1$. The black hole itself can be interpreted as a long string in flat space, where the Rindler energy $E_R$ and the oscillation number $N$ of this long string (not the worldsheet CFT of the Rindler string!) are related as $E_R = 2\sqrt{N}$, such that $E_{\text{sh}} = \frac{2}{\ell_s}\sqrt{N}$ \cite{Maldacena:1996ya}\cite{Halyo:1996vi}. Hence the added entropy is related to an added oscillation number of the long string making up the black hole. It can be written as
\begin{equation}
\delta S = 4\pi \delta \sqrt{N} = 2\pi \sqrt{\frac{c}{6}} 2 \delta \sqrt{N},
\end{equation}
immediately leading to $c=6$. \\
The different Hagedorn temperatures for flat space versus Rindler space are what ultimately cause the $c=12 \to c=6$ shift of the central charge. \\
\emph{The universality of $c=6$ is hence directly related to the universality of $T_H = T_{\text{Hawking}}$.}

\subsection{Relevance for Microscopic Interpretation}
Let us finally turn to a more detailed discussion of the microstructure implied by the above story.\footnote{We thank H. Verlinde for emphasizing the importance of this discussion.}
\subsubsection{Long strings and equilibration}
The above derivation correctly reproduces the Bekenstein-Hawking entropy. But is this not true for any matter falling in? After all, if one throws in some matter of energy $\delta E$ and entropy $\delta S$, then after a long time, the black hole's mass will have increased by $\delta E$ and its entropy by $\delta S$ where both quantities are related through the area law. This is basically the Zurek-Thorne argument where the infalling information that falls behind the stretched horizon is reinterpreted as the black hole entropy \cite{Zurek:1985gd}. \\
The main difference with our approach is that it is in general impossible to realize thermodynamical equilibrium of the infalling matter (with fixed energy and entropy) on its own. The first law of thermodynamics of the infalling matter, living in the heat bath of the black hole, states that
\begin{equation}
dE = T_{\text{Hawking}} dS,
\end{equation}
but on the other hand, the microcanonical temperature of the infalling gas is defined as
\begin{equation}
\frac{1}{T(E)} = \frac{dS}{dE}(E),
\end{equation}
and hence, in general, $T(E) \neq T_{\text{Hawking}}$, so equilibrium is not realized immediately: the infalling matter either needs to deposit energy in the thermal bath or absorb energy from it before it can achieve equilibrium. \\
Of course, as discussed above as well, matter is simply expected to fall in indefinitely and no sensible equilibrium configuration of the infalling gas is achieved in any case (ignoring diffusion into the stretched horizon). \\

\noindent For long strings, we have $S \sim E$ (a Hagedorn density of states) and the microcanonical temperature is energy-independent. Hence, unless $T_H = T_{\text{Hawking}}$, it is \emph{impossible} to achieve thermodynamical equilibrium for the infalling long string gas. \\

\noindent The upshot is that only long strings (having a Hagedorn density of states) with $T_H = T_{\text{Hawking}}$ can equilibrate with the black hole already present, without interacting with it in a significant way. The long strings can hence be deposited on top of the existing black hole one by one, each one in equilibrium with the already present black hole. \\

\noindent Taking a step back, one can think of the black hole realized in this way as a set of equilibrated long strings. \\
Hence the black hole microstructure is realized in a quite natural fashion by these long strings.

\subsubsection{Black hole microstructure}
There are several ways in which the microstructure of the black hole is realized in string theory. Let us recapitulate these and point out some open questions here. \\

\noindent Firstly, as suggested by Susskind and Uglum and discussed earlier \cite{Susskind:1994sm}, the tree-level spherical amplitude is expected to contain the tree-level black hole entropy. Microscopically, it corresponds to open strings whose endpoints are fixed on the horizon. This idea is quite general and is expected to apply to any black hole. The intriguing part of this perspective is that it gives a property of the background itself (the tree-level black hole entropy) from a stringy perturbative computation (the spherical amplitude is computed in the fixed black hole geometry). Unfortunately, the spherical amplitude proves to be quite difficult to compute. \\

\noindent A second set of approaches is to view the black hole itself as being built up by some stringy building blocks.\footnote{It should be mentioned here that an alternative approach is the fuzzball program (see e.g. \cite{Mathur:2005zp} and references therein) where the black hole horizon itself is absent in string theory but only arises in an effective coarse-grained way. This approach has received increased interest lately due to its implications on the firewall paradox. We will not discuss this approach further. } Two main routes have been explored here. \\
The first viewpoint is that of long strings. Susskind argued in the past \cite{Susskind:1993ws} that long strings really can account for the black hole entropy of uncharged black holes. \\
The second route, pioneered by Strominger and Vafa \cite{Strominger:1996sh}, uses the beautiful idea that D-branes and their open string excitations can precisely account for the degrees of freedom in certain extremal and near-extremal black holes. Varying the string coupling constant for these BPS objects, allows a direct interpretation in terms of the D-brane building blocks of the supergravity black hole solution. Of course, these considerations in general need RR-charged black holes and a high degree of supersymmetry. \\
These two different ideas of building the black hole through stringy building blocks, are linked by the work of Horowitz and Polchinski in their correspondence principle \cite{Horowitz:1996nw}. They argue that uncharged and NS-NS charged black holes are built from long fundamental strings, whereas RR-charged black holes are built from D-brane states and their excitations. \\

\noindent In our approach as given above, we are studying uncharged black holes and hence expect long strings to carry the entropy. This is indeed borne out in the previous discussion of the entropy. \\

\noindent A very important open problem (that has been open for two decades now) is to relate the Susskind-Uglum picture of the open strings stuck at the horizon to the D-brane picture of open strings moving on the brane worldvolume. As far as the author can tell, no significant progress has been made on this front yet.

\section{Summary}
\label{summary}
Let us summarize the results obtained in this chapter.
\begin{itemize}
\item{The thermal scalar energy-momentum tensor captures the energy-momentum tensor of an average of highly excited Lorentzian strings.}
\item{The $U(1)$ charge symmetry of the thermal scalar action leads to a classical Noether current which is the same as the (time-averaged) string charge tensor of an average of highly excited Lorentzian strings.}
\item{One can readily extend these results to the canonical thermal average of the energy-momentum tensor or the string charge. We find expressions with the same spatial distribution. Our main result is equation (\ref{mainresult}) for the energy-momentum tensor in the canonical ensemble.}
\item{A special class of correlators can also be computed using these methods, where each individual stress tensor is time-averaged. The near-Hagedorn dominant behavior (for a fully compact space) is simply the product of the expectation values.}
\item{We have demonstrated these results on three spacetimes. We briefly looked at flat spacetime, demonstrating the presence of a pressureless state of matter at the Hagedorn temperature. $AdS_3$ showed that a non-trivial string charge is possible for backgrounds including Kalb-Ramond fields. Finally we looked at Rindler spacetime, where a negative-energy state of matter was found living close to the event horizon.}
\item{Finally, we presented a short demonstration that $T_H = T_{\text{Hawking}}$ implies that long strings can indeed account fully for the tree-level black hole entropy.}
\end{itemize}
In general we conclude that quantum properties of the long string are translated into classical properties of the thermal scalar.

\section{*Total energy of matter in a stationary spacetime}
\label{stationary}
\subsection{Local analysis}
Consider an observer in a stationary spacetime moving along a trajectory tangential to $\frac{\partial}{\partial t}$. Hence his 4-velocity equals $u^{\mu} = \frac{1}{\sqrt{-G_{00}}} \frac{\partial}{\partial t}$, where the norm is fixed by requiring $u^{\mu}u_{\mu} = -1$. The energy density as measured by such an observer equals $T^{\mu\nu}u_{\mu}u_{\nu}$. In writing these expressions we are assuming that this vector is globally timelike in the globally hyperbolic section of the spacetime onto which we are focusing; this excludes for instance generic Kerr-Newman black holes. Note though that there exist rotating black holes which do have globally timelike Killing vectors (such as a subclass of Kerr-AdS black holes \cite{Hawking:1999dp}\cite{Winstanley:2001nx}). Before continuing, we make the transition to the Euclidean signature manifold. This (Euclidean) energy density can be expanded as
\begin{equation}
\label{total}
T^{\mu\nu}u_{\mu}u_{\nu} = T^{\tau\tau}G_{\tau\tau} + 2 T^{\tau i} G_{i\tau} + T^{ij}\frac{G_{i\tau}G_{j\tau}}{G_{\tau\tau}}.
\end{equation}
and it is related to the Lorentzian energy density by a sign change. \\

\noindent First we compute only $T^{\tau\tau}G_{\tau\tau} + T^{\tau i} G_{i\tau} = T^{\tau}_{\tau}$. We will show that this is precisely the total energy. Thus we consider
\begin{align}
-\int \left\langle T_{\tau,e}^{\tau}\right\rangle_E \sqrt{G} dV &= -\int \left(G_{\tau\tau}\left\langle T^{\tau\tau}_{e}\right\rangle_E + G_{\tau k}\left\langle T^{\tau k}_{e}\right\rangle_E\right) \sqrt{G} dV.
\end{align}
This expression has three different contributions. Part of the first term is the same as in the static case and yields the total energy:
\begin{align}
\frac{E}{\beta_H^{2}} \frac{\int dV \sqrt{G} G_{\tau\tau}\frac{\beta_H^2 TT^*}{4\pi^2\alpha'^2} }{\left.\frac{\partial \lambda_0}{\partial \beta^2}\right|_{\beta = \beta_H}} = E. 
\end{align}
For the other terms we also need
\begin{align}
\frac{e^{2\Phi}}{\sqrt{G}}\frac{\delta \lambda_0}{\delta G_{\tau i}} &= -\left( G^{ki} G^{\tau l} + (k \leftrightarrow l)\right)\left(\frac{\partial_k T^* \partial_{l} T + \partial_{l}T^* \partial_{k} T}{4}\right) \nonumber \\
&+ \frac{1}{2}G^{\tau i}\left[G^{kl}\partial_k T \partial_l T^* + \frac{\beta^2G_{\tau\tau}-\beta_{H,\text{flat}}^2}{4\pi^2\alpha'^2}TT^*\right].
\end{align}
A second term (proportional to the Lagrangian), vanishes after using $G_{\tau\tau}G^{\tau\tau} + G_{\tau i}G^{\tau i} = 1$ due to the fact that the thermal scalar is a zero-mode. \\
Finally, the remaining term equals
\begin{align}
\frac{E}{\beta_H^{2}} \frac{\int dV \sqrt{G} \left[ - G_{\tau\tau}G^{\tau i}G^{\tau j}\partial_i T \partial_j T^* - \frac{G_{\tau k}}{2}\left(G^{ki}G^{\tau j} + G^{j k}G^{\tau i}\right)\partial_i T \partial_j T^*\right]}{\left.\frac{\partial \lambda_0}{\partial \beta^2}\right|_{\beta = \beta_H}}.
\end{align}
Upon using $G_{\tau k}G^{i k}+ G_{\tau\tau}G^{\tau i} =0$, one readily finds this term vanishes as well. In all, one finds that $-\left\langle T_{\tau,e}^{\tau}\right\rangle_E$ represents the energy density. \\

\noindent Next we show that the remainder of (\ref{total}) vanishes. One readily obtains
\begin{equation}
\frac{\delta \bar{G}^{ij}}{\delta G_{kl}} = - \frac{G^{ik} G^{jl} + (i \leftrightarrow j)}{2}.
\end{equation}
Using this result we obtain
\begin{align}
\frac{e^{2\Phi}}{\sqrt{G}}\frac{\delta \lambda_0}{\delta G_{kl}} &= -\left( G^{ik} G^{j l} + (i \leftrightarrow j)\right)\left(\frac{\partial_i T^* \partial_{j} T + \partial_{j}T^* \partial_{i} T}{4}\right) \nonumber \\
&+ \frac{1}{2}G^{kl}\left[G^{ij}\partial_i T \partial_j T^* + \frac{\beta^2G_{\tau\tau}-\beta_{H,\text{flat}}^2}{4\pi^2\alpha'^2}TT^*\right].
\end{align}
The term on the second line of this expression combines with the second term of $T^{\tau i} G_{i\tau}$, with the factor in front of the square brackets yielding
\begin{equation}
G_{\tau i}G^{\tau i} + G^{kl}\frac{G_{k\tau}G_{l\tau}}{G_{\tau\tau}} = G_{\tau i}G^{\tau i} - G^{\tau k}G_{\tau k} = 0.
\end{equation}
The other terms combine into
\begin{align}
\frac{E}{\beta_H^{2}} \frac{\int dV \sqrt{G} \left[ - \frac{G_{\tau k}}{2}\left(G^{ki}G^{\tau j} + G^{j k}G^{\tau i}\right)\partial_i T \partial_j T^* - \frac{G_{\tau k}G_{\tau l}}{2G_{\tau\tau}}\left(G^{ik}G^{jl} + G^{j k}G^{il}\right)\partial_i T \partial_j T^*\right]}{\left.\frac{\partial \lambda_0}{\partial \beta^2}\right|_{\beta = \beta_H}}.
\end{align}
The prefactors (for fixed $i$ and $j$) can be seen to vanish by again using $G_{\tau l}G^{jl} + G_{\tau\tau}G^{\tau j} = 0$.

\subsection{Komar integral}
\begin{figure}[h]
\centering
\includegraphics[width=0.5\textwidth]{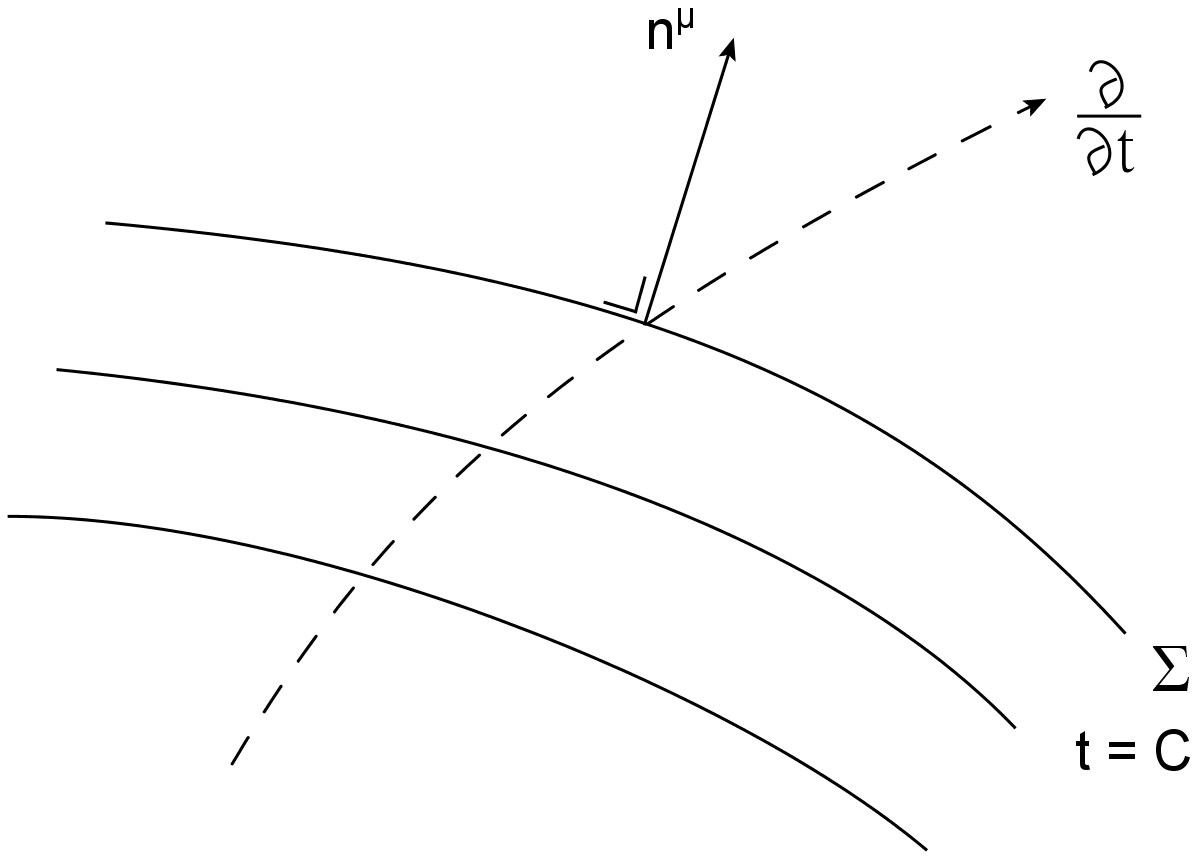}
\caption{Constant time hypersurfaces and their normal 4-vector. Also shown is the flow of the timelike Killing vector. For a non-static spacetime, these vectors are not parallel.}
\label{curvilinear}
\end{figure}

There exists an alternative method of determining the total matter energy according to the asymptotic observer, the Komar integral:
\begin{equation}
\label{Kom}
E= \int_{\Sigma} T^{\mu\nu}\xi_{\mu}n_{\nu} d\Sigma,
\end{equation}
where we integrate over a spacelike hypersurface with normal $n^{\mu}$ and where $\xi^{\mu}$ is the timelike Killing vector. For static spacetimes it is readily seen that this definition coincides with the description used above, but for stationary spacetimes this seems not immediate. Let us consider this case in detail. A defining feature of non-static (but stationary) spacetimes is the fact that the normal vector to spatial slices is not parallel to the tangent of the timelike Killing vector. This implies that one can define two different notions of being constant in time. A sketch of this situation is given in figure \ref{curvilinear}. 
The 4-vector $n^{\mu}$ is normal to the constant time hypersurfaces and is hence of the form:
\begin{equation}
n^{\mu} = \frac{\nabla^{\mu} \tau}{\left(\nabla^{\nu} \tau \nabla_{\nu} \tau\right)^{1/2}} = \frac{G^{\tau\mu}}{\sqrt{G^{\tau\tau}}}.
\end{equation}
Its covariant components are readily found to be
\begin{align}
n_\tau &= \frac{G_{\tau\tau}G^{\tau\tau}}{\sqrt{G^{\tau\tau}}} + \frac{G_{\tau i}G^{\tau i}}{\sqrt{G^{\tau\tau}}} = \frac{1}{\sqrt{G^{\tau\tau}}}, \\
n_i &= \frac{G_{i\tau}G^{\tau\tau}}{\sqrt{G^{\tau\tau}}} + \frac{G_{i j}G^{\tau j}}{\sqrt{G^{\tau\tau}}} = 0.
\end{align}
The other ingredient is the timelike Killing vector. Its covariant components are
\begin{align}
\xi_\tau &= G_{\tau\tau}, \\
\xi_i &= G_{\tau i}.
\end{align}
The spacelike slice $\Sigma$ is integrated over with the pull-back metric $G_{ij}$ (setting $dt=0$ in the full metric). \\
With these formulas, the integrand of (\ref{Kom}) equals
\begin{align}
\sqrt{G_{ij}} T^{\mu\nu}\xi_{\mu}n_{\nu} = \frac{\sqrt{G_{ij}}}{\sqrt{G^{\tau\tau}}}\left[ T^{\tau\tau}G_{\tau\tau} + T^{\tau i}G_{\tau i}\right].
\end{align}
To proceed, we need the following identities:
\begin{align}
\sqrt{G} = \sqrt{G_{ij}}\sqrt{G_{\tau\tau} - G_{\tau k}G_{\tau l}\bar{G}^{kl}}, \\
G^{\tau\tau}\left(G_{\tau\tau} - G_{\tau k}G_{\tau l}\bar{G}^{kl}\right) = 1.
\end{align}
In these formulas, we have written 
\begin{equation}
\bar{G}^{kl} = G^{kl} - \frac{G^{\tau k}G^{\tau l}}{G^{\tau\tau}},
\end{equation}
which is the matrix inverse of the purely spatial matrix $G_{ij}$. \\

\noindent We hence have 
\begin{align}
\sqrt{G_{ij}} T^{\mu\nu}\xi_{\mu}n_{\nu} = \sqrt{G} T^{\tau}_{\tau},
\end{align}
precisely the same as before. \\

\noindent The physical interpretation of the previous result is that we should redshift the local energy density $T_{\mu\nu}u^{\mu}u^{\nu}$ on the spatial slices $\Sigma$. For stationary spacetimes, the redshift factor is $\sqrt{G_{\tau\tau}}$ which combines with the spatial metric as defined by observers moving along $\frac{\partial}{\partial t}$ to yield the total background metric:
\begin{equation}
\sqrt{G} = \sqrt{G_{\tau\tau}}\sqrt{G_{ij}-\frac{G_{\tau i}G_{\tau j}}{G_{\tau\tau}}}.
\end{equation}
We conclude that also for stationary spacetimes we have that
\begin{equation}
\boxed{
E = \int_{\Sigma} T^{\mu\nu}\xi_{\mu}n_{\nu} \sqrt{G_{ij}} d\Sigma = \int_{\Sigma} T^{\mu\nu}u_{\mu}u_{\nu} \sqrt{G} d\Sigma}.
\end{equation}
The first equality is the global Komar analysis, whereas the second expression is interpreted as summing (and redshifting) the local energy density over the entire spatial slice using the spatial metric as constructed by the radar definition of the stationary observers. \\
This story applies to string theory, in spite of the fact that the dominant near-Hagedorn behavior implies non-local long strings. The crucial point is that this long string is effectively treated as a Euclidean quantum field whose energy density can be viewed as local in space.

\chapter{Conclusion}

High-temperature string thermodynamics remains a mysterious phase of nature. In the eighties, physicists showed that a critical temperature, the Hagedorn temperature, emerges in the canonical ensemble, indicating either a phase transition or precluding further heating of the string gas. Close to this temperature, strings tend to coalesce and form long highly excited strings, describing a random walk in space. The main piece of evidence for this random walk picture is that the high energy density of states precisely matches the number of random walks of fixed length $L$, proportional to the energy of the string. An alternative description can be given on the thermal manifold. In field theory, one describes the thermal ensemble as a vacuum amplitude on this thermal manifold. In string theory, two possibilities emerge. Firstly, one can describe string theory as a sum of fields and one is then naturally led to considering the genus one amplitude on the thermal manifold on the modular strip with only one wrapping number. The Hagedorn divergence arises from the small $\tau_2$ region from wrapping number one. Alternatively, one can really consider string theory on the thermal manifold and compute the torus amplitude on the modular fundamental domain with both wrapping numbers, manifestly exhibiting modular invariance. The Hagedorn divergence arises in this language from the large $\tau_2$ region as a thermal state, singly wound around the Euclidean time direction, becoming massless. This state is called the thermal scalar. It turns out however, that both descriptions of thermodynamics are actually identical.\\
Our main goal in this work was to find out to what extent this story can be generalized to curved spacetimes. \\
The random walk picture was obtained in chapter \ref{chth} where we combined first quantized and second quantized descriptions to obtain a description of the random walk. It turned out that the one-loop amplitude of the thermal scalar field on the spatial submanifold precisely gives the dominant contribution to the free energy of the string gas, showing that the trajectory of the thermal scalar particle in its loop should be interpreted as the spatial form of the random walking string. \\
We further analyzed how the random walk path integral operated by looking into the linear dilaton space, toroidal models of flat space and a simple $\mathbb{Z}_2$-orbifold model. \\

\noindent Black holes provide one of the most intriguing objects in theoretical physics and they are still ill-understood at the quantum level. This was made very explicit by the discovery of the so-called firewall paradox a few years ago. \\
\noindent Using conventional wisdom, in describing the infalling vacuum according to fiducial observers, one miraculously finds a thermal population of the energy levels (Unruh effect). The locally measured temperature increases as one approaches the horizon and goes all the way to infinity. Naively, one would expect some Hagedorn phenomenon to occur very close to the black hole horizon. \\
In chapter \ref{chri} we set out to investigate this matter from the thermal scalar perspective. We found the important result that $\beta_H = \beta_{\text{Hawking}}$ such that the string gas around the black hole behaves precisely the same as the Hagedorn gas in flat space. Note that the Hagedorn temperature being only black hole scale (and not string scale) is not a contradiction. The Hawking temperature is measured at infinity, but the locally measured temperature is arbitrarily high. Thus locally string-scale temperatures are indeed reached at the Hawking temperature. \\

\noindent In chapter \ref{chwzw} we investigated whether our starting point (the thermal scalar state predicts $T_H$ and is the dominant high-temperature mode) is valid in a curved background. The $AdS_3$ WZW model has the advantage of being a group manifold for which the string path integral at genus one has been computed explicitly. We determined the thermal spectrum and indeed found the thermal scalar was responsible for the Hagedorn divergence. However, infinitely many states turned out to become marginal simultaneously at $T_H$ (in agreement with the form of the high-energy density of states). For the BTZ WZW model, a more puzzling situation arises: no wound states are present at all! To further analyze these peculiarities, we also analyzed $\mathbb{Z}_N$ orbifold models (related to the correct analytical continuation of Poisson's summation formula) and chemical potentials in $AdS_3$ (related to the entire $SL(2,\mathbb{Z})$ family of black holes in the Farey Tail story). \\

\noindent The last few chapters contain the conclusions we drew from these example models. In chapter \ref{chgentd} we discussed some generalities on the thermal scalar action and in particular we suggested that for type II superstrings, no corrections to this action appear to be present. \\
For black holes, the fact that $T_H = T_{\text{Hawking}}$ has some very deep consequences. Aspects of this and related lessons on black holes were given in chapter \ref{chgenbh}. Firstly, one cannot neglect the massive string modes around the black hole when computing thermodynamical quantities. After all, these will give the random walker around the horizon. Secondly, the perturbation series in $g_s$ appears to break down at the Hawking temperature, suggesting one should turn to non-perturbative methods to analyze the problem further. 
We also provided some details on the different path integrals one can define on a cigar-shaped manifold. In particular this is closely related to Susskind's and Uglum's strange open strings stuck on the horizon. We illustrated how these predictions might be manifested in concrete computations. Then we speculated how these developments can be related to a firewall-like paradox. We argued that string theory both provides a firewall and then demonstrates how to destroy it again and restore the equivalence principle. \\

\noindent Finally in chapter \ref{chrel} we examined whether the thermal scalar is of any use beyond simply determing the dominant thermodynamics. It turns out that also the high temperature stress tensor and string charge can be written in terms of the (classical!) stress tensor and string charge of the thermal scalar. We provided three examples to test these formulas: the flat space case where a pressureless fluid was obtained, $AdS_3$ where a non-zero string charge is present and finally Rindler space where we obtained a zone of negative energy density close to the horizon. We applied these formulas to motivate a derivation of the Bekenstein-Hawking entropy based on equilibrating long strings near the black hole. It turned out that, because $T_H = T_{\text{Hawking}}$, the long strings carry precisely enough entropy to account for the black hole entropy. \\

\noindent Many important issues still require further research. Throughout the chapters we already presented several open avenues for future research. Here we only mention a few broader goals. \\
The relation with the firewall problem we presented in chapter \ref{chgenbh} is interesting to study further, especially in the context of the different partition functions that appear to be possible in cigar-shaped backgrounds. It would be interesting to have a better handle on these partition functions and how their field content is precisely encoded. \\
The Rindler problem is not fully solved and several important questions and checks still await solving. For instance it would be interesting to analyze in somewhat more detail the way the discrete states are incorporated in the continuum to obtain the 2d flat partition function in the end. Also, a full analysis of the Lorentzian Rindler case would be a highly valuable result. Unfortunately, our understanding is more rudimentary here as for instance no succesful path integral computation has been performed in the literature. The Lorentzian Rindler spectrum would also allow us to compute the high energy density of states and confirm the Hagedorn temperature from the Lorentzian perspective. \\
Another interesting point is the relation to fluid dynamics on the stretched horizon. In particular we would like to extract in some way the Brown-York boundary stress tensor directly from the highly excited string on the stretched horizon. Research is underway in line with this goal. \\
Finally, the higher genus results suggest string perturbation theory, as a method of computing thermodynamical quantities, seems to break down in black hole geometries. This leads us to non-perturbative approaches, such as matrix models, $AdS/CFT$ and M-theory to further analyze the string gas around black holes. There exist some results on this (see e.g. \cite{Kazakov:2000pm} and \cite{Russo:2001vh}) but more extensive research into this is needed. \\

\noindent As a final remark, it is clear that black hole horizons are not (completely) understood at the quantum level and a sizeable portion of the coming years in theoretical physics will undoubtably be devoted towards unraveling some of these mysteries that lurk behind the horizon.

\bibliography{bibfile}

\providecommand{\href}[2]{#2}\begingroup\raggedright\begin{thebibliography}{100}

\bibitem{Mertens:2013pza}
T.~G. Mertens, H.~Verschelde, and V.~I. Zakharov, ``{Near-Hagedorn
  Thermodynamics and Random Walks: a General Formalism in Curved
  Backgrounds},'' \href{http://dx.doi.org/10.1007/JHEP02(2014)127}{{\em JHEP}
  {\bfseries 1402} (2014) 127},
\href{http://arxiv.org/abs/1305.7443}{{\ttfamily arXiv:1305.7443 [hep-th]}}.

\bibitem{Mertens:2013zya}
T.~G. Mertens, H.~Verschelde, and V.~I. Zakharov, ``{Random Walks in Rindler
  Spacetime and String Theory at the Tip of the Cigar},''
  \href{http://dx.doi.org/10.1007/JHEP03(2014)086}{{\em JHEP} {\bfseries 1403}
  (2014) 086},
\href{http://arxiv.org/abs/1307.3491}{{\ttfamily arXiv:1307.3491 [hep-th]}}.

\bibitem{Mertens:2014nca}
T.~G. Mertens, H.~Verschelde, and V.~I. Zakharov, ``{The thermal scalar and
  random walks in $AdS_3$ and $BTZ$},''
  \href{http://dx.doi.org/10.1007/JHEP06(2014)156}{{\em JHEP} {\bfseries 1406}
  (2014) 156},
\href{http://arxiv.org/abs/1402.2808}{{\ttfamily arXiv:1402.2808 [hep-th]}}.

\bibitem{Mertens:2014cia}
T.~G. Mertens, H.~Verschelde, and V.~I. Zakharov, ``{Near-Hagedorn
  Thermodynamics and Random Walks - Extensions and Examples},''
  \href{http://dx.doi.org/10.1007/JHEP11(2014)107}{{\em JHEP} {\bfseries 1411}
  (2014) 107},
\href{http://arxiv.org/abs/1408.6999}{{\ttfamily arXiv:1408.6999 [hep-th]}}.

\bibitem{Mertens:2014dia}
T.~G. Mertens, H.~Verschelde, and V.~I. Zakharov, ``{On the Relevance of the
  Thermal Scalar},'' \href{http://dx.doi.org/10.1007/JHEP11(2014)157}{{\em
  JHEP} {\bfseries 1411} (2014) 157},
\href{http://arxiv.org/abs/1408.7012}{{\ttfamily arXiv:1408.7012 [hep-th]}}.

\bibitem{Mertens:2014saa}
T.~G. Mertens, H.~Verschelde, and V.~I. Zakharov, ``{Perturbative String
  Thermodynamics near Black Hole Horizons},''
\href{http://arxiv.org/abs/1410.8009}{{\ttfamily arXiv:1410.8009 [hep-th]}}.

\bibitem{Mertens:2015hia}
T.~G. Mertens, H.~Verschelde, and V.~I. Zakharov, ``{The long string at the
  stretched horizon and the entropy of large non-extremal black holes},''
\href{http://arxiv.org/abs/1505.04025}{{\ttfamily arXiv:1505.04025 [hep-th]}}.

\bibitem{Dudal:2014jfa}
D.~Dudal and T.~G. Mertens, ``{The melting of charmonium in a magnetic field
  from an effective AdS/QCD model},''
\href{http://arxiv.org/abs/1410.3297}{{\ttfamily arXiv:1410.3297 [hep-th]}}.

\bibitem{Mitchell:1987hr}
D.~Mitchell and N.~Turok, ``{Statistical Mechanics of Cosmic Strings},''
\href{http://dx.doi.org/10.1103/PhysRevLett.58.1577}{{\em Phys.Rev.Lett.}
  {\bfseries 58} (1987) 1577}.

\bibitem{Mitchell:1987th}
D.~Mitchell and N.~Turok, ``{Statistical Properties of Cosmic Strings},''
\href{http://dx.doi.org/10.1016/0550-3213(87)90626-2}{{\em Nucl.Phys.}
  {\bfseries B294} (1987) 1138}.

\bibitem{Bowick:1989us}
M.~J. Bowick and S.~B. Giddings, ``{High Temperature Strings},''
\href{http://dx.doi.org/10.1016/0550-3213(89)90500-2}{{\em Nucl.Phys.}
  {\bfseries B325} (1989) 631}.

\bibitem{Deo:1989bv}
N.~Deo, S.~Jain, and C.-I. Tan, ``{String Statistical Mechanics Above Hagedorn
  Energy Density},''
\href{http://dx.doi.org/10.1103/PhysRevD.40.2626}{{\em Phys.Rev.} {\bfseries
  D40} (1989) 2626}.

\bibitem{Atick:1988si}
J.~J. Atick and E.~Witten, ``{The Hagedorn Transition and the Number of Degrees
  of Freedom of String Theory},''
\href{http://dx.doi.org/10.1016/0550-3213(88)90151-4}{{\em Nucl.Phys.}
  {\bfseries B310} (1988) 291--334}.

\bibitem{Horowitz:1997jc}
G.~T. Horowitz and J.~Polchinski, ``{Selfgravitating fundamental strings},''
  \href{http://dx.doi.org/10.1103/PhysRevD.57.2557}{{\em Phys.Rev.} {\bfseries
  D57} (1998) 2557--2563},
\href{http://arxiv.org/abs/hep-th/9707170}{{\ttfamily arXiv:hep-th/9707170
  [hep-th]}}.

\bibitem{Barbon:2004dd}
J.~Barbon and E.~Rabinovici, ``{Touring the Hagedorn ridge},''
\href{http://arxiv.org/abs/hep-th/0407236}{{\ttfamily arXiv:hep-th/0407236
  [hep-th]}}.

\bibitem{Polchinski:1998rq}
J.~Polchinski, {\em {String theory. Vol. 1: An introduction to the bosonic
  string}}.
\newblock Cambridge University Press,
1998.
\newblock

\bibitem{Zwiebach:2004tj}
B.~Zwiebach, {\em {A first course in string theory}}.
\newblock Cambridge University Press,
2004.
\newblock

\bibitem{Polchinski:1995mt}
J.~Polchinski, ``{Dirichlet Branes and Ramond-Ramond charges},''
  \href{http://dx.doi.org/10.1103/PhysRevLett.75.4724}{{\em Phys.Rev.Lett.}
  {\bfseries 75} (1995) 4724--4727},
\href{http://arxiv.org/abs/hep-th/9510017}{{\ttfamily arXiv:hep-th/9510017
  [hep-th]}}.

\bibitem{Maldacena:1997re}
J.~M. Maldacena, ``{The Large N limit of superconformal field theories and
  supergravity},'' \href{http://dx.doi.org/10.1023/A:1026654312961}{{\em
  Int.J.Theor.Phys.} {\bfseries 38} (1999) 1113--1133},
\href{http://arxiv.org/abs/hep-th/9711200}{{\ttfamily arXiv:hep-th/9711200
  [hep-th]}}.

\bibitem{Witten:1995ex}
E.~Witten, ``{String theory dynamics in various dimensions},''
  \href{http://dx.doi.org/10.1016/0550-3213(95)00158-O}{{\em Nucl.Phys.}
  {\bfseries B443} (1995) 85--126},
\href{http://arxiv.org/abs/hep-th/9503124}{{\ttfamily arXiv:hep-th/9503124
  [hep-th]}}.

\bibitem{Banks:1996vh}
T.~Banks, W.~Fischler, S.~Shenker, and L.~Susskind, ``{M theory as a matrix
  model: A Conjecture},''
  \href{http://dx.doi.org/10.1103/PhysRevD.55.5112}{{\em Phys.Rev.} {\bfseries
  D55} (1997) 5112--5128},
\href{http://arxiv.org/abs/hep-th/9610043}{{\ttfamily arXiv:hep-th/9610043
  [hep-th]}}.

\bibitem{Dijkgraaf:1997vv}
R.~Dijkgraaf, E.~P. Verlinde, and H.~L. Verlinde, ``{Matrix string theory},''
  \href{http://dx.doi.org/10.1016/S0550-3213(97)00326-X}{{\em Nucl.Phys.}
  {\bfseries B500} (1997) 43--61},
\href{http://arxiv.org/abs/hep-th/9703030}{{\ttfamily arXiv:hep-th/9703030
  [hep-th]}}.

\bibitem{Callan:1985ia}
J.~Callan, Curtis~G., E.~Martinec, M.~Perry, and D.~Friedan, ``{Strings in
  Background Fields},''
\href{http://dx.doi.org/10.1016/0550-3213(85)90506-1}{{\em Nucl.Phys.}
  {\bfseries B262} (1985) 593}.

\bibitem{Buscher:1987sk}
T.~Buscher, ``{A Symmetry of the String Background Field Equations},''
\href{http://dx.doi.org/10.1016/0370-2693(87)90769-6}{{\em Phys.Lett.}
  {\bfseries B194} (1987) 59}.

\bibitem{Buscher:1987qj}
T.~Buscher, ``{Path Integral Derivation of Quantum Duality in Nonlinear Sigma
  Models},''
\href{http://dx.doi.org/10.1016/0370-2693(88)90602-8}{{\em Phys.Lett.}
  {\bfseries B201} (1988) 466}.

\bibitem{Rocek:1991ps}
M.~Rocek and E.~P. Verlinde, ``{Duality, quotients, and currents},''
  \href{http://dx.doi.org/10.1016/0550-3213(92)90269-H}{{\em Nucl.Phys.}
  {\bfseries B373} (1992) 630--646},
\href{http://arxiv.org/abs/hep-th/9110053}{{\ttfamily arXiv:hep-th/9110053
  [hep-th]}}.

\bibitem{Tseytlin:1992ri}
A.~A. Tseytlin, ``{Effective action of gauged WZW model and exact string
  solutions},'' \href{http://dx.doi.org/10.1016/0550-3213(93)90511-M}{{\em
  Nucl.Phys.} {\bfseries B399} (1993) 601--622},
\href{http://arxiv.org/abs/hep-th/9301015}{{\ttfamily arXiv:hep-th/9301015
  [hep-th]}}.

\bibitem{Tseytlin:1993my}
A.~A. Tseytlin, ``{Conformal sigma models corresponding to gauged
  Wess-Zumino-Witten theories},''
  \href{http://dx.doi.org/10.1016/0550-3213(94)90461-8}{{\em Nucl.Phys.}
  {\bfseries B411} (1994) 509--558},
\href{http://arxiv.org/abs/hep-th/9302083}{{\ttfamily arXiv:hep-th/9302083
  [hep-th]}}.

\bibitem{Kaloper:1997ux}
N.~Kaloper and K.~A. Meissner, ``{Duality beyond the first loop},''
  \href{http://dx.doi.org/10.1103/PhysRevD.56.7940}{{\em Phys.Rev.} {\bfseries
  D56} (1997) 7940--7953},
\href{http://arxiv.org/abs/hep-th/9705193}{{\ttfamily arXiv:hep-th/9705193
  [hep-th]}}.

\bibitem{Garousi:2013gea}
M.~R. Garousi, A.~Ghodsi, T.~Houri, and G.~Jafari, ``{T-duality of D-brane
  action at order $\alpha'$ in bosonic string theory},''
  \href{http://dx.doi.org/10.1007/JHEP10(2013)103}{{\em JHEP} {\bfseries 1310}
  (2013) 103},
\href{http://arxiv.org/abs/1308.4609}{{\ttfamily arXiv:1308.4609 [hep-th]}}.

\bibitem{Hagedorn:1965st}
R.~Hagedorn, ``{Statistical thermodynamics of strong interactions at
  high-energies},''
{\em Nuovo Cim.Suppl.} {\bfseries 3} (1965) 147--186.

\bibitem{Frautschi:1971ij}
S.~C. Frautschi, ``{Statistical bootstrap model of hadrons},''
\href{http://dx.doi.org/10.1103/PhysRevD.3.2821}{{\em Phys.Rev.} {\bfseries D3}
  (1971) 2821--2834}.

\bibitem{Carlitz:1972uf}
R.~D. Carlitz, ``{Hadronic matter at high density},''
\href{http://dx.doi.org/10.1103/PhysRevD.5.3231}{{\em Phys.Rev.} {\bfseries D5}
  (1972) 3231--3242}.

\bibitem{Abel:1999rq}
S.~Abel, J.~Barbon, I.~Kogan, and E.~Rabinovici, ``{String thermodynamics in
  D-brane backgrounds},''
  \href{http://dx.doi.org/10.1088/1126-6708/1999/04/015}{{\em JHEP} {\bfseries
  9904} (1999) 015},
\href{http://arxiv.org/abs/hep-th/9902058}{{\ttfamily arXiv:hep-th/9902058
  [hep-th]}}.

\bibitem{Manes:2004nd}
J.~L. Manes, ``{Portrait of the string as a random walk},''
  \href{http://dx.doi.org/10.1088/1126-6708/2005/03/070}{{\em JHEP} {\bfseries
  0503} (2005) 070},
\href{http://arxiv.org/abs/hep-th/0412104}{{\ttfamily arXiv:hep-th/0412104
  [hep-th]}}.

\bibitem{Athanasiu:1988st}
G.~Athanasiu and J.~Atick, ``{On the nature of the Hagedorn temperature},''
\href{http://dx.doi.org/10.1016/0920-5632(89)90017-0}{{\em
  Nucl.Phys.Proc.Suppl.} {\bfseries 11} (1989) 304--315}.

\bibitem{Polchinski:1985zf}
J.~Polchinski, ``{Evaluation of the One Loop String Path Integral},''
\href{http://dx.doi.org/10.1007/BF01210791}{{\em Commun.Math.Phys.} {\bfseries
  104} (1986) 37}.

\bibitem{McClain:1986id}
B.~McClain and B.~D.~B. Roth, ``{Modular Invariance for Interacting Bosonic
  Strings at Finite Temperature},''
\href{http://dx.doi.org/10.1007/BF01219073}{{\em Commun.Math.Phys.} {\bfseries
  111} (1987) 539}.

\bibitem{O'Brien:1987pn}
K.~O'Brien and C.~Tan, ``{Modular Invariance of Thermopartition Function and
  Global Phase Structure of Heterotic String},''
\href{http://dx.doi.org/10.1103/PhysRevD.36.1184}{{\em Phys.Rev.} {\bfseries
  D36} (1987) 1184}.

\bibitem{Alvarez:1986sj}
E.~Alvarez and M.~Osorio, ``{Superstrings at Finite Temperature},''
\href{http://dx.doi.org/10.1103/PhysRevD.36.1175}{{\em Phys.Rev.} {\bfseries
  D36} (1987) 1175}.

\bibitem{Liu:2014nva}
L.~Liu, ``{Lecture notes on thermodynamics of ideal string gases and its
  application in cosmology},''
\href{http://arxiv.org/abs/1412.2059}{{\ttfamily arXiv:1412.2059 [hep-th]}}.

\bibitem{Giveon:2013ica}
A.~Giveon and N.~Itzhaki, ``{String theory at the tip of the cigar},''
  \href{http://dx.doi.org/10.1007/JHEP09(2013)079}{{\em JHEP} {\bfseries 1309}
  (2013) 079},
\href{http://arxiv.org/abs/1305.4799}{{\ttfamily arXiv:1305.4799 [hep-th]}}.

\bibitem{Gross:1982cv}
D.~Gross, M.~Perry, and L.~Yaffe, ``{Instability of Flat Space at Finite
  Temperature},''
\href{http://dx.doi.org/10.1103/PhysRevD.25.330}{{\em Phys.Rev.} {\bfseries
  D25} (1982) 330--355}.

\bibitem{Barbon:2001di}
J.~Barbon and E.~Rabinovici, ``{Closed string tachyons and the Hagedorn
  transition in AdS space},''
  \href{http://dx.doi.org/10.1088/1126-6708/2002/03/057}{{\em JHEP} {\bfseries
  0203} (2002) 057},
\href{http://arxiv.org/abs/hep-th/0112173}{{\ttfamily arXiv:hep-th/0112173
  [hep-th]}}.

\bibitem{Abel:1999dy}
S.~Abel, J.~Barbon, I.~Kogan, and E.~Rabinovici, ``{Some thermodynamical
  aspects of string theory},''
\href{http://arxiv.org/abs/hep-th/9911004}{{\ttfamily arXiv:hep-th/9911004
  [hep-th]}}.

\bibitem{Barbon:1998ix}
J.~Barbon and E.~Rabinovici, ``{Extensivity versus holography in anti-de Sitter
  spaces},'' \href{http://dx.doi.org/10.1016/S0550-3213(98)00824-4}{{\em
  Nucl.Phys.} {\bfseries B545} (1999) 371--384},
\href{http://arxiv.org/abs/hep-th/9805143}{{\ttfamily arXiv:hep-th/9805143
  [hep-th]}}.

\bibitem{Barbon:1998cr}
J.~Barbon, I.~Kogan, and E.~Rabinovici, ``{On stringy thresholds in SYM / AdS
  thermodynamics},''
  \href{http://dx.doi.org/10.1016/S0550-3213(98)00868-2}{{\em Nucl.Phys.}
  {\bfseries B544} (1999) 104--144},
\href{http://arxiv.org/abs/hep-th/9809033}{{\ttfamily arXiv:hep-th/9809033
  [hep-th]}}.

\bibitem{Kleban:2013wba}
M.~Kleban, A.~Lawrence, M.~M. Roberts, and S.~Storace, ``{Metastability and
  instability in holographic gauge theories},''
  \href{http://dx.doi.org/10.1007/JHEP06(2014)152}{{\em JHEP} {\bfseries 1406}
  (2014) 152},
\href{http://arxiv.org/abs/1312.1312}{{\ttfamily arXiv:1312.1312 [hep-th]}}.

\bibitem{Hawking:1982dh}
S.~Hawking and D.~N. Page, ``{Thermodynamics of Black Holes in anti-De Sitter
  Space},''
\href{http://dx.doi.org/10.1007/BF01208266}{{\em Commun.Math.Phys.} {\bfseries
  87} (1983) 577}.

\bibitem{Kruczenski:2005pj}
M.~Kruczenski and A.~Lawrence, ``{Random walks and the Hagedorn transition},''
  \href{http://dx.doi.org/10.1088/1126-6708/2006/07/031}{{\em JHEP} {\bfseries
  0607} (2006) 031},
\href{http://arxiv.org/abs/hep-th/0508148}{{\ttfamily arXiv:hep-th/0508148
  [hep-th]}}.

\bibitem{Susskind:1993ws}
L.~Susskind, ``{Some speculations about black hole entropy in string theory},''
\href{http://arxiv.org/abs/hep-th/9309145}{{\ttfamily arXiv:hep-th/9309145
  [hep-th]}}.

\bibitem{Susskind:2005js}
L.~Susskind and J.~Lindesay, {\em {An introduction to black holes, information
  and the string theory revolution: The holographic universe}}.
\newblock World Scientific,
2005.
\newblock

\bibitem{Thorne:1986iy}
K.~S. Thorne, R.~Price, and D.~Macdonald, {\em {Black Holes: The Membrane
  Paradigm}}.
\newblock Yale University Press,
1986.
\newblock

\bibitem{Kutasov:2005rr}
D.~Kutasov, ``{Accelerating branes and the string/black hole transition},''
\href{http://arxiv.org/abs/hep-th/0509170}{{\ttfamily arXiv:hep-th/0509170
  [hep-th]}}.

\bibitem{Kutasov:2000jp}
D.~Kutasov and D.~Sahakyan, ``{Comments on the thermodynamics of little string
  theory},'' \href{http://dx.doi.org/10.1088/1126-6708/2001/02/021}{{\em JHEP}
  {\bfseries 0102} (2001) 021},
\href{http://arxiv.org/abs/hep-th/0012258}{{\ttfamily arXiv:hep-th/0012258
  [hep-th]}}.

\bibitem{Dabholkar:2001if}
A.~Dabholkar, ``{Tachyon condensation and black hole entropy},''
  \href{http://dx.doi.org/10.1103/PhysRevLett.88.091301}{{\em Phys.Rev.Lett.}
  {\bfseries 88} (2002) 091301},
\href{http://arxiv.org/abs/hep-th/0111004}{{\ttfamily arXiv:hep-th/0111004
  [hep-th]}}.

\bibitem{Adams:2001sv}
A.~Adams, J.~Polchinski, and E.~Silverstein, ``{Don't panic! Closed string
  tachyons in ALE space-times},''
  \href{http://dx.doi.org/10.1088/1126-6708/2001/10/029}{{\em JHEP} {\bfseries
  0110} (2001) 029},
\href{http://arxiv.org/abs/hep-th/0108075}{{\ttfamily arXiv:hep-th/0108075
  [hep-th]}}.

\bibitem{Okawa:2004rh}
Y.~Okawa and B.~Zwiebach, ``{Twisted tachyon condensation in closed string
  field theory},'' \href{http://dx.doi.org/10.1088/1126-6708/2004/03/056}{{\em
  JHEP} {\bfseries 0403} (2004) 056},
\href{http://arxiv.org/abs/hep-th/0403051}{{\ttfamily arXiv:hep-th/0403051
  [hep-th]}}.

\bibitem{Almheiri:2012rt}
A.~Almheiri, D.~Marolf, J.~Polchinski, and J.~Sully, ``{Black Holes:
  Complementarity or Firewalls?},''
  \href{http://dx.doi.org/10.1007/JHEP02(2013)062}{{\em JHEP} {\bfseries 1302}
  (2013) 062},
\href{http://arxiv.org/abs/1207.3123}{{\ttfamily arXiv:1207.3123 [hep-th]}}.

\bibitem{Braunstein:2009my}
S.~L. Braunstein, S.~Pirandola, and K.~Życzkowski, ``{Better Late than Never:
  Information Retrieval from Black Holes},''
  \href{http://dx.doi.org/10.1103/PhysRevLett.110.101301}{{\em Phys.Rev.Lett.}
  {\bfseries 110} no.~10, (2013) 101301},
\href{http://arxiv.org/abs/0907.1190}{{\ttfamily arXiv:0907.1190 [quant-ph]}}.

\bibitem{Giveon:2012kp}
A.~Giveon and N.~Itzhaki, ``{String Theory Versus Black Hole
  Complementarity},'' \href{http://dx.doi.org/10.1007/JHEP12(2012)094}{{\em
  JHEP} {\bfseries 1212} (2012) 094},
\href{http://arxiv.org/abs/1208.3930}{{\ttfamily arXiv:1208.3930 [hep-th]}}.

\bibitem{Brandenberger:1988aj}
R.~H. Brandenberger and C.~Vafa, ``{Superstrings in the Early Universe},''
\href{http://dx.doi.org/10.1016/0550-3213(89)90037-0}{{\em Nucl.Phys.}
  {\bfseries B316} (1989) 391}.

\bibitem{Deo:1988jj}
N.~Deo, S.~Jain, and C.-I. Tan, ``{Strings at High-energy Densities and Complex
  Temperature},''
\href{http://dx.doi.org/10.1016/0370-2693(89)90024-5}{{\em Phys.Lett.}
  {\bfseries B220} (1989) 125}.

\bibitem{Maldacena:2000kv}
J.~M. Maldacena, H.~Ooguri, and J.~Son, ``{Strings in AdS(3) and the SL(2,R)
  WZW model. Part 2. Euclidean black hole},''
  \href{http://dx.doi.org/10.1063/1.1377039}{{\em J.Math.Phys.} {\bfseries 42}
  (2001) 2961--2977},
\href{http://arxiv.org/abs/hep-th/0005183}{{\ttfamily arXiv:hep-th/0005183
  [hep-th]}}.

\bibitem{Maldacena:2000hw}
J.~M. Maldacena and H.~Ooguri, ``{Strings in AdS(3) and SL(2,R) WZW model 1.:
  The Spectrum},'' \href{http://dx.doi.org/10.1063/1.1377273}{{\em
  J.Math.Phys.} {\bfseries 42} (2001) 2929--2960},
\href{http://arxiv.org/abs/hep-th/0001053}{{\ttfamily arXiv:hep-th/0001053
  [hep-th]}}.

\bibitem{Ferrer:1990na}
E.~Ferrer, E.~Fradkin, and V.~de~la Incera, ``{Effect of a background electric
  field of the Hagedorn temperature},''
\href{http://dx.doi.org/10.1016/0370-2693(90)90293-F}{{\em Phys.Lett.}
  {\bfseries B248} (1990) 281--287}.

\bibitem{Braaten:1995jr}
E.~Braaten and A.~Nieto, ``{Free energy of QCD at high temperature},''
  \href{http://dx.doi.org/10.1103/PhysRevD.53.3421}{{\em Phys.Rev.} {\bfseries
  D53} (1996) 3421--3437},
\href{http://arxiv.org/abs/hep-ph/9510408}{{\ttfamily arXiv:hep-ph/9510408
  [hep-ph]}}.

\bibitem{Braaten:1995cm}
E.~Braaten and A.~Nieto, ``{Effective field theory approach to high temperature
  thermodynamics},'' \href{http://dx.doi.org/10.1103/PhysRevD.51.6990}{{\em
  Phys.Rev.} {\bfseries D51} (1995) 6990--7006},
\href{http://arxiv.org/abs/hep-ph/9501375}{{\ttfamily arXiv:hep-ph/9501375
  [hep-ph]}}.

\bibitem{Gross:1985fr}
D.~J. Gross, J.~A. Harvey, E.~J. Martinec, and R.~Rohm, ``{Heterotic String
  Theory. 1. The Free Heterotic String},''
\href{http://dx.doi.org/10.1016/0550-3213(85)90394-3}{{\em Nucl.Phys.}
  {\bfseries B256} (1985) 253}.

\bibitem{Alvarez:1985fw}
E.~Alvarez, ``{Strings At Finite Temperature},''
\href{http://dx.doi.org/10.1016/0550-3213(86)90514-6}{{\em Nucl.Phys.}
  {\bfseries B269} (1986) 596}.

\bibitem{Khandekar:1986ib}
D.~Khandekar and S.~Lawande, ``{Feynman Path Integrals: Some Exact Results and
  Applications},''
\href{http://dx.doi.org/10.1016/0370-1573(86)90029-3}{{\em Phys.Rept.}
  {\bfseries 137} (1986) 115--229}.

\bibitem{Jones:1971kk}
A.~Jones and G.~Papadopoulos, ``{On the exact propagator},''
\href{http://dx.doi.org/10.1088/0305-4470/4/5/019}{{\em J.Phys.} {\bfseries A4}
  (1971) L86--L89}.

\bibitem{Gao-Feng:2008}
W.~Gao-Feng, L.~Chao-Yun, L.~Zheng-Wen, and Q.~Shui-Jie, ``{Exact solution to
  two-dimensional isotropic charged harmonic oscillator in uniform magnetic
  field in non-commutative phase space},'' {\em Chinese Physics C (HEP and NP)}
  {\bfseries 32, No. 4.} (2008) .

\bibitem{Dijkgraaf:1991ba}
R.~Dijkgraaf, H.~L. Verlinde, and E.~P. Verlinde, ``{String propagation in a
  black hole geometry},''
\href{http://dx.doi.org/10.1016/0550-3213(92)90237-6}{{\em Nucl.Phys.}
  {\bfseries B371} (1992) 269--314}.

\bibitem{Schulgin:2011zb}
W.~Schulgin and J.~Troost, ``{The heterotic string at high temperature (or with
  strong supersymmetry breaking)},''
  \href{http://dx.doi.org/10.1007/JHEP10(2011)047}{{\em JHEP} {\bfseries 1110}
  (2011) 047},
\href{http://arxiv.org/abs/1107.5316}{{\ttfamily arXiv:1107.5316 [hep-th]}}.

\bibitem{Abers:1973qs}
E.~Abers and B.~Lee, ``{Gauge Theories},''
\href{http://dx.doi.org/10.1016/0370-1573(73)90027-6}{{\em Phys.Rept.}
  {\bfseries 9} (1973) 1--141}.

\bibitem{Weinberg:1995mt}
S.~Weinberg, {\em {The Quantum theory of fields. Vol. 1: Foundations}}.
\newblock Cambridge University Press,
1995.
\newblock

\bibitem{Weinberg:1996kr}
S.~Weinberg, {\em {The Quantum theory of fields. Vol. 2: Modern applications}}.
\newblock Cambridge University Press,
1996.
\newblock

\bibitem{Grignani:2001ik}
G.~Grignani, M.~Orselli, and G.~W. Semenoff, ``{The Target space dependence of
  the Hagedorn temperature},''
  \href{http://dx.doi.org/10.1088/1126-6708/2001/11/058}{{\em JHEP} {\bfseries
  0111} (2001) 058},
\href{http://arxiv.org/abs/hep-th/0110152}{{\ttfamily arXiv:hep-th/0110152
  [hep-th]}}.

\bibitem{Grignani:2001hb}
G.~Grignani, M.~Orselli, and G.~W. Semenoff, ``{Matrix strings in a B field},''
  \href{http://dx.doi.org/10.1088/1126-6708/2001/07/004}{{\em JHEP} {\bfseries
  0107} (2001) 004},
\href{http://arxiv.org/abs/hep-th/0104112}{{\ttfamily arXiv:hep-th/0104112
  [hep-th]}}.

\bibitem{Gubser:2000mf}
S.~S. Gubser, S.~Gukov, I.~R. Klebanov, M.~Rangamani, and E.~Witten, ``{The
  Hagedorn transition in noncommutative open string theory},''
  \href{http://dx.doi.org/10.1063/1.1372176}{{\em J.Math.Phys.} {\bfseries 42}
  (2001) 2749--2764},
\href{http://arxiv.org/abs/hep-th/0009140}{{\ttfamily arXiv:hep-th/0009140
  [hep-th]}}.

\bibitem{Sugawara:2012ag}
Y.~Sugawara, ``{Thermodynamics of Superstring on Near-extremal NS5 and
  Effective Hagedorn Behavior},''
  \href{http://dx.doi.org/10.1007/JHEP10(2012)159}{{\em JHEP} {\bfseries 1210}
  (2012) 159},
\href{http://arxiv.org/abs/1208.3534}{{\ttfamily arXiv:1208.3534 [hep-th]}}.

\bibitem{Dienes:2012dc}
K.~R. Dienes, M.~Lennek, and M.~Sharma, ``{Strings at Finite Temperature:
  Wilson Lines, Free Energies, and the Thermal Landscape},''
  \href{http://dx.doi.org/10.1103/PhysRevD.86.066007}{{\em Phys.Rev.}
  {\bfseries D86} (2012) 066007},
\href{http://arxiv.org/abs/1205.5752}{{\ttfamily arXiv:1205.5752 [hep-th]}}.

\bibitem{Burgess:1988qs}
C.~Burgess, N.~Hambli, and A.~Kshirsagar, ``{Strings, Strong Fields and
  Boundaries},''
\href{http://dx.doi.org/10.1088/0264-9381/6/10/016}{{\em Class.Quant.Grav.}
  {\bfseries 6} (1989) 1473}.

\bibitem{Kleinert:2004ev}
H.~Kleinert, {\em {Path Integrals in Quantum Mechanics, Statistics, Polymer
  Physics, and Financial Markets}}.
\newblock World Scientific, 2004.

\bibitem{Weinberg:2008zzc}
S.~Weinberg, {\em {Cosmology}}.
\newblock Oxford University Press,
2008.
\newblock

\bibitem{Gradshteyn:2007}
I.~Gradshteyn and I.~Ryzhik, {\em {Table of Integrals, Series, and Products,
  (Edited by A. Jeffrey and D. Zwillinger)}}.
\newblock Academic Press, seventh~ed., 2007.

\bibitem{Witten:1981gj}
E.~Witten, ``{Instability of the Kaluza-Klein Vacuum},''
\href{http://dx.doi.org/10.1016/0550-3213(82)90007-4}{{\em Nucl.Phys.}
  {\bfseries B195} (1982) 481}.

\bibitem{Aspinwall:1993nu}
P.~S. Aspinwall, B.~R. Greene, and D.~R. Morrison, ``{Calabi-Yau moduli space,
  mirror manifolds and space-time topology change in string theory},''
  \href{http://dx.doi.org/10.1016/0550-3213(94)90321-2}{{\em Nucl.Phys.}
  {\bfseries B416} (1994) 414--480},
\href{http://arxiv.org/abs/hep-th/9309097}{{\ttfamily arXiv:hep-th/9309097
  [hep-th]}}.

\bibitem{Witten:1993yc}
E.~Witten, ``{Phases of N=2 theories in two-dimensions},''
  \href{http://dx.doi.org/10.1016/0550-3213(93)90033-L}{{\em Nucl.Phys.}
  {\bfseries B403} (1993) 159--222},
\href{http://arxiv.org/abs/hep-th/9301042}{{\ttfamily arXiv:hep-th/9301042
  [hep-th]}}.

\bibitem{Tseytlin:1988tv}
A.~A. Tseytlin, ``{Mobius Infinity Subtraction and Effective Action in $\sigma$
  Model Approach to Closed String Theory},''
\href{http://dx.doi.org/10.1016/0370-2693(88)90421-2}{{\em Phys.Lett.}
  {\bfseries B208} (1988) 221}.

\bibitem{Susskind:1994sm}
L.~Susskind and J.~Uglum, ``{Black hole entropy in canonical quantum gravity
  and superstring theory},''
  \href{http://dx.doi.org/10.1103/PhysRevD.50.2700}{{\em Phys.Rev.} {\bfseries
  D50} (1994) 2700--2711},
\href{http://arxiv.org/abs/hep-th/9401070}{{\ttfamily arXiv:hep-th/9401070
  [hep-th]}}.

\bibitem{Balasubramanian:2013rqa}
V.~Balasubramanian, B.~Czech, B.~D. Chowdhury, and J.~de~Boer, ``{The entropy
  of a hole in spacetime},''
  \href{http://dx.doi.org/10.1007/JHEP10(2013)220}{{\em JHEP} {\bfseries 1310}
  (2013) 220},
\href{http://arxiv.org/abs/1305.0856}{{\ttfamily arXiv:1305.0856 [hep-th]}}.

\bibitem{Carlip:1993sa}
S.~Carlip and C.~Teitelboim, ``{The Off-shell black hole},''
  \href{http://dx.doi.org/10.1088/0264-9381/12/7/011}{{\em Class.Quant.Grav.}
  {\bfseries 12} (1995) 1699--1704},
\href{http://arxiv.org/abs/gr-qc/9312002}{{\ttfamily arXiv:gr-qc/9312002
  [gr-qc]}}.

\bibitem{'tHooft:1984re}
G.~'t~Hooft, ``{On the Quantum Structure of a Black Hole},''
\href{http://dx.doi.org/10.1016/0550-3213(85)90418-3}{{\em Nucl.Phys.}
  {\bfseries B256} (1985) 727}.

\bibitem{Barbon:1994ej}
J.~Barbon, ``{Horizon divergences of fields and strings in black hole
  backgrounds},'' \href{http://dx.doi.org/10.1103/PhysRevD.50.2712}{{\em
  Phys.Rev.} {\bfseries D50} (1994) 2712--2718},
\href{http://arxiv.org/abs/hep-th/9402004}{{\ttfamily arXiv:hep-th/9402004
  [hep-th]}}.

\bibitem{Emparan:1994qa}
R.~Emparan, ``{Heat kernels and thermodynamics in Rindler space},''
  \href{http://dx.doi.org/10.1103/PhysRevD.51.5716}{{\em Phys.Rev.} {\bfseries
  D51} (1995) 5716--5719},
\href{http://arxiv.org/abs/hep-th/9407064}{{\ttfamily arXiv:hep-th/9407064
  [hep-th]}}.

\bibitem{Susskind:1993ki}
L.~Susskind, ``{String theory and the principles of black hole
  complementarity},'' \href{http://dx.doi.org/10.1103/PhysRevLett.71.2367}{{\em
  Phys.Rev.Lett.} {\bfseries 71} (1993) 2367--2368},
\href{http://arxiv.org/abs/hep-th/9307168}{{\ttfamily arXiv:hep-th/9307168
  [hep-th]}}.

\bibitem{Susskind:1994uu}
L.~Susskind and J.~Uglum, ``{Black holes, interactions, and strings},''
\href{http://arxiv.org/abs/hep-th/9410074}{{\ttfamily arXiv:hep-th/9410074
  [hep-th]}}.

\bibitem{Susskind:1993aa}
L.~Susskind, ``{Strings, black holes and Lorentz contraction},''
  \href{http://dx.doi.org/10.1103/PhysRevD.49.6606}{{\em Phys.Rev.} {\bfseries
  D49} (1994) 6606--6611},
\href{http://arxiv.org/abs/hep-th/9308139}{{\ttfamily arXiv:hep-th/9308139
  [hep-th]}}.

\bibitem{Gibbons:1976es}
G.~Gibbons and M.~Perry, ``{Black Holes in Thermal Equilibrium},''
\href{http://dx.doi.org/10.1103/PhysRevLett.36.985}{{\em Phys.Rev.Lett.}
  {\bfseries 36} (1976) 985}.

\bibitem{Fulling:1972md}
S.~A. Fulling, ``{Nonuniqueness of canonical field quantization in Riemannian
  space-time},''
\href{http://dx.doi.org/10.1103/PhysRevD.7.2850}{{\em Phys.Rev.} {\bfseries D7}
  (1973) 2850--2862}.

\bibitem{Davies:1974th}
P.~Davies, ``{Scalar particle production in Schwarzschild and Rindler
  metrics},''
\href{http://dx.doi.org/10.1088/0305-4470/8/4/022}{{\em J.Phys.} {\bfseries A8}
  (1975) 609--616}.

\bibitem{Unruh:1976db}
W.~Unruh, ``{Notes on black hole evaporation},''
\href{http://dx.doi.org/10.1103/PhysRevD.14.870}{{\em Phys.Rev.} {\bfseries
  D14} (1976) 870}.

\bibitem{Callan:1994py}
J.~Callan, Curtis~G. and F.~Wilczek, ``{On geometric entropy},''
  \href{http://dx.doi.org/10.1016/0370-2693(94)91007-3}{{\em Phys.Lett.}
  {\bfseries B333} (1994) 55--61},
\href{http://arxiv.org/abs/hep-th/9401072}{{\ttfamily arXiv:hep-th/9401072
  [hep-th]}}.

\bibitem{Parikh:1999mf}
M.~K. Parikh and F.~Wilczek, ``{Hawking radiation as tunneling},''
  \href{http://dx.doi.org/10.1103/PhysRevLett.85.5042}{{\em Phys.Rev.Lett.}
  {\bfseries 85} (2000) 5042--5045},
\href{http://arxiv.org/abs/hep-th/9907001}{{\ttfamily arXiv:hep-th/9907001
  [hep-th]}}.

\bibitem{Hartle:1976tp}
J.~Hartle and S.~Hawking, ``{Path Integral Derivation of Black Hole
  Radiance},''
\href{http://dx.doi.org/10.1103/PhysRevD.13.2188}{{\em Phys.Rev.} {\bfseries
  D13} (1976) 2188--2203}.

\bibitem{Sciama:1981hr}
D.~Sciama, P.~Candelas, and D.~Deutsch, ``{Quantum Field Theory, Horizons and
  Thermodynamics},''
\href{http://dx.doi.org/10.1080/00018738100101457}{{\em Adv.Phys.} {\bfseries
  30} (1981) 327--366}.

\bibitem{Dowker:1994fi}
J.~Dowker, ``{Remarks on geometric entropy},''
  \href{http://dx.doi.org/10.1088/0264-9381/11/4/001}{{\em Class.Quant.Grav.}
  {\bfseries 11} (1994) L55--L60},
\href{http://arxiv.org/abs/hep-th/9401159}{{\ttfamily arXiv:hep-th/9401159
  [hep-th]}}.

\bibitem{Frolov:1989jh}
V.~P. Frolov and K.~Thorne, ``{Renormalized Stress - Energy Tensor Near the
  Horizon of a Slowly Evolving, Rotating Black Hole},''
\href{http://dx.doi.org/10.1103/PhysRevD.39.2125}{{\em Phys.Rev.} {\bfseries
  D39} (1989) 2125--2154}.

\bibitem{Howard:1984qp}
K.~Howard and P.~Candelas, ``{Quantum Stress Tensor In Schwarzschild
  Space-time},''
\href{http://dx.doi.org/10.1103/PhysRevLett.53.403}{{\em Phys.Rev.Lett.}
  {\bfseries 53} (1984) 403--406}.

\bibitem{Candelas:1980zt}
P.~Candelas, ``{Vacuum Polarization in Schwarzschild Space-Time},''
\href{http://dx.doi.org/10.1103/PhysRevD.21.2185}{{\em Phys.Rev.} {\bfseries
  D21} (1980) 2185--2202}.

\bibitem{Page:1982fm}
D.~N. Page, ``{Thermal Stress Tensors in Static Einstein Spaces},''
\href{http://dx.doi.org/10.1103/PhysRevD.25.1499}{{\em Phys.Rev.} {\bfseries
  D25} (1982) 1499}.

\bibitem{Dowker:1977zj}
J.~Dowker, ``{Quantum Field Theory on a Cone},''
\href{http://dx.doi.org/10.1088/0305-4470/10/1/023}{{\em J.Phys.} {\bfseries
  A10} (1977) 115--124}.

\bibitem{Troost:1977dw}
W.~Troost and H.~Van~Dam, ``{Thermal Effects for an Accelerating Observer},''
\href{http://dx.doi.org/10.1016/0370-2693(77)90764-X}{{\em Phys.Lett.}
  {\bfseries B71} (1977) 149}.

\bibitem{Troost:1978yk}
W.~Troost and H.~van Dam, ``{Thermal Propagators and Accelerated Frames of
  Reference},''
\href{http://dx.doi.org/10.1016/0550-3213(79)90091-9}{{\em Nucl.Phys.}
  {\bfseries B152} (1979) 442}.

\bibitem{Parentani:1989gq}
R.~Parentani and R.~Potting, ``{The Accelerating Observer and the Hagedorn
  Temperature},''
\href{http://dx.doi.org/10.1103/PhysRevLett.63.945}{{\em Phys.Rev.Lett.}
  {\bfseries 63} (1989) 945}.

\bibitem{Cohen:1985sm}
A.~G. Cohen, G.~W. Moore, P.~C. Nelson, and J.~Polchinski, ``{An Off-Shell
  Propagator for String Theory},''
\href{http://dx.doi.org/10.1016/0550-3213(86)90148-3}{{\em Nucl.Phys.}
  {\bfseries B267} (1986) 143}.

\bibitem{Susskind:1994vu}
L.~Susskind, ``{The World as a hologram},''
  \href{http://dx.doi.org/10.1063/1.531249}{{\em J.Math.Phys.} {\bfseries 36}
  (1995) 6377--6396},
\href{http://arxiv.org/abs/hep-th/9409089}{{\ttfamily arXiv:hep-th/9409089
  [hep-th]}}.

\bibitem{Halyo:1996vi}
E.~Halyo, A.~Rajaraman, and L.~Susskind, ``{Braneless black holes},''
  \href{http://dx.doi.org/10.1016/S0370-2693(96)01544-4}{{\em Phys.Lett.}
  {\bfseries B392} (1997) 319--322},
\href{http://arxiv.org/abs/hep-th/9605112}{{\ttfamily arXiv:hep-th/9605112
  [hep-th]}}.

\bibitem{Maldacena:1996ds}
J.~M. Maldacena and L.~Susskind, ``{D-branes and fat black holes},''
  \href{http://dx.doi.org/10.1016/0550-3213(96)00323-9}{{\em Nucl.Phys.}
  {\bfseries B475} (1996) 679--690},
\href{http://arxiv.org/abs/hep-th/9604042}{{\ttfamily arXiv:hep-th/9604042
  [hep-th]}}.

\bibitem{Halyo:1996xe}
E.~Halyo, B.~Kol, A.~Rajaraman, and L.~Susskind, ``{Counting Schwarzschild and
  charged black holes},''
  \href{http://dx.doi.org/10.1016/S0370-2693(97)00357-2}{{\em Phys.Lett.}
  {\bfseries B401} (1997) 15--20},
\href{http://arxiv.org/abs/hep-th/9609075}{{\ttfamily arXiv:hep-th/9609075
  [hep-th]}}.

\bibitem{Halyo:2001us}
E.~Halyo, ``{Universal counting of black hole entropy by strings on the
  stretched horizon},''
  \href{http://dx.doi.org/10.1088/1126-6708/2001/12/005}{{\em JHEP} {\bfseries
  0112} (2001) 005},
\href{http://arxiv.org/abs/hep-th/0108167}{{\ttfamily arXiv:hep-th/0108167
  [hep-th]}}.

\bibitem{Halyo:2003bt}
E.~Halyo, ``{Gravitational entropy and string bits on the stretched horizon},''
\href{http://arxiv.org/abs/hep-th/0308166}{{\ttfamily arXiv:hep-th/0308166
  [hep-th]}}.

\bibitem{Sen:2004dp}
A.~Sen, ``{How does a fundamental string stretch its horizon?},''
  \href{http://dx.doi.org/10.1088/1126-6708/2005/05/059}{{\em JHEP} {\bfseries
  0505} (2005) 059},
\href{http://arxiv.org/abs/hep-th/0411255}{{\ttfamily arXiv:hep-th/0411255
  [hep-th]}}.

\bibitem{Giveon:2006pr}
A.~Giveon and D.~Kutasov, ``{Fundamental strings and black holes},''
  \href{http://dx.doi.org/10.1088/1126-6708/2007/01/071}{{\em JHEP} {\bfseries
  0701} (2007) 071},
\href{http://arxiv.org/abs/hep-th/0611062}{{\ttfamily arXiv:hep-th/0611062
  [hep-th]}}.

\bibitem{BatoniAbdalla:2007zv}
M.~Batoni~Abdalla, M.~Botta~Cantcheff, and D.~L. Nedel, ``{Strings in horizons,
  dissipation and a simple interpretation of the Hagedorn temperature},''
  \href{http://dx.doi.org/10.1140/epjc/s10052-007-0513-2}{{\em Eur.Phys.J.}
  {\bfseries C54} (2008) 311--317},
\href{http://arxiv.org/abs/hep-th/0703064}{{\ttfamily arXiv:hep-th/0703064
  [HEP-TH]}}.

\bibitem{Sasai:2010pz}
Y.~Sasai and A.~Zahabi, ``{Shear viscosity of a highly excited string and the
  black hole membrane paradigm},''
  \href{http://dx.doi.org/10.1103/PhysRevD.83.026002}{{\em Phys.Rev.}
  {\bfseries D83} (2011) 026002},
\href{http://arxiv.org/abs/1010.5380}{{\ttfamily arXiv:1010.5380 [hep-th]}}.

\bibitem{Halyo:2015ffa}
E.~Halyo, ``{Black Holes as Conformal Field Theories on Horizons},''
\href{http://arxiv.org/abs/1502.01979}{{\ttfamily arXiv:1502.01979 [hep-th]}}.

\bibitem{Giveon:2015cma}
A.~Giveon, N.~Itzhaki, and D.~Kutasov, ``{Stringy Horizons},''
\href{http://arxiv.org/abs/1502.03633}{{\ttfamily arXiv:1502.03633 [hep-th]}}.

\bibitem{Dabholkar:1994ai}
A.~Dabholkar, ``{Strings on a cone and black hole entropy},''
  \href{http://dx.doi.org/10.1016/0550-3213(95)00050-3}{{\em Nucl.Phys.}
  {\bfseries B439} (1995) 650--664},
\href{http://arxiv.org/abs/hep-th/9408098}{{\ttfamily arXiv:hep-th/9408098
  [hep-th]}}.

\bibitem{Lowe:1994ah}
D.~A. Lowe and A.~Strominger, ``{Strings near a Rindler or black hole
  horizon},'' \href{http://dx.doi.org/10.1103/PhysRevD.51.1793}{{\em Phys.Rev.}
  {\bfseries D51} (1995) 1793--1799},
\href{http://arxiv.org/abs/hep-th/9410215}{{\ttfamily arXiv:hep-th/9410215
  [hep-th]}}.

\bibitem{Emparan:1994bt}
R.~Emparan, ``{Remarks on the Atick-Witten behavior and strings near black hole
  horizons},''
\href{http://arxiv.org/abs/hep-th/9412003}{{\ttfamily arXiv:hep-th/9412003
  [hep-th]}}.

\bibitem{McGuigan:1994tg}
M.~McGuigan, ``{Finite black hole entropy and string theory},''
  \href{http://dx.doi.org/10.1103/PhysRevD.50.5225}{{\em Phys.Rev.} {\bfseries
  D50} (1994) 5225--5231},
\href{http://arxiv.org/abs/hep-th/9406201}{{\ttfamily arXiv:hep-th/9406201
  [hep-th]}}.

\bibitem{Giveon:2014hfa}
A.~Giveon, N.~Itzhaki, and J.~Troost, ``{Lessons on Black Holes from the
  Elliptic Genus},''
\href{http://arxiv.org/abs/1401.3104}{{\ttfamily arXiv:1401.3104 [hep-th]}}.

\bibitem{Calcagni:2013eua}
G.~Calcagni and L.~Modesto, ``{Nonlocality in string theory},''
  \href{http://dx.doi.org/10.1088/1751-8113/47/35/355402}{{\em J.Phys.}
  {\bfseries A47} no.~35, (2014) 355402},
\href{http://arxiv.org/abs/1310.4957}{{\ttfamily arXiv:1310.4957 [hep-th]}}.

\bibitem{Rangamani:2007fz}
M.~Rangamani and S.~F. Ross, ``{Winding tachyons in BTZ},''
  \href{http://dx.doi.org/10.1103/PhysRevD.77.026010}{{\em Phys.Rev.}
  {\bfseries D77} (2008) 026010},
\href{http://arxiv.org/abs/0706.0663}{{\ttfamily arXiv:0706.0663 [hep-th]}}.

\bibitem{Witten:1991yr}
E.~Witten, ``{On string theory and black holes},''
\href{http://dx.doi.org/10.1103/PhysRevD.44.314}{{\em Phys.Rev.} {\bfseries
  D44} (1991) 314--324}.

\bibitem{Tseytlin:1991ht}
A.~A. Tseytlin, ``{On the form of the black hole solution in D = 2 theory},''
\href{http://dx.doi.org/10.1016/0370-2693(91)90800-6}{{\em Phys.Lett.}
  {\bfseries B268} (1991) 175--178}.

\bibitem{Jack:1992mk}
I.~Jack, D.~Jones, and J.~Panvel, ``{Exact bosonic and supersymmetric string
  black hole solutions},''
  \href{http://dx.doi.org/10.1016/0550-3213(93)90239-L}{{\em Nucl.Phys.}
  {\bfseries B393} (1993) 95--110},
\href{http://arxiv.org/abs/hep-th/9201039}{{\ttfamily arXiv:hep-th/9201039
  [hep-th]}}.

\bibitem{Giveon:1999px}
A.~Giveon and D.~Kutasov, ``{Little string theory in a double scaling limit},''
  \href{http://dx.doi.org/10.1088/1126-6708/1999/10/034}{{\em JHEP} {\bfseries
  9910} (1999) 034},
\href{http://arxiv.org/abs/hep-th/9909110}{{\ttfamily arXiv:hep-th/9909110
  [hep-th]}}.

\bibitem{Bars:1992sr}
I.~Bars and K.~Sfetsos, ``{Conformally exact metric and dilaton in string
  theory on curved space-time},''
  \href{http://dx.doi.org/10.1103/PhysRevD.46.4510}{{\em Phys.Rev.} {\bfseries
  D46} (1992) 4510--4519},
\href{http://arxiv.org/abs/hep-th/9206006}{{\ttfamily arXiv:hep-th/9206006
  [hep-th]}}.

\bibitem{Taylor:2003gn}
W.~Taylor and B.~Zwiebach, ``{D-branes, tachyons, and string field theory},''
\href{http://arxiv.org/abs/hep-th/0311017}{{\ttfamily arXiv:hep-th/0311017
  [hep-th]}}.

\bibitem{Brigante:2007jv}
M.~Brigante, G.~Festuccia, and H.~Liu, ``{Hagedorn divergences and tachyon
  potential},'' \href{http://dx.doi.org/10.1088/1126-6708/2007/06/008}{{\em
  JHEP} {\bfseries 0706} (2007) 008},
\href{http://arxiv.org/abs/hep-th/0701205}{{\ttfamily arXiv:hep-th/0701205
  [hep-th]}}.

\bibitem{Giveon:1993hm}
A.~Giveon, E.~Rabinovici, and A.~A. Tseytlin, ``{Heterotic string solutions and
  coset conformal field theories},''
  \href{http://dx.doi.org/10.1016/0550-3213(93)90583-B}{{\em Nucl.Phys.}
  {\bfseries B409} (1993) 339--362},
\href{http://arxiv.org/abs/hep-th/9304155}{{\ttfamily arXiv:hep-th/9304155
  [hep-th]}}.

\bibitem{Sfetsos:1993bh}
K.~Sfetsos and A.~A. Tseytlin, ``{Chiral gauged WZNW models and heterotic
  string backgrounds},''
  \href{http://dx.doi.org/10.1016/0550-3213(94)90069-8}{{\em Nucl.Phys.}
  {\bfseries B415} (1994) 116--154},
\href{http://arxiv.org/abs/hep-th/9308018}{{\ttfamily arXiv:hep-th/9308018
  [hep-th]}}.

\bibitem{Aldazabal:2013sca}
G.~Aldazabal, D.~Marques, and C.~Nunez, ``{Double Field Theory: A Pedagogical
  Review},'' \href{http://dx.doi.org/10.1088/0264-9381/30/16/163001}{{\em
  Class.Quant.Grav.} {\bfseries 30} (2013) 163001},
\href{http://arxiv.org/abs/1305.1907}{{\ttfamily arXiv:1305.1907 [hep-th]}}.

\bibitem{Aharony:2004xn}
O.~Aharony, A.~Giveon, and D.~Kutasov, ``{LSZ in LST},''
  \href{http://dx.doi.org/10.1016/j.nuclphysb.2004.05.015}{{\em Nucl.Phys.}
  {\bfseries B691} (2004) 3--78},
\href{http://arxiv.org/abs/hep-th/0404016}{{\ttfamily arXiv:hep-th/0404016
  [hep-th]}}.

\bibitem{Martinec:2001cf}
E.~J. Martinec and W.~McElgin, ``{String theory on AdS orbifolds},''
  \href{http://dx.doi.org/10.1088/1126-6708/2002/04/029}{{\em JHEP} {\bfseries
  0204} (2002) 029},
\href{http://arxiv.org/abs/hep-th/0106171}{{\ttfamily arXiv:hep-th/0106171
  [hep-th]}}.

\bibitem{Son:2001qm}
J.~Son, ``{String theory on AdS(3) / Z(N)},''
\href{http://arxiv.org/abs/hep-th/0107131}{{\ttfamily arXiv:hep-th/0107131
  [hep-th]}}.

\bibitem{Hanany:2002ev}
A.~Hanany, N.~Prezas, and J.~Troost, ``{The Partition function of the
  two-dimensional black hole conformal field theory},''
  \href{http://dx.doi.org/10.1088/1126-6708/2002/04/014}{{\em JHEP} {\bfseries
  0204} (2002) 014},
\href{http://arxiv.org/abs/hep-th/0202129}{{\ttfamily arXiv:hep-th/0202129
  [hep-th]}}.

\bibitem{Barbon:2002nw}
J.~Barbon and E.~Rabinovici, ``{Remarks on black hole instabilities and closed
  string tachyons},'' \href{http://dx.doi.org/10.1023/A:1022823926674}{{\em
  Found.Phys.} {\bfseries 33} (2003) 145--165},
\href{http://arxiv.org/abs/hep-th/0211212}{{\ttfamily arXiv:hep-th/0211212
  [hep-th]}}.

\bibitem{Giveon:2005jv}
A.~Giveon and D.~Kutasov, ``{The Charged black hole/string transition},''
  \href{http://dx.doi.org/10.1088/1126-6708/2006/01/120}{{\em JHEP} {\bfseries
  0601} (2006) 120},
\href{http://arxiv.org/abs/hep-th/0510211}{{\ttfamily arXiv:hep-th/0510211
  [hep-th]}}.

\bibitem{Horowitz:1996nw}
G.~T. Horowitz and J.~Polchinski, ``{A Correspondence principle for black holes
  and strings},'' \href{http://dx.doi.org/10.1103/PhysRevD.55.6189}{{\em
  Phys.Rev.} {\bfseries D55} (1997) 6189--6197},
\href{http://arxiv.org/abs/hep-th/9612146}{{\ttfamily arXiv:hep-th/9612146
  [hep-th]}}.

\bibitem{Johnson:1994jw}
C.~V. Johnson, ``{Exact models of extremal dyonic 4-D black hole solutions of
  heterotic string theory},''
  \href{http://dx.doi.org/10.1103/PhysRevD.50.4032}{{\em Phys.Rev.} {\bfseries
  D50} (1994) 4032--4050},
\href{http://arxiv.org/abs/hep-th/9403192}{{\ttfamily arXiv:hep-th/9403192
  [hep-th]}}.

\bibitem{Johnson:1994kv}
C.~V. Johnson, ``{Heterotic Coset Models},''
  \href{http://dx.doi.org/10.1142/S0217732395000582}{{\em Mod.Phys.Lett.}
  {\bfseries A10} (1995) 549--560},
\href{http://arxiv.org/abs/hep-th/9409062}{{\ttfamily arXiv:hep-th/9409062
  [hep-th]}}.

\bibitem{Johnson:2004zq}
C.~V. Johnson and H.~G. Svendsen, ``{An Exact string theory model of closed
  time-like curves and cosmological singularities},''
  \href{http://dx.doi.org/10.1103/PhysRevD.70.126011}{{\em Phys.Rev.}
  {\bfseries D70} (2004) 126011},
\href{http://arxiv.org/abs/hep-th/0405141}{{\ttfamily arXiv:hep-th/0405141
  [hep-th]}}.

\bibitem{Svendsen:2005gy}
H.~G. Svendsen, ``{Global properties of an exact string theory solution in two
  and four dimensions},''
  \href{http://dx.doi.org/10.1103/PhysRevD.73.064032}{{\em Phys.Rev.}
  {\bfseries D73} (2006) 064032},
\href{http://arxiv.org/abs/hep-th/0511289}{{\ttfamily arXiv:hep-th/0511289
  [hep-th]}}.

\bibitem{Green:1984sg}
M.~B. Green and J.~H. Schwarz, ``{Anomaly Cancellation in Supersymmetric D=10
  Gauge Theory and Superstring Theory},''
\href{http://dx.doi.org/10.1016/0370-2693(84)91565-X}{{\em Phys.Lett.}
  {\bfseries B149} (1984) 117--122}.

\bibitem{Metsaev:1986yb}
R.~Metsaev and A.~A. Tseytlin, ``{Curvature Cubed Terms in String Theory
  Effective Actions},''
\href{http://dx.doi.org/10.1016/0370-2693(87)91527-9}{{\em Phys.Lett.}
  {\bfseries B185} (1987) 52}.

\bibitem{Foakes:1987bn}
A.~Foakes, N.~Mohammedi, and D.~Ross, ``{Effective Action and beta Function for
  the Heterotic String},''
\href{http://dx.doi.org/10.1016/0370-2693(88)91262-2}{{\em Phys.Lett.}
  {\bfseries B206} (1988) 57}.

\bibitem{Dunne:1998qy}
G.~V. Dunne, ``{Aspects of Chern-Simons theory},''
\href{http://arxiv.org/abs/hep-th/9902115}{{\ttfamily arXiv:hep-th/9902115
  [hep-th]}}.

\bibitem{Giveon:2004rw}
A.~Giveon, A.~Konechny, E.~Rabinovici, and A.~Sever, ``{On thermodynamical
  properties of some coset CFT backgrounds},''
  \href{http://dx.doi.org/10.1088/1126-6708/2004/07/076}{{\em JHEP} {\bfseries
  0407} (2004) 076},
\href{http://arxiv.org/abs/hep-th/0406131}{{\ttfamily arXiv:hep-th/0406131
  [hep-th]}}.

\bibitem{Eguchi:2010cb}
T.~Eguchi and Y.~Sugawara, ``{Non-holomorphic Modular Forms and SL(2,R)/U(1)
  Superconformal Field Theory},''
  \href{http://dx.doi.org/10.1007/JHEP03(2011)107}{{\em JHEP} {\bfseries 1103}
  (2011) 107},
\href{http://arxiv.org/abs/1012.5721}{{\ttfamily arXiv:1012.5721 [hep-th]}}.

\bibitem{Sugawara:2011vg}
Y.~Sugawara, ``{Comments on Non-holomorphic Modular Forms and Non-compact
  Superconformal Field Theories},''
  \href{http://dx.doi.org/10.1007/JHEP01(2012)098}{{\em JHEP} {\bfseries 1201}
  (2012) 098},
\href{http://arxiv.org/abs/1109.3365}{{\ttfamily arXiv:1109.3365 [hep-th]}}.

\bibitem{Giveon:2005mi}
A.~Giveon, D.~Kutasov, E.~Rabinovici, and A.~Sever, ``{Phases of quantum
  gravity in AdS(3) and linear dilaton backgrounds},''
  \href{http://dx.doi.org/10.1016/j.nuclphysb.2005.04.015}{{\em Nucl.Phys.}
  {\bfseries B719} (2005) 3--34},
\href{http://arxiv.org/abs/hep-th/0503121}{{\ttfamily arXiv:hep-th/0503121
  [hep-th]}}.

\bibitem{Parnachev:2005qr}
A.~Parnachev and D.~A. Sahakyan, ``{On non-critical superstring/black hole
  transition},'' \href{http://dx.doi.org/10.1103/PhysRevD.73.086008}{{\em
  Phys.Rev.} {\bfseries D73} (2006) 086008},
\href{http://arxiv.org/abs/hep-th/0512075}{{\ttfamily arXiv:hep-th/0512075
  [hep-th]}}.

\bibitem{Nakayama:2005pk}
Y.~Nakayama, S.-J. Rey, and Y.~Sugawara, ``{D-brane propagation in
  two-dimensional black hole geometries},''
  \href{http://dx.doi.org/10.1088/1126-6708/2005/09/020}{{\em JHEP} {\bfseries
  0509} (2005) 020},
\href{http://arxiv.org/abs/hep-th/0507040}{{\ttfamily arXiv:hep-th/0507040
  [hep-th]}}.

\bibitem{Barbon:2007za}
J.~L. Barbon, C.~A. Fuertes, and E.~Rabinovici, ``{Deconstructing the little
  Hagedorn holography},''
  \href{http://dx.doi.org/10.1088/1126-6708/2007/09/055}{{\em JHEP} {\bfseries
  0709} (2007) 055},
\href{http://arxiv.org/abs/0707.1158}{{\ttfamily arXiv:0707.1158 [hep-th]}}.

\bibitem{Hawking:1999dp}
S.~Hawking and H.~Reall, ``{Charged and rotating AdS black holes and their CFT
  duals},'' \href{http://dx.doi.org/10.1103/PhysRevD.61.024014}{{\em Phys.Rev.}
  {\bfseries D61} (2000) 024014},
\href{http://arxiv.org/abs/hep-th/9908109}{{\ttfamily arXiv:hep-th/9908109
  [hep-th]}}.

\bibitem{Winstanley:2001nx}
E.~Winstanley, ``{On classical superradiance in Kerr-Newman - anti-de Sitter
  black holes},'' \href{http://dx.doi.org/10.1103/PhysRevD.64.104010}{{\em
  Phys.Rev.} {\bfseries D64} (2001) 104010},
\href{http://arxiv.org/abs/gr-qc/0106032}{{\ttfamily arXiv:gr-qc/0106032
  [gr-qc]}}.

\bibitem{Rangamani:2001ir}
M.~Rangamani, ``{Little string thermodynamics},''
  \href{http://dx.doi.org/10.1088/1126-6708/2001/06/042}{{\em JHEP} {\bfseries
  0106} (2001) 042},
\href{http://arxiv.org/abs/hep-th/0104125}{{\ttfamily arXiv:hep-th/0104125
  [hep-th]}}.

\bibitem{Buchel:2001dg}
A.~Buchel, ``{On the thermodynamic instability of LST},''
\href{http://arxiv.org/abs/hep-th/0107102}{{\ttfamily arXiv:hep-th/0107102
  [hep-th]}}.

\bibitem{Khandekar:1975}
D.~Khandekar and S.~Lawande, ``{Exact propagator for a time-dependent harmonic
  oscillator with and without a singular perturbation},'' {\em J. Math. Phys}
  {\bfseries 16} (1975) 384.

\bibitem{Deo:1991mp}
N.~Deo, S.~Jain, O.~Narayan, and C.-I. Tan, ``{The Effect of topology on the
  thermodynamic limit for a string gas},''
\href{http://dx.doi.org/10.1103/PhysRevD.45.3641}{{\em Phys.Rev.} {\bfseries
  D45} (1992) 3641--3650}.

\bibitem{Jatkar:1992np}
D.~P. Jatkar, ``{The Spectrum of SL(2,R) / U(1) black hole conformal field
  theory},'' \href{http://dx.doi.org/10.1016/0550-3213(93)90213-9}{{\em
  Nucl.Phys.} {\bfseries B395} (1993) 167--184},
\href{http://arxiv.org/abs/hep-th/9203063}{{\ttfamily arXiv:hep-th/9203063
  [hep-th]}}.

\bibitem{Nappi:1993ie}
C.~R. Nappi and E.~Witten, ``{A WZW model based on a nonsemisimple group},''
  \href{http://dx.doi.org/10.1103/PhysRevLett.71.3751}{{\em Phys.Rev.Lett.}
  {\bfseries 71} (1993) 3751--3753},
\href{http://arxiv.org/abs/hep-th/9310112}{{\ttfamily arXiv:hep-th/9310112
  [hep-th]}}.

\bibitem{DiFrancesco:1997nk}
P.~Di~Francesco, P.~Mathieu, and D.~Senechal,
``{Conformal field theory},''.

\bibitem{Hemming:2002kd}
S.~Hemming, E.~Keski-Vakkuri, and P.~Kraus, ``{Strings in the extended BTZ
  space-time},'' \href{http://dx.doi.org/10.1088/1126-6708/2002/10/006}{{\em
  JHEP} {\bfseries 0210} (2002) 006},
\href{http://arxiv.org/abs/hep-th/0208003}{{\ttfamily arXiv:hep-th/0208003
  [hep-th]}}.

\bibitem{Balog:1988jb}
J.~Balog, L.~O'Raifeartaigh, P.~Forgacs, and A.~Wipf, ``{Consistency of String
  Propagation on Curved Space-Times: An SU(1,1) Based Counterexample},''
\href{http://dx.doi.org/10.1016/0550-3213(89)90380-5}{{\em Nucl.Phys.}
  {\bfseries B325} (1989) 225}.

\bibitem{Petropoulos:1989fc}
P.~Petropoulos, ``{Comments On SU(1,1) String Theory},''
\href{http://dx.doi.org/10.1016/0370-2693(90)90819-R}{{\em Phys.Lett.}
  {\bfseries B236} (1990) 151}.

\bibitem{Hwang:1990aq}
S.~Hwang, ``{No ghost theorem for SU(1,1) string theories},''
\href{http://dx.doi.org/10.1016/0550-3213(91)90177-Y}{{\em Nucl.Phys.}
  {\bfseries B354} (1991) 100--112}.

\bibitem{Hwang:1991ana}
S.~Hwang, ``{Cosets as gauge slices in SU(1,1) strings},''
  \href{http://dx.doi.org/10.1016/0370-2693(92)91666-W}{{\em Phys.Lett.}
  {\bfseries B276} (1992) 451--454},
\href{http://arxiv.org/abs/hep-th/9110039}{{\ttfamily arXiv:hep-th/9110039
  [hep-th]}}.

\bibitem{Bars:1995mf}
I.~Bars, ``{Ghost - free spectrum of a quantum string in SL(2,R) curved
  space-time},'' \href{http://dx.doi.org/10.1103/PhysRevD.53.3308}{{\em
  Phys.Rev.} {\bfseries D53} (1996) 3308--3323},
\href{http://arxiv.org/abs/hep-th/9503205}{{\ttfamily arXiv:hep-th/9503205
  [hep-th]}}.

\bibitem{Bars:1999ik}
I.~Bars, C.~Deliduman, and D.~Minic, ``{String theory on AdS(3) revisited},''
\href{http://arxiv.org/abs/hep-th/9907087}{{\ttfamily arXiv:hep-th/9907087
  [hep-th]}}.

\bibitem{deBoer:1998pp}
J.~de~Boer, H.~Ooguri, H.~Robins, and J.~Tannenhauser, ``{String theory on
  AdS(3)},'' \href{http://dx.doi.org/10.1088/1126-6708/1998/12/026}{{\em JHEP}
  {\bfseries 9812} (1998) 026},
\href{http://arxiv.org/abs/hep-th/9812046}{{\ttfamily arXiv:hep-th/9812046
  [hep-th]}}.

\bibitem{Giveon:1998ns}
A.~Giveon, D.~Kutasov, and N.~Seiberg, ``{Comments on string theory on
  AdS(3)},'' {\em Adv.Theor.Math.Phys.} {\bfseries 2} (1998) 733--780,
\href{http://arxiv.org/abs/hep-th/9806194}{{\ttfamily arXiv:hep-th/9806194
  [hep-th]}}.

\bibitem{Kutasov:1999xu}
D.~Kutasov and N.~Seiberg, ``{More comments on string theory on AdS(3)},''
  \href{http://dx.doi.org/10.1088/1126-6708/1999/04/008}{{\em JHEP} {\bfseries
  9904} (1999) 008},
\href{http://arxiv.org/abs/hep-th/9903219}{{\ttfamily arXiv:hep-th/9903219
  [hep-th]}}.

\bibitem{Evans:1998qu}
J.~M. Evans, M.~R. Gaberdiel, and M.~J. Perry, ``{The no ghost theorem for
  AdS(3) and the stringy exclusion principle},''
  \href{http://dx.doi.org/10.1016/S0550-3213(98)00561-6}{{\em Nucl.Phys.}
  {\bfseries B535} (1998) 152--170},
\href{http://arxiv.org/abs/hep-th/9806024}{{\ttfamily arXiv:hep-th/9806024
  [hep-th]}}.

\bibitem{Berkooz:2007fe}
M.~Berkooz, Z.~Komargodski, and D.~Reichmann, ``{Thermal AdS(3), BTZ and
  competing winding modes condensation},''
  \href{http://dx.doi.org/10.1088/1126-6708/2007/12/020}{{\em JHEP} {\bfseries
  0712} (2007) 020},
\href{http://arxiv.org/abs/0706.0610}{{\ttfamily arXiv:0706.0610 [hep-th]}}.

\bibitem{Lin:2007gi}
F.-L. Lin, T.~Matsuo, and D.~Tomino, ``{Hagedorn Strings and Correspondence
  Principle in AdS(3)},''
  \href{http://dx.doi.org/10.1088/1126-6708/2007/09/042}{{\em JHEP} {\bfseries
  0709} (2007) 042},
\href{http://arxiv.org/abs/0705.4514}{{\ttfamily arXiv:0705.4514 [hep-th]}}.

\bibitem{Sundborg:1999ue}
B.~Sundborg, ``{The Hagedorn transition, deconfinement and N=4 SYM theory},''
  \href{http://dx.doi.org/10.1016/S0550-3213(00)00044-4}{{\em Nucl.Phys.}
  {\bfseries B573} (2000) 349--363},
\href{http://arxiv.org/abs/hep-th/9908001}{{\ttfamily arXiv:hep-th/9908001
  [hep-th]}}.

\bibitem{Aharony:2003sx}
O.~Aharony, J.~Marsano, S.~Minwalla, K.~Papadodimas, and M.~Van~Raamsdonk,
  ``{The Hagedorn - deconfinement phase transition in weakly coupled large N
  gauge theories},'' \href{http://dx.doi.org/10.4310/ATMP.2004.v8.n4.a1}{{\em
  Adv.Theor.Math.Phys.} {\bfseries 8} (2004) 603--696},
\href{http://arxiv.org/abs/hep-th/0310285}{{\ttfamily arXiv:hep-th/0310285
  [hep-th]}}.

\bibitem{Argurio:2000tb}
R.~Argurio, A.~Giveon, and A.~Shomer, ``{Superstrings on AdS(3) and symmetric
  products},'' \href{http://dx.doi.org/10.1088/1126-6708/2000/12/003}{{\em
  JHEP} {\bfseries 0012} (2000) 003},
\href{http://arxiv.org/abs/hep-th/0009242}{{\ttfamily arXiv:hep-th/0009242
  [hep-th]}}.

\bibitem{Dixon:1986qv}
L.~J. Dixon, D.~Friedan, E.~J. Martinec, and S.~H. Shenker, ``{The Conformal
  Field Theory of Orbifolds},''
\href{http://dx.doi.org/10.1016/0550-3213(87)90676-6}{{\em Nucl.Phys.}
  {\bfseries B282} (1987) 13--73}.

\bibitem{Parsons:2009si}
J.~Parsons and S.~F. Ross, ``{Strings in extremal BTZ black holes},''
  \href{http://dx.doi.org/10.1088/1126-6708/2009/04/134}{{\em JHEP} {\bfseries
  0904} (2009) 134},
\href{http://arxiv.org/abs/0901.3044}{{\ttfamily arXiv:0901.3044 [hep-th]}}.

\bibitem{Giddings:1993wn}
S.~B. Giddings, J.~Polchinski, and A.~Strominger, ``{Four-dimensional black
  holes in string theory},''
  \href{http://dx.doi.org/10.1103/PhysRevD.48.5784}{{\em Phys.Rev.} {\bfseries
  D48} (1993) 5784--5797},
\href{http://arxiv.org/abs/hep-th/9305083}{{\ttfamily arXiv:hep-th/9305083
  [hep-th]}}.

\bibitem{Natsuume:1996ij}
M.~Natsuume and Y.~Satoh, ``{String theory on three-dimensional black holes},''
  \href{http://dx.doi.org/10.1142/S0217751X98000585}{{\em Int.J.Mod.Phys.}
  {\bfseries A13} (1998) 1229--1262},
\href{http://arxiv.org/abs/hep-th/9611041}{{\ttfamily arXiv:hep-th/9611041
  [hep-th]}}.

\bibitem{Maldacena:2001km}
J.~M. Maldacena and H.~Ooguri, ``{Strings in AdS(3) and the SL(2,R) WZW model.
  Part 3. Correlation functions},''
  \href{http://dx.doi.org/10.1103/PhysRevD.65.106006}{{\em Phys.Rev.}
  {\bfseries D65} (2002) 106006},
\href{http://arxiv.org/abs/hep-th/0111180}{{\ttfamily arXiv:hep-th/0111180
  [hep-th]}}.

\bibitem{Teschner:1997ft}
J.~Teschner, ``{On structure constants and fusion rules in the SL(2,C) / SU(2)
  WZNW model},'' \href{http://dx.doi.org/10.1016/S0550-3213(99)00072-3}{{\em
  Nucl.Phys.} {\bfseries B546} (1999) 390--422},
\href{http://arxiv.org/abs/hep-th/9712256}{{\ttfamily arXiv:hep-th/9712256
  [hep-th]}}.

\bibitem{Kutasov:1990sv}
D.~Kutasov and N.~Seiberg, ``{Number of degrees of freedom, density of states
  and tachyons in string theory and CFT},''
\href{http://dx.doi.org/10.1016/0550-3213(91)90426-X}{{\em Nucl.Phys.}
  {\bfseries B358} (1991) 600--618}.

\bibitem{Israel:2003ry}
D.~Israel, C.~Kounnas, and M.~P. Petropoulos, ``{Superstrings on NS5
  backgrounds, deformed AdS(3) and holography},''
  \href{http://dx.doi.org/10.1088/1126-6708/2003/10/028}{{\em JHEP} {\bfseries
  0310} (2003) 028},
\href{http://arxiv.org/abs/hep-th/0306053}{{\ttfamily arXiv:hep-th/0306053
  [hep-th]}}.

\bibitem{Israel:2004ir}
D.~Israel, C.~Kounnas, A.~Pakman, and J.~Troost, ``{The Partition function of
  the supersymmetric two-dimensional black hole and little string theory},''
  \href{http://dx.doi.org/10.1088/1126-6708/2004/06/033}{{\em JHEP} {\bfseries
  0406} (2004) 033},
\href{http://arxiv.org/abs/hep-th/0403237}{{\ttfamily arXiv:hep-th/0403237
  [hep-th]}}.

\bibitem{Maldacena:1998bw}
J.~M. Maldacena and A.~Strominger, ``{AdS(3) black holes and a stringy
  exclusion principle},''
  \href{http://dx.doi.org/10.1088/1126-6708/1998/12/005}{{\em JHEP} {\bfseries
  9812} (1998) 005},
\href{http://arxiv.org/abs/hep-th/9804085}{{\ttfamily arXiv:hep-th/9804085
  [hep-th]}}.

\bibitem{Gawedzki:1991yu}
K.~Gawedzki, ``{Noncompact WZW conformal field theories},''
\href{http://arxiv.org/abs/hep-th/9110076}{{\ttfamily arXiv:hep-th/9110076
  [hep-th]}}.

\bibitem{Gawedzki:1988nj}
K.~Gawedzki and A.~Kupiainen, ``{Coset Construction from Functional
  Integrals},''
\href{http://dx.doi.org/10.1016/0550-3213(89)90015-1}{{\em Nucl.Phys.}
  {\bfseries B320} (1989) 625}.

\bibitem{Maloney:2007ud}
A.~Maloney and E.~Witten, ``{Quantum Gravity Partition Functions in Three
  Dimensions},'' \href{http://dx.doi.org/10.1007/JHEP02(2010)029}{{\em JHEP}
  {\bfseries 1002} (2010) 029},
\href{http://arxiv.org/abs/0712.0155}{{\ttfamily arXiv:0712.0155 [hep-th]}}.

\bibitem{Dijkgraaf:2000fq}
R.~Dijkgraaf, J.~M. Maldacena, G.~W. Moore, and E.~P. Verlinde, ``{A Black hole
  Farey tail},''
\href{http://arxiv.org/abs/hep-th/0005003}{{\ttfamily arXiv:hep-th/0005003
  [hep-th]}}.

\bibitem{deBoer:2006vg}
J.~de~Boer, M.~C. Cheng, R.~Dijkgraaf, J.~Manschot, and E.~Verlinde, ``{A Farey
  Tail for Attractor Black Holes},''
  \href{http://dx.doi.org/10.1088/1126-6708/2006/11/024}{{\em JHEP} {\bfseries
  0611} (2006) 024},
\href{http://arxiv.org/abs/hep-th/0608059}{{\ttfamily arXiv:hep-th/0608059
  [hep-th]}}.

\bibitem{Forste:1994wp}
S.~Forste, ``{A Truly marginal deformation of SL(2, R) in a null direction},''
  \href{http://dx.doi.org/10.1016/0370-2693(94)91340-4}{{\em Phys.Lett.}
  {\bfseries B338} (1994) 36--39},
\href{http://arxiv.org/abs/hep-th/9407198}{{\ttfamily arXiv:hep-th/9407198
  [hep-th]}}.

\bibitem{Giveon:1993ph}
A.~Giveon and E.~Kiritsis, ``{Axial vector duality as a gauge symmetry and
  topology change in string theory},''
  \href{http://dx.doi.org/10.1016/0550-3213(94)90460-X}{{\em Nucl.Phys.}
  {\bfseries B411} (1994) 487--508},
\href{http://arxiv.org/abs/hep-th/9303016}{{\ttfamily arXiv:hep-th/9303016
  [hep-th]}}.

\bibitem{Ray:1973sb}
D.~Ray and I.~Singer, ``{Analytic torsion for complex manifolds},''
\href{http://dx.doi.org/10.2307/1970909}{{\em Annals Math.} {\bfseries 98}
  (1973) 154--177}.

\bibitem{Teschner:1997fv}
J.~Teschner, ``{The Minisuperspace limit of the sl(2,C) / SU(2) WZNW model},''
  \href{http://dx.doi.org/10.1016/S0550-3213(99)00071-1}{{\em Nucl.Phys.}
  {\bfseries B546} (1999) 369--389},
\href{http://arxiv.org/abs/hep-th/9712258}{{\ttfamily arXiv:hep-th/9712258
  [hep-th]}}.

\bibitem{David:2009xg}
J.~R. David, M.~R. Gaberdiel, and R.~Gopakumar, ``{The Heat Kernel on AdS(3)
  and its Applications},''
  \href{http://dx.doi.org/10.1007/JHEP04(2010)125}{{\em JHEP} {\bfseries 1004}
  (2010) 125},
\href{http://arxiv.org/abs/0911.5085}{{\ttfamily arXiv:0911.5085 [hep-th]}}.

\bibitem{Spiegelglas:1988hr}
M.~Spiegelglas, ``{String Thermal Tachyons as Multiparticle Instabilities},''
\href{http://dx.doi.org/10.1016/0370-2693(89)90893-9}{{\em Phys.Lett.}
  {\bfseries B220} (1989) 391}.

\bibitem{Denef:2009yy}
F.~Denef, S.~A. Hartnoll, and S.~Sachdev, ``{Quantum oscillations and black
  hole ringing},'' \href{http://dx.doi.org/10.1103/PhysRevD.80.126016}{{\em
  Phys.Rev.} {\bfseries D80} (2009) 126016},
\href{http://arxiv.org/abs/0908.1788}{{\ttfamily arXiv:0908.1788 [hep-th]}}.

\bibitem{Kabat:1995eq}
D.~N. Kabat, ``{Black hole entropy and entropy of entanglement},''
  \href{http://dx.doi.org/10.1016/0550-3213(95)00443-V}{{\em Nucl.Phys.}
  {\bfseries B453} (1995) 281--299},
\href{http://arxiv.org/abs/hep-th/9503016}{{\ttfamily arXiv:hep-th/9503016
  [hep-th]}}.

\bibitem{Kabat:1995jq}
D.~N. Kabat, S.~Shenker, and M.~Strassler, ``{Black hole entropy in the O(N)
  model},'' \href{http://dx.doi.org/10.1103/PhysRevD.52.7027}{{\em Phys.Rev.}
  {\bfseries D52} (1995) 7027--7036},
\href{http://arxiv.org/abs/hep-th/9506182}{{\ttfamily arXiv:hep-th/9506182
  [hep-th]}}.

\bibitem{Kabat:2012ns}
D.~Kabat and D.~Sarkar, ``{Cosmic string interactions induced by gauge and
  scalar fields},'' \href{http://dx.doi.org/10.1103/PhysRevD.86.084021}{{\em
  Phys. Rev.} {\bfseries D86} (2012) 084021},
\href{http://arxiv.org/abs/1206.5642}{{\ttfamily arXiv:1206.5642 [hep-th]}}.

\bibitem{Donnelly:2012st}
W.~Donnelly and A.~C. Wall, ``{Do gauge fields really contribute negatively to
  black hole entropy?},''
  \href{http://dx.doi.org/10.1103/PhysRevD.86.064042}{{\em Phys.Rev.}
  {\bfseries D86} (2012) 064042},
\href{http://arxiv.org/abs/1206.5831}{{\ttfamily arXiv:1206.5831 [hep-th]}}.

\bibitem{Donnelly:2014fua}
W.~Donnelly and A.~C. Wall, ``{Entanglement entropy of electromagnetic edge
  modes},''
\href{http://arxiv.org/abs/1412.1895}{{\ttfamily arXiv:1412.1895 [hep-th]}}.

\bibitem{He:2014gva}
S.~He, T.~Numasawa, T.~Takayanagi, and K.~Watanabe, ``{Notes on Entanglement
  Entropy in String Theory},''
  \href{http://dx.doi.org/10.1007/JHEP05(2015)106}{{\em JHEP} {\bfseries 05}
  (2015) 106},
\href{http://arxiv.org/abs/1412.5606}{{\ttfamily arXiv:1412.5606 [hep-th]}}.

\bibitem{Mertens:2015adr}
T.~G. Mertens, H.~Verschelde, and V.~I. Zakharov, ``{String partition functions
  in Rindler space and maximal acceleration},''
\href{http://arxiv.org/abs/1511.00560}{{\ttfamily arXiv:1511.00560 [hep-th]}}.

\bibitem{Verlinde:2012cy}
E.~Verlinde and H.~Verlinde, ``{Black Hole Entanglement and Quantum Error
  Correction},'' \href{http://dx.doi.org/10.1007/JHEP10(2013)107}{{\em JHEP}
  {\bfseries 1310} (2013) 107},
\href{http://arxiv.org/abs/1211.6913}{{\ttfamily arXiv:1211.6913 [hep-th]}}.

\bibitem{Harlow:2013tf}
D.~Harlow and P.~Hayden, ``{Quantum Computation vs. Firewalls},''
  \href{http://dx.doi.org/10.1007/JHEP06(2013)085}{{\em JHEP} {\bfseries 1306}
  (2013) 085},
\href{http://arxiv.org/abs/1301.4504}{{\ttfamily arXiv:1301.4504 [hep-th]}}.

\bibitem{Almheiri:2013hfa}
A.~Almheiri, D.~Marolf, J.~Polchinski, D.~Stanford, and J.~Sully, ``{An
  Apologia for Firewalls},''
  \href{http://dx.doi.org/10.1007/JHEP09(2013)018}{{\em JHEP} {\bfseries 1309}
  (2013) 018},
\href{http://arxiv.org/abs/1304.6483}{{\ttfamily arXiv:1304.6483 [hep-th]}}.

\bibitem{Bousso:2013wia}
R.~Bousso, ``{Firewalls from double purity},''
  \href{http://dx.doi.org/10.1103/PhysRevD.88.084035}{{\em Phys.Rev.}
  {\bfseries D88} no.~8, (2013) 084035},
\href{http://arxiv.org/abs/1308.2665}{{\ttfamily arXiv:1308.2665 [hep-th]}}.

\bibitem{Silverstein:2014yza}
E.~Silverstein, ``{Backdraft: String Creation in an Old Schwarzschild Black
  Hole},''
\href{http://arxiv.org/abs/1402.1486}{{\ttfamily arXiv:1402.1486 [hep-th]}}.

\bibitem{Larjo:2012jt}
K.~Larjo, D.~A. Lowe, and L.~Thorlacius, ``{Black holes without firewalls},''
  \href{http://dx.doi.org/10.1103/PhysRevD.87.104018}{{\em Phys.Rev.}
  {\bfseries D87} no.~10, (2013) 104018},
\href{http://arxiv.org/abs/1211.4620}{{\ttfamily arXiv:1211.4620 [hep-th]}}.

\bibitem{Polchinski:1988jq}
J.~Polchinski, ``{Factorization of Bosonic String Amplitudes},''
\href{http://dx.doi.org/10.1016/0550-3213(88)90522-6}{{\em Nucl.Phys.}
  {\bfseries B307} (1988) 61}.

\bibitem{Giveon:1999tq}
A.~Giveon and D.~Kutasov, ``{Comments on double scaled little string theory},''
  \href{http://dx.doi.org/10.1088/1126-6708/2000/01/023}{{\em JHEP} {\bfseries
  0001} (2000) 023},
\href{http://arxiv.org/abs/hep-th/9911039}{{\ttfamily arXiv:hep-th/9911039
  [hep-th]}}.

\bibitem{Liu:2004vy}
H.~Liu, ``{Fine structure of Hagedorn transitions},''
\href{http://arxiv.org/abs/hep-th/0408001}{{\ttfamily arXiv:hep-th/0408001
  [hep-th]}}.

\bibitem{AlvarezGaume:2005fv}
L.~Alvarez-Gaume, C.~Gomez, H.~Liu, and S.~Wadia, ``{Finite temperature
  effective action, AdS(5) black holes, and 1/N expansion},''
  \href{http://dx.doi.org/10.1103/PhysRevD.71.124023}{{\em Phys.Rev.}
  {\bfseries D71} (2005) 124023},
\href{http://arxiv.org/abs/hep-th/0502227}{{\ttfamily arXiv:hep-th/0502227
  [hep-th]}}.

\bibitem{Aharony:2003vk}
O.~Aharony, B.~Fiol, D.~Kutasov, and D.~A. Sahakyan, ``{Little string theory
  and heterotic / type II duality},''
  \href{http://dx.doi.org/10.1016/j.nuclphysb.2003.11.041}{{\em Nucl.Phys.}
  {\bfseries B679} (2004) 3--65},
\href{http://arxiv.org/abs/hep-th/0310197}{{\ttfamily arXiv:hep-th/0310197
  [hep-th]}}.

\bibitem{Damour:1999aw}
T.~Damour and G.~Veneziano, ``{Selfgravitating fundamental strings and black
  holes},'' \href{http://dx.doi.org/10.1016/S0550-3213(99)00596-9}{{\em
  Nucl.Phys.} {\bfseries B568} (2000) 93--119},
\href{http://arxiv.org/abs/hep-th/9907030}{{\ttfamily arXiv:hep-th/9907030
  [hep-th]}}.

\bibitem{Berkooz:2000mz}
M.~Berkooz and M.~Rozali, ``{Near Hagedorn dynamics of NS five-branes, or a new
  universality class of coiled strings},''
  \href{http://dx.doi.org/10.1088/1126-6708/2000/05/040}{{\em JHEP} {\bfseries
  0005} (2000) 040},
\href{http://arxiv.org/abs/hep-th/0005047}{{\ttfamily arXiv:hep-th/0005047
  [hep-th]}}.

\bibitem{Brandenberger:2006xi}
R.~H. Brandenberger, A.~Nayeri, S.~P. Patil, and C.~Vafa, ``{Tensor Modes from
  a Primordial Hagedorn Phase of String Cosmology},''
  \href{http://dx.doi.org/10.1103/PhysRevLett.98.231302}{{\em Phys.Rev.Lett.}
  {\bfseries 98} (2007) 231302},
\href{http://arxiv.org/abs/hep-th/0604126}{{\ttfamily arXiv:hep-th/0604126
  [hep-th]}}.

\bibitem{Brandenberger:2006vv}
R.~H. Brandenberger, A.~Nayeri, S.~P. Patil, and C.~Vafa, ``{String gas
  cosmology and structure formation},''
  \href{http://dx.doi.org/10.1142/S0217751X07037159}{{\em Int.J.Mod.Phys.}
  {\bfseries A22} (2007) 3621--3642},
\href{http://arxiv.org/abs/hep-th/0608121}{{\ttfamily arXiv:hep-th/0608121
  [hep-th]}}.

\bibitem{Sen:2004nf}
A.~Sen, ``{Tachyon dynamics in open string theory},''
  \href{http://dx.doi.org/10.1142/S0217751X0502519X}{{\em Int.J.Mod.Phys.}
  {\bfseries A20} (2005) 5513--5656},
\href{http://arxiv.org/abs/hep-th/0410103}{{\ttfamily arXiv:hep-th/0410103
  [hep-th]}}.

\bibitem{Bassett:2003ck}
B.~A. Bassett, M.~Borunda, M.~Serone, and S.~Tsujikawa, ``{Aspects of string
  gas cosmology at finite temperature},''
  \href{http://dx.doi.org/10.1103/PhysRevD.67.123506}{{\em Phys.Rev.}
  {\bfseries D67} (2003) 123506},
\href{http://arxiv.org/abs/hep-th/0301180}{{\ttfamily arXiv:hep-th/0301180
  [hep-th]}}.

\bibitem{Takamizu:2006sy}
Y.-i. Takamizu and H.~Kudoh, ``{Thermal equilibrium of string gas in Hagedorn
  universe},'' \href{http://dx.doi.org/10.1103/PhysRevD.74.103511}{{\em
  Phys.Rev.} {\bfseries D74} (2006) 103511},
\href{http://arxiv.org/abs/hep-th/0607231}{{\ttfamily arXiv:hep-th/0607231
  [hep-th]}}.

\bibitem{Frolov:1986ut}
V.~P. Frolov and A.~Zelnikov, ``{Vacuum Polarization Of The Electromagnetic
  Field Near A Rotating Black Hole},''
\href{http://dx.doi.org/10.1103/PhysRevD.32.3150}{{\em Phys.Rev.} {\bfseries
  D32} (1985) 3150--3163}.

\bibitem{Visser:1996ix}
M.~Visser, ``{Gravitational vacuum polarization. 3: Energy conditions in the
  (1+1) Schwarzschild space-time},''
  \href{http://dx.doi.org/10.1103/PhysRevD.54.5123}{{\em Phys.Rev.} {\bfseries
  D54} (1996) 5123--5128},
\href{http://arxiv.org/abs/gr-qc/9604009}{{\ttfamily arXiv:gr-qc/9604009
  [gr-qc]}}.

\bibitem{Carlson:2003ub}
E.~D. Carlson, W.~H. Hirsch, B.~Obermayer, P.~R. Anderson, and P.~B. Groves,
  ``{Stress energy tensor for the massless spin 1/2 field in static black hole
  space-times},'' \href{http://dx.doi.org/10.1103/PhysRevLett.91.051301}{{\em
  Phys.Rev.Lett.} {\bfseries 91} (2003) 051301},
\href{http://arxiv.org/abs/gr-qc/0305045}{{\ttfamily arXiv:gr-qc/0305045
  [gr-qc]}}.

\bibitem{Zurek:1985gd}
W.~Zurek and K.~S. Thorne, ``{Statistical mechanical origin of the entropy of a
  rotating, cha rged black hole},''
\href{http://dx.doi.org/10.1103/PhysRevLett.54.2171}{{\em Phys.Rev.Lett.}
  {\bfseries 54} (1985) 2171}.

\bibitem{Haddad:2013tha}
N.~Haddad, ``{Hawking Radiation from Small Black Holes at Strong Coupling and
  Large N},'' \href{http://dx.doi.org/10.1088/0264-9381/30/19/195002}{{\em
  Class.Quant.Grav.} {\bfseries 30} (2013) 195002},
\href{http://arxiv.org/abs/1306.0086}{{\ttfamily arXiv:1306.0086 [hep-th]}}.

\bibitem{Figueras:2013jja}
P.~Figueras and S.~Tunyasuvunakool, ``{CFTs in rotating black hole
  backgrounds},'' \href{http://dx.doi.org/10.1088/0264-9381/30/12/125015}{{\em
  Class.Quant.Grav.} {\bfseries 30} (2013) 125015},
\href{http://arxiv.org/abs/1304.1162}{{\ttfamily arXiv:1304.1162 [hep-th]}}.

\bibitem{Maldacena:1996ya}
J.~M. Maldacena, ``{Statistical entropy of near extremal five-branes},''
  \href{http://dx.doi.org/10.1016/0550-3213(96)00368-9}{{\em Nucl.Phys.}
  {\bfseries B477} (1996) 168--174},
\href{http://arxiv.org/abs/hep-th/9605016}{{\ttfamily arXiv:hep-th/9605016
  [hep-th]}}.

\bibitem{Tseytlin:1996qg}
A.~A. Tseytlin, ``{Extremal black hole entropy from conformal string sigma
  model},'' \href{http://dx.doi.org/10.1016/0550-3213(96)00383-5}{{\em
  Nucl.Phys.} {\bfseries B477} (1996) 431--448},
\href{http://arxiv.org/abs/hep-th/9605091}{{\ttfamily arXiv:hep-th/9605091
  [hep-th]}}.

\bibitem{Mathur:2005zp}
S.~D. Mathur, ``{The Fuzzball proposal for black holes: An Elementary
  review},'' \href{http://dx.doi.org/10.1002/prop.200410203}{{\em Fortsch.
  Phys.} {\bfseries 53} (2005) 793--827},
\href{http://arxiv.org/abs/hep-th/0502050}{{\ttfamily arXiv:hep-th/0502050
  [hep-th]}}.

\bibitem{Strominger:1996sh}
A.~Strominger and C.~Vafa, ``{Microscopic origin of the Bekenstein-Hawking
  entropy},'' \href{http://dx.doi.org/10.1016/0370-2693(96)00345-0}{{\em
  Phys.Lett.} {\bfseries B379} (1996) 99--104},
\href{http://arxiv.org/abs/hep-th/9601029}{{\ttfamily arXiv:hep-th/9601029
  [hep-th]}}.

\bibitem{Kazakov:2000pm}
V.~Kazakov, I.~K. Kostov, and D.~Kutasov, ``{A Matrix model for the
  two-dimensional black hole},''
  \href{http://dx.doi.org/10.1016/S0550-3213(01)00606-X}{{\em Nucl.Phys.}
  {\bfseries B622} (2002) 141--188},
\href{http://arxiv.org/abs/hep-th/0101011}{{\ttfamily arXiv:hep-th/0101011
  [hep-th]}}.

\bibitem{Russo:2001vh}
J.~G. Russo, ``{Free energy and critical temperature in eleven-dimensions},''
  \href{http://dx.doi.org/10.1016/S0550-3213(01)00128-6}{{\em Nucl.Phys.}
  {\bfseries B602} (2001) 109--131},
\href{http://arxiv.org/abs/hep-th/0101132}{{\ttfamily arXiv:hep-th/0101132
  [hep-th]}}.

\end{thebibliography}\endgroup
\bibliographystyle{utphys}   
\end{document}